%% file: these.tex
\documentclass[a4paper,11pt,onecolumn]{book}
\usepackage{pdfsync}
\usepackage[latin1]{inputenc}
\usepackage[T1]{fontenc}
\usepackage[normalem]{ulem}
\usepackage[french]{babel}
\usepackage{verbatim}
\usepackage{graphicx}
\usepackage {amsfonts}
\usepackage{amssymb}
\usepackage{amsmath}
\usepackage {txfonts}
\usepackage{french}
\usepackage{deleq}
\usepackage{setspace}
\usepackage {hyperref}
\usepackage[small]{caption}
\usepackage[francais]{minitoc}
\usepackage[Sonny]{fncychap}
\usepackage{mathrsfs}
\usepackage{fancyhdr}
\usepackage[left=3.5cm , right=3cm, top=4cm, bottom=4cm]{geometry}
\usepackage{epsfig}
\usepackage{float}
\usepackage{chngpage}
\usepackage{wasysym}
\usepackage{ragged2e}
\usepackage{wrapfig}
\usepackage{rotating}
\usepackage{epigraph}

\newenvironment{changemargin}[2]{\begin{list}{}{%
\setlength{\topsep}{0pt}%
\setlength{\leftmargin}{0pt}%
\setlength{\rightmargin}{0pt}%
\setlength{\listparindent}{\parindent}%
\setlength{\itemindent}{\parindent}%
\setlength{\parsep}{0pt plus 1pt}%
\addtolength{\leftmargin}{#1}%
\addtolength{\rightmargin}{#2}%
}\item }{\end{list}}

\newcommand{\ket}{\rangle}
\newcommand{\bra}{\langle}
\newcommand{\hata}{\hat{a}}
\newcommand{\hatad}{\hat{a}^\dag}

\fancypagestyle{monstyle}{
	\fancyhf{}
	\fancyhead[LE,RO]{\thepage}  
	\fancyhead[LO]{\rightmark}   
	\fancyhead[RE]{\leftmark}
	}
\DeclareTextSymbol{\degre}{OT1}{23}

\fancypagestyle{intro}{
	\fancyhf{}                            
	\fancyhead[LE,RO]{\thepage}  
	\fancyhead[LO]{ Introduction}   
	\fancyhead[RE]{ Introduction}}
	
\fancypagestyle{conclu}{
	\fancyhf{}                            
	\fancyhead[LE,RO]{\thepage}  
	\fancyhead[LO]{ Conclusion}   
	\fancyhead[RE]{ Conclusion}}

\fancypagestyle{annexe}{
	\fancyhf{}                            
	\fancyhead[LE,RO]{\thepage}  
	\fancyhead[LO]{ Annexes}   
	\fancyhead[RE]{ Annexes}}

\hypersetup{
colorlinks=true, 
breaklinks=true, 
urlcolor= blue, 
linkcolor= black, 
bookmarksopen=true, 
citecolor=red
}

\begin{document}

\bibliographystyle{these.bst}

\include{pagedegarde}

\include{introductionv5}

\pagestyle{monstyle}

\part{Outils théoriques et expérimentaux}

		\include{chapitre1v5}

		\include{chapitre2v5}

\part{Modèle théorique du mélange à 4 ondes}	

\include{chapitre3v5}

		\include{chapitre4v5}

\part{Réalisations expérimentales}
\include{chapitre5v5}

\include{chapitre6v5}

\chapter*{Conclusion}\label{conclusion:sec}
\addcontentsline{toc}{chapter}{Conclusion}
\pagestyle{conclu}
\begin{flushright}
\normalsize \textit{Comment expliquer autrement le mystère incompréhensible de ses perpétuelles fluctuations ?\\}
\textbf{Honoré de Balzac}, Histoire des treize : la Duchesse de Langeais.
\end{flushright}
\large

\vspace{1cm}

Ce travail de thèse avait pour but initial de diversifier les sources d'états comprimés vers les longueur d'ondes visibles les plus courtes et notamment de développer une source proche de transition $5S^{1/2}\to5P^{1/2}$ avec les ions $^{88}$Sr$^+$ à 422 nm.
La réalisation des expériences à 795 nm sur la raie D1 du $^{85}$Rb et l'exploration de l'espace des paramètres était donc initialement prévu comme la première étape de cette étude.
Nous nous sommes vite rendu compte que, pour pouvoir espérer adresser une autre transition, il était indispensable de développer un modèle (qui n'existait pas dans la littérature) pour pouvoir trouver les paramètres adéquats accessibles expérimentalement.
Ce développement théorique nous a permis de comprendre les expériences réalisées à 795 nm, d'identifier un régime original, et de conclure négativement sur la faisabilité de cette expérience à 422 nm.

\paragraph*{Approche théorique}
Pour donner une compréhension simple du processus de mélange à 4 ondes, nous avons développé une approche phénoménologique de l'amplification paramétrique dans un milieu de susceptibilité non-linéaire $\chi^{(3)}$ dans deux configurations : l'amplification sensible et insensible à la phase.
Pour le premier cas, nous avons montré qu'il était possible d'amplifier ou de déamplifier un champ sonde et ainsi de générer des états comprimés sur une quadrature arbitrairement choisie.
Pour le second, nous avons vérifié que de corrélations quantiques sous la limite quantique standard pouvaient être mises en évidence entre un champ sonde et son conjugué.\\
Par la suite, nous avons développé un modèle microscopique de l'interaction lumière--matière, basé sur des atomes immobiles ayant une structure de niveaux d'énergie en double--$\Lambda$, afin de donner un contenu physique au coefficient $\chi^{(3)}$ dans le cas de l'amplificateur insensible à la phase.
A l'aide de ce modèle nous avons notamment pu prévoir, pour la première fois, la génération de faisceaux intriqués intenses à 795 nm dans un milieu constitué d'atomes de $^{85}$Rb froids.\\
Dans un second temps, nous avons étendu ce modèle microscopique au cas d'une vapeur atomique ``chaude'' en démontrant que les résultats obtenus pour les atomes immobiles pouvaient être appliqués en prenant simplement en compte la densité d'atome ainsi que le taux de décohérence des niveaux fondamentaux que l'on observe dans des vapeurs atomiques.
Nous avons alors étudié théoriquement la possibilité de générer des faisceaux corrélés à 422 nm sur la transition $5S^{1/2}\to 6P^{1/2}$ du $^{85}$Rb et nous avons montré qu'il n'était pas possible de dégager des conditions réalistes expérimentalement similaires à celles de la transition $5S^{1/2}\to 5P^{1/2}$.\\

\paragraph*{Corrélations quantiques à 795 nm}
Mon travail de thèse s'est concentré autour de l'étude de la génération de corrélations quantiques sur la raie D1 du  $^{85}$Rb, bien que la plupart des résultats que nous avons obtenus sont généralisables à d'autres transitions de cet atome (la ligne D2 en particulier) ou à d'autres espèces atomiques.
Afin d'interpréter les expériences de mélange à 4 ondes dans une vapeur atomique et pour prendre en compte le temps d'interaction fini des atomes avec le faisceau pompe dans ce cas, nous avons étudié l'établissement du régime stationnaire et nous avons démontré qu'une part significative des atomes pouvaient ne pas avoir atteint cet état lorsqu'ils interagissaient avec le faisceau sonde, ce que nous avons observé expérimentalement.\\
D'autre part, à l'aide d'une étude expérimentale systématique de l'espace des paramètres pertinents, nous avons pu optimiser les corrélations quantiques observées entre les faisceaux sonde et conjugué dans l'expérience de mélange à 4 ondes que nous avons mise en place durant ce travail de thèse.
Nous avons notamment mis en évidence le rôle essentiel joué par un éventuel excès de bruit technique sur le faisceau sonde avant le processus d'amplification et trouvé une configuration expérimentale qui permettait de minimiser ce bruit.
Ceci fournit des renseignements précieux sur le type de sources laser à utiliser et en particulier sur l'impossibilité d'employer une diode laser (nécessairement trop bruitée) pour générer le faisceau sonde.\\
Nous avons mesuré à une fréquence d'analyse de 1 MHz, jusqu'à 9.2 dB de réduction du bruit sous la limite quantique standard sur la différence d'intensité entre ces deux faisceaux intenses.
Cette valeur fait partie des plus hautes actuellement répertoriées dans la littérature pour la génération de corrélations quantiques.\\

\paragraph*{Lame séparatrice quantique}
A l'aide de notre modèle microscopique, nous avons pu, de plus, mettre en évidence un nouveau régime qui permet la production de corrélations quantiques entre les faisceaux sonde et conjugué sans augmentation du nombre total de photons.
Nous présentons dans ce manuscrit la première observation expérimentale de ce phénomène.
Ainsi, pour un faisceau sonde incident, nous avons obtenu deux faisceaux présentant jusqu'à 1 dB de corrélations sous la limite quantique standard et dont l'intensité totale était inférieure à l'intensité du champ incident.
De par sa ressemblance avec le comportement d'une lame séparatrice, nous avons appelé ce nouveau régime : ``lame séparatrice quantique''.

\paragraph*{Perspectives}
La réalisation d'une source de lumière non-classique à 422 nm résonante avec la transition $5S^{1/2}\to5P^{1/2}$ du $^{88}$Sr$^+$ reste une question ouverte.
Si nous avons montré que les expériences sur la raie D1 du  $^{85}$Rb n'étaient pas transposable à l'identique sur la raie $5S^{1/2}\to 6P^{1/2}$, de nombreuses pistes restent à explorer, notamment l'étude du modèle microscopique dans le cas de l'amplification sensible à la phase reste à faire.
On peut imaginer que ce processus sensible à la phase soit alors plus favorable à 422 nm qu'à 795 nm car le processus insensible à la phase étant très faible dans le bleu, il ne perturbera pas le phénomène, ce qui est actuellement un problème de ce genre d'expériences dans l'infrarouge.\\
C'est dans ce cadre que nous avons réalisé des expériences préliminaires sur la transition $5S^{1/2}\to 6P^{1/2}$ du $^{85}$Rb et notamment nous avons observé pour la première fois le phénomène la transparence électromagnétiquement sur cette transition.\\
Une autre voie à explorer (qui nécessite l'emploi d'un second laser de pompe) consiste à utiliser les deux niveaux excités $5P^{1/2}$ et $6P^{1/2}$ du $^{85}$Rb afin de générer des corrélations quantiques entre deux faisceaux laser de différentes couleurs.\\
Enfin, nous pouvons dégager trois points supplémentaires qui mériteraient d'être approfondis dans des travaux ultérieurs.
Premièrement, il pourrait être intéressant de prendre en compte de façon plus complète le profil transverse du faisceau pompe dans l'étude des phénomènes de décohérence en résolvant les équations d'évolutions des atomes sans faire l'hypothèse d'avoir atteint un état stationnaire.
En effet, en simulant la trajectoire des atomes dans les faisceaux pompe et sonde, on peut connaitre l'état de chaque atome lorsqu'il interagit avec le faisceau sonde et ainsi rendre compte plus fidèlement des effets d'absorption résiduelle.
D'autre part, la mesure expérimentale de très fort taux d'intrication reste un problème ouvert.
En effet, contrairement à ce que prédit notre modèle, les niveaux d'anti-corrélations de phase mesurés et reportés dans littérature sont généralement bien moins importants que sur les corrélations d'intensité (et cela pas uniquement dans les expériences de mélange à 4 ondes).
La mesure des anti-corrélations de phase nécessite l'utilisation d'un oscillateur local et d'un montage de détection homodyne, on peut donc se demander si les observations expérimentales ne sont pas affectées par une erreur systématique.
Une piste serait alors l'amélioration de ces techniques expérimentales afin de mesurer des niveaux comparables sur ces deux quadratures.\\
Enfin, contrairement aux expériences de génération d'états non-classiques utilisant des cavités, le mélange à 4 ondes dans une vapeur atomique est une technique intrinsèquement multimode.
Il a déjà été démontré qu'il était possible de générer des images corrélés quantiquement par le processus d'amplification insensible à la phase.
Par ailleurs, par le processus de l'amplificateur sensible à la phase il est vraisemblablement possible de réaliser l'amplification d'images quantiques, et donc de produire des états comprimés multimodes.

\appendix
\part*{Annexes}
\addcontentsline{toc}{chapter}{Annexes}
\pagestyle{annexe}
\section{Transformée de Fourier}
On définit la transformée de Fourier d'un opérateur $\hat{a}(t)$ de la manière suivante :
\begin{equation}
\hat{a}(\omega)=\int_{-\infty}^\infty \hat{a}(t)\ e^{i\omega t}\ dt.
\end{equation}
La notation pour la transformée de Fourier de l'opérateur conjugué $\hat{a}^\dag(t)$ est plus ambigüe. On définit cette quantité de la manière suivante :
\begin{equation}\label{def_conjuguee2}
\hat{a}^\dag(\omega)=\int_{-\infty}^\infty \hat{a}^\dag(t)\ e^{i\omega t}\ dt.
\end{equation}
Mais dans ce cas il faut noter que la conjugaison s'exprime par
\begin{eqnarray}\label{conjuguee}
[\hat{a}^\dag(\omega)]^\dag &=&\int_{-\infty}^\infty \hat{a}(t)\ e^{-i\omega t}\ dt.\\
&=&\hat{a}(-\omega)
\end{eqnarray}
On fera donc particulièrement attention à ne pas confondre la conjuguée de la transformée de Fourier :
\begin{equation}
[\hat{a}(\omega)]^\dag=\left[\int_{-\infty}^\infty \hat{a}(t)\ e^{i\omega t}\ dt\right]^\dag = \int_{-\infty}^\infty \hat{a}^\dag(t)\ e^{-i\omega t}\ dt = \hat{a}^\dag(-\omega).
\end{equation}
et la transformée de Fourier de la conjuguée :
\begin{equation}
\hat{a}^\dag(\omega)=\int_{-\infty}^\infty \hat{a}^\dag(t)\ e^{i\omega t}\ dt.
\end{equation}
\clearpage
\section{Calculs des coefficients de diffusion}\label{Annexe_diffu}
Les coefficients de diffusion pour les forces de Langevin que nous avons introduits à l'équation \eqref{452} peuvent être calculés à l'aide de l'équation d'Einstein généralisée \cite{CohenTannoudji:1996p4732}.
La méthode de calcul est détaillée dans \cite{Davidovich:1996p1958}.\\
Comme nous l'avons vu à l'équation \eqref{evol_coherence}, les équations de Heisenberg-Langevin s'écrivent de façon générale sous la forme :
\begin{equation}
\frac{d}{dt}\tilde\sigma_{uv}=\mathfrak{E}_{uv}[\tilde\sigma_{23},\tilde\sigma_{41},\tilde\sigma_{43},\tilde\sigma_{21}]+\tilde f_{uv}
\end{equation}
pour $uv\in\{23,41,43,21\}$, où $\mathfrak{E}_{uv}$ décrit la combinaison linéaire des opérateurs.\\
La relation d'Einstein généralisée s'écrit alors :
\begin{equation}
2D_{uv,u'v'}=\bra \frac{d}{dt}(\tilde\sigma_{uv} \tilde\sigma_{u'v'})-\mathfrak{E}_{uv}\tilde\sigma_{u'v'}-\tilde\sigma_{uv}\mathfrak{E}_{u'v'}\ket 
\end{equation}
Nous pouvons alors calculer les coefficients de diffusion à l'aide des équations \eqref{evol_coherence}.
Nous pouvons définir les deux matrices de diffusion $[D_1]$ et $[D_2]$ sous la forme :
\begin{deqarr}
[D_1]  2 \delta(t-t')\delta(z-z')&=&\langle|F_1(z,t)] [F_1^\dag(z,t')|\rangle,\\
\left[D_2\right] 2\delta(t-t')\delta(z-z')&=& \langle|F_1^\dag(z,t)] [F_1(z,t')|\rangle.
\end{deqarr}
On rappelle que :
\begin{equation}
|F_1(z,t)]=\left |
\begin{array}{c}
\tilde{f}_{23}(z,t)\\
\tilde{f}_{41}(z,t)\\
\tilde{f}_{43}(z,t)\\
\tilde{f}_{21}(z,t)
\end{array}
\right]\text{ et }
|F_1^\dag(z,t)]=\left |
\begin{array}{c}
\tilde{f}_{32}(z,t)\\
\tilde{f}_{14}(z,t)\\
\tilde{f}_{34}(z,t)\\
\tilde{f}_{12}(z,t)
\end{array}
\right]
\end{equation}
On obtient alors pour les matrices  $[D_1]$ et $[D_2]$ :
\begin{scriptsize}
\begin{eqnarray}
\nonumber [D_1]&=&\frac{1}{2\tau}\left[
\begin{array}{cccc}
\Gamma  \left(\Gamma ^2+4 \Delta ^2+2 \Omega ^2+8 \Delta  \omega_0+4 \omega_0^2\right) & 0 & i \Gamma  \Omega  (\Gamma +2 i (\Delta +\omega_0)) & 0 \\
0 & 0 & 0 & -i \gamma  \Omega  (\Gamma -2 i (\Delta +\omega_0)) \\
-i \Gamma  \Omega  (\Gamma -2 i (\Delta +\omega_0)) & 0 & \Gamma  \Omega ^2 & 0 \\
0 & i \gamma  \Omega  (\Gamma +2 i (\Delta +\omega_0)) & 0 & \Gamma  \Omega ^2+2 \gamma  \left(\Gamma ^2+4 \Delta ^2+\Omega ^2+8 \Delta  \omega_0+4 \omega_0^2\right)
\end{array}\right]\\
~[D_2]&=&\frac{1}{2\tau}\left[
\begin{array}{cccc}
0 & 0 & 0 & -i \gamma  (\Gamma -2 i \Delta ) \Omega  \\
0 & \Gamma  \left(\Gamma ^2+4 \Delta ^2+2 \Omega ^2\right) & i \Gamma  (\Gamma +2 i \Delta ) \Omega  & 0 \\
0 & -i \Gamma  (\Gamma -2 i \Delta ) \Omega  & \Gamma  \Omega ^2 & 0 \\
i \gamma  (\Gamma +2 i \Delta ) \Omega  & 0 & 0 & \Gamma  \Omega ^2+2 \gamma  \left(\Gamma ^2+4 \Delta ^2+\Omega ^2\right)
\end{array}
\right]
\end{eqnarray}
\end{scriptsize}
avec  $\tau=2 \Gamma ^2+4\Omega ^2+4\omega_0^2+8 \Delta ^2+8 \Delta  \omega_0$.\\
La matrice des coefficients de diffusion symétrisés $[D]$ est donnée par :
\begin{equation}
[D]=\frac{[D_1]+[D_2]}{2}
\end{equation}
\clearpage
\section{Niveaux d'énergie du rubidium}\label{Annexe_Rb} 
Nous rappelons ici les principales données correspondant aux propriétés physiques ainsi qu'aux transitions optiques étudiées pour le $^{85}$Rb.
Ces données sont issues de \cite{Steck08}.

\begin{table}[h!]\centering\begin{tabular}{|c|c|}
\hline Numéro atomique & 37 \\ 
\hline  Nombre de masse & 85 \\ 
\hline Abondance naturelle & 72$\ \%$ \\ 
\hline Point de fusion & 39.3$^{\circ}$C \\ 
\hline Point d'ébullition & 688$^{\circ}$C \\ 
\hline Pression de vapeur saturante à 25$^{\circ}$C & $3.92\times 10^{-7}$ Torr \\ 
\hline Spin du noyeau & 5/2 \\ 
\hline 
\end{tabular}
\caption{Principales propriétés physique du $^{85}$Rb.}
\end{table}
Dans un deuxième temps nous allons donner les grandeurs associées à la transition optique D1 du $^{85}$Rb ($5S^{1/2}\to 5P^{1/2}$).
\begin{table}[h!]\centering\begin{tabular}{|c|c|}
\hline Fréquence & $2\pi\ 377.107$ THz \\ 
\hline  Longueur d'onde (dans le vide) & 794.979 nm \\ 
\hline Temps de vie du niveau excité & 27.68 ns \\ 
\hline Taux de relaxation (largeur naturelle - FWHM) & $36.13\times 10^6 $s$^{-1}=2\pi\  5.75$ MHz\\ 
\hline 
\end{tabular}
\caption{Principales grandeurs associées à la transition optique D1 du $^{85}$Rb ($5S^{1/2}\to 5P^{1/2}$).}
\end{table}

\begin{figure}[h!]
 \centering
 	\includegraphics[width=15cm]{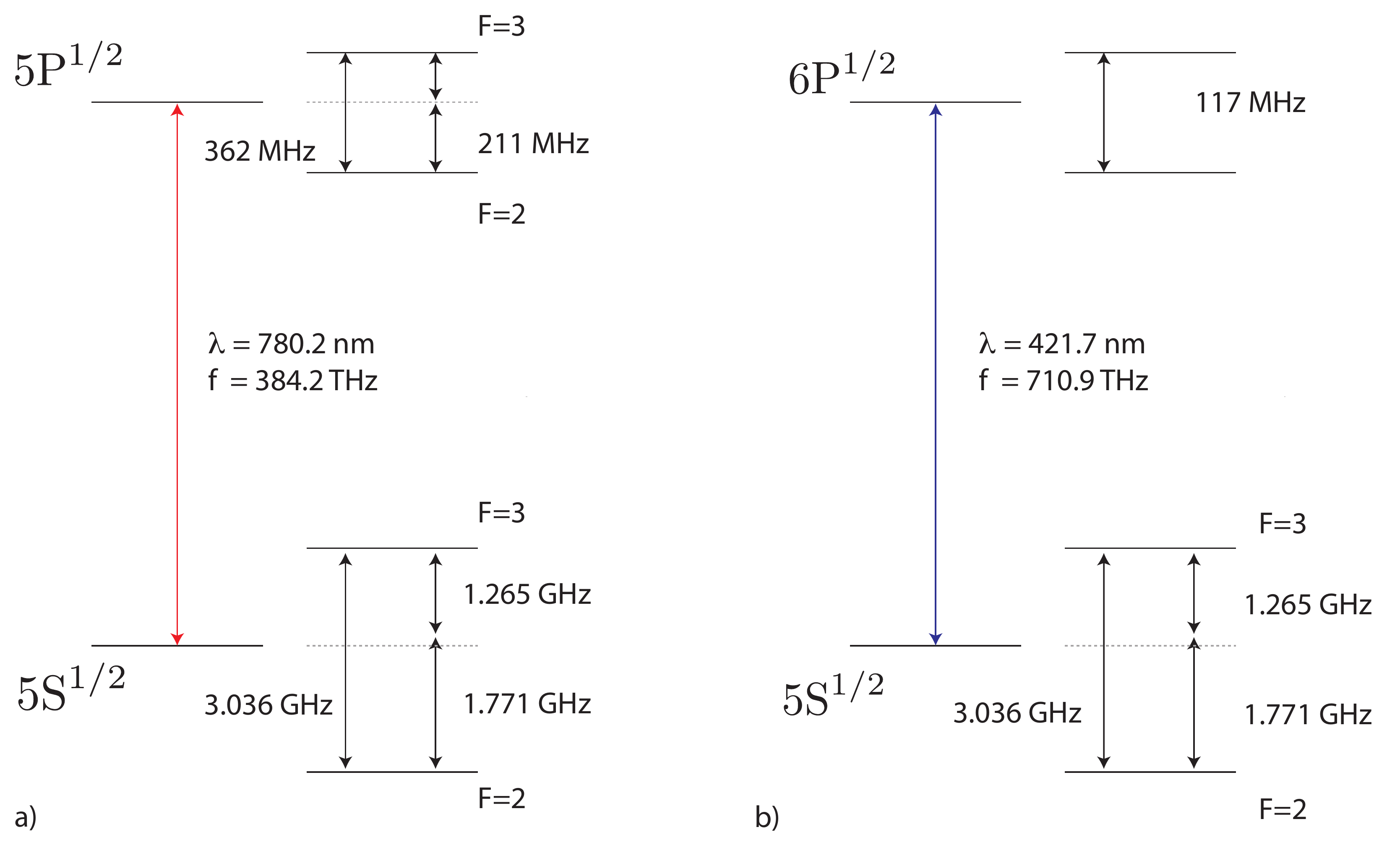} 
 		\caption[Schéma de niveaux équivalent]{Schémas de niveaux pour les transitions D1 et $5S^{1/2}\to 6P^{1/2}$ du $^{85}$Rb \label{nivooo}}	
 \end{figure}

\noindent De même, nous donnons les grandeurs associés à la transition $5S^{1/2}\to 6P^{1/2}$ du $^{85}$Rb.
Ces données sont reprises sur le schéma de niveau simplifié présenté sur la figure \ref{nivooo}.
\begin{table}[h!]\centering\begin{tabular}{|c|c|}
\hline Fréquence & $2\pi\ 710.96$ THz \\ 
\hline  Longueur d'onde (dans le vide) & 421.55 nm \\ 
\hline Temps de vie du niveau excité & 4.19 $\mu$s \\ 
\hline Taux de relaxation (largeur naturelle - FWHM) & $1.50\times 10^6 $s$^{-1}=2\pi\  0.238$ MHz\\ 
\hline 
\end{tabular}
\caption{Principales grandeurs associées à la transition optique $5S^{1/2}\to 6P^{1/2}$ du $^{85}$Rb.}
\end{table}

\noindent On peut alors calculer un certain nombre de grandeurs que nous avons utilisé tout au long de ce manuscrit.

\begin{table}[h!]\centering\begin{tabular}{|c|c|}
\hline Fréquence de Rabi & $0.59 \sqrt{\frac{\text{P}}{R^2}}$ MHz\\ 
\hline Moment dipolaire de la transition au carré & 6.41$\times10^{-58}$ SI \\ 
\hline Section efficace de la transition & 1$\times10^{-9}$ cm$^{2}$ \\ 
\hline 
\end{tabular}
\caption{Principales grandeurs associées à la transition optique $5S^{1/2}\to 5P^{1/2}$ du $^{85}$Rb.}
\end{table}

\begin{table}[h!]\centering\begin{tabular}{|c|c|}
\hline Fréquence de Rabi & $0.0467 \sqrt{\frac{\text{P}}{R^2}}$ MHz\\ 
\hline Moment dipolaire de la transition au carré & 3.99$\times10^{-60}$ SI \\ 
\hline Section efficace de la transition & 2.83$\times10^{-10}$ cm$^{2}$ \\ 
\hline 
\end{tabular}
\caption{Principales grandeurs associées à la transition optique $5S^{1/2}\to 6P^{1/2}$ du $^{85}$Rb.}
\end{table}
\noindent Dans les tableaux qui suivent nous introduisons $P$, la puissance du laser de pompe (en W) et $R$ son rayon a $1/e^2$ (en m).

\vspace{2cm}
\clearpage

\section{Modèle en double-$\Lambda$ et $^{85}$Rb }\label{Annexe_Lambda} 
Dans ce manuscrit nous avons, à de nombreuses reprises, utilisé un modèle atomique simplifié en double-$\Lambda$ pour décrire l'interaction des lasers avec le $^{85}$Rb.
Dans cette annexe nous allons voir pourquoi ce modèle est justifié en première approximation pour des polarisations linéaires sur les faisceaux pompe et sonde.
Dans un premier temps, nous rappelons la définition des différents moments cinétiques et des couplages responsables de la levée de dégénérescence des niveaux d'énergie pour des atomes de $^{85}$Rb.
Par la suite, nous établirons la matrice de passage entre la base des états propres des opérateurs associés aux polarisations circulaires et ceux associé aux polarisations linéaires.
Enfin, nous utiliserons cette matrice de passage sur les matrice de couplage afin de voir comment elles  s'écrivent dans la nouvelle base associée aux polarisations linéaires.
\subsection{Moments cinétiques}
La transition $5S^{1/2}\to 5P^{1/2}$ du $^{85}$Rb comprend 24 sous--niveaux Zeeman repartis entre quatre niveaux hyperfins.\\
On définit les moments cinétiques suivants :
\begin{itemize}
\item $\mathbf{\hat L}$ est le moment cinétique orbital.
\item $\mathbf{\hat S}$ est le moment cinétique de spin.
\item $\mathbf{\hat J}=\mathbf{\hat L}+\mathbf{\hat S}$ est le moment cinétique global lorsque l'on prend en compte le couplage spin-ortbite (couplage $\mathbf{\hat L}.\mathbf{\hat S}$). Ce terme fait apparaître  la structure fine.
\item $\mathbf{ \hat I}$ est le moment cinétique du noyau.
\item $\mathbf{ \hat F}=\mathbf{\hat J}+\mathbf{\hat I}$ est le moment cinétique global lorsque l'on prend en compte le couplage  $\mathbf{\hat I}.\mathbf{\hat S}$. Ce terme fait apparaître la structure hyperfine.
\end{itemize}
\subsection{Etats propres}
Les niveaux hyperfins $|F\ket$ sont les états propres de l'opérateur $\mathbf{ \hat F}^2$ associés à la valeur propre $F$.
Les sous--niveaux Zeeman $|m_F\ket$ sont les $2F+1$ états propres de l'une des composantes d'un opérateur moment cinétique que l'on va appeler $\hat F_z$ associés à la valeur propre $m_F$.
Ces états propres définissent une base de l'espace de définition de $\hat F_z$,, que l'on notera $\mathcal{B}_z^{(2F+1)}$.
On peut définir les opérateurs $ \hat F_+$ et $\hat F_-$ par leur action sur les vecteurs de cette base :
\begin{deqarr}
\hat F_+|F,m_F\ket= \sqrt{(F(F+1)-m_F(m_F+1)}\ |F,m_F+1\ket,\\
\hat F_-|F,m_F\ket= \sqrt{(F(F+1)-m_F(m_F-1)}\ |F,m_F-1\ket.
\end{deqarr} 
On définit alors un opérateur $\hat F_V$ qui est la superposition linéaire des deux opérateurs $ \hat F_+$ et $\hat F_-$ :
\begin{equation}
\hat F_V=\frac{\hat F_+ +\hat F_-}{2}.
\end{equation}
Les états $|F,m_F\ket$ ne sont pas des états propres de $\hat F_V$.
On notera $ F_{V2}$ la matrice $5\times 5$ de l'opérateur $\hat F_V$ pour $F=2$ sur la base $\mathcal{B}_z^{(5)}$ et $F_{V3}$ la matrice $7\times 7$ de l'opérateur $\hat F_V$ pour $F=3$ sur la base $\mathcal{B}_z^{(7)}$.
On obtient ainsi aisément :
\begin{deqn}
F_{V2}=\left(
\begin{array}{ccccc}
 0 & 1 & 0 & 0 & 0 \\
 1 & 0 & \sqrt{\frac{3}{2}} & 0 & 0 \\
 0 & \sqrt{\frac{3}{2}} & 0 & \sqrt{\frac{3}{2}} & 0 \\
 0 & 0 & \sqrt{\frac{3}{2}} & 0 & 1 \\
 0 & 0 & 0 & 1 & 0
\end{array}
\right),
\end{deqn}
\begin{ddeqn}
\text{ et }
 F_{V3}=\left(
\begin{array}{ccccccc}
 0 & \sqrt{\frac{3}{2}} & 0 & 0 & 0 & 0 & 0 \\
 \sqrt{\frac{3}{2}} & 0 & \sqrt{\frac{5}{2}} & 0 & 0 & 0 & 0 \\
 0 & \sqrt{\frac{5}{2}} & 0 & \sqrt{3} & 0 & 0 & 0 \\
 0 & 0 & \sqrt{3} & 0 & \sqrt{3} & 0 & 0 \\
 0 & 0 & 0 & \sqrt{3} & 0 & \sqrt{\frac{5}{2}} & 0 \\
 0 & 0 & 0 & 0 & \sqrt{\frac{5}{2}} & 0 & \sqrt{\frac{3}{2}} \\
 0 & 0 & 0 & 0 & 0 & \sqrt{\frac{3}{2}} & 0
\end{array}
\right).
\end{ddeqn}
On peut alors diagonaliser ces matrices afin d'obtenir la matrice de passage de la base $\mathcal{B}_z^{(2F+1)}$ à la base $\mathcal{B}_V^{(2F+1)}$ qui sera la base propre de $\hat F_V$.
Les matrices de passage $Q_2$ et $Q_3$ sont composées des vecteurs propres de $\hat F_V$ écrits en colonne :
\begin{deqarr}\arrlabel{passm}
Q_2&=&\left(
\begin{array}{ccccc}
 1 & 1 & -1 & -1 & 1 \\
 -2 & 2 & 1 & -1 & 0 \\
 \sqrt{6} & \sqrt{6} & 0 & 0 & -\sqrt{\frac{2}{3}} \\
 -2 & 2 & -1 & 1 & 0 \\
 1 & 1 & 1 & 1 & 1
\end{array}
\right),\\
\text{ et }
 Q_{3}&=&\left(
 \begin{array}{ccccccc}
  1 & 1 & -1 & -1 & 1 & 1 & -1 \\
  -\sqrt{6} & \sqrt{6} & 2 \sqrt{\frac{2}{3}} & -2 \sqrt{\frac{2}{3}} & -\sqrt{\frac{2}{3}} & \sqrt{\frac{2}{3}} & 0 \\
  \sqrt{15} & \sqrt{15} & -\sqrt{\frac{5}{3}} & -\sqrt{\frac{5}{3}} & -\frac{1}{\sqrt{15}} & -\frac{1}{\sqrt{15}} & \sqrt{\frac{3}{5}} \\
  -2 \sqrt{5} & 2 \sqrt{5} & 0 & 0 & \frac{2}{\sqrt{5}} & -\frac{2}{\sqrt{5}} & 0 \\
  \sqrt{15} & \sqrt{15} & \sqrt{\frac{5}{3}} & \sqrt{\frac{5}{3}} & -\frac{1}{\sqrt{15}} & -\frac{1}{\sqrt{15}} & -\sqrt{\frac{3}{5}} \\
  -\sqrt{6} & \sqrt{6} & -2 \sqrt{\frac{2}{3}} & 2 \sqrt{\frac{2}{3}} & -\sqrt{\frac{2}{3}} & \sqrt{\frac{2}{3}} & 0 \\
  1 & 1 & 1 & 1 & 1 & 1 & 1
 \end{array}
 \right).
\end{deqarr}\clearpage

\subsection{Matrices de couplage}
L'opérateur moment dipolaire $\hat D$ peut se décomposer sur ses différentes composantes $\hat D_x$, $\hat D_y$, $\hat D_z$.
On peut définir les opérateurs $\hat D_+$, $\hat D_-$, respectivement associés à des photons de polarisation $\sigma_+$ et $\sigma_+$ par :
\begin{equation}
\hat D_+=\hat D_x+i\hat D_y,\text{ et } \hat D_-=\hat D_x-i\hat D_y.
\end{equation}
Le théorème de Wigner-Eckart permet d'écrire en fonction de l'élément de matrice réduit $\bra F||\hat D|| F'\ket$ les éléments de la matrice de couplage pour les opérateurs $\hat D_+$, $\hat D_-$ :
\begin{equation}
\bra F,m|\hat D_\pm| F',m'\ket=\bra F',m';1,\pm 1\ket \bra F||\hat D|| F'\ket
\end{equation}
où $\bra F',m';1,\pm 1\ket$ est le coefficient de Clebsch-Gordan associé à la transition de $F,m \to~F',m'$ par un photon  $\sigma_+$ ou $\sigma_+$.\\
On écrit donc les quatre matrices de couplage $M_{ij}^{\sigma_+}$, pour les photons polarisés $\sigma_+$, correspondantes aux transitions $F=i\to F=j$, avec $i\text{ et } j\in\{2,3\}$ et les quatre matrices  $M_{ij}^{\sigma_-}$, pour les photons polarisés $\sigma_-$ : 
\begin{deqn}
M_{22}^{\sigma_+}=\left(
\begin{array}{ccccc}
 0 & 0 & 0 & 0 & 0 \\
 -\frac{1}{\sqrt{3}} & 0 & 0 & 0 & 0 \\
 0 & -\frac{1}{\sqrt{2}} & 0 & 0 & 0 \\
 0 & 0 & -\frac{1}{\sqrt{2}} & 0 & 0 \\
 0 & 0 & 0 & -\frac{1}{\sqrt{3}} & 0
\end{array}
\right)\text{ et } 
M_{22}^{\sigma_-}=\left(
\begin{array}{ccccc}
 0 & \frac{1}{\sqrt{3}} & 0 & 0 & 0 \\
 0 & 0 & \frac{1}{\sqrt{2}} & 0 & 0 \\
 0 & 0 & 0 & \frac{1}{\sqrt{2}} & 0 \\
 0 & 0 & 0 & 0 & \frac{1}{\sqrt{3}} \\
 0 & 0 & 0 & 0 & 0
\end{array}
\right),
\end{deqn}
\begin{ddeqn}
M_{33}^{\sigma_+}=\left(
\begin{array}{ccccccc}
 0 & 0 & 0 & 0 & 0 & 0 & 0 \\
 -\frac{1}{2} & 0 & 0 & 0 & 0 & 0 & 0 \\
 0 & -\frac{\sqrt{\frac{5}{3}}}{2} & 0 & 0 & 0 & 0 & 0 \\
 0 & 0 & -\frac{1}{\sqrt{2}} & 0 & 0 & 0 & 0 \\
 0 & 0 & 0 & -\frac{1}{\sqrt{2}} & 0 & 0 & 0 \\
 0 & 0 & 0 & 0 & -\frac{\sqrt{\frac{5}{3}}}{2} & 0 & 0 \\
 0 & 0 & 0 & 0 & 0 & -\frac{1}{2} & 0
\end{array}
\right),
M_{33}^{\sigma_-}=\left(
\begin{array}{ccccccc}
 0 & \frac{1}{2} & 0 & 0 & 0 & 0 & 0 \\
 0 & 0 & \frac{\sqrt{\frac{5}{3}}}{2} & 0 & 0 & 0 & 0 \\
 0 & 0 & 0 & \frac{1}{\sqrt{2}} & 0 & 0 & 0 \\
 0 & 0 & 0 & 0 & \frac{1}{\sqrt{2}} & 0 & 0 \\
 0 & 0 & 0 & 0 & 0 & \frac{\sqrt{\frac{5}{3}}}{2} & 0 \\
 0 & 0 & 0 & 0 & 0 & 0 & \frac{1}{2} \\
 0 & 0 & 0 & 0 & 0 & 0 & 0
\end{array}
\right),
\end{ddeqn}

\begin{ddeqn}
M_{23}^{\sigma_+}=\left(
\begin{array}{ccccc}
 0 & 0 & 0 & 0 & 0 \\
 0 & 0 & 0 & 0 & 0 \\
 \frac{1}{\sqrt{15}} & 0 & 0 & 0 & 0 \\
 0 & \frac{1}{\sqrt{5}} & 0 & 0 & 0 \\
 0 & 0 & \sqrt{\frac{2}{5}} & 0 & 0 \\
 0 & 0 & 0 & \sqrt{\frac{2}{3}} & 0 \\
 0 & 0 & 0 & 0 & 1
\end{array}
\right),
M_{23}^{\sigma_-}=\left(
\begin{array}{ccccc}
 1 & 0 & 0 & 0 & 0 \\
 0 & \sqrt{\frac{2}{3}} & 0 & 0 & 0 \\
 0 & 0 & \sqrt{\frac{2}{5}} & 0 & 0 \\
 0 & 0 & 0 & \frac{1}{\sqrt{5}} & 0 \\
 0 & 0 & 0 & 0 & \frac{1}{\sqrt{15}} \\
 0 & 0 & 0 & 0 & 0 \\
 0 & 0 & 0 & 0 & 0
\end{array}
\right)
\end{ddeqn}
et
\begin{ddeqn}
M_{32}^{\sigma_+}=\left(
\begin{array}{ccccccc}
 \sqrt{\frac{5}{7}} & 0 & 0 & 0 & 0 & 0 & 0 \\
 0 & \sqrt{\frac{10}{21}} & 0 & 0 & 0 & 0 & 0 \\
 0 & 0 & \sqrt{\frac{2}{7}} & 0 & 0 & 0 & 0 \\
 0 & 0 & 0 & \frac{1}{\sqrt{7}} & 0 & 0 & 0 \\
 0 & 0 & 0 & 0 & \frac{1}{\sqrt{21}} & 0 & 0
\end{array}
\right),
M_{32}^{\sigma_-}=\left(
\begin{array}{ccccccc}
 0 & 0 & \frac{1}{\sqrt{21}} & 0 & 0 & 0 & 0 \\
 0 & 0 & 0 & \frac{1}{\sqrt{7}} & 0 & 0 & 0 \\
 0 & 0 & 0 & 0 & \sqrt{\frac{2}{7}} & 0 & 0 \\
 0 & 0 & 0 & 0 & 0 & \sqrt{\frac{10}{21}} & 0 \\
 0 & 0 & 0 & 0 & 0 & 0 & \sqrt{\frac{5}{7}}
\end{array}
\right).
\end{ddeqn}
On peut alors écrire la matrice de couplage pour les différentes polarisations linéaires à l'aide d'une combinaison linéaire des $M_{ij}^{\sigma_+}$ et $M_{ij}^{\sigma_-}$.
Si désormais, on souhaite connaitre les couplages pour deux polarisations linéaires orthogonales, que l'on notera $H$ et $V$, dans les bases $\mathcal{B}_V^{(2F+1)}$, on peut utiliser les matrices de passage que l'on a définies à l'équation \eqref{passm}.
On peut choisir par exemple pour les deux polarisations $H$ et $V$ de prendre les superpositions :
\begin{equation}
M_{ij}^{H}=\frac{M_{ij}^{\sigma_+}-M_{ij}^{\sigma_-}}{2i} \text{ et } M_{ij}^{V}=\frac{M_{ij}^{\sigma_+}+M_{ij}^{\sigma_-}}{2}.
\end{equation} 
On peut donc écrire la relation de passage d'une base à l'autre :
\begin{equation}
N_{ij}^{H}(\mathcal{B}_V^{(2F+1)})=Q_j^{-1} M_{ij}^{H}(\mathcal{B}_z^{(2F+1)}) Q_i.
\end{equation}
On obtient alors les matrices de couplage pour les polarisations $H$ et $V$ dans la base $\mathcal{B}_V^{(2F+1)}$ que l'on va noter $N_{ij}^{H}$ :
\small
\begin{deqn}
N_{22}^{H}=\left(
\begin{array}{ccccc}
 \frac{1}{i\sqrt{3}} & 0 & 0 & 0 & 0 \\
 0 & -\frac{1}{i\sqrt{3}} & 0 & 0 & 0 \\
 0 & 0 & \frac{1}{2 i\sqrt{3}} & 0 & 0 \\
 0 & 0 & 0 & -\frac{1}{2i \sqrt{3}}& 0 \\
 0 & 0 & 0 & 0 & 0
\end{array}
\right)\text{ et } 
N_{22}^{V}=\left(
\begin{array}{ccccc}
 0 & 0 & -\frac{1}{4 \sqrt{3}} & 0 & 0 \\
 0 & 0 & 0 & \frac{1}{4 \sqrt{3}} & 0 \\
 \frac{1}{\sqrt{3}} & 0 & 0 & 0 & -\frac{1}{2 \sqrt{3}} \\
 0 & -\frac{1}{\sqrt{3}} & 0 & 0 & \frac{1}{2 \sqrt{3}} \\
 0 & 0 & \frac{\sqrt{3}}{4} & -\frac{\sqrt{3}}{4} & 0
\end{array}
\right),
\end{deqn}

\begin{ddeqn}
N_{33}^{H}=\left(
\begin{array}{ccccccc}
 \frac{\sqrt{\frac{3}{2}}}{2i} & 0 & 0 & 0 & 0 & 0 & 0 \\
 0 & -\frac{\sqrt{\frac{3}{2}}}{2i} & 0 & 0 & 0 & 0 & 0 \\
 0 & 0 & \frac{1}{i\sqrt{6}} & 0 & 0 & 0 & 0 \\
 0 & 0 & 0 & -\frac{1}{i\sqrt{6}} & 0 & 0 & 0 \\
 0 & 0 & 0 & 0 & \frac{1}{2 i\sqrt{6}} & 0 & 0 \\
 0 & 0 & 0 & 0 & 0 & -\frac{1}{2 i\sqrt{6}} & 0 \\
 0 & 0 & 0 & 0 & 0 & 0 & 0
\end{array}
\right),
N_{33}^{V}=\left(
\begin{array}{ccccccc}
 0 & 0 & -\frac{1}{4 \sqrt{6}} & 0 & 0 & 0 & 0 \\
 0 & 0 & 0 & \frac{1}{4 \sqrt{6}} & 0 & 0 & 0 \\
 \frac{\sqrt{\frac{3}{2}}}{2} & 0 & 0 & 0 & -\frac{1}{2 \sqrt{6}} & 0 & 0 \\
 0 & -\frac{\sqrt{\frac{3}{2}}}{2} & 0 & 0 & 0 & \frac{1}{2 \sqrt{6}} & 0 \\
 0 & 0 & \frac{5}{4 \sqrt{6}} & 0 & 0 & 0 & -\frac{\sqrt{\frac{3}{2}}}{4} \\
 0 & 0 & 0 & -\frac{5}{4 \sqrt{6}} & 0 & 0 & \frac{\sqrt{\frac{3}{2}}}{4} \\
 0 & 0 & 0 & 0 & \frac{1}{\sqrt{6}} & -\frac{1}{\sqrt{6}} & 0
\end{array}
\right),
\end{ddeqn}

\begin{figure}
	\includegraphics[width=15cm]{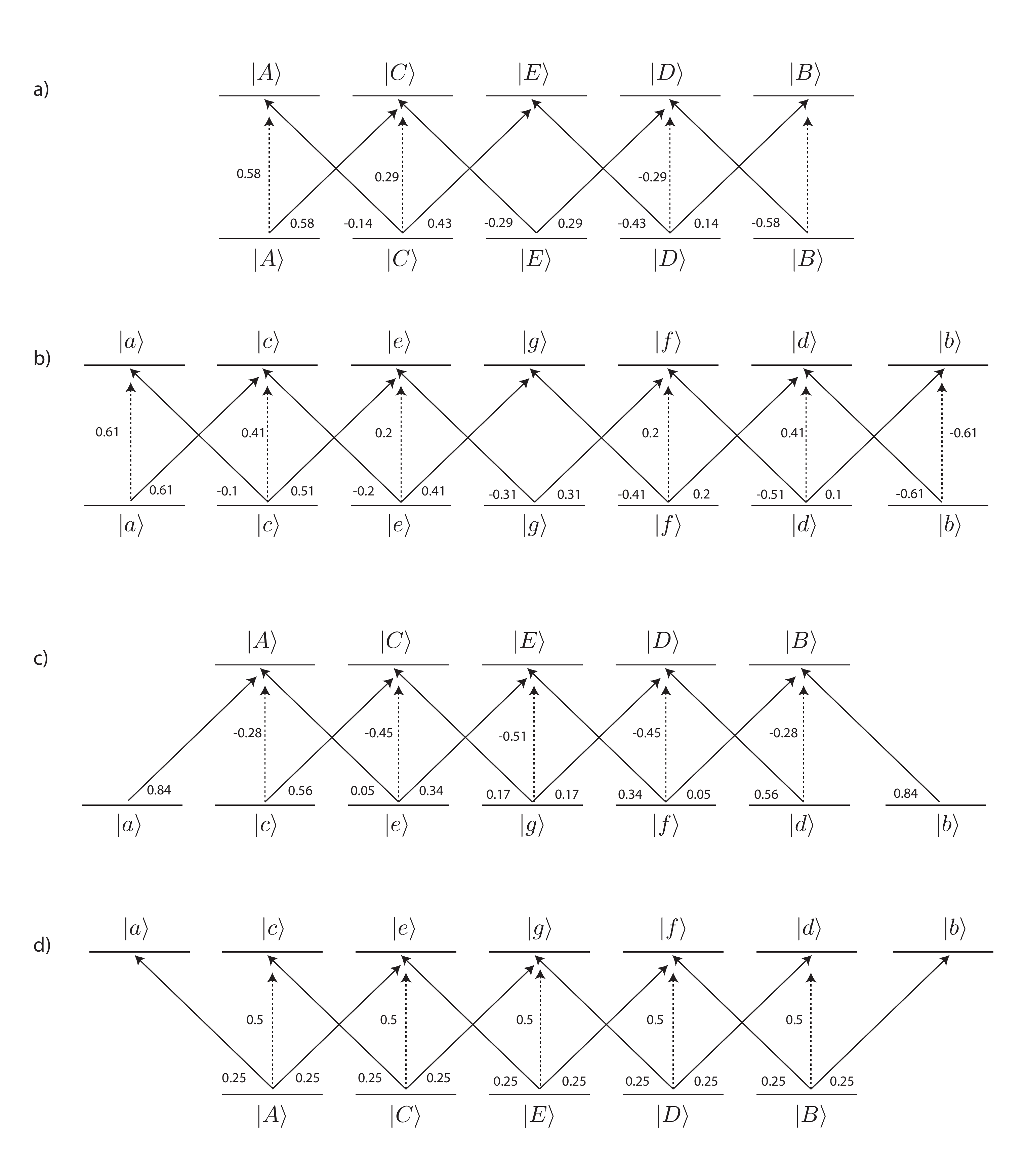} 
		\caption[Schéma de niveaux équivalent]{Schémas de niveaux équivalents pour la transition D1 du $^{85}$Rb pour deux polarisations linéaires croisées $H$ (en pointillés) et $V$ (en traits pleins).\\
		a) $F=2\to F=2$, \\
		b) $F=3\to F=3$, \\
		c) $F=3\to F=2$, \\
		d) $F=2\to F=3$. \label{annexniv}}		
\end{figure}

\begin{ddeqn}
N_{23}^{H}=\left(
\begin{array}{ccccc}
 0 & 0 & 0 & 0 & 0 \\
 0 & 0 & 0 & 0 & 0 \\
 \frac{1}{2i} & 0 & 0 & 0 & 0 \\
 0 & \frac{1}{2i} & 0 & 0 & 0 \\
 0 & 0 & \frac{1}{2i} & 0 & 0 \\
 0 & 0 & 0 & \frac{1}{2i} & 0 \\
 0 & 0 & 0 & 0 & \frac{1}{2i}
\end{array}
\right),
N_{23}^{V}=\left(
\begin{array}{ccccc}
 \frac{1}{4} & 0 & 0 & 0 & 0 \\
 0 & \frac{1}{4} & 0 & 0 & 0 \\
 0 & 0 & \frac{1}{4} & 0 & 0 \\
 0 & 0 & 0 & \frac{1}{4} & 0 \\
 \frac{1}{4} & 0 & 0 & 0 & \frac{1}{4} \\
 0 & \frac{1}{4} & 0 & 0 & \frac{1}{4} \\
 0 & 0 & \frac{1}{4} & \frac{1}{4} & 0
\end{array}
\right)
\end{ddeqn}

et
\begin{ddeqn}
N_{32}^{H}=\left(
\begin{array}{ccccccc}
 0 & 0 & -\frac{\sqrt{\frac{5}{7}}}{3i} & 0 & 0 & 0 & 0 \\
 0 & 0 & 0 & -\frac{\sqrt{\frac{5}{7}}}{3i} & 0 & 0 & 0 \\
 0 & 0 & 0 & 0 & -\frac{8}{3 i\sqrt{35}} & 0 & 0 \\
 0 & 0 & 0 & 0 & 0 & -\frac{8}{3 i\sqrt{35}} & 0 \\
 0 & 0 & 0 & 0 & 0 & 0 & -\frac{3}{i\sqrt{35}}
\end{array}
\right),
N_{32}^{V}=\left(
\begin{array}{ccccccc}
 \sqrt{\frac{5}{7}} & 0 & 0 & 0 & \frac{1}{3 \sqrt{35}} & 0 & 0 \\
 0 & \sqrt{\frac{5}{7}} & 0 & 0 & 0 & \frac{1}{3 \sqrt{35}} & 0 \\
 0 & 0 & \frac{2 \sqrt{\frac{5}{7}}}{3} & 0 & 0 & 0 & \frac{1}{\sqrt{35}} \\
 0 & 0 & 0 & \frac{2 \sqrt{\frac{5}{7}}}{3} & 0 & 0 & \frac{1}{\sqrt{35}} \\
 0 & 0 & 0 & 0 & \frac{2}{\sqrt{35}} & \frac{2}{\sqrt{35}} & 0
\end{array}
\right).
\end{ddeqn}
\large
On note la base $\mathcal{B}_V^{(5)}$ sous la forme :
\begin{deqn}
\mathcal{B}_V^{(5)}=\{|A\ket,|B\ket,|C\ket,|D\ket,|E\ket\},
\end{deqn}
et la base $\mathcal{B}_V^{(7)}$ sous la forme :
\begin{ddeqn}
\mathcal{B}_V^{(7)}=\{|a\ket,|b\ket,|c\ket,|d\ket,|e\ket,|f\ket,|g\ket\}.
\end{ddeqn}
On peut donc résumer les couplages pour les polarisations $H$ et $V$ dans ces bases sous la forme des schémas de la figure \ref{annexniv}.\\
Cette représentation nous permet de justifier pourquoi nous avons utilisé un modèle en double-$\Lambda$ pour décrire l'interaction des atomes.
En effet le système peut se décomposer en de nombreux sous-systèmes en double-$\Lambda$.
Prenons l'exemple de la transition $F=2\to F'=2$ (schéma  \ref{annexniv} a)).
Dans ce cas, on obtient un premier système en double--$\Lambda$ pour les niveaux $|A\ket$ et $|C\ket$ de l'état fondamental et de l'état excité.
Une polarisation linéaire (la pompe par exemple), couple les deux niveaux $|A\ket$ et les deux niveaux $|C\ket$ (traits pointillés), tandis que la polarisation linéaire orthogonale (la sonde par exemple) couple les niveaux $|A\ket$ du fondamental et $|C\ket$ de l'état excité et réciproquement $|C\ket$ du fondamental et $|A\ket$ de l'état excité (traits pleins).
Il s'agit bien d'un système en double-$\Lambda$ comme nous l'avons décrit dans le chapitre 4.\\
Par contre, il est important de noter, qu'il s'agit bien sur d'un schéma simplifié car on peut le voir sur la figure \ref{annexniv}, il est possible pour les atomes de se désexciter de l'état $|C\ket$  vers l'état fondamental $|E\ket$.
Dans ce cas l'atome sort du schéma en double-$\Lambda$ et doit être repompé pour participer à nouveau au processus de mélange à ondes.
Par ce repompage, il peut retourner dans le double-$\Lambda$ composé des niveaux $|A\ket$ et $|C\ket$ ou basculer de l'autre côté dans le second sous-système en double-$\Lambda$ composé des niveaux $|B\ket$ et $|D\ket$.\\
Une des perspectives de ce travail de thèse serait donc d'étendre notre étude théorique à un modèle microscopique plus complet en prenant en compte l'ensemble des sous niveaux Zeeman. 

\small
\listoffigures
\large
\pagestyle{monstyle}
\addcontentsline{toc}{chapter}{Bibliographie}
\bibliography{biblio} 	

\end{document}

%% file: pagedegarde.tex
\thispagestyle{empty}

\begin{wrapfigure}{l}[300pt]{6.5cm}
\includegraphics[width=2.3cm]{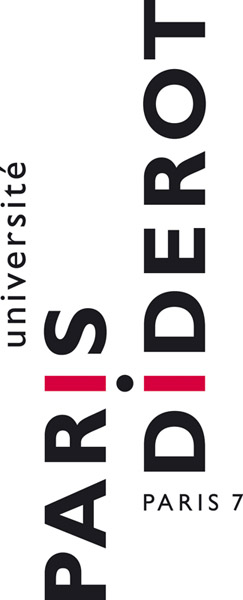}
\end{wrapfigure}

 \centering
\Large 
\textbf{
 LABORATOIRE\\ MATÉRIAUX ET PHÉNOMÈNES QUANTIQUES} \\
 CNRS UMR 7162\\
 \vskip 1 cm

 THÈSE DE DOCTORAT DE \\L'UNIVERSITÉ PARIS DIDEROT - PARIS 7 \\
 \vskip 0.6 cm
     \textbf{}\\
     Spécialité : \textit{Physique Quantique }
     
     \vskip 1.2 cm
     présentée par  \\[1em]

\LARGE
 \textbf{QUENTIN GLORIEUX}\\  \vskip 1 cm
\large
Pour obtenir le grade de\\ docteur de l'Université Paris Diderot
 \vskip 1 cm
  \Large
 \textbf{ETUDE THEORIQUE ET EXPERIMENTALE DES CORRELATIONS QUANTIQUES OBTENUES PAR MELANGE A QUATRE ONDES DANS UNE VAPEUR ATOMIQUE}

\large
\vskip 1 cm
Soutenue, le 19 novembre 2010 devant la commission d'examen composée de :
\vskip 0.5 cm
\bf \large

\begin{tabular}{p{1cm} p{6cm} p{6cm}}


&M. Ennio ARIMONDO & Membre invité \hspace{2cm}\\
&M. Philippe BOUYER & Examinateur\\
&M. Vincent BOYER & Rapporteur\\
&M. Thomas COUDREAU & Directeur de thèse\\
&M. Cristiano CIUTI & Président\\
&M. Claude FABRE & Rapporteur\\

 \end{tabular}

\clearpage
\thispagestyle{empty}
\chapter*{Résumé}
\addcontentsline{toc}{chapter}{Résumé}

\justifying
\normalfont
\paragraph*{Résumé}
\thispagestyle{empty}
Ce travail de thèse se consacre à l'étude théorique et expérimentale de la génération de corrélations quantiques dans le régime des variables continues par mélange à 4 ondes dans une vapeur atomique.

Deux approches théoriques complémentaires sont développées.
D'une part, nous étudions le mélange à 4 ondes sous le point de vue de l'optique non-linéaire "classique" afin d'obtenir les équations d'évolution d'un amplificateur idéal pour un milieu de susceptibilité diélectrique $\chi^{(3)}$.
D'autre part, nous présentons un modèle microscopique à 4 niveaux en double-$\Lambda$ permettant de calculer le coefficient $\chi^{(3)}$ pour une vapeur atomique en présence de laser, et ainsi obtenir théoriquement les spectres de bruit en intensité des grandeurs observées.

La mise en oeuvre expérimentale de ce processus sur la raie D1 du rubidium 85 est ensuite détaillée.
Nous présentons notamment un taux de compression sous la limite quantique standard de -9.2dB, ainsi qu'un régime original permettant la génération de corrélations quantiques sans amplification.

L'ensemble de ces résultats a des applications très importantes dans le domaine de l'optique quantique multimode. Citons par exemple l'imagerie quantique ou les mémoires quantiques multimodes.
\paragraph*{Mots clés}
fluctuations quantiques, réduction du bruit quantique, corrélations quantiques, vapeur atomique, mélange à 4 ondes, variable continue, intrication.
\clearpage
\paragraph*{Abstract}
We study both theoretically and experimentally the generation of quantum correlations in the continuous variable regime by way of four-wave mixing in a hot atomic vapor.

Two theoretical approaches have been developed.
On one side, we study the four-wave mixing under the "classical" non-linear optics point of view.
In such a way we obtain the evolution equation for an ideal linear amplifier in a  $\chi^{(3)}$ medium.
On the other side, we present a microscopic model with  4 levels in the double-$\Lambda$ configuration to calculate the 
$\chi^{(3)}$ coefficient in a atomic vapor dressed with a laser.
This calculation allows us to derive the spectra of intensity noise for interesting parameters.

The experimental part of this work describes the demonstration of this effect on the D1 line of rubidium 85.
We present a measurement of relative intensity squeezing as high as -9.2dB below the standard quantum limit, and an original regime where quantum correlations have been measured without amplification.

 These results have broad applications in the field of multimode quantum optics, e.g. quantum imaging or the multimode quantum memories.
 \paragraph*{Keywords}
quantum fluctuations, squeezing, quantum correlations, atomic vapor, four wave mixing, continuous variable, entanglement.

\clearpage
\thispagestyle{empty}
\chapter*{Remerciements}
\addcontentsline{toc}{chapter}{Remerciements}

Me voila arrivé à la page des remerciements...
ll faut croire que c'est en quelque sorte un aboutissement dans la rédaction d'une thèse !\\

Avant tout, je tiens à remercier très sincèrement Thomas Coudreau, mon directeur de thèse, pour son soutien constant et sa grande disponibilité en dépit de ses nombreuses activités. J'ai été particulièrement touché par son aide dans les moments de doute ainsi que par son ouverture. Je garde un souvenir ému de la nuit de manip à chercher le signal de mélange à 4 ondes "bleu", ainsi que des nombreuses discussions extra-scientifiques que nous avons pu avoir.
Je remercie également très chaleureusement Luca Guidoni dont la patience et la grande rigueur scientifique m'ont fait énormément progresser durant ces trois années. 
Le travail présenté dans ce manuscrit doit beaucoup à Thomas et à Luca qui ont relu avec opiniâtreté les versions successives (malgré les mauvaises blagues de la Poste) et je leur en suis profondément reconnaissant.

Je remercie également l'ensemble des membres permanents de l'équipe IPIQ, Samuel Guibal avec qui j'ai partagé les joies et les peines du travail de recherche ainsi que les soirées au siège du CNRS et Jean-Pierre Likforman qui m'a initié à la physique expérimentale des lasers femto et avec qui je partage une passion commune pour la gastronomie, qu'elle soit universitaire ou valdôtaine.
Un grand merci également à Nicolas Sangouard et Perola Milman avec qui j'ai pu discuter de nombreuses questions théoriques de ce travail.\\

Mes remerciements les plus sincères vont aussi aux membres du jury qui ont accepté la lourde tache de relire ce manuscrit et en premier lieux à Claude Fabre et Vincent Boyer qui ont bien voulu rapporter ce travail. Je remercie également Philippe Bouyer pour avoir traverser l'Atlantique pour cette soutenance ainsi que Cristiano Ciuti qui a accepté de présider le jury.
Je tiens à remercier particulièrement chaleureusement Ennio Arimondo, qui accepté de participer à ce jury en tant que membre et avec qui j'ai pu collaborer de manière très fructueuse durant ma thèse.\\

Durant ces trois ans j'ai été amené à travailler avec un certain nombre d'autres thésards au sein de l'équipe et je me dois de commencer par remercier Romain Dubessy dont le calme et la très (très) grande compétence scientifique, m'ont permis de me sortir de situations qui semblaient parfois désespérées. J'ai eu beaucoup de plaisir à travailler avec lui, que ce soit durant "la nuit de calculs" au mois d'août dans un labo vide ou en salle de manip à confronter nos goûts musicaux.
Je tiens à remercier également Sebastien Rémoville (ou "sremo2") et Brice Dubost qui, dans des styles très différents, ont su me faire profiter de leurs qualités et de leur rigueur de normaliens.
Simon Pigeon a aussi le droit à une ligne (ou deux) de remerciement dans la catégorie travail pour avoir pris patiemment le temps d'écouter mes plaintes contre les forces de Langevin et autres bêtes sauvages de l'optique quantique et m'avoir même parfois donné des pistes pour en découdre !\\

Je remercie aussi les nombreux stagiaires ont fait partie de l'équipe : Marie-Blandine, Pu, Olivier, Daria, Benjamin, Quentin qui m'a donné un coup de main sur le réglage des acoustos et Aurélie qui a passé ses soirées à prendre des données avec moi sur les expériences d'EIT.\\

Une pensée également pour tous les thésard-e-s que j'ai pu croiser durant ces trois années dans le laboratoire MPQ. 
Que ce soit mes partenaires de courses à pied et de ski tout d'abord : Nicolas, Xavier et Alex, mes adversaires de Molky : Ludivine, Pauline, David et Wilfried ou mes acolytes du Shannon : Carole, Jean, Seb et Aurore.
Je vous souhaite à tou-te-s une très bonne continuation.\\

Je tiens à remercier également les nombreuses personnes avec qui j'ai pu collaborer ou simplement discuter durant ces trois années et qui ont rendu cette expérience très enrichissante : Patrick Lepert de l'atelier de mécanique ainsi que Alain Cangemi et Marc Apfel de l'atelier de mécanique et Alain Roger qui m'ont tous bien aidé lorsque j'en avais besoin.
Yann Girard et Edouard Boulat, dont l'intérêt pour les conditions de travail des étudiants m'a beaucoup touché, ainsi qu'évidemment Lynda Silva, Anne Servouze et Joelle Mercier pour leur gentillesse et leur grand professionnalisme. 
De manière générale, un grand merci à l'ensemble des membres de MPQ pour la bonne ambiance qui règne dans le laboratoire.\\

Un grand merci à Nicolas Treps (et ses chaussons d'escalade) pour ses photodiodes tout d'abord et pour  avoir pris le temps de me conseiller avec Claude Fabre, Julien Laurat et Gaétan Messin sur mon choix de post-doc.\\

Quelques lignes pour remercier Valérie P. d'avoir permis l'organisation des toutes ces longues randonnées (et même parfois en ronde) durant ma deuxième année de thèse ainsi qu'à tout ceux et celles qui m'y ont accompagné (Marie, Lucien et son karaoké, John et son méga, Giulia, Morgane, Léa et les jicreux, Manue R, Démos, Damien, Caroline, Kyan, Lucile, Nadir et même Vincent B.)\\

Enfin, il me semble évident que l'ensemble de mes amis ont contribué par leur présence à la réussite de ce travail de thèse et je les en remercie très sincèrement. Il serait difficile d'essayer de les citer tous et toutes ici mais je pense qu'ils ou elles se reconnaitront dans ces quelques lignes et notamment les courageux spectateurs de ma soutenance qui se sont consciencieusement concentrés pour comprendre ce dont je parlais.

Je voudrais dire enfin un grand merci à Julien pour être comme il est, à Marine et son degré d'abstraction pour m'avoir inscrit de force en prépa (et pour tout le reste) ainsi qu'à Eve à qui mon moral durant la fin de ma thèse doit beaucoup.

Je ne peux pas finir sans remercier tout particulièrement toute ma famille pour son soutien durant ces trois ans et les 24 années qui les ont précédées avec trois petites pensées spéciales à ma maman qui a toujours été là quand il fallait, à Olivier et à ses voisins, ainsi qu'à ma grand mère qui s'est même mise à internet pour pouvoir communiquer avec son petit fils...

And last but not least : un immense MERCI à Elise pour m'avoir supporté avant, pendant et après cette thèse (et même durant la rédaction !)

\thispagestyle{empty}
\justifying
\normalfont
\tableofcontents \markboth{TABLES DE MATI\`{E}RES}{Table des matières}

%% file: introductionv5.tex
\pagestyle{intro}
\chapter*{Introduction}
\addcontentsline{toc}{chapter}{Introduction}

\paragraph*{Contexte général}
L'information quantique  \cite{Bennett:2000p16750,Bouwmeester:2001p16436,Cerf07} est un domaine de recherche qui a pour but d'utiliser les propriétés quantiques de la lumière et de la matière afin de développer une nouvelle approche dans le traitement de l'information.
Dès 1982, Richard Feynman propose d'utiliser des systèmes quantiques simples et contrôlés pour simuler des propriétés de systèmes plus complexes \cite{Feynman:1982p16738}.
A la même période, les premiers protocoles de distribution de clé publique sont publiés \cite{Bennett:1984p18774}.
Ces travaux ont ouvert la voie à une nouveau domaine, qui s'est développé de manière très importante durant les 25 dernières années : les communication quantiques\\

Le domaine des \textit{communications quantiques} regroupe les travaux visant à transmettre des états quantiques entre deux points ou plus \cite{Kimble:2008p17197}.
Basés sur la nature quantique du champ électromagnétique, des protocoles de \textit{cryptographie quantique} ont été proposés afin de garantir la confidentialité des communications, non pas grâce à la complexité du cryptage mais grâce aux lois de la physique quantique.
Un état \textit{intriqué} est défini par la présence de corrélations non locales entre deux sous-systèmes qui le constituent.
Par conséquent, la fonction d'onde d'un tel état ne pourra pas s'écrire comme le produit tensoriel de chacun des deux sous-systèmes. 
Cette propriété a été mise en évidence dans un article de Einstein, Podolsky et Rosen en 1935 et porte le nom de paradoxe EPR \cite{Einstein:1935p3706}.
En effet, le caractère non-local de l'intrication s'oppose au principe de réalisme local adopté par les auteurs, qui suggèrent que la description en terme de fonction d'onde de la physique quantique est incomplète.
En 1964, le mathématicien John Bell, utilise l'hypothèse de l'existence de variables cachées pour dériver les inégalités qui portent son nom.
Ainsi, pour tout état décrit par une théorie respectant le réalisme local, ces inégalités doivent être respectées\cite{Bell:1964p17065}.
Les premières preuves de la violation des inégalités de Bell a été obtenues dans les années 1970 \cite{Freedman:1972p19572,Kasday:1975p19364}.
Quelques années plus tard, en 1981, les expériences d'Orsay démontrent la violation de ces inégalités en utilisant les photons émis par cascade radiative dans un jet calcium \cite{Aspect:1981p17124}.
En parallèle de ces travaux sur la non-localité de la physique quantique, de nombreux travaux se sont concentrés sur la production d'états non-classiques de la lumière.

\paragraph*{Etats comprimés en variables continues}
En mécanique quantique, \textit{les états cohérents} ont été introduits par Erwin Schrödinger pour définir les états propres de l'oscillateur harmonique. 
Les travaux de Roy Glauber ont permis d'appliquer ce formalisme au domaine de l'optique quantique, où chaque mode du champ électromagnétique est assimilé à un oscillateur harmonique\cite{Glauber:1963p1343}.
Dans un régime impliquant un très grand nombre de photons, le champ électromagnétique est décrit dans le formalisme des variables continues par deux observables conjuguées que l'on appelle des \textit{quadratures du champ}.
Le produit des variances de deux quadratures est donc naturellement borné par la relation d'inégalité d'Heisenberg.\\
Pour un état cohérent, aussi appelé \textit{état quasi-classique}, le rôle de chacune des quadratures est identique, ce qui signifie que les fluctuations de chacune des observables possèdent une borne inférieure que l'on appelle \textit{la limite quantique standard.}
Au prix de l'augmentation des fluctuations sur l'une des quadratures, il est possible de réduire sous la limite quantique standard les fluctuations de la quadrature conjuguée.
On appelle ces états, des \textit{états non-classiques} de la lumière.
Ces états permettent notamment d'améliorer la sensibilité des interféromètres dans la perspective de la détection des ondes gravitationnelles \cite{Mehmet:2010p12073}.
Depuis une trentaine d'années de nombreux travaux se sont concentrés sur la production de ce type d'état.
Pour des modes non-vides du champ électromagnétique, nous allons distinguer deux catégories d'états non-classiques  en fonction du nombre de modes mis en jeu (un ou deux modes).
Les états non-classiques à un mode du champ sont appelés \textit{états comprimés}.
Les états non-classiques à deux modes du champ peuvent être mis en évidence par la mesure de corrélations quantiques de l'une des quadratures, entre les deux modes.\\
Généralement le champ directement produit par un laser peut être assimilé à un état cohérent.
Pour produire des états non-classiques du champ, il est nécessaire d'utiliser des phénomènes de l'optique non-linéaire.
La première démonstration expérimentale de la production d'états non-classiques (état comprimé à un mode) a été réalisée en 1985 dans une expérience de mélange à 4 ondes à l'aide d'un jet atomique de sodium \cite{Slusher:1985p5993}.
Plus récemment, le mélange à 4 ondes dans un système à trois niveaux en $\Lambda$ a été étudié théoriquement en vue de produire des états comprimés \cite{Shahriar:1998p17972,Lukin:1999p1647}.
De plus le phénomène de transparence électromagnétiquement induite (Electromagnetically Induced Transparency ou EIT) s'est révélé un atout utile afin de permettre un gain paramétrique important, pour les expériences de mélange à 4 ondes dans un système atomique modélisé par des atomes à quatre niveaux en double--$\Lambda$ \cite{Zibrov:1999p1361}.
En effet, le phénomène d'EIT permet d'observer la transparence, en présence d'un champ de contrôle, d'un milieu initialement opaque en l'absence de champ \cite{Harris:1990p9025,Boller:1991p9541,Harris:1997p9203}.
Suite à ces travaux, le mélange à 4 ondes a été utilisé expérimentalement pour produire des corrélations quantiques (état comprimé à deux modes) dans une vapeur atomique de rubidium
\cite{McCormick:2007p652,Boyer:2008p1401,McCormick:2008p6669,Pooser:2009p9625}. 
Dans ces expériences, un faisceau pompe intense interagit avec un faisceau sonde en présence d'une vapeur atomique de $^{85}$Rb. Cette interaction permet de générer un faisceau conjugué, qui peut posséder des corrélations quantiques avec le faisceau sonde.
Dans cette situation, aucune cavité optique n'est nécessaire pour amplifier le phénomène et le processus est par conséquent intrinsèquement multimode, ce qui ouvre la voie à la production de corrélations point à point entre des images quantiques \cite{Marino:2009p1400}.\\

Bien que la plupart des propositions théoriques en information quantique aient été initialement écrites dans le régime des variables discrètes (polarisation d'un photon, spin individuel d'un atome ou d'un ion unique par exemple), une large majorité de ces propositions a été étendue aux variables continues \cite{Braunstein:2005p5169}.
Par exemple, le protocole de téléportation quantique en variable discrète de \cite{Bennett:1993p18466} a été presque immédiatement adapté aux variables continues \cite{Vaidman:1994p18380}.
Le principal intérêt de travailler avec des variables continues vient de la dimension de l'espace de Hilbert sous-jacent qui est de dimension infinie : il contient bien plus d'information potentielle qu'un état dans un espace de dimension deux.
De plus, ces états sont plus faciles à exploiter expérimentalement.
Par exemple, la mesure des quadratures d'un état en variables continues repose simplement sur une détection homodyne réalisée à l'aide d'une lame séparatrice et de photodiodes commerciales relativement peu couteuses.
En comparaison des détecteurs de photons individuels sont nécessaire pour travailler dans le régime des variables discrètes avec des photons uniques.
Cependant, les états comprimés en variables continues sont très sensibles aux pertes (sur le chemin optique ou lors de la détection).
Alors que dans le cas des variables discrètes, les pertes ne diminuent que le débit des transmissions, dans ce régime elles réduisent le caractère quantique des états non-classiques utilisés.
Ainsi, à cause des pertes, il est souhaitable de disposer de sources d'états très intriqués pour distribuer une grande quantité d'information sur de longues distances.
De plus, ces états doivent être susceptibles d'interagir avec des mémoires quantiques, qui sont la brique de base pour accroitre les distance de transmission par la méthode des \textit{``répéteurs quantiques'' }\cite{Duan:2001p1089}.
La diversification des sources d'états non-classiques, et notamment la recherche de sources à des longueurs d'onde compatibles avec les mémoires quantiques (proche de résonnance des transitions atomiques) est un domaine de recherche très actif.


\paragraph*{Contexte dans l'équipe IPIQ}
C'est dans ce contexte que l'équipe IPIQ du laboratoire Matériaux et Phénomènes Quantiques s'intéresse à la réalisation d'une mémoire quantique en variables continues dans un nuage d'ions strontium refroidis par laser.
Ce projet qui a débuté fin 2004 par la construction d'un piège à ions de dimensions centimétriques, s'est poursuivi durant le thèse de Sebastien Rémoville qui a mis en évidence le piégeage et le refroidissement des ions $^{88}$Sr$^+$ \cite{Removille:2009p19653}.
Dans le but final de disposer d'une source de photons corrélés à la longueur d'onde de la transition $5S_{1/2} \to 5P_{1/2}$ du $^{88}$Sr$^+$, j'ai travaillé sur la mise en place d'une expérience de mélange à 4 ondes dans une vapeur de rubidium.
En effet, il existe une quasi-coïncidence entre la transition $5S_{1/2} \to 5P_{1/2}$ du Sr$^+$ et la transition $5S_{1/2} \to 6P_{1/2}$ du $^{85}$Rb \cite{Madej:1998p13999}.
De plus, l'utilisation du mélange à 4 ondes dans une vapeur atomique est une voie pour contourner les problèmes techniques de mise en oeuvre des méthodes utilisant la conversion paramétrique.
Le mélange à 4 ondes n'ayant jamais été observé sur cette transition, la première partie de mon travail de thèse a consisté à mettre en évidence les corrélations quantiques  sur la raie D1 du $^{85}$Rb à 795 nm, en mettant en place une expérience basée sur le montage décrit dans \cite{McCormick:2007p652}.
En parallèle, j'ai étudié théoriquement les phénomènes mis en jeu dans cette expérience en utilisant un modèle microscopique à quatre niveaux.
Cette étude m'a permis de mettre en évidence théoriquement un régime original permettant la production de corrélations quantiques entre le faisceau sonde et conjugué sans amplification, que nous avons pu observer expérimentalement pour la première fois durant cette thèse.
Enfin, l'étude théorique ayant montré qu'il n'était pas envisageable de transposer à l'identique ces expériences sur la raie $5S_{1/2} \to 6P_{1/2}$ du $^{85}$Rb, nous avons utilisé les ressources expérimentales disponibles pour mettre en évidence l'effet de transparence électromagnétiquement induite sur cette transition, ce qui, à notre connaissance, n'avait jamais été étudié auparavant.

\paragraph*{Organisation du manuscrit}
Ce manuscrit est composé de trois parties.
La première partie permet d'introduire les éléments théoriques et expérimentaux qui seront utilisés au cours des chapitres suivants.
Dans le \textbf{premier chapitre}, nous donnons une description quantique du champ électromagnétique en présentant deux catégories d'états : les états cohérents et les états comprimés.
Nous introduisons ensuite, le formalisme de la représentation de Wigner et les notions de bruit quantique et de photodétection.
Les corrélations quantiques en variables continues sont ensuite décrites de façon synthétique.
Enfin nous présentons brièvement les équations de Heisenberg-Langevin, qui permettent de traiter de l'interaction lumière-matière de façon entièrement quantique.\\

Le \textbf{chapitre \ref{ch2}}, nous permet de décrire les différentes techniques expérimentales utilisées dans la suite de ce manuscrit.
Nous commençons par une revue de la littérature sur les méthodes de production des états non-classiques de la lumière.
Nous présentons plus en détails le mélange à 4 ondes qui est la technique qui a été utilisée durant cette thèse.
Dans ces expériences, un faisceau pompe interagit avec un faisceau sonde dans une vapeur de rubidium et permet de générer un faisceau conjugué.
Nous détaillons donc les caractéristiques des sources utilisées pour produire les différents faisceaux et nous décrivons le milieu atomique, afin d'établir une loi empirique permettant de déterminer la densité de rubidium en phase vapeur dans la cellule utilisée.
Enfin, nous donnons les caractéristiques des différents éléments de la chaine de photodétection (photodiodes, analyseur de spectres) et nous étudions l'effet des imperfections expérimentales sur la mesure de corrélations quantiques.\\

La seconde partie de ce manuscrit est consacrée à l'étude théorique du mélange à 4 ondes.
Dans le \textbf{chapitre \ref{ch3}}, nous en présentons une approche phénoménologique.
Dans un premier temps, nous écrivons les équations de l'optique non-linéaire qui décrivent le processus de mélange à 4 ondes.
Puis, nous étudions le phénomène d'amplification paramétrique dans deux situations : d'une part l'amplification des faisceaux sonde et conjugué dans une configuration insensible à la phase du faisceau pompe, d'autre part l'amplification ou la déamplification du faisceau sonde dans une  configuration sensible à la phase des faisceaux de pompe.
Dans un second temps, nous utilisons le modèle de l'amplificateur linéaire idéal pour calculer les valeurs moyennes de l'intensité ainsi que les fluctuations quantiques associées aux différents faisceaux (intensité du faisceau pompe ou différence d'intensité entre le faisceau sonde et conjugué, selon le cas).\\

La susceptibilité  non--linéaire  $\chi^{(3)}$, introduite dans le chapitre 3 de manière phénoménologique, trouve son contenu physique dans le modèle microscopique que nous développons dans le \textbf{chapitre \ref{ch4}}.
Après avoir décrit brièvement le phénomène de transparence électromagnétiquement induite dans un système à 3 niveaux en $\Lambda$, nous étudions le processus de mélange à 4 ondes à l'aide d'un système à quatre niveaux en double-$\Lambda$ \cite{Lukin:2000p1648}.
Nous présentons, la méthode de résolution des équations de Heisenberg-Langevin qui nous a permis de dériver l'expression des corrélations quantiques d'intensité et des anti-corrélations de phase dans ce type de système en présence de deux champs pompes et d'un champ sonde.
A l'aide de ce modèle, nous mettons en évidence pour la première fois la possibilité de générer un haut niveau de corrélations quantiques dans un milieu constitué d'atomes froids par mélange à 4 ondes. \\
Dans une seconde partie, nous discutons de l'extension de ce modèle à une vapeur atomique ``chaude'' ce qui permet de comparer les prédictions à nos résultats expérimentaux.
C'est dans cette section également que nous présentons un régime qui a été proposé pour la première fois durant cette thèse, permettant de réaliser une \textit{lame séparatrice quantique}, c'est-à-dire générer deux faisceaux corrélés quantiquement à partir d'un état cohérent et sans amplification.
Enfin, nous étudions théoriquement la possibilité de transposer ces expériences sur la transition $5S_{1/2} \to 6P_{1/2}$ du $^{85}$Rb et nous pouvons conclure que dans les régimes que nous avons étudiés, une telle expérience n'est pas réalisable.\\

La troisième partie de ce manuscrit est consacrée aux résultats expérimentaux obtenus durant mon travail de thèse.
Dans le\textbf{ chapitre \ref{ch5}}, nous présentons la génération d'états comprimés à deux modes du champ à 795 nm sur la transition D1 du $^{85}$Rb.
Après avoir détaillé le dispositif expérimental, nous donnons une caractérisation du milieu atomique en présence d'un champ pompe.
Puis, nous effectuons une étude des différents paramètres expérimentaux qui affectent le gain du processus et les corrélations obtenues.
Chacune de ces expériences comparée aux prédictions théoriques.
Enfin, nous présentons la mise en évidence expérimentale du régime  de  la \textit{lame séparatrice quantique}, que nous avons proposé théoriquement.\\

Pour finir, le \textbf{chapitre 6}, est consacré à lune étude expérimentale préliminaire du phénomène de transparence électromagnétiquement induite (EIT) sur la transition $5S_{1/2} \to 6P_{1/2}$ du $^{85}$Rb. Il s'agit de la première démonstration expérimentale du phénomène d'EIT observée sur cette transition.

\clearpage
\thispagestyle{empty}



%% file: chapitre1v5.tex
\chapter{Optique quantique en variables continues}\label{ch1}\minitoc
\vspace{2cm}
L'objectif de ce chapitre est de présenter les outils théoriques utilisés dans ce manuscrit au cours de l'étude théorique qui sera détaillée dans les chapitres \ref{ch3} et \ref{ch4}.
Au cours du texte, la traduction anglaise des notions les plus souvent utilisées est donnée en italique.\\
Après une présentation générale de la notion de variable continue (\textit{continuous variable}), nous étudions comment cette notion s'applique dans le cas d'un champ électromagnétique quantifié.
Différents états faisant intervenir un seul mode du champ sont ensuite présentés, notamment nous introduisons deux types d'états gaussiens : les états cohérents (\textit{coherent states}) et  les états comprimés de la lumière (\textit{squeezed states}).\\
Nous étendons notre étude aux états du champ à deux modes et nous décrivons les propriétés de corrélations quantiques entre ces modes.
En présence de corrélations en intensité entre les deux modes, les états sont appelés états comprimés à deux modes (\textit{twin beams}).
Si dans le même temps, on observe des anti--corrélations sur l'observable conjuguée, on parlera alors d'états intriqués (\textit{entangled states}).
Dans le régime des variables continues, il s'agit de l'analogue des états intriqués introduits dans le célèbre article EPR de 1935 \cite{Einstein:1935p3706,Zeilinger:1999p5174,Reid:2009p5180}.

\section{Quantification du champ et états quantiques de la lumière.}
Historiquement les variables dites ``discrètes'' ont joué un rôle prédominant dans l'étude des propriétés quantiques de la lumière.
En effet il s'agit d'un très bon système modèle car un grand nombre de phénomènes peuvent être décrits dans un espace de Hilbert de faible dimension et plus particulièrement dans un espace de dimension deux.
De nombreuses propositions théoriques et démonstrations expérimentales utilisant ces systèmes ont été réalisées ces dernières années.
On peut citer notamment : la réalisation de portes logiques destinée à l'ordinateur quantique, la cryptographie quantique ou les protocoles de téléportation quantique \cite{Bouwmeester:1997p2392,Gisin:2002p4832,Duan:2010p5192}.\\
Dans cette section nous laissons de côté les variables discrètes pour nous intéresser au formalisme des variables dites ``continues'' \cite{Braunstein:2005p5169}.

\subsection{Description classique du champ}
Les équations qui régissent la propagation d'une onde électromagnétique classique dans le vide sont les équations de Maxwell et s'écrivent :
\begin{equation}
\nabla \mathbf{E}(\mathbf{r},t)-\frac{1}{c}\frac{\partial^2}{\partial t^2}\mathbf{E}(\mathbf{r},t)=0.
\end{equation}
En supposant  le champ électrique $\mathbf{E}(\mathbf{r},t)$ se propageant sur l'axe $z$ et polarisé selon $\mathbf{p}$, alors  le champ s'écrit  $\mathbf{E}(\mathbf{r},t)=E(z,t)\ \mathbf{p}$ avec :
\begin{equation}
E(z,t)=|E_0|\left(\alpha\ e^{i k z}e^{-i \omega t}+\alpha^*\ e^{-i k z}e^{i \omega t}\right).
\end{equation}
où $\alpha$ est un nombre complexe de module 1 et $|E_0|$ l'amplitude du champ $E$.
On peut alors écrire cette expression en faisant apparaitre les quadratures du champ $X$ et $Y$ :
\begin{equation}
E(z,t)=|E_0|\left(X\cos\omega t+Y\sin\omega t\right) ,
\end{equation}
où les quadratures s'écrivent :
\begin{deqarr}
X&=&\alpha\ e^{i k z}+\alpha^*\ e^{-i k z},\\
Y&=&i\left(\alpha\ e^{i k z}-\alpha^*\ e^{-i k z}\right).
\end{deqarr}
On s'intéresse à l'évolution temporelle du champ.
Pour une position donnée on pourra écrire simplement les quadratures sous la forme :
\begin{deqarr}
X&=&2 \cos \phi,\\
Y&=&2 \sin\phi.
\end{deqarr}
avec 
\begin{equation}
\phi=\text{tan}^{-1}\frac{Y}{X}
\end{equation}
Cette description classique du champ peut être représentée dans le repère de Fresnel par une flèche de longueur $|E_0|$ et une phase $\phi$ comme le montre la figure 1.1 a).

\subsection{Description quantique du champ}
Par analogie avec la description de l'oscillateur harmonique, où la position et l'impulsion peuvent être quantifiées, on peut donner une description quantique du champ électromagnétique qui devient alors une observable notée $\hat{E}(t)$ \cite{FabreHouches}.
Pour une onde plane, monomode de fréquence $\omega$ et d'amplitude $E_0$, on exprime le champ en un point donné de l'espace sous la forme  \cite{Meystre:2007p4122}:
\begin{equation}\label{champquantifie}
\hat{E}(t)=\mathcal{E}_0\left(\hat{X} \cos \omega t +\hat{Y} \sin \omega t\right),
\end{equation}
où l'on a introduit la constante de normalisation $\mathcal{E}_0$ qui correspond au champ électrique associé à un photon et qui a pour expression :
\begin{equation}
\mathcal{E}_0=\sqrt{\frac{\hbar \omega}{2 \epsilon_0 V}},
\end{equation}
où $V$ est le volume de quantification.
A l'aide de la relation (\ref{champquantifie}), on définit les opérateurs de création et d'annihilation d'un photon à la fréquence $\omega$ :
\begin{equation}
\hat{a}_\omega=\frac{\hat{X}+i\hat{Y}}{2},\hspace{1cm}\hat{a}^\dag_\omega=\frac{\hat{X}-i\hat{Y}}{2},
\end{equation}
ce qui nous permet d'écrire la relation (\ref{champquantifie}) sous la forme :
\begin{equation}\label{champquantifie2}
\hat{E}(t)=\mathcal{E}_0\left(\hat{a} e^{ - i\omega t} +\hat{a}^\dag  e^{  i\omega t}\right).
\end{equation}
Réciproquement pour les quadratures $\hat{X}$ et $\hat{Y}$ on a :
\begin{equation}\label{defquadrature}
\hat{X}=\hat{a}_\omega+\hat{a}^\dag_\omega,\hspace{1cm}\hat{Y}=-i\left(\hat{a}_\omega-\hat{a}^\dag_\omega\right).
\end{equation}

 \subsection{Inégalité d'Heisenberg et fluctuations quantiques du champ}
 	\begin{figure}	
 			\centering
 			\includegraphics[width=13.2cm]{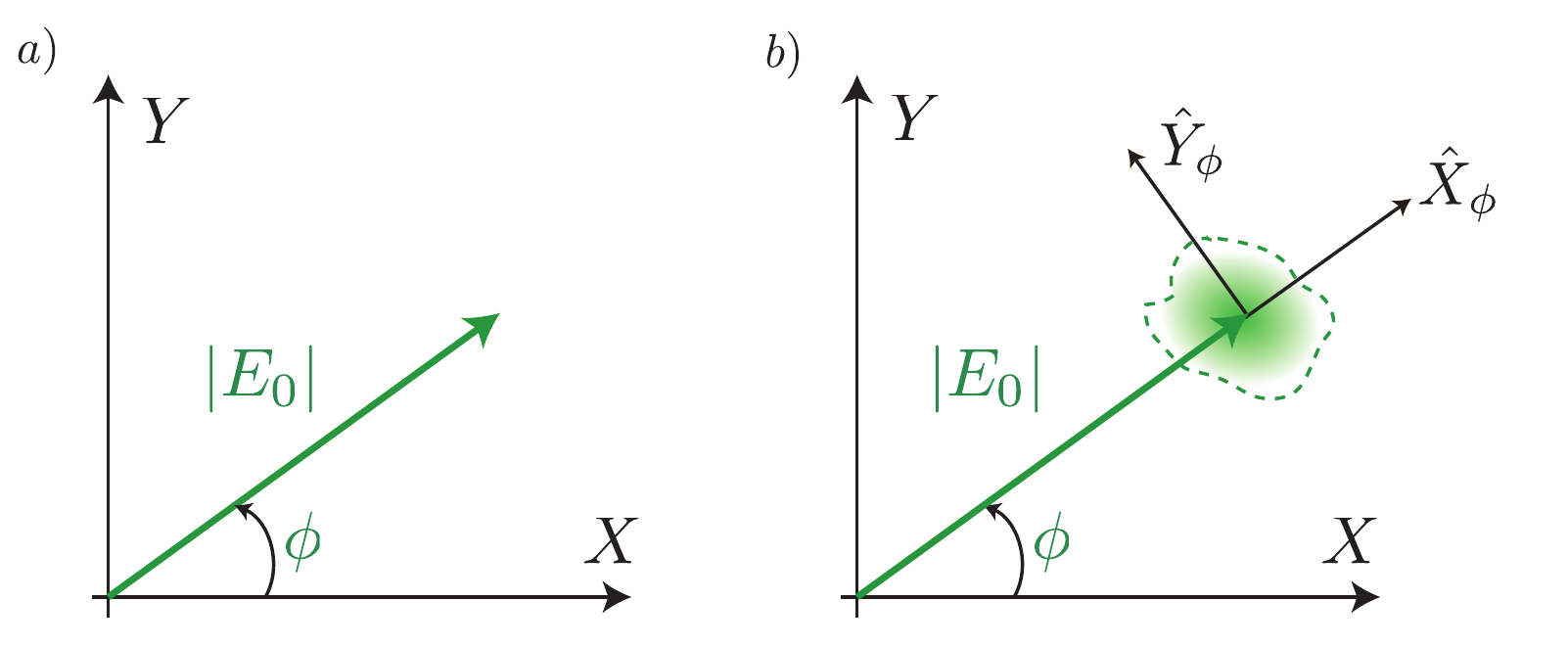}
 			\caption[Champ électromagnétique dans le repère de Fresnel]{Description d'un champ électromagnétique d'amplitude $|E_0|$ et de phase $\phi$ dans le repère de Fresnel. a) description classique. b) description quantique incluant les quadratures $\hat{X}_\phi$ et $\hat{Y}_\phi$ et la zone où l'écart par rapport à la valeur moyenne du champ est inférieure à l'écart quadratique moyen.}		\label{fresnel_fig}	
 			\end{figure}
D'après la définition que nous venons de donner des quadratures d'un champ électromagnétique quantifié, nous pouvons faire l'analogie avec les opérateurs position et impulsion de l'oscillateur harmonique.
Au même titre que ces opérateurs, les opérateurs $\hat{X}$ et $\hat{Y}$ sont un couple d'opérateurs conjugués, par conséquent ils ne commutent pas :
\begin{equation}
\left[\hat{X},\hat{Y}\right]=2 i.
\end{equation}
 Cette non-commutation signifie que l'on ne peut pas mesurer simultanément avec une précision arbitraire ces deux observables.
 Elle se traduit par la relation d'inégalité d'Heisenberg qui donne la borne inférieure du produit des variances de ces deux observables 
 \begin{equation}\label{heisenberg}
 \Delta \hat{X}^2\Delta \hat{Y}^2\geq 1,
 \end{equation}
 où  l'on a défini la variance par :
\begin{equation}
\Delta \hat{X}^2=\langle\hat{X}^2\rangle-\langle\hat{X}\rangle^2.
\end{equation}
 La mesure de ces quadratures ne pourra donc pas se faire simultanément avec une précision infinie et nous allons nous intéresser aux fluctuations des valeurs obtenues par des réalisations successives, que l'on va qualifier de bruit quantique de la lumière (\textit{quantum noise}).
 Comme le choix d'un couple de quadratures est arbitraire, on étend la notation de quadrature (relation (\ref{defquadrature})) à celle de quadrature généralisée que l'on note pour une phase $\theta$ quelconque :
 \begin{equation}\label{defquadraturegene}
 \hat{X}_\theta=\hat{a}\ \text{e}^{-i\theta}+\hat{a}^\dag \ \text{e}^{i\theta},\hspace{1cm}\hat{Y}_\theta=-i\left(\hat{a} \ \text{e}^{-i\theta}-\hat{a}^\dag \ \text{e}^{i\theta}\right).
 \end{equation}
On peut écrire la valeur moyenne de l'opérateur $\hat{a}$ sous la forme $\langle\hat{a}\rangle=|\alpha| e^{i\phi}$.\\
Dans le cas particulier où l'on choisit $\theta=\phi$, avec $\phi$ la phase du champ moyen de valeur moyenne non nulle ($\alpha \neq 0$), alors les quadratures $ \hat{X}_\theta$ et $ \hat{Y}_\theta$ sont respectivement appelées les quadratures d'intensité et de phase du champ.
La figure \ref{fresnel_fig}b) donne une représentation du champ dans un repère de Fresnel.
On a ajouté sur cette figure, par rapport à la description classique, une surface a priori quelconque autour de la valeur moyenne qui correspond à l'aire des fluctuations, car ces quadratures doivent respecter la relation  (\ref{heisenberg}).
L'aire des fluctuations sera définie rigoureusement dans la section \ref{Rwigner}.
On peut la voir, pour l'instant, comme la zone où l'écart par rapport à la valeur moyenne du champ est inférieure à une certaine valeur (l'écart quadratique moyen par exemple).

\subsection{Etats nombres, états cohérents et états comprimés}
Dans la théorie quantique de la lumière, il existe plusieurs façons de décrire l'état d'un champ électromagnétique.
Nous présentons ici celles qui nous serons utiles dans ce manuscrit \cite{Knight:1995p3906}.
\subsubsection{Etats nombres}
Les états nombres sont utilisés en optique quantique dans le régime des variables discrètes.
Un état nombre $|n\rangle$, aussi connu sous le nom d'état de Fock, est vecteur propre de l'opérateur nombre $\hat{N}=\hat{a}^\dag\hat{a}$. La valeur propre associée à ce vecteur propre est $n$ :
\begin{equation}
\hat{N}|n\rangle=n|n\rangle.
\end{equation}
L'état fondamental, c'est-à-dire l'état de plus basse énergie, est noté $|0\rangle$.
Cet état correspond au vide électromagnétique.
Bien que son appellation soit trompeuse, le vide ne correspond pas à une énergie nulle pour le système.
En effet le Hamiltonien du champ pour un mode s'écrit 
\begin{equation}
\hat{H}=\frac 12 \hbar \omega (\hat{a}^\dagger\hat{a}+\hat{a}\hat{a}^\dagger).
\end{equation}
La relation de commutation $[\hat{a},\hat{a}^\dagger] =1$ permet d'écrire le Hamiltonien sous la forme :
\begin{equation}
\hat{H}= \hbar \omega \left(\hat{N}+\frac 12\right).
\end{equation}
Les états de Fock sont donc des vecteurs propres du Hamiltonien du champ et les valeurs propres associées valent $(n+\frac 12) \hbar \omega$.
L'énergie du vide pour un mode de fréquence $\omega$ vaut donc~$\frac 12 \hbar \omega$.\\
D'autre part l'action des opérateurs d'annihilation $\hata$ ou de création $\hatad$ sur un état de Fock permet respectivement de diminuer ou d'augmenter d'une unité la valeur propre de l'opérateur nombre associé à l'état qui en résulte :
\begin{equation}
\hata |n\rangle=\sqrt{n}|n-1\rangle,\hspace{1.5cm}\hatad |n\rangle=\sqrt{n+1}|n+1\rangle.
\end{equation}
De plus, on peut décrire un état nombre à l'aide d'un produit de l'opérateur de création $\hatad$ appliqué au vide sous la forme:
\begin{equation}\label{Fock1}
|n\rangle=\frac{(\hatad)^n}{\sqrt{n!}}|0\rangle.
\end{equation}
Dans le cas des états de Fock, on connait avec une précision infinie le nombre de photons dans le mode, et donc l'intensité.
Par conséquent la phase est connue avec une précision nulle, ou plus simplement la phase n'est pas définie pour un état de Fock.
Actuellement, les plus grands états de Fock qui peuvent être générés sont de l'ordre de la dizaine de photons \cite{Haroche:1998p16411}.
Pour un très grand nombre de photons, la description en terme d'état de Fock devient compliquée dès lors que le système subit des pertes.
Ainsi, ces états ne sont pas adaptés pour la description de faisceaux relativement intenses.

\subsubsection{Etats cohérents} 
 	
\begin{figure}	
\centering
\includegraphics[width=13.2cm]{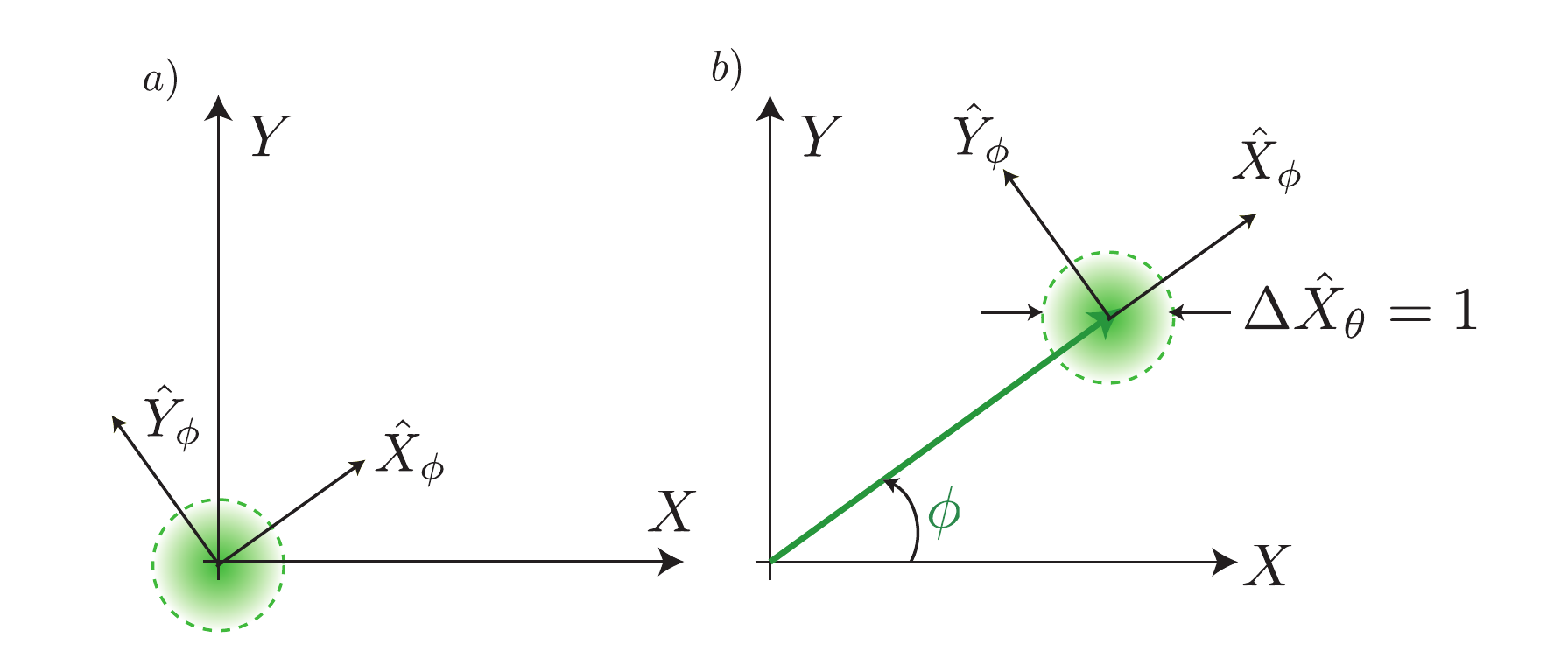}
\caption[Etat cohérent et état vide.]{a) Etat vide et b) état cohérent de moyenne non nulle dans le repère de Fresnel. Les fluctuations sont données par un disque de rayon 1 autour de la valeur moyenne. La variance des quadratures $\hat{X}_\theta$ et $\hat{Y}_\theta$ vaut 1 quelque soit $\theta$.}		\label{fresnelcoherent}	
 \end{figure}
Les états cohérents ou états quasi-classiques ont été introduits en optique quantique par Glauber \cite{Glauber:1963p1343} pour décrire des modes contenant un très grand nombre de photons.
Ils sont utilisés pour rendre compte des états générés par un laser.
Un état cohérent que l'on notera $|\alpha\rangle$ est vecteur propre de l'opérateur d'annihilation associé à la valeur propre $\alpha$ :
\begin{equation}
\hata |\alpha\rangle = \alpha |\alpha\rangle.
\end{equation}
Ces états sont normés :
\begin{equation}
 \langle \alpha | \alpha \rangle =1 ,
\end{equation}
et la valeur moyenne du nombre de photons dans ce mode est donnée par :
\begin{equation}
\langle \hatad \hata \rangle =  |\alpha|^2 .
\end{equation}
On retrouve ici la forme que prendrait la valeur moyenne de l'intensité du champ dans une description classique pour un champ d'amplitude $\alpha$.
Par ailleurs, il est intéressant de noter qu'un état cohérent n'est pas vecteur propre de l'opérateur nombre et par conséquent du Hamiltonien du champ.
Cela veut dire que la quantité d'énergie d'un état cohérent n'est pas parfaitement définie.
On peut comprendre cela comme une conséquence du fait qu'un état cohérent peut être écrit dans la base des états de Fock comme une combinaison linéaire d'états à nombre de photons différents c'est-à-dire d'énergies différentes :
\begin{equation}
|\alpha\ket = \sum_{n=0}^\infty \bra n|\alpha\ket |n\ket.
\end{equation}
A l'aide de la relation (\ref{Fock1}) on peut écrire :
\begin{equation}\label{coherentfock}
|\alpha\ket =  \sum_{n=0}^\infty \frac{\alpha^n}{\sqrt{n!}} \exp\left(-\frac{1}{2}|\alpha|^2\right)  |n\ket.
\end{equation}
Cette forme donne immédiatement la probabilité de trouver $n$ photons dans un état cohérent :
\begin{equation}
p_n=|\bra n | \alpha\ket |^2 =\exp -|\alpha|^2 \frac{\alpha^{2n}}{n!}
\end{equation}
Cette probabilité est appelée une distribution de Poisson.\\
Une propriété importante des états cohérent réside dans ses fluctuations autour de sa valeur moyenne.
Nous l'avons vu précédemment, la limite ultime de la précision sur la mesure d'observables conjugués, en mécanique quantique, est donnée par l'inégalité d'Heisenberg.
Un état cohérent est un \textit{état minimal}, c'est à dire, qu'il sature la relation d'Heisenberg  :
\begin{equation}
\Delta \hat{X}_\theta^2\Delta \hat{Y}_\theta^2 =1.
\end{equation}
De plus, il n'y a pas de quadrature privilégiée et donc pour tout angle $\theta$ :
\begin{equation}
\Delta \hat{X}_\theta^2=\Delta \hat{Y}_\theta^2 =1.
\end{equation}
Il s'agit de ce que l'on appelle la\textbf{ limite quantique standard.} (\textit{standard quantum limit}, SQL).
Les fluctuations seront représentées par un disque de rayon 1 autour de la valeur moyenne (figure \ref{fresnelcoherent}).
Le bruit associé à une mesure de ces quadratures sera donc fixé par ces fluctuations autour de la valeur moyenne.
On appelle ce bruit, le bruit quantique standard, ou\textbf{ bruit de grenaille} (\textit {shot-noise}).
La seconde appellation est dûe à la vision corpusculaire du photon, dans laquelle la présence de fluctuations sur la mesure d'une quadrature revient à décrire le flux de photons par une \textbf{statistique poissonienne}, qui se caractérise par une variance du nombre de photons égale à la valeur moyenne :
 \begin{equation}
\Delta \hat{N}^2=\bra \hat{N}\ket.
\end{equation}
Une des conséquences de la nature  poissonienne du flux de photons est que si l'on sépare en deux parties à l'aide d'une lame séparatrice un état cohérent, les deux états produits sont deux états cohérents.\\
Nous l'avons vu précédemment, le vide peut être décrit comme un état de Fock.
Il peut aussi être vu comme un état cohérent et donc possède les mêmes fluctuations que les autres états cohérents.
Un état cohérent non vide peut donc être décrit comme un \textit{déplacement} de l'état vide de valeur moyenne nulle vers un état de valeur moyenne $|\alpha|^2$.\\
\subsubsection{Etats gaussiens}
Un état gaussien est un état pour lequel la distribution de probabilité de deux quadratures orthogonales sont des gaussiennes.
De plus, pour ces états, l'inégalité de Heisenberg est saturée de telle manière que les états gaussiens soient des états minimaux.
Les états cohérents respectent ces deux conditions et par conséquent sont des états gaussiens.\\
Par ailleurs, on peut noter que des états gaussiens sont entièrement caractérisés par la donnée du moment d'ordre un (la valeur moyenne) et du moment d'ordre deux (la variance).
La valeur moyenne sera identifiée à une grandeur classique, et les propriétés purement quantiques de ces états seront donc contenues dans la description en terme de variance.
Tout au long de ce manuscrit, lorsque l'on étudiera les propriétés quantiques d'états gaussiens, on s'intéressera donc aux moments d'ordre deux.

\subsubsection{Etats comprimés}
   	\begin{figure}	
   	   	   	\centering
   	   	   	\includegraphics[width=11.5cm]{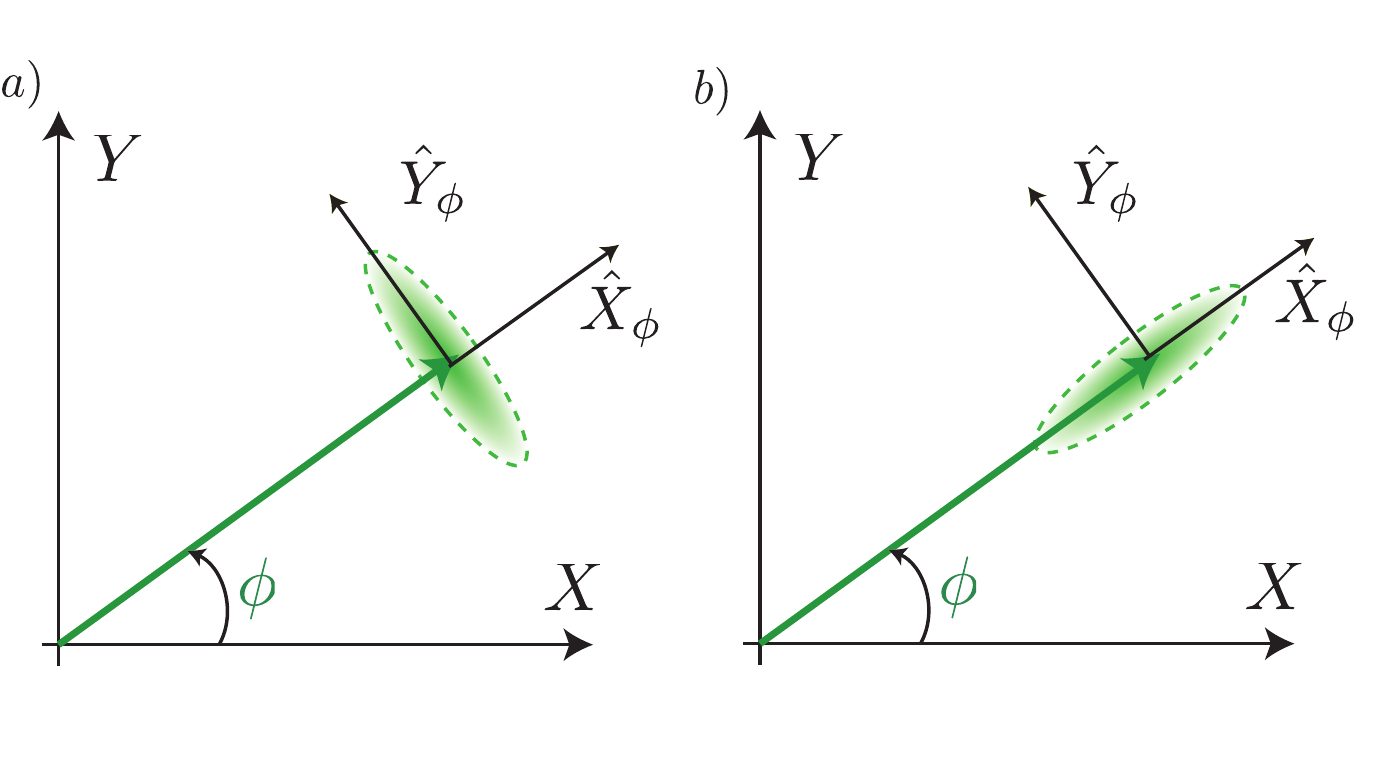}
   	   	   	\caption[Etat comprimé.]{a) Etat comprimé en intensité et b) état comprimé en phase dans le repère de Fresnel. Les fluctuations sont données par une ellipse  autour de la valeur moyenne. Le produit des variances des quadratures $\hat{X}_\phi$ et $\hat{Y}_\phi$ vaut 1	dans le cas d'états comprimés minimaux et est supérieur à 1 sinon.}\label{fresnelsq}	
\end{figure}
Un autre type d'état gaussien, sont les états comprimés à un mode du champ (\textit{single mode squeezed states}).
Pour ce type d'état, les fluctuations autour de la valeur moyenne du champ ne se décrivent pas par un disque mais par une ellipse. 
Dans ce cas, toutes les quadratures ne sont plus équivalentes, et tout en respectant l'inégalité d'Heisenberg, la variance de l'une des quadratures peut être inférieure à 1 au prix d'une augmentation de la variance sur la quadrature conjuguée.
Sur la figure \ref{fresnelsq}, deux cas particuliers des ces états sont représentés.
La figure a) décrit une compression des fluctuations en intensité c'est-à-dire :
\begin{equation}
\Delta\hat{X}_\theta^2<1,\text{ et }  \Delta\hat{Y}_\theta^2>1.
\end{equation}
Tandis que  la figure b) montre une compression des fluctuations en phase c'est-à-dire :
\begin{equation}
\Delta\hat{X}_\theta^2>1,\text{ et }   \Delta\hat{Y}_\theta^2<1.
\end{equation}
La réalisation expérimentale de tels états est un enjeu majeur en optique quantique depuis les trois dernières décennies \cite{Bachor:2004p4500}. De nombreuses méthodes ont été utilisées et nous en décrirons quelques unes dans le chapitre \ref{ch2}.
Comme nous l'avons vu pour des états cohérents, une description corpusculaire, c'est-à-dire en terme de statistique du flux de photons dans le mode peut être utile.
Dans le cas d'un état comprimé en intensité la statistique sera sub--poissonienne et le bruit de mesure sera donc inférieur au bruit quantique standard :
\begin{equation}
\Delta \hat{N}^2<\bra \hat{N}\ket.
\end{equation}

\subsubsection{Représentation des états du champ et mesure du taux de compression.}

\begin{figure}
\centering
\includegraphics[width=14.5cm]{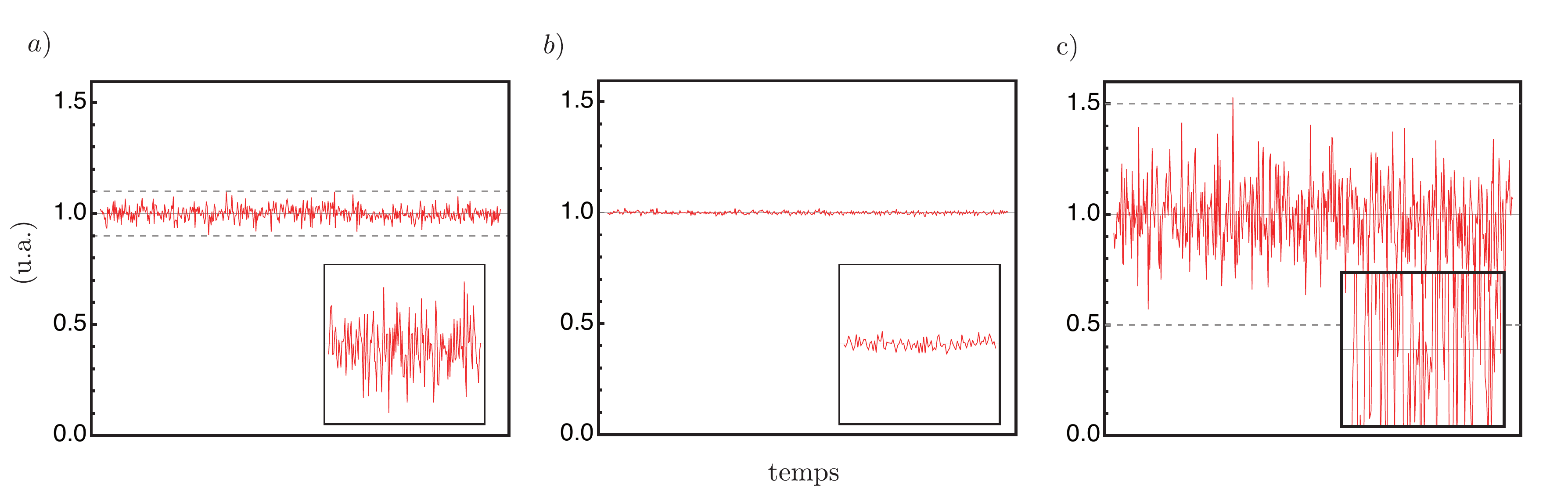}
\caption[Fluctuations pour un état cohérent et pour un état comprimé]{a) Fluctuations pour un état cohérent, b) pour un état comprimé (-7dB) en intensité et c) pour un état comprimé en phase (excès de bruit de +7dB en intensité). Les encarts montrent un agrandissement de ce signal à la même échelle pour les trois figures.\label{shotft}	}

\includegraphics[width=14.5cm]{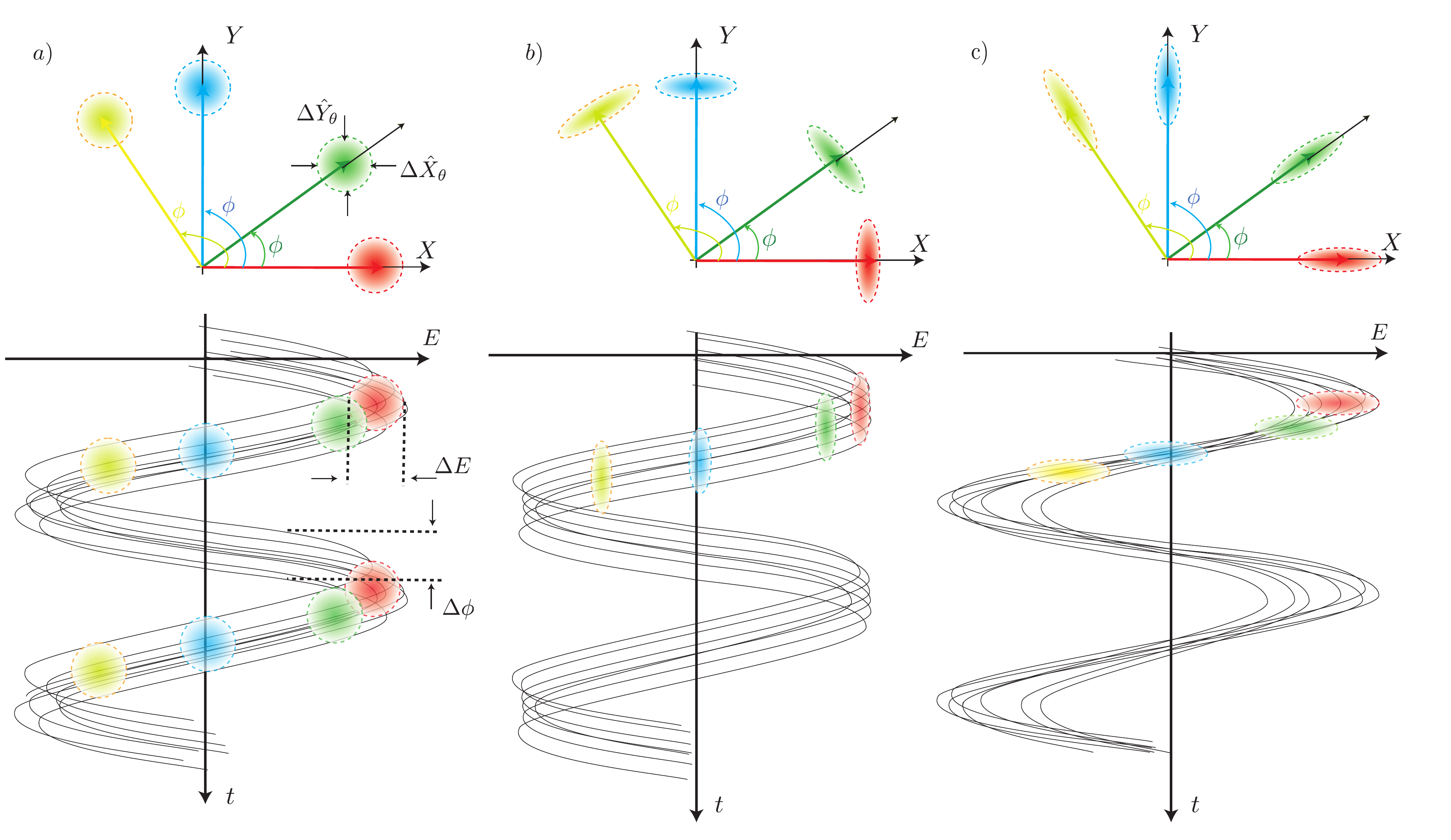}
\caption[Espace des phases et champ électrique associé.]{Espace des phases et champ électrique associé dans le cas a) d'un état cohérent, b) d'un état comprimé en intensité et c) d'un état comprimé en phase. Les fluctuations sont indiquées par une zone d'incertitude  autour de la valeur moyenne dans l'espace des phases qui correspondent aux valeurs que peut prendre le champ électrique lors de différentes mesures.\label{fresnelsq2}}	
\end{figure}

Différentes représentations graphiques permettent de décrire les états du champ en variables continues \cite{Lam:1998p11060}. 
Nous donnons ici ces multiples représentations ainsi qu'un critère quantitatif pour mesurer le niveau de compression d'une quadrature. \\
Sur la figure \ref{shotft}, nous avons simulé les fluctuations d'un flux de photons suivant une statistique poissonienne de valeur moyenne 1000 (état cohérent), d'un second suivant une statistique sub--poissonienne de même valeur moyenne (état comprimé en intensité) et d'un troisième suivant une statistique sur--poissonienne (état comprimé en phase) .\\
Une autre représentation des états du champ, consiste à en donner la valeur moyenne et les fluctuations dans l'espace des phases.
A cette description dans l'espace des phases, on peut mettre en regard l'évolution temporelle du champ électrique.
Sur la figure \ref{fresnelsq2}, on donne ces deux représentations dans le cas d'un état cohérent et dans celui de deux états comprimé (en intensité et en phase).

Pour déterminer l'importance de la compression sur une des quadratures, il faut définir une référence.
Ce sont les fluctuations de l'état cohérent, c'est à dire la limite quantique standard, qui sont choisies pour jouer ce rôle.
Ainsi, les mesures de bruit décrites dans ce manuscrit seront toujours comparées  à cette limite.\\
Dans la littérature, deux échelles sont utilisées pour quantifier le niveau de compression.
Une échelle linéaire, qui définit le taux de compression $S$ par le rapport entre la variance d'une quadrature donnée de l'état étudié et la variance dans le cas d'un état cohérent :
\begin{equation}\label{ajjt}
S_\theta=\frac{\Delta \hat{X}_\theta^2}{\Delta \hat{X}^2_{\ coherent}}.
\end{equation}
Comme la statistique d'un flux de photon pour un état cohérent est poissonienne, cette définition est identique à celle du facteur de Fano de l'état étudié \cite{Fano:1947p10925}.
Ce paramètre vaut donc 1 pour un état cohérent et 0 pour un état infiniment comprimé selon la quadrature $\theta$.\\
On peut définir une seconde échelle basée sur le logarithme de ce rapport :
\begin{equation}
S_{dB}=10 \log \left(\frac{\Delta \hat{X}_\theta^2}{\Delta \hat{X}^2_{\ coherent}}\right).
\end{equation}
Cette représentation a l'avantage de décrire avec une meilleure dynamique les taux de compression entre 0.5 (-3 dB) et 0.1 (-10 dB), c'est-à-dire les niveaux que l'on observe typiquement dans la littérature \cite{Lambrecht:1996p6121,Silberhorn:2001p11006,Bachor:2004p4500}.

\section{Quasi--probabilité}
\subsection{Etats mixtes et opérateur densité.}
Il existe des états qui ne sont pas des états purs, c'est-à-dire qui ne peuvent pas être décrits par un vecteur d'état $|\psi\ket $, contrairement à ceux que nous venons d'étudier.
Pour les décrire  il est nécessaire d'utiliser une somme statistique de plusieurs états purs.
On appelle ces états, des états mixtes du champ.
Pour décrire un état mixte de façon générale, on utilise l'opérateur densité, qui s'écrit sous la forme :
\begin{equation}
\rho=\sum_\psi P_\psi |\psi\ket\bra\psi|,
\end{equation}
où $P_\psi $ est la probabilité d'être dans l'état $|\psi\ket $ de telle sorte que l'on ait : $\sum P_\psi =1$.\\
Nous l'avons déjà mentionné, les états de Fock forment une base de l'espace des états.
On peut donc écrire une décomposition sur les états de Fock de l'opérateur densité :
\begin{equation}
\rho=\sum_m \sum_n  |n\ket\bra n|\rho  |m\ket\bra m| = \sum_m \sum_n \rho_{nm}|n\ket\bra m|.
\end{equation}
Pour un opérateur quelconque $\hat{A}$, la valeur moyenne dans un état pur $|\psi\ket$ est donnée par :
\begin{equation}
\bra \hat{A}\ket = \bra\psi |\hat{A}|\psi\ket.
\end{equation}
Pour un état mixte, décrit par l'opérateur densité $\rho$, on peut écrire la moyenne d'ensemble sous la forme :
\begin{equation}
\bra \hat{A}\ket= \mbox{Tr}(\rho \hat{A}).
\end{equation}
La base des états de Fock n'est pas adaptée à la description des états mixtes.
En effet, plus un état mixte s'éloigne d'une description sous la forme d'un état pur, plus la taille de l'opérateur densité et de sa décomposition sur la base des états de Fock augmentent.
Or dans une expérience d'optique quantique, le manque de connaissances sur l'état initial, les pertes et plus généralement les phénomènes de décohérence vont éloigner les états étudiés d'états purs.
C'est pourquoi des représentations basées sur une description en terme d'états cohérents ont été introduites.
Il s'agit des représentations dites $P$ et $W$ \cite{Meystre:2007p4122}.

\subsection{Représentation $P$}
Contrairement aux états de Fock, les états cohérents forment une base sur-complète de l'espace des états.
Les états cohérents forment une famille génératrice (non libre) et ne sont pas orthogonaux.
Le caractère générateur de  cet ensemble  va permettre de décrire l'opérateur densité comme une somme d'éléments diagonaux : \begin{equation}\label{rho}
\rho = \int P(\alpha)|\alpha\ket\bra \alpha | \mbox{d}^2 \alpha,
\end{equation}
avec $\mbox{d}^2\alpha=\mbox{d Re}(\alpha)\mbox{d Im}(\alpha)$.\\
On a introduit à l'équation \eqref{rho}, la fonction $P(\alpha)$.
On appelle cette fonction la représentation $P$.
Cette fonction contient toutes les informations sur l'état du système.
L'intérêt d'une telle fonction est de pouvoir passer d'un calcul difficile sur des opérateurs quantiques qui ne commutent pas à un calcul simple sur des nombres complexes à l'aide de la méthode qui suit.\\
Dans un premier temps il faut, grâce aux règles de commutation, écrire l'opérateur étudié $\hat {A}$ sous la forme d'une combinaison linéaire de terme en $(\hat{a}^{\dag})^{n}\hata^m$.
Cette écriture, où les opérateurs de création sont à gauche des opérateurs d'annihilation, est appelée l'ordre normal \cite{Glauber:1963p1343}.
Sous cette forme, le calcul de la valeur moyenne de l'opérateur $\hat {A}$ se réduit à un calcul avec des nombres complexes : 
\begin{eqnarray}
\nonumber \bra \hat{A}\ket &=& \text{Tr}(\rho \hat{A})=\sum_n \bra n | \int P(\alpha)|\alpha\ket\bra \alpha | \hat{A} |n\ket \mbox{d}^2 \alpha\\
\nonumber &=& \int   P(\alpha) \sum_n \bra \alpha | \hat{A} |n\ket \bra n|\alpha\ket\mbox{d}^2 \alpha\\
\nonumber &=& \int   P(\alpha)  \bra \alpha | \hat{A} |\alpha\ket\mbox{d}^2 \alpha\\
&=& \int   P(\alpha)\  \alpha^{*n} \alpha^m\ \mbox{d}^2 \alpha\label{tracederhoa}.
\end{eqnarray}
Par exemple, si l'on s'intéresse à l'opérateur nombre, c'est-à-dire le cas $n=1$ et $m=1$ on a :
\begin{equation}\label{normal_aad}
\bra \hat{a}^{\dag}\hata\ket =\int   P(\alpha)\  \alpha^{*} \alpha\ \mbox{d}^2 \alpha=\int   P(\alpha)\  | \alpha |^{2}\ \mbox{d}^2 \alpha
\end{equation}
On retrouve ainsi un résultat de l'optique classique : le nombre moyen de photons est égal à la valeur moyenne du module carré de l'amplitude.\\
$P(\alpha)$ peut être vu comme l'analogue d'une distribution de probabilité pour les valeurs de $\alpha$.
Comme nous l'avons souligné, les états cohérents ne sont pas orthogonaux.
Ainsi dans le cas général, $P(\alpha)$ n'est pas une véritable distribution de probabilité.
Par exemple, pour certains états du champ, $P(\alpha)$ pourra prendre une valeur négative.
Dans ce cas la distribution $P(\alpha)$ ne pourra évidemment pas être interprétée comme une probabilité classique et ce sera la signature d'un état non classique.
On appellera donc $P(\alpha)$ une quasi--probabilité. 

\subsection{Représentation de Wigner}\label{Rwigner}
 	\begin{figure}	
 	\centering
 	\includegraphics[width=14.5cm]{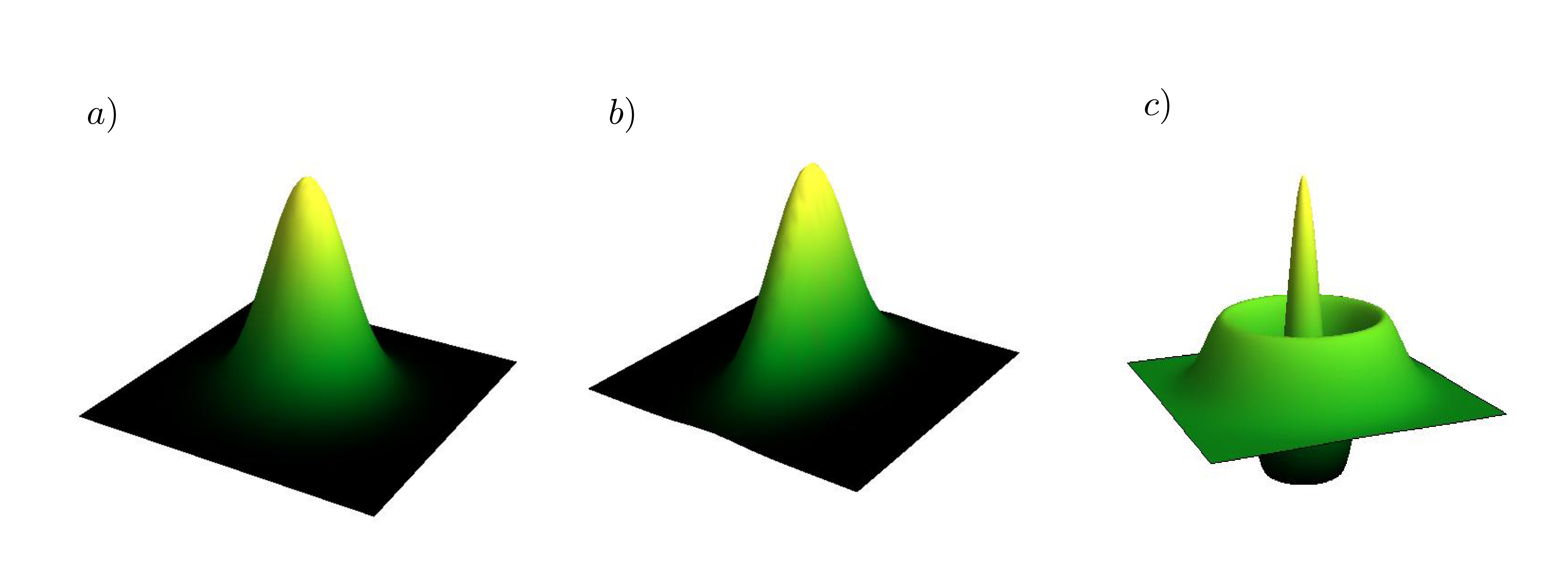}
 	\caption[Représentation de de Wigner]{Représentation de Wigner pour trois états du champs. De gauche à droite, a) un état cohérent, b) un état comprimé et c) un état nombre (n=2)  \cite{OrvilScully:1997p3797}. }		\label{wigner}	
 \end{figure}

Nous venons de le voir, la représentation $P$ est adaptée au calcul des produits d'opérateurs de création et d'annihilation  dans l'ordre normal.
Lorsque l'on étudie les quadratures $\hat{X}$ et $\hat{Y}$ du champ, chaque terme produit $(\hat{a}^{\dag})^{n}\hata^m$ dans l'ordre normal est couplé à un terme similaire dans l'ordre anti-normal  $\hata^m(\hat{a}^{\dag})^{n}$.
Dans ce cas, il est plus simple d'utiliser la représentation de Wigner qui est adaptée à cet ordre dit ``symétrique'' \cite{Tatarskii:1983p4201,Gardiner:1991p4207,FabreHouches}.\\

\noindent Dans un premier temps, nous allons introduire l'opérateur déplacement $\hat D$ pour un champ monomode $\hata$ :
\begin{equation}
\hat D(\eta) =e^{\eta \hatad -\eta^* \hata}.
\end{equation}
En séparant la partie réelle et la partie imaginaire de $\eta=u+i v$, l'opérateur déplacement peut s'écrire à l'aide des quadratures $\hat X$ et $\hat Y$ sous la forme :
\begin{equation}
\hat D(u,v) =e^{i(v \hat X -u \hat Y)}.
\end{equation}
La fonction de Wigner est la transformée de Fourier de la valeur moyenne de l'opérateur déplacement :
\begin{equation}
W(\alpha)=\frac {1}{\pi^2}\int \text{Tr}[\rho \hat D(\eta)]\ \mbox{e}^{\eta \alpha^*-\eta^* \alpha} \mbox{d}^2\eta,
\end{equation}
ou écrit autrement :
\begin{equation}
W(u,v)=\frac {1}{(2\pi)^2}\int \text{Tr}[\rho \hat D(u,v)]\ \mbox{e}^{i(v  X +u  Y)}\mbox{d}u\mbox{d}v.
\end{equation}
Les propriétés de cette fonction sont décrites en détails dans \cite{FabreHouches}.\\
%

\begin{figure}	
 	 	\centering
 	 	\includegraphics[width=5cm]{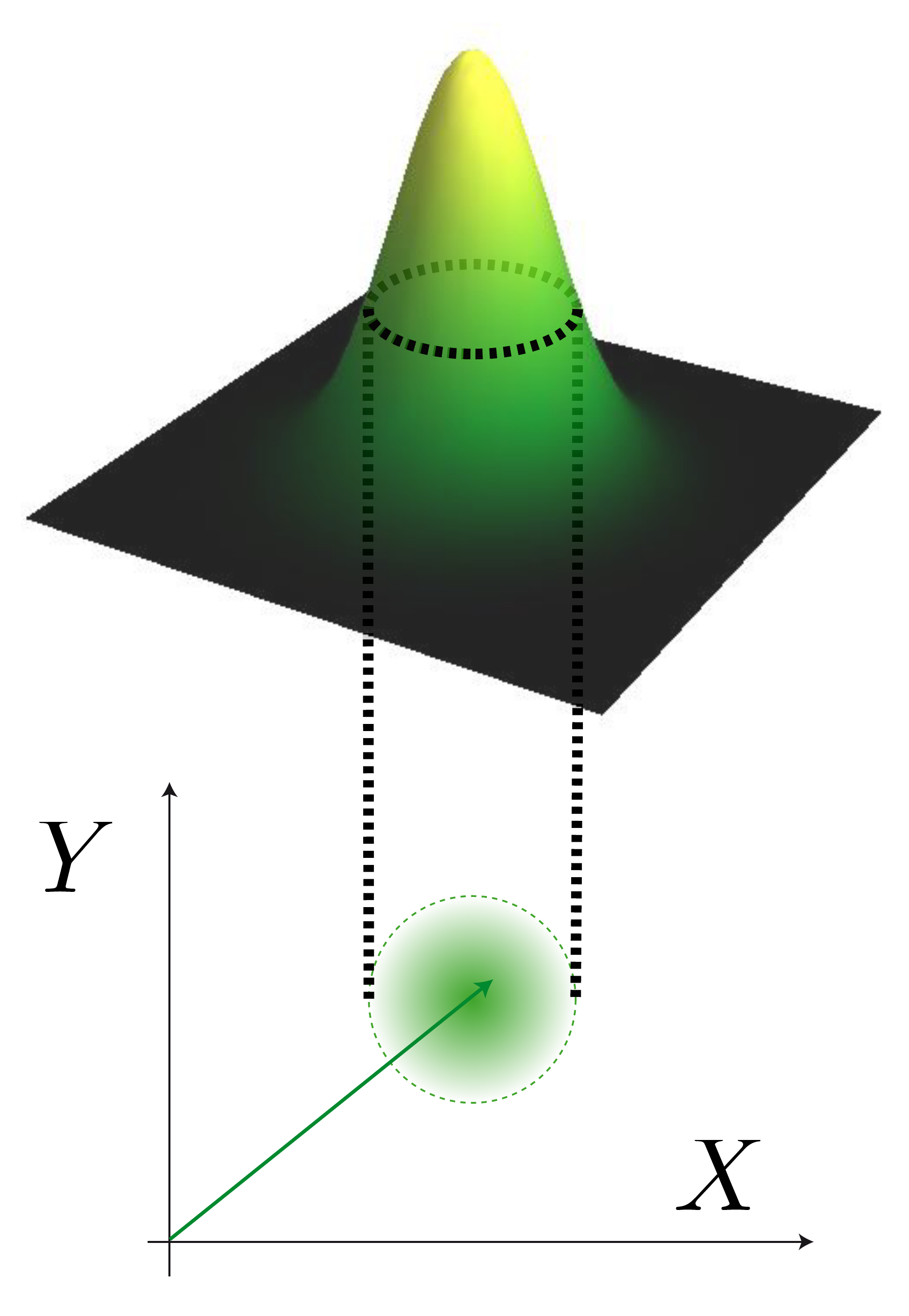}
 	 	\caption[Représentation de de Wigner]{Correspondance entre la représentation de Wigner et  le diagramme de Fresnel pour un état cohérent	}		\label{wig}	
 	 \end{figure}

Jusqu'à présent, nous avions utilisé, pour décrire le champ dans le repère de Fresnel, l'image classique d'un segment pour décrire la valeur moyenne de l'amplitude (sa longueur) et d'un disque pour décrire les fluctuations \ref{fresnelcoherent}.
A l'aide de la représentation de Wigner, on peut donner une définition rigoureuse de ce disque.
Il peut être décrit comme une ligne de niveau de la fonction de Wigner (par exemple 1/e).
La figure \ref{wig} illustre cette définition.
Il est intéressant de noter au sujet de la fonction de Wigner que toute fonction de Wigner positive est une fonction gaussienne (par conséquent associée à un état gaussien).
Elle peut alors être assimilée à une distribution de probabilité classique.
\subsection{Linéarisation des opérateurs d'annihilation}
Les méthodes de résolution des problèmes d'optique quantique sont décrites de manière détaillée dans de nombreuses références, notamment \cite{Gardiner:1991p4207,FWalls:2008p11264} et par conséquent ne seront pas étudiées ici.
Par contre, nous allons justifier le fait de pouvoir linéariser les opérateurs de création et d'annihilation.
En effet, dans les problèmes étudiés dans ce manuscrit, le bruit quantique lié aux fluctuations est bien plus faible que les valeurs moyennes de l'amplitude du champ.
Dans ce cas, nous pouvons utiliser le formalisme introduit dans le domaine de l'optique quantique par \cite{Lugiato:1982p11849} et utilisé notamment dans \cite{Yurke:1984p11396,Reynaud:1989p11268} :
\begin{equation}\label{linearisation}
\hat{a}(t)\simeq \alpha +\delta \hata(t),
\end{equation}
où $\alpha$ est un nombre complexe représentant l'amplitude moyenne du champ et $ \delta \hata(t)$ est un opérateur dépendant du temps représentant les fluctuations de $\hata$.
Pour écrire l'opérateur annihilation sous cette forme, nous avons fait l'hypothèse que la valeur moyenne de $ \delta \hata(t)$ est nulle et que son module est très petit devant le module de $\alpha$.
La seconde hypothèse, nous permettra par la suite de faire une approximation au premier ordre en $ \delta \hata(t)$ dans les grandeurs étudiées.
On peut noter que dans ce cas, les termes non négligés ne contiennent pas de produit d'opérateurs et que par conséquent tous les termes commutent.
\section{Photodétection}
La plupart des mesures en optique quantique se concentrent sur une détection de photons à l'aide de photodétecteurs, typiquement une ou plusieurs photodiodes.
En effet, la détection d'un photon passe, la plupart du temps, par sa conversion en un photo-courant qui sera alors analysé.\\
La lumière visible est une onde électromagnétique oscillante à plusieurs centaines de THz.
Il est clair qu'aucune photodiode ne peut répondre suffisamment vite pour rendre compte de ces oscillations de façon directe.
La mesure la plus simple sera donc une mesure de l'intensité lumineuse c'est-à-dire de l'enveloppe lentement variable du champ.
Évidemment cette description ne contient aucune information sur la phase du champ mais uniquement sur son amplitude.\\
Pour décrire certains phénomènes, il peut être intéressant de disposer d'une mesure sensible à la phase du champ.
Dans ce cas, il sera nécessaire d'utiliser des techniques dites de détection homodyne ou hétérodyne.\\

\subsection{Densité spectrale de bruit}
		On définit l'opérateur photocourant $\hat{i}$ proportionnel à l'opérateur nombre ${\hat{N}_a=\hat{a}^\dag\hat{a}}$.\\
		On définit la fonction d'auto-corrélation pour l'opérateur photocourant, comme la variance à deux temps de celui-ci :
		\begin{equation}
		C_i(t,t')=\langle\hat{i}(t)\hat{i}(t')\rangle-\langle\hat{i}(t)\rangle\langle\hat{i}(t')\rangle.
		\end{equation}
		En linéarisant l'opérateur $\hat{i}$ c'est-à-dire en écrivant $\hat{i}(t)=\langle \hat{i}\rangle +\delta \hat{i}(t)$, on peut montrer aisément que :
		\begin{equation}
		C_i(t,t')=\langle\delta\hat{i}(t)\delta\hat{i}(t')\rangle.
		\end{equation}
		Pour un processus stationnaire, $C_i$ dépend uniquement de la différence temporelle ${\tau=t'-t}$ :
		\begin{equation}
		C_i(\tau)=\langle\delta\hat{i}(t)\delta\hat{i}(t+\tau)\rangle.
			\end{equation}
		D'après le théorème de Wiener-Khintchine, la densité spectrale de bruit $S_i(\omega)$ est donnée par la transformée de Fourier
			de la fonction d'auto-corrélation \cite{Fabre:1997p4333} : 
					\begin{equation}\label{DSP}
					S_i(\omega)=\int_{-\infty}^\infty C_i(\tau)\ e^{i \omega \tau}\ d\tau.
					\end{equation}
Dans le cas particulier où le système de détection peut être modélisé par un filtre passe-bande très fin de bande passante $\delta f$ et de fréquence centrale $\omega_c$, on peut montrer  \cite{Fabre:1997p4333} que la variance du photocourant $\Delta_i^2$ est reliée à la densité spectrale de bruit par :
\begin{equation}
\Delta_i^2=2\ \delta f S_i(\omega_c)
\end{equation}
Or, comme nous le verrons au chapitre \ref{ch2}, un analyseur de spectre se comporte comme tel un filtre.
On peut donc avoir accès expérimentalement à la grandeur $S_i(\omega_c)$ qui pourra être comparée aux prédictions théoriques.

\subsection{Photo-courant}

Le photo-courant est proportionnel au flux d'électrons $F_{el}$  produit par une photodiode en réponse à un flux de photons incident $ F_{ph}$.
Dans le cas idéal, une photodiode a un taux de conversion, que l'on nomme efficacité quantique (\textit{quantum efficiency}) de 1 : chaque photon incident génère un électron.
On introduira le paramètre $\eta$ pour quantifier cette efficacité :
\begin{equation}
F_{el}=\eta F_{ph}.
\end{equation}
Le flux d'électrons possède la même statistique que le flux de photons uniquement dans le cas $\eta=1$.
Une efficacité quantique inférieure à 1 sera donc interprétée comme un terme de pertes dans l'analyse du signal.
Ces pertes, comme celles qui interviennent sur le flux de photons, sont des phénomènes aléatoires et par conséquent font tendre, dans la limite des pertes importantes, toute statistique vers une statistique poissonienne (voir paragraphe \ref{parag_pertes}).

Le flux de photons est donné, de manière générale, par la puissance optique $P$ incidente sur le détecteur :
\begin{equation}
 F_{ph}=\frac{P}{h\nu},
\end{equation}
où $h\nu$ est l'énergie d'un photon.\\
On exprime la valeur moyenne $\overline{i}$ du photo-courant sous la forme :
\begin{equation}
\overline{i}=\eta e \frac{P}{h\nu} 
\end{equation}
Rappelons que le bruit quantique standard ou shot-noise correspond à une statistique poissonienne de photons incidents et donc d'électrons.
Pour un nombre moyen d'électrons $n_e$, on a :
\begin{equation}
\Delta n_e^2= n_e.
\end{equation}
On en déduit la variance du photo-courant $\Delta i(t)^2$ pour un temps $\Delta t$ :
\begin{equation}
\Delta i(t)^2=\frac{e^2\Delta n_e^2}{\Delta t^2}=\frac{e^2 n_e}{\Delta t^2}.
\end{equation} 
Comme on peut écrire $n_e=\overline{i}\Delta t/e$, on a :
 \begin{equation}
 \Delta i(t)^2=\frac{\overline{i} e}{\Delta t}.
 \end{equation}
Le terme $\Delta t$ est relié au processus de détection, il correspond à l'intervalle de temps durant lequel dure la détection.
En termes fréquentiels, il s'agit de la bande passante $B=1/2\Delta t$, la plus étroite de l'ensemble des appareils de mesure.
Typiquement, on peut confondre $B$ avec la bande-passante de résolution (RBW, resolution bandwidth) de l'analyseur de spectre, comme nous le verrons au chapitre \ref{ch2}.
On pourra écrire  :
\begin{equation}
 \Delta i(t)^2=2 \overline{i} e B.
\end{equation}
La variance du photocourant pour un état cohérent dépend donc uniquement de sa valeur moyenne ainsi que de la bande passante de l'appareil de mesure.
Ce bruit sera étudié et mesuré expérimentalement dans le chapitre \ref{ch2}.
\subsection{Détection d'intensité}
	\begin{figure}	
		 	\centering
		 	\includegraphics[width=11.5cm]{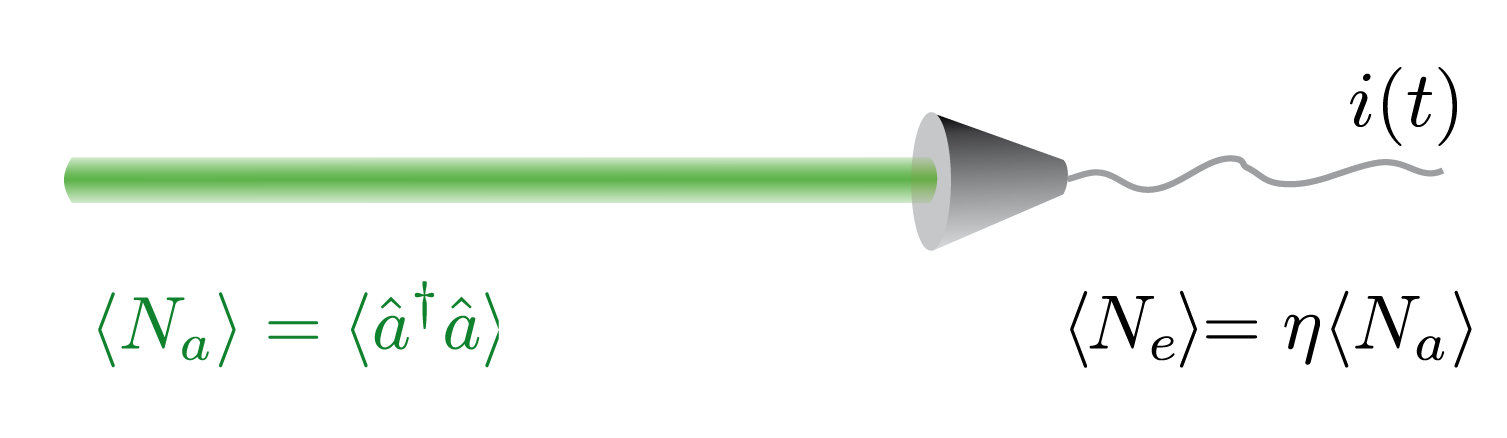}
		 	\caption[Photodétection en intensité]{Photodétection en intensité : flux de photons arrivant sur une photodiode. Le photocourant $i(t)$ est proportionnel au nombre de photons incidents par unité de temps}		\label{detection}	
		 	 \end{figure}
La détection d'intensité est le cas le plus simple de photodétection.
Dans cette situation, on mesure le flux de photons qui arrivent sur un photodétecteur et on l'analyser pour connaitre la valeur moyenne du nombre de photons ainsi que les fluctuations associées.
On étudie donc l'opérateur nombre : $\hat{N}_a=\hatad\hata$.\\
En linéarisant les opérateurs à l'aide de la relation \eqref{linearisation} on peut écrire l'opérateur nombre sous la forme :
\begin{equation}
\hat{N}_a=\bra \hata^\dag\ket\bra \hata\ket+\bra \hata^\dag\ket\delta \hat{a} +\bra \hata\ket \delta \hat{a}^\dag +\delta \hat{a}\delta \hata^\dag.
\end{equation}
Au premier ordre en $\delta \hat{a}$ et en notant les valeurs moyennes $\bra \hata\ket$ et $\bra \hata^\dag\ket$ respectivement $|\alpha| e^{i\phi}$ et $|\alpha| e^{-i\phi}$ on a :
\begin{deqarr}\arrlabel{linearisation2}
\bra \hat{N}_a \ket &=&|\alpha|^2\\
\delta\hat{N}_a &=& |\alpha| e^{i\phi} \delta \hata^\dag +|\alpha| e^{-i\phi} \delta\hata= |\alpha|\ \delta\hat{X}_a^\phi.\label{deltaNa}
\end{deqarr}
Les fluctuations sur l'opérateur nombre sont directement reliées aux fluctuations de la quadrature $\hat{X}_a^\phi$ où $\phi$ est la valeur moyenne de la phase du champ $\hata$.
Le taux de compression est alors donné d'après la relation \eqref{ajjt} par :
\begin{equation}
S=\frac{\langle \delta\hat{N}_a \delta\hat{N}_a\rangle}{\langle \hat{N}_a\rangle}
\end{equation}

\subsection{Détection balancée}
\begin{figure}	
		 	\centering
		 	\includegraphics[width=15cm]{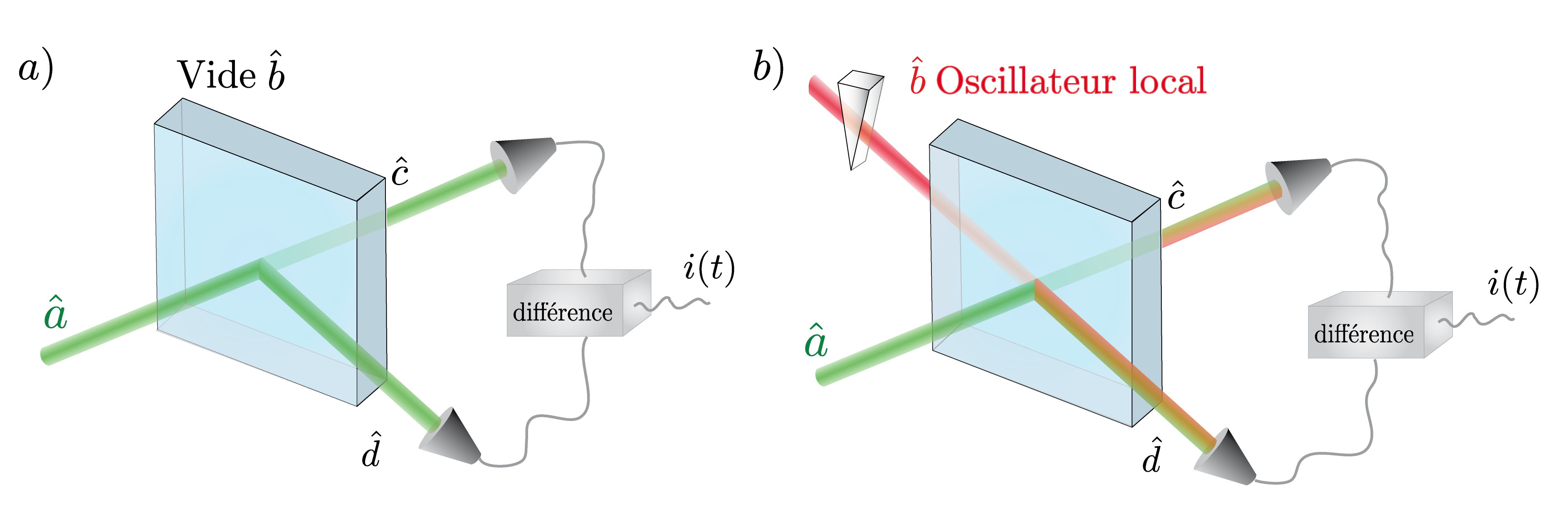}
		 	\caption[Détection homodyne]{Schémas de détection homodyne. Sur la figure a) le vide est homodyné par le champ $\hata$, sur la figure b) le champ $\hata$ est homodyné par l'oscillateur local. }		\label{homodyne}	
		 	 \end{figure}
La détection balancée (voir figure \ref{homodyne}.a) est utilisée pour s'affranchir du bruit classique, c'est-à-dire du bruit technique, souvent présent expérimentalement.
Dans ce type de montage, on sépare sur une lame semi--réfléchissante le faisceau que l'on souhaite étudier et on observe la somme et la différence d'intensité à l'aide de deux photodiodes dans les deux voies de sortie de la lame.
D'un coté, sur la sortie ``somme'', on mesure le bruit total du faisceau.
De l'autre, sur la sortie ``différence'', on ne mesure que le bruit quantique.
En effet le bruit technique du faisceau est corrélé entre les deux voies de sortie de la lame séparatrice et l'opération de différence (dans le cas d'un montage parfaitement balancé) va donc supprimer la contribution de ce bruit.\\
Pour un état cohérent sans excès de bruit technique, on mesure donc le même niveau de bruit sur les deux voies.
On peut ainsi mesurer le bruit quantique standard sur la voie différence et détecter tout excès de bruit technique sur la voie somme.\\

\noindent Ce montage peut être considéré comme une détection homodyne pour le vide.
En effet, le champ du vide (noté $\hat{b}$) qui intervient dans le processus de photodétection par la deuxième face de la lame est ``homodyné'' par un champ relativement intense (noté $\hat{a}$).
On obtient donc après la lame sur les deux voies $\hat c$ et $\hat d$ :
\begin{deqarr}
 \hat c&=& \frac {1}{\sqrt 2}(\bra \hata \ket+\delta \hat{a} +\delta \hat{b})\\
  \hat d&=&\frac {1}{\sqrt 2}(\bra \hata \ket+\delta \hat{a} -\delta \hat{b}).
\end{deqarr}
A l'aide de cette écriture linéarisée, nous allons pouvoir calculer les opérateurs différence $\hat N_-$ et somme d'intensité $\hat N_+$ en sortie de la lame.

\subsubsection{Différence d'intensité}
\label{diffint}
En utilisant les mêmes notations que dans l'équation \eqref{linearisation2}, on peut écrire l'opérateur différence d'intensité sous la forme :
\begin{eqnarray}
\nonumber \hat N_-&=&\frac{1}{\sqrt 2}(|\alpha| e^{-i\phi_a}+\delta \hat{a}^\dag +\delta \hat{b}^\dag) (|\alpha |e^{i\phi_a}+\delta \hat{a} +\delta \hat{b})-  \frac{1}{\sqrt 2}(|\alpha| e^{-i\phi_a} +\delta \hat{a}^\dag -\delta \hat{b}^\dag) (|\alpha |e^{i\phi_a}+\delta \hat{a} -\delta \hat{b})\\
\nonumber &=&|\alpha |e^{i\phi_a} \delta \hat{b}^\dag +|\alpha |e^{-i\phi_a} \delta \hat{b}= |\alpha|\ \delta \hat{X}_{b}^{\phi_a} .
\end{eqnarray}
Le vide étant un état cohérent, toutes les quadratures sont identiques et la phase ne joue aucun rôle.
La différence d'intensité mesurée dans une détection balancée est donc égal au bruit quantique standard du vide multiplié par l'amplitude du champ incident.\\
Comme nous l'avons dit, cette méthode permet donc de calibrer la valeur de la limite quantique standard pour une intensité donnée.

\subsubsection{Somme d'intensité}
Si désormais on s'intéresse à la somme de l'intensité mesurée dans les deux bras on obtient le bruit du champ $\hata$.
En effet, on peut écrire au premier ordre :
\begin{eqnarray}\label{balancee}
\nonumber \hat N_+&=&\frac{1}{\sqrt 2}(|\alpha| e^{-i\phi_a}+\delta \hat{a}^\dag +\delta \hat{b}^\dag) (|\alpha| e^{i\phi_a}+\delta \hat{a} +\delta \hat{b})+  \frac{1}{\sqrt 2}(|\alpha| e^{-i\phi_a}+\delta \hat{a}^\dag -\delta \hat{b}^\dag) (|\alpha |e^{i\phi_a}+\delta \hat{a} -\delta \hat{b})\\
\nonumber &=&|\alpha|^2+|\alpha |e^{i\phi_a} \delta \hat{a}^\dag +|\alpha |e^{-i\phi_a} \delta \hat{a}=|\alpha|^2+|\alpha |\delta \hat{X}_{a}^{\phi_a}.
\end{eqnarray}
Ainsi la variance de $\hat N+$ s'exprime par :
\begin{equation}
(\Delta \hat N_+)^2=|\alpha |^2\bra (\delta \hat{X}_{a}^{\phi_a})^2\ket=|\alpha |^2(\Delta \hat{X}_{a}^{\phi_a})^2,
\end{equation}
Lorsque l'on mesure les fluctuations de $\hat{N}_+$, on obtient un signal proportionnel à la valeur moyenne du champ et aux fluctuations de la quadrature $ \hat{X}_{a}^{\phi_a}$.
Dans ce cas, la phase est fixée par la valeur moyenne de la phase du champ $\phi_a$.
On mesure donc directement par cette méthode la quadrature intensité du champ $\hata$.

\subsection{Détection homodyne}
\begin{figure}	
\centering
\includegraphics[width=8cm]{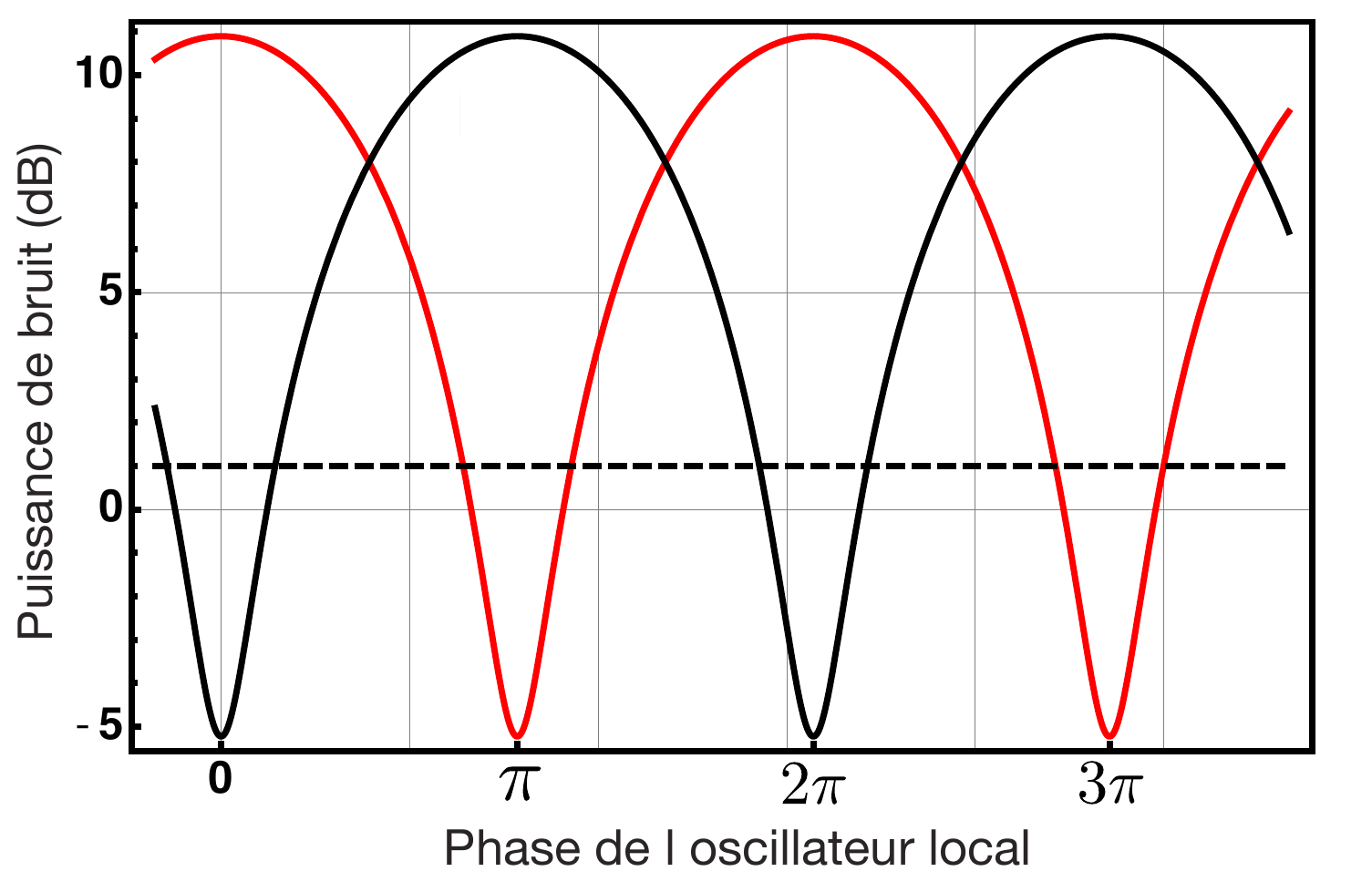}
\caption[Puissance de bruit détectée dans une détection homodyne]{Puissance de bruit par rapport à la limite quantique standard détectée dans une détection homodyne pour un état comprimé. La courbe pointillée représente le bruit quantique standard (vide en entrée de la détection homodyne), les courbes rouges et noires sont le bruit sur les deux quadratures $ \hat{X}_{a}^\phi$ et $ \hat{Y}_{a}^\phi$ en fonction de la phase de l'oscillateur local. On notera que lorsque l'on a une réduction du bruit sous le bruit quantique standard (ici -5dB), on observe simultanément un excès de bruit sur la quadrature conjuguée (ici +11dB). Ce n'est donc pas un état minimal. }		\label{olocal}	
\end{figure}

Pour connaitre les fluctuations d'un champ non vide noté (noté $\hat{a}$), on doit réaliser un montage de détection homodyne avec un oscillateur local intense (noté $E_{LO}$).
On peut alors écrire sous forme linéarisée ces deux champs :  $\hat{a}=\alpha+\delta \hat{a}$ et $E_{LO}=|E_{LO}|\ e^{i\phi}+\delta E_{LO}$.
Le montage de détection homodyne est représenté sur la figure \ref{homodyne}.b.
La phase du champ intense pourra être modifiée, ce qui permet d'explorer toutes les quadratures du champ $\hat{a}$ (on choisit par exemple le champ $\hat{a}$ comme référence de phase, soit $\alpha$ réel).
On se place alors dans les conditions telles que l'oscillateur soit suffisamment intense et que ses fluctuations ne soient pas grandes devant la valeur moyenne du champ $\hata$ : $\alpha \delta E_{LO} \ll E_{LO}\delta \hata$.\\
On obtient alors :
\begin{eqnarray}
\nonumber\hat{ N}_-&=&\frac{1}{\sqrt 2}(\alpha +\delta \hat{a}^\dag +|E_{LO}|\ e^{-i\phi}+\delta E_{LO}) (\alpha +\delta \hat{a} +|E_{LO}|\ e^{i\phi}+\delta E_{LO})- \\
 \nonumber &&\frac{1}{\sqrt 2}(\alpha +\delta \hat{a}^\dag -|E_{LO}|\ e^{-i\phi}-\delta E_{LO}) (\alpha +\delta \hat{a} -|E_{LO}|\ e^{i\phi}-\delta E_{LO})\\
 &\simeq&|E_{LO}|(\alpha\  e^{i\phi}+\alpha \ e^{-i\phi})+| E_{LO}|(\delta \hat{a}^\dag \  e^{i\phi}+\delta \hat{a}\ e^{-i\phi})\simeq|E_{LO}|\bra \hat{X}_{a}^\phi \ket + |E_{LO}|\ \delta\hat{X}_{a}^\phi.\hspace{1cm}
\end{eqnarray}
On voit donc que le signal de la détection homodyne peut se décomposer en deux termes :
d'une part la valeur moyenne proportionnelle à la valeur moyenne de la quadrature $ \hat{X}_{a}^\phi$ et d'autre part les fluctuations qui sont proportionnelles aux fluctuations de cette même quadrature. 
En pratique c'est la méthode qui sera utilisée pour mesurer le niveau de bruit d'une quadrature arbitraire.
Expérimentalement, on comparera les fluctuations de $\hata$ aux fluctuations du vide.
Le vide étant un état cohérent, son aire de fluctuations est circulaire et aucune direction $\phi$ n'est privilégiée.
Une mesure des fluctuations du vide par détection homodyne donnera donc un niveau de bruit identique pour toutes les phases de l'oscillateur local.
Par contre,  lorsque l'on observe une compression sur une quadrature $ \hat{X}_{a}^\phi$, on pourra vérifier la présence d'un excès de bruit sur la quadrature conjuguée $ \hat{Y}_{a}^\phi$ (voir figure \ref{olocal}).
On notera que lorsque l'on se place dans les conditions $\alpha \delta E_{LO} \ll E_{LO}\delta \hata$, alors le bruit sur l'oscillateur local n'entre pas en compte lors de la mesure de bruit par détection homodyne.

\subsection{Effet des pertes}\label{parag_pertes}

\begin{figure}	[h]
\centering
\includegraphics[width=8cm]{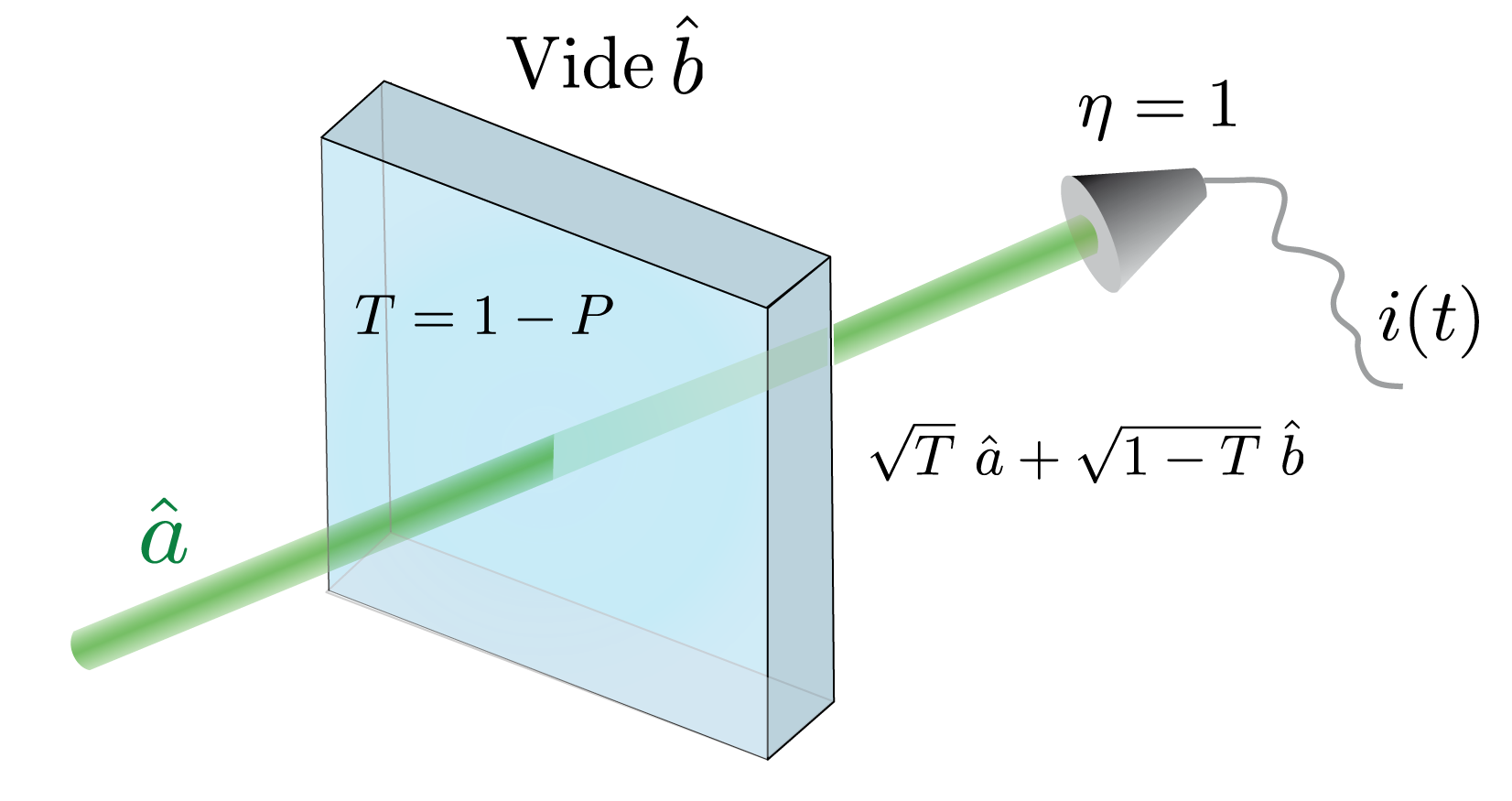}
\caption[Modélisation des pertes dans une mesure d'optique quantique]{Modélisation des pertes dans une mesure d'optique quantique. L'ensemble des pertes $P$ de la chaine de détection est remplacé par une lame séparatrice de transmission $T=1-P$ et un détecteur d'efficacité quantique $\eta=1$. }		\label{pertes}	
 \end{figure}

\clearpage
\begin{figure}	[h]
\centering
\includegraphics[width=8cm]{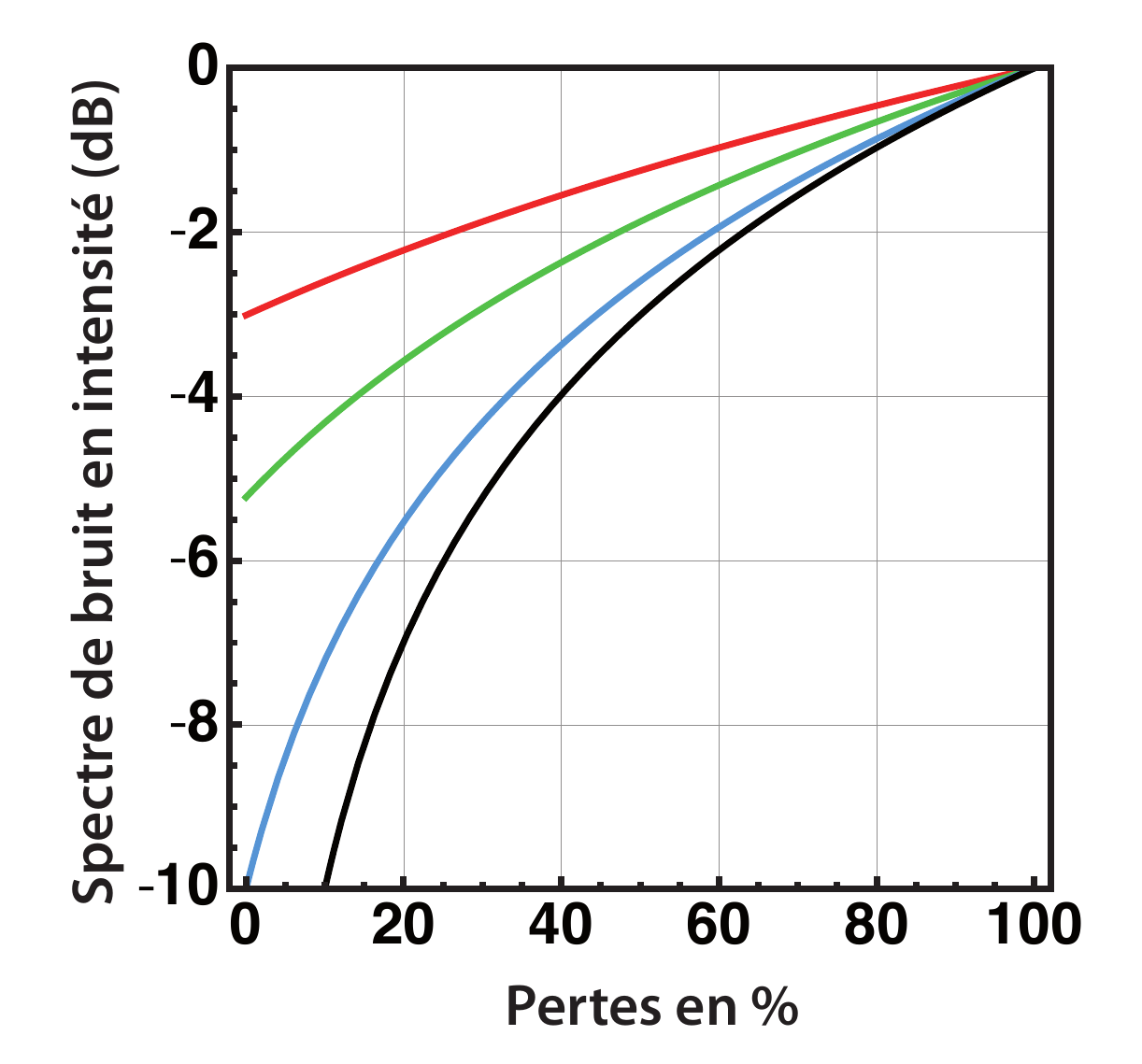}
\caption[Effet des pertes dans une mesure d'optique quantique]{Effet des pertes sur la détection d'un état comprimé en intensité de -3 dB (rouge), -5.2	dB (vert) et -10 dB (bleu) sous le bruit quantique standard et pour un état infiniment comprimé (noir).}		\label{pertes2}	
\end{figure}
\noindent Lorsque l'on mesure les fluctuations d'un champ électromagnétique, une attention toute particulière doit être portée aux pertes introduites par le système de détection.
Ces pertes ont principalement deux origines : les pertes optiques sur la propagation du faisceau et l'efficacité quantique inférieure à 1 de la photodiode.
Ces deux sources peuvent être traitées de la même façon, c'est-à-dire modélisées par une lame séparatrice ayant un coefficient de transmission $T=1-P$, où $P$ quantifie les pertes de la chaine de détection \cite{Bachor:2004p4500}.\\
Dans ce cas on a la relation entrée--sortie suivante pour la lame :
\begin{equation}
\hata_{out}=\sqrt{T}\ \hat{a}_{in}+\sqrt{1-T}\ \hat{b}
\end{equation}
Le champ $\hat{b}$ étant le vide, sa valeur moyenne est nulle. On va donc écrire $\hata_{out}$ de manière linéarisée :
\begin{equation}
\bra{\hata_{out}}\ket = \sqrt{T}\ |\alpha|\ ,\ \delta \hata_{out}=\sqrt{T}\ \delta\hat{a}_{in}+\sqrt{1-T}\ \delta\hat{b}
\end{equation}
On peut donc donner à l'aide de la relation (\ref{balancee}) la valeur moyenne et les fluctuations du champ $\hata$ en présence de pertes $P$ :
\begin{equation}
\bra{\hat{N}_{+,out}}\ket =(1-P) |\alpha|^2\ , \ \delta\hat{N}_{+,out} = \sqrt{1-P}\ \delta \hat{X}_{a,in}+\sqrt{P}\ \delta \hat{X}_{b,in}
\end{equation}
Ainsi, on obtient le spectre de bruit en intensité après la lame en fonction du spectre entrant sous la forme :
\begin{equation}
S\hat{N}_{+,out}=(1-P)\ S\hat{N}_{+,in}+P
\end{equation}
Pour un état comprimé en intensité, l'effet des pertes dépend du niveau de compression initial.
 Par exemple, 10 \% de pertes pour un état infiniment comprimé le ramènera à une compression  de -10 dB, tandis que pour un état de -3 dB les mêmes pertes le ramèneront à une compression de -2.6 dB (voir figure \ref{pertes2}).

\section{Corrélations quantiques en variables continues. }
\begin{figure}	[h]
		 	\centering
		 	\includegraphics[width=12cm]{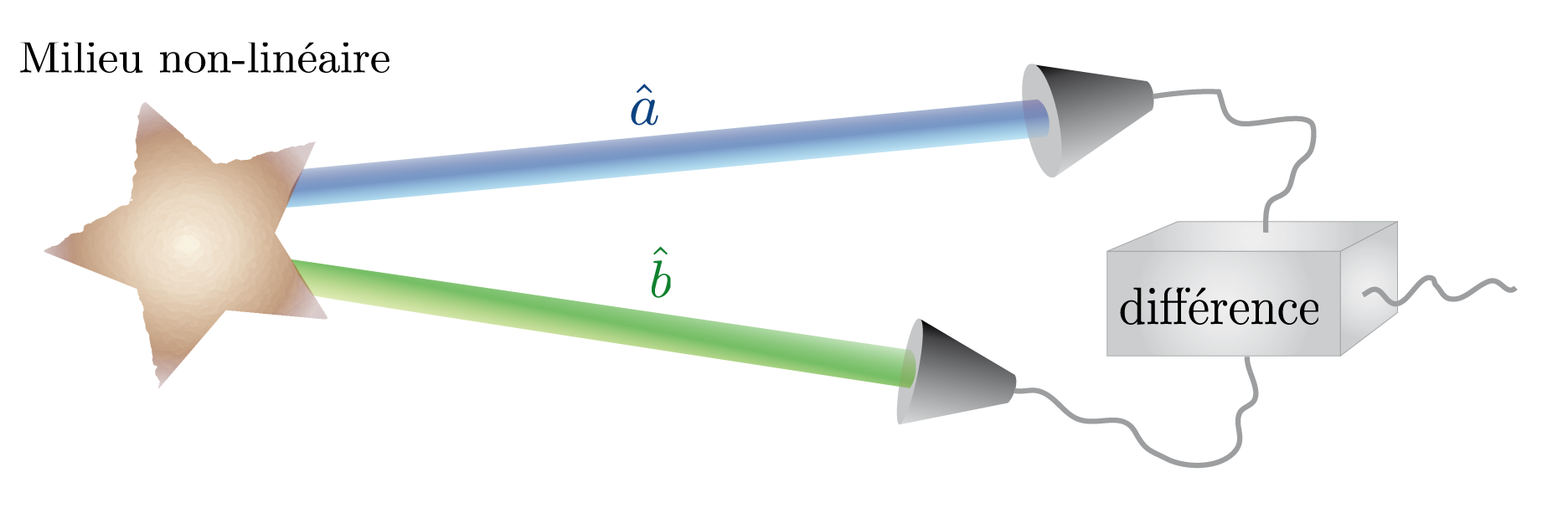}
		 	\caption[Photodétection en intensité pour la mesure de corrélations.]{Photodétection en intensité pour la mesure de corrélations en intensité : flux de photons arrivant sur une photodiode. Le photocourant $i(t)$ est proportionnel au nombre de photons incidents par unité de temps. le signal mesuré est la différence des deux photocourants.}		\label{2modesfig}	
		 	 \end{figure}
\subsection{Mesure de corrélations en variables continues} \label{corel_ch1}
Dans ce manuscrit, on s'intéresse aux corrélations entre deux faisceaux  $\hat{a}$ et  $\hat{b}$ (voir figure \ref{2modesfig}).
On va donc introduire les mesures de corrélations en variables continues et définir les états comprimés à deux modes du champ.
En reprenant les notations précédentes pour $\hata$ et en notant le champ $\hat{b}=|\beta|\  e^{i\psi}+\delta \hat{b}$, on a :
\begin{eqnarray}
\nonumber \hat N_-&=&(|\alpha|\  e^{-i\phi}+\delta \hat{a}^\dag)(|\alpha|\  e^{i\phi}+\delta \hat{a})-(|\beta|\  e^{-i\psi}+\delta \hat{b}^\dag)(|\beta|\  e^{i\psi}+\delta \hat{b})\\
&=&|\alpha|^2-|\beta|^2+(|\alpha|\ \delta\hat{X}_{a}^\phi-| \beta|\ \delta\hat{X}_{b}^\psi).
\end{eqnarray}
La valeur moyenne $\bra N_-\ket = |\alpha|^2-|\beta|^2$ nous renseigne sur la différence d'intensité moyenne entre les deux faisceaux.
Dans le cas balancé, c'est-à-dire si cette différence est nulle, on pourra écrire les fluctuations de la différence sous la forme :
\begin{equation}\label{deltaN-}
\delta N_- = |\alpha| (\delta\hat{X}_{a}^\phi-  \delta\hat{X}_{b}^\psi)
\end{equation}
Dans ce cas on peut faire l'analogie avec les fluctuations de l'état à un mode, décrites par l'équation~(\ref{balancee}) en remplaçant la quadrature intensité du champ par la différence des deux quadratures intensité associées à chacun des deux modes.
Le cas où $\bra N_-\ket \neq 0$ nécessite une étude plus détaillée.
On note en particulier que, dans ce cas, les bruits quantiques individuels des deux modes jouent un rôle sur la mesure des corrélations. 
On pourra consulter par exemple \cite{Treps:2004p1389}.
\subsection{Matrice de covariance}
La relation (\ref{DSP}) relie la densité spectrale de puissance à la transformée de Fourier de la fonction d'auto-corrélation.
Pour déterminer les spectres de bruit, il est donc nécessaire de calculer les moments d'ordre deux des opérateurs différence et somme d'intensité que l'on vient d'introduire dans différentes configurations de photodétection.
Pour cela on définit la matrice de covariance $[V(\omega)]$ pour un opérateur $\hat a$ de la manière suivante \cite{Hilico:1992p2633,Dantan05}:
\begin{equation}
[V(\omega)]2\pi\ \delta(\omega+\omega')=\langle|\hat{A}(\omega)][\hat{A}^\dag(\omega')|\rangle
\end{equation}
où 
\begin{equation}
|\hat{A}(\omega)]=\left |
\begin{array}{c}
\hat{a}(\omega)\\
 \hat{a}^\dag(\omega)
\end{array}\right]
\text{ et }
[\hat{A}^\dag(\omega')|=
\left[
\hat{a}^\dag(\omega')\ \hat{a}(\omega') \right |.
\end{equation}
De manière générale on notera $|\ ]$ les vecteurs colonnes, $[\ |$ les vecteurs lignes et $[\ ]$ les matrices.
En linéarisant l'opérateur $\hat{a}$ de façon à faire apparaitre la valeur moyenne que l'on note $\langle\hat{a} \rangle=\alpha$ et les fluctuations (opérateur $\delta\hat{a}(t)$), on peut écrire :

\begin{deqarr}\arrlabel{delta1}
\langle \delta \hat{a} (t) \delta \hat{a}(t')\rangle&=&\langle  \hat{a} (t) \hat{a}(t')\rangle-\alpha^2\\
\langle \delta \hat{a}^\dag (t) \delta \hat{a}^\dag(t')\rangle&=&\langle  \hat{a}^\dag (t) \hat{a}^\dag(t')\rangle-\alpha^{*2}\\
\langle \delta \hat{a} (t) \delta \hat{a}^\dag(t')\rangle&=&\langle  \hat{a} (t) \hat{a}^\dag(t')\rangle-|\alpha|^2\\
\langle \delta \hat{a}^\dag (t) \delta \hat{a}(t')\rangle&=&\langle  \hat{a}^\dag (t) \hat{a}(t')\rangle-|\alpha|^2.
\end{deqarr}
Nous avons vu à l'équation \eqref{normal_aad} que l'on pouvait calculer les valeurs moyennes des moments d'ordre 2 des opérateurs $\hata$ et $\hatad$ en les remplaçant par des nombres complexes lorsque ceux-ci sont exprimés dans l'ordre normal (c'est-à-dire dans la représentation P).
Après avoir passé l'équation (\ref{delta1}c) dans l'ordre normal à l'aide de la relation de commutation $[ \hat{a} (t), \hat{a}^\dag (t')]=\delta(t-t')$, on obtient à l'aide de \eqref{normal_aad} les relations suivantes  
\begin{deqarr}\arrlabel{delta2}
\langle \delta \hat{a} (t) \delta \hat{a}(t')\rangle &=&0\\
\langle \delta \hat{a}^\dag (t) \delta \hat{a}^\dag(t')\rangle&=&0\\
\langle \delta \hat{a} (t) \delta \hat{a}^\dag(t')\rangle &=&\delta(t-t')\\
\langle \delta \hat{a}^\dag (t) \delta \hat{a}(t')\rangle &=&0.
\end{deqarr}
On peut donc obtenir par transformée de Fourier sur $t$ et sur $t'$, les valeurs moyennes dans le domaine fréquentiel. 
On a, dans le cas d'un état cohérent \cite{Hilico:1992p2633} :
	\begin{equation}\label{delta_etat_coherent}
[V(\omega)]=\left[
\begin{array}{cc}
1 & 0\\ 
0 & 0
\end{array} \right].
\end{equation}
Cette forme de la matrice de covariance est utile pour les calculs d'optique quantique faisant intervenir l'équation de Fokker Planck \cite{Gardiner:1985p9724,Gardiner:1985p3576}.
Il a été démontré  pour l'oscillateur paramétrique dans \cite{Reynaud:1989p11268,Fabre:1990p2839}, qu'il était équivalent d'utiliser une méthode dite ``semi-classique'' pour calculer les fluctuations du champ dans ce système.
Dans cette méthode ce sont des quantités complexes qui sont employées et non plus des opérateurs.
Il est alors nécessaire d'utiliser des grandeurs symétrisées\footnote{Dans l'ordre symétrique on a la relation : $\langle \delta \hat{a} (t) \delta \hat{a}^\dag(t')\rangle_\mathcal{S}=\frac 12 \langle \delta \hat{a} (t) \delta \hat{a}^\dag(t')\rangle+\frac 12\langle \delta \hat{a}^\dag (t) \delta \hat{a}(t')\rangle$.} que l'on indiquera par l'indice $\mathcal{S}$ .
La matrice de covariance dans l'ordre symétrique s'écrit donc :
	\begin{equation}\label{delta_etat_coherent_sym}
[V_\mathcal{S}(\omega)]=\left[
\begin{array}{cc}
\frac 12 & 0\\
0 & \frac 12 
\end{array} \right].
\end{equation}
Les résultats physiques étant indépendants de l'ordre choisi, nous utiliserons dans ce manuscrit de préférence l'ordre symétrique et donc la matrice de covariance donnée par la relation \eqref{delta_etat_coherent_sym}.

\section{Interaction lumière-matière. }
Pour rendre compte des phénomènes d'interaction entre la lumière et la matière étudié dans ce manuscrit, il est nécessaire d'utiliser le formalisme de la mécanique quantique, aussi bien pour les atomes que pour le champ.
Le début de ce chapitre nous a permis d'introduire ce formalisme pour le champ électromagnétique.
Pour une description quantique des atomes, nous suivons dans ce chapitre l'approche présentée par exemple dans \cite{JFoot:2005p6575}

\subsection{Hamiltonien de Jaynes Cummings}
Le Hamiltonien de Jaynes Cummings permet de décrire les interactions lumière--matière entre des systèmes dont les niveaux d'énergie et les modes du champ électromagnétique sont quantifiés \cite{OrvilScully:1997p3797}.\\
Nous allons rappeler ici les points essentiels de ce modèle dans le cas d'un atome à deux niveaux $|g\ket$ (fondamental) et $ |e\ket$ (excité d'énergie $\hbar \omega_0$ ) couplé avec un seul mode du champ $\hata$ de fréquence $\omega_L$.
Le Hamiltonien total du système $\hat{\mathcal{H}}$ se décompose en trois termes :
\begin{equation}
\hat{\mathcal{H}}=\hat{\mathcal{H}_{at}}+\hat{\mathcal{H}_{ch}}+\hat{\mathcal{H}_{int}},
\end{equation}
où l'on a introduit le Hamiltonien des atomes seuls 
\begin{deqn}
\hat{\mathcal{H}_{at}}=\hbar \omega_0 |e\ket\bra e|	 ,
\end{deqn} 
le Hamiltonien du champ seul 
\begin{ddeqn}
\hat{\mathcal{H}_{ch}}=\hbar \omega_L (\hatad \hata +\frac 12),
\end{ddeqn} 
et le Hamiltonien d'interaction  
\begin{ddeqn}
\hat{\mathcal{H}_{int}}=-\vec{D}\cdot\vec{E}.
\end{ddeqn}
Le Hamiltonien d'interaction écrit ainsi correspond au terme le plus bas des termes d'interaction à savoir le terme dipolaire électrique où $\vec{D}$ est le dipôle atomique et $\vec{E}$ le champ électrique.
On peut l'écrire dans le cas d'un atome à deux niveaux couplé avec un mode du champ :
\begin{ddeqn}
\hat{\mathcal{H}_{int}}=i \hbar g \left(\hatad |g\ket\bra e| + \hata |e\ket\bra g| \right).
\end{ddeqn}
avec la constante de couplage $g= \frac{\mu \mathcal{E}}{\hbar}$ où $\mu$ est le moment de dipôle de la transition et $\mathcal{E}$ le champ électrique d'un photon.

\subsection{Représentations de Schrödinger et de Heisenberg}
De manière générale, la mécanique quantique utilise deux types d'objets mathématiques pour décrire les systèmes étudiés.
Il s'agit des vecteurs d'états d'une part et des observables d'autre part.
Les premiers étant des vecteurs dans un espace de Hilbert tandis que les seconds sont des opérateurs hermitiens agissant sur les éléments (vecteurs d'états) de cet espace.\\
Deux représentations peuvent alors être adoptées pour décrire l'évolution du système :
\begin{itemize}
\item La représentation de Schrödinger qui consiste à considérer indépendants du temps et donc invariants temporellement les observables, et de décrire l'évolution du système par l'évolution des vecteurs d'état $|\Psi(t)\ket$. Cette évolution est alors régie par l'équation de Schrödinger que l'on peut écrire pour un Hamiltonien $ \hat{\mathcal{H}}(t)$ :
\begin{equation}\label{eq-Sch}
i\hbar\frac{d}{dt}|\Psi(t)\ket=\hat{\mathcal{H}}(t)\ |\Psi(t)\ket. 
\end{equation}
\item Dans la représentation de Heisenberg, on décrit le système par un vecteur d'état indépendant du temps et des observables qui évoluent temporellement. Cette représentation est très souvent adoptée en optique quantique du fait de la grande similarité que l'on observe entre les équations d'évolution des opérateurs champ et les équations de Maxwell de l'électromagnétisme classique. Cette évolution est régie par l'équation de Heisenberg qui donne l'évolution temporelle des opérateurs :
\end{itemize}
%
\begin{equation}\label{eq-heisenberg}
\frac{d}{dt}\hat{A}(t)=\frac{1}{i\hbar }[\hat{A}(t),\hat{\mathcal{H}}].
\end{equation} 
\subsection{Equations de Heisenberg Langevin}
Comme le système qui nous intéresse n'est pas un système isolé mais un système quantique couplé à l'environnement (réservoir), on observe des phénomènes de relaxation du système vers l'extérieur.
Un des effets du réservoir est donc de produire des événements aléatoires sur le système et ainsi y introduire des fluctuations.
Ce phénomène est une des facettes du phénomène de  ``fluctuation--dissipation''  \cite{CohenTannoudji:2001p6627} et peut être décrit par les équations d'Heisenberg--Langevin que nous présentons ici à partir de la référence \cite{Davidovich:1996p1958}.\\

En mécanique classique, les équations de Langevin ont été introduites pour décrire une force fluctuante de valeur moyenne nulle de même type que celle que l'on retrouve dans le cas du mouvement Brownien.
Prenons l'exemple d'une particule de masse $m$ soumis à une force de frottement fluide $-\gamma\ \vec{v}$, son mouvement sera décrit par :
\begin{equation}\label{Langevinclassique}
m\frac{d\vec{v}}{dt}=m\vec{g}-\gamma\ \vec{v} +\vec{F}_L(t).
\end{equation}
 L'interaction de la particule avec le réservoir (le liquide qui l'entoure) est la résultante de deux contributions.
 D'une part la force moyenne qui s'applique sur le système (la force de friction) et d'autre part une force $\vec{F}_L(t)$ de valeur moyenne nulle qui va induire des fluctuations sur le système et que l'on appellera force de Langevin.\\
Cette force aléatoire est caractérisé par une ``mémoire'' qui tend vers zéro\footnote{C'est l'approximation de mémoire courte ou de Markov \cite{LeBellac:2007p11866}.} et par le coefficient de diffusion $D$ :
 \begin{deqarr}
\overline{\vec{F}_L(t)}&=&0,\\
\overline{\vec{F}_L(t)\vec{F}_L(t')}&= &2D \ \delta(t-t').
 \end{deqarr}
 Dans le cas d'un système quantique, l'équation de Heisenberg permet d'avoir accès à l'évolution des opérateurs du système.
 De manière générale la connaissance du Hamiltonien du système permet, dans le formalisme de Heisenberg, d'écrire les équations de Bloch optiques, et ainsi connaitre l'évolution de la valeur moyenne d'une observable quelconque $\hat{A}$.
  Il s'agit du théorème d'Ehrenfest.
 En présence de relaxation, que l'on quantifiera par un taux $\Gamma$, l'équation d'évolution de la valeur moyenne de $\hat{A}$ s'écrit sous la forme :
 \begin{equation}
\frac{d}{dt}\bra \hat{A}\ket=\frac{1}{  i\hbar}\bra[\hat{A},\hat{\mathcal{H}}]\ket-\Gamma\bra \hat{A}\ket,
 \end{equation}
où  $\hat{\mathcal{H}}$ est le Hamiltonien du système.\\

\noindent Si on s'intéresse maintenant aux fluctuations d'une observable quelconque que l'on note $\delta\hat{A}$, le théorème d'Ehrenfest ne permet pas de déterminer $\delta\hat{A}$, car il ne donne accès qu'aux valeurs moyennes.
En présence de dissipation, on peut vérifier que l'équation d'Heisenberg telle que nous l'avons écrite à la relation (\ref{eq-heisenberg}) ne décrit pas convenablement le système :
 \begin{equation}
\frac{d}{dt} \delta\hat{A}\neq \frac{1}{  i\hbar}[\delta\hat{A},\hat{\mathcal{H}}]-\Gamma\ \delta\hat{A}.
  \end{equation}
  En effet, dans ce cas, lorsque deux opérateurs quelconques vérifient cette équation, leur commutateur tend vers zéro pour des temps longs, ce qui n'a pas de sens physique.
  Pour conserver les commutateurs au cours de l'évolution il faut ajouter un terme $\hat{F}_A(t)$ de valeur moyenne nulle qui est l'analogue quantique du terme de Langevin de l'équation (\ref{Langevinclassique}). 
  La forme correcte de l'équation qui décrit l'évolution des fluctuations d'un opérateur quelconque, dite de Heisenberg-Langevin, s'écrit donc :

   \begin{equation}
\frac{d}{dt} \delta\hat{A}=\frac{1}{  i\hbar} [\delta\hat{A},\hat{\mathcal{H}}]-\Gamma \delta\hat{A}+\hat{F}_A(t).
  \end{equation}
  Comme dans le cas classique la valeur moyenne de $\hat{F}_A(t)$ est nulle et on définit le coefficient de diffusion $ D_{A,A}$:
  \begin{deqarr}
\bra\hat{F}_A(t)\ket&=&0,\\
\bra\hat{F}_A(t)\hat{F}_A(t')\ket&= &2D_{A,A} \ \delta(t-t')
   \end{deqarr}
   La méthode de calcul de ces coefficients de diffusion dépend du système atomique.
   Elle est explicitée dans \cite{CohenTannoudji:2001p6627} et sera détaillée au chapitre \ref{ch4} dans un cas particulier.

\subsection{Equations de Maxwell Langevin}
Pour décrire l'interaction d'un état du champ au cours de sa propagation dans le formalisme de Heisenberg, nous devons introduire les équation de Maxwell Langevin.
En effet en utilisant l'équation de Heisenberg, on peut démontrer \cite{CohenTannoudji:1996p4732} que les équations d'évolution des opérateurs champs s'obtiennent en remplaçant les champs classiques par des opérateurs quantiques dans les équations de Maxwell.
On obtient alors :
\begin{equation}
\left( \frac{\partial }{\partial t}+c\frac{\partial}{\partial z}\right)\hat{a}(z,t)=\frac{1}{i\hbar}\left[\hat{a}(z,t),\hat{\mathcal{H}}\right].
\end{equation}
Le terme de droite dans cette équation est un terme de source qui va modifier l'opérateur $\hata$ au cours de la propagation.
Cette contribution des atomes à la propagation du champ peut être un simple effet linéaire (absorption) ou un effet non linéaire qui peut être à l'origine de la production d'état non classiques comme nous le verrons dans le chapitre \ref{ch2}.
Lorsque le terme de droite tend vers zéro, les atomes n'interagissent plus avec le champ et la lumière se propage à la vitesse $c$.



\section*{Conclusion du chapitre}
Dans ce chapitre nous avons introduit le formalisme permettant d'étudier les interactions lumière--matière dans le domaine des variables continues.
Les différents états à un mode du champ, et en particulier les états cohérents et les états comprimés ont été présentés.
Une extension pour les états comprimés à deux modes du champ, c'est-à-dire les corrélations quantiques entre deux modes a aussi été discutée.\\
La question de la photodétection a été abordée d'un point de vue théorique et sera détaillée pour les questions expérimentales dans le chapitre suivant.
Nous avons donné les bases qui nous permettront de faire une étude théorique du mélange à 4 ondes comme source d'état comprimés aux chapitres \ref{ch3} et \ref{ch4}.

%% file: chapitre2v5.tex
\chapter{Techniques expérimentales}\minitoc\label{ch2}
\vspace{2cm}
La génération d'états non--classiques intenses de la lumière est un domaine de recherche expérimentale très actif depuis le début des années 80 \cite{Polzik:2008p13199}.
Si on étudie l'histoire de ce domaine, des expériences pionnières de 1984 \cite{Slusher:1985p3642} au récent record de -12.7 dB sous la limite quantique \cite{Mehmet:2010p12073}, on peut noter deux éléments principaux.
Dans un premier temps on assiste à une diversification des techniques expérimentales employées, puis par la suite on peut constater les nombreux progrès dans les outils expérimentaux disponibles.
Au cours de ce chapitre nous allons, dans un premier temps, donner un panorama des méthodes les plus couramment utilisées pour produire des états non--classiques du champ.
La configuration choisie pour les expériences réalisées dans le cadre de ce travail de thèse sera présentée.\\
Dans un second temps, nous décrirons les outils expérimentaux de l'optique non--linéaire et de l'optique quantique que nous avons utilisés pour ce travail.

\newpage
\section{Revue des méthodes de production des états non classiques de la lumière} 
\label{genesqueez}
	\begin{figure}		
	\centering
		\includegraphics[width=11cm]{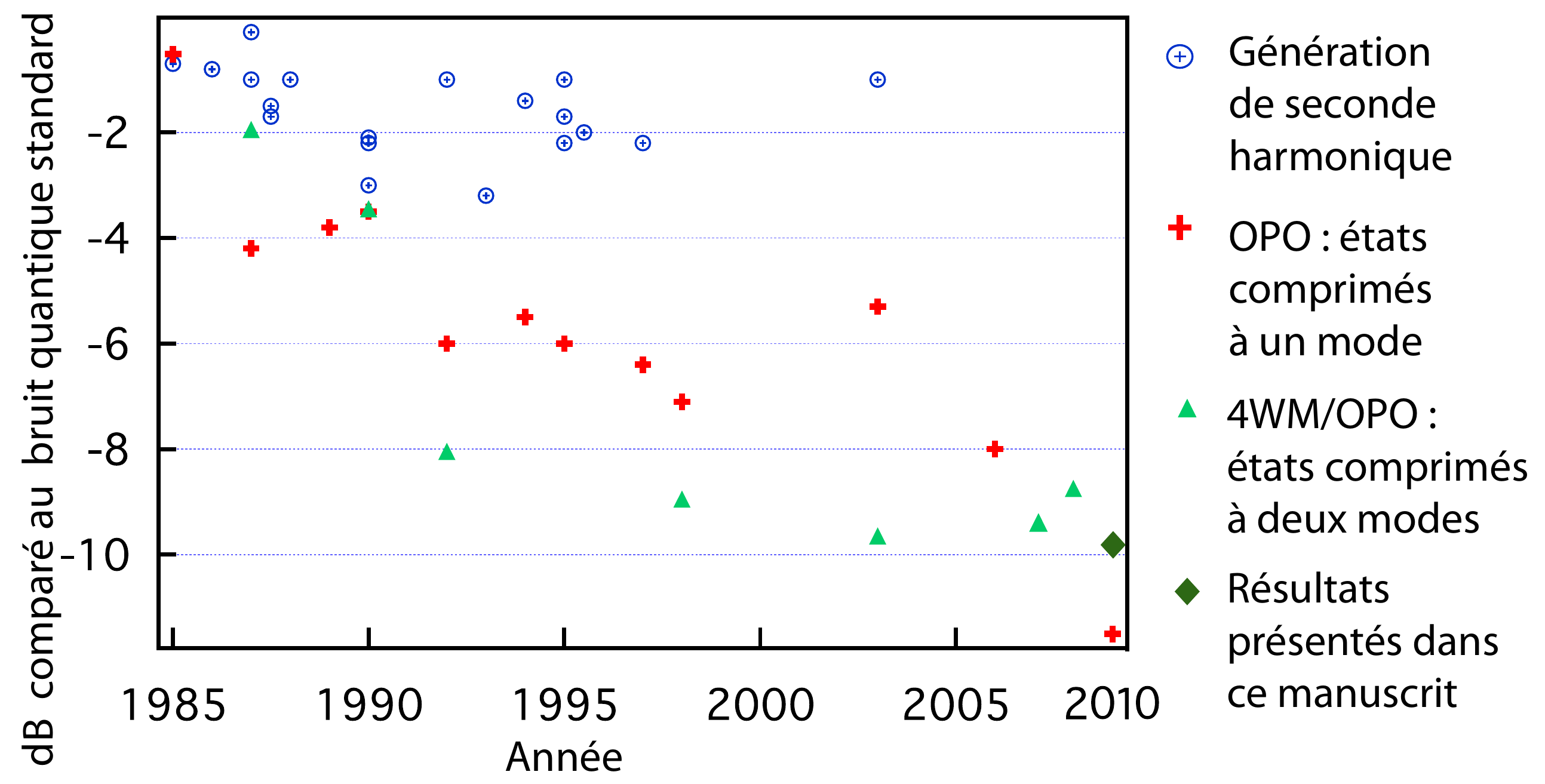} 
	\caption[Panorama historique de la production d'états comprimés]{Panorama historique de la production d'états comprimés depuis 1985. Les ronds bleus correspondent aux expériences de génération de seconde harmonique ou d'effet Kerr. Les croix rouges  représentent les expériences de compression à un mode avec des OPO, tandis que les triangles verts indiquent la compression à deux mode avec des OPO ou du mélange à quatre ondes \cite{Bachor:2004p4500}. Nos résultats expérimentaux sont inclus dans ce diagramme et sont identifiés par le losange vert.\label{history}}		
	\end{figure}
Depuis les trois dernières décennies de très nombreuses méthodes de production d'états non--classiques ont été proposées et réalisées expérimentalement \cite{Bachor:2004p4500}. 
Nous revenons ici sur les principaux systèmes expérimentaux qui permettent d'obtenir des états comprimés.
On peut distinguer deux catégories pour les états comprimés, selon qu'ils mettent en jeu un ou deux modes du champ.\\
La plupart des méthodes utilisées sont basées sur une interaction non-linéaire avec un milieu de susceptibilité non--linéaire $\chi^{(2)}$ ou $\chi^{(3)}$ qui permet de briser la symétrie entre les quadratures du champ\footnote{Une autre méthode utilisant une pompe régulière pour produire une statistique sub--poissonienne dans l'émission de photons a pu permettre de générer des états non-classiques. Une mise en oeuvre dans les diodes lasers alimentées par un courant régulier et stabilisé est décrite dans \cite{Marin:1995p6360}.}.\\
 Pour donner un aperçu de cette histoire la figure \ref{history} présente un grand nombre de résultats expérimentaux depuis 1984, en séparant les expériences de compression à un et deux modes du champ.

\subsection{Mélange à 4 ondes}
	\begin{figure}		
		\centering
			\includegraphics[width=7cm]{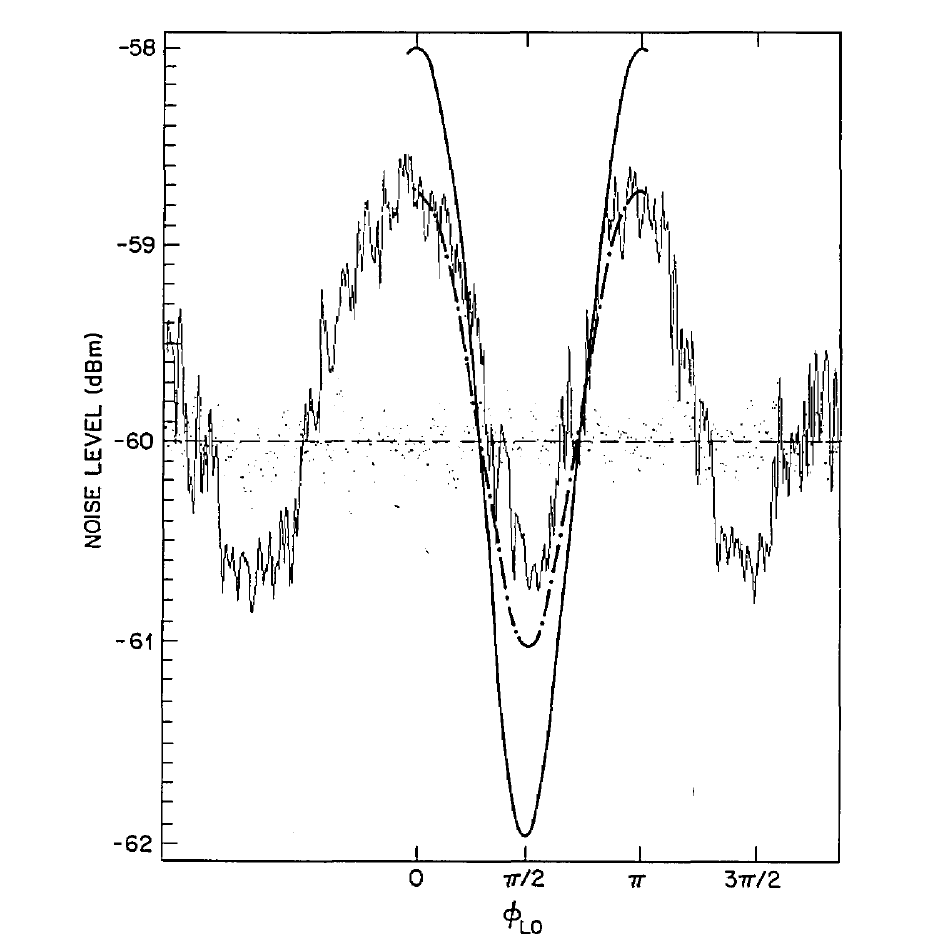} 
		\caption[Mesure de -1dB de compression dans le mélange à 4 ondes]{Réalisation expérimentale de compression sous la limite quantique standard à l'aide de mélange à 4 ondes. En pointillés, l'ajustement des données expérimentales en prenant en compte les pertes, et en trait plein, la prédiction théorique dans cette situation. Données \cite{Slusher:1985p3642}.\label{slusher}}		
		\end{figure}
Le mélange à 4 ondes (\textit{four-wave-mixing}) a été identifié très tôt comme une technique intéressante pour la production d'états non-classiques de la lumière \cite{Bondurant:1984p10551}.
Les expériences pionnières de Slusher  au Bell Labs en 1984 ont été la première démonstration expérimentale d'un état comprimé sous le bruit quantique standard \cite{Slusher:1985p3642}.\\
L'expérience était basée sur un jet atomique de sodium éclairé de façon transverse par un laser de pompe désaccordé de 2 GHz par rapport à la transition atomique.
Ce faisceau était rétro--réfléchi de manière à interagir une seconde fois avec les atomes.
Un faisceau sonde interagissait avec les mêmes atomes, mais avec un léger angle par rapport à la pompe.
Un faisceau conjugué était alors généré, colinéaire et co--propageant avec la sonde.
Pour augmenter l'importance de l'effet, une cavité résonnante avec les deux faisceaux (sonde et conjugué) était placée autour du jet atomique.
Le faisceau sortant de la cavité présentait des fluctuations sous le bruit quantique standard (voir figure \ref{slusher}) ; il s'agit du premier état non-classique intense observé expérimentalement.\\
	\begin{figure}	
			\centering
				\includegraphics[width=12cm]{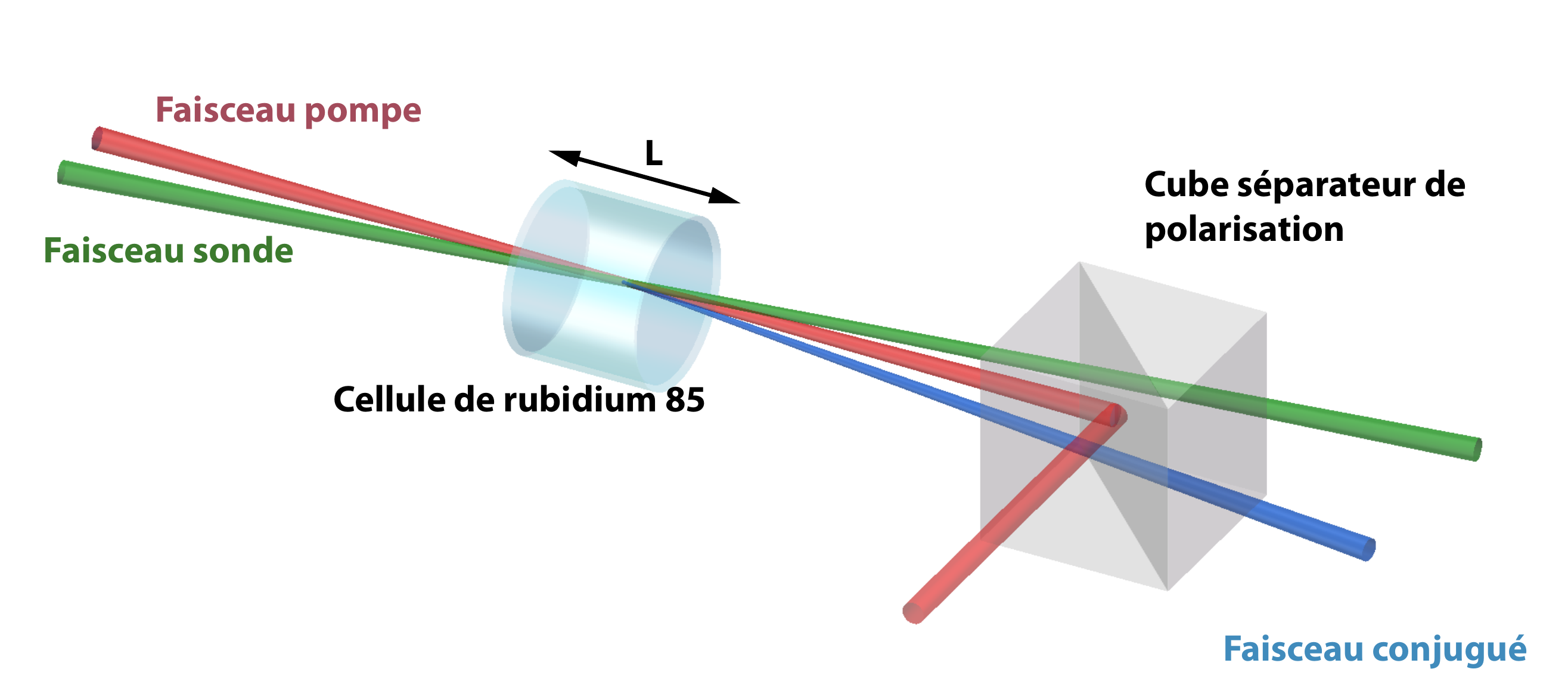} 
			\caption[Montage expérimental du mélange à 4 ondes.]{Schéma de la zone d'interaction dans une expérience de mélange à 4 ondes dans une vapeur atomique. Un faisceau pompe interagit avec un faisceau sonde dans une cellule de rubidium pour générer un faisceau conjugué.\label{figlett1}}		
			\end{figure}
			
Plus récemment, cette technique a connu un regain d'intérêt grâce aux expériences décrites dans \cite{McCormick:2007p652,McCormick:2008p6669,Boyer:2008p1401,Pooser:2009p9625,Akulshin:2009p10219,Camacho:2009p10696}.
En effet -8.8 dB de corrélations sous la limite quantique standard ont été observées entre deux modes du champ à 795 nm.
Dans ces expériences, le jet atomique de sodium est remplacé par une vapeur atomique de rubidium 85 dans une cellule chauffée à une centaine de degrés.
Un faisceau pompe de plusieurs centaines de mW polarisé linéairement et désaccordé d'environ 1 GHz par rapport à la transition est superposé au sein de la cellule avec un angle faible à un faisceau sonde de polarisation orthogonale et décalé en fréquence de 3GHz par rapport au faisceau pompe.
Cet écart correspond à l'écart hyperfin entre les deux niveaux fondamentaux du rubidium 85.
Un faisceau conjugué, de même polarisation que le faisceau sonde, est alors généré durant la propagation dans le milieu.
En sortie, le faisceau pompe est filtré par un cube séparateur de polarisation (voir figure \ref{figlett1}) et les deux faisceaux sonde et conjugué sont détectés sur les deux voies d'une détection balancée.
La sortie différence d'intensité est étudiée à l'analyseur de spectre afin d'observer les éventuelles corrélations sous la limite quantique standard.\\
L'expérience que nous avons développée durant ce travail de thèse reprend cette configuration.
Ainsi, les différents outils qui sont nécessaires pour ce montage expérimental vont être décrits dans la seconde partie de ce chapitre.
Il s'agit d'un faisceau pompe asservi sur la transition D1 du rubidium 85, un faisceau sonde décalé de 3 GHz par rapport à la pompe, une cellule de rubidium, un montage de photodétection ainsi qu'un analyseur de spectre.
\subsection{Amplificateur paramétrique}
	\begin{figure}		
			\centering
				\includegraphics[width=13cm]{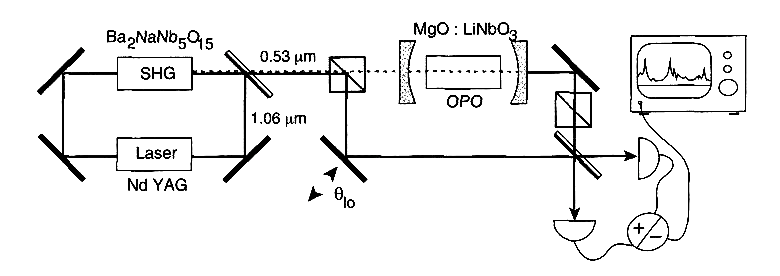} 
			\caption[Schéma d'une expérience de génération d'état comprimé avec un OPO]{Schéma d'une expérience de génération d'un état vide comprimé à l'aide d'un OPO sous le seuil. Un cristal non linéaire MgO:LiNbO$_3$ est pompé par un laser à 0.53 $\mu$m. L'oscillateur local est un faisceau intense à 1.06 $\mu$m. Données \cite{Wu:1987p6008}.\label{OPO1}}		
			\end{figure}
Dans un milieu de susceptibilité non--linéaire $\chi^{(2)}$, une onde incidente que l'on appelle faisceau pompe de valeur moyenne $\alpha_P$ peut donner naissance à deux ondes appelées ``signal'' et ``complémentaire''.
Dans le cas dégénéré, la fréquence de la pompe est égale au double des fréquences du signal et du complémentaire.
Le processus paramétrique qui va donner naissance au champ signal de valeur moyenne $\alpha_S$ va s'écrire sous la forme :
\begin{equation}
\frac{\partial}{\partial z}\alpha_S=-g^{(2)}\alpha_P\alpha_S^* ,
\end{equation}
où $g^{(2)}$ est une constante, a priori complexe, qui quantifie la réponse non--linéaire du milieu.
On voit  que le champ signal pourra être amplifié ou desamplifié selon sa phase par rapport au champ pompe.
Ainsi, certaines quadratures vont donc pouvoir être amplifiées tandis que d'autres seront atténuées.
Pour un choix de phase tel que $\alpha_P$ soit réel positif,  on voit que le champ moyen et la quadrature $\hat{X}_S$ sont désamplifiés tandis que la  quadrature $\hat{Y}_S$ est amplifiée :
\begin{equation}
\frac{\partial}{\partial z}\hat{X}_S=-g^{(2)}\alpha_P\hat{X}_S\ ,\ \frac{\partial}{\partial z}\hat{Y}_S=+g^{(2)}\alpha_P\hat{Y}_S.
\end{equation}
Dans ce cas, pour un champ cohérent en entrée, le champ en sortie sera comprimé suivant la quadrature $\hat{X}_S$. 
Le processus paramétrique a ainsi permis de briser la symétrie entre les deux quadratures du champ.\\
En pratique, le milieu de susceptibilité non--linéaire $\chi^{(2)}$ est placé dans une cavité pour renforcer l'interaction non--linéaire \cite{Wu:1986p6016,Wu:1987p6008}.
L'ensemble ``cavité + milieu non--linéaire'' forme ce que l'on appelle un amplificateur paramétrique optique (OPA) (figure \ref{OPO1}).
Il existe alors un seuil d'oscillation pour le champ signal dans la cavité.
En dessous de ce seuil, le champ généré par un tel dispositif est un état vide comprimé \cite{Hetet:2007p3123}.\\
C'est dans une expérience basée sur un OPA qu'a été obtenu le plus haut taux de compression, à ce jour \cite{Mehmet:2010p12073}.

\subsection{Génération de seconde harmonique}
La génération de seconde harmonique dans un milieu de susceptibilité non--linéaire $\chi^{(2)}$ a aussi été utilisée pour produire des états comprimés.
On peut donner un description qualitative du phénomène dans le cas d'un simple passage à travers le milieu.
En effet pour un flux de photons incident suivant un statistique poissonienne, c'est à dire un état cohérent, l'écart temporel entre deux photons est aléatoire. 
Si le processus de génération de seconde harmonique était indépendant de l'intensité incidente, le flux de photons produit aurait lui aussi la statistique poissonienne d'un état cohérent.
Or ce processus a une probabilité de conversion, non pas constante, mais proportionnelle à $I^2$, où $I$ est l'intensité du faisceau de pompe.
Ainsi la probabilité de  conversion augmente avec le nombre de photons par unité de temps.
La statistique du flux de photons générés sera donc fortement influencée par la statistique des paires de photons dans le faisceau incident, c'est à dire une statistique sub-poissonienne.
De la même manière la statistique des photons restants dans la pompe ne sera plus affectée par les paires de photons et sera par conséquent plus régulière que la statistique initiale.\\
Dans les réalisations expérimentales, il est souvent nécessaire d'amplifier le phénomène en ajoutant une cavité autour du milieu pour le mode de la pompe \cite{Pereira:1988p6513,Kurz:1993p12232,Paschotta:1994p12167}.
\subsection{Effet Kerr}
	\begin{figure}		
				\centering
					\includegraphics[width=6cm]{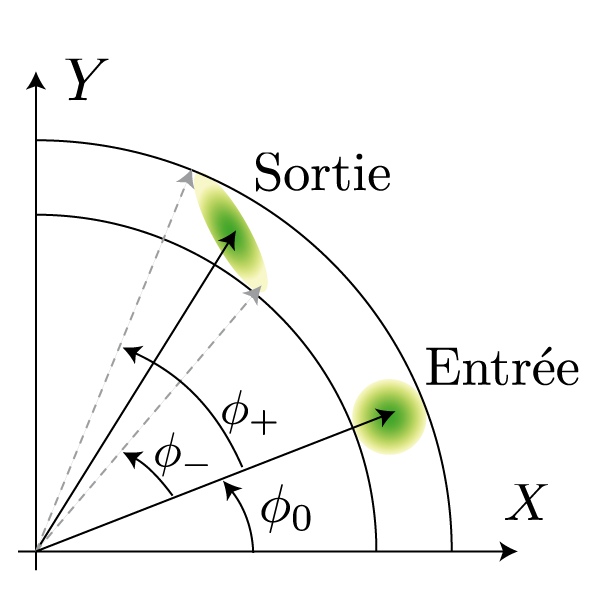} 
				\caption[Génération d'états comprimés par effet Kerr]{Génération d'états comprimés par effet Kerr. Le champ entrant possède une phase moyenne $\phi_0$ et un disque de fluctuations qui correspond à celui d'un état cohérent. La phase accumulée durant la propagation diffèrent selon l'intensité du champ dans le milieu Kerr. Pour le champ d'intensité $\bra I\ket+\delta I$ la phase accumulée vaut $\phi_+$, réciproquement pour le champ d'intensité $\bra I\ket-\delta I$ la phase accumulée vaut $\phi_-$. L'état final est un état comprimé.	\label{kerr_fig}}
\end{figure}
L'effet Kerr est un effet non--linéaire que l'on observe dans les milieux de susceptibilité non--linéaire $\chi^{(3)}$.
L'indice de réfraction d'un tel milieu dépend de l'intensité lumineuse $I$ qui l'éclaire :
\begin{equation}
n=n_0+n_2 I.
\end{equation}
On peut comprendre comment l'effet Kerr permet de briser la symétrie entre les quadratures en considérant un état cohérent à l'entrée de ce milieu.
Lors de la propagation, les fluctuations d'intensité de cet état vont induire une modulation de l'indice, qui va en retour introduire des fluctuations supplémentaires sur la phase du champ.
La phase $\phi$ accumulée lors de la propagation sur une longueur $L$, pour une longueur d'onde $\lambda$, s'exprime sous la forme :
\begin{equation}
\phi_{milieu}=\frac{2\pi  L}{\lambda}(n_0+n_2 I).
\end{equation}
Ainsi les fluctuations en sortie de la quadrature $\hat Y$  vont être modifiées par la non--linéarité du milieu.
La propagation dans un milieu Kerr transforme donc le disque de fluctuations des états comprimés en une ellipse.
Pour un milieu Kerr parfait, l'aire des fluctuations est conservée et l'état en sortie est un état comprimé (voir figure \ref{kerr_fig}).\\
Pour obtenir un effet important et donc une grande compression, il est nécessaire de disposer d'un milieu présentant une non--linéarité $\chi^{(3)}$ importante.\\
Différentes pistes ont été explorées parmi les milieux Kerr existants, que l'on peut classer en deux catégories : les milieux non résonants et les milieux résonants.\\
Parmi les premiers, l'idée est de compenser une non-linéarité relativement faible par une grande longueur de propagation.
Les fibres optiques ont par conséquent été utilisées comme milieu Kerr pour la production d'états comprimés \cite{Levenson:1985p10498,Sizmann:1999p6422}.\\
La seconde approche est basée sur une interaction avec des milieux atomiques à résonance ou proche de résonance.
Dans ce cas la linéarité peut être très forte, mais le taux d'absorption par le milieu introduit des pertes qui diminuent fortement la compression attendue.
Cette méthode a été largement employée pour produire des états comprimés avec des atomes froids \cite{Lambrecht:1996p6121}\footnote{Même si des expériences similaires ont aussi été publiées dans une vapeur atomique \cite{Ries:2003p6186}, il est important de remarquer qu'un article récent \cite{Johnsson:2006p6202} démontre que la production d'états comprimés avec des atomes chauds ne semble pas envisageable.}.

		\section{Source laser et asservissement}
Lorsque nous avons présenté les expériences de mélange à 4 ondes dans une vapeur atomique, nous avons vu qu'elles nécessitaient deux faisceaux laser proches d'une résonance atomique (la transition D1 du $^{85}$Rb à 795 nm dans les expériences présentées dans ce manuscrit).
Il s'agit d'une part, d'un faisceau intense ($P\simeq1$W) appelé faisceau pompe et d'autre part d'un faisceau  de plus faible intensité ($P<0.5$mW) stabilisé en phase avec le premier et décalé en fréquence de l'équivalent de l'écart hyperfin entre les niveaux fondamentaux de l'atome utilisé (3GHz pour le rubidium 85).
Ce second faisceau est appelé faisceau sonde.
Nous allons voir comment ces deux champs sont générés expérimentalement.\\

\noindent Le laser utilisé pour générer le faisceau pompe est un laser titane-saphir continu de la société Coherent\footnote{http://www.coherent.com/}.
Il s'agit du modèle MBR-110 sans la cavité de référence pour la stabilisation.
Le modèle commercial a donc été légèrement modifié pour permettre une stabilisation sur une référence externe, à savoir une raie d'absorption du rubidium.
Nous présentons ici ce laser et nous donnons les détails techniques qui permettent de l'asservir sur une transition atomique.
	\begin{figure}	
		\centering
		\includegraphics[width=14cm]{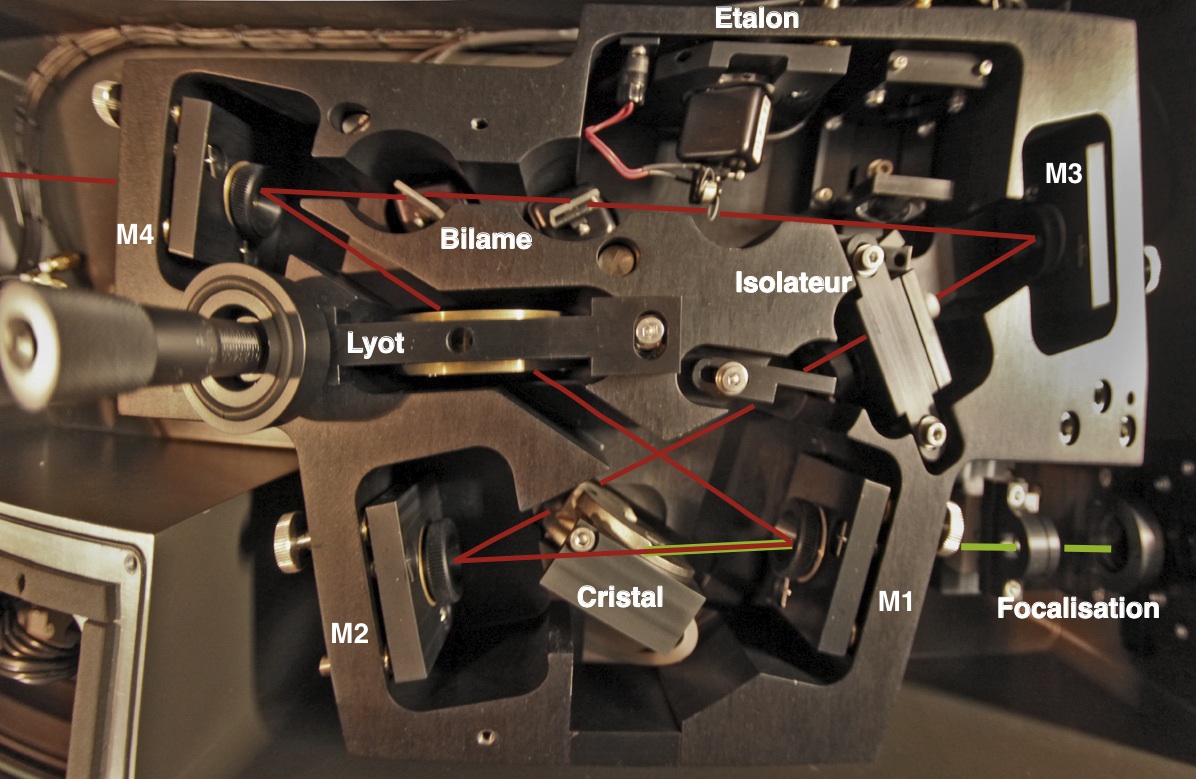}
		\caption[Cavité titane-saphir]{Cavité titane-saphir. Les miroirs de la cavité sont indiqués de M1 à M4. Le laser de pompe est indiqué en vert, et le trajet du laser dans la cavité est indiqué en rouge. Voir le texte pour les détails.}		\label{lenspump}	
		\end{figure}
					\subsection{Laser titane-saphir}
					\subsubsection{Cavité laser}

Le laser MBR est composé d'une cavité monobloc et d'un cristal de saphir dopé aux ions titane Ti$^{3+}$.
Ce cristal est pompé par un laser Nd:YVO$_4$ doublé à 532nm de 18W (Verdi V18  de la société Coherent).
Selon le jeu de miroirs utilisé pour la cavité laser, il est possible d'obtenir l'effet laser une gamme de longueurs d'onde allant de 700~nm à 990~nm grâce à la très grande plage de gain du milieu amplificateur. 
Dans cette thèse, nous avons utilisé les miroirs dits ``mid-wave'' ou ``MW'' qui permettent l'émission laser mono--mode de 780 à 870~nm.\\
Le laser de pompe est focalisé dans le cristal à l'aide d'un système de deux lentilles dont la position peut être réglée finement afin d'optimiser les performances du laser (voir la zone de focalisation sur la figure \ref{lenspump}).\\
Afin de s'assurer du fonctionnement mono--mode longitudinal du laser, il faut supprimer les effets de \textit{''hole burning''} dans le milieu amplificateur présents, par exemple, pour une cavité linéaire.
Pour ce faire, la cavité du MBR est une cavité en noeud-papillon dotée d'un isolateur optique afin de ne supporter qu'une seule direction de propagation (voir figure \ref{lenspump}).
La cavité est donc composée de quatre miroirs.
Les miroirs M1 et M2 sont des miroirs sphériques de rayon 100~mm, couverts d'un traitement $R_{max}$ (réflectivité maximale), large bande autour de la longueur d'onde d'émission.
Comme le faisceau de pompe est injecté à travers le miroir M1, il joue un rôle de lentille pour ce faisceau. Ainsi sa position est doublement critique pour un bon fonctionnement du laser.\\
Le miroir M3 est un miroir plan, traité $R_{max}$, monté sur transducteur piézo--électrique (PZT) qui permet de régler finement la longueur de la cavité afin d'asservir le laser.
Le miroir plan M4 est le coupleur de sortie.
Nous disposons de deux coupleurs de sortie différents, adaptés à une puissance de pompe de 10 et 18~W.
			\subsubsection{Réglage de la fréquence}
			\begin{figure}	
				\centering
				\includegraphics[width=8.5cm]{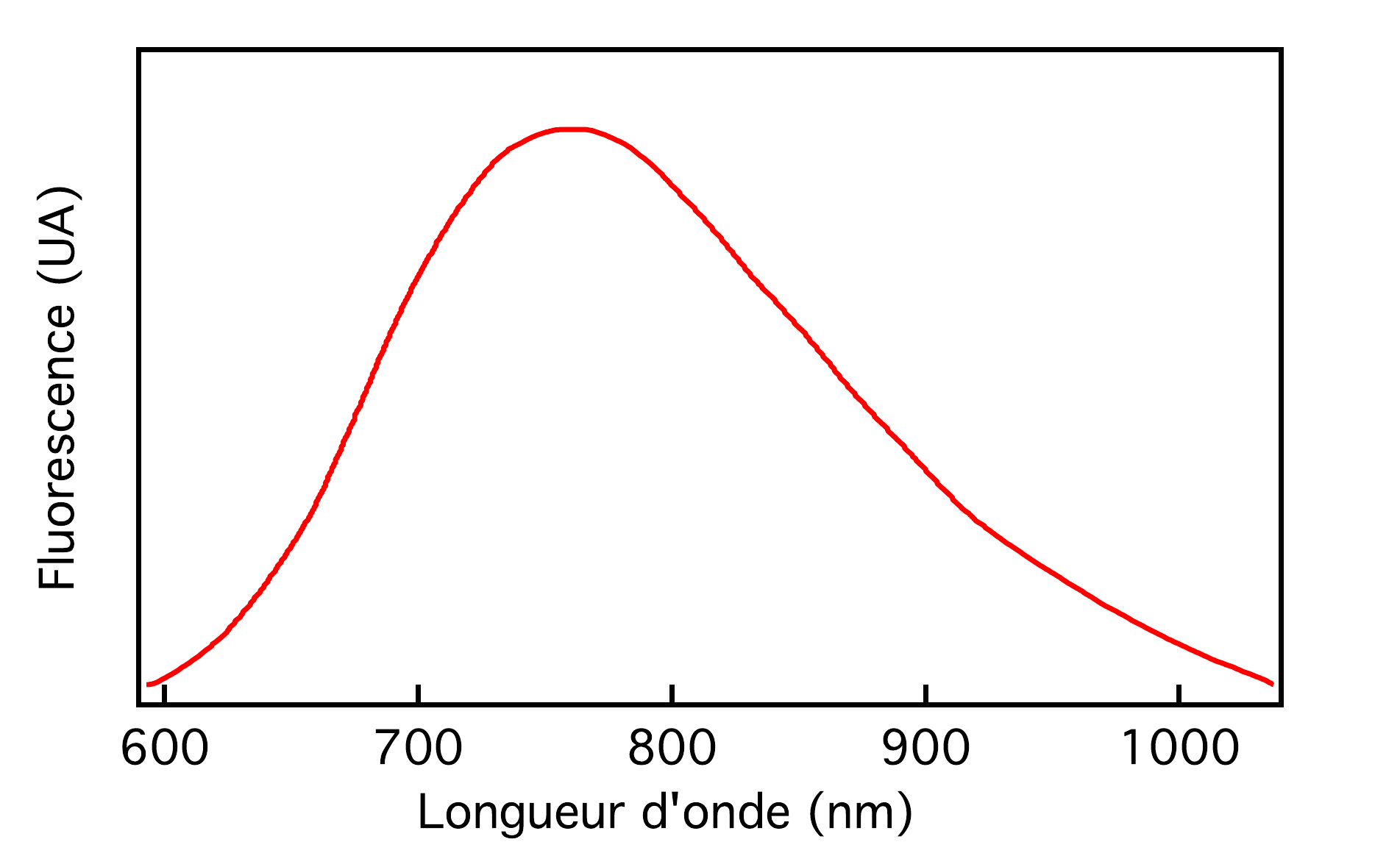}
				\caption[Fluorescence d'un cristal titane-saphir pompé à 532 nm]{Fluorescence d'un cristal titane-saphir pompé à 532 nm. Données \cite{Albers:1986p7043}.}	\label{tisaph}	
				\end{figure}
\begin{figure}	
\centering
\includegraphics[width=11cm]{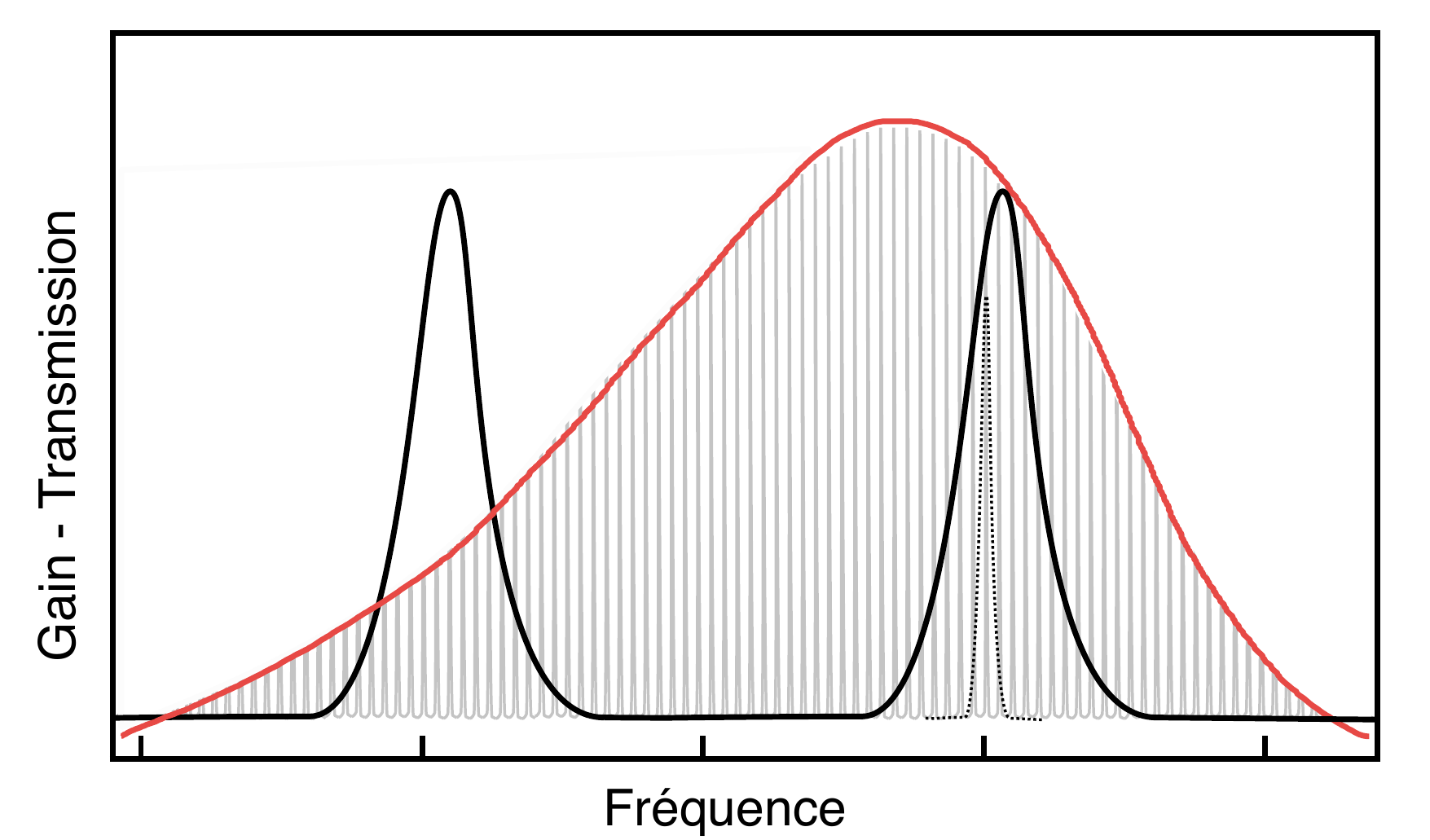}
\caption[Schéma de principe du fonctionnement mono--mode du laser titane saphir]{Schéma de principe du fonctionnement mono--mode du laser titane saphir. Le gain de la cavité est donné en rouge. Il couvre plusieurs centaines de THz. le peigne de mode longitudinaux de la cavité est représenté en gris. Notons que l'échelle horizontale n'est pas respectée et que le nombre de modes longitudinaux dans la zone de gain devrait être bien plus important (typiquement $10^5$). Le filtre de Lyot (courbe de transmission en noir) effectue une première sélection d'un nombre réduit de modes dans ce peigne. Enfin la transmission de l'étalon  (courbe en pointillés) dont la position est asservie, assure le fonctionnement mono--mode du laser. }		\label{asserv}	
\end{figure}

Comme nous l'avons mentionné précédemment, la courbe de gain d'un cristal de titane saphir est très large (figure \ref{tisaph}).
En effet, dans les lasers à cristaux dopés par des métaux de transition, il existe une forte interaction entre les niveaux électroniques et les modes de vibration du réseau cristallin \cite{Schwendimann:1988p12520}.
Cette interaction induit un élargissement homogène important et par conséquent une large bande passante de gain \cite{Moulton:1986p12334,Pollnau:2007p12283}.\\
Ainsi, un très grand nombre de modes longitudinaux de la cavité peuvent vérifier la condition d'un gain supérieur ou égal aux pertes de la cavité.
Pour permettre au laser de fonctionner de manière mono--mode, c'est à dire sur un seul mode longitudinal, il est nécessaire de placer des filtres dans la cavité.
Dans le cas du MBR, deux filtres sont utilisés.
Un filtre biréfringent (filtre de Lyot) permet un premier réglage grossier du mode par sauts de $225$ GHz, tandis qu'une fine lame de verre (étalon Fabry Perot de 500 microns d'épaisseur) permet des réglages plus fins à l'intérieur de cette gamme.
L'asservissement de l'angle de l'étalon permet de maintenir le fonctionnement mono--mode du laser en suivant un mode de la cavité (voir figure \ref{asserv}).
Pour faire varier continûment la longueur d'onde du laser, on dispose de deux lames de verre à l'angle de Brewster placées à l'intérieur de la cavité et montées sur des moteurs galvanométriques.
La rotation de ce bilame permet de modifier le chemin optique sans modifier le trajet du faisceau dans la cavité. 
La longueur d'onde du laser peut être ajustée de manière contrôlée par ce biais.
Cette méthode peut être utilisée pour compenser les dérives lentes du laser (typiquement inférieures au GHz par seconde), alors que le PZT permet, en principe, de compenser des dérives à plus haute fréquence.
Dans la configuration décrite dans cette thèse le miroir M3 est fixe et seules les dérives lentes sont compensées par l'asservissement.
			\subsubsection{Électronique}
Afin de pouvoir stabiliser le laser sur une référence externe, nous avons dû modifier légèrement  l'électronique de contrôle du MBR.
En effet, dans sa version commerciale, il est conçu pour être asservi sur une cavité de référence interne.
Pour asservir la fréquence du laser sur une référence atomique, nous avons utilisé l'entrée \textit{External Scan Input} du boitier de contrôle.
Cette entrée accepte une tension entre -10 V et +10 V.
Sur cette plage, la fréquence du laser est modifiée à l'aide du bilame.
La gamme de fréquence est réglée sur le boitier de contrôle externe du laser (\textit{Scan Width}), ainsi que la fréquence centrale (\textit{Scan offset}).
La plage maximale de balayage est de 40 GHz à une vitesse de 8 GHz/s.
			
\subsection{Asservissement}
Afin d'asservir en fréquence le laser et d'éviter les sauts de modes, deux techniques sont utilisées conjointement.
L'asservissement de l'étalon a déjà été présenté dans la section précédente et permet d'éviter les sauts entre les différents modes longitudinaux de la cavité.
Par contre, les dérives de la cavité laser elle-même ne sont pas compensées par cet asservissement.
Pour ce faire il est nécessaire d'asservir la cavité sur une référence atomique.
Dans ce manuscrit, il s'agit selon les cas, de la raie D1 ou de la raie $5S_{1/2} \rightarrow 6P_{1/2} $ du $^{85}$Rb.

\subsubsection{Élargissement inhomogène par effet Doppler}
Pour la stabilisation du laser nous utilisons une vapeur de $^{85}$Rb dans une cellule en verre chauffée à 60$^\circ$C.
A cette température, l'élargissement Doppler doit être pris en compte.
La valeur moyenne de la projection du module de la vitesse des atomes sur l'axe de propagation du faisceau laser est donnée par :
\begin{equation}\label{rouge}
v_D=\sqrt{\frac{k_B T}{m}},
\end{equation}
où $k_B$ est la constante de Boltzmann, $T$ la température des atomes et $m$ leur masse .
On peut en déduire simplement l'élargissement en fréquence par effet Doppler d'une raie atomique infiniment étroite à la longueur d'onde $\lambda$ :
\begin{equation}
\Delta_{Dop}=\frac 2\lambda \sqrt{\frac{k_B T}{m}}.
\end{equation}
A 60$^\circ$C pour le $^{85}$Rb sur la raie D1,  la largeur Doppler de la transition est de 450~MHz, ce qui est plus large que les 5.7~MHz de la largeur naturelle (détails de la raie D1 en annexe \ref{Annexe_Rb}).
Cette largeur est aussi plus grande que l'écart hyperfin des niveaux excités F=2 et F=3 (360 MHz).
Ainsi le signal observé en sortie d'une cellule de rubidium chaude ne permet pas de séparer les deux niveaux excités et on observe un creux d'absorption unique  large d'environ 700 MHz.
Il n'est donc pas possible d'asservir le laser précisément sur ce signal.
\subsubsection{Absorption saturée}
			\begin{figure}	
	\centering
	\includegraphics[width=7.5cm]{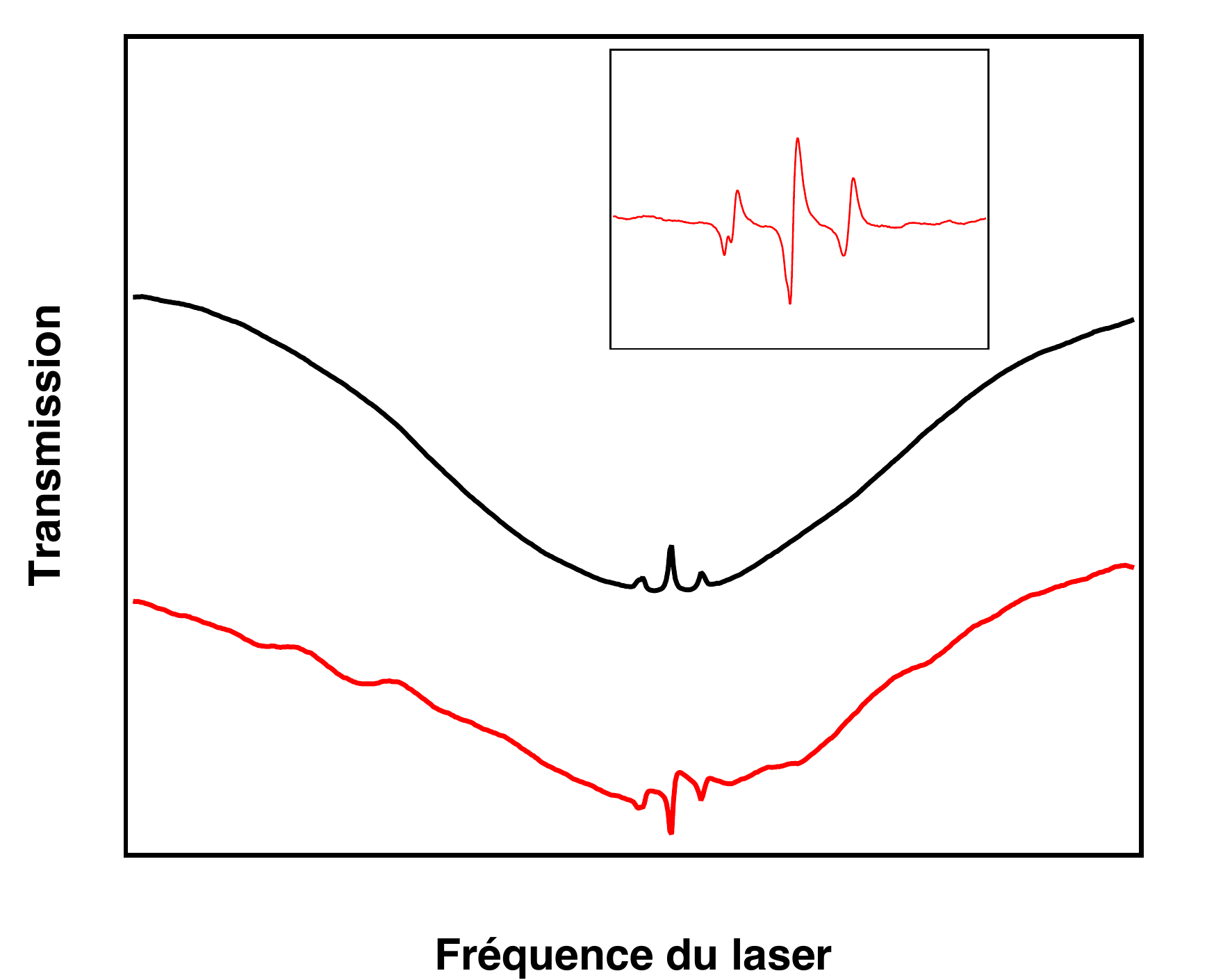}
	\caption[Spectre d'absorption saturée.]{Spectre d'absorption saturée sur la raie $5S_{1/2},F=2 \rightarrow 6P_{1/2} $  du $^{85}$Rb. La courbe noire est le signal de transmission. La courbe rouge est le signal démodulé qui sert de signal d'erreur pour l'asservissement du laser. En encart, il s'agit d'un détail du signal d'erreur.}		\label{abs}	
\end{figure}
Une méthode bien connue pour résoudre ce problème est d'utiliser un montage d'absorption saturée \cite{Siegman:1986p20060}.
Un premier faisceau est utilisé pour saturer la transition (l'intensité de saturation vaut $Isat =$ 4.4 mW/cm$^2$, voir annexe \ref{Annexe_Rb}).
Un second faisceau contra--propageant vient sonder le milieu ainsi préparé. 
Seuls les atomes, dont la projection de la vitesse sur l'axe de propagation des lasers est nulle, interagissent sans décalage Doppler avec les deux faisceaux laser.
On observe alors un pic de transmission pour chaque fréquence correspondant à l'écart d'énergie entre un niveau de l'état fondamental et un niveau de l'état excité.
S'il existe plusieurs niveaux hyperfins dans l'état excité (deux dans le cas de la raie D1 du $^{85}$Rb), des pics ne correspondant pas à une transition atomique, dits ``pics de croisement de niveaux'', apparaissent à la fréquence médiane de chaque couple de niveaux hyperfins. 
Le montage d'absorption saturée permet ainsi d'obtenir des pics de transmission beaucoup plus fins que la largeur Doppler\footnote{La largeur des pics est égale, typiquement, à la somme de la largeur naturelle et de la largeur de raie du laser.} (voir figure \ref{abs} qui présente une courbe d'absorption saturée pour la transition "bleue" $5S_{1/2},F=2 \rightarrow 6P_{1/2} $  du $^{85}$Rb).
Par contre, le signal n'est pas utilisable en l'état comme signal d'erreur.
\subsubsection{Détection synchrone}
Pour augmenter le rapport signal à bruit et générer un signal d'erreur utilisable, on utilise une détection synchrone.
Le modèle utilisée au laboratoire est le LIA-MV-200-H de la société Femto\footnote{http://www.femto.de/}.
Le laser MBR est modulé en interne à une fréquence de 89.3 kHz, pour réaliser l'asservissement de l'étalon.
On utilise donc aussi cette modulation comme fréquence de référence pour l'asservissement sur la raie atomique.
Afin de récupérer la référence de modulation pour la détection synchrone, l'électronique interne du laser a été modifiée.
Dans cette version modifiée, le signal de modulation est disponible sur la sortie \textit{Fast External Lock} du boitier de contrôle du laser que nous n'utilisions pas.
Pour obtenir le signal de la figure \ref{abs}, la constante de temps et la sensibilité  de la détection synchrone sont réglées sur 1~ms et 30~mV, respectivement.
La phase est ajustée afin de maximiser le signal.
Un correcteur PID (proportionnel, intégrateur, différentiel) est utilisé afin de fermer la boucle de l'asservissement.
Dans ces conditions, le laser peut être asservi sur l'une des raies du rubidium, et l'excursion en fréquence résiduelle du laser est inférieure à la largeur naturelle de la transition.
\subsection {Bruit technique du laser}
Pour caractériser les performances du laser nous avons réalisé des mesures de bruit en intensité.
Ces mesures sont réalisées à l'aide d'une photodiode d'efficacité quantique $\eta=0.85$ en atténuant le laser de 3 W à 3 mW. 
Le photocourant est ensuite observé sur l'analyseur de spectre décrit à la section \ref{Agilent}, réglé sur une bande passante de résolution RBW = 100 kHz et une bande passante vidéo VBW = 10 Hz.
Le bruit du laser est comparé au bruit quantique standard, mesuré à l'aide d'une détection équilibrée.
Lors de ces mesures, nous avons pu vérifier les données du constructeur, à savoir qu'au delà de 500~kHz, le bruit technique du laser est négligeable devant le bruit quantique standard.

\section{Génération de seconde harmonique}
Il existe une quasi-coïncidence (environ $500$ MHz d'écart  \cite{Madej:1998p13999}), entre la transition $5S_{1/2}~\rightarrow~5P_{1/2} $ de l'ion $^{88}$Sr$^+$ et  la transition $5S_{1/2},F=2 \rightarrow 6P_{1/2} $ du $^{85}$Rb.
Afin de réaliser des expériences d'optique quantique dans un nuage d'ions Sr$^+$ refroidis par laser, qui est une des thématiques de l'équipe dans laquelle j'ai réalisé ce travail de thèse, il est utile de disposer d'une source laser à cette longueur d'onde.
De plus, dans le chapitre \ref{ch6} de ce manuscrit, nous décrivons des expériences de transparence électromagnétiquement induite sur la transition $5S_{1/2} \rightarrow 6P_{1/2} $  du $^{85}$Rb à 422 nm.
Nous donnons ici les informations relatives à la réalisation expérimentale d'une telle source laser.
\subsection{Principe de la génération de seconde harmonique}
Nous avons utilisé la génération de seconde harmonique (doublage de fréquence) à partir du laser titane saphir réglé à 844 nm, pour produire un champ laser intense à 422 nm.
La génération de seconde harmonique est un processus d'optique non--linéaire dans les milieux de susceptibilité non--linéaire $\chi^{(2)}$ \cite{Armstrong:1962p13945}.
Nous en rappelons ici brièvement l'idée générale.
Pour des raisons de simplicité, nous faisons l'hypothèse que le champ électrique ainsi que la polarisation dans le milieu sont des quantités scalaires.
Ainsi la susceptibilité non--linéaire $\chi^{(2)}$, qui dans un traitement vectoriel est un tenseur d'ordre 3, devient simplement une grandeur scalaire \cite{WBoyd:2008p7258}.
Lors de l'interaction d'un champ électrique $E(z,t)$ avec un tel milieu, la polarisation $P(z,t)$ induite par ce champ s'écrit dans l'approximation que nous venons de faire :
\begin{equation}
P(z,t)=\epsilon_0\ \left( \chi^{(1)}E(z,t)+\chi^{(2)}E(z,t)^2\right).
\end{equation}
La polarisation induite dans le milieu agit comme un terme source dans l'équation de propagation :
\begin{equation}
\frac{\partial^2}{\partial z^2}E(z,t)-\frac {n^2}{c^2}\frac{\partial^2}{\partial t^2}E(z,t)=\frac 1{\epsilon_0 c^2}\frac{\partial^2}{\partial t^2}P^{NL}(z,t),
\end{equation}
avec $P^{NL}(z,t)=\epsilon_0\chi^{(2)}E(z,t)^2$ la polarisation non-linéaire du milieu et $n$ l'indice linéaire usuel.\\
Ainsi, la polarisation non--linéaire permet de faire apparaitre des termes à la fréquence angulaire $2\omega$ dans l'équation de propagation, caractéristiques de la génération de seconde harmonique, .
Un traitement plus complet \cite{RShen:2003p7271}, permet de donner une condition nécessaire pour obtenir une efficacité de conversion importante : l'accord de phase.
Pour des vecteurs d'onde $k_\omega$ et $k_{2\omega}$ respectivement associés au champ à la fréquence $\omega$ et $2\omega$, cette condition impose :
\begin{equation}
2k_{\omega}-k_{2\omega}=0.
\end{equation} 
\subsection{Quasi accord de phase et mise en oeuvre expérimentale}
\begin{figure}	
\centering
\includegraphics[width=8cm]{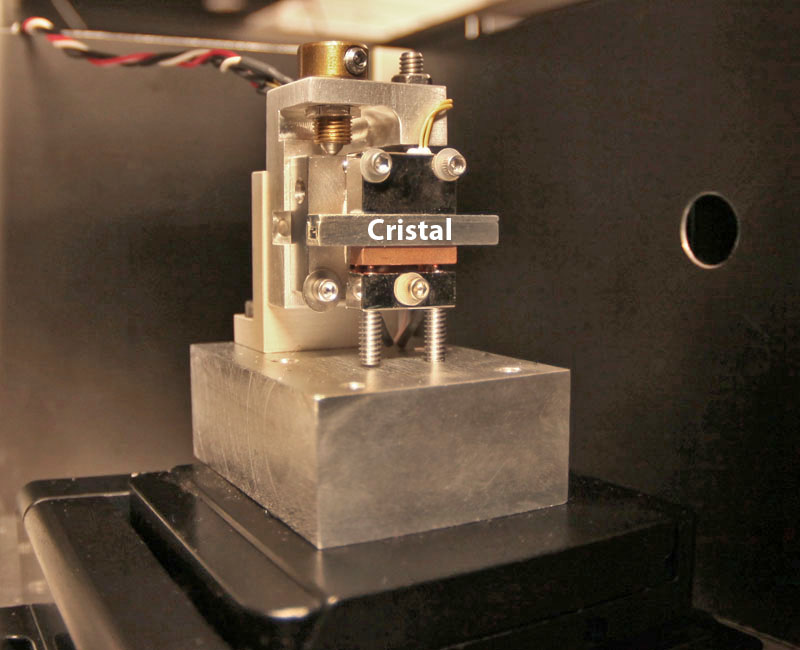}
\caption[Cristal de PPKTP.]{Cristal de PPKTP dans le four qui permet de contrôler la température. \label{four_raicol}	}		
\end{figure}
\begin{figure}	
\centering
\includegraphics[width=11.5cm]{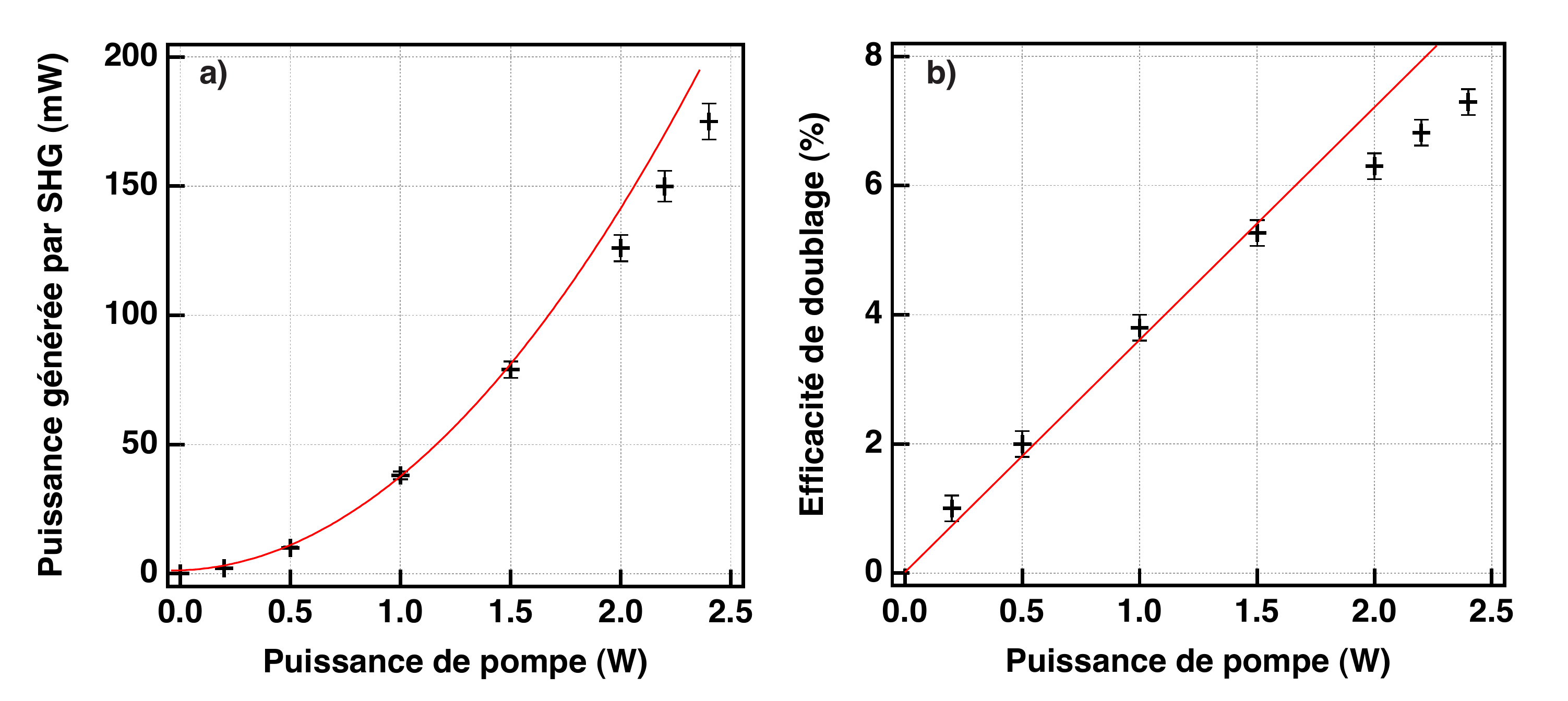}
\caption[Rendement du SHG.]{a) Puissance de signal à la fréquence double en fonction de la puissance incidente. b) Rendement de conversion du doublage de fréquence. En rouge, les courbes d'ajustement respectivement quadratique a) et linéaire b). D'après la figure b), on peut voir que le rendement vaut 3,6$\%$/W \label{SHG}	}		
\end{figure}

Expérimentalement, la mise en oeuvre d'un doublage de fréquence nécessite un cristal ayant un tenseur de susceptibilité non--linéaire $\chi^{(2)}$ non nul et un laser de pompe intense (plusieurs centaines de mW), à la fréquence du fondamental.
De plus, un processus de génération de seconde harmonique n'est efficace que si la condition d'accord de phase est satisfaite.
Pour réaliser la condition d'accord de phase dans les cristaux, il n'est pas toujours possible d'utiliser l'accord de phase par biréfringence \cite{WBoyd:2008p7258}.\\
Dans ce cas, sans annuler complètement le désaccord, une des techniques employées est de le mettre à zéro périodiquement grâce à un matériau ``périodiquement retourné'' (\textit{periodically poled}).
En effet, en renversant periodiquement le signe du coefficient non--linéaire, on peut compenser la différence de vitesse de phase entre le champ pompe et la seconde harmonique.
On appelle cette condition : la condition de quasi-accord de phase \cite{GRosenman:1997p13600,Pasiskevicius:1998p13967,Pierrou:1999p13520}.\\
Le cristal que nous avons utilisé est un cristal de PP-KTP (Periodically Poled Potassium Titanyl Phosphate) qui mesure $1\times 2\times 30$ mm$^3$ et qui a été produit par la société Raicol Crystals.
Afin d'assurer un bon rendement de conversion, il est nécessaire de bien maitriser l'indice du cristal de PP-KTP et donc sa température.
Le cristal est donc placé dans un four présenté sur la figure \ref{four_raicol}.
Un asservissement en température de ce four permet de stabiliser sa température à $0.05~$K de précision autour de 310 K.\\
Il est important de noter que la température de consigne du four doit être ajustée en fonction de la puissance du faisceau pompe.
En effet, le faisceau incident est focalisé sur une taille de 400 microns de rayon (waist à $1/e^2$) au centre du cristal, ce qui provoque des effets thermiques non négligeables.\\
En ajustant la température de consigne pour chaque point, on peut obtenir un signal généré à $2\omega$, proportionnel au carré de la puissance du faisceau pompe incident, et donc un rendement de conversion proportionnel à la puissance du faisceau incident. 
On peut constater sur la figure \ref{SHG}, que les valeurs de conversion et de rendement pour des puissances de pompe grandes (supérieures à 1.5W) sont légèrement plus basses que celles attendues pour les courbes théoriques.
En effet, à de telles puissances de pompe, le signal généré devient important (près de 200~mW pour 2.4 W de pompe) et l'hypothèse de non-dépletion de la pompe (qui a permis d'établir la dépendance quadratique de la puissance du signal en fonction de la puissance incidente) n'est plus valable.
\clearpage
\section{Génération du faisceau sonde}
Pour réaliser l'expérience de mélange à 4 ondes décrite précédemment, il est nécessaire de disposer d'un faisceau de faible intensité décalé de 3 GHz par rapport à la pompe et stabilisé en phase avec la pompe.
La méthode que nous avons choisie pour cela est d'utiliser un modulateur acousto-optique (MAO) à 1.5 GHz dans une configuration de double passage.
Nous allons détailler dans cette section le principe de fonctionnement d'un MAO et les mesures de bruit que nous avons effectuées sur le faisceau ainsi généré.

\subsection{Principe de fonctionnement d'un modulateur acousto-optique}
Un MAO est un composant opto-électronique qui utilise l'effet acousto--optique pour diffracter et décaler en fréquence un faisceau laser.
Un PZT collé sur un cristal de TeO$_2$ permet de générer une onde acoustique progressive (de fréquence 1.5 GHz dans notre cas) dans ce cristal.
Une modulation sinusoïdale de l'indice est ainsi produite dans le cristal et un faisceau laser peut être diffracté par diffraction de Bragg dans les différents ordres $m$, à l'angle $\theta_m$ donné par :
\begin{equation}
\sin\theta_m = \frac{ m\lambda}{2\Lambda},
\end{equation}
avec $\lambda$ la longueur d'onde optique et $\Lambda$ la longueur d'onde acoustique.
Dans notre cas, on obtient un angle de diffraction de 140 mrad à 800 nm et 75 mrad à 422 nm pour le premier ordre diffracté (la vitesse de propagation de l'onde dans le  TeO$_2$ est de 4200~$m.s^{-1}$).\\
Contrairement à la diffraction de Bragg classique, la modulation d'indice due à l'onde acoustique, se propage dans le milieu.
La fréquence de l'onde diffractée dans l'ordre~+1 sera donc décalée par effet Doppler de l'équivalent de la fréquence acoustique.
C'est cette propriété qui nous intéresse, en vue de générer un faisceau sonde décalé de 3 GHz par rapport au faisceau pompe.
\subsection{Efficacité de diffraction}
\begin{figure}	
	\centering
	\includegraphics[width=14cm]{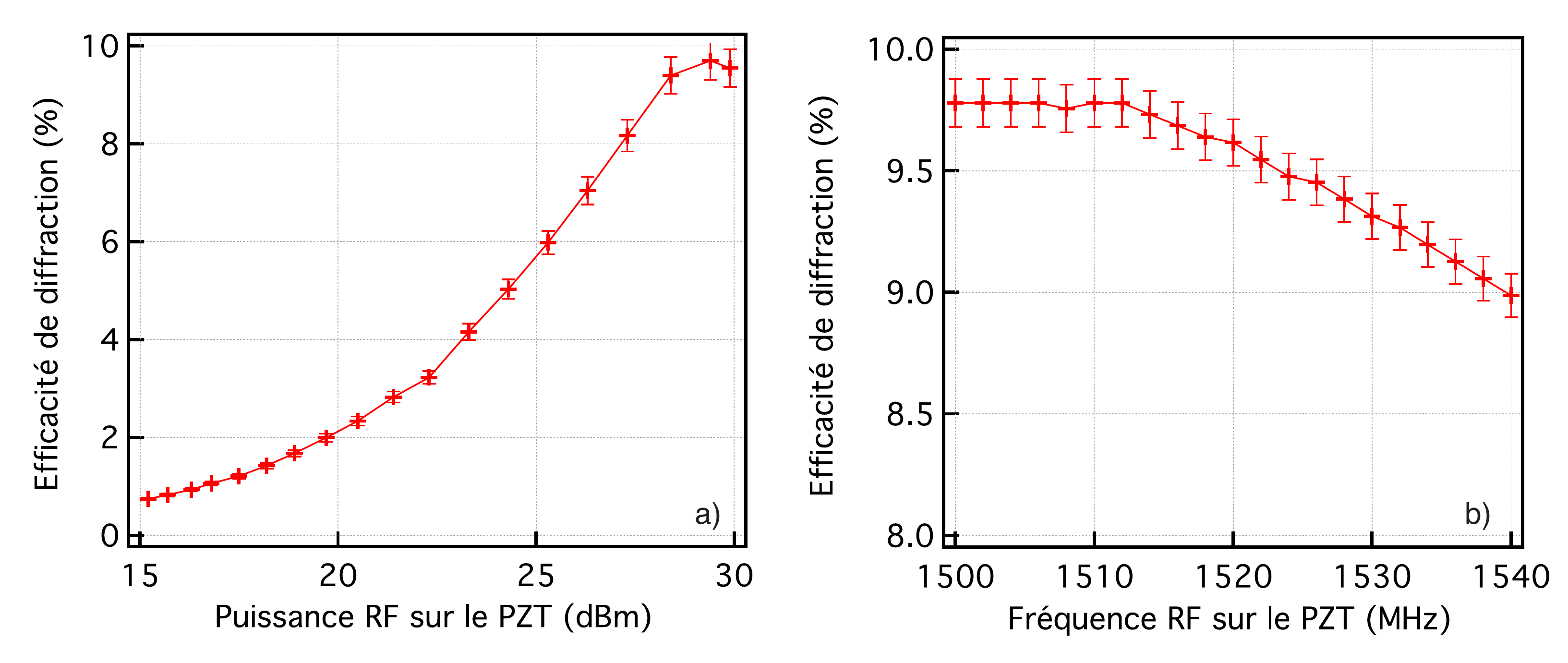}
	\caption[Efficacité de diffraction du MAO]{Efficacité de diffraction du MAO en fonction de la puissance RF a) et de la fréquence RF b).}\label{calibMAO}	
	\end{figure}
Pour augmenter l'efficacité de diffraction, il est nécessaire de focaliser le faisceau incident sur la zone active du cristal (onde acoustique).
Dans notre cas, pour le modèle TEF-1500-200-422 de la société Brimrose\footnote{http://www.brimrose.com/}, cette zone mesure 75 microns.
Le cristal est traité anti--reflet et le traitement ne supporte pas une intensité lumineuse supérieure à 5 W.mm$^{-2}$.
La puissance maximale incidente est donc limitée au seuil de dommage : $50~$mW.
A 800 nm, nous avons obtenu un maximum de 10\% d'efficacité de diffraction en simple passage pour un total de 1\% en double passage.
On dispose donc d'un maximum de 0.5 mW pour le faisceau sonde, ce qui sera amplement suffisant.
A 422 nm, nous avons obtenu un maximum de 15\% d'efficacité de diffraction en simple passage pour un total de 2\% en double passage.
Les courbes de calibration du MAO à 800 nm sont présentées sur la figure \ref{calibMAO}.

\subsection{Mesure du bruit technique sur le faisceau diffracté}
Pour produire une modulation d'indice importante dans le cristal, il est nécessaire d'utiliser une puissance élevée (ici de l'ordre de 1W) pour alimenter le PZT.
Deux étages sont nécessaires pour produire un signal radio fréquence (RF) de cette puissance : une source RF suivie d'un amplificateur. 
Nous allons comparer différentes méthodes de génération et d'amplification par rapport à un paramètre très important : le bruit en intensité introduit sur le faisceau diffracté.
Comme nous le verrons dans le chapitre \ref{ch5}, l'excès de bruit sur le faisceau diffracté doit être minimisé pour mesurer des fluctuations sous la limite quantique standard dans les expériences de mélange à 4 ondes dans une vapeur atomique.
\subsubsection{Source RF}
			\begin{figure}	
				\centering
				\includegraphics[width=8cm]{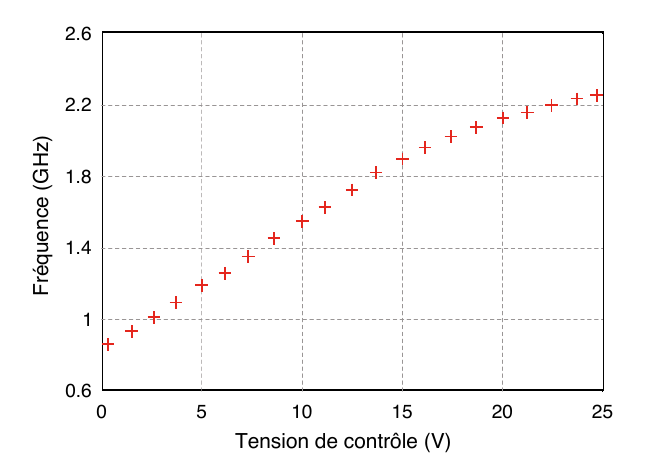}
				\caption[Calibration du VCO.]{Courbe de calibration du VCO ZX95-2150VW-S à 20$^\circ$C.\label{figvco}	}		
				\end{figure}
Pour synthétiser un signal RF à 1.5 GHz nous avons testé deux dispositifs : un oscillateur contrôlé par une tension électrique (VCO) et un synthétiseur de signaux.\\
Un VCO est un oscillateur électronique dont la fréquence d'oscillation dépend de la tension de contrôle appliquée à ses bornes.
Le modèle que nous avons utilisé est le ZX95-2150VW-S+ de la société Mini-circuits\footnote{http://www.minicircuits.com/}.
Ce modèle permet de générer un signal de $+4$ dBm entre 800 et 2300 MHz (voir la figure \ref{figvco} pour la calibration en fréquence).\\
La seconde méthode que nous avons utilisée, est basée sur le synthétiseur de signaux SMBV100 de la marque Rohde \& Schwarz\footnote{http://www2.rohde-schwarz.com/}.
Ce générateur délivre des signaux jusqu'à 3.2 GHz pour une puissance maximale de $+18$ dBm.\\
Ces deux générateurs ne délivrent pas assez de puissance pour piloter le PZT du MAO et un amplificateur est donc nécessaire.
\subsubsection{Amplificateur RF}
		 		\begin{figure}	
							\centering
							\includegraphics[width=13cm]{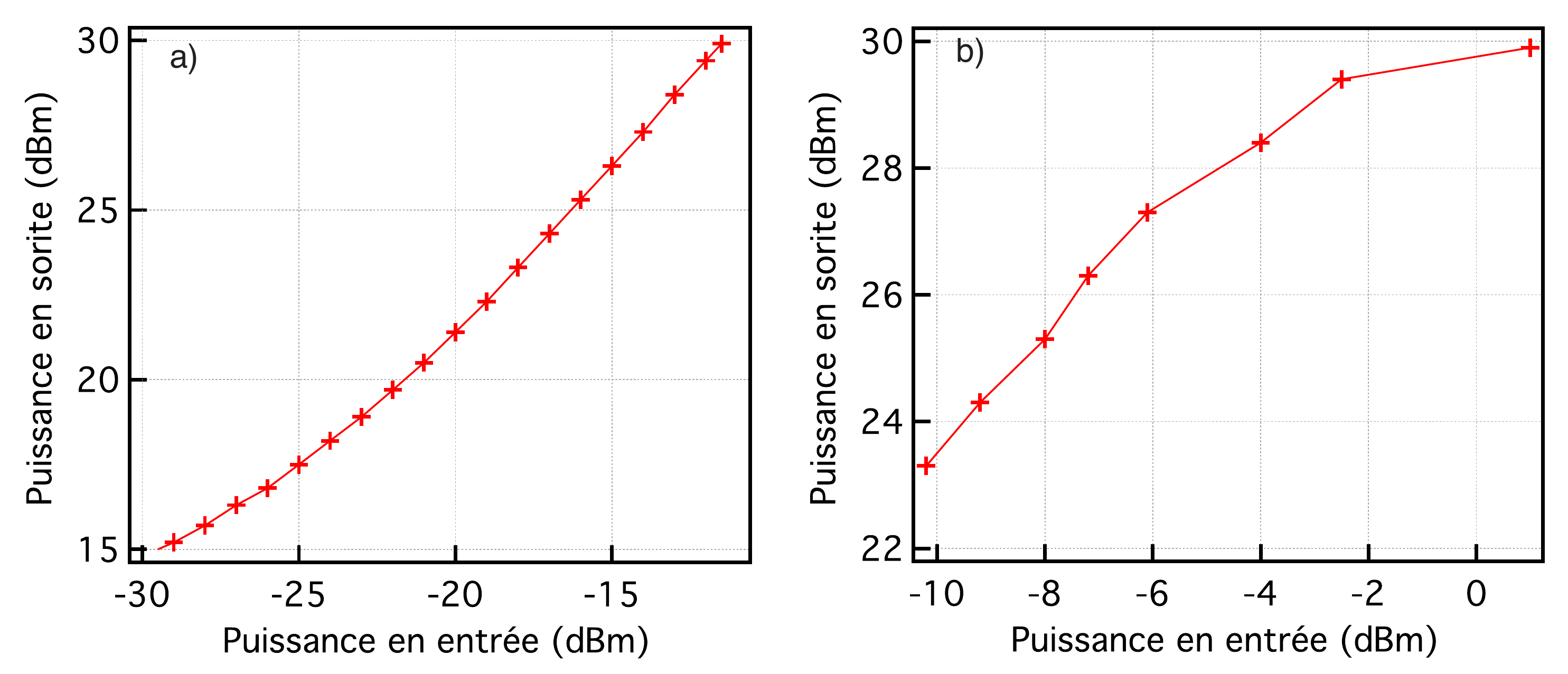}
							\caption[Réponse de l'amplificateur RF.]{Réponse de l'amplificateur RF. a) sans atténuation, c'est à dire en régime non-saturé. b) atténué de 7dB en sortie, c'est à dire saturé pour une puissance d'entrée de 1dBm.\label{ampli}	}		
							\end{figure}
L'amplificateur dont nous disposons au laboratoire est un Mini-circuits ZHL-5W-2GX.
Cet amplificateur a un gain nominal de 50 dB et une puissance de sortie de maximale de +37 dBm.
La puissance d'entrée maximale acceptable par l'amplificateur est de +1 dBm.
Dans le cas d'un fonctionnement en mode saturé, nous devons atténuer le signal de sortie de l'amplificateur pour ne pas endommager le MAO.
Nous utilisons un atténuateur de -6 dB avec un radiateur pour dissiper la chaleur (Mini-circuits MCL BW-S6W5).
Nous ajoutons un filtre passe haut Mini-circuits VHF-1200+, dont la bande passante à 2 dB est 1200-4600 MHz, afin de couper le bruit basse fréquence éventuellement présent sur la sortie de l'amplificateur.
A 1.5 GHz, la sortie de ce filtre est atténuée de 0.8 dB environ par rapport à l'entrée.
Les courbes de gain de l'amplificateur sont données sur la figure \ref{ampli} avec et sans atténuation, c'est à dire dans le régime saturé et non saturé (respectivement).
\subsubsection{Bruit technique}
\begin{figure}	
\centering
\includegraphics[width=10cm]{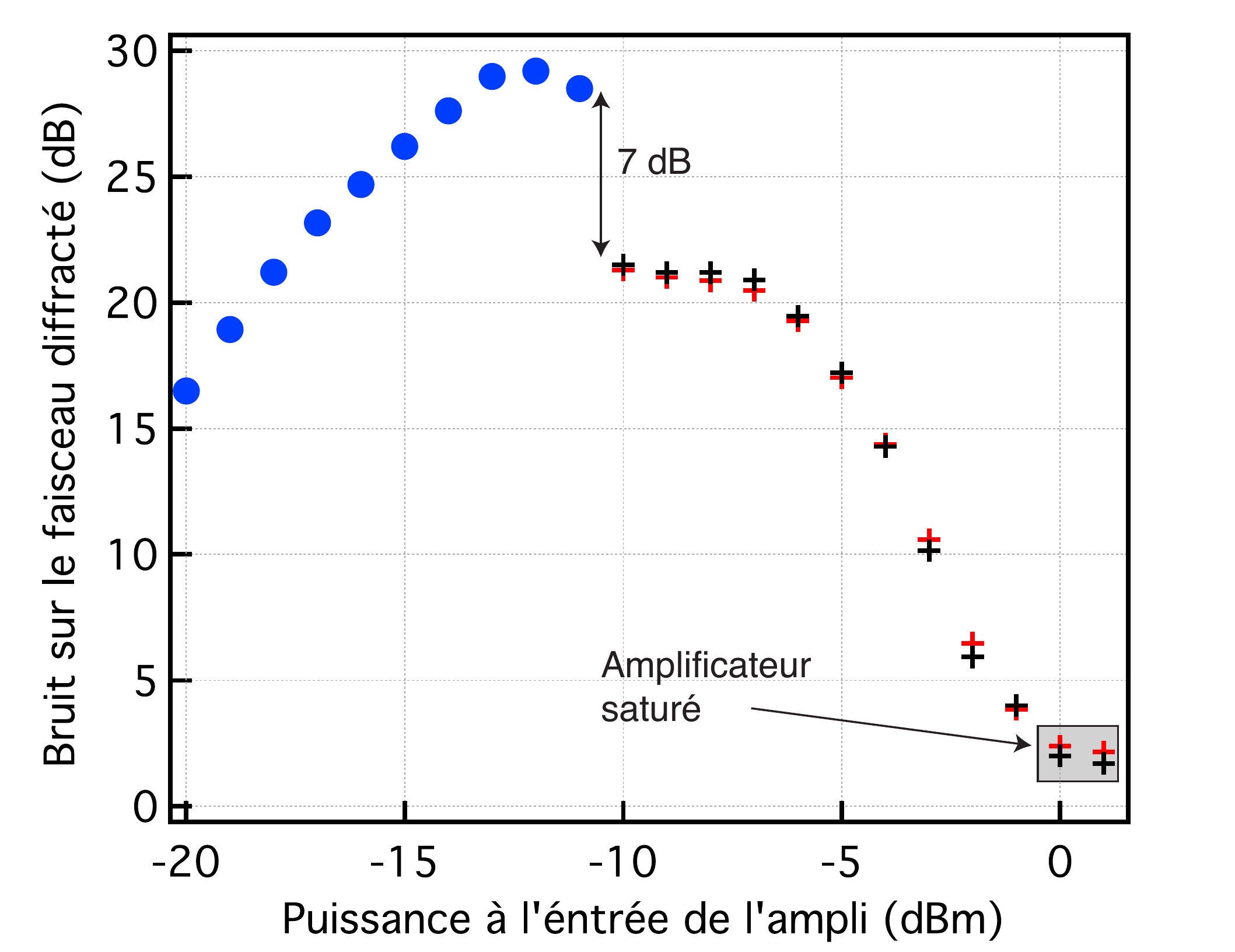}
\caption[Effet de la saturation de l'amplificateur sur le bruit du faisceau diffracté]{Effet de la saturation de l'amplificateur sur le bruit du faisceau diffracté. Les croix rouges et noires donnent le cas d'un amplificateur atténué en sortie respectivement sans et avec un filtre passe haut. Les points bleus ont été réalisés pour l'amplificateur sans atténuation. On peut constater que l'écart entre les deux courbes est de -7 dB, ce qui correspond à l'atténuation de sortie de l'amplificateur. \label{effetsat}	}		

\includegraphics[width=10cm]{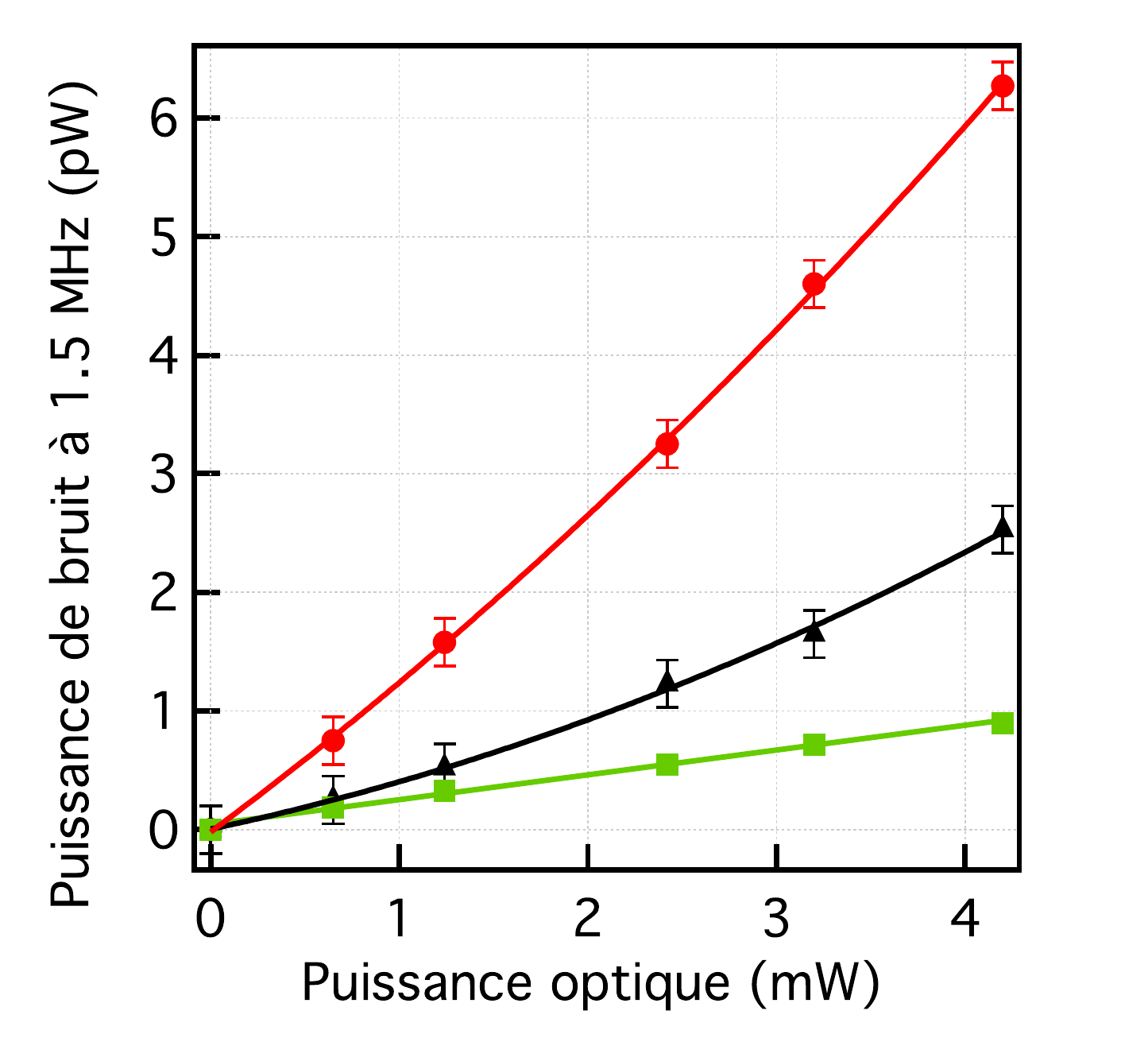}
	\caption[Excès de bruit sur le faisceau diffracté.]{Bruit sur le faisceau diffracté dans différentes configurations. En vert : le bruit quantique standard qui sert de référence. En noir, la source est le générateur de signaux et en rouge la source est un VCO. Notons que l'ajustement pour le bruit quantique est linéaire (vert), alors qu'il est quadratique pour les bruits techniques (rouge et noir).
   \label{bruitMAO}	}		
\end{figure}

Pour comparer les différents montages nous mesurons le bruit d'un faisceau diffracté par le MAO piloté par trois sources différentes : un VCO suivi d'un amplificateur saturé, un synthétiseur suivi d'un amplificateur non-saturé et un synthétiseur suivi d'un amplificateur saturé.
Pour ce faire nous utilisons un faisceau de 10 mW et une détection balancée (paragraphe \ref{detbal}) et nous mesurons le bruit à 1.2~MHz.\\
Premièrement, nous pouvons constater sur la figure \ref{effetsat} que l'atténuation du signal RF en sortie de l'amplificateur permet de réduire d'environ 7 dB le bruit sur le faisceau diffracté par rapport au cas où le signal est atténué en entrée (saut de 7 dB entre la série de points bleus et de croix rouges).
Cette atténuation très importante du bruit technique sur le faisceau diffracté, correspond aux 6.8 dB d'atténuation électronique sur la sortie de l'amplificateur.
Nous nous placerons donc par la suite dans cette configuration.\\
D'autre part, nous pouvons voir sur la figure \ref{effetsat}, que le bruit sur le faisceau diffracté diminue à mesure que l'on se rapproche de la saturation de l'amplificateur.
Le régime saturé, qui correspond au deux derniers points de la figure \ref{effetsat}, est donc le plus favorable en terme de bruit technique ajouté.
De façon complémentaire, on peut noter que l'ajout du filtre passe-bande permet une légère amélioration du bruit de 0.2 dB dans le régime saturé.\\

Dans un second temps nous avons étudié le rôle du générateur RF sur le bruit technique du faisceau diffracté en comparant un VCO et un synthétiseur de signaux.
On peut conclure grâce aux données de la figure \ref{bruitMAO}  que l'utilisation d'un synthétiseur de signaux comme source RF est préférable à un VCO.
En effet l'excès de bruit mesuré est de l'ordre de 1 dB pour le premier et de 4 dB pour le second à 3 mW.\\

Nous avons observé que le fonctionnement en régime saturé de l'amplificateur atténué en sortie améliore de façon significative le bruit sur le faisceau diffracté.
De plus pour la source RF, il est préférable d'utiliser un synthétiseur de signaux plutôt qu'un VCO.
Nous utiliserons donc la configuration optimale, c'est-à-dire un synthétiseur saturant l'amplificateur atténué en sortie.

		 	
\newpage
\section{Cellule de Rubidium}
\begin{figure}[]	
		\centering
		\includegraphics[width=4cm]{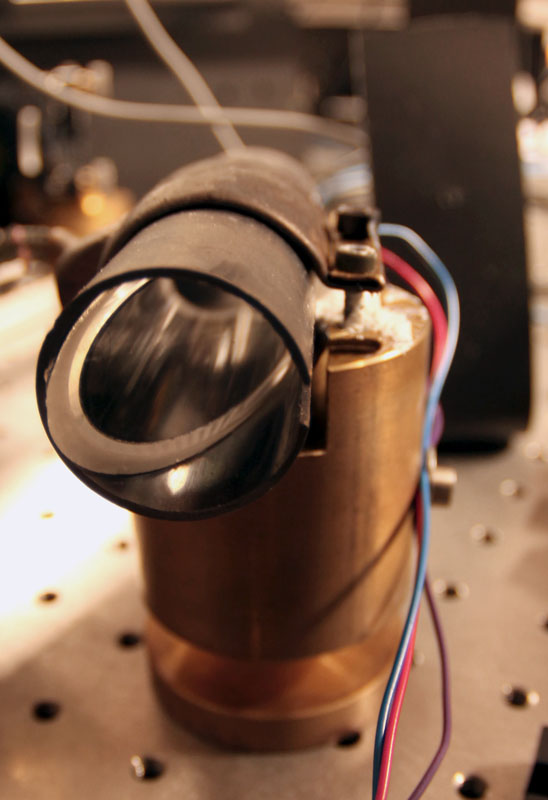}
			\caption[Cellule de rubidium.]{Cellule de rubidium  de 5 cm de long et dont les facettes sont non traitées et orientées à l'angle de Brewster. La cellule est dans un four placé sur un pied en laiton. Les câbles de la sonde de température PT-100 sont visibles sur la droite.\label{celluleRB}}		
		\end{figure}
	Le milieu dans lequel nous avons réalisé les expériences de mélanges à 4 ondes décrites dans ce manuscrit est une vapeur de $^{85}$Rb isotopique.
Cette vapeur est contenue dans une cellule en verre dont les facettes ont étés traitées anti-reflet\footnote{Nous disposons de deux cellules, l'une de 12.5 mm dont les faces sont traitées anti-reflets, l'autre de 5 cm de long et dont les facettes sont non traitées et orientées à l'angle de Brewster (voir figure \ref{celluleRB}).}.
La cellule est placée dans une bague qui supporte une résistance chauffante, alimentée en 220~V.
La bague est montée sur un pied en laiton qui est isolé de son support et de la table par une rondelle de 1 cm de Macor (isolant thermique).
Un capteur de température (PT-100) est placé sur le queusot de la cellule et un contrôleur de température permet de stabiliser l'ensemble avec une précision de l'ordre de $\pm 3K$.
A chaud ($T>100^\circ$C), les pertes optiques sur les faces de la cellule à 795 nm sont mesurées en incidence normale à 2.5\% $\pm 0.5\%$ pour chacun des deux hublots.

\subsection{Densité atomique}
La densité d'atomes en interaction est un paramètre qu'il faut contrôler pour les expériences de mélange à 4 ondes. 
Comme nous le verrons aux chapitres \ref{ch4} et \ref{ch5}, il est possible de calculer la densité théorique d'atomes en phase vapeur $n_{at}$ dans la cellule à l'aide de la loi des gaz parfaits et de la relation de Clausius-Clapeyron :
\begin{equation}
  \label{eq:11}
  n_{at} = \frac{1}{k_B T} 10^{A-  B/T}     
\end{equation}
où $T$ est la température en K et $k_B$ est la constante de Boltzmann.
Les coefficients $A=-9.138$ et $B=4040 K^{-1}$ sont reportés dans \cite{Alcock:1984p1928,Alcock:2001p1913}.
Cette relation fait l'hypothèse d'un gaz parfait à l'équilibre thermodynamique, ce qui n'est pas nécessairement le cas ici.
Pour affiner cette estimation, nous avons donc évalué quantitativement la densité d'atomes en phase vapeur dans la cellule.
Pour cela, nous avons étudié le profil d'une raie d'absorption d'un faisceau laser (de 100 $\mu$W focalisé sur 5 mm$^2$) dans la cellule de rubidium chauffée à différentes températures.
La mesure de l'absorption permet de déduire la fraction des atomes dans la cellule qui sont en phase vapeur par rapport à l'estimation donnée par la relation \ref{eq:11}.\\
	\begin{figure}	
			\centering
			\includegraphics[width=9.5cm]{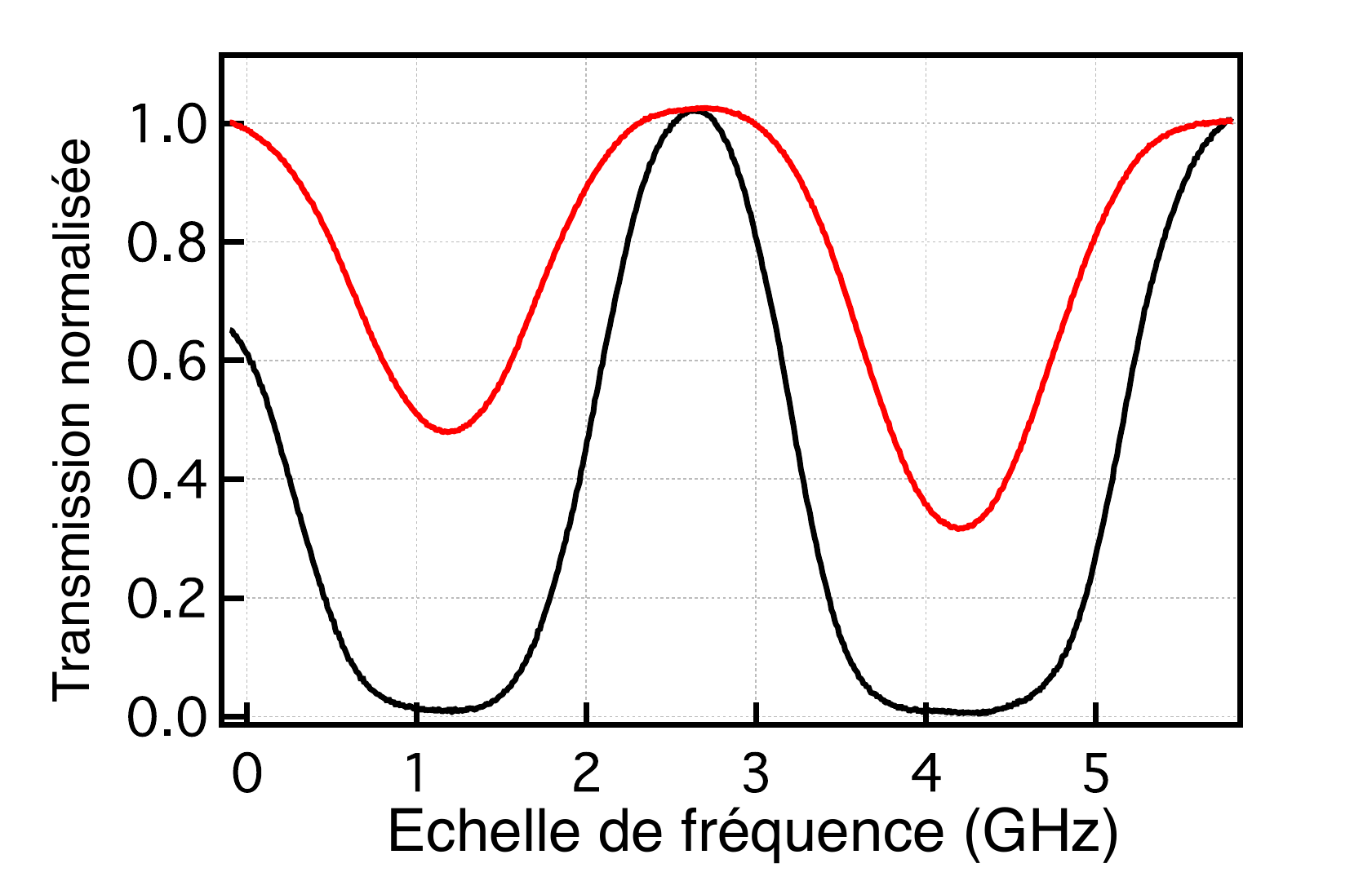}
				\caption[Absorption dans une cellule de rubidium.]{Spectre d'absorption de la raie $5S_{1/2} \rightarrow 6P_{1/2} $  dans une cellule de $^{85}$Rb à 100$^\circ$C (rouge) et 145$^\circ$C (noir) sur la raie .\label{absss}}		
			\end{figure}
			
Aux températures que nous étudions (au delà de 100$^\circ$C), il n'est pas simple de mesurer précisément l'absorption tant elle est importante.
En effet, on peut le voir sur la figure \ref{absss}, si à 100$^\circ$C la mesure est encore aisée, à 145$^\circ$C, le spectre est plat sur près de 1 GHz et la valeur du maximum d'absorption ne peut pas être mesurée simplement.
Pour résoudre ce problème et afin de déterminer la densité d'atomes, nous avons utilisé non pas la valeur du maximum d'absorption mais un ajustement sur l'ensemble du profil en prenant en compte l'élargissement Doppler.

\begin{figure}	
\centering
\includegraphics[width=9cm]{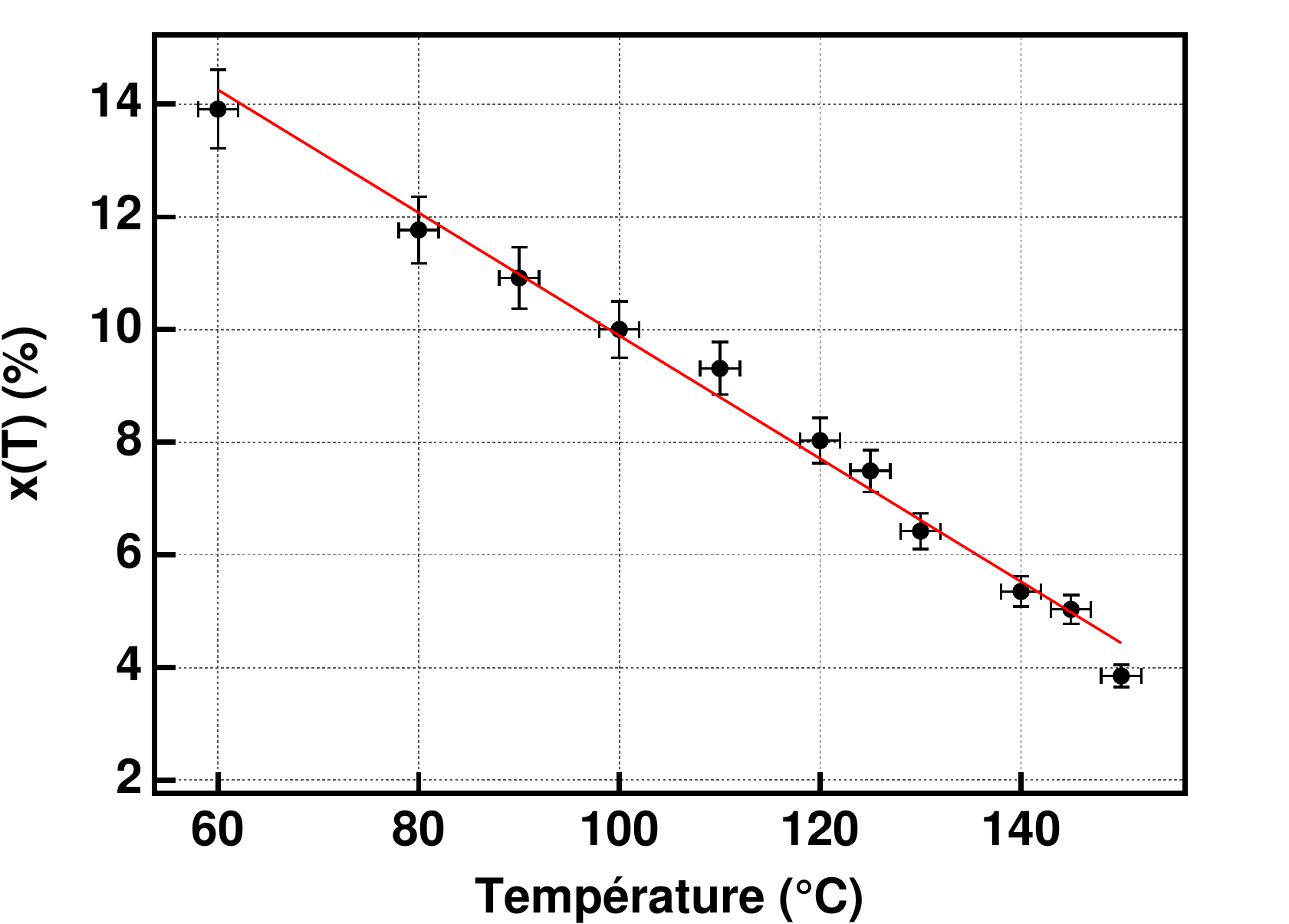}
\caption[Pourcentage d'atomes en phase vapeur par rapport à la densité prévue par le modèle thermodynamique]{Pourcentage d'atomes en phase vapeur par rapport à la densité prévue par le modèle thermodynamique. Ces points ont été obtenus par des mesures d'absorption sur la raie $5S_{1/2} \rightarrow 6P_{1/2} $  du $^{85}$Rb à 422 nm. En noir les points expérimentaux, en rouge l'ajustement linéaire.\label{abss}}	
\end{figure}

\begin{figure}	
\centering
\includegraphics[width=14cm]{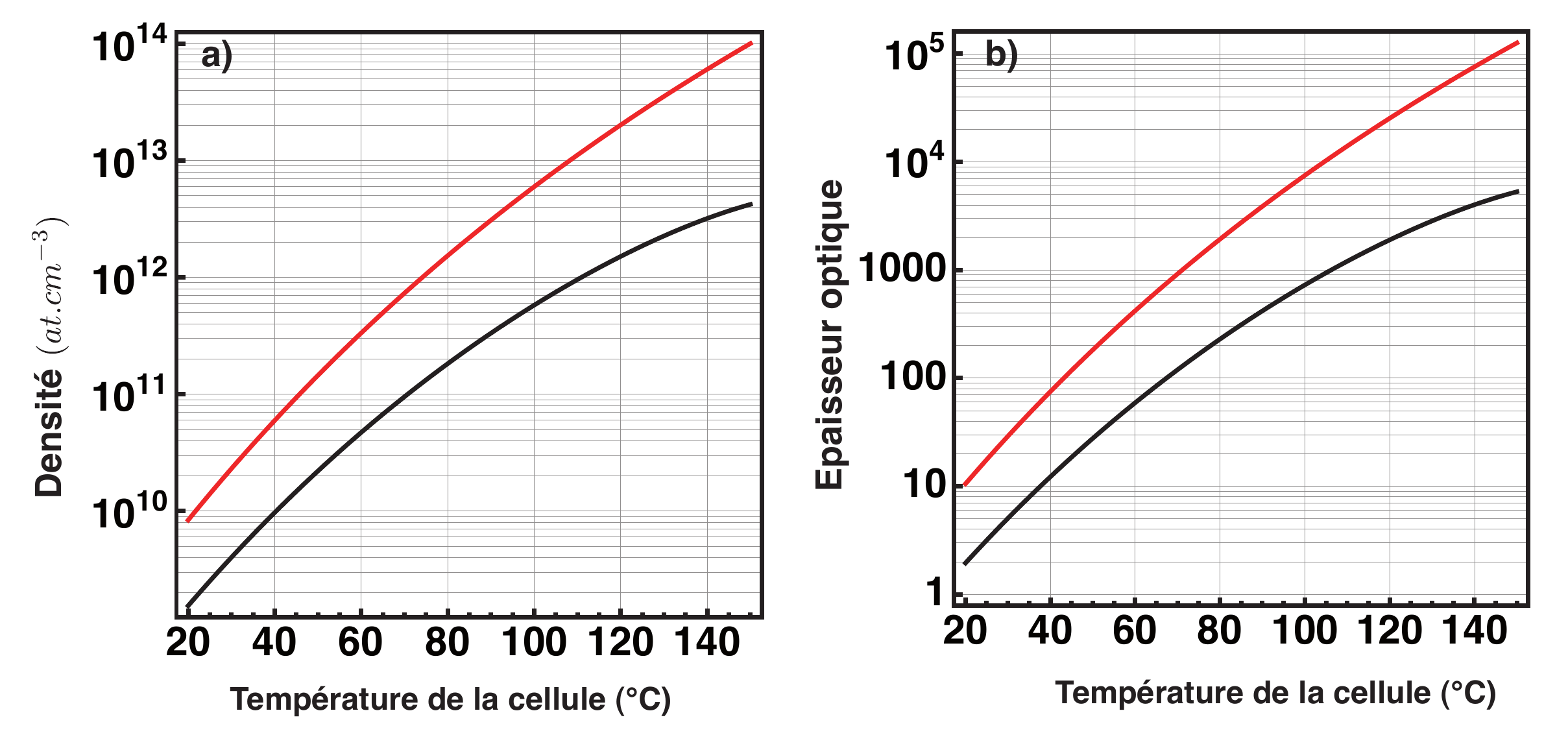}
\caption[Densité atomique et épaisseur optique en fonction de la température pour la raie D1 du $^{85}$Rb]{a) Densité atomique et b) épaisseur optique en fonction de la température pour la raie D1 du $^{85}$Rb. En rouge les valeurs données par la thermodynamique, en noir en prenant en compte la correction pour la part d'atomes en phase vapeur.\label{figdensite}}		
\end{figure}

Dans une vapeur atomique, la densité d'atomes $dn$ dans la classe de vitesse $d{\rm v}$ est donnée par la distribution de Maxwell-Boltzmann :
\begin{equation}\label{eq215}
dn=\frac{n}{\sqrt{2\pi}\sigma_{\rm v}}e^{-\rm v^2/\sigma_{\rm v}^2}d\rm v,
\end{equation}
avec $\sigma_{\rm v}^2=k_B T/m$, où $k_B$ est la constante de Boltzmann, $m $ la masse d'un atome et $T$ la température de la cellule.\\
En présence d'élargissement inhomogène l'intensité $I$, transmise à travers un milieu de longueur $L$ s'exprime sous la forme :
\begin{equation}
I(L)=I(0)\ e^{-\kappa L},
\end{equation}
avec
\begin{equation}\label{eq216}
\kappa=\frac{h \nu \mathcal{N}\ \Gamma/I_{sat}}{2\sigma_{v}}\sqrt{\frac{\pi}{2}}\ e^{-(\nu-\nu_0)^2/2\sigma_\nu^2} \text{\ \ et\ \ }\sigma_\nu=\nu_0\sqrt{k_B T/mc^2}
\end{equation}
En utilisant ce résultat, nous avons pu faire un ajustement de chacun des spectres d'absorption (pour différentes températures de la cellule) et déterminer la densité d'atomes en phase vapeur dans chaque cas.
Ces mesures ont été réalisées sur la raie $5S_{1/2} \rightarrow 6P_{1/2} $  du $^{85}$Rb, dont le dipôle est plus faible que pour  la raie D, afin d'obtenir des absorptions plus faibles.\\
Nous avons ensuite comparé ces données à la densité prévue par le modèle thermodynamique (relation \eqref{eq:11}) afin de définir un coefficient $x(T)$ tel que l'on ait :
\begin{equation}
  \label{eq:12}
  n_{at} = x(T)\frac{1}{k_B T} 10^{A-  B/T}
\end{equation}
La figure \ref{abss} présente cette comparaison et permet de déterminer  le coefficient $x(T)$ qui dépend de la température :
\begin{equation}\label{eq174d}
x(T)=20.7-0.11\ T\ (^\circ C)
\end{equation}
On notera que plus la température augmente, plus la densité mesurée s'éloigne de la densité calculée par la relation \eqref{eq:11}.
On notera de plus que la relation (\ref{eq174d}) est une relation empirique valable uniquement pour la cellule considérée et dans la gamme de températures étudiée.
Cette formule permet de tracer la densité atomique en fonction de la température pour cette cellule de rubidium (voir figure \ref{figdensite}) pour le modèle thermodynamique d'une part et en prenant en compte ce coefficient d'autre part.
L'épaisseur optique $\alpha L$ est définie par la relation :
\begin{equation}
\alpha L = n_{at} \sigma L,
\end{equation}
où $L$ est la longueur du milieu et $\sigma$ la section efficace de la transition (voir annexe \ref{Annexe_Rb}).
Dans la gamme de températures étudiée, nous pouvons donc constater qu'il y a à peu près un ordre de grandeur entre la densité estimée par la thermodynamique et la valeur mesurée expérimentalement.
	\subsection{Auto-focalisation par effet Kerr optique}
\begin{figure}	
\centering
\includegraphics[width=12cm]{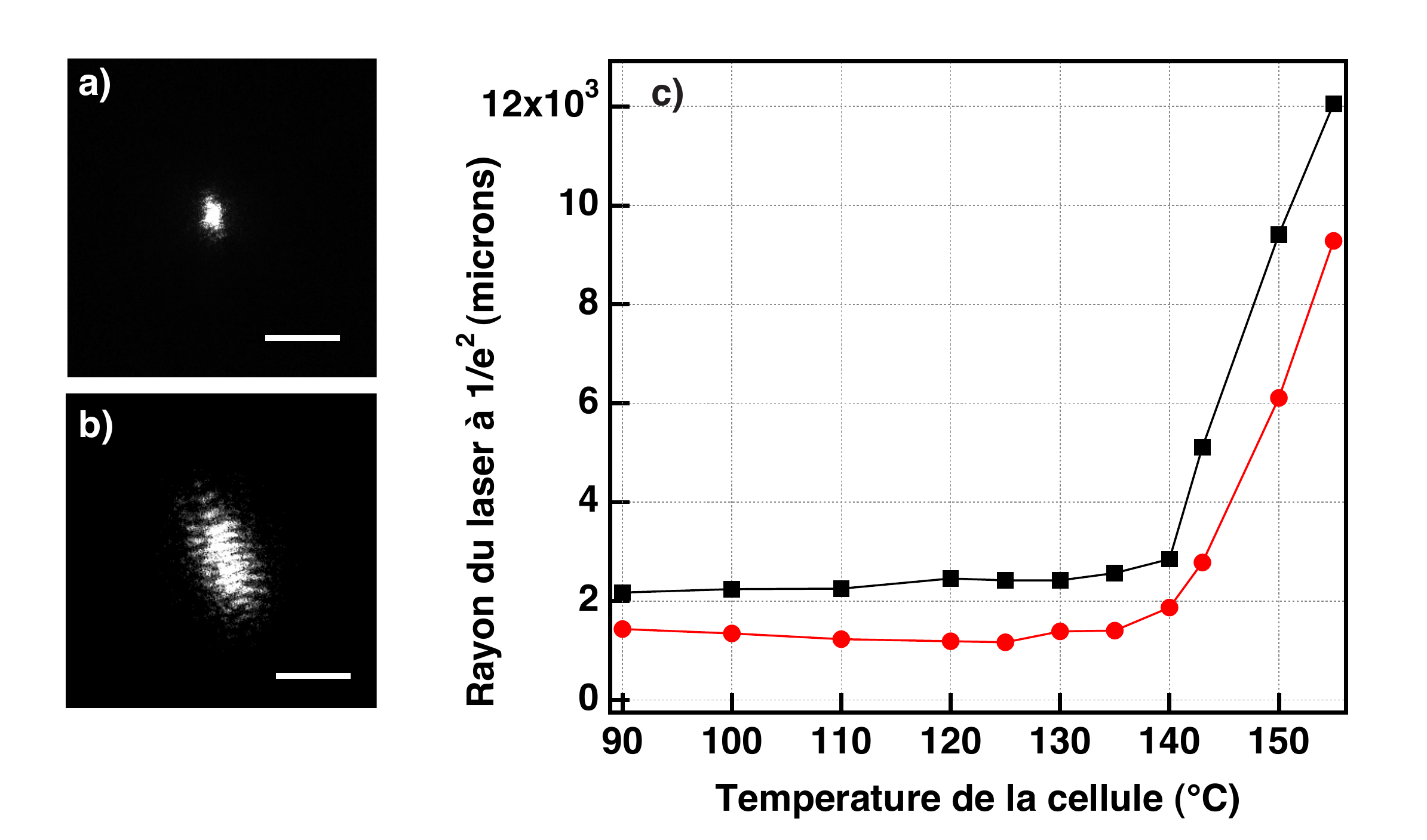}
\caption[Auto--focalisation par effet Kerr]{Auto--focalisation par effet Kerr. Profils transverses du laser (pour une puissance de 1.5 W et un rayon de 650 microns au niveau du waist dans la cellule)  pour une cellule chauffée à 100$^\circ$C a) et à 145$^\circ$C b).
Sur les figures a) et b), l'échelle est indiquée par le trait blanc qui correspond à 1 mm. Sur la figure c) les points noirs et les ronds rouges correspondent respectivement au rayon du faisceau laser à 1/e$^2$ dans l'axe Y et X mesuré dans le plan du capteur de la caméra CCD à 1.3 m de la cellule, en fonction de la température. \label{kerr_auto}}		
\end{figure}

Dans les milieux présentant une susceptibilité non--linéaire $\chi^{(3)}$, on peut observer le phénomène d'auto-focalisation par effet Kerr.
L'effet Kerr décrit une modification de l'indice dans un milieu $\chi^{(3)}$ sous l'effet d'un champ électrique.
La variation d'indice (l'indice non--linéaire) est proportionnelle à l'intensité du champ excitateur.
Pour un faisceau laser TEM00, le profil d'intensité est gaussien, et l'intensité au centre du faisceau est plus grande que sur les bords.
L'indice sera donc différent entre le centre du faisceau et ses bords.
Le milieu agit alors comme une lentille pour le faisceau laser excitateur, c'est pour cela que l'on parle d'auto--focalisation.\\
Nous avons observé expérimentalement l'auto-focalisation dans une vapeur atomique de rubidium 85 proche de résonance (800 MHz de désaccord) et étudié l'effet de la densité d'atomes (i.e. de la température de la cellule) sur ce phénomène (voir figure \ref{kerr_auto}).
Le faisceau laser utilisé a une puissance de 1.5 W et il est focalisé sur un rayon de 650 microns au niveau de la cellule, ce qui correspond à une distance de Rayleigh de 1.6 m.
En plaçant la caméra qui nous permet de mesurer la taille du faisceau à 1.3 m, on peut considérer que le faisceau est collimaté en l'absence d'effet d'auto--focalisation.
On peut alors noter qu'au delà de 145$^\circ$C, l'auto-focalisation n'est plus négligeable et que le profil transverse du faisceau est modifié.
Cela nous donne donc une borne supérieure de la température (et donc de la densité atomique) que l'on pourra explorer dans les expériences de mélange à 4 ondes décrites dans ce manuscrit au chapitre 5.
D'après les résultats de la section précédente, cette limite correspond à une épaisseur optique autour de 5000 pour la raie D1 du $^{85}$Rb. 

\section{Photodétection}
Nous avons décrit au chapitre 1 le champ électromagnétique à l'aide de ses quadratures.
Nous avons associé à ces quadratures des observables, ainsi que des états du champ.
L'objectif de cette section est de comprendre comment sont mesurées, expérimentalement, ces observables.
\subsection{Efficacité quantique}
Les photodétecteurs qui nous permettent de mesurer les quadratures décrivant le champ dans le régime des variables continues, sont des photodiodes.
Deux montages vont concentrer notre attention : d'une part la détection équilibrée qui va permettre de mesurer le bruit quantique standard en s'affranchissant du bruit technique et d'autre part la mesure de corrélations d'intensité que nous allons utiliser pour caractériser le caractère non-classique des faisceaux générés par mélange à 4 ondes.\\
Comme nous l'avons vu au chapitre 1, le courant généré par une photodiode $i(t)$ dépend du nombre de photons $N_{ph}$ incidents par intervalle de temps $\Delta t$.
En effet, les photons incidents sont convertis en électrons avec un rendement $\eta$ inférieur à 1 que l'on appelle efficacité quantique.
On obtient :
\begin{equation}
i(t)=\frac{\eta N_{ph}(t) e}{\Delta t}=\frac{ \eta P(t) e }{\hbar \omega},
\end{equation}
où l'on a introduit $P(t)$ la puissance optique du faisceau étudié, $\hbar \omega$ l'énergie d'un photon à la fréquence $\omega$ et $e$ la charge de l'électron.\\

\subsubsection{Détection équilibrée\label{detbal}}
Le montage de détection équilibrée permet de mesurer le bruit quantique standard en s'affranchissant du bruit technique du laser.
Le formalisme de ce système de détection a été introduit au paragraphe \ref{diffint} du chapitre \ref{ch1}.
Le faisceau laser que l'on souhaite étudier est envoyé sur une lame séparatrice semi-réfléchissante.
Les deux faisceaux en sortie de la lame sont détectés dans deux photodiodes indépendantes.
On mesure ensuite la différence et la somme d'intensité.
Comme nous l'avons vu, la différence donne directement accès au bruit quantique standard pour une puissance moyenne donnée.
La somme permet de mesurer un éventuel excès ou réduction de bruit par rapport au bruit quantique standard.

\subsubsection{Corrélation d'intensité\label{correl}}
Le montage pour la détection de corrélation d'intensité est très proche du précédent (voir paragraphe \ref{corel_ch1} du chapitre \ref{ch1}).
Cette fois les deux faisceaux qui sont détectés par les photodiodes ne proviennent pas d'un même faisceau scindé en deux par une lame 50/50, mais du processus non linéaire de conversion par mélange à 4 ondes.
La différence d'intensité, nous renseigne donc sur les corrélations entre les deux modes du champ détectés par les deux photodiodes.
Pour connaitre le bruit individuel d'un des deux faisceaux, il suffit de bloquer la deuxième voie et de mesurer le bruit détecté par une seule des deux photodiodes.

\subsection{Caractérisation des photo--détecteurs utilisés}
	\begin{figure}	
		\centering
		\includegraphics[width=10cm]{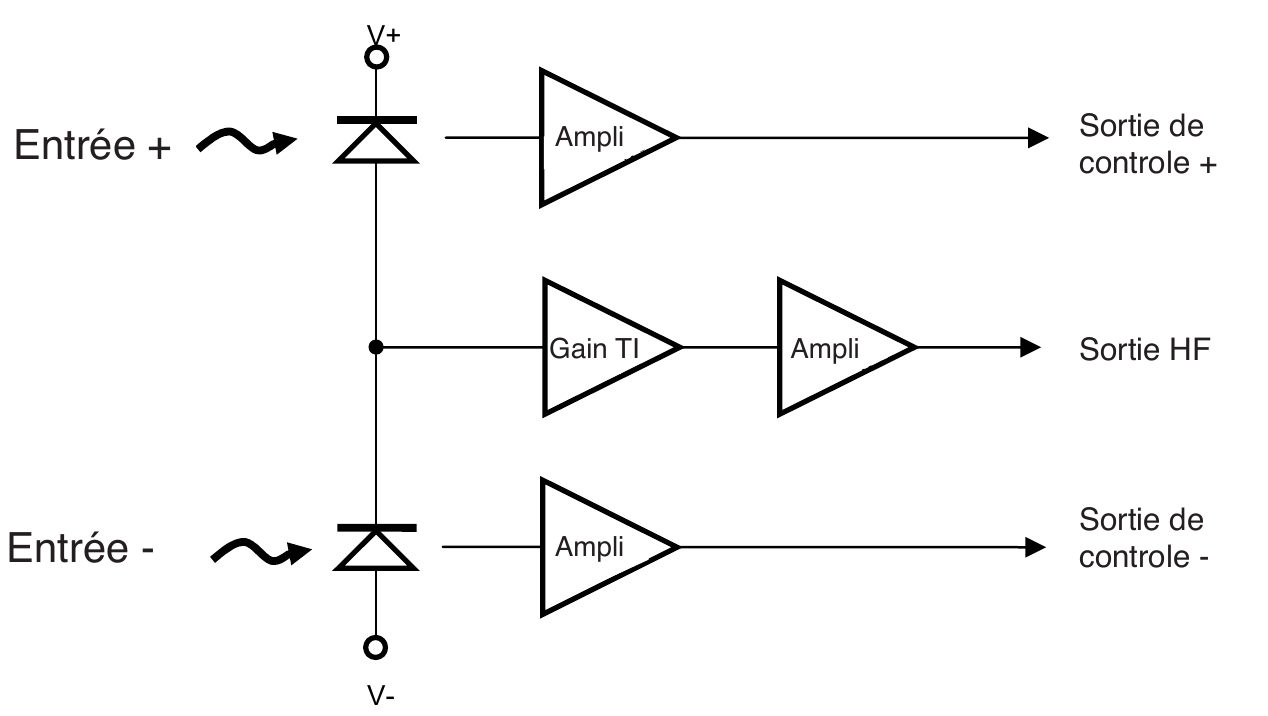}
			\caption[Schéma de principe du bloc de photodiode PDB 150]{Schéma de principe du bloc de photodiode PDB 150. Les deux entrées correspondent aux deux photodiodes. Deux voies basse-fréquence permettent de surveiller les signaux sur les deux entrées. Le gain de ces voies n'est pas réglable. Une voie haute fréquence permet de détecter le bruit sur la différence d'intensité des deux champs incidents. Le gain transimpédance (Ampli TI) est réglable.\label{pd1}}		
		\end{figure}
\begin{figure}	
			\centering
			\includegraphics[width=9cm]{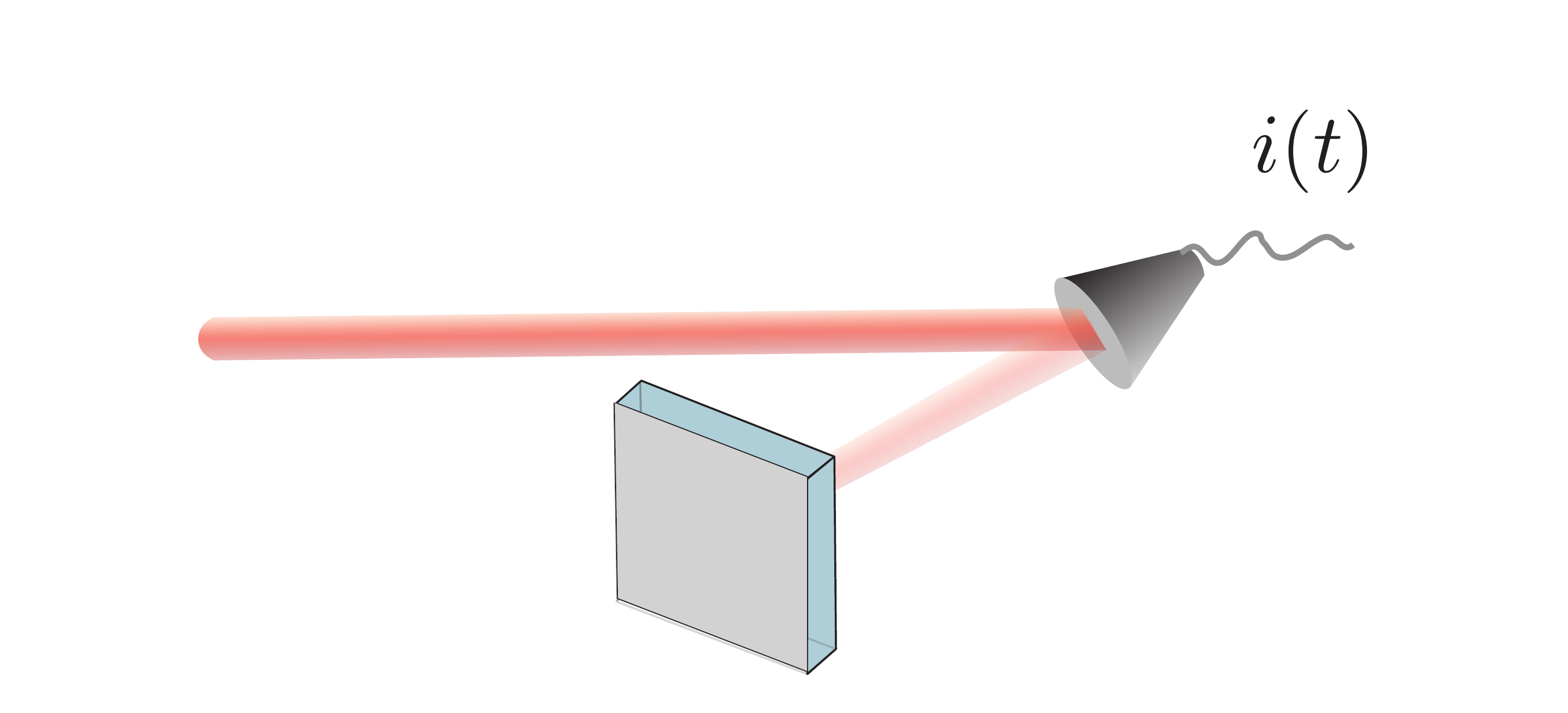}
				\caption[Montage d'une photodiode à l'incidence de Brewster.]{Montage d'une photodiode à l'incidence de Brewster, avec rétro-réflexions des réflexions résiduelles.\label{pd2}}		
			\end{figure}

Les photo--détecteurs que nous avons utilisés sont basés sur le montage de détection équilibrée PDB150 de la société Thorlabs\footnote{http://www.thorlabs.de/}.
Il s'agit d'un montage électronique d'amplification et de deux photodiodes en silicium ayant un rendement théorique donné par le constructeur de 0.53~$A/W$ à 795 nm, ce qui correspond à une efficacité quantique de 83\%.
Les deux photodiodes sont montées de façon inversée l'une par rapport à l'autre (voir figure \ref{pd1}) de manière à ce que l'électronique n'amplifie que la différence de courant entre les deux voies\footnote{Dans ce montage, il n'est donc pas possible de mesurer la somme des photo-courants entre les deux voies.
Lorsque nous avions besoin de cette mesure, nous avons utilisé deux photodiodes S3883 de la société Hamamatsu (http://www.hamamatsu.com/) et des circuits d'amplification bas-bruit fabriqués au laboratoire.}.
Le gain transimpédance de la voie haute fréquence (HF) du montage est réglable par décades entre $10^3$ et $10^7\ V/A$.
Il est important de noter que le gain transimpédance fourni par le fabricant est réduit d'un facteur 2 lorsque la sortie est terminée sur 50 Ohms.
Le gain des voies de contrôle basse fréquence (BF) est donné par le constructeur et vaut 10V/mW à 800 nm.
Un des paramètres importants pour quantifier les performances d'une détection équilibrée est le taux de rejection (\textit{common mode rejection}).
Nous avons mesuré ce taux en appliquant une modulation sur le laser, et on obtient une valeur de 34 dB de rejection à 1.5 MHz.\\
La fréquence de coupure basse de la voie HF dépend du gain utilisé.
Typiquement pour un gain de $10^3\ V/A$ la fréquence de coupure à -3 dB vaut 100 MHz et pour un gain de $10^5\ V/A$ que nous utiliserons par la suite la fréquence de coupure est de 8 MHz.\\
Le bruit électronique en l'absence de champ (bruit de fond des détecteurs) à 1.5 MHz a une valeur de -80 dBm (pour une bande passante de résolution RBW = 100 kHz).\\
Ce montage a donc de bonnes propriétés de bruit de fond et de taux de rejection mais l'efficacité quantique des photodiodes est insuffisante pour mesurer avec précision des niveaux importants de compression sous le bruit quantique standard.
En effet, nous l'avons vu au chapitre \ref{ch1}, les pertes font tendre les mesures de bruit vers la limite quantique standard.
Nous avons donc changé les photodiodes de série pour les remplacer par des photodiodes S3883 produites par la société Hamamatsu, dont nous avons de plus retiré le verre de protection.
En plaçant ces photodiodes à l'incidence de Brewster et en rétro--réfléchissant les réflexions résiduelles (voir figure \ref{pd2}), nous avons estimé leur efficacité quantique totale à 95 \% $\pm 2$\% en comparant les valeurs mesurées à celles données par un détecteur calibré.
\subsection{Effet des imperfections expérimentales}

\subsubsection{Effet des pertes sur la mesure de compression}\label{xdfr}
L'effet des pertes sur les mesures en optique quantique a été introduit au paragraphe \ref{parag_pertes} du chapitre \ref{ch1}.
On rappelle que la densité spectrale de bruit mesuré $S_{m}$ pour une efficacité globale du processus $\eta_T$ est donné par :
\begin{equation}\label{efpertes}
S_{m}=\eta_T S+(1-\eta_T).
\end{equation}
où $S$ est la densité spectrale de bruit idéale (c'est à dire en l'absence de pertes).\\
Dans les expériences décrites dans ce manuscrit, les pertes sont de plusieurs origines :
\begin{itemize}
\item les réflexions sur la face de sortie de la cellule de rubidium mesurées à 2.5~\%~$\pm 0.5\%$;
\item les pertes introduites par la transmission d'un cube séparateur de polarisation (de la société Fichou) mesurée à 2.5~\%~$\pm 0.5\%$;
\item l'efficacité quantique de la photodiode S3883 à l'incidence de Brewster, avec rétro-réflexions des réflexions résiduelles, estimée à $\eta=95$~\%~$\pm 2\%$
\end{itemize}
L'efficacité globale de collection est donc estimée à $\eta_T= 90 \pm 3\%$.
Cette efficacité va donc limiter notre détectivité à $90 \pm 3\%$ de compression soit entre $-9$ dB et $-12$ dB mesuré pour une compression parfaite.

\subsubsection{Effet des pertes sur la mesure de corrélations}
\begin{figure}	
\centering
\includegraphics[width=8cm]{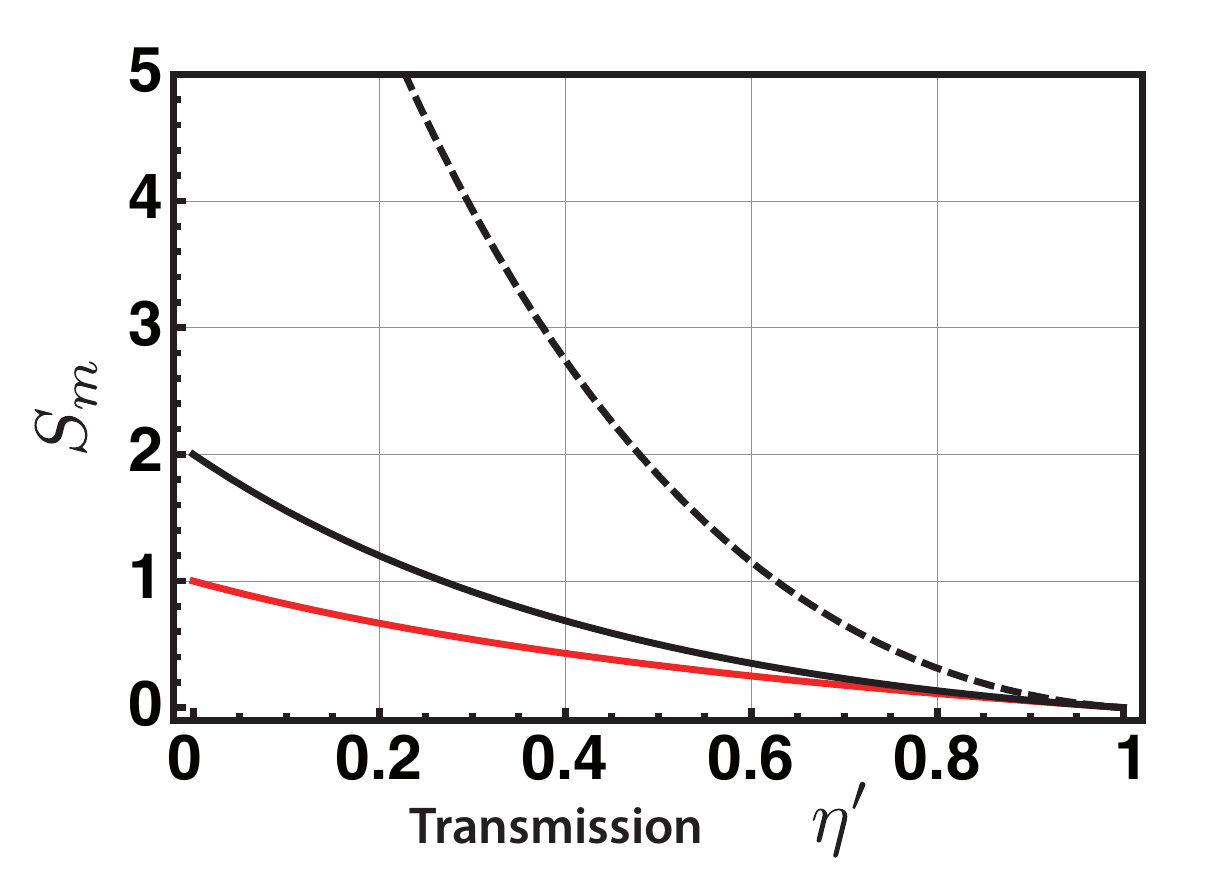}
	\caption[Effet de $\eta$ sur la densité spectrale de bruit pour la différence d'intensité de deux modes.]{Effet de la transmission du mode $\hata$ notée $\eta'$ sur la densité spectrale de bruit pour la différence d'intensité de deux modes pour différentes valeurs d'excès de bruit sur les modes $\hata$ et $\hat b$ (en rouge 0dB, en noir plein +3dB, en noir pointillés +10dB) dans le cas de faisceaux parfaitement corrélés.\label{pd3}}		
\end{figure}
Lorsque l'on mesure les corrélations entre deux modes, les pertes ne sont pas nécessairement identiques sur les deux voies.
Dans ce cas, il faut décomposer le problème en deux étapes.
Dans une premier temps, on identifie les pertes qui sont présentes sur les deux faisceaux et non corrélées (typiquement l'efficacité quantique des photodiodes inférieure à 1).
A l'aide de l'équation \eqref{deltaN-}, on peut démontrer que ces pertes modifient la mesure du bruit de la même façon que celles décrites dans la section \ref{xdfr}.
On traitera donc ces pertes de manière séparée.\\
Le problème se résume alors à une situation déséquilibrée, où les pertes se concentrent sur l'un des deux faisceaux et peuvent être considérées comme nulles sur l'autre.
Une étude détaillée de ce problème est faite au chapitre 3.
Nous donnons ici  un résultat simple dans le cas où l'intensité moyenne des deux champs est identique (avant les pertes).
Dans ce cas, on peut montrer que la densité spectrale de bruit mesurée est égale à :
\begin{equation}\label{228}
S_{m}=\frac{1}{1+\eta'}\left(\eta'^2 S_a +S_b -2 \eta' \bra\delta\hat X_a\delta \hat X_b\rangle +\eta' (1-\eta')\right),
\end{equation} 
où l'on a introduit $S_a$ et $S_b$ respectivement  la densité spectrale de bruit sur le mode $\hata$ et sur le mode $\hat b$ et $\eta'$ qui correspond à la transmission du faisceau $\hat a$.
Pour des corrélations parfaites, on peut faire l'hypothèse que $S_a=S_b$ (cela revient, dans le cas de corrélations parfaites, à supposer que le bruit sur les deux champs pris individuellement sont égaux).
Dans ce cas l'équation (\ref{228}) se simplifie en :
\begin{equation}
S_{m}=\frac{(1-\eta')^2}{1+\eta'}S_a+\frac{(1-\eta')\eta'}{1+\eta'}.
\end{equation}
On voit donc, que la mesure du spectre de bruit, dans ce cas, va dépendre du bruit sur les deux champs pris individuellement.
La figure \ref{pd3} donne l'effet de la transmission du mode $\hata$ sur la densité spectrale de bruit pour la différence d'intensité de deux modes pour différentes valeurs de bruit individuel sur les modes $\hata$ et $\hat b$.
Pour des corrélations supposées parfaites, $S_{m}$ tend vers 0 lorsque la transmission est proche de 1.
Si les modes $\hata$ et $\hat b$ sont considérés comme étant individuellement à la limite quantique standard (courbe rouge de la figure \ref{pd3}), alors le spectre $S_{m}$ tend vers 1 lorsque les pertes augmentent.
De manière générale, $S_{m}$ tend vers $S_a$ lorsque la transmission du champ $\hata$ devient faible.
Expérimentalement, on sait \cite{McCormick:2007p652} qu'il est possible d'observer des corrélations entre $\hata$ et $\hat b$ sous la limite quantique standard malgré un bruit individuel sur les faisceaux supérieur à cette limite.
On donne donc sur la figure \ref{pd3} (courbe en pointillés) le cas qui correspond à un excès de bruit de 10 dB sur chacun des faisceaux $\hata$ et $\hat b$ par rapport à la limite quantique standard.
On peut voir sur cette courbe qu'une diminution de la transmission du champ $\hata$ va dégrader les corrélations mesurées, et qu'en dessous de 65\% de transmission on ne mesure plus de corrélations sous  la limite quantique standard dans ce cas.

\subsubsection{Effet du bruit électronique sur la mesure de compression}
Le signal des mesures de corrélations que nous souhaitons détecter et mesurer est souvent très faible et donc difficile à discriminer d'autres sources de bruit.
Les bruits des différents appareils électroniques de la chaine de détection s'ajoutent au signal (le bruit du faisceaux lumineux) et peuvent fausser sa détection.
Les deux bruits que nous devons prendre en compte sont le bruit de l'analyseur de spectre, et le bruit de l'électronique de détection (photodiode et amplificateur).\\
Le bruit de l'analyseur de spectre est très bas (typiquement inférieur à -100 dBm).
Le bruit électronique est mesuré à -80 dBm pour une bande passante de résolution de l'analyseur de spectre, RBW=100 kHz 
Ces bruits sont indépendants et décorrélés des observables du champ, ils vont simplement s'ajouter au signal pour donner $S_{global}$  la densité spectrale de bruit globale réellement mesurée :  
\begin{equation}
S_{global}=S_{m}+S_{elec},
\end{equation}
où $S_{elec}$ est la densité spectrale de bruit de la chaine de détection.
Comme $S_{elec}$  est facilement mesurable en bloquant tous les champs lumineux, on peut calibrer le bruit électronique et le soustraire aux mesures.
On présentera donc dans ce manuscrit, le bruit mesuré brut et le bruit corrigé du bruit électronique.

\section{Calibration du bruit quantique standard}
Comme nous l'avons déjà introduit, les mesures de bruit (corrélations) doivent être rapportées au bruit quantique standard.
Il est donc essentiel de calibrer de manière très précise ce bruit.
Pour cela nous avons utilisé une détection balancée introduite au chapitre 1 et décrite expérimentalement à la section \ref{detbal} de ce chapitre.

\subsection{Analyseur de spectre}\label{Agilent}
L'appareil de mesure que nous avons utilisé est un analyseur de spectre N1996A de la société Agilent\footnote{www.agilent.com/}.
Généralement, les gammes de fréquence que nous avons étudiées se situent entre 500 kHz et 6 MHz.
La limite basse est la limite au delà de laquelle le bruit technique du laser devient important.
La limite haute, quant à elle correspond à la bande passante à -3dB des photodiodes pour un gain de $10^5$~A/W.\\
Il est important de donner quelques précisions sur le fonctionnement d'un analyseur de spectre.
Comme nous l'avons souligné au chapitre 1, la puissance de bruit mesurée dépend de la bande passante de l'appareil de détection.
La bande passante est fixée par le réglage de la bande passante de résolution (RBW) sur l'analyseur de spectre.
Une modification de la RBW va donc induire une modification sur le niveau de bruit mesuré.
Il est donc impératif de toujours donner la RBW lorsque l'on présente des spectres de bruit.
D'autre part, la bande passante vidéo (VBW) permet de moyenner (à l'affichage) les spectres, mais n'influe pas, en principe, sur la valeur mesurée. Il faut néanmoins faire attention au type d'opération effectué, à savoir si la moyenne est réalisée avant ou après la conversion de la puissance en échelle logarithmique.
\subsection{Mesures}
\begin{figure}	
				\centering
				\includegraphics[width=12cm]{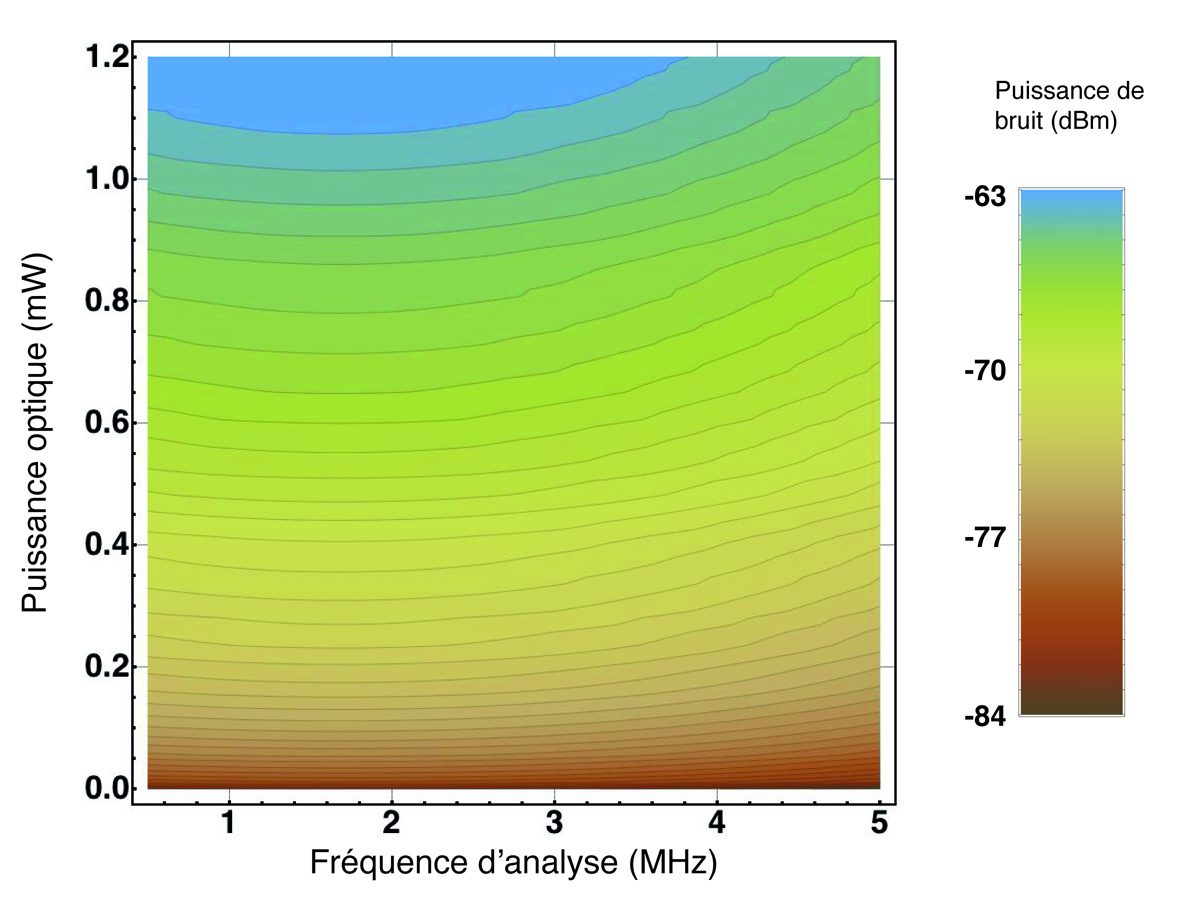}
					\caption[Bruit quantique standard en fonction de la puissance et de la fréquence d'analyse.]{Bruit quantique standard en fonction de la puissance et de la fréquence d'analyse .\label{bruit2}}		
				\end{figure}

Le bruit quantique standard est un un bruit blanc, c'est à dire qu'il ne dépend pas de la fréquence à laquelle on le mesure.
Théoriquement, il pourrait donc suffire de faire une mesure pour une fréquence fixée afin de déterminer la densité spectrale de bruit pour le bruit quantique standard.
En pratique, la chaîne de détection n'a pas une réponse plate en fréquence et l'on devra acquérir un spectre (et non juste un point) pour déterminer la valeur du bruit quantique standard à toutes les fréquences.\\
Nous avons donc réalisé une série de spectres de bruit entre 0 et 1.2 mW (le gain des photodiodes était réglé sur $10^5$~A/W, la RBW = 100 kHz et la VBW= 10 Hz).
De là, il a donc été possible d'extrapoler les points pour obtenir carte 2D du bruit quantique standard en fonction de la puissance et de la fréquence d'analyse (voir figure \ref{bruit2}).
Dans la suite lorsque nous comparons nos données au bruit quantique standard, c'est à l'ajustement 2D de cette carte que nous nous referons.

\section*{Conclusion du chapitre}

Dans ce chapitre nous avons présenté les différents outils expérimentaux permettant de générer des états non-classiques du champ.
A l'aide une brève revue de la littérature, nous avons introduit les différentes techniques utilisées, et principalement la technique que nous avons retenue, c'est à dire le mélange à 4 ondes dans une vapeur atomique.
Dans un second temps nous avons donné les caractéristiques techniques des différents instruments utilisés au cours de ce travail de thèse, à savoir un laser titane saphir, un modulateur acousto-optique, une cellule de rubidium, un système de photodétection ainsi qu'un analyseur de spectres.
Deux résultats nouveaux ont été obtenus.
\begin{itemize}
\item Nous avons mis en évidence le régime dans lequel devait fonctionner une source RF et un amplificateur pour minimiser le bruit technique sur le faisceau diffracté par un MAO alimenté par cette source.
\item Nous avons démontré que la densité d'atomes dans notre cellule était surévaluée par la relation de Clausius-Clapeyron et nous avons obtenu une loi empirique pour déterminer cette densité en fonction de la température de la cellule.
\end{itemize}

%% file: chapitre3v5.tex
\chapter{Approche phénoménologique}\label{ch3}
\setcounter{minitocdepth}{2}
\minitoc
\section{Mélange à 4 ondes en optique non--linéaire }
Les processus d'amplification sensible et insensible à la phase sont deux processus paramétriques d'optique non-linéaire largement étudiés \cite{Mollow:1967p14234,Baumgartner:1979p14191,Collett:1984p14218,Holm:1987p10558,Gigan:2006p14361}.
On peut, entre autres, observer ces processus dans un milieu de susceptibilité non--linéaire $\chi^{(3)}$ par mélange à 4 ondes \cite{OrvilScully:1997p3797,RShen:2003p7271,WBoyd:2008p7258}.
Nous allons nous intéresser dans ce chapitre aux configurations qui permettent d'obtenir ces deux effets.
Nous les traiterons tout d'abord dans le cadre du formalisme de l'optique non--linéaire afin de déterminer l'expression générale des champs en sortie du milieu en fonction de la susceptibilité  non--linéaire $\chi^{(3)}$ et de l'amplitude du champ pompe.
Dans cette étude, nous nous limiterons à une description scalaire du champ électrique et de la polarisation ($\chi^{(3)}$ scalaire) et à des milieux  non--linéaires non dissipatifs ($n$ et $\chi^{(3)}$  réels).
Cela nous permettra de simplifier les expressions obtenues, afin de donner une description simple des phénomènes mis en jeu et d'introduire dans une seconde partie le modèle de l'amplificateur linéaire idéal.
A l'aide de ce modèle, nous pourrons alors écrire les relations qui relient l'état quantique d'entrée à l'état de sortie pour les deux processus.
Alors qu'une approche classique permet de déterminer uniquement des grandeurs telles que le gain sur la valeur moyenne, ce modèle nous permettra d'étudier les fluctuations d'intensité.

\subsection {Différentes configurations permettant le processus mélange à 4 ondes}
L'utilisation d'un milieu atomique proche de résonance permet d'atteindre des non-linéarités du troisième ordre importantes \cite{Yariv:1977p6885,Abrams:1978p6900}.
Les premières expériences d'amplification par mélange à 4 ondes utilisant la réponse non-linéaire d'un milieu atomique ont été réalisées en 1981 et sont décrites dans l'approximation de l'atome à deux niveaux dans \cite{Boyd:1981p7611,Agarwal:1986p9913}.
De nombreuses configurations sont envisageables \cite{Kolchin:2006p9848,McCormick:2007p652,Becerra:2008p10775,Akulshin:2009p10219,Agha:2010p13052}.
De plus, nous l'avons vu au chapitre \ref{ch2}, le mélange à 4 ondes permet de générer des états non--classiques \cite{Yuen:1979p7810,Kumar:1984p7847,Reid:1985p3606,Slusher:1985p5993}.
Plus récemment, de nombreuses descriptions théoriques se sont intéressées à des ensembles atomiques constitués d'atomes décrits par un modèle à 3 niveaux en simple-$\Lambda$ et à 4 niveaux en double-$\Lambda$ comme milieu pour les expériences de mélange à 4 ondes \cite{Fleischhauer95,Lukin:1999p1647,Zibrov:1999p1361,Lukin:2000p1648}.
La configuration décrite dans ce manuscrit au chapitre 4 est celle d'un milieu atomique à 4 niveaux en double-$\Lambda$.
Dans cette configuration, deux processus vont nous intéresser particulièrement.\\
L'un impliquera un champ intense appelé champ pompe et deux champs faibles appelés sonde et conjugué (voir figure \ref{fig31}).
Ce sont alors les corrélations en intensité entre les deux faisceaux qui vont nous intéresser. 
Ce phénomène est associé à un processus d'amplification insensible à la phase.\\
L'autre impliquera deux champs intenses (pompes) et un champ faible (sonde).
On étudiera dans ce cas le bruit d'une des quadratures du faisceau sonde qui pourra être réduit sous la limite quantique standard.
Ce phénomène est observé en présence d'une amplification sensible à la phase.

\subsection {Amplification insensible à la phase}\label{PIA:sec}
\begin{figure}		
		\centering
			\includegraphics[width=10 cm]{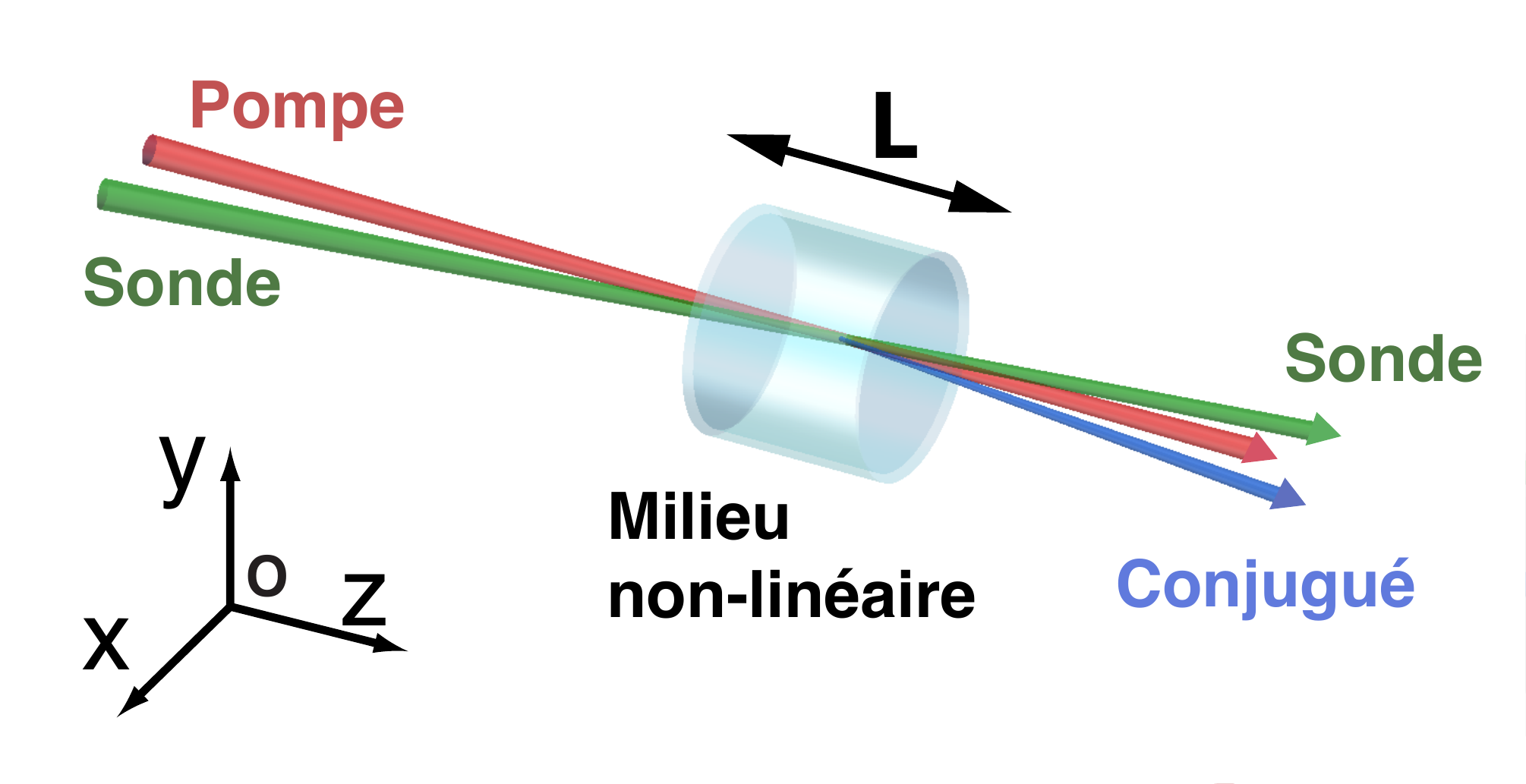} 
		\caption[Géométrie du mélange à 4 ondes]{Géométrie du mélange à 4 ondes décrite dans l'amplification insensible à la phase. Le champ pompe est se propage sur l'axe à $z$. Le champ sonde est dans le plan xOz. Le champ conjugué généré par mélange à 4 ondes respecte l'accord de phase ($2\textbf{k}_p-\textbf{k}_a-\textbf{k}_b=0$).\label{fig31}}	
		\end{figure}
Nous nous intéressons à la géométrie décrite sur la figure \ref{fig31}.
Un champ pompe intense $E_p$ à la fréquence angulaire $\omega_p$ et un champ sonde peu intense $E_a$  à la fréquence angulaire $\omega_a$ sont injectés à l'abscisse $z=0$ d'un milieu de longueur $L$ caractérisé par une susceptibilité non--linéaire d'ordre 3 ($\chi^{(3)}$).
Dans cette configuration un champ conjugué $E_b$  à la fréquence angulaire $\omega_b$ peut être généré par le processus de mélange à 4 ondes.
Ce processus peut intervenir si la conservation d'énergie est assurée.
Ainsi, $\omega_b$ doit respecter la relation suivante sur les fréquences angulaires :
\begin{equation}
2\omega_p-\omega_a-\omega_b=0.
\end{equation}
Pour des raisons de simplicité, nous prenons les champ $E_p$, $E_a$ et $E_b$ polarisés linéairement et parallèlement.
De plus, on suppose qu'ils se propagent tous selon $z$ et qu'ils sont stabilisés en phase (c'est-à-dire qu'il n'y pas d'évolution temporelle de la phase relative entre les lasers).
Les champs peuvent alors s'écrire sous la forme :
\begin{equation}
E_j(z,t)=\frac 12 \mathscr{E}_j(z)e^{i(k_j z-\omega_j t)}+c.c.,
\end{equation}
où $j \in \{a,b,p\}$, $\mathscr{E}_j(z)$ est l'enveloppe lentement variable du champ à priori complexe et $k_j$ la projection du vecteur d'onde du champ $E_j$ sur l'axe $Oz$.
Pour étudier les champs sonde et conjugué dans le milieu non-linéaire, on écrit l'équation de propagation 
\begin{equation}\label{propae}
\frac{\partial^2 E(z,t)}{\partial z^2}-\frac 1{c^2}\frac{\partial^2 E(z,t)}{\partial t^2}=\frac{1}{\epsilon_0 c^2}\frac{\partial^2 P(z,t)}{\partial t^2},
\end{equation}
où $E(z,t)$ est le champ total dans le milieu :
\begin{equation}
E(z,t)=E_p(z,t)+E_a(z,t)+E_b(z,t),
\end{equation}
et $P(z,t)$ est la polarisation non-linéaire du milieu qui est donnée, avec les approximations\footnote{Nous avons supposé que le champ $E$ était décrit par une quantité scalaire et non par un vecteur. Ainsi, dans cette approche, la susceptibilité non--linéaire $\chi^{(3)}$ est un scalaire et non pas un tenseur d'ordre 3.} que nous avons faites, par \cite{OrvilScully:1997p3797,WBoyd:2008p7258} :
\begin{equation}\label{pola}
P(z,t)=\chi^{(3)}E(z,t)^3.
\end{equation}
A l'aide de  l'approximation sur l'enveloppe lentement variable, on peut écrire l'équation de propagation \eqref{propae} sous la forme :
\begin{equation}\label{eq38}
i\ k_j \frac{\partial \mathscr{E}_j(z)}{\partial z}  e^{i(k_jz -\omega_j t)}=\frac{1}{\epsilon_0 c^2}\frac{\partial^2 P(z,t)}{\partial t^2}.
\end{equation}
En ne prenant en compte que la polarisation non--linéaire dans l'équation de propagation, on a omis le terme lié à l'indice linéaire $n$ du milieu.
Comme nous avons supposé le milieu non dissipatif ($n$ réel), cela revient simplement à redéfinir la référence de phase en prenant en compte la phase acquise au cours de la propagation sur une longueur $L$ : $e^{i k_j L}$.
Au vu de l'équation \eqref{pola}, le terme de polarisation dans le membre de droite de l'équation de propagation va contenir dix termes.
Parmi ceux-ci, on cherche les termes qui contiennent une dépendance spatio-temporelle identique au membre de gauche $ e^{i(k_jz -\omega_j t)}$.\\
On définit alors les deux polarisations $P_a(z,t)$ et $P_b(z,t)$, dont les dépendances spatiale et temporelle respectives sont $ e^{i(\textbf{k}_a\textbf{r}-\omega_a t)}$ et $ e^{i(\textbf{k}_b\textbf{r}-\omega_b t)}$.
En ne conservant que les termes qui vérifient la condition d'accord de phase ($2\textbf{k}_p-\textbf{k}_a-\textbf{k}_b=0$), on peut alors écrire les équations de propagation à l'aide de l'équation \eqref{pola} sous la forme:
\begin{deqarr}\label{39}
k_a\frac{ \partial \mathscr{E}_a (z)}{\partial z} e^{i(k_az-\omega_a t)}&=&\frac{i\omega_a^2}{\epsilon_0 c^2}P_a(z,t),\\
k_b \frac{ \partial \mathscr{E}_b(z) }{\partial z}  e^{i(k_bz-\omega_b t)}&=&\frac{i\omega_b^2}{\epsilon_0 c^2}P_b(z,t),
\end{deqarr}
avec 
\begin{deqarr}\label{Pa}
P_a(z,t)&=&\frac{3\chi^{(3)}}{8}\left(\mathscr{E}_a^2\mathscr{E}_a^*+2\mathscr{E}_a\mathscr{E}_b\mathscr{E}_b^*+2\mathscr{E}_a\mathscr{E}_p\mathscr{E}_p^*+2\mathscr{E}_p\mathscr{E}_p \mathscr{E}_b^*   \right)e^{i(k_az-\omega_a t)},\\
P_b(z,t)&=&\frac{3\chi^{(3)}}{8}\left(\mathscr{E}_b^2\mathscr{E}_b^*+2\mathscr{E}_b\mathscr{E}_a\mathscr{E}_a^*+2\mathscr{E}_b\mathscr{E}_p\mathscr{E}_p^*+2\mathscr{E}_p\mathscr{E}_p \mathscr{E}_a^*   \right)e^{i(k_bz-\omega_b t)}.
\end{deqarr}
Pour un champ pompe intense, on peut simplifier les expressions précédentes en négligeant les termes d'ordre 2 en champ sonde et conjugué devant les termes de champ pompe.
On ne conserve donc que les deux derniers termes de ces équations.
Par souci de simplicité on suppose que $\omega_a=\omega_b=\omega$ ; on peut ainsi écrire l'équation \eqref{eq38} sous la forme :
\begin{deqarr}\arrlabel{eqnabla}
k_a \frac{ \partial \mathscr{E}_a (z)}{\partial z} &=&i\frac \omega c \left(\kappa \mathscr{E}_a(z)+\eta \mathscr{E}_b^*(z)\right),\\
k_b\frac{ \partial \mathscr{E}_b (z)}{\partial z} &=&i\frac \omega c \left(\kappa \mathscr{E}_b(z)+\eta \mathscr{E}_a^*(z)\right),
\end{deqarr}
avec 
\begin{deqarr}\arrlabel{312}
\kappa &=&\frac{3\chi^{(3)}\omega}{4\epsilon_0 c}| \mathscr{E}_p|^2,\\
\eta &=& \frac{3\chi^{(3)}\omega}{4\epsilon_0 c}\mathscr{E}_p^2.
	\end{deqarr}
On a donc obtenu un jeu de deux équations non--linéaires couplées qui peuvent être résolues exactement en précisant les conditions aux limites.
Afin de ne pas compliquer le formalisme, on fera l'hypothèse d'une pompe très intense et de non déplétion, pour que l'on puisse considérer son amplitude comme constante au cours de la propagation.
Dans cette approximation les coefficients $\kappa$ et $\eta$ sont alors indépendants de z.
On effectue alors le changement de variables suivant :
\begin{equation}\label{chgt}
\tilde{\mathscr{E}}_j=\mathscr{E}_j e^{-i\kappa z}.
	\end{equation}
	Comme nous avons supposé $\chi^{(3)}$ (et donc $\kappa$) réel, cela revient simplement à un changement de référence de phase pour les champs $a$ et $b$.
On peut alors écrire les équations \eqref{eqnabla} sous forme matricielle :
\begin{equation}\label{eqnabla2}
\frac {\partial} {\partial z}\begin{bmatrix}
\tilde{\mathscr{E}}_a \\ 
\tilde{\mathscr{E}}_b^*
\end{bmatrix} =
\begin{bmatrix}
0 & i\eta\\ 
-i\eta^* & 0
\end{bmatrix}
\begin{bmatrix}
\tilde{\mathscr{E}}_a \\ 
\tilde{\mathscr{E}}_b^*
\end{bmatrix}.
\end{equation}
Les solutions de ce système
s'écrivent sous la forme :
\begin{equation}\label{PIA_ONL}
\begin{bmatrix}
\tilde{\mathscr{E}}_a(z) \\ 
\tilde{\mathscr{E}}_b^*(z)
\end{bmatrix} =
\begin{bmatrix}
\text{cosh}\ |\eta|z & i\frac{\eta}{|\eta|}\text{sinh}\  |\eta|z \\ 
-i\frac{\eta^*}{|\eta|}\text{sinh}\  |\eta| z& \text{cosh}\  |\eta|z 
\end{bmatrix}
\begin{bmatrix}
\tilde{\mathscr{E}}_a(0) \\ 
\tilde{\mathscr{E}}_b^*(0)
\end{bmatrix}.
\end{equation}
	
Avec les conditions initiales suivantes $\tilde{\mathscr{E}}_a(0)=\tilde{\mathscr{E}}_{in}$ et $\tilde{\mathscr{E}}_b(0)=0$, on obtient :
\begin{deqarr}\arrlabel{eqnabla3}
\tilde{\mathscr{E}}_a(z)&=&\tilde{\mathscr{E}}_{in}\ \text{cosh}\ | \eta |z,\\
\tilde{\mathscr{E}}_b(z)&=&i\frac {\eta} {|\eta|}\tilde{\mathscr{E}}_{in}^* \ \text{sinh}\ |\eta |z.
\end{deqarr}
Nous étudions l'intensité moyenne des champs sonde et conjugué en sortie du milieu.
L'étude des fluctuations sera faite dans la section \ref{PIA_OQ_bruit}.
Comme le changement de variable \eqref{chgt} n'influe pas sur l'intensité car il ajoute uniquement un terme de phase, on peut écrire en sortie d'un milieu de longueur $L$ :
\begin{deqarr}\arrlabel{eqnabla4}
|\mathscr{E}_a(L) |^2&=& |\mathscr{E}_{in}|^2\ \text{cosh}^2\left(|\eta  |L\right),\\
|\mathscr{E}_b(L)|^2 &=& |\mathscr{E}_{in}|^2\ \text{sinh}^2\left(|\eta  |L\right),
\end{deqarr}
que l'on peut mettre sous la forme : 
\begin{deqarr}\arrlabel{eqnabla5}
|\mathscr{E}_a(L) |^2&=& G |\mathscr{E}_{in}|^2,\\
|\mathscr{E}_b(L)|^2 &=&  (G-1) |\mathscr{E}_{in}|^2.
\end{deqarr}
avec 
\begin{equation}\arrlabel{eqnabla6}
G=\text{cosh}^2\left(|\eta  |L\right)
\end{equation}
Ainsi, on obtient à la sortie du milieu une amplification du champ sonde et la génération du champ conjugué.
Le coefficient $|\eta  |L$ qui quantifie le gain est piloté par trois paramètres qui sont la longueur du milieu $L$, le coefficient non--linéaire $\chi^{(3)}$ et l'intensité $|\mathscr{E}_p|^2$ du champ pompe.
Pour un gain élevé, c'est-à-dire des valeurs de  $|\eta  |L\gg 1$, on a :
\begin{equation}
|\mathscr{E}_a(L)|\simeq|\mathscr{E}_b(L)| \simeq |\mathscr{E}_{in}|\ e^{|\eta  |L}.
\end{equation}
Ce calcul d'optique non--linéaire décrit l'amplification du champ sonde et la génération d'un champ conjugué dans un milieu de susceptibilité non--linéaire $\chi^{(3)}$.
Ce processus sera décrit dans la suite comme une amplification insensible à la phase (PIA).
Dans la section \ref{PIA:OQ}, on verra que l'on retrouve les équations \eqref{eqnabla5} dans un modèle d'amplificateur parfait insensible à la phase.
Les fluctuations quantiques dans le cas de la PIA seront étudiées dans cette section.

	\subsection {Amplification sensible à la phase}\label{PSA:sec}
Une autre configuration intéressante pour le mélange à 4 onde est à l'origine de l'amplification sensible à la phase \cite{Marhic:1991p14967,Hansryd:2002p14966,Tang:2008p14950,Marino:2010p15163}.
Dans cette configuration dégénérée, deux pompes intenses $E_{p1}$ et $E_{p2}$, de fréquence angulaire respective $\omega_{p1}$ et $\omega_{p2}$, interagissent avec un champ sonde $E_a$ de fréquence angulaire $\omega_a$.
Deux photons sonde sont mis en jeu pour un photon de la pompe $E_{p1}$ et un photon de la pompe $E_{p2}$.
C'est la configuration qui échange le rôle des faisceaux intenses et faibles par rapport à celle de la section précédente.
On peut écrire la conservation de l'énergie sous la forme :
\begin{equation}
\omega_{p1}+\omega_{p2}-2\omega_a=0.
\end{equation}
Nous reprenons les notations et les hypothèses de la section \ref{PIA:sec} pour les champs pompe et sonde.
Le formalisme pour déterminer l'évolution du champ sonde est identique à celui développé dans la section \ref{PIA:sec}.
Notamment, les équations de propagation \eqref{eq38} restent valables pour le champ sonde.
Le champ total $E $ dans le milieu est donné par :
\begin{equation}
E(z,t)=E_{p1}(z,t)+E_{p2}(z,t)+E_a(z,t).
\end{equation}
Ainsi l'équation de propagation donnée à l'équation \eqref{39} reste valable, en définissant la polarisation $P_a(z,t)$ par :
\begin{equation}\label{polaPSA}
P_a(z,t)=\frac{3\chi^{(3)}}{8}\left(\mathscr{E}_a^2\mathscr{E}_a^*+2\mathscr{E}_a\mathscr{E}_{p1}\mathscr{E}_{p1}^*+2\mathscr{E}_a\mathscr{E}_{p2}\mathscr{E}_{p2}^*+2\mathscr{E}_{p1}\mathscr{E}_{p2} \mathscr{E}_a^*   \right)e^{i(k_az-\omega_a t)}.
\end{equation}
En négligeant dans l'équation \eqref{polaPSA} les termes qui ne contiennent pas de champs pompe, on obtient l'équation de propagation du champ sonde sous la forme :
\begin{equation}
\frac{\partial \mathscr{E}_a(z)}{\partial z}=i(\kappa \mathscr{E}_a(z)+\eta \mathscr{E}_a^*(z)),
\end{equation}
avec 
\begin{deqarr}\arrlabel{324}
\kappa &=&\frac{3\chi^{(3)}\omega}{4\epsilon_0 c}(| \mathscr{E}_{p1}|^2+| \mathscr{E}_{p2}|^2),\\
\eta &=& \frac{3\chi^{(3)}\omega}{4\epsilon_0 c}\mathscr{E}_{p1}\mathscr{E}_{p2}.
	\end{deqarr}
	Pour résoudre cette équation différentielle (et l'équation conjuguée), on peut les écrire sous forme matricielle\footnote{Comme on a fait l'hypothèse que $\chi^{(3)}$ était réel, alors $\kappa$ l'est aussi. }:
	\begin{equation}
	\frac {\partial} {\partial z}\begin{bmatrix}
	\mathscr{E}_a(z) \\ 
	\mathscr{E}_a^*(z) 
	\end{bmatrix} =i
	\begin{bmatrix}
		\kappa & \eta\\ 
		-\eta^* & -\kappa
		\end{bmatrix}
		\begin{bmatrix}
			\mathscr{E}_a (z) \\ 
			\mathscr{E}_a^*(z) 
			\end{bmatrix}.
	\end{equation}
	On utilise la condition initiale $\mathscr{E}_a(0)=\mathscr{E}_{in}$ afin de résoudre cette équation sous la forme :
	\begin{equation}
		\begin{bmatrix}
		\mathscr{E}_a(z) \\ 
		\mathscr{E}_a^*(z)
		\end{bmatrix} =e^{M z}
			\begin{bmatrix}
				\mathscr{E}_{in} \\ 
				\mathscr{E}_{in}^*
				\end{bmatrix}.
		\end{equation}
		Le matrice M est définie par :
		\begin{equation}
		M=i
			\begin{bmatrix}
				\kappa & \eta\\ 
				-\eta^* & -\kappa
				\end{bmatrix}.
		\end{equation}
On a donc à l'abscisse $z=L$ :
\begin{equation}
e^{M L}=\frac{1}{\Delta}
	\begin{bmatrix}
	\Delta\	\text{cos}\Delta L+i\ \kappa\ \text{sin}\Delta L & i\ \eta\ \text{sin}\Delta L \\ 
		-i\ \eta^*\ \text{sin}\Delta L & \Delta\ \text{cos}\Delta  L-i\ \kappa\ \text{sin}\Delta L
		\end{bmatrix},
\end{equation}
où l'on a posé $\Delta =\sqrt{|\kappa|^2-|\eta|^2}$.
On pourra noter d'après les équations \eqref{324}, que le coefficient $|\kappa|^2-|\eta|^2$ est réel positif.
Ainsi, on décrit l'évolution du champ $E_a$ par son enveloppe lentement variable :
\begin{equation}\label{331}
\mathscr{E}_a(L)=(\text{cos}\Delta L+i\frac{ \kappa}{\Delta}\text{sin}\Delta L)\ \mathscr{E}_{in}+i\frac{ \eta}{\Delta} \text{sin}\Delta L\ \mathscr{E}_{in}^*.
\end{equation}
On peut écrire cette équation sous la forme suivante :
\begin{equation}\label{PSA_ONL2}
\mathscr{E}_a(L)=\sqrt{G}\ e^{i\phi_1}\ \mathscr{E}_{in}+\sqrt{G-1}\ e^{i\phi_2} \ \mathscr{E}_{in}^*,
\end{equation}
où le gain $G$ s'écrit sous la forme :
\begin{equation}
G=\text{cos}^2\Delta L+\frac{ \kappa^2}{\Delta^2}\text{sin}^2\Delta L=1+\frac{ |\eta|^2}{\Delta^2} \text{sin}^2\Delta L.
\end{equation}
Les phases $\phi_1$ et $\phi_2$ peuvent être dérivées à partir de l'expression \eqref{331}.
L'expression $\label{PSA_ONL}$ est alors de la forme de celle que nous utiliserons pour décrire le modèle de l'amplificateur parfait sensible à la phase dans la section \ref{PSA:OQ} (Eq. \ref{PSA_eq}).\\
On peut écrire l'intensité du champ en sortie du milieu sous la forme suivante :
\begin{eqnarray}\label{334}
\nonumber |\mathscr{E}_a(L)|^2&&=(\text{cos}^2\Delta L+\frac{ \kappa^2}{\Delta^2}\text{sin}^2\Delta L+\frac{ |\eta|^2}{\Delta^2} \text{sin}^2\Delta L)|\mathscr{E}_{in}|^2\\
&&+2\text{Re}\left(\frac{\kappa \eta}{\Delta^2}\text{sin}^2\Delta L\ \mathscr{E}_{in}^{*2} +i\frac{\eta}{\Delta}\text{cos}\Delta L\ \text{sin}\Delta L\ \mathscr{E}_{in}^{*2}\right).
\end{eqnarray}
Cette expression nous montre que l'intensité en sortie est sensible à la phase.
En effet le terme de la seconde ligne de l'équation \eqref{334} va dépendre de la phase de $\eta$ et de celle du champ d'entrée $\mathscr{E}_{in}$.
Comme nous n'avons pas fixé de référence de phase, c'est donc la différence entre ces deux phases qui va jouer un rôle.\\
\clearpage
\section{Modèle phénoménologique de l'amplificateur linéaire idéal}

Nous reprenons ici en le détaillant un modèle introduit dans \cite{McCormick:2008p6669}.
On considère ici le mélange à quatre ondes comme un processus d'amplification idéale.
Dans le cas d'un amplificateur idéal insensible à la phase, chaque photon généré dans le mode de la sonde aura son homologue dans le mode du conjugué.
Ainsi le taux de compression sur la différence d'intensité peut être évalué quantitativement en connaissant  le gain du processus.
Pour l'amplificateur idéal sensible à la phase, selon sa phase chaque quadrature pourra être amplifiée ou desamplifiée.
Ainsi il existe une situation où la valeur moyenne du champ est amplifiée et une des quadratures est desamplifiée sans ajout de bruit, ce qui permet une réduction du bruit sous la limite quantique standard.  \\
Dans un second temps, une approche phénoménologique des pertes au cours de la propagation va permettre d'améliorer ce modèle.
Enfin nous nous intéressons au processus d'amplificateur idéal sensible à la phase pour mettre en évidence la capacité d'un tel système à produire des états comprimés à un mode du champ.
Les équations entrée-sortie qui décrivent ces processus sont compatibles avec les relations \eqref{PIA_ONL} pour le PIA et \eqref{PSA_ONL2} pour le PSA, obtenues dans le cadre de l'optique non--linéaire.\\

\subsection{Amplification idéale insensible à la phase}\label{PIA:OQ}
L'amplificateur linéaire idéal \cite{Caves:1982p6739} est un modèle qui permet de décrire les processus d'amplification paramétrique de façon simple.
Dans le cadre de l'optique quantique, il a été utilisé par exemple dans les thèses de \cite{Bencheikh:1996p15359} et de \cite{Gigan:2004p8300}.
Dans ce modèle le signal en sortie est relié au signal d'entrée par des relations linéaires.
Il n'y a donc ni de bruit classique ajouté, ni de pertes et la bande passante est considérée comme infinie.
Après avoir présenté ce modèle, nous verrons comment il décrit les valeurs moyennes des intensités des champs sonde et conjugué.
Dans un second temps, nous étudierons les fluctuations quantiques des faisceaux à la sortie d'un tel système.
\subsubsection{Valeurs moyennes}
Dans un processus d'amplification linéaire idéale à deux modes, on peut écrire les relations entrée-sortie des opérateurs $\hata$ et $\hat{b}$ pour un gain $G\geq 1$, sous la forme suivante \cite{OrvilScully:1997p3797} :
\begin{deqarr} \arrlabel{ampli_ideale}
\hata_{out}=\sqrt{G}\ \hata_{in}+\sqrt{G-1}\ \hat{b}_{in}^\dag,\\
\hat{b}^\dag_{out}=\sqrt{G}\ \hat{b}_{in}^\dag+\sqrt{G-1}\ \hata_{in}.
\end{deqarr}
On s'intéresse aux états à deux modes du champ qui ont été introduits au paragraphe \ref{corel_ch1} du chapitre~\ref{ch1}, c'est à dire aux corrélations entre les mode $\hat a$ et $\hat b$, et particulièrement aux corrélations d'intensité.
A partir des équations (\ref{ampli_ideale}), on peut dériver simplement l'expression de la valeur moyenne des opérateurs nombres en sortie.
En entrée du milieu,on injecte un état cohérent $|\alpha\rangle$ sur le mode $\hata$ et le vide sur le mode $\hat{b}$.
On obtient donc \cite{Gigan:2004p8300} :
\begin{deqarr}\arrlabel{gain_PIA}
\bra\hat{N}_{a,out}\ket&=& G\ |\alpha|^2,\\
\bra\hat{N}_{b,out}\ket&=&( G-1)\  |\alpha|^2,
\end{deqarr}
avec $|\alpha|^2= \bra\alpha|\hat{N}_{a,in}|\alpha\ket$.\\
On voit que pour un amplificateur idéal, on a en sortie $G\ |\alpha|^2$ photons dans le mode $\hata$ et $( G-1)\  |\alpha|^2$ dans le mode $\hat{b}$.
Ainsi les valeurs moyennes des opérateurs somme et différence de photons s'écrivent :
\begin{deqarr}
\bra\hat{N}_{+,out}\ket&=& (2G-1)\  |\alpha|^2,\\
\bra\hat{N}_{-,out}\ket&=& |\alpha|^2.
\end{deqarr}
On peut constater que la différence d'intensité n'est pas modifiée par la propagation dans le cas d'un amplificateur idéal.
\subsubsection{Spectres de bruit}\label{PIA_OQ_bruit}
Dans un processus d'amplification linéaire idéale, le spectre de bruit  de la différence d'intensité est donné par :
\begin{equation}\label{335}
S(N_-)=\frac{1}{2G-1}.
\end{equation}
On peut démontrer ce résultat, en appliquant les relations (\ref{deltaNa} et \ref{deltaN-})  à l'état de sortie.
En effet, on peut écrire des relations entrée-sortie pour les fonctions de corrélations :
\begin{deqarr}
\bra\delta \hat X_a \delta \hat X_a\ket_{out}&=& G\ \bra\delta \hat X_a \delta\hat X_a\ket_{in}+(G-1)\ \bra\delta \hat X_b\delta\hat X_b\ket_{in}+2\sqrt{G(G-1)}\ \bra\delta \hat X_a\delta\hat X_b\ket_{in}\\
\bra\delta \hat X_b \delta \hat X_b\ket_{out}&=& (G-1)\ \bra\delta \hat X_a \delta\hat X_a\ket_{in}+G\ \bra\delta \hat X_b\delta\hat X_b\ket_{in}+2\sqrt{G(G-1)}\ \bra\delta \hat X_a\delta\hat X_b\ket_{in}\\
\bra\delta \hat X_a \delta \hat X_b\ket_{out}&=& \sqrt{G(G-1)} \left( \bra\delta \hat X_a \delta\hat X_a\ket_{in}+\bra\delta \hat X_b\delta\hat X_b\ket_{in}\right)+(2G-1)\ \bra\delta \hat X_a\delta\hat X_b\ket_{in}.\hspace{1.8cm}
\end{deqarr}
	\begin{figure}	[]	
				\centering
					\includegraphics[width=9cm]{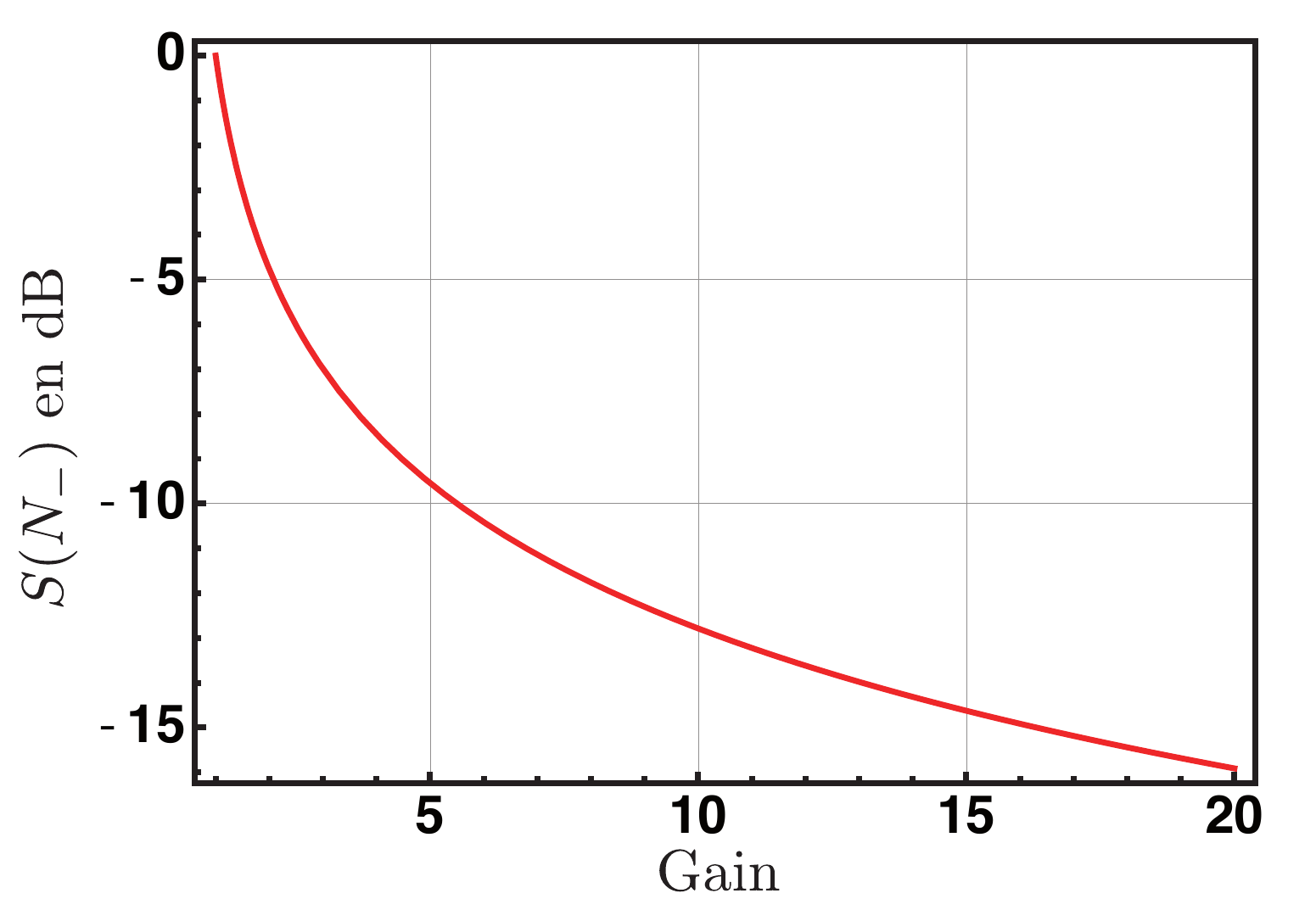} 
				\caption[Bruit de la différence d'intensité  en fonction du gain]{Bruit de la différence d'intensité  en fonction du gain pour deux modes d'un amplificateur parfait (Eq. \ref{335}).\label{fig_ampliparfait}}	
				\end{figure}
Les  spectres de bruit en intensité du mode $\hata$ et de la différence d'intensité, normalisés par le bruit quantique standard, sont donnés par :
\begin{deqarr}\arrlabel{spectres_1}
S(N_a)&=&\bra\delta \hat X_a \delta \hat X_a\ket_{out},\\
S(N_-)&=&\frac{G \bra\delta \hat X_a \delta \hat X_a\ket_{out}+(G-1)\bra\delta \hat X_b \delta \hat X_b\ket_{out}-2\sqrt{G (G-1)}\bra\delta \hat X_a \delta \hat X_b\ket_{out}}{2G-1}.\hspace{1.5cm}
\end{deqarr}
Le champ $\hata$ en entrée est un état cohérent. Le champ $\hat{b}$ en entrée est le vide donc aussi un état cohérent.
Les corrélations entre ces deux champs sont nulles.
Les fonctions de corrélations pour les quadratures des champs entrants s'écrivent donc:
\begin{deqarr}
\bra\delta \hat X_a \delta \hat X_a\ket_{in}&=&1,\\
\bra\delta \hat X_b \delta \hat X_b\ket_{in}&=&1,\\
\bra\delta \hat X_a \delta \hat X_b\ket_{in}&=&0.
\end{deqarr}
Les expressions (\ref{spectres_1}) pour  les spectres de bruit en sortie, normalisés par le bruit quantique standard s'écrivent alors :
\begin{deqarr}
S(N_a)&=&2G-1,\\
S(N_-)&=& \frac{1}{2G-1}.\label{339b}
\end{deqarr}
On voit donc que le bruit d'intensité du champ $\hata$ est amplifié par rapport au bruit quantique standard, alors que le bruit de la différence d'intensité est comprimé sous la limite quantique standard pour $G>1$ ce qui est caractéristique d'une amplification linéaire parfaite  \cite{Caves:1982p6739,Gigan:2004p8300}.
Dans ce modèle, des paires de photons parfaitement corrélées entre les deux modes sont générées.
C'est pourquoi les fluctuations sont alors inférieures à la limite quantique standard comme le montre la figure \ref{fig_ampliparfait}.
Sur cette figure on a tracé en échelle logarithmique la relation \eqref{339b}.
\subsubsection{Effet des pertes}
\begin{figure}		
		\centering
			\includegraphics[width=9cm]{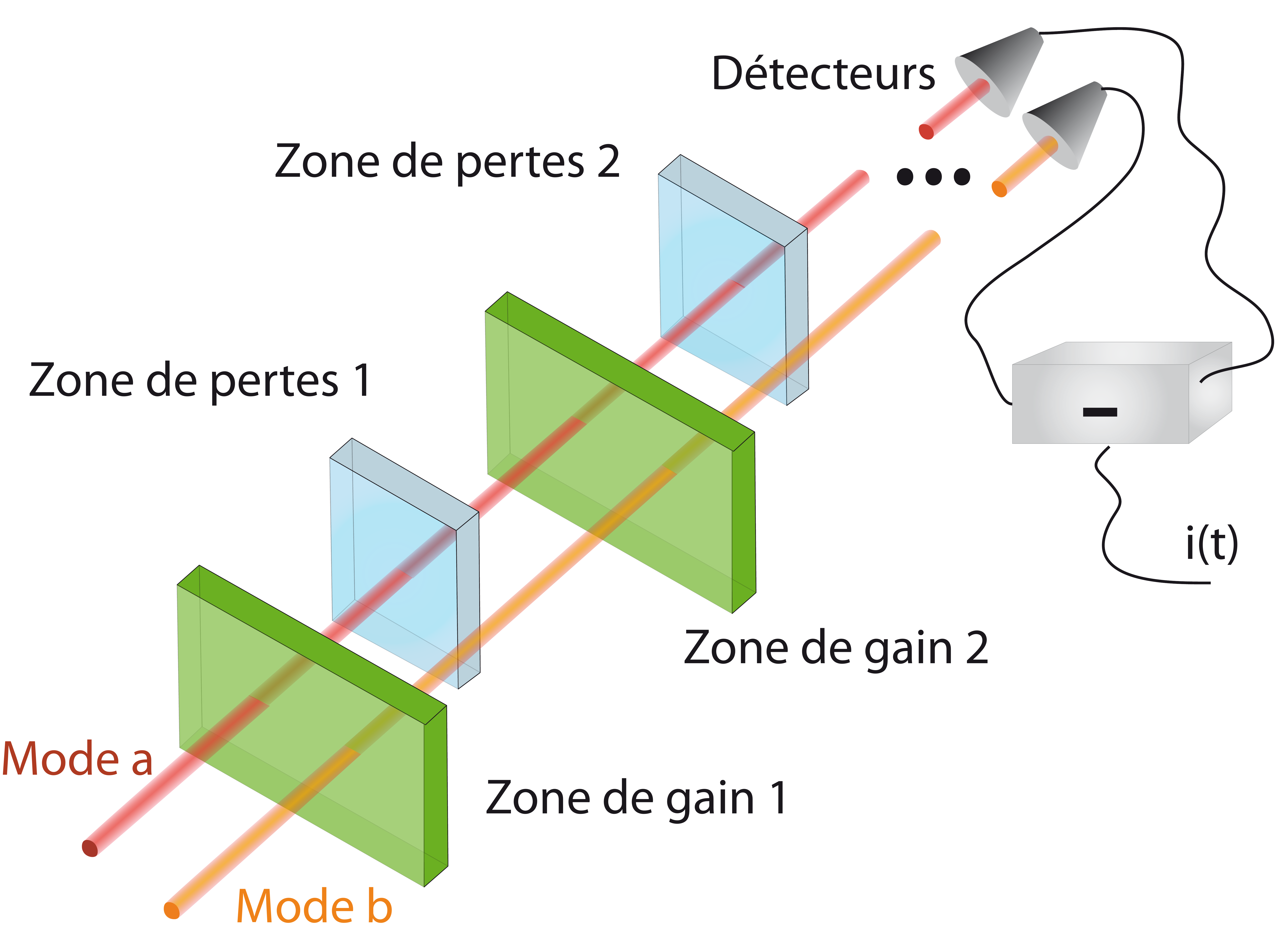} 
		\caption[Modèle phénoménologique pour le mélange à 4 ondes]{Modèle phénoménologique pour décrire les expériences de mélange à 4 ondes dans une vapeur atomique. Les modes $\hata$ et $\hat{b}$ sont amplifiés dans  des zones d'amplification linéaire idéale et le mode $\hata$ subit une absorption linéaire dans des zones de pertes. Le photocourant mesuré est proportionnel à la différence d'intensité entre les modes $\hata$ et $\hat{b}$. \label{modele}}	
		\end{figure}
Pour rendre le modèle de l'amplificateur linéaire plus réaliste, on peut ajouter, de manière phénoménologique, des pertes sur un ou deux des modes du champ.
Ce type d'approche a été introduit dans \cite{Jeffers:1993p15522} et développé par \cite{Fiorentino:2002p15564} et plus récemment par \cite{McCormick:2008p6669}  pour décrire les expériences de mélange à 4 ondes dans une vapeur atomique \cite{McCormick:2007p652}.
Dans ces expériences la non--linéarité du troisième ordre est produite par une interaction quasi-résonnante avec une vapeur atomique.
Comme nous le montrerons dans la suite, il suffit d'étudier l'effet des pertes uniquement sur le faisceau sonde.\\
Les pertes que nous introduisons correspondent à des pertes linéaires au cours de la propagation du mode $\hata$ dans le milieu.
Le milieu non--linéaire va donc être décrit comme une succession de $N$ zones de gain (milieu amplificateur idéal avec un gain $g>1$) et de  $N$ zones de pertes (lames séparatrices de transmission $t<1$).\\
Les équations d'entrée-sortie pour chaque zone de gain sont les équations  (\ref{ampli_ideale}); où le gain total $G$ est remplacé par le gain d'une tranche g.
On introduit le champ vide $\hat{c}$ et la transmission  d'une tranche $t$.
Les équations entrée--sortie pour chaque zone d'absorption sont identiques à celle d'une lame séparatrice et s'écrivent alors :
\begin{deqarr} \arrlabel{zone_pertes}
\hata_{out}&=&\sqrt{t}\ \hata_{in}+\sqrt{1-t}\ \hat{c},\\
\hat{b}^\dag_{out}&=&\hat{b}_{in}^\dag.
\end{deqarr}

\begin{figure}		
		\centering
			\includegraphics[width=15cm]{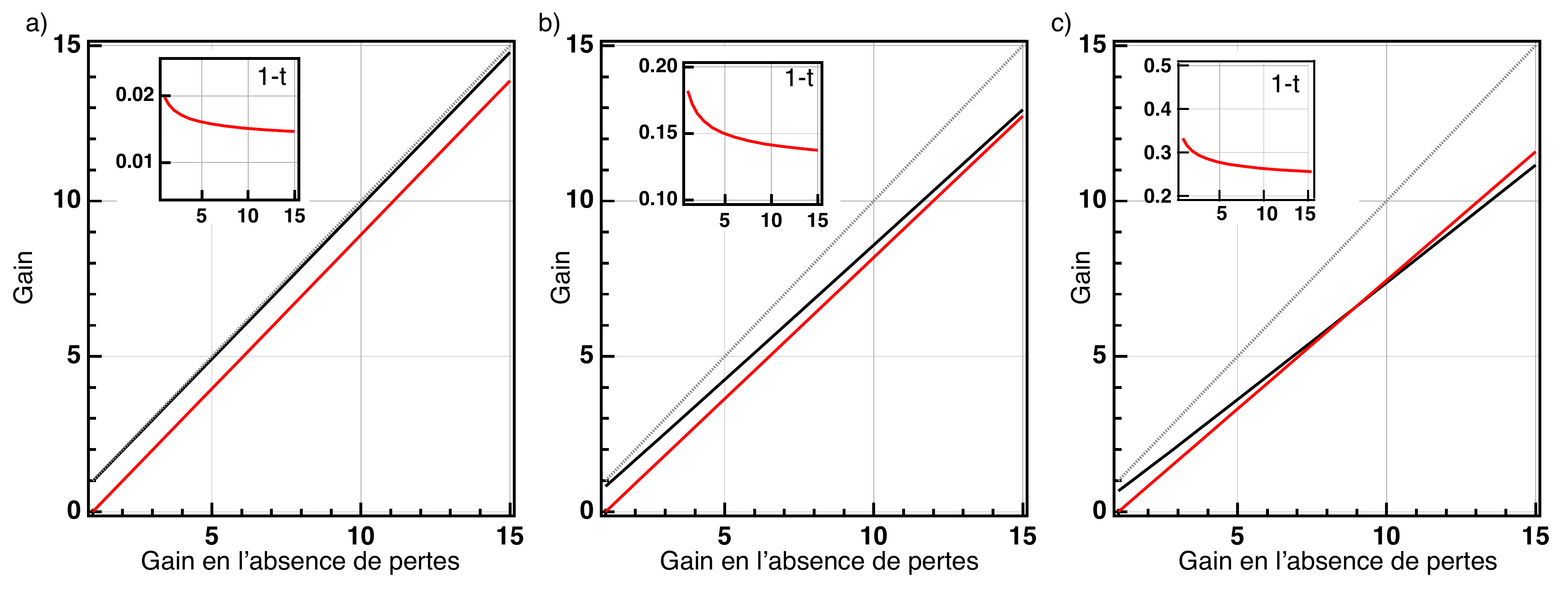} 
		\caption[Gain pour l'amplificateur idéal en présence de pertes]{Gain pour l'amplificateur idéal en présence de pertes. Le gain pour le mode $\hata$ est donné en noir. Le mode $\hat{b}$ étant vide en entrée, le gain (en rouge sur la figure) est défini comme l'intensité de sortie divisée par l'intensité d'entrée du mode $\hata$. Le gain en absence de pertes pour le mode $\hata$ est donné en pointillé gris. Les figures a), b) et c) correspondent à des valeurs différentes de pertes, respectivement $(1-t)\simeq 2\%,15\%,30\%$. Comme les pertes totales varient légèrement en fonction du gain, on les a indiquées en fonction de l'intensité de sortie dans les encarts. \label{gainmodele}}	
		\vspace{2cm}
				\centering
					\includegraphics[width=15cm]{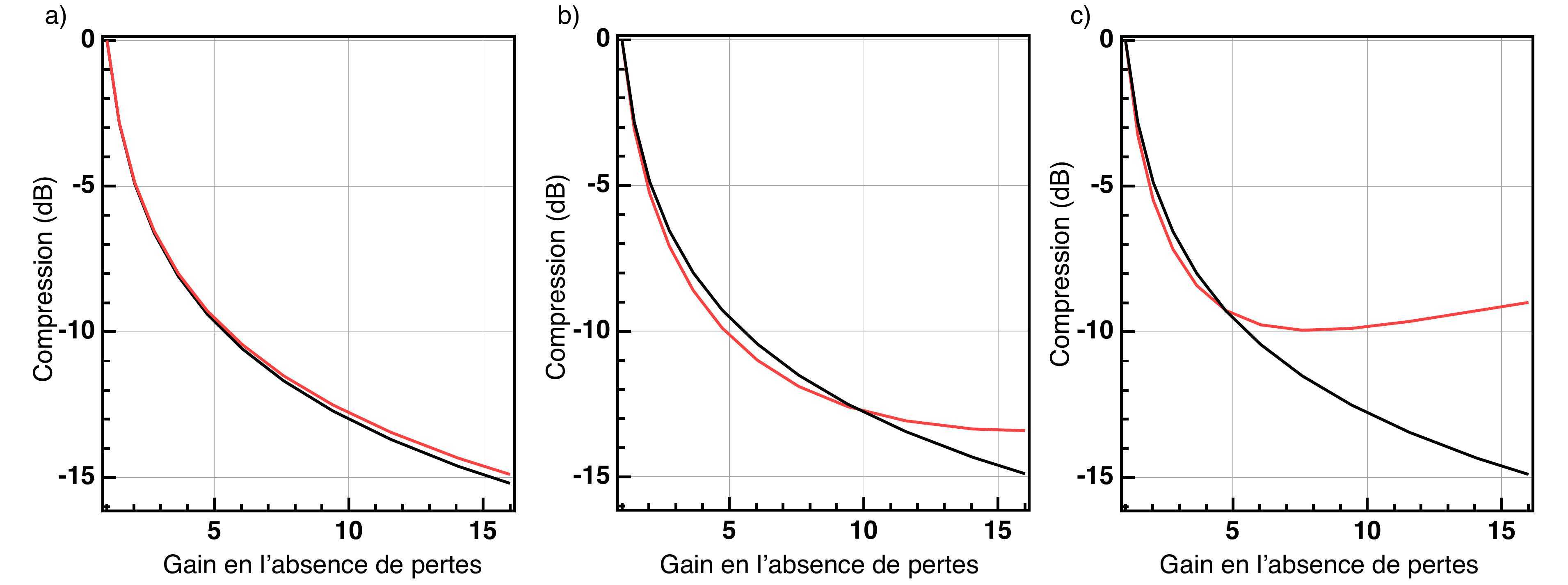} 
				\caption[Compression théorique de la différence d'intensité pour le modèle de l'amplificateur idéal en présence de pertes]{Compression théorique de la différence d'intensité pour le modèle de l'amplificateur idéal en présence de pertes. Le taux de compression en l'absence de pertes est donné en noir et en présence de pertes en rouge. Les figures a), b) et c) correspondent aux valeurs différentes de pertes qui sont indiquées sur la figure \ref{gainmodele},  respectivement $(1-t)\simeq 2\%,15\%,30\%$. \label{sqmodele}}	
				\end{figure}
\begin{figure}		
	\centering
\includegraphics[width=11cm]{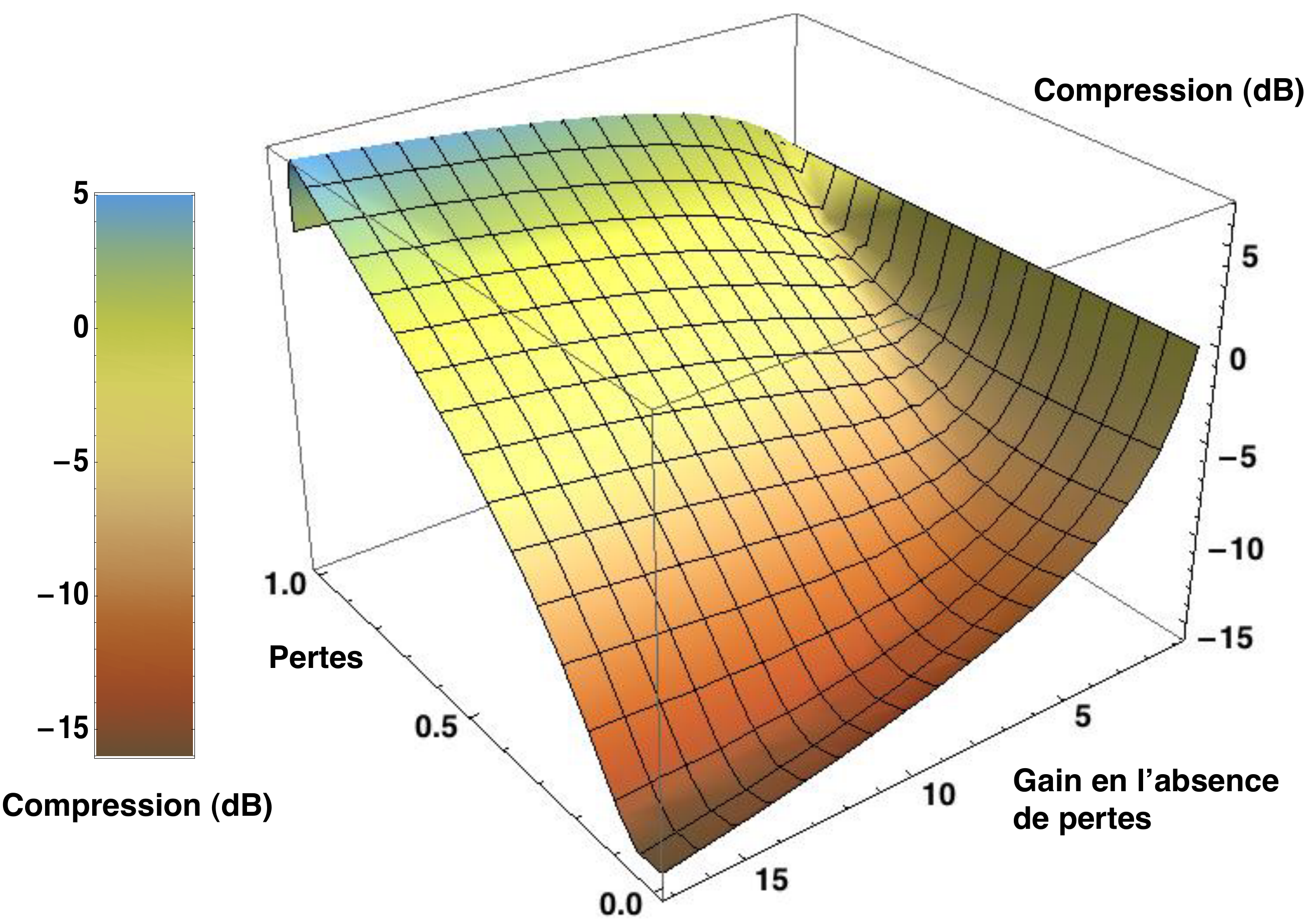} 
\caption[Compression théorique pour le modèle de l'amplificateur idéal en présence de pertes]{Compression théorique pour le modèle de l'amplificateur idéal en présence de pertes. \label{3dmodele}}	
\end{figure}
\noindent On souhaite étudier les corrélations en sortie du milieu entre les modes $\hata$ et $\hat{b}$.
On peut simuler la propagation dans ce milieu pour un nombre fini de zones.
On obtient la formule de récurrence suivante pour passer d'une tranche $n$ à la suivante :
\begin{deqarr}
\hat a_{n + 1}&=&\sqrt t \left(\sqrt g\ a_n + \sqrt{g - 1}\ \hat b^\dag_n\right) + \sqrt{1 - t}\ \hat c,\\ 
\hat b_{n + 1}&=& \sqrt{g}\ \hat  b_n + \sqrt{g - 1}\ \hat a_n.
\end{deqarr}
On écrit les relations entrée--sortie pour chaque tranche pour les fonctions de corrélation. On obtient les relations de récurrence suivantes :
\begin{deqarr}
\bra\delta \hat X_a \delta \hat X_a\ket_{n+1}&=& gt\ \bra\delta \hat X_a \delta\hat X_a\ket_{n}+(g-1)t\ \bra\delta \hat X_b\delta\hat X_b\ket_{n}+2t\sqrt{g(g-1)}\ \bra\delta \hat X_a\delta\hat X_b\ket_{n},\\
\nonumber &&+ (1-t)\bra\delta \hat X_c \delta\hat X_c\ket_n\\
\bra\delta \hat X_b \delta \hat X_b\ket_{n+1}&=& (g-1)\ \bra\delta \hat X_a \delta\hat X_a\ket_{n}+g\ \bra\delta \hat X_b\delta\hat X_b\ket_{n}+2\sqrt{g(g-1)}\ \bra\delta \hat X_a\delta\hat X_b\ket_{n},\\
\bra\delta \hat X_a \delta \hat X_b\ket_{n+1}&=& \sqrt{tg(g-1)} \left( \bra\delta \hat X_a \delta\hat X_a\ket_{n}+\bra\delta \hat X_b\delta\hat X_b\ket_{n}\right)+(2g-1)\sqrt{t}\ \bra\delta \hat X_a\delta\hat X_b\ket_{n}.\hspace{1.5cm}
\end{deqarr}
Dans ces relations nous n'avons pas écrit les termes de corrélations mixtes entre les champs ($\hata$ ou $\hat b$) et le vide $\hat c$ qui sont nuls.
De plus les corrélations entre deux modes $\hat c$ correspondant à des tranches successives sont également nulles. Ainsi pour chaque tranche les seuls termes dépendant de $\hat c$ et non nuls sont $\bra\delta \hat X_c \delta\hat X_c\ket_n=1$. 
On peut résoudre ce système d'équations de récurrence afin de déterminer une relation d'entrée sortie pour les spectres en utilisant les relations \eqref{spectres_1}.\\

\noindent Les figures \eqref{gainmodele}, \eqref{sqmodele}  et \eqref{3dmodele} présentent les résultats d'un calcul numérique pour $N=100$.
Ce choix de $N$ est dicté  par un compromis entre le temps de calcul et la convergence du modèle de discrétisation du milieu.
En effet les résultats des simulations semblent tendre vers une limite lorsque $N$ augmente ; en pratique, cela signifie que le nombre de tranches influence peu le résultat pour $N\gtrsim8 0$.
On peut voir sur la figure \eqref{gainmodele} que, en présence de pertes sur le mode $\hata$, l'intensité de sortie sur le mode $\hat b$ diminue aussi mais moins rapidement que celle du mode $\hata$. Dans certaines conditions, on a donc la même puissance en sortie sur les deux modes.
De plus, on constate que l'effet des pertes n'est pas forcément négatif pour la mesure de la compression sur le bruit de la différence d'intensité.
En effet, dans des situations très déséquilibrées d'intensité entre le mode $\hata$ et $\hat b$, c'est à dire à très faible gain, des pertes faibles peuvent améliorer légèrement le taux de compression figure \ref{sqmodele}.
Ceci peut se comprendre par le fait que dans cette situation, la part de photons n'étant pas issus du processus de mélange à 4 ondes (c'est-à-dire les photons incidents) est grand devant les photons générés.
Ainsi en ajoutant des pertes sur le mode $\hata$, à la fois des photons générés par mélange à 4 ondes (et donc parfaitement corrélés) mais surtout des photons incidents non amplifiés (et donc non corrélés) sont détruits.\\
Pour des gains importants ($G\gg 1$) et pour des transmissions faibles ($t\simeq 0$), l'effet des pertes devient négatif et le bruit sur la différence augmente.

\subsection{Amplification idéale sensible à la phase}\label{PSA:OQ}
Il est intéressant de comparer l'amplification insensible à la phase que nous venons d'étudier au cas de l'amplification sensible à la phase.
On utilise de même un modèle d'amplificateur linéaire idéal.
L'amplification sensible à la phase a été largement étudiée pour le mélange à 4 ondes dans le régime classique \cite{Abrams:1978p6900}.
Nous présentons ici un modèle simple pour décrire les propriétés de bruit du faisceau généré dans un tel processus.\\

\subsubsection{Modèle et valeurs moyennes}\label{PSA_OQ_moy}
Le processus d'amplification sensible à la phase peut être décrit, dans le cas idéal, par les relations entrée--sortie suivantes :
\begin{equation}\label{PSA_eq}
\hat a_{out}=\sqrt G \  e^{i\phi_1}\hat a_{in}+\sqrt {G-1} \ e^{ i\phi_2}\  \hat a_{in}^\dag,
\end{equation}
avec $G$ réel et $\phi_{\{1,2\}}$ des phases.
Dans le processus de mélange à 4 ondes permettant l'amplification sensible à la phase, trois faisceaux sont nécessaires en entrée : 2 pompes et une sonde. Pour décrire les phases relatives entre ces faisceaux, deux paramètres sont donc nécessaires et suffisants. 
On peut noter que l'on dispose d'une phase relative de plus par rapport au cas de l'amplification sensible à la phase dans les milieu les milieux $\chi^{(2)}$  \cite{Gigan:2004p8300}.\\

\noindent Comme précédemment nous allons écrire le champ $\hata$ sous la forme linéarisée : $$\hata =\bra \hata \ket +\delta \hata.$$
La valeur moyenne de $\hata$ en entrée sera notée $\bra \hata \ket_{in}=  |\alpha|_{in} e^{i\varphi}$ pour faire apparaître spécifiquement la phase $\varphi$ du champ incident.
L'intensité du champ en sortie est donc donnée par :
\begin{equation}
\bra\hat{N}_{a,out}\ket=|\alpha_{in}|^2 \left(2 G-1+2 \sqrt{G(G-1) }\ \text{cos}[2\varphi +(\phi_1 -\phi_2)]\right).
\end{equation}
On définit la phase $\theta=2\varphi +(\phi_1 -\phi_2)$.
On peut alors écrire le gain $\mathcal{G}$ de ce processus en fonction de $\theta$ :
\begin{equation}\label{eqtheta}
\mathcal{G}=2 G-1+2 \sqrt{G(G-1) }\ \text{cos}\ \theta.
\end{equation}
Le gain est ainsi sensible à la phase.
La figure \ref{pascourbes} a) représente le gain en fonction de $\theta$ et montre que l'on peut réaliser une amplification $\mathcal{G}>1$ ou une déamplification $\mathcal{G}<1$ selon la valeur de la phase  $\theta$.
Pour avoir un gain compris en 0 et 10, nous avons choisi pour les figures \ref{pascourbes}, une valeur de $G=3$.

\subsubsection{Spectres de bruit}\label{PSA_OQ_bruit}
\begin{figure}		
	\centering
\includegraphics[width=14.5cm]{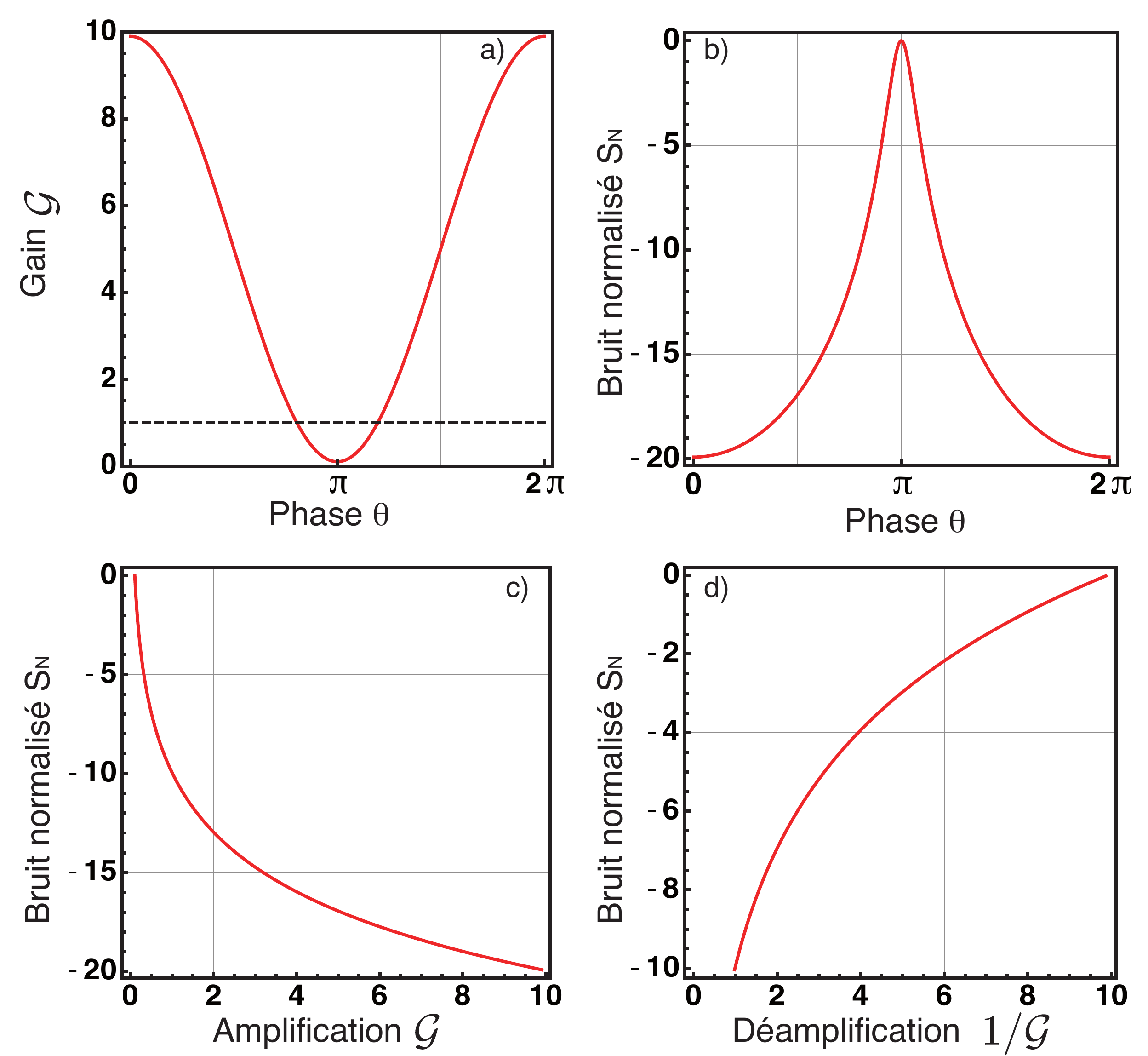} 
\caption[Gain et bruit pour un processus sensible à la phase.]{Processus sensible à la phase pour une valeur de $G=3$.
a) Gain en fonction de la phase $\theta$ en rouge et limite entre l'amplification et la déamplification en pointillés.
Différentes phases entrent en jeu mais un seul degré de liberté est pertinent lorsque l'on étudie la valeur moyenne de l'intensité (le paramètre $\theta$). Ce paramètre est introduit à l'équation \eqref{eqtheta}.
b) Bruit normalisé sur la quadrature présentant le bruit le plus faible ($\Theta=\pi/2$) en fonction de la phase $\theta$.
c) Bruit normalisé sur la quadrature présentant le bruit le plus faible ($\Theta=\pi/2$) en fonction du facteur d'amplification $\mathcal{G}$.
d) Bruit normalisé sur la quadrature présentant le bruit le plus faible ($\Theta=\pi/2$) en fonction du facteur de déamplification $1/\mathcal{G}$.
\label{pascourbes}}	
\end{figure}

On souhaite déterminer les spectres de bruit pour les différentes quadratures $\hat X^{\theta_{LO}}$, où $\theta_{LO}$ est la phase variable d'un oscillateur local utilisé pour une détection homodyne (voir chapitre \ref{ch1}).
On rappelle que : $$\delta \hat X^{\theta_{LO}}=\delta \hat a e^{-i \theta_{LO}} +\delta \hat a^\dag e^{i \theta_{LO}}.$$
A l'aide de la relation \eqref{PSA_eq}, on obtient en sortie :

\begin{equation}
\delta\hat X_{out}^{\theta_{LO}}=\sqrt G\ \delta\hat X_{in}^{(\theta_{LO}-\phi_1)}+\sqrt {G-1}\ \delta\hat X_{in}^{(\phi_2-\theta_{LO})}.
\end{equation}
On peut alors calculer le spectre de bruit de cette quadrature :
\begin{equation}
S (\hat X^{\theta_{LO}}_{out})=2 G-1 +2 \sqrt{G(G-1)} \bra \delta\hat X_{in}^{(\theta_{LO}-\phi_1)}\delta\hat X_{in}^{(\theta_{LO}-\phi_1)}\ket.
\end{equation}
On rappelle que pour un état cohérent on a :
\begin{deqarr}
\bra \delta\hat X_{in}^{\phi_a}\delta\hat X_{in}^{\phi_b}\ket = \cos (\phi_a-\phi_b).
\end{deqarr}
On en déduit que l'on peut écrire le spectre de la quadrature $\hat X^{\theta_{LO}}$ sous la forme :
\begin{equation}\label{eq352}
S=2 G-1+2 \sqrt{G(G-1) }\ \text{cos}\ \Theta,
\end{equation}
avec $\Theta=2\theta_{LO}-(\phi_1+\phi_2)$.\\
Pour une phase  $\Theta=0$, on voit que le bruit est maximum et vaut :
\begin{equation}
S_{\max}=2 G-1+2 \sqrt{G(G-1) }.
\end{equation}
A l'inverse pour une phase $\Theta=\pi/2$, le bruit est minimum : 
\begin{equation}
S_{\min}=2 G-1-2 \sqrt{G(G-1) }.
\end{equation}
Pour étudier les fluctuations, il est intéressant de normaliser les spectres de bruit par le bruit quantique standard.
Dans ce cas, il faut diviser l'expression \eqref{eq352} par le gain donné par l'expression \eqref{eqtheta}.
On obtient donc le spectre de bruit normalisé $S_N$:
\begin{equation}
S_N=\frac{2 G-1+2 \sqrt{G(G-1) }\ \text{cos}\ \Theta}{2 G-1+2 \sqrt{G(G-1) }\ \text{cos}\ \theta}.
\end{equation}
\begin{figure}		
	\centering
\includegraphics[width=12cm]{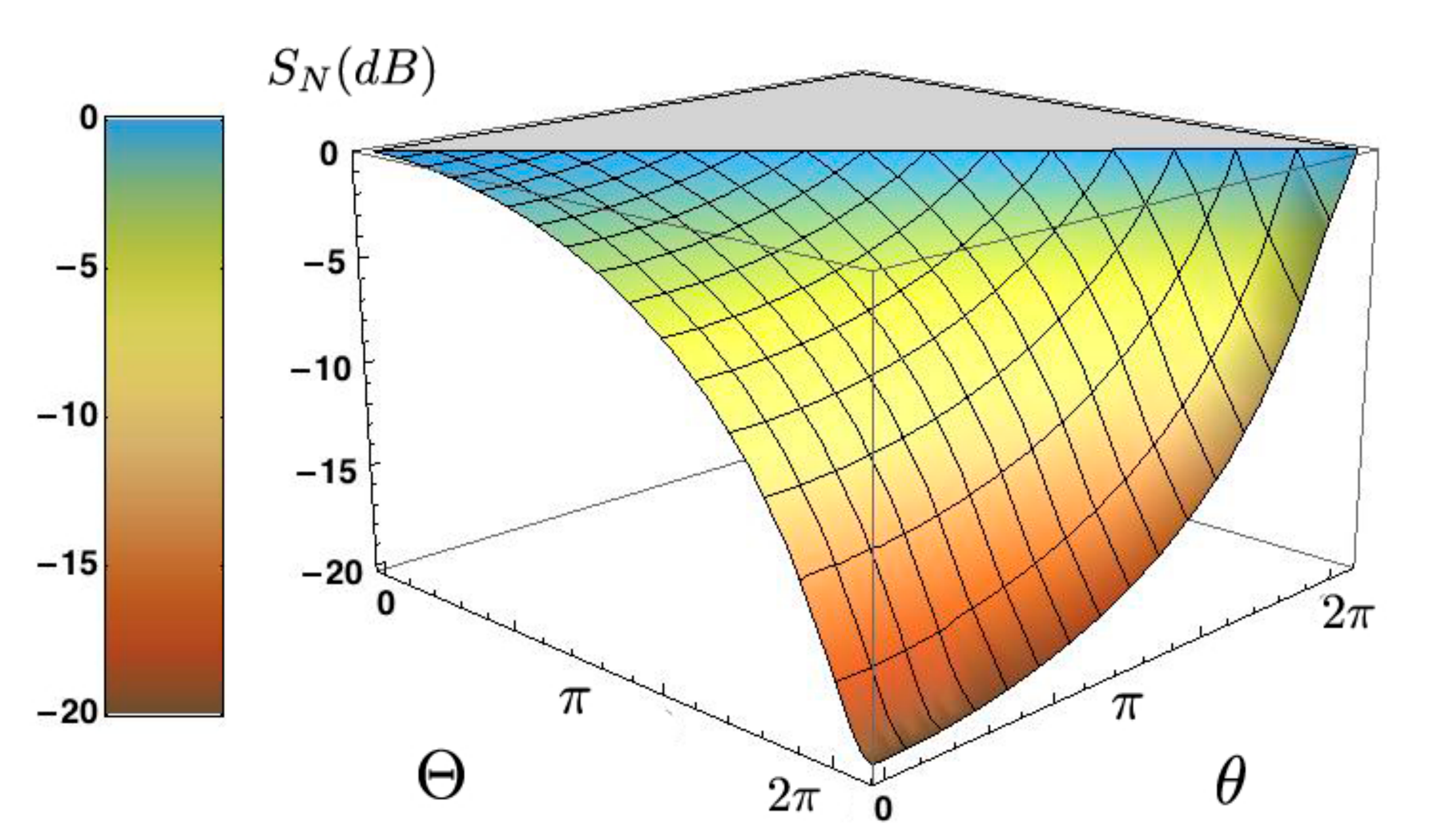} 
\caption[Compression pour un processus sensible à la phase.]{Compression pour un processus sensible à la phase pour une valeur de $G=3$ en fonction de $\theta$ et $\Theta$. On constate que l'on peut atteindre des compressions jusqu'à -20 dB dans le cas optimal. \label{psa3d}}	
\end{figure}
Revenons brièvement sur les différentes phases que nous avons introduites.
Nous prenons comme référence de phase le champ sonde.
Dans ce modèle simple, on a donc deux phénomènes sensibles à la phase, que l'on peut piloter indépendamment. 
D'une part, l'amplification ou la déamplification de la valeur moyenne du champ, qui est contrôlée par le paramètre $\theta$ et qui dépend de la différence de phase relative entre les deux pompes.
Et d'autre part, le spectre de bruit de la quadrature effectivement mesurée via la détection homodyne $\hat X^\theta_{LO}$ qui dépend de la phase de l'oscillateur local et de la somme de la phase des deux pompes donnée par le paramètre $\Theta$.
Ces deux paramètres ($\theta$ et $\Theta$) peuvent être modifiés de façon indépendante (même s'il est important de noter que les phase que nous avons introduites $\phi_1$ et $\phi_2$ ne sont pas les phases des pompes 1 et 2, elles peuvent être obtenues à l'aide du formalisme de l'optique non-linéaire présenté au début de ce chapitre.
)
On peut donc atteindre un grand nombre de régimes différents (voir figures \ref{pascourbes} et \ref{psa3d}).
D'après ce modèle la différence de phase $\theta$ peut prendre n'importe quelle valeur et on obtient donc au choix une amplification ou une déamplification de la valeur moyenne du champ associée à une réduction du bruit sous la limite quantique standard.

\section{Conclusion du chapitre}
Dans ce chapitre nous avons tout d'abord rappelé le formalisme de l'optique non--linéaire pour le mélange à 4 ondes.
Dans ce formalisme, nous avons utilisé la susceptibilité non--linéaire $\chi^{(3)}$ sans détailler l'origine de ce terme.
Deux configurations ont été étudiées qui conduisent d'une part à l'amplification idéale insensible à la phase et d'autre part à l'amplification/déamplification idéale sensible à la phase.
Nous avons détaillé comment ces processus permettent de générer des états comprimés de la lumière, respectivement à deux et un mode du champ.
A l'aide d'un modèle discret d'amplificateurs linéaires idéaux et de pertes linéaires nous avons pu donner un ordre de grandeur des taux de compression atteignables dans ces milieux.
Nous verrons au chapitre \ref{ch4} qu'une structure atomique en double $\Lambda$ permet d'envisager la réalisation de ces deux processus par mélange à 4 ondes.
La susceptibilité  non--linéaire  $\chi^{(3)}$, introduite dans ce chapitre de manière phénoménologique, trouvera alors son contenu physique.

%% file: chapitre4v5.tex
\setcounter{minitocdepth}{1}
\chapter{Modèle microscopique et traitement quantique}\minitoc\label{ch4}
\vspace{2cm}
Dans ce chapitre, nous présentons en détail le modèle théorique qui permet de rendre compte des résultats expérimentaux qui seront présentés dans la partie~III.\\
Dans un premier temps, nous rappelons des résultats  sur la \textit{transparence électromagnétiquement induite} (EIT) pour un schéma de niveaux atomiques en $\Lambda$.\\
Puis, nous introduisons le modèle microscopique basé sur un schéma des niveaux en \textit{double}--$\Lambda$ et nous détaillons la résolution des équations de Heisenberg-Langevin dans ce cas.
Les résultats obtenus pour des atomes immobiles sont alors présentés.
Dans cette situation, qui correspond à des expériences dans un ensemble d'atomes froids, nous distinguons deux régimes de fonctionnement en fonction du gain supérieur ou inférieur à 1.\\
Dans une dernière partie, une extension du modèle à une vapeur atomique c'est-a-dire un milieu où les atomes ne sont plus immobiles (\textit{atomes chauds}) est discutée.\\

	\section{Transparence électromagnétiquement induite. Modèle de l'atome en $\Lambda$}
Nous présentons ici un modèle microscopique à 3 niveaux en \textit{simple}-$\Lambda$, dans lequel nous allons décrire le phénomène d'EIT.
Les processus d'interactions entre les photons et les atomes sont décrits à l'aide des équations d'Heisenberg-Langevin \cite{Marangos:1998p9901,Fleischhauer:2000p4594,Andre05,Dantan05}.
L'objectif ici n'est pas d'étudier le phénomène d'EIT en détail \cite{Mishina:2008p8741,Kupriyanov:2010p8743}, mais de rappeler le formalisme utilisé et d'introduire les effets que l'on retrouvera dans la partie \ref{Modele4WM} lors de l'étude du système, plus complexe, en \textit{double--$\Lambda$}.

\subsection{Schéma énergétique à 3 niveaux en simple Lambda }
\begin{figure}
\centering
\includegraphics[scale=0.75]{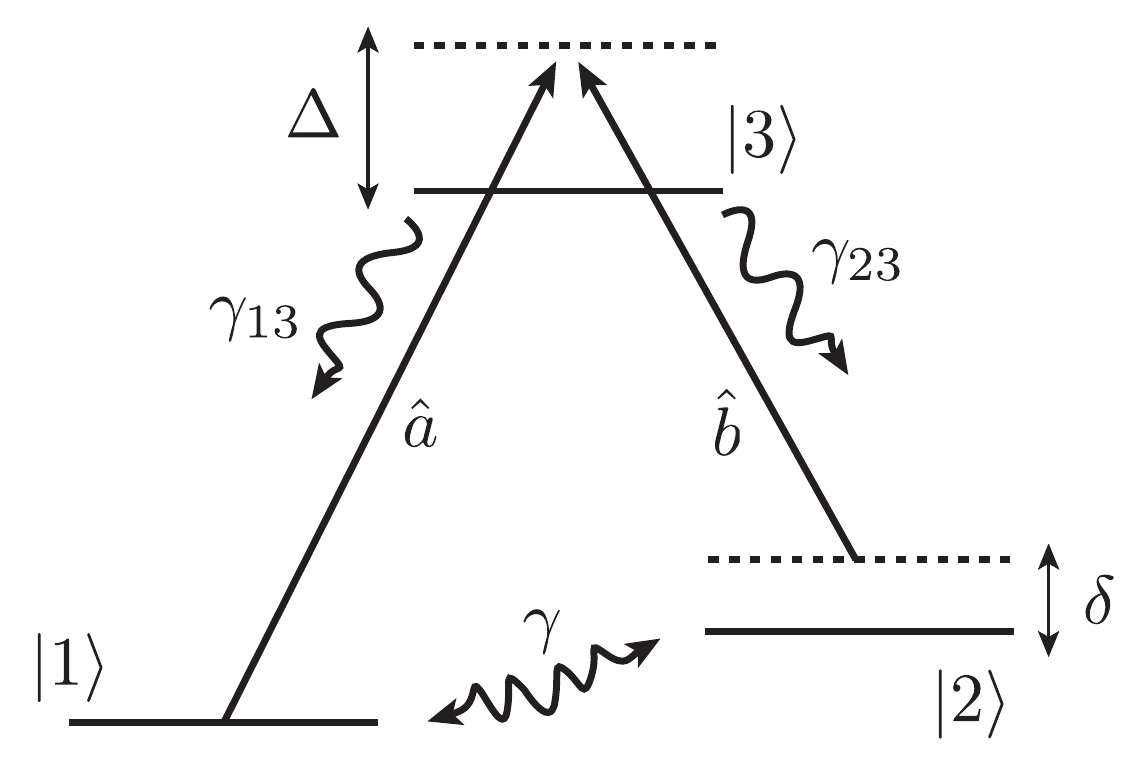} 
\caption{Schéma d'interaction lumière-matière à trois niveaux en simple $\Lambda$.}		
\label{fig_3niveaux}
\end{figure}

Dans cette partie, nous allons nous intéresser à un schéma d'interaction lumière-matière à trois niveaux et deux champs électromagnétiques dit en $\Lambda$.
Ce schéma est décrit sur la figure \ref{fig_3niveaux}.
Il s'agit de deux niveaux fondamentaux $|1\rangle$ et $|2\rangle$ et d'un niveau excité $|3\rangle$, couplés par deux champs notés $\hat{a}$ et $\hat{b}$.\\
Les notations de la figure \ref{fig_3niveaux} sont détaillées ci-dessous :
\begin{itemize}
\item $\Delta$ est le désaccord \textit{à un photon} qui correspond à $\Delta=\omega_{a}-\omega_{31}$, c'est-à-dire la différence entre la fréquence du champ $\hat{a}$ et celle correspondant à la transition $|3\rangle \leftrightarrow |1\rangle$.\\
\item $\delta$ est le désaccord \textit{à deux photons} qui s'écrit : $\delta=\omega_{32}+\Delta-\omega_{b}$, avec $\omega_{32}$ la fréquence de la transition $|3\rangle \leftrightarrow |2\rangle$ et $\omega_b$ celle du champ $\hat{b}$.\\
\item $\gamma_{i3}$ désigne le taux de relaxation du niveau $|3\rangle$ vers le fondamental $|i\rangle$.
On prendra : $\gamma_{13}=\gamma_{23}=\frac{\Gamma}{2}$ avec $\Gamma$ la largeur du niveau excité.
Par souci de simplification, nous faisons ici l'hypothèse qu'il n'y a pas de désexcitation depuis le niveau $|3\rangle$ vers d'autres niveaux.\\
\item On note $\gamma$ le taux de décohérence de la cohérence atomique entre les niveaux $|1\rangle$ et $|2\ket$.
\end{itemize}
~~\\
	Dans ce modèle simple il n'y a pas de taux de relaxation ni de pompage des populations pour les niveaux  $|1\rangle$ et $|2\rangle$ depuis l'extérieur.
Il s'agit d'un modèle microscopique fermé à trois niveaux.

\subsection{Equations d'Heisenberg-Langevin }\label{Hamiltonien_3niveaux}

Pour décrire l'évolution de ce système atome-champ et les fluctuations quantiques des observables du champ, nous avons utilisé le formalisme des équations dites de Heisenberg-Langevin. Le milieu atomique considéré est un ensemble de $N$ atomes contenus dans un volume cylindrique $V$ défini par la surface transverse $S$ du faisceau laser et la longueur du milieu $L$ selon l'axe de propagation $z$.\\
Le champ $\hat{a}$  (respectivement $\hat{b}$) s'écrit dans une description quantique:

\begin{equation}\label{def_a}
\hat{E_a}(z,t)= \mathcal{E}_a \left(\hat{a}(z,t)e^{i(k_az-\omega_a t)}+\hat{a}^\dag(z,t)e^{-i(k_az-\omega_a t)}\right),
\end{equation}
avec $ \mathcal{E}_a =\sqrt{\frac{\hbar\omega_a}{2\epsilon_0 V}}$, le champ électrique d'un photon.\\

Pour les variables atomiques, nous utiliserons les opérateurs atomiques collectifs tels que proposés dans ~\cite{Fleischhauer95,Lukin00,Dantan05}. 
Les opérateurs collectifs sont définis sur une tranche $\Delta z$ contenant un grand nombre d'atomes $N_z\gg 1$ :

\begin{equation}\label{def_sigma_tilde}
\tilde \sigma_{uv}(z,t) = \frac{1}{N_z} \sum_{j=1}^{N_z} |u_j  \rangle \langle v_j | e^{(- i \tilde \omega_{uv} t+i k_{uv} z)}.
\end{equation}
Dans cette expression on a défini : $\tilde\omega_{31}=\omega_a$, $\tilde\omega_{32}=\omega_b$ et  $\tilde\omega_{21}=\omega_b-\omega_a$. 
$ k_{uv}$ est la projection sur l'axe $z$ de $\vec k_{uv}$, et $\vec k_{31}$ est le vecteur d'onde correspondant au champ $\hat{a}$, $\vec k_{32}$ est le vecteur d'onde correspondant au  champ $\hat{b}$ et $\vec k_{21}=\vec k_{23}-\vec k_{13}$.
Dans cette notation on a pris la convention : $\vec{k}_{uv}=-\vec{k}_{vu}$.\\
Les termes $\tilde\sigma_{11},\tilde\sigma_{22},\tilde\sigma_{33}$ sont appelés des populations.
Les termes $\tilde\sigma_{13},\tilde\sigma_{23},\tilde\sigma_{12}$ sont appelés des cohérences.\\

Dans l'approximation de l'onde tournante, c'est-à-dire en négligeant les termes évoluant à une fréquence de l'ordre de $2\omega_a$, le hamiltonien dipolaire électrique s'écrit :

\begin{equation}\label{eq_ham_3}
H_{int}=-\frac{\hbar N}{L}\int^L_0 \Delta \tilde{\sigma}_{33}(z,t)+ \delta \tilde{\sigma}_{22}(z,t)+\left( g_a\hat{a}(z,t)\tilde{\sigma}_{31}(z,t)+g_b\hat{b}(z,t)\tilde{\sigma}_{32}(z,t) + H.c.\right)dz,
\end{equation}
avec $g_i=\frac{\wp_i  \mathcal{E}_i}{\hbar}$ pour $i\in\{a,b\}$, H.c. désignant l'hermitien conjugué et $\wp_i$ l'élément de dipôle de la transition concernée.\\
On peut alors écrire l'évolution hamiltonienne de la manière suivante :

\begin{equation}\label{evol_hamiltonienne}
\frac{\partial}{\partial t}\tilde{\sigma}_{uv}=\frac{i}{\hbar}[H_{int},\tilde{\sigma}_{uv}(z,t)].
\end{equation}
	On obtient alors le système suivant :
\begin{deqarr}\arrlabel{eq_evolution_eit}
\frac{\partial}{\partial t}\tilde{\sigma}_{11}&=&-i\left( g_a \hat{a}\tilde{\sigma}_{31}-g_a^* \hat{a}^\dag\tilde{\sigma}_{13}\right)+\frac{\Gamma}{2}\tilde{\sigma}_{33}\\
\frac{\partial}{\partial t}\tilde{\sigma}_{22}&=&-i\left( g_b \hat{b}\tilde{\sigma}_{32}-g_b^* \hat{b}^\dag\tilde{\sigma}_{23}\right)+\frac{\Gamma}{2}\tilde{\sigma}_{33}\\
\frac{\partial}{\partial t}\tilde{\sigma}_{33}&=&-i\left(g_a^* \hat{a}^\dag\tilde{\sigma}_{13}+g_b^* \hat{b}^\dag\tilde{\sigma}_{23}-g_a \hat{a}\tilde{\sigma}_{31}-g_b \hat{b}\tilde{\sigma}_{32}\right)-\Gamma\tilde{\sigma}_{33}\\
\frac{\partial}{\partial t}\tilde{\sigma}_{31}&=&-i\left(\Delta\tilde{\sigma}_{31}+g_a^*\hat{a}^\dag(\tilde{\sigma}_{11}-\tilde{\sigma}_{33})+g_b^*\hat{b}^\dag\tilde{\sigma}_{21}\right)-\frac{\Gamma}{2}\tilde{\sigma}_{31}\\
\frac{\partial}{\partial t}\tilde{\sigma}_{23}&=&-i\left((\delta-\Delta)\tilde{\sigma}_{23}+g_b\hat{b}(\tilde{\sigma}_{33}-\tilde{\sigma}_{22})-g_a\hat{a}\tilde{\sigma}_{21}\right)-\frac{\Gamma}{2}\tilde{\sigma}_{23}\\
	\frac{\partial}{\partial t}\tilde{\sigma}_{21}&=&-i\left(\delta\tilde{\sigma}_{21}+g_b\hat{b}\tilde{\sigma}_{31}-g_a^*\hat{a}^\dag\tilde{\sigma}_{23}\right)-\gamma\tilde{\sigma}_{21}.
\end{deqarr}
Dans ces équations, les termes non-hamiltoniens ne sont naturellement pas présents.
En effet, dans cette partie, nous allons nous intéresser uniquement à l'évolution des valeurs moyennes des opérateurs.
Dans ce cas, l'équation de Heisenberg--Langevin et l'équation d'Ehrenfest coïncident comme nous l'avons vu au chapitre 1.
L'étude du rôle de ces termes sur les fluctuations quantiques sera faite dans la section \ref{Lange}.\\

Pour décrire l'évolution des champs $\hat{a}$ et $\hat{b}$, il faut ajouter au groupe d'équations \ref{eq_evolution_eit}, les deux équations de Maxwell dans le milieu atomique pour les enveloppes lentement variables des champs.
\begin{deqarr}
\left(\frac{\partial}{\partial t}+c\frac{\partial}{\partial z}\right)\hat{a}(z,t)=ig_a\tilde{\sigma}_{13}(z,t),\\
\left(\frac{\partial}{\partial t}+c\frac{\partial}{\partial z}\right)\hat{b}(z,t)=ig_b\tilde{\sigma}_{23}(z,t).
\end{deqarr}

\subsection{Régime stationnaire}
Nous allons désormais nous placer dans la situation où le champ $\hat{a}$ est très intense devant le champ $\hat{b}$.
Le champ $\hat{a}$ sera appelé champ de contrôle et sera traité classiquement ; le champ $\hat{b}$ sera appelé champ sonde.  
Comme nous nous intéressons, pour l'instant, aux valeurs moyennes du champ sonde, il pourra lui aussi être traité comme un champ classique.
On remplacera donc dans les équations (\ref{eq_evolution_eit}) les termes $g_a\hat{a}$ par $\frac{\Omega_c}{2}$ et $g_b\hat{b}$ par $\frac{\Omega_s}{2}$, où $\Omega_c$ est la pulsation de Rabi du champ pompe et $\Omega_s$ la pulsation de Rabi du champ sonde.
Ce système s'écrit alors dans le régime stationnaire :
\begin{deqarr}\arrlabel{eq_evolution_eit_stat}
	0&=&-i\left( \frac{\Omega_c}{2}\tilde{\sigma}_{31}-\frac{\Omega_c^*}{2}\tilde{\sigma}_{13}\right)+\frac{\Gamma}{2}\tilde{\sigma}_{33}\\
	0&=&-i\left( \frac{\Omega_s}{2}\tilde{\sigma}_{32}-\frac{\Omega_s^*}{2}\tilde{\sigma}_{23}\right)+\frac{\Gamma}{2}\tilde{\sigma}_{33}\\
0&=&-i\left(\frac{\Omega_c^*}{2}\tilde{\sigma}_{13}+\frac{\Omega_s^*}{2}\tilde{\sigma}_{23}-\frac{\Omega_c}{2}\tilde{\sigma}_{31}-\frac{\Omega_s}{2}\tilde{\sigma}_{32}\right)-\Gamma\tilde{\sigma}_{33}\\
	0&=&-i\left(\Delta\tilde{\sigma}_{31}+\frac{\Omega_c^*}{2}(\tilde{\sigma}_{11}-\tilde{\sigma}_{33})+\frac{\Omega_s^*}{2}\tilde{\sigma}_{21}\right)-\frac{\Gamma}{2}\tilde{\sigma}_{31}\\
	0&=&-i\left((\delta-\Delta)\tilde{\sigma}_{23}+\frac{\Omega_s^*}{2}(\tilde{\sigma}_{33}-\tilde{\sigma}_{22})-\frac{\Omega_c}{2}\tilde{\sigma}_{21}\right)-\frac{\Gamma}{2}\tilde{\sigma}_{23}\\
 	0&=&-i\left(\delta\tilde{\sigma}_{21}+\frac{\Omega_s}{2}\tilde{\sigma}_{31}-\frac{\Omega_c^*}{2}\tilde{\sigma}_{23}\right)-\gamma\tilde{\sigma}_{21},
	\end{deqarr}

\subsection{Calcul de la susceptibilité diélectrique}
\begin{figure} [h!]
\centering
\includegraphics[width=11.5	cm]{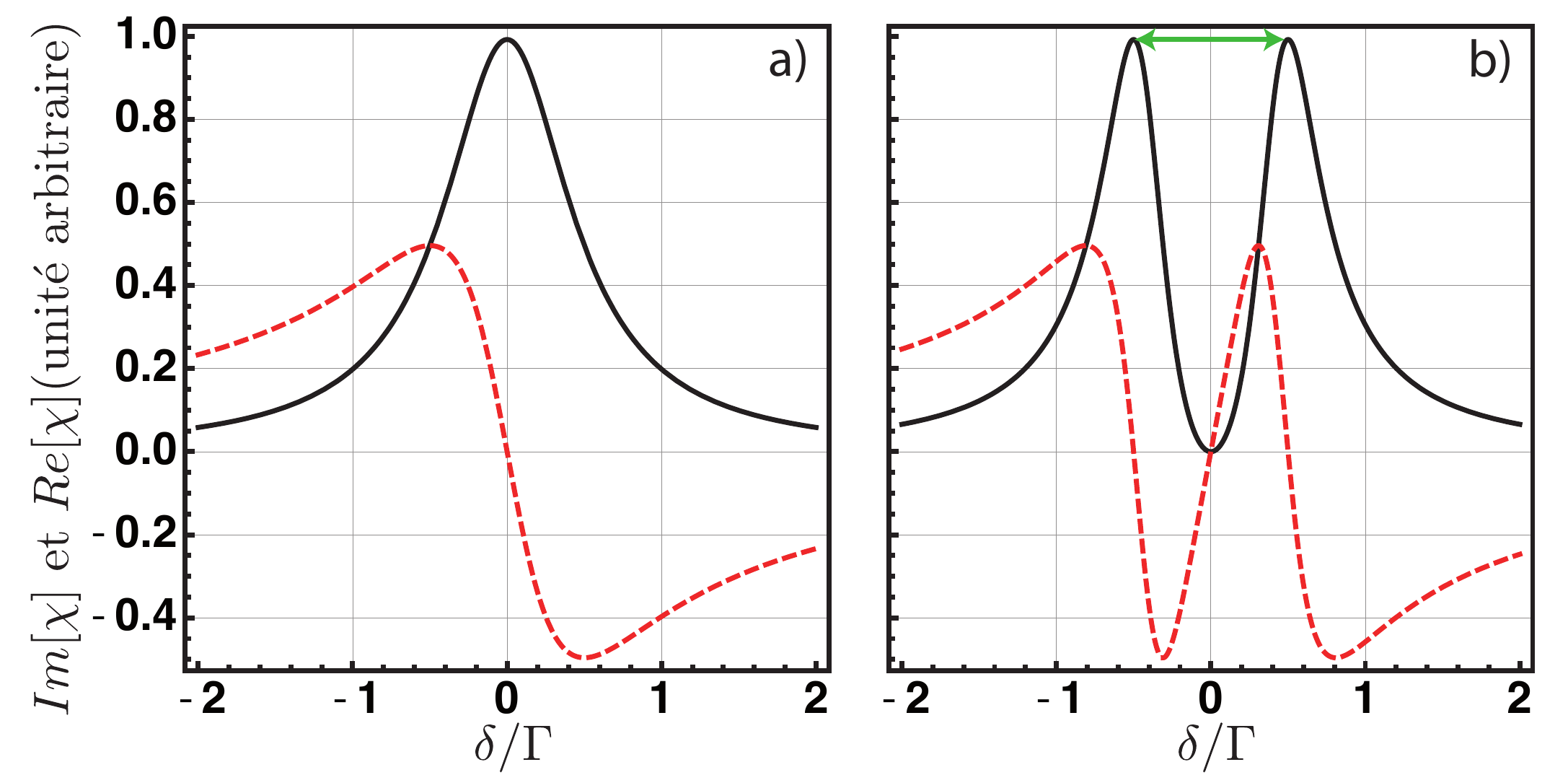}
\caption[Partie réelle et imaginaire de $\chi$ en présence ou en l'absence de champ de contrôle.]{Partie réelle (rouge pointillé) et imaginaire (noir plein) de la susceptibilité atomique  $\chi$ pour le ~rubidium~85 en l'absence a) ou en présence b) du champ de contrôle. La flèche verte correspond à la largeur de la fenêtre EIT, telle que nous l'avons définie. Paramètres utilisés : la pulsation de Rabi $\Omega_c=\Gamma$, le taux de décohérence des niveaux 1 et 2 : $\gamma=0$ et le désaccord à un photon $\Delta=0$. La fenêtre de transparence est donnée par la flèche verte sur la figure b).}	
\label{fig_EIT1}	
\end{figure}

	Dans l'hypothèse d'un champ de contrôle intense ($\Omega_c \gg \Omega_s$), on peut faire l'approximation que la population est majoritairement pompée dans le niveau $|2\rangle$.
On peut alors résoudre le système \eqref{eq_evolution_eit_stat} ainsi simplifié afin de déterminer la susceptibilité linéaire $\chi$ du milieu atomique vue par le champ sonde. Elle est donnée par la relation :
\begin{equation}
N\wp_{23}\ \tilde{\sigma}_{23}^{(1)} = \epsilon_0\ \chi E_s ,
\end{equation}
où $\tilde{\sigma}_{23}^{(1)}$ correspond à la solution à l'ordre 1 en champ sonde pour la cohérence entre les niveaux $|2\rangle$ et $|3\rangle$ et $E_s=\frac{\hbar \Omega_s/2}{\wp_{23}}$ est l'amplitude du champ électrique pour le faisceau sonde.\\
On obtient alors simplement la susceptibilité linéaire pour le champ sonde \cite{Ortalo09} :
\begin{equation}
\chi  =\frac{N\wp_{23}^2}{\hbar\epsilon_0 }\frac{2 (\gamma +i\delta)}{2 (\gamma +i \delta) (  2 (\delta - \Delta)-i\Gamma) - i \Omega_c^2}.
\end{equation}
La partie réelle de la susceptibilité, que l'on note $\chi_r$, détermine l'indice de réfraction du milieu pour le faisceau sonde.
La partie imaginaire, notée $\chi_i$, donne accès à la dissipation du champ sonde par le milieu c'est-à-dire à l'absorption \cite{Fleischhauer:2005p1645}.
Les profils théoriques, en présence et en absence de champ de contrôle, de $\chi_r$ et $\chi_i$  sont présentés dans la figure \ref{fig_EIT1}.
Ces courbes sont donnés pour la ligne D1 des atomes de Rb$^{85}$, dont les grandeurs importantes sont rappelées dans l'annexe \ref{Annexe_Rb}.

\subsection{Influence des paramètres sur la fenêtre de transparence}
Le phénomène dit de transparence électromagnétiquement induite correspond à l'annulation de $\chi_i$ en présence de pompe pour un désaccord à 2 photons nul.
On définit la fenêtre de transparence comme l'écart entre les deux pics de la partie imaginaire en présence de pompe (voir figure \ref{fig_EIT1}.b).
C'est la zone du spectre où l'absorption est significativement modifiée par rapport à la situation en absence de champ de contrôle. Il est intéressant de regarder l'effet des différents paramètres sur la largeur et le contraste de cette fenêtre. 
\subsubsection{Effet de la pulsation de Rabi du champ de contrôle}
La figure \ref{fig_EIT_omega} présente la modification de la largeur de la fenêtre de transparence en fonction de la pulsation de Rabi du champ de contrôle.
Dans le cas simple de trois niveaux non élargis par effet Doppler (atomes froids), on peut noter que la fenêtre de transparence est donnée la valeur de $\Omega_c$.

\subsubsection{Effet du taux de décohérence}
\begin{figure}[h!]
\centering
	\includegraphics[width=14.4cm]{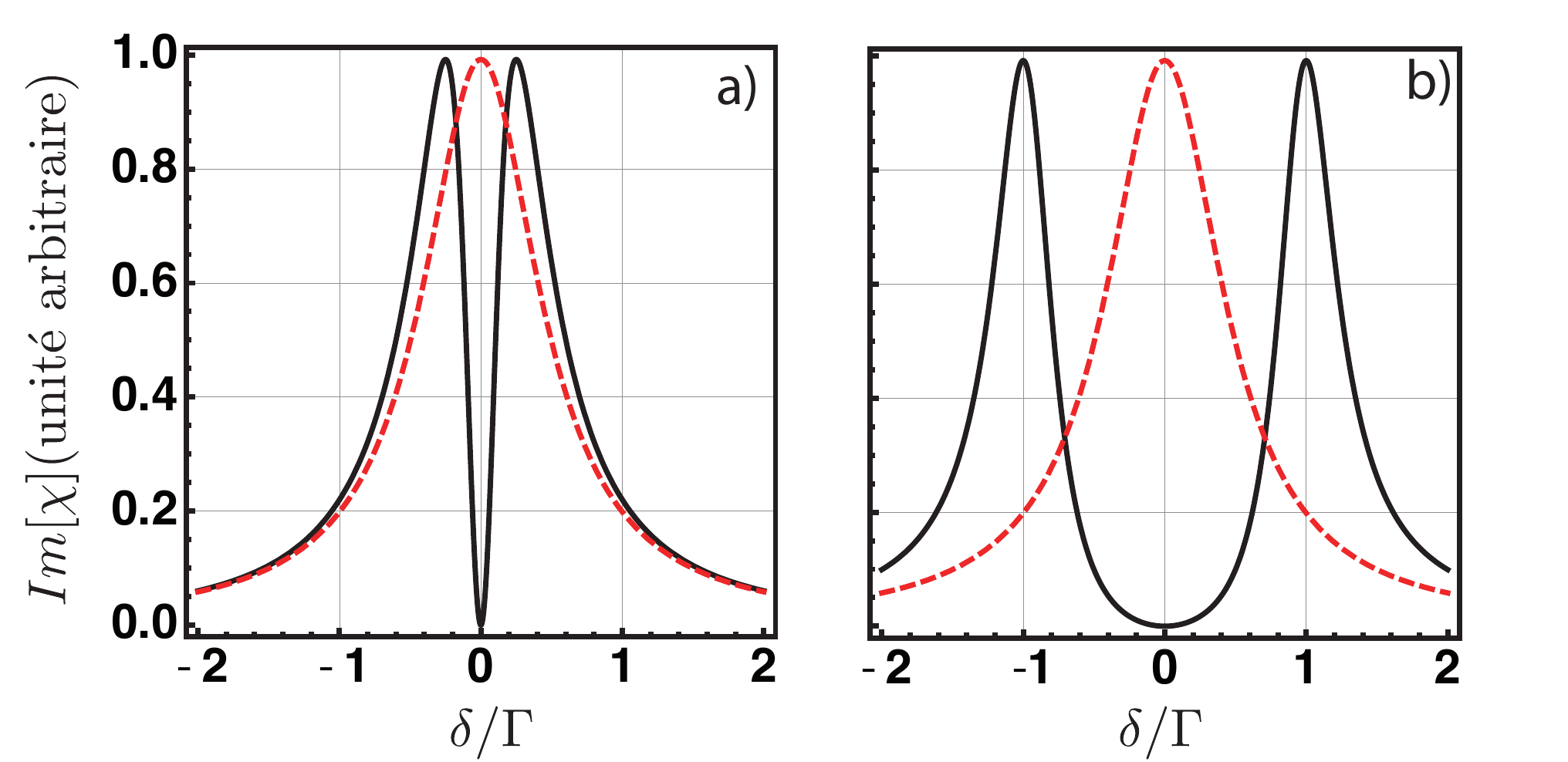}
	\caption[Partie imaginaire de $\chi$ en présence ou en l'absence de champ de contrôle. Effet de $\Omega_c$]{Partie imaginaire de  $\chi$  en l'absence (rouge pointillé) ou en présence (noir plein) du champ de contrôle. Paramètres utilisés : la pulsation de Rabi a) $\Omega_c=0.5\ \Gamma$ et b) $\Omega_c=2\ \Gamma$. Le taux de décohérence des niveaux 1 et 2 et le désaccord à un photon sont pris nuls.}	
 \label{fig_EIT_omega}	
	 	\includegraphics[width=14.4cm]{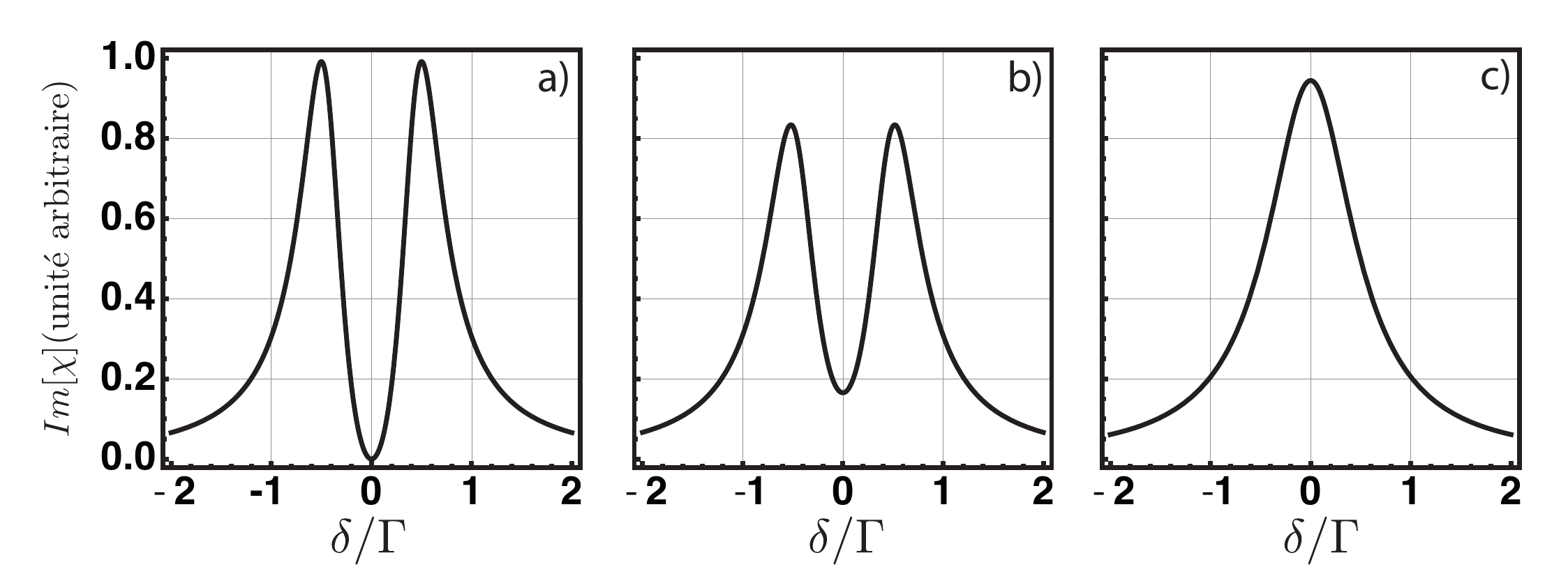}
	 		\caption[Partie imaginaire de $\chi$ en présence ou en l'absence de champ de contrôle. Effet de $\gamma$]{Partie imaginaire de  $\chi$  en présence du champ de contrôle. Paramètres utilisés : la pulsation de Rabi $\Omega_c=\Gamma$ . Le taux de décohérence des niveaux 1 et 2 vaut a) $\gamma=0$, b) $\gamma=0.1\ \Gamma$ et c) $\gamma=10\ \Gamma$. Le désaccord à un photon est pris nul.}	
 		 \label{fig_EIT_gamma}	

\end{figure}

Un paramètre important pour l'observation du phénomène d'EIT et plus généralement de tous les phénomènes reposant sur la préparation cohérente d'un milieu atomique est le taux de décohérence \cite{Lukin00,Lukin:1999p1647}.
Dans le cas étudié ici, il s'agit du taux de décohérence de l'opérateur $\tilde{\sigma}_{12}$, noté $\gamma$. 
La figure \ref{fig_EIT_gamma} présente l'effet de ce paramètre sur la valeur de l'absorption à résonance. 
L'augmentation de $\gamma$ va avoir tendance a réduire, voire à supprimer totalement, l'effet de transparence lorsqu'il devient grand devant $\Gamma$.
Cet effet de la décohérence sera donc une différence importante entre les différentes configurations pour ce genre d'expérience.
Les paramètres qui influents sur la valeur de $\gamma$ seront discutés dans la partie \ref{param_froids}.

\subsubsection{Autres effets}
Nous verrons, au chapitre 6, une démonstration expérimentale de l'EIT dans une vapeur atomique.
Dans ce cas, outre la pulsation de Rabi et le taux de décohérence, l'élargissement inhomogène va jouer un rôle sur la largeur de la fenêtre de transmission \cite{Field:1991p9519,Li:2004p8962}.
De même, comme cela a été démontré dans \cite{Ortalo09}, la prise en compte de la structure hyperfine de l'atome en dépassant le modèle de l'atome à 3 niveaux, permet de rendre compte plus précisément des profils de transmission observés expérimentalement.

	\section{Mélange à 4 ondes. Modèle microscopique en double--$\Lambda$}	\label{Modele4WM}
Nous avons montré dans le chapitre précédent qu'il est possible de produire des états non classiques du champ à l'aide d'amplificateurs sensible ou insensible à la phase.
Ce type d'amplification peut, entre autres, être réalisées dans un milieu atomique.
Nous venons de voir comment la susceptibilité d'un milieu atomique était modifiée en présence de champ électromagnétique dans le cas simple d'atomes décrit par un modèle à trois niveaux.
Nous allons nous intéresser maintenant au modèle microscopique qui permet de rendre compte de ces phénomènes.
Les atomes seront décrits par un modèle à 4 niveaux en double-$\Lambda$ en présence de deux champs de contrôle.
Cette étude sera réalisée tout d'abord pour des atomes froids, puis pour une vapeur atomique en prenant en compte les effets de l'élargissement inhomogène. \\
Seul le cas de l'amplification insensible à la phase sera traité dans ce manuscrit car il correspond à la situation expérimentale que nous avons étudiée dans le chapitre \ref{ch4}.
Le modèle microscopique pour l'amplification sensible à la phase par mélange à 4 ondes sera détaillé dans \cite{Glorieux11}.

Pour décrire le plus complètement possible l'interaction entre la lumière et la matière dans les expériences présentées dans ce manuscrit, nous avons utilisé un modèle microscopique à quatre niveaux dit en double $\Lambda$. 
Nous décrivons donc tout d'abord ce modèle, puis nous dérivons les équations d'évolution afin d'étudier les propriétés quantiques des faisceaux générés.

\subsection{Schéma énergétique à 4 niveaux en double $\Lambda$}
\begin{figure}
\centering
\includegraphics[width=10.5cm]{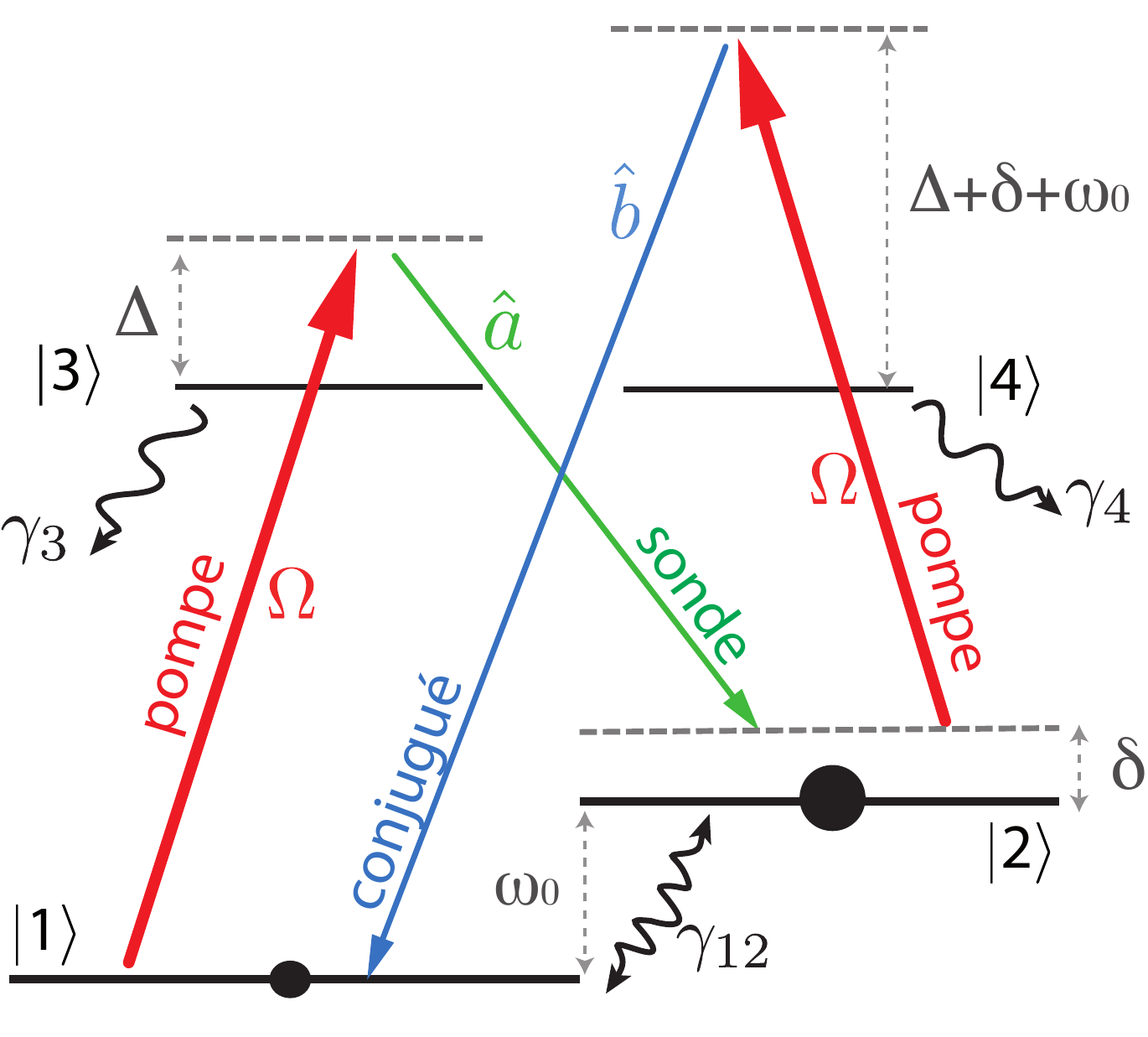} 
\caption{Schéma d'interaction lumière-matière à quatre niveaux en double $\Lambda$.	ON verra plus loin que dans l'état stationnaire, la population est essentiellement dans le niveau $|2\rangle$, ce que l'on schématise par la taille des points dans les niveaux  $|1\rangle$ et  $|2\rangle$.
	\label{fig_4niveaux}}		
\end{figure}

La figure \ref{fig_4niveaux} présente notre modèle microscopique à quatre niveaux.
Les atomes interagissent avec quatre champs électromagnétiques.\\
Deux champs intenses que l'on appellera champ pompe et dont l'interaction avec le milieu atomique sera traitée de manière semi-classique.\\
Deux champs $\hat{a}$ et $\hat{b}$ que l'on appellera respectivement champs sonde et conjugué.
Ces champs seront, eux, traités de manière quantique à l'aide d'opérateurs en représentation d'Heisenberg.
En effet, nous nous intéressons non seulement aux valeurs moyennes de ces opérateurs qui sont des grandeurs accessibles classiquement, mais aussi à leurs variances, ce qui nécessite un traitement quantique.\\

Le système atomique est décrit par deux niveaux fondamentaux $|1\rangle$ et $|2\rangle$ et de deux niveaux excités $|3\rangle$ et $|4\rangle$. Les transitions $|1\rangle \leftrightarrow|2\rangle$ et $|3\rangle \leftrightarrow |4\rangle$ sont supposées interdites par couplage dipolaire.
Les notations de la figure \ref{fig_4niveaux} sont détaillées ci-dessous :
\begin{itemize}
\item $\omega_0$ est la différence en fréquence entre les niveaux $|1\rangle$ et $|2\rangle$. Il s'agira typiquement de l'écart hyperfin entre deux niveaux fondamentaux.\\
\item $\Omega$ est la pulsation de Rabi du champ pompe définie par $\Omega=\frac{2\wp E}{\hbar}$, avec $\wp$ l'élément de dipôle pris identique pour les transitions $|1\rangle \leftrightarrow|3\rangle$ et $|2\rangle \leftrightarrow |4\rangle$ et $E$ le champ électrique.\\
\item $\Delta$ est le désaccord \textit{à un photon} qui correspond à $\Delta=\omega_{p}-\omega_{13}$, c'est-à-dire la différence entre la fréquence du champ pompe et celle de la transition $|1\rangle \leftrightarrow |3\rangle$.\\
\item $\delta$ est le désaccord \textit{à deux photons} qui s'écrit : $\delta=\omega_{23}+\Delta-\omega_{b}$, avec $\omega_{23}$ la fréquence de la transition $|2\rangle \leftrightarrow |3\rangle$ et $\omega_b$ celle du champ $\hat{b}$.\\
\item $\gamma_{k}$ désigne le taux de relaxation du niveau $|k\rangle$.
Pour $k=3,4$ ce taux vaut $\gamma_{k}=\Gamma$ la largeur du niveau excité. Pour $k=1,2$ ce taux est supposé nul : nous faisons ici l'hypothèse qu'il n'y a pas de désexcitation vers l'extérieur du système ni de phénomène de relaxation des populations entre les deux niveaux de l'état fondamental. \\
On considère de plus que la desexcitation spontanée depuis les niveaux de l'état excité se réalise de manière isotrope vers les deux niveaux fondamentaux, ce qui implique que les taux de  relaxation de $|3\rangle$ à $ |1\rangle$ et de $|4\rangle $ à $ |2\rangle$ soient identiques et égaux ~à~$\frac{\Gamma}{2}$\\
\item On note $\gamma$ le taux de décohérence de la cohérence atomique entre les niveaux $|1\rangle$ et $| 2\ket$.
\end{itemize}
	Dans ce modèle il n'y a pas de taux de relaxation ni de pompage des populations pour les niveaux  $|1\rangle$ et $|2\rangle$ depuis l'extérieur.
Il s'agit d'un modèle microscopique fermé à quatre niveaux.
De plus, seul les couplages représentés sur la figure  \ref{fig_4niveaux}  sont pris en compte. Par exemple, on négligera le couplage du champ $\hata$ avec la transition $|2\ket\to |4\ket$ pour des raisons de règles de sélection qui seront détaillées dans l'annexe \ref{Annexe_Lambda}.

\subsection{Equations d'Heisenberg-Langevin}\label{Hamiltonien_4niveaux}
L'évolution des opérateurs atomiques est décrite par les équations d'Heisenberg.
Pour prendre en compte les termes de fluctuations introduits par la dissipation, il est nécessaire d'ajouter à l'évolution hamiltonienne des équations d'Heisenberg des termes dit de \textit{forces de Langevin.}
L'ensemble de ces deux contributions est décrit par les équations de Heisenberg-Langevin que nous allons obtenir dans cette section dans le cadre de notre modèle \cite{Lezama:2008p12938}.
		\subsubsection{Hamiltonien d'interaction}
De manière similaire au paragraphe \ref{Hamiltonien_3niveaux}, on établit le hamiltonien effectif d'interaction dans sa forme continue pour le système étudié:
\begin{equation}
\hat{H}_{int}=-\frac{\hbar N}{L}\int^L_0 \Delta_0\ \tilde{\sigma}_{44}+\Delta\tilde{\sigma}_{33}+\delta\tilde{\sigma}_{22}+\left[g_a\hat{a}\ \tilde{\sigma}_{32} +g_b \hat{b}\ \tilde{\sigma}_{41}+\frac{\Omega}{2}(\tilde{\sigma}_{31}+\tilde{\sigma}_{42})+H.c.\right]dz,
\end{equation} 
avec $\Delta_0=\omega_0+\Delta+\delta$.\\
Dans cette équation les termes $\tilde{\sigma}_{uv}$, $\hat{a}$ et $\hat{b}$ dépendent à priori de $z$ et de $t$.\\
Les $\tilde{\sigma}_{uv}$ sont les enveloppes lentement variables des opérateurs atomiques collectifs définis dans l'équation~(\ref{def_sigma_tilde}) et les opérateurs $\hat{a}$ et $\hat{b}$ sont les enveloppes lentement variables des champs électriques sonde et conjugué définis dans l'équation~(\ref{def_a}).\\
Dans l'équation ~(\ref{def_sigma_tilde}) nous avons introduit $\tilde \omega_{31} = \tilde \omega_{42}= \omega_p $, $\tilde \omega_{32}=\omega_a$, $\tilde \omega_{41}=\omega_b$ et  $\tilde \omega_{21}=~\tilde \omega_{43}=~\omega_{p}~-~\omega_{a}$.\\
$\vec k_{31}$ et $\vec k_{42}$ sont les vecteurs d'onde du champ pompe, $\vec k_{32}$ est  le vecteur d'onde du champ $\hata$, et $\vec k_{41}$ est  le vecteur d'onde du champ $\hat b$. Enfin, on note : $\vec k_{43}=\vec k_{42} - \vec k_{32}$.\\

\noindent	Pour déterminer l'évolution hamiltonienne des populations $\tilde{\sigma}_{ii}$, on utilise la relation~(\ref{evol_hamiltonienne}).
En ajoutant ensuite les contributions non hamiltonienne on obtient  : 

\begin{deqarr}\arrlabel{eq_evolution_pop_4WM}
\frac{\partial}{\partial t}\tilde{\sigma}_{11}&=&-i\left( g_b \hat{b}\tilde{\sigma}_{41}-g_b^* \hat{b}^\dag\tilde{\sigma}_{14}+\frac{\Omega}{2}(\tilde{\sigma}_{31}-\tilde{\sigma}_{13})\right)+\frac{\Gamma}{2}(\tilde{\sigma}_{33}+\tilde{\sigma}_{44})+\tilde{f}_{11}\ \\
	\frac{\partial}{\partial t}\tilde{\sigma}_{22}&=&-i\left( g_a \hat{a}\tilde{\sigma}_{32}-g_a^* \hat{a}^\dag\tilde{\sigma}_{23}+\frac{\Omega}{2}(\tilde{\sigma}_{42}-\tilde{\sigma}_{24})\right)+\frac{\Gamma}{2}(\tilde{\sigma}_{33}+\tilde{\sigma}_{44})+\tilde{f}_{22}\ \\
\frac{\partial}{\partial t}\tilde{\sigma}_{33}&=&-i\left( g_a^* \hat{a}^\dag\tilde{\sigma}_{23}-g_a \hat{a}\tilde{\sigma}_{32}+\frac{\Omega}{2}(\tilde{\sigma}_{13}-\tilde{\sigma}_{31})\right)-\gamma_3\tilde{\sigma}_{33}+\tilde{f}_{33}\\
	\frac{\partial}{\partial t}\tilde{\sigma}_{44}&=&-i\left( g_b^* \hat{b}^\dag\tilde{\sigma}_{14}-g_b \hat{b}\tilde{\sigma}_{41}+\frac{\Omega}{2}(\tilde{\sigma}_{24}-\tilde{\sigma}_{42})\right)-\gamma_4\tilde{\sigma}_{44}+\tilde{f}_{44}.
	\end{deqarr}
où l'on a introduit les opérateurs de Langevin $\tilde f_{uv}$ caractérisés par \cite{Davidovich:1996p1958,RaymondOoi:2007p15590} :
   \begin{equation}
  \bra\hat f_{uv}(z,t)\ket=0,
  \end{equation}
On définit les coefficients de diffusion$D_{uv,u'v'}$ de ces opérateurs sous la forme :
  \begin{equation}
\bra\hat F_{uv}^\dag(z,t)\hat{F}_{u'v'}(z',t')\ket=2D_{uv,u'v'}\delta(t-t')\delta(z-z').
 \label{diffu}
  \end{equation}
\subsubsection{Solutions stationnaires pour les populations}
On cherche les solutions de ce système, en négligeant la contribution des champs sonde et conjugué devant le champ pompe supposé beaucoup plus intense.
Dans ce cas, pour obtenir un système fermé d'équations, il faut ajouter les équations d'évolution des cohérences $\tilde{\sigma}_{31}$ et $\tilde{\sigma}_{42}$ (ainsi que celle des opérateurs adjoints) :
\begin{ddeqar}\arrlabel{eq_evolution_coh1_4WM}
\frac{\partial}{\partial t}\tilde{\sigma}_{31}&=&-i\left( g_a^* \hat{a}^\dag\tilde{\sigma}_{21}- g_b^* \hat{b}^\dag\tilde{\sigma}_{34}+\frac{\Omega}{2}(\tilde{\sigma}_{11}-\tilde{\sigma}_{33})\right)-\left(\frac{\Gamma}{2}+i\Delta\right)\tilde{\sigma}_{31}+\tilde{f}_{31},\\
\frac{\partial}{\partial t}\tilde{\sigma}_{42}&=&-i\left( g_b^* \hat{b}^\dag\tilde{\sigma}_{12}- g_a^* \hat{a}^\dag\tilde{\sigma}_{43}+\frac{\Omega}{2}(\tilde{\sigma}_{22}-\tilde{\sigma}_{44})\right)-\left(\frac{\Gamma}{2}+i \Delta +i\omega_0\right)\tilde{\sigma}_{42}+\tilde{f}_{42}.\hspace*{1.4cm}
	\end{ddeqar}
La conservation du nombre d'atomes dans le système donne la relation de fermeture :
\begin{equation}
\tilde{\sigma}_{11}+\tilde{\sigma}_{22}+\tilde{\sigma}_{33}+\tilde{\sigma}_{44}=1.
\end{equation}

Si le temps d'interaction des atomes avec le laser de pompe est suffisamment long par rapport aux temps caractéristiques d'évolutions du système\footnote{Ce point sera étudié plus en détail dans la section \ref{extension}. }, alors le système (\ref{eq_evolution_pop_4WM}) peut être résolu en supposant qu'il a atteint l'état stationnaire.\\
Pour simplifier la lecture, nous allons désormais utiliser un formalisme matriciel pour décrire ce système d'équations linéaires.
En effet, on peut réécrire le système (\ref{eq_evolution_pop_4WM})  en négligeant les termes proportionnels aux champs sonde et conjugué, sous la forme suivante :
\begin{deqn}\label{415}
\left(i[\mathbf{1}]\frac{\partial}{\partial{t}}+[M_0]\right)|\Sigma_0]=|S_0]+i|F_0],
\end{deqn}
avec 

\begin{ddeqar}
\nonumber &[M_0]=\left[
\setstretch{1.1}\begin{array}{ccccccc}
i \frac{\Gamma }{2} & i \frac{\Gamma }{2} & 0 & -\frac{\Omega }{2} & \frac{\Omega }{2} & 0 & 0 \\
i \frac{\Gamma }{2} & i \frac{\Gamma }{2} & 0 & 0 & 0 & -\frac{\Omega }{2} & \frac{\Omega }{2} \\
0 & 0 & i \Gamma  & \frac{\Omega }{2} & -\frac{\Omega }{2} & 0 & 0 \\
-\frac{\Omega }{2} & 0 & \frac{\Omega }{2} & -\Delta +i \frac{\Gamma }{2} & 0 & 0 & 0 \\
\frac{\Omega }{2} & 0 & -\frac{\Omega }{2} & 0 & \Delta +i \frac{\Gamma }{2} & 0 & 0 \\
-\frac{\Omega }{2} & -\Omega  & -\frac{\Omega }{2} & 0 & 0 & -\Delta -\omega_0 +i \frac{\Gamma }{2} & 0 \\
\frac{\Omega }{2} & \Omega  & \frac{\Omega }{2} & 0 & 0 & 0 & \Delta +\omega_0+i \frac{\Gamma }{2}
\end{array}
\right]&,\\
&|\Sigma_0]=\left |
\begin{array}{c}
\tilde{\sigma}_{11}\\
\tilde{\sigma}_{22}\\
\tilde{\sigma}_{33}\\
\tilde{\sigma}_{31}\\
\tilde{\sigma}_{13}\\
\tilde{\sigma}_{42}\\
\tilde{\sigma}_{24}
\end{array}
\right],\ 
|S_0]=\frac{1}{2}\left |
\begin{array}{c}
i \Gamma  \\
i \Gamma  \\
0 \\
0 \\
0 \\
-\Omega  \\
\Omega 
\end{array}
\right],\ 
|F_0]=\left |
\begin{array}{c}
\tilde{f}_{11}\\
\tilde{f}_{22}\\
\tilde{f}_{33}\\
\tilde{f}_{31}\\
\tilde{f}_{13}\\
\tilde{f}_{42}\\
\tilde{f}_{24}
\end{array}
\right].&
\end{ddeqar}
On obtient alors la solution stationnaire de ce système sous la forme : 
\begin{equation}
|\langle\Sigma_0\rangle]=[M_0]^{-1}|S_0].
\end{equation}

On introduit le coefficient $D=\Gamma^2+2(\Omega^2+\Delta^2+(\Delta+\omega_0)^2)$.
On trouve alors les solutions suivantes pour les valeurs moyennes des populations:
\begin{equation}\label{solS0}
|\langle\Sigma_0\rangle]=\frac{1}{2D}\left |
\begin{array}{c}
\Gamma^2+\Omega^2+4\Delta^2\\
\Gamma^2+\Omega^2+4(\Delta+\omega_0)^2\\
\Omega^2\\
-\Omega(2\Delta+i\Gamma)\\
-\Omega(2\Delta-i\Gamma)\\
-\Omega(2(\Delta+\omega_0)+i\Gamma)\\
-\Omega(2(\Delta+\omega_0)-i\Gamma)
\end{array}
\right].
\end{equation}
La figure présente l'effet de la pulsation de Rabi $\Omega$ et du désaccord à un photon $\Delta$ dans le cas de la raie D1 du rubidium~85. Les données de cette transition sont détaillées dans l'annexe \ref{Annexe_Rb}.\\
On vérifie aisément que dans les conditions expérimentales décrites dans ce manuscrit ($\Omega\leq 2\pi\times 2\text{GHz}$ et $ 0\leq\Delta\leq  2 \pi\times 1.5\text{GHz}$) la population dans l'état stationnaire est essentiellement dans le niveau $|2\rangle$. On schématise ce résultat sur la figure \ref{fig_4niveaux} par la taille des points dans les niveaux  $|1\rangle$ et  $|2\rangle$.

\begin{figure}
\centering
\includegraphics[width=13.5cm]{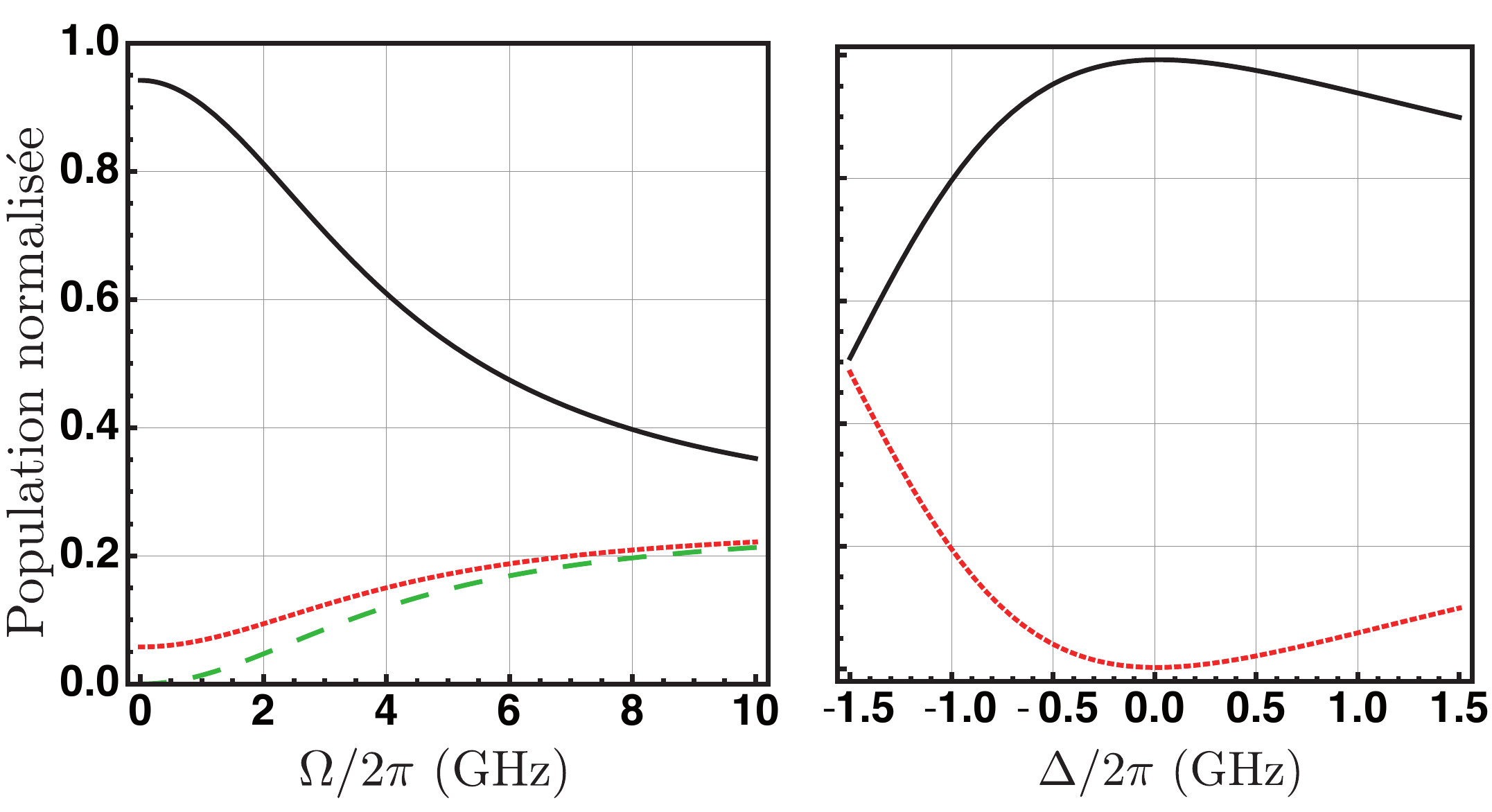} 
\caption[ Population des niveaux d'un modèle en double $\Lambda$]		{Effet  a) de  la pusaltion de Rabi $\Omega$  et b) du désaccord $\Delta$  sur la population des niveaux $|1\rangle$ (rouge pointillés), $|2\rangle$ (noir plein), $|3\rangle$ (vert tirets) d'un modèle en double $\Lambda$. Le désaccord à un deux photons et le taux de décohérence  $\gamma$ sont pris nuls. Pour la figure a) $\Delta =  2\pi\times 1\text{GHz}$. Dans la figure b) $\Omega\leq 2\pi\times 0.3 \text{GHz}$. Dans ce cas la population du niveau $|3\rangle$ est constante et inférieure à 0.5\%}
	\label{pop_4niveaux}
\end{figure}

\subsubsection{Evolution des cohérences}
A l'aide de la relation~(\ref{evol_hamiltonienne}), on obtient simplement l'évolution des cohérences $\tilde{\sigma}_{23},\tilde{\sigma}_{41},\tilde{\sigma}_{43} \text{ et } \tilde{\sigma}_{21}$ :
\begin{deqarr}\arrlabel{evol_coherence}
i\frac{\partial}{\partial t}  \tilde{\sigma}_{23}&=&(\delta-\Delta-i\frac{\Gamma}{2})\tilde{\sigma}_{23}-\frac{\Omega}{2}(\tilde{\sigma}_{21}-\tilde{\sigma}_{43})+ g \hat{a} (\tilde{\sigma}^0_{33}-\tilde{\sigma}^0_{22})+i\tilde{f}_{23}\\
i\frac{\partial}{\partial t} \tilde{\sigma}_{41}&=&(\delta+\Delta+\omega_0 -i\frac{\Gamma}{2})\tilde{\sigma}_{41}-\frac{\Omega}{2}(\tilde{\sigma}_{43}-\tilde{\sigma}_{21})- g \hat{b}^\dag (\tilde{\sigma}^0_{44}-\tilde{\sigma}^0_{11})+i\tilde{f}_{41}\hspace*{1cm}\\
i\frac{\partial}{\partial t} \tilde{\sigma}_{43}&=&(\delta+\omega_0 -i\Gamma)\tilde{\sigma}_{43}-\frac{\Omega}{2}(\tilde{\sigma}_{41}-\tilde{\sigma}_{23})+ g (\hat{b}^\dag\tilde{\sigma}^0_{13}-\hat{a} \tilde{\sigma}^0_{42})+i\tilde{f}_{43}\\
i\frac{\partial}{\partial t} \tilde{\sigma}_{21}&=&(\delta-i\gamma)\tilde{\sigma}_{21}-\frac{\Omega}{2}(\tilde{\sigma}_{23}-\tilde{\sigma}_{41})- g (\hat{b}^\dag\tilde{\sigma}^0_{24}-\hat{a} \tilde{\sigma}^0_{31})+i\tilde{f}_{21},
\end{deqarr}
ainsi que des opérateurs adjoints.
Pour résoudre ce système, nous avons ré-injecté les solutions stationnaires obtenues au paragraphe précédent. 
Elles sont notées $\tilde{\sigma}^0_{uv}$ dans les équations (\ref{evol_coherence}).\\
Pour simplifier, nous avons pris $g_a$ et $g_b$, les constantes de couplage, identiques et égales à $g$.\\
Nous allons écrire ces équations sous forme matricielle :
\begin{deqn}\label{mat_evol_coherence}
\left(i[\mathbf{1}]\frac{\partial}{\partial{t}}+[M_1]\right)|\hat{\Sigma_1}(z,t)]=g[S_1]|\hat{A}(z,t)]+i|F_{1}(z,t)]
\end{deqn}
avec
\begin{ddeqar}\label{mat_evol_coherence_def}\nonumber
\setstretch{1.1}
&[M_1]=\left[\begin{array}{cccc}
i \frac{\Gamma }{2}+(\Delta -\delta)  & 0 & -\frac{\Omega }{2} & \frac{\Omega }{2} \\
0 & i \frac{\Gamma }{2}-(\Delta +\delta +\omega_0) & \frac{\Omega }{2} & -\frac{\Omega }{2} \\
-\frac{\Omega }{2} & \frac{\Omega }{2} &i\Gamma -(\delta +\omega_0) & 0 \\
\frac{\Omega }{2} & -\frac{\Omega }{2} & 0 &  i \gamma -\delta 
\end{array}
\right]&,\\
\nonumber ~\\
&|\Sigma_1(z,t)]=\left|
\begin{array}{c}
\tilde\sigma_{23}(z,t)\\
\tilde\sigma_{41}(z,t)\\
\tilde\sigma_{43}(z,t)\\
\tilde\sigma_{21}(z,t)
\end{array}
\right],\ 
[S_1]=\left[
\begin{array}{cc}\setstretch{1.2}
\tilde\sigma^0_{33}-\tilde\sigma^0_{22}&0 \\
0 &\tilde\sigma^0_{11}-\tilde\sigma^0_{44}\\
-\tilde\sigma^0_{42}&\tilde\sigma^0_{13} \\
\tilde\sigma^0_{31} &-\tilde\sigma^0_{24}
\end{array}
\right]&,\\
\nonumber ~\\
\nonumber &|F_1(z,t)]=\left |
\begin{array}{c}
\tilde{f}_{23}(z,t)\\
\tilde{f}_{41}(z,t)\\
\tilde{f}_{43}(z,t)\\
\tilde{f}_{21}(z,t)
\end{array}
\right],\ 
|\hat{A}(z,t)]=\left |
\begin{array}{c}
\hat{a}(z,t)\\
\hat{b}^\dag(z,t)
\end{array}
\right].&
\end{ddeqar}

\subsubsection{Equations de propagation}
Les champs sonde et conjugué sont choisis, pour des raisons de simplicité, colinéaires et co--propageants avec le champ pompe. De plus, nous supposons que l'accord de phase est vérifié, $2 \overrightarrow{k_p}-\overrightarrow{k_a}-\overrightarrow{k_b}=\overrightarrow{0}$.
On définit l'axe $z$ comme la direction de propagation.
Les équations de Maxwell pour les champs sonde et conjugué donnent les équations de propagation :
\begin{deqarr}\arrlabel{propagation}
\left(\frac{\partial}{\partial t}+c\frac{\partial}{\partial z}\right) \hat a(z,t)&=&i g N \tilde\sigma_{23}(z,t),\\
\left(\frac{\partial}{\partial t}+c\frac{\partial}{\partial z}\right) \hat{b}^\dag(z,t)&=& - i g N \tilde\sigma_{41}(z,t).
\end{deqarr}
Les équations (\ref{propagation}) s'écrivent sous forme matricielle :
\begin{equation}\label{mat_propagation}
\left(\frac{\partial}{\partial t}+c\frac{\partial}{\partial z}\right)|\hat{A}(z,t)]=i g N [T]|\Sigma_1(z,t)],
\end{equation}
avec $[T]=\left[
\begin{array}{cccc}
1 &0&0&0\\
0&-1&0&0
\end{array}
\right]$.\\

\noindent Avant de résoudre ce système, nous allons simplifier les équations (\ref{propagation}) et (\ref{mat_propagation}) en négligeant la dérivée temporelle.
En effet pour des longueurs d'interaction inférieures à 10 cm, le terme $\frac{\partial}{\partial t}$ est négligeable devant $c\frac{\partial}{\partial z}$ pour des évolutions à des fréquences inférieures à 1 GHz.
Les grandeurs qui nous intéressent, c'est-à-dire les spectres de bruit dans la bande passante des détecteurs usuels, sont à des fréquences très inférieures au GHz.
		\subsection{Résolution}
Afin de résoudre l'équation de propagation (\ref{propagation}), nous allons dans un premier temps passer dans l'espace des fréquences par transformée de Fourier, puis extraire des équations d'évolution des cohérences (\ref{evol_coherence}) les solutions $\tilde\sigma_{23}$ et $\tilde\sigma_{41}$.\\
En notant $[M_1'(\omega)]=\omega [\mathbf{1}] + [M_1]$, l'équation (\ref{mat_evol_coherence}) s'écrit dans l'espace de Fourier temporel :
\begin{equation}\label{mat_evol_coherence_f}
[M_1'(\omega)]|\Sigma_1(z,\omega)]=g[S_1]|\hat{A}(z,\omega)]+i|F_{1}^\mathcal{N}(z,\omega)],
\end{equation}
et	peut alors se résoudre sous la forme : 
\begin{equation}
|\Sigma_1(z,\omega)]=g[M_1'(\omega)]^{-1}[S_1]|\hat{A}(z,\omega)]+i[M_1'(\omega)]^{-1}|F_1(z,\omega)].
\end{equation}
Cela nous permet d'écrire l'équation de propagation (\ref{mat_propagation}) dans l'espace de Fourier :
\begin{deqn}\label{eq_evol_2}
\frac{\partial}{\partial z}|\hat{A}(z,\omega)]=[M(\omega)]|\hat{A}(z,\omega)]+ [M_F(\omega)]|F_1(z,\omega)],
\end{deqn}
avec 
\begin{ddeqn}
\nonumber [M(\omega)]=i\ \frac{g^2 N}{c} [T][M_1'(\omega)]^{-1}[S_1] \text{ et }
[M_F(\omega)]=-\frac{g N}{c} [T][M_1'(\omega)]^{-1}.
\end{ddeqn}
Les coefficients de la matrice $[M(\omega)]$ sont notés, par analogie avec l'équation \eqref{312} :
\begin{equation}\label{424}
[M(\omega)]=i\left[
\begin{array}{cc}
\kappa_a  (\omega)& \eta_a (\omega)\\
\eta_b  (\omega)& \kappa_b  (\omega)
\end{array}
\right].
\end{equation}
	\begin{figure}\centering
\includegraphics[width=14cm]{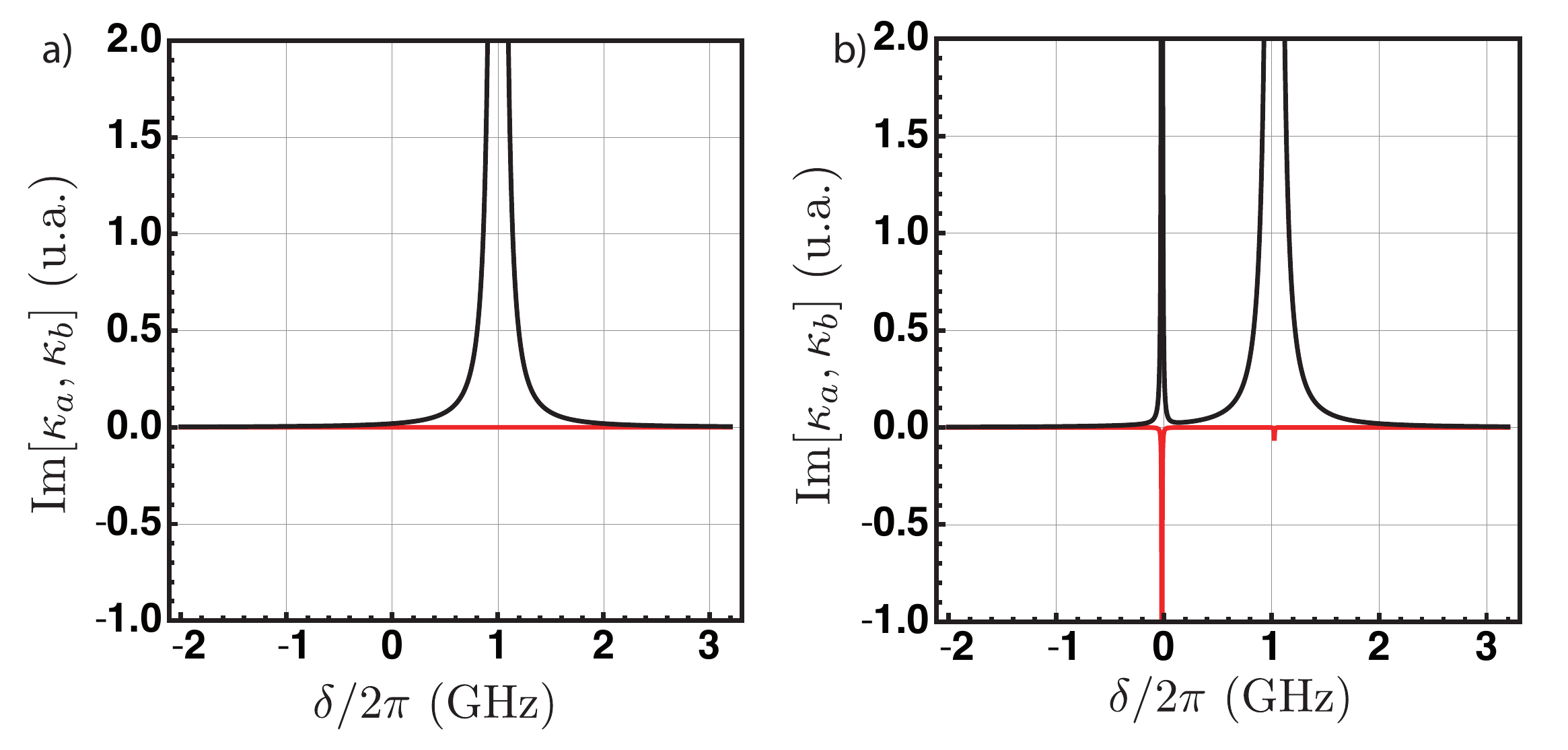} 	\caption[Profil spectral pour les champ sonde et conjugué]		{Profil spectral la partie imaginaire de $\kappa_a$ (noir) et de $\kappa_b$ (rouge) en fonction du désaccord à deux photons. a) $\Omega/2\pi=0$ GHz et b)  $\Omega/2\pi=0.5$ GHz. Paramètres utilisés : $\gamma=1$ kHz, $\Delta=1$ GHz.}
	\label{fig_Re_kappa}
\end{figure}
En effet, le modèle microscopique que nous venons de détailler permet d'obtenir le coefficient $\chi^{(3)}$ que nous avions introduit pour modéliser l'interaction lumière matière.
Il est intéressant de noter que les coefficients $\kappa_a$ et $\kappa_b$ ne sont pas identiques a priori, ce qui complique notablement la résolution.\\
On peut donner une image physique des différents termes de la matrice $[M]$.
Sous la forme de la relation \eqref{424}, la partie réelle des éléments de la matrice est reliée à la dispersion du milieu, tandis que la partie imaginaire est reliée à la dissipation ou à l'amplification dans le milieu.
Sur la figure \ref{fig_Re_kappa}, on présente les profils de $\kappa_a$ et $\kappa_b$ en fonction de $\delta$.
Pour $\delta\simeq \Delta$, on observe un pic positif sur la partie imaginaire de $\kappa_a$ (absorption).
Cela correspond à l'absorption du faisceau sonde lorsqu'il passe à résonance avec la transition. 
La largeur de ce pic dépend donc principalement de $\Gamma$.
La position du pic qui est exactement à la valeur $\delta=\Delta$ en l'absence de pompe, est modifiée par déplacement lumineux pour $\Omega\neq 0$. 
Dans ce cas, le pic sera déplacé vers  $\delta>\Delta$.\\
Autour de $\delta\simeq 0$, on observe en présence de pompe, des pics correspondants à de l'absorption pour le champ sonde et à du gain pour le champ conjugué.
Ces pics décrivent deux processus Raman. 
L'un effectue un transfert de photons du champ sonde vers le champ pompe et l'autre du champ pompe vers le champ conjugué.
La dissymétrie nait du fait que la population atomique est principalement dans le niveau $|2\ket$ et cela favorise ce sens de conversion.\\
Ces effets (absorption et transfert Raman) ne peuvent donc pas être à l'origine de corrélations entre les champs sonde et conjugué car ils ne correspondent jamais à la création ou à la destruction simultanée d'un photon dans chacun des deux modes $\hat a$ et $\hat b$.
Ce sont les coefficients $\eta_a$ et $\eta_b$ qui vont nous renseigner sur le couplage entre les deux champs.
Or, il est plus difficile d'avoir une interprétation en terme d'indice du milieu pour ces termes.
On va donc résoudre l'équation (\ref{eq_evol_2}) afin de relier directement les champs en entrée aux champs en sortie du milieu.
L'équation (\ref{eq_evol_2}) est une équation différentielle matricielle linéaire du premier ordre.
Pour une longueur d'interaction $L$, on peut la résoudre de la façon suivante :
\begin{equation}\label{sol_propa}
| \hat{A}(L,\omega)]=e^{[M(\omega)].L} \left(|\hat{A}(0,\omega)]+L\int_0^1  e^{-[M(\omega)]Lz} [M_F(\omega)]|F_1(z,\omega)]dz\right).
\end{equation}
Le premier terme du membre de droite correspond à une matrice de transfert dans le formalisme entrée-sortie décrit dans le chapitre \ref{ch1}.
A l'intérieur de la parenthèse, le premier terme décrit donc les champs en entrée du milieu. 
La description de l'état d'entrée dépend du choix du vecteur $|\hat{A}(0,\omega)]$.
En pratique, comme on s'intéresse à des fonctions de corrélations à 2 points, c'est la matrice de covariance   $|\hat{A}(0,\omega)] [\hat{A}^\dag(0,\omega')|$ en non le vecteur $|\hat{A}(0,\omega)]$ que l'on doit choisir pour définir l'état d'entrée \cite{Dantan05}.\\
Le second terme de la parenthèse est un terme de dissipation.
Il s'agit des forces de Langevin atomiques intégrées sur la longueur de zone d'interaction.
Ce terme fera donc apparaitre les coefficients de diffusion lors de l'étude des fonctions de corrélations à deux points.

\subsection{Valeurs moyennes}
Dans un premier temps, nous allons étudier les résultats à fréquence nulle, c'est-à-dire les valeurs moyennes des opérateurs.
\subsubsection{Gain}

On s'intéresse aux valeurs moyennes de l'intensité des champs $\hat{a}$ et $\hat{b}$, c'est-à-dire aux valeurs moyennes de l'opérateur nombre $\hat{N}_a$ ou $\hat{N}_b$ :
\begin{equation}\label{nombre}
\langle \hat{N}_a\rangle= \langle \hat{a}^\dag\hat{a}\rangle.
\end{equation}
On note les valeurs moyennes du champ $\hat{a}$ : $\alpha= \langle\hat{a}\rangle$ et du champ $\hat{b}$ : $\beta= \langle\hat{b}\rangle$, avec $\alpha$ et $\beta$ a priori complexes.\\
L'équation (\ref{nombre}) va donc s'écrire au premier ordre sous la forme :
\begin{equation}\label{nombre_simplifié}
\langle \hat{N}_a\rangle= | \alpha | ^2  \text{ et } \langle \hat{N}_b\rangle=0.
\end{equation}	
On définit alors le gain pour les champs $\hat{a}$ et  $\hat{b}$ comme une quantité classique de la manière suivante :
\begin{equation}
G_a=\frac{\langle \hat{a}^\dag(L)\hat{a}(L)\rangle}{\langle \hat{a}^\dag(0)\hat{a}(0)\rangle}=\frac{|\alpha_{out}|^2}{|\alpha_{in}|^2},\ G_b=\frac{\langle \hat{b}^\dag(L)\hat{b}(L)\rangle}{\langle \hat{a}^\dag(0)\hat{a}(0)\rangle}=\frac{|\beta_{out}|^2}{|\alpha_{in}|^2}.
\end{equation}
Pour déterminer la valeur du gain, on utilise la relation (\ref{sol_propa}).
Prendre la valeur moyenne de $|\langle \hat{A}(z,\omega)\rangle]$ revient à fixer $\omega=0$ et à ne pas prendre en compte les forces de Langevin dont la valeur moyenne est nulle.
On obtient alors :
\begin{equation}
|\langle \hat{A}(z=L,\omega=0)\rangle]=e^{[M(\omega=0)].L}|\langle \hat{A}(z=0,\omega=0)\rangle].
\end{equation}

On peut écrire de façon simple cette équation dans le formalisme entrée--sortie  :

\begin{equation}\label{entree_sortie_moyenne}
|\langle \hat{A}_{out}\rangle]=\left[
\begin{array}{cc}
A(0) & B(0) \\ 
C(0) & D(0)
\end{array} \right] |\langle \hat{A}_{in}\rangle],
\end{equation}
où $\left[\begin{array}{cc}
A(0) & B(0) \\ 
C(0) & D(0)
\end{array} \right]=e^{[M(0)].L}.$
On obtient alors :

\begin{equation}\label{gain_def}
G_a=|A(0)|^2,\ G_b=|C(0)|^2.
\end{equation}
Le gain sur les modes  $\hata$ et  $\hat b$ se déduisent donc simplement du module carré des termes $A$ et $C$ de l'exponentielle de la matrice de transfert à fréquence nulle.
\subsubsection{Phase}
On souhaite déterminer la valeur moyenne de la phase $\langle {\phi_a}_{out}\rangle$ et $\langle {\phi_b}_{out}\rangle$ des champ $\hat{a}$ et $\hat{b}$ en sortie du milieu.
Comme pour le gain, il s'agit d'une quantité classique.
Ainsi en utilisant la relation \ref{entree_sortie_moyenne}, on peut voir que la phase du champ $\hat{a}$ va s'écrire :
\begin{equation}
\langle {\phi_a}_{out}\rangle=\arctan \frac{\text{Im} \ A(0) }{\text{Re} \ A(0) }.
\end{equation}
De même pour $\hat{b}$ :
\begin{equation}
\langle {\phi_b}_{out}\rangle=\arctan \frac{\text{Im} \ C(0) }{\text{Re} \ C(0) }.
\end{equation}
\subsection{Fluctuations quantiques}
Nous venons de dériver les équations entrée-sortie pour obtenir l'expression des valeurs moyennes de l'amplitude et de la phase des champs quantiques $\hat{a}$ et  $\hat{b}$.
Pour des champs quantiques, toute l'information ne réside pas dans les valeurs moyennes des opérateurs et on peut s'intéresser également aux fluctuations de ces opérateurs autour des valeurs moyennes, notamment à leur variance (ce qui est suffisant pour les états gaussiens).
Les fluctuations doivent être étudiées à fréquence non nulle.
Comme nous l'avons introduit au chapitre 2, les mesures des fluctuations sont donc réalisées à l'aide d'un analyseur de spectre.
La fréquence $\omega$ telle que nous l'avons définie dans ce manuscrit correspond la fréquence d'analyse de l'appareil de mesure.
\subsubsection{Fluctuations quantiques d'intensité à un mode}
Étudions dans un premier temps les fluctuations quantiques d'intensité à un mode pour le champ sonde.
On rappelle que l'on note la valeur moyenne de l'opérateur annihilation de la manière suivante :

\begin{equation}
\alpha= |\alpha | e^{i \phi}.
\end{equation}
On cherche à déterminer les fluctuations $\delta \hat{N}_a$ de l'opérateur nombre $\hat{N_a}$, défini à l'équation (\ref{nombre}).
En linéarisant son expression on trouve au premier ordre:
\begin{equation}\label{def_Na}
\delta \hat{N}_a = |\alpha |\ \delta\hat{a} e^{-i\phi}+ | \alpha |\ \delta\hat{a}^\dag e^{i\phi}=| \alpha |\delta \hat{X}_{\phi},
\end{equation}
avec  les fluctuations  $\delta\hat{X_\phi}$ de la quadrature  $\hat{X_\phi}$ définie dans le chapitre \ref{ch1} par :
\begin{equation}\label{def_Xa}
\delta \hat{X}_{\phi}=\delta\hat{a}\ e^{-i\phi}+\delta\hat{a}^\dag\ e^{i\phi}.
\end{equation}
	La quantité mesurée expérimentalement  est la densité spectrale de bruit $S_{N_a}(\omega)$, c'est-à-dire la transformée de Fourier de la fonction d'auto-corrélation.
On définit la transformée de Fourier de $\hat{a}(t)$  par  :
\begin{equation}
\hat{a}(\omega)=\int_{-\infty}^\infty \hat{a}(t)\ e^{i\omega t}\ dt.
\end{equation}
La notation pour la transformée de Fourier de $\hat{a}^\dag(t)$ est plus ambiguë. On définit $\hat{a}^\dag(\omega)$ de la manière suivante\footnote{Dans ce cas il faut noter que la conjugaison s'exprime par $
[\hat{a}^\dag(\omega)]^\dag =\int_{-\infty}^\infty \hat{a}(t)\ e^{-i\omega t}\ dt=\hat{a}(-\omega).$
Ainsi, on ferra donc particulièrement attention à ne pas confondre la conjuguée de la transformée de Fourier 
$
[\hat{a}(\omega)]^\dag=\left[\int_{-\infty}^\infty \hat{a}(t)\ e^{i\omega t}\ dt\right]^\dag = \int_{-\infty}^\infty \hat{a}^\dag(t)\ e^{-i\omega t}\ dt = \hat{a}^\dag(-\omega), 
$
et la transformée de Fourier de la conjuguée $
\hat{a}^\dag(\omega)=\int_{-\infty}^\infty \hat{a}^\dag(t)\ e^{i\omega t}\ dt  .
$
 Pour plus de détails on se reportera à l'annexe A à la fin de ce manuscrit.} :
\begin{equation}\label{def_conjuguee}
\hat{a}^\dag(\omega)=\int_{-\infty}^\infty \hat{a}^\dag(t)\ e^{i\omega t}\ dt.
\end{equation}
On peut donc écrire la densité spectrale de bruit sous la forme :
\begin{equation}\label{def_spectre2}
S_{N_a}(\omega)\ 2\pi\ \delta(\omega+\omega')=\langle \delta N_a(\omega)\ \delta N_a^\dag(\omega')\rangle.
\end{equation}
A l'aide de (\ref{def_Na}) et (\ref{def_Xa}) on peut alors l'exprimer à partir de la fonction de corrélation de la quadrature $X_\phi$ :
\begin{equation}\label{def_spectre3}
S_{N_a}(\omega)\ 2\pi\ \delta(\omega+\omega')=|\alpha |^2 \langle \delta \hat{X}_\phi(\omega)\ \delta \hat{X}_\phi^\dag (\omega ')\rangle.
\end{equation}
\paragraph{Calcul du spectre de bruit d'intensité à un mode}~\\
		L'équation (\ref{sol_propa}) peut être linéarisée afin de séparer les valeurs moyennes des fluctuations $ |\delta\hat{A}(\omega)]$.
		 Dans le formalisme entrée-sortie, en introduisant la matrice de transfert ABCD à la manière de (\ref{entree_sortie_moyenne}) et un vecteur $|F(L,\omega)]=\left[
\begin{array}{c}
F_a(L,\omega)  \\ 
F_{b^\dag}(L,\omega) 
\end{array} \right]$ pour tenir compte de la dissipation, on peut écrire : 
	 \begin{deqn}
|\delta\hat{A}_{out}(\omega)]=\left[
\begin{array}{cc}
A(\omega) & B(\omega) \\ 
C(\omega) & D(\omega)
\end{array} \right]\left( |\delta\hat{A}_{in}(\omega)]+
|F(L,\omega)]\right),
\end{deqn}
		avec 	 \begin{ddeqn}\label{def_F}
	 \left[
\begin{array}{cc}
A(\omega) & B(\omega) \\ 
C(\omega) & D(\omega)
\end{array} \right]=e^{[M(\omega)].L}\text{ et } |F(L,\omega)]=L\int_0^1  e^{-[M(\omega)]Lz} [M_F(\omega)]|F_1(z,\omega)]dz
.\end{ddeqn}
De même pour le vecteur adjoint : 
\begin{deqn}\label{adjoint}
|\delta\hat{A}^\dag_{out}(\omega)]=\left[
\begin{array}{c}
\delta a^\dag(L,\omega)  \\ 
\delta{b}(L,\omega) 
\end{array} \right]=\left[
\begin{array}{cc}
A^*(-\omega) & B^*(-\omega) \\ 
C^*(-\omega) & D^*(-\omega)
\end{array} \right]\left( |\delta\hat{A}^\dag_{in}(\omega)]+
|F^\dag(L,\omega)]\right),
\end{deqn}
		avec 	 \begin{ddeqn}\label{def_Fdag}
		|F^\dag(L,\omega)]=\left[
\begin{array}{c}
F_{a^\dag}(L,\omega)  \\ 
F_{b}(L,\omega) 
\end{array} \right]=L\int_0^1  e^{-[M^*(-\omega)]Lz} [M_F^*(-\omega)]|F_1^\dag(z,\omega)]dz,
\end{ddeqn} où $|F_1^\dag(z,\omega)]=\left |
\begin{array}{c}
\tilde{f}_{23}^\dag(z,\omega)\\
\tilde{f}_{41}^\dag(z,\omega)\\
\tilde{f}_{43}^\dag(z,\omega)\\
\tilde{f}_{21}^\dag(z,\omega)
\end{array}
\right]=\left |
\begin{array}{c}
\tilde{f}_{32}(z,\omega)\\
\tilde{f}_{14}(z,\omega)\\
\tilde{f}_{34}(z,\omega)\\
\tilde{f}_{12}(z,\omega)
\end{array}
\right]$.\\

\noindent La densité spectrale de bruit en intensité à la sortie du milieu en fonction des fluctuations des champs $\hat{a}$ et $\hat{b}$ en entrée s'écrit alors :
\begin{eqnarray}\label{SNa1}
\nonumber S_{N_{a,out}}(\omega)\ 2\pi\ \delta(\omega+\omega')&=&|\alpha_{out}|^2 \langle [A(\omega)(\delta \hat{a}_{in}(\omega) +F_a(L,\omega))e^{-i\theta}+B(\omega)(\delta \hat{b}_{in}^\dag(\omega)+F_{b^\dag}(L,\omega)) e^{-i\theta}\\
\nonumber&&+A^*(-\omega)(\delta \hat{a}_{in}^\dag(\omega)+F_{a^\dag}(L,\omega)) e^{i\theta}+B^*(-\omega )(\delta \hat{b}_{in}(\omega)+F_b(L,\omega)) e^{i\theta}]\\
\nonumber&&\times[A^*(-\omega')(\delta \hat{a}_{in}^\dag(\omega')+F_{a^\dag}(L,\omega')) e^{i\theta}+B^*(-\omega')(\delta \hat{b}_{in}(\omega')+F_b(L,\omega')) e^{i\theta}\\
\nonumber&&+ A(\omega ')(\delta \hat{a}_{in}(\omega ') +F_a(L,\omega'))e^{-i\theta}+B(\omega ')(\delta \hat{b}_{in}^\dag(\omega ') +F_{b^\dag}(L,\omega'))e^{-i\theta}]\rangle.\\
~ \end{eqnarray}

On a donc obtenu une expression qui relie la densité spectrale de bruit en sortie aux fonctions de corrélation à deux points du champ en entrée ainsi qu'aux coefficients de diffusion des forces de Langevin.
Pour calculer une grandeur scalaire à partir d'opérateurs, on est amené à choisir un ordre pour les termes des équations matricielles.
Nous allons voir dans ce qui suit, comment obtenir les coefficients de diffusion des forces de Langevin dans le système atomique que nous avons considéré.\\
Notons que les fluctuations du champ ainsi que les termes de force de Langevin étant indépendamment nuls en valeur moyenne, leur produit l'est aussi.
\paragraph{Fonctions de corrélation dans l'ordre symétrique}~\\
Comme nous l'avons vu, pour calculer les fonctions de corrélation, une possibilité est d'utiliser l'ordre symétrique pour les opérateurs \cite{Fabre:1990p2839}.
La méthode de transformation d'opérateurs vers des nombres complexes est décrite en détail dans les références \cite{Davidovich:1996p1958,Hilico:1992p2633}.
Dans la représentation symétrique, les produits ordonnés d'opérateurs $\hat{a}\hat{a}^\dag$ et  $\hat{a}^\dag\hat{a}$ sont remplacés par la demi-somme de ces deux produits.
En suivant cette transformation on obtient pour un état cohérent :
\begin{deqarr}\label{cov_sym}
\langle\delta \hat{a}(\omega)\delta \hat{a}^\dag(\omega')\rangle^{\mathcal{S}}=\langle\delta \hat{a}^\dag(\omega)\delta \hat{a}(\omega')\rangle^{\mathcal{S}}&=&
\frac 12\langle\delta \hat{a}(\omega)\delta \hat{a}^\dag(\omega')+\delta \hat{a}^\dag(\omega)\delta \hat{a}(\omega')\rangle\hspace{0.5cm}\\
&=&\frac 12 2\pi\ \delta(\omega+\omega').
\end{deqarr}
On obtient donc pour l'expression (\ref{SNa1}), en prenant un champ cohérent non vide en entrée sur le mode $\hat a$ et le vide sur le mode $\hat b$.
\begin{eqnarray}\label{SNa2}
&& S_{N_{a,out}}(\omega)\ 2\pi\ \delta(\omega+\omega')=\\
&&\nonumber \frac {|\alpha_{out}|^2}{2}   \left(A(\omega)A^*(-\omega')+A^*(-\omega)A(\omega')\right. +\left.(B(\omega)B^*(-\omega')+B^*(-\omega)B(\omega')\right)  \times \ 2\pi\ \delta(\omega+\omega')\\
\nonumber &+&|\alpha_{out}|^2 (A(\omega)A^*(-\omega')\langle F_a(L,\omega)F_{a^\dag}(L,\omega')\rangle^{\mathcal{S}}+A^*(-\omega)A(\omega')\langle F_{a^\dag}(L,\omega)F_a(L,\omega')\rangle)^{\mathcal{S}}\\
\nonumber &+&|\alpha_{out}|^2 (B(\omega)B^*(-\omega')\langle F_{b^\dag}(L,\omega)F_{b}(L,\omega')\rangle^{\mathcal{S}}+B^*(-\omega)B(\omega') \langle F_b(L,\omega)F_{b^\dag}(L,\omega')\rangle^{\mathcal{S}})\\
\nonumber &+&|\alpha_{out}|^2 (A(\omega)B^*(-\omega')\langle F_{a}(L,\omega)F_{b}(L,\omega')\rangle^{\mathcal{S}}+B(\omega)A^*(-\omega') \langle F_{b^\dag}(L,\omega)F_{a^\dag}(L,\omega')\rangle^{\mathcal{S}}\\
\nonumber &&\hspace{1cm}+A^*(-\omega)B(\omega') \langle F_{a^\dag}(L,\omega)F_{b^\dag}(L,\omega')\rangle^{\mathcal{S}}+B^*(-\omega)A(\omega') \langle F_b(L,\omega)F_{a}(L,\omega')\rangle^{\mathcal{S}}).
\end{eqnarray}

L'utilisation de l'ordre symétrique garantit, par sa structure même, la parité en fonction de $\omega$ du terme : $A(\omega)A^*(-\omega')+A^*(-\omega)A(\omega')+B(\omega)B^*(-\omega')+B^*(-\omega)B(\omega')$.
Cela veut dire que $ S_{N_{a,out}}(\omega)$ est une grandeur paire, même si l'on néglige les contributions des forces de Langevin.

\paragraph{Coefficient de diffusion des forces de Langevin dans l'ordre symétrique}~\\
A priori, les forces de Langevin ne peuvent pas être négligées dans \eqref{SNa2}.
Voyons comment calculer les termes qui apparaissent dans cette équation.
On peut exprimer la valeur moyenne d'un produit de deux forces de Langevin $\tilde f_{uv}(z,t)$ et $\tilde{f}_{u'v'}(z',t')$ à l'aide du coefficient de diffusion $D_{uv,u'v'}$ défini par :
\begin{equation}\label{452}
\bra{\tilde f_{uv}(z,t)\tilde{f}_{u'v'}(z',t')}\ket=2D_{uv,u'v'}\delta(t-t')\delta(z-z').\end{equation}
Le calcul de chacun des coefficients dans le cas particulier d'un système à 4 niveaux est fait dans l'annexe \ref{Annexe_diffu}.
Dans notre système, ces coefficients de diffusion doivent être intégrés pour prendre en compte la propagation.
On va définir les coefficients de diffusion après intégration pour simplifier l'écriture des équations entrée sortie.
A l'aide des relations (\ref{def_F}) et (\ref{def_Fdag}) on obtient pour les coefficients de diffusion quantiques \cite{CohenTannoudji:1996p4732,Davidovich:1996p1958,OrvilScully:1997p3797} :
\begin{eqnarray}
\nonumber\langle F_{a}(\omega)F_{a^\dag}(\omega ')\rangle= L^2&&[1\ 0|\bra\int_0^1  e^{-[M(\omega)]Lz} [M_F(\omega)]|F_1(Lz,\omega)]dz\\
\times &&[1\ 0|\int_0^1  e^{-[M^*(-\omega')]Lz'}[M_{F}^*(-\omega')]|F_1^\dag(Lz',\omega')]dz'\rangle,\hspace{0.5cm}
\end{eqnarray}
que l'on peut écrire aussi :
\begin{eqnarray}
\nonumber\langle F_{a}(\omega)F_{a^\dag}(\omega ')\rangle= L^2&&[1\ 0|\bra\int_0^1\int_0^1  e^{-[M(\omega)]Lz} [M_F(\omega)]|F_1(Lz,\omega)]\\
\times && [F_1^\dag(Lz',\omega')|\ ^t[M_{F}^*(-\omega')]e^{-^t[M^*(-\omega')]Lz'}dz dz'\rangle|1\ 0].\hspace{0.5cm}
\end{eqnarray}
La delta-corrélation en $z$ des forces de Langevin permet de réduire le problème de la propagation à une seule intégrale sur $z$ : 
\begin{eqnarray}\label{455}
\nonumber\langle F_{a}(\omega)F_{a^\dag}(\omega ')\rangle =L^2 &&[1\ 0|\langle\int_0^1  e^{-[M(\omega)]Lz}[M_{F}(\omega)]|F_1(Lz,\omega)]\\
\times &&  [F_1^\dag(Lz,\omega')|\ ^t[M_{F}^*(-\omega')]e^{-^t[M^*(-\omega')]Lz}dz\rangle|1\ 0].
\end{eqnarray}
Comme nous avons adopté la représentation symétrique pour calculer les fonction de corrélations de bruit, il est nécessaire de remplacer la matrice $\langle|F_1(Lz,\omega)]\times  [F_1^\dag(Lz,\omega')\rangle|$ qui contient les coefficients de diffusion $D_{ij,kl}$ pour les équations quantiques   avec $ij\in\{23,41,43,21\}$ et $kl\in\{32,14,34,12\}$ par la demi somme de cette matrice et de la matrice $\langle|F_1^\dag(Lz,\omega)]\times  [F_1(Lz,\omega')|\rangle$ qui contient les coefficients de diffusion $D_{ij,kl}$ pour les équations quantiques   avec $ij\in\{32,14,34,12\}$ et $kl\in\{23,41,43,21\}$.\\
En effet c'est l'équivalent pour les forces de Langevin de la transformation appliquée dans la relation (\ref{cov_sym}).
En notant $[D]$ cette matrice de diffusion et en explicitant la delta-corrélation en $\omega$ on peut écrire pour l'ordre symétrique  :
\begin{equation}
\langle|F_1(Lz,\omega)]  [F_1^\dag(Lz,\omega')|\rangle^{\mathcal{S}}=[D]\ 2\pi\ \delta(\omega+\omega').
\end{equation}
On peut alors écrire l'équation \eqref{455} sous la forme :
\begin{deqn}
\langle F_{a}(\omega)F_{a^\dag}(\omega ')\rangle =D_{aa^\dag}(\omega,-\omega') 2\pi\ \delta(\omega+\omega'),
\end{deqn}
avec 
\begin{ddeqn}
D_{aa^\dag}(\omega,-\omega')=L^2 [1\ 0|\int_0^1  \bra e^{-[M(\omega)]Lz}[M_{F}(\omega)][D]\ ^t[M_{F}^*(-\omega')]e^{-^t[M^*(-\omega')]Lz}dz\rangle|1\ 0].
\end{ddeqn}
On écrit de la même manière les termes $\langle F_{a^\dag}(L,\omega)F_a(L,\omega')\rangle$, $\langle F_{b^\dag}(\omega)F_{b}(\omega ')\rangle$ et $\langle F_{b}(\omega)F_{b^\dag}(\omega ')\rangle$ sous la forme :
\begin{deqarr}
\langle F_{a^\dag}(\omega)F_{a}(\omega ')\rangle =D_{a^\dag a}(-\omega,\omega') 2\pi\ \delta(\omega+\omega'),\\
\langle F_{b^\dag}(\omega)F_{b}(\omega ')\rangle =D_{b^\dag b}(\omega,-\omega') 2\pi\ \delta(\omega+\omega'),\\
\langle F_{b}(\omega)F_{b^\dag}(\omega ')\rangle =D_{bb^\dag}(\omega,-\omega') 2\pi\ \delta(\omega+\omega').
\end{deqarr}
où l'on a défini $D_{a^\dag a}(-\omega,\omega')$, $D_{b^\dag b}(\omega,-\omega')$ et $D_{bb^\dag }(-\omega,\omega')$ par :
\begin{deqarr}
D_{a^\dag a}(-\omega,\omega')=L^2 [1\ 0|\int_0^1  e^{-[M^*(-\omega)]Lz}[M^*_{F}(-\omega)][D]\ ^t[M_{F}(\omega')]e^{-^t[M(\omega')]Lz}dz\rangle|1\ 0],\hspace{1cm}\\
 D_{b^\dag b}(\omega,-\omega')=L^2 [0\ 1|\int_0^1  e^{-[M(\omega)]Lz}[M_{F}(\omega)][D]\ ^t[M_{F}^*(-\omega')]e^{-^t[M^*(-\omega')]Lz}dz\rangle|0\ 1],\hspace{1cm}\\
D_{bb^\dag }(-\omega,\omega')=L^2 [0\ 1|\int_0^1  e^{-[M^*(-\omega)]Lz}[M^*_{F}(-\omega)][D]\ ^t[M_{F}(\omega')]e^{-^t[M(\omega')]Lz}dz\rangle|0\ 1].\hspace{1cm}
\end{deqarr}
Le spectre de bruit en intensité pour un champ $\hat{a}$ s'écrit donc de façon simplifiée sous la forme :
\begin{eqnarray}\label{SNa3}
\nonumber S_{N_{a,out}}(\omega)&=&\frac{|\alpha_{out}|^2}{2}\left( |A(\omega)|^2(1+2D_{aa^\dag}(\omega,\omega))+|A(-\omega)|^2(1+2D_{a^\dag a}(-\omega,-\omega))\right.\\
\nonumber &&+ \left. |B(\omega)|^2(1+2D_{b^\dag b}(\omega,\omega))+|B(-\omega)|^2(1+2D_{bb^\dag}(-\omega,-\omega))\right)\\
\nonumber&&+\frac{|\alpha_{out}|^2}{2}\left(A(\omega)B^*(\omega) 2D_{a b}(\omega,\omega)+A^*(\omega)B(\omega) 2D_{b^\dag a^\dag+}(\omega,\omega)\right.\\
&&\left.+A^*(-\omega)B(-\omega) 2D_{a^\dag b^\dag}(-\omega,-\omega)+B^*(-\omega)A(-\omega) 2D_{b a}(-\omega,-\omega)\right).\hspace{1cm}
\end{eqnarray}
Notons que l'on peut vérifier numériquement que dans les cas traités dans ce manuscrit, cette équation peut être évaluée de manière simplifiée sous la forme suivante :
\begin{eqnarray}\label{SNa3b}
\nonumber S_{N_{a,out}}(\omega)&\simeq&\frac{|\alpha_{out}|^2}{2}\left( |A(\omega)|^2(1+D_{aa^\dag}(\omega,\omega))+|A(-\omega)|^2(1+D_{a^\dag a}(-\omega,-\omega))\right.\\
&&+ \left. |B(\omega)|^2(1+D_{b^\dag b}(\omega,\omega))+|B(-\omega)|^2(1+D_{bb^\dag}(-\omega,-\omega))\right).
\end{eqnarray}

\subsubsection{Fluctuations quantiques de phase à un mode}
Afin de déterminer la densité spectrale de bruit de phase pour un mode du champ nous allons utiliser le formalisme introduit par  \cite{Pegg:1988p2364,Pegg:1989p2351} de l'opérateur phase $\phi_a$, qui s'exprime sous la forme suivante :
\begin{equation}
\hat{a}=\hat{e}^{i\hat\phi_a}\ \hat{N}^\frac{1}{2}.
\end{equation}
Cette expression est adaptée au formalisme des variables continues, c'est àd ire des états avec un grand nombre de photons.
Pour un opérateur nombre qui ne tend pas vers 0, on peut alors écrire cette relation sous la forme :
\begin{equation}\label{def_operateur_phase}
\hat{e}^{i\hat\phi_a}=\hat{a}\ \hat{N_a}^{-\frac{1}{2}}.
\end{equation}
%
On utilise alors cette expression pour écrire les fluctuations de manière linéarisée :
\begin{deqn}
e^{i\langle\hat\phi_a\rangle}e^{i\delta \hat\phi_a}=(\alpha+\delta\hat{a})\ \left((\alpha^*+\delta\hat{a}^\dag)(\alpha+\delta\hat{a})\right)^{-\frac{1}{2}}.
\end{deqn}
En utilisant un développement limité au premier ordre en $\frac{\delta a}{\alpha}$, on a :
\begin{ddeqn}
e^{i\langle\hat\phi_a\rangle}(1+{i\delta\hat{\phi}_a})=\sqrt{\frac{\alpha}{\alpha^*}}\left(1+\frac{1}{2}\frac{\delta \hat{a}}{\alpha}\right)\left(1-\frac{1}{2}\frac{\delta \hat{a}^\dag}{\alpha^*}\right).
\end{ddeqn}
Cela nous permet d'écrire les fluctuations de la phase en fonction de la quadrature $\hat{Y}_{ \phi_a}$ :
\begin{equation}\label{fluctuation_phase}
\delta \hat{\phi}_a=\frac{-i}{2|\alpha|}\left(\delta \hat{a}\ e^{-i\phi_a}-\delta \hat{a}^\dag e^{i\phi_a}\right)=\frac{\delta \hat{Y}_{ \phi_a}}{2|\alpha|}.
\end{equation}
La densité spectrale de bruit est ainsi donnée par :
		
\begin{equation}\label{def_spectre_phi}
S_{\phi_a}(\omega)\ 2\pi\ \delta(\omega+\omega')=\langle \delta\hat{\phi}_a(\omega)\ \delta \hat{\phi}_a^\dag(\omega')\rangle .
\end{equation}
En utilisant les équations  (\ref{fluctuation_phase}) et (\ref{def_spectre_phi}), on trouve le spectre de bruit de phase :
\begin{equation}\label{def_spectre_phi2}
S_{\phi_a}(\omega)\ 2\pi\ \delta(\omega+\omega')=\frac{1}{4|\alpha|^2} \langle \delta \hat{Y}_{\phi_a }(\omega) \delta \hat{Y}_{\phi_a }^\dag (\omega ')\rangle .
\end{equation}
On retrouve donc un résultat connu, à savoir le bruit sur la phase est donné par la fonction de corrélations de la quadrature $\hat{Y}$ associée à la phase moyenne du champ c'est-à-dire la quadrature : $\hat{Y}_{\phi_a } $.

\paragraph*{Calcul du spectre de bruit de phase à un mode}~\\
On peut calculer la relation  (\ref{def_spectre_phi}), avec la même méthode que pour les spectres de bruit en intensité, ce qui donne pour le spectre en sortie du milieu :
\begin{eqnarray}\label{Sphia1}
\nonumber S_{\phi_{a,out}}(\omega)\ 2\pi\ \delta(\omega+\omega')&=&\frac{-1}{4|\alpha_{out}|^2} \langle [A(\omega)(\delta \hat{a}_{in}(\omega) +F_a(L,\omega))e^{-i\theta}+B(\omega)(\delta \hat{b}_{in}^\dag(\omega)+F_{b^\dag}(L,\omega)) e^{-i\theta}\\
\nonumber&&-A^*(-\omega)(\delta \hat{a}_{in}^\dag(\omega)+F_{a^\dag}(L,\omega)) e^{i\theta}+B^*(-\omega )(\delta \hat{b}_{in}(\omega)+F_b(L,\omega)) e^{i\theta}]\\
\nonumber&&\times[A(\omega ')(\delta \hat{a}_{in}(\omega ') +F_a(L,\omega'))e^{-i\theta}+B(\omega ')(\delta \hat{b}_{in}^\dag(\omega ') +F_{b^\dag}(L,\omega'))e^{-i\theta}\\
\nonumber&&- A^*(-\omega')(\delta \hat{a}_{in}^\dag(\omega')+F_{a^\dag}(L,\omega')) e^{i\theta}+B^*(-\omega')(\delta \hat{b}_{in}(\omega')+F_b(L,\omega')) e^{i\theta}]\rangle .\\
~ \end{eqnarray}
On se place dans l'ordre symétrique comme précédemment et on écrit le spectre pour un état cohérent en entrée sur le mode $\hat{a}$ et le vide sur le mode $\hat{b}$ :
\begin{eqnarray}\label{Sphia2}
\nonumber S_{\phi_{a,out}}(\omega)\ 2\pi\ \delta(\omega+\omega')&=&\frac{1}{4|\alpha_{out}|^2} \left(A(\omega)A^*(-\omega')+A^*(-\omega)A(\omega')\right.\\
\nonumber &+&\left.(B(\omega)B^*(-\omega')+B^*(-\omega)B(\omega')\right)  \times\frac 12  \ 2\pi\ \delta(\omega+\omega')\\
\nonumber &+&\frac{1}{4|\alpha_{out}|^2}  (A(\omega)A^*(-\omega')\langle F_a(L,\omega)F_{a^\dag}(L,\omega')\rangle^{\mathcal{S}}\\
\nonumber &&\hspace{1cm}+A^*(-\omega)A(\omega')\langle F_{a^\dag}(L,\omega)F_a(L,\omega')\rangle)^{\mathcal{S}}\\
\nonumber &+&\frac{1}{4|\alpha_{out}|^2}  (B(\omega)B^*(-\omega')\langle F_{b^\dag}(L,\omega)F_{b}(L,\omega')\rangle^{\mathcal{S}}\\
\nonumber \hspace{1cm}&&+B^*(-\omega)B(\omega') \langle F_b(L,\omega)F_{b^\dag}(L,\omega')\rangle^{\mathcal{S}})\\
&+&\text{termes croisés}.
\end{eqnarray}
Les termes croisés correspondent aux deux dernières lignes de l'équation \eqref{SNa2} que nous avons omis pour simplifier (un peu) la lecture.
A l'aide d'une approximation identique à celle effectuée au paragraphe précédent pour passer de l'équation \eqref{SNa3} à \eqref{SNa3b}, on peut écrire le spectre de bruit de phase pour un champ $\hat{a}$ sous la forme :
\begin{eqnarray}\label{Sphia3}
\nonumber S_{\phi_{a,out}}(\omega)&\simeq&\frac{1}{4|\alpha_{out}|^2}\left( |A(\omega)|^2(1+D_{aa^\dag}(\omega,\omega))+|A(-\omega)|^2(1+D_{a^\dag a}(-\omega,-\omega))\right.\\
&&+ \left. |B(\omega)|^2(1+D_{b^\dag b}(\omega,\omega))+|B(-\omega)|^2(1+D_{bb^\dag}(-\omega,-\omega))\right).\end{eqnarray}
Pour comparer ce spectre à un état cohérent (la limite quantique standard) il faut le normaliser par $\frac{1}{4|\alpha|^2}$.
On obtient alors le spectre de bruit normalisé : 
\begin{eqnarray}\label{Sphia4}
\nonumber S_{\phi_{a,out}}^\mathcal{N}(\omega)&=&\left( |A(\omega)|^2(1+D_{aa^\dag}(\omega,\omega))+|A(-\omega)|^2(1+D_{a^\dag a}(-\omega,-\omega))\right.\\
&&+ \left. |B(\omega)|^2(1+D_{b^\dag b}(\omega,\omega))+|B(-\omega)|^2(1+D_{bb^\dag}(-\omega,-\omega))\right).\end{eqnarray}
On peut alors noter qu'après normalisation de (\ref{SNa3b}), on obtient une expression identique pour $S_{\phi_{a,out}}^N(\omega)$ et $S_{N_{a,out}}^N(\omega)$
Le fluctuations sur les quadratures identités et phase sont donc identiques.
On peut déduire de ce résultat très simple que notre système ne pourra pas générer un état comprimé à un mode, ce qui est attendu pour un amplificateur insensible à la phase

\subsection{Corrélations quantiques}
\subsubsection{Corrélations quantiques d'intensité}
\textit{Dans ce paragraphe nous avons aussi omis l'écriture des "termes croisés", afin de ne pas surcharger très lourdement l'écriture. Les résultats obtenus sont par conséquent des résultats approchés, mais qui pour les paramètres numériques explorés dans ce manuscrit en sont une très bonne approximation.}\\
On calcule le spectre de bruit de la différence d'intensité des deux modes $\hat{a}$ et $\hat{b}$ en sortie du milieu.
On définit l'opérateur différence d'intensité $\hat{N}_-$ par :
\begin{equation}
\hat{N}_-=\hat{N}_a-\hat{N}_b.
\end{equation}
En notant les valeurs moyennes des opérateurs $\hat{a}$ et $\hat{b}$ de la façon suivante : $\langle \hat{a}\rangle = |\alpha | e^{i \phi_a}$ et $\langle \hat{b}\rangle = |\beta | e^{i \phi_b} $, on obtient pour les fluctuations de l'opérateur $\hat{N}_-$ la relation :
\begin{equation}\label{defN-}
\delta \hat{N}_-(\omega)=|\alpha | \delta \hat{X}_{a,\phi_a}(\omega)- |\beta |\delta \hat{X}_{b,\phi_b}(\omega).
\end{equation}
Le spectre de bruit s'écrit, comme précédemment, comme la transformée de Fourier de la fonction de corrélation :
\begin{equation}\label{def_spectre_N-}
S_{N_-}(\omega)2\pi\ \delta(\omega+\omega')=\langle \delta N_-(\omega)\delta N_-^\dag(\omega')\rangle.
\end{equation}
En substituant la relation (\ref{defN-}) dans (\ref{def_spectre_N-}) on peut écrire :
\begin{eqnarray}\label{sn-1}
\nonumber S_{N_-}(\omega)2\pi\ \delta(\omega+\omega')&=& \langle \left( |\alpha | \delta \hat{X}_{a,\phi_a}(\omega)- |\beta |\delta \hat{X}_{b,\phi_b}(\omega)\right) \left( |\alpha | \delta \hat{X}_{a,\phi_a}(\omega ')- |\beta |\delta \hat{X}_{b,\phi_b}(\omega ')\right)\rangle\\
\nonumber &=&\langle |\alpha |^2 \delta \hat{X}_{a,\phi_a}(\omega) \delta \hat{X}_{a,\phi_a}(\omega ')+ |\beta |^2 \delta \hat{X}_{b,\phi_b}(\omega)\delta \hat{X}_{b,\phi_b}(\omega ')\\
&&- |\alpha \beta| \left(\delta \hat{X}_{b,\phi_b}(\omega)\delta \hat{X}_{a,\phi_a}(\omega ')+\delta \hat{X}_{a,\phi_a}(\omega)\delta \hat{X}_{b,\phi_b}(\omega ')\right)\rangle.
\end{eqnarray}
Le terme $ |\alpha |^2\langle \delta \hat{X}_{a,\phi_a}(\omega) \delta \hat{X}_{a,\phi_a}(\omega ')\rangle$  est identique à celui défini pour le à l'équation (\ref{def_spectre3}) et calculé en (\ref{SNa3}).
De même le terme $|\beta |^2\langle \delta \hat{X}_{b,\phi_b}(\omega)\delta \hat{X}_{b,\phi_b}(\omega ')\rangle$ correspond à un calcul similaire pour le champ $\hat b$.\\
Comme précédemment on ne s'intéresse qu'aux termes dont la valeur moyenne est non nulle (cf. \ref{cov_sym}).
A l'aide des relations entrée-sortie on écrit le terme $\langle\delta \hat{X}_{a,\phi_a}(\omega)\delta \hat{X}_{b,\phi_b}(\omega ')\rangle$ en sortie dans l'ordre symétrique sous la forme :
\begin{eqnarray}
\nonumber \langle\delta \hat{X}_{a,\phi_a}(\omega)\delta \hat{X}_{b,\phi_b}(\omega ')\rangle^{\mathcal{S}} & =  &
A(\omega)C^*(-\omega ')e^{-i(\phi_{a}^{out}+\phi_{b}^{out})}\left(\langle\delta\hat{a}_{in}(\omega) \delta\hat{a}_{in}^\dag(\omega')\rangle^{\mathcal{S}} +\langle F_a(\omega) F_{a^\dag}(\omega')\rangle^{\mathcal{S}} \right) \\
\nonumber & + & B(\omega)D^*(-\omega ')e^{-i(\phi_{a}^{out}+\phi_{b}^{out})}\left(\langle\delta\hat{b}_{in}^\dag(\omega) \delta\hat{b}_{in}(\omega')\rangle^{\mathcal{S}} +\langle F_{b^\dag}(\omega) F_b(\omega')\rangle^{\mathcal{S}} \right)\\
\nonumber  &+&	A^*(-\omega)C(\omega ')e^{i(\phi_{a}^{out}+\phi_{b}^{out})}\left(\langle\delta\hat{a}^\dag_{in}(\omega) \delta\hat{a}_{in}(\omega')\rangle^{\mathcal{S}}  +\langle F_{a^\dag}(\omega) F_a(\omega')\rangle^{\mathcal{S}} \right)\\
\nonumber &+&	B^*(-\omega)D(\omega ')e^{i(\phi_{a}^{out}+\phi_{b}^{out})}\left(\langle\delta\hat{b}_{in}(\omega) \delta\hat{b}_{in}^\dag(\omega')\rangle^{\mathcal{S}} +\langle F_b(\omega) F_{b^\dag}(\omega')\rangle^{\mathcal{S}} \right) .\\
\end{eqnarray}
En utilisant les relations que l'on a obtenues dans la section précédente sur la valeur moyenne de l'amplitude et de la phase en sortie du milieu on peut écrire : 
\begin{equation}
 e^{i(\phi_{a}^{out}+\phi_{b}^{out})} =\frac{ |\alpha_{in}|^2 A(0)C(0)^*}{|\alpha_{out} \beta_{out}| }.
\end{equation}
On peut alors écrire le terme $|\alpha_{out} \beta_{out}| \langle\delta \hat{X}_{a,\phi_a}(\omega)\delta\hat{X}_{b,\phi_b}(\omega ')\rangle$ de l'équation (\ref{sn-1}) sous la forme : 
\begin{eqnarray}
\nonumber|\alpha_{out} \beta_{out}| \langle\delta \hat{X}_{a,\phi_a}(\omega)\delta \hat{X}_{b,\phi_b}(\omega ')\rangle&=&\frac{|\alpha_{in}|^2}{2}\  2\pi\delta(\omega+\omega')\times\\
\nonumber&&\left[ A(0)C(0)^*\left(A^*(-\omega)C(\omega ')+B^*(-\omega)D(\omega ')\right)\right.\\
\nonumber&+& \left. A(0)^*C(0) \left(A(\omega)C^*(-\omega ')+ B(\omega)D^*(-\omega ')\right)\right]\\
\nonumber&+& A(0)C(0)^*\left(A^*(-\omega)C(\omega ')\langle F_{a^\dag}(\omega) F_a(\omega')\rangle^{\mathcal{S}}\right.\\
\nonumber && \hspace{1cm}\left.+B^*(-\omega)D(\omega ')\langle F_b(\omega) F_{b^\dag}(\omega')\rangle^{\mathcal{S}}\right)\\
\nonumber&+& \left. A(0)^*C(0) \left(A(\omega)C^*(-\omega ')\langle F_a(\omega)F_{a^\dag}(\omega')\rangle^{\mathcal{S}}\right.\right. \\
&& \hspace{1cm}+\left. B(\omega)D^*(-\omega ')\langle F_{b^\dag}(\omega) F_b(\omega')\rangle^{\mathcal{S}}\right).
\end{eqnarray}
Le spectre de bruit sur la différence d'intensité des deux champs s'écrit donc:
\begin{eqnarray}
S_{N_-}(\omega)&=&\frac{|\alpha_{in}\ A(0)|^2}{2}\left( |A(\omega)|^2(1+D_a(\omega))+|A(-\omega)|^2(1+D_a(\omega))\right.\\
\nonumber&&\hspace{1.5cm}+ \left. |B(\omega)|^2(1+D_b(\omega))+|B(-\omega)|^2(1+D_b(\omega))\right)\\
\nonumber &+&\frac{|\alpha_{in}\ C(0)|^2}{2}\left( |C(\omega)|^2(1+D_a(\omega))+|C(-\omega)|^2(1+D_a(\omega))\right.\\
\nonumber&&\hspace{1.5cm}+  \left. |D(\omega)|^2(1+D_b(\omega))+|D(-\omega)|^2(1+D_b(\omega))\right)\\
\nonumber &-&|\alpha_{in}|^2 \mathbf{Re}\left[ A(0)C(0)^*\left( C(\omega)A^*(\omega  )(1+Da(\omega))+D(\omega)B^*(\omega )(1+Db(\omega))\right.\right.\\
\nonumber&&\hspace{2cm}+\left.\left.A^*(-\omega)C(-\omega )(1+Da(\omega))+B^*(-\omega)D(-\omega )(1+Db(\omega))\right)\right].
\end{eqnarray}
Une fois ce terme normalisé par le bruit d'intensité d'un faisceau cohérent contenant la somme du nombre de photons sur $\hat a$ et $\hat b$, $|\alpha_{in}\ A(0)|^2+|\alpha_{in}\ C(0)|^2$, on obtient :

\begin{eqnarray}\label{eq:sol_a}
S^\mathcal{N}_{N_-}(\omega)&=&
\frac{1}{2(|A(0)|^2+|C(0)|^2)}\times\\
\nonumber&&\left(  |A(0)^* A(\omega)-C(0)^* C(\omega)|^2(1+D_{aa^\dag}(\omega))\right.\\
\nonumber &&+|A(0) A(-\omega)^*-C(0) C(-\omega)^*|^2(1+D_{a^\dag a}(-\omega))\\
\nonumber&&+|A(0)^* B(\omega)-C(0)^* D(\omega)|^2(1+D_{b^\dag b}(\omega))\\
\nonumber&&\left.+ |A(0) B(-\omega)^*-C(0) D(-\omega)^*|^2(1+D_{bb^\dag}(-\omega))\right).\end{eqnarray}
Il est intéressant de noter que, contrairement au spectre de bruit du mode $\hat a$ seul qui est une somme de termes tous positifs (équation \eqref{SNa3} ), il n'est pas nécessaire que les coefficients $A,B,C,D$ de l'équation \eqref{eq:sol_a} soient nuls pour que $S^\mathcal{N}_{N_-}(\omega)$ tende vers 0.\\
En effet, pour un système se comportant comme un amplificateur idéal de bande passante infinie décrit au chapitre 3, on a :
$|A(\omega)|^2=|D(\omega)|^2=G$ et $|B(\omega)|^2=|C(\omega)|^2=G-1$ ainsi que des coefficients de diffusions nuls : $D_{uv}=0$.
On retrouve alors les résultats du chapitre 3 pour l'amplificateur insensible à la phase :
\begin{equation}S^\mathcal{N}_{N_-} = \frac{1}{2G-1}.\end{equation}
\subsubsection{Anti--corrélations quantiques de phase}
Pour démontrer l'intrication de deux faisceaux en variables continues, on peut utiliser un critère basé sur l'inséparabilité qui sera détaillé dans la section \ref{froid}.
L'inséparabilité est définie par:
\begin{equation}\label{insep}
\mathcal{I}(\omega)=\frac 12 (S^\mathcal{N}_{N^-} + S^\mathcal{N}_{\phi^+}).
\end{equation}
Pour calculer la valeur de l'inséparabilité, il est donc nécessaire de déterminer les anti--corrélations de phase $\hat{\phi}_+$.
C'est ce que nous allons étudier dans ce paragraphe en suivant la même méthode que précédemment.
On écrit l'opérateur de somme des phases :
\begin{equation}
\hat{\phi}_+=\hat{\phi}_a+\hat{\phi}_b.
\end{equation}
Puis on utilise (\ref{fluctuation_phase}) pour écrire :
\begin{equation}
\delta \hat{\phi_+}=\frac{\delta \hat{Y}_{a,\phi_a}}{2|\alpha|}+\frac{\delta \hat{Y}_{b,\phi_b}}{2|\beta|}.
\end{equation}
On peut alors calculer le spectre :
\begin{eqnarray}
\nonumber S_{\phi +}(\omega)2\pi\ \delta(\omega+\omega')&=& \frac{\langle\delta \hat{Y}_{a,\phi_a}(\omega)\delta \hat{Y}_{a,\phi_a}(\omega ')\rangle}{4|\alpha |^2 }+ \frac{\langle\delta \hat{Y}_{b,\phi_b}(\omega)\delta \hat{Y}_{b,\phi_b}(\omega ')\rangle}{4|\beta |^2 }\\
&&+ \frac{\langle\delta \hat{Y}_{a,\phi_a}(\omega)\delta \hat{Y}_{b,\phi_b}(\omega ')+\delta \hat{Y}_{b,\phi_b}(\omega)\delta \hat{Y}_{a,\phi_a}(\omega ')\rangle}{4|\alpha \beta| }.
\end{eqnarray}
A nouveau les termes de la première ligne découlent directement du calcul de bruit à un champ (équation \ref{Sphia2}).
Le terme de la seconde ligne doit lui être calculé.
En ne gardant que les termes non nuls pour notre état d'entrée, on a :
\begin{eqnarray}
\langle\delta \hat{Y}_{a\ \phi_a}(\omega)\delta \hat{Y}_{b\ \phi_b}(\omega ')\rangle&=&
\nonumber				-A(\omega)C^*(-\omega ')e^{-i(\phi_{a}^{out}+\phi_{b}^{out})}\left(\langle\delta\hat{a}_{in} (\omega)\delta \hat{a}_{in}^\dag(\omega')\rangle +\langle F_a(\omega) F_{a^\dag}(\omega')\rangle^{\mathcal{S}} \right)\\
\nonumber & - & B(\omega)D^*(-\omega ')e^{-i(\phi_{a}^{out}+\phi_{b}^{out})}\left(\langle\delta\hat{b}_{in} ^\dag(\omega)\delta\hat{b}_{in}(\omega')\rangle+\langle F_{b^\dag}(\omega)F_b (\omega')\rangle^{\mathcal{S}} \right)\\
\nonumber  &-&	A^*(-\omega)C(\omega ')e^{i(\phi_{a}^{out}+\phi_{b}^{out})}\left(\langle\delta\hat{a}_{in}^\dag(\omega) \delta\hat{a}_{in}(\omega')\rangle+\langle F_{a^\dag}(\omega) F_a(\omega')\rangle^{\mathcal{S}} \right) \\
\nonumber  &-&	B^*(-\omega)D(\omega ')e^{i(\phi_{a}^{out}+\phi_{b}^{out})}\left(\langle\delta\hat{b}_{in} (\omega)\delta\hat{b}_{in}^\dag(\omega')\rangle+\langle F_b(\omega) F_{b^\dag}(\omega')\rangle^{\mathcal{S}} \right) .\hspace{5mm}\\
\end{eqnarray}
On obtient donc finalement le spectre de bruit des anti-corrélations de phase sous la forme générale :
\begin{small}
\begin{eqnarray}\label{sphi+}
\nonumber S_{\phi_+}(\omega)&=&\frac{1}{2}\ \frac{1}{4|\alpha_{out}|^2}\left( |A(\omega)|^2(1+D_a(\omega))+|A(-\omega)|^2(1+D_a(\omega))\right.+ \left. |B(\omega)|^2(1+D_b(\omega))+|B(-\omega)|^2(1+D_b(\omega))\right)\\
\nonumber &+&\frac{1}{2}\ \frac{1}{4|\beta_{out}|^2}\left( |C(\omega)|^2(1+D_a(\omega))+|C(-\omega)|^2(1+D_a(\omega))\right.+  \left. |D(\omega)|^2(1+D_b(\omega))+|D(-\omega)|^2(1+D_b(\omega))\right)\\
\nonumber &-&\frac{|\alpha_{in}|^2}{4|\alpha_{out}\ \beta_{out}|^2} \mathbf{Re}\left[ A(0)C(0)^*\left( C(\omega)A^*(\omega  )(1+Da(\omega))+D(\omega)B^*(\omega )(1+Db(\omega))\right.\right.\\
&&\hspace{2cm}+\left.\left.A^*(-\omega)C(-\omega )(1+Da(\omega))+B^*(-\omega)D(-\omega )(1+Db(\omega))\right)\right].
\end{eqnarray}
\end{small}
Comme pour les spectres de corrélations il faut normaliser ce terme par le bruit de phase d'un faisceau cohérent pour le comparer au bruit quantique standard : 
\begin{equation}
S{_\phi}=\frac{|\alpha_{out}|^2+|\beta_{out}|^2}{4|\alpha_{out}\ \beta_{out}|^2}.
\end{equation}
L'expression (\ref{sphi+}) peut alors s'écrire sous la forme :
\begin{eqnarray}\label{eq:sol_b}
     S_{\phi_+}^\mathcal{N}(\omega)&=&
      \frac{1}{2(|A(0)|^2+|C(0)|^2)}\times\\
       \nonumber&&\left(  |A(0) C(\omega)-C(0) A(\omega)|^2(1+D_{aa^\dag}(\omega))\right.\\
     \nonumber &&+|A(0) C(-\omega)-C(0) A(-\omega)|^2(1+D_{a^\dag a}(-\omega))\\
      \nonumber&&+|A(0) D(\omega)-C(0) B(\omega)|^2(1+D_{b^\dag b}(\omega))\\
      \nonumber&&\left.+ |A(0) D(-\omega)-C(0) B(-\omega)|^2(1+D_{bb^\dag}(-\omega))\right).\end{eqnarray}
Dans la référence \cite{Glorieux:2010p9415}, les équations \eqref{eq:sol_a} et \eqref{eq:sol_b} ont été obtenues en utilisant un formalisme légèrement différent, à l'aide des quadratures $\hat x$ et $\hat p$.
On pourra vérifier que ces deux formalismes mènent à un résultat identique.\\

Dans le modèle de l'amplificateur linéaire idéal de bande passante infinie (
$|A(\omega)|^2=|D(\omega)|^2=G$ et $|B(\omega)|^2=|C(\omega)|^2=G-1$ et $D_{uv}=0$), on peut donner l'expression des anticorrélations d'intensité sous la forme :
\begin{equation}     S_{\phi_+}^\mathcal{N}= \frac{1}{2G-1}.\end{equation}

Ainsi dans ce cas l'inséparabilité s'écrit de la même manière :
\begin{equation}     \mathcal{I}= \frac{1}{2G-1}.\end{equation}
Dans, le cas de l'amplificateur linéaire idéal de bande passante infinie, on obtient donc que lorsque le gain est plus grand que 1, alors le système génère à la fois des corrélations d'intensité et des anti-corrélations de phase sous la limite quantique standard.
Ainsi le système produit des faisceaux intriqués.\\
Nous allons maintenant nous intéresser aux différents paramètres que nous avons introduit dans le modèle microscopique pour dépasser  l'amplificateur linéaire idéal  et prédire les résultats expérimentaux.
\newpage 
	
\section{Paramètres}\label{param_froids}
Les différentes grandeurs physiques associées à la raie D1 du rubidium 85, que nous utiliseront dans ce paragraphe, sont décrites en détail dans l'annexe \ref{Annexe_Rb}.
En effet, comme nous en donnons l'explication dans l'annexe \ref{Annexe_Lambda}, la raie D1 ($5S_{1/2} \to 5P_{1/2}$) ou la transition à 422 nm ($5S_{1/2} \to 6P_{1/2}$) du rubidium 85 peuvent être modélisées comme un système en double--$\Lambda$, lorsque les champ pompe et sonde sont polarisés linéairement et de façon orthogonale.

\subsection{Constante de couplage}	
On rappelle que le la solution de l'équation de propagation \ref{sol_propa} s'écrit :
\begin{equation*}
| \hat{A}(L,\omega)]=e^{[M(\omega)].L} \left(|\hat{A}(0,\omega)]+L\int_0^1  e^{-[M(\omega)]Lz} [M_F(\omega)]|F_1(z,\omega)]dz\right),
\end{equation*}
avec  $
[M(\omega)]=i\ \frac{g^2 N}{c} [T][M_1'(\omega)]^{-1}[S_1]. $\\
Le terme de couplage $g$  \cite{Hilborn:2002p1355} est défini dans l'équation (\ref{eq_ham_3}) par : 
\begin{equation}\label{def_alphaL}
g=\frac{\wp\ \varepsilon}{\hbar}\ \text{avec}\ \varepsilon=\sqrt{\frac{\hbar\omega_L}{2\epsilon_0 V}}\ \text{et}\ \wp=\sqrt{\frac{ \epsilon_0\ c\ \hbar\  \Gamma/2\  \sigma}{\omega_L}}.
\end{equation}
Par simplicité, ce terme de couplage $g$ est pris égal pour les deux champs $\hat a$ et $\hat b$.
On peut donc écrire la matrice $[M]$ de l'équation de propagation (\ref{sol_propa}) en introduisant la densité d'atomes dans la zone d'interaction  $\mathcal{N}$ sous la forme :
\begin{equation}
[M(\omega)]=i \mathcal{N} \ \sigma\  L\ \frac{\Gamma}{4}[T][M_1'(\omega)]^{-1}[S_1].
\end{equation}
avec $ \sigma$ la section efficace de la transition.
Par analogie avec la manière dont on traite les effets d'absorption linéaire dans le modèle de l'atome à deux niveaux, on notera le produit $\alpha L= \mathcal{N} \ \sigma\ L$ et on appellera ce terme \textit{l'épaisseur optique du milieu}.

	\subsection{Décohérence}
	
Le taux de décohérence $\gamma$ de la cohérence atomique $\tilde \sigma_{12}$ correspond dans notre étude au taux de décohérence entre les deux niveaux hyperfins de l'état fondamental.
Ce paramètre  a été étudié de manière approfondie dans le cadre des expériences de pompage optique et pour des applications aux magnétomètres et à la mesure des standards de fréquence \cite{Happer:1972p2002,Oreto:2004p2032}.\\
La question du déphasage dû aux collisions induisant un changement de vitesse de l'atome sans en modifier l'état interne a été étudiée dans \cite{Ghosh:2009p2129}.
Pour un milieu constitué d'atomes froids cet effet sera négligé car la distribution de vitesse est très étroite. \\
Le taux de décohérence s'écrit alors sous la forme :
\begin{equation}
\gamma=\gamma_t+\gamma_{col}+\gamma_B,
\end{equation}
où nous avons introduit les taux suivants :
\begin{itemize}
\item $\gamma_{col}$ est le taux de décohérence collisionel, déterminé par les collisions rubidium-rubidium (ou les collisions rubidium-gaz tampon dans le cas d'une cellule contenant un gaz tampon) \cite{Oreto:2004p2032}.
\item Le terme $\gamma_B$ correspond au déphasage introduit par une inhomogénéité du champ magnétique. Typiquement, dans la littérature, il est usuel de trouver des valeurs de  $\gamma_B\simeq 1$ kHz.
\item $\gamma_t$ provient du temps de transit des atomes dans un faisceau de diamètre $R$ à une vitesse $\overline{v}$. S'il est nul pour un milieu constitué d'atomes froids, ce terme sera le terme dominant dans le cas d'une vapeur atomique. Il est donné en l'absence d'un gaz tampon par \cite{Oreto:2004p2032}:
\begin{equation}
\gamma_{\rm t}= \frac{{\bar{v}}}{R}.
\end{equation}  
\end{itemize}
En conclusion, pour un milieu constitué d'atomes froids, le terme dominant est $\gamma_B$ et on a donc $$\gamma\simeq 1\text{ kHz}.$$\\
Pour une vapeur atomique \textit{``chaude''} le terme dominant est $\gamma_t$.
Nous reviendrons en détail sur ce terme lorsque nous aborderons l'extension du modèle aux vapeurs atomiques.
On peut noter que pour $\overline{v}\simeq 300\ \text{m.s}^{-1},\ T= 100^\circ$C et $R=1$ mm le taux de décohérence est de l'ordre de  : $$\gamma\simeq\gamma_t\simeq 500\text{ kHz}.$$\\

\subsection{Pulsation de Rabi}
La pulsation de Rabi d'un laser $\Omega$ quantifie le couplage lumière matière pour une transition atomique donnée.
Elle s'écrit sous la forme :
\begin{equation}
\Omega=\frac{\wp E}{\hbar},
\end{equation}
où $\wp$ est le dipole de la transition atomique et $E$ le champ électrique correspondant au faisceau laser.
Ce champ $E$ s'exprime en fonction de l'intensité $I$ du laser sous la forme :
\begin{equation}
E=\sqrt{\frac{2 I}{\epsilon_0 c}}.
\end{equation}
Expérimentalement, les grandeurs accessibles sont la puissance $P$ du laser et son waist $w$ (rayon à $1/e^2$).
L'intensité s'écrit donc :
\begin{equation}
I=\frac{P}{\pi w^2}.
\end{equation}
On peut donc donner un ordre de grandeur des pulsations de Rabi accessibles expérimentalement :
\begin{itemize}
\item Sur la transition D1 ($5S_{1/2} \to 5P_{1/2}$) à 795 nm du rubidium 85, une puissance de laser de 1W et une focalisation sur un waist de 600 microns donnent une pulsation de Rabi $\Omega = 2\pi \times 1$ GHz.
\item Sur la transition à 422 nm ($5S_{1/2} \to 6P_{1/2}$) du rubidium 85, une puissance de laser de 300 mW et une focalisation sur un waist de 200 microns donnent une pulsation de Rabi $\Omega = 2\pi \times 0.15$ GHz.
\end{itemize}

\section{Intrication en variables continues dans un milieu d'atomes froids}\label{froid}
Nous allons maintenant présenter les résultats obtenus lors de cette étude pour des atomes à vitesse nulle correspondant à un milieu constitué d\textit{'atomes froids.}
Nous utiliserons des paramètres accessibles dans ce genre d'expériences et en particulier pour la valeur de l'épaisseur optique ($\alpha L=150$).
Nous verrons par la suite que les processus que nous présentons ici permettent tout de même de produire des corrélations (certes moins importantes) pour des épaisseurs optiques plus faibles.
Nous présenterons l'effet de ces différents paramètres sur les corrélations d'intensité générées.
Comme nous le verrons, nos calculs démontrent qu'un milieu d'atomes froids permet de générer par mélange à 4 ondes des faisceaux intriqués intenses.
\subsection{Inséparabilité}
Nous avons déjà introduit les corrélations d'intensité $S_{N}^-$ (équation \eqref{SNa3}) et les anti-corrélations de phase $S_{\phi}^+$  (équation \eqref{Sphia4}), ainsi que l'inséparabilité $\mathcal{I}(\omega)$.
Pour détecter la présence d'intrication, nous utiliserons le critère suffisant  introduit par \cite{Duan:2000p15733,Simon:2000p15824}
:\begin{equation}
\mathcal{I}(\omega)< 1,
\end{equation} où l'inséparabilité $\mathcal{I}(\omega)$ est définie à l'équation \eqref{insep}.
Ainsi, il s'agit d'un critère différent de la simple présence de corrélations d'intensité (ou d'anti-corrélations de phase).
\subsection{Limites expérimentales}
Nous allons appliquer le modèle que nous venons de décrire à la transition $5S_{1/2}\to 5P_{1/2}$ du rubidium 85.
Les paramètres que nous utiliserons ont été introduits dans la section~\ref{param_froids}.
La pulsation de Rabi et le désaccord à un photon seront choisis semblables à ceux que l'on trouve dans les expériences de mélange à 4 ondes dans une vapeur atomique  \cite{Boyer:2007p1404,McCormick:2007p652,Boyer:2008p1401,Marino:2009p1400,GlorieuxSPIE2010} ; soit $0.7<\Delta /2\pi  <3$~GHz et $0.3<\Omega/2\pi<2$~GHz.\\
On utilisera pour la valeur de l'épaisseur optique $\alpha L=150$, ce qui est supérieur mais compatible avec ce qui a été obtenu dans des expériences utilisant un piège magnéto-optique très anisotrope \cite{Du:2008p8781,Lin:2008p16593}.
\subsection{Spectre de gain}\label{parag:spec}
\begin{figure}
\centering
\includegraphics[width=13.5cm]{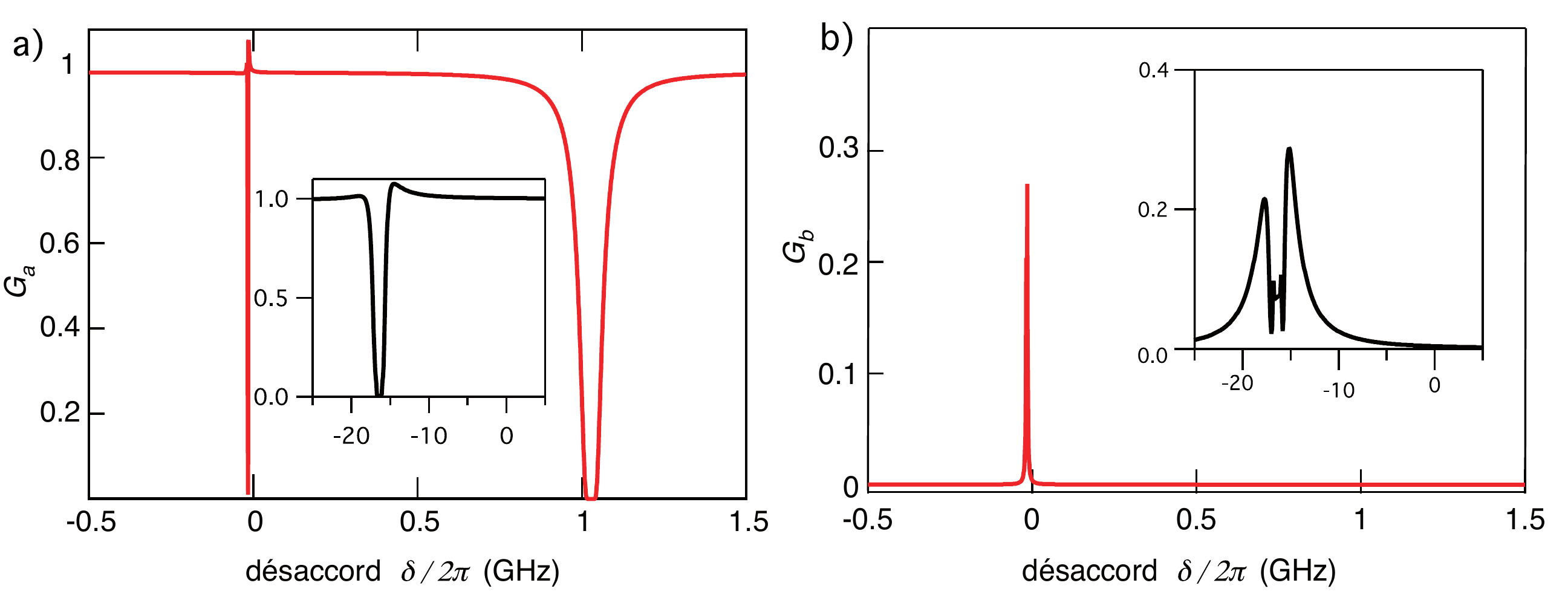} 
\caption[Spectre de gain pour la sonde et le conjugué.]	{Spectre de gain sur la sonde a) et sur le conjugué b) en fonction du désaccord à 2 photons $\delta$. Paramètres : $\gamma/2\pi=10$~kHz; $\Omega/2\pi=0.3$~GHz;
$\Delta/2\pi=1$~GHz.
Les encarts sont des vue de détails de la zone proche de $\delta=0$.}
	\label{fig2}
\end{figure} 
Dans un premier temps, on s'intéresse aux quantités classiques.
Le gain sur la sonde $G_a$ et sur le conjugué $G_b$ sont tracés en fonction du désaccord à 2 photons sur la figure \ref{fig2}.
Différents processus élémentaires contribuent au profil observé \cite{Lukin:2000p1648}.
Lorsque la condition de résonance à 4 photons est remplie ($\delta=0$), on observe une redistribution cohérente des photons de la pompe vers la sonde et le conjugué.
C'est le processus de mélange à 4 ondes proprement dit.\\
De plus, dans cette situation, une transition Raman impliquant un photon sonde et un photon pompe peut avoir lieue (condition d'accord à 2 photons).
Ainsi, le spectre de gain du conjugué peut être compris en terme du seul processus de mélange à 4 ondes, alors que celui de la sonde fait apparaître une combinaison entre les processus de mélange à 4 ondes et de transition Raman.
Sur l'encadré de la figure \ref{fig2} a), on peut donc attribuer l'amplification du champ sonde au mélange à 4 ondes, alors que l'absorption pour $\delta\simeq 0$ est due au processus Raman impliquant un photon sonde et un photon pompe.
D'autre part, la zone d'absorption pour $\delta\simeq \Delta$ est simplement une  conséquence du passage à résonance de la sonde sur la transition $|2\ket\to|3\ket$.
Pour un milieu constitué d'atomes froids, la largeur de ce creux est donnée par la largeur naturelle, $\Gamma$, ainsi que par l'épaisseur optique pour prendre en compte la propagation.\\
Si on considère le sous système en simple $\Lambda$  : $| 1 \ket ,|2\ket ,|3\ket$, on peut voir qu'il s'agit d'une configuration d'EIT un peu particulière car le faisceau pompe est hors résonance.
Dans ce cas, on observe, un profil asymétrique caractéristique des profils de Fano décrit dans les expériences de \cite{Alzetta:1979p16248,Kaivola:1985p16290,Zhu:1997p16306,Zhu:1996p16300} et étudié théoriquement dans \cite{Lounis:1992p16335}.
Le gain de la sonde autour de $\delta=0$ est similaire aux profils décrits dans les références précédentes, à la différence du pic de gain qui n'est pas présent dans les profil d'EIT et qui provient, comme nous l'avons dit, du processus de mélange à 4 ondes.
On peut vérifier que la position exacte de ce pic est donnée par le déplacement lumineux (\textit{AC Stark shift}) induit par le faisceau pompe \cite{Wei:1998p8801}.\\
Enfin, sur le spectre du mode conjugué, on peut observer une réduction du gain dans la zone $\delta\simeq 0$ dont la largeur est compatible avec celle du creux d'absorption sur le mode sonde.
De manière qualitative, on peut comprendre cela par le fait qu'en absence de faisceau sonde (c'est-à-dire lorsqu'il est absorbé), il n'y a pas de création de photons dans le mode conjugué.
Le taux de génération du faisceau conjugué diminue donc lorsque la sonde est absorbée au cours de la propagation.

\subsection{Corrélations d'intensité et anti-corrélations de phase}
\begin{figure}
	\centering
	\includegraphics[width=14.5cm]{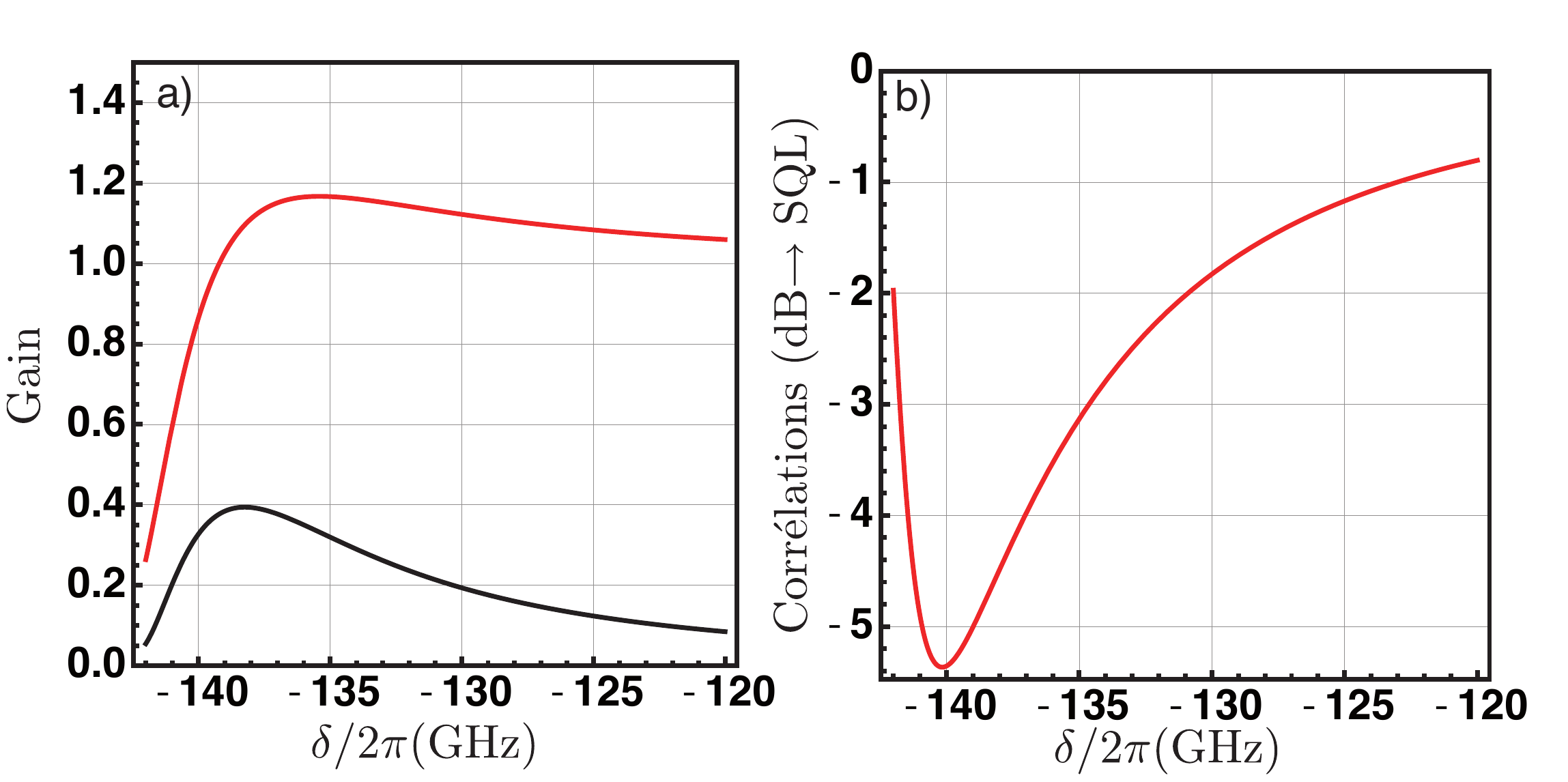} 
	\caption[Gain pour la sonde et le conjugué et corrélations en fonction de $\delta$]	{ a) Spectre de gain sur la sonde $G_a$ (rouge) et sur le conjugué $G_b$ (noir) en fonction du désaccord à 2 photons $\delta$. b) Corrélations quantiques en intensité à 1 MHz en fonction de $\delta$ normalisé au bruit quantique standard. 
	Paramètres : $\gamma/2\pi=10$~kHz; $\Omega/2\pi=1$~GHz;
 $\Delta/2\pi=1$~GHz.}
		\label{fig2.5}
	\end{figure}
\begin{figure}
	\centering
	\includegraphics[width=13cm]{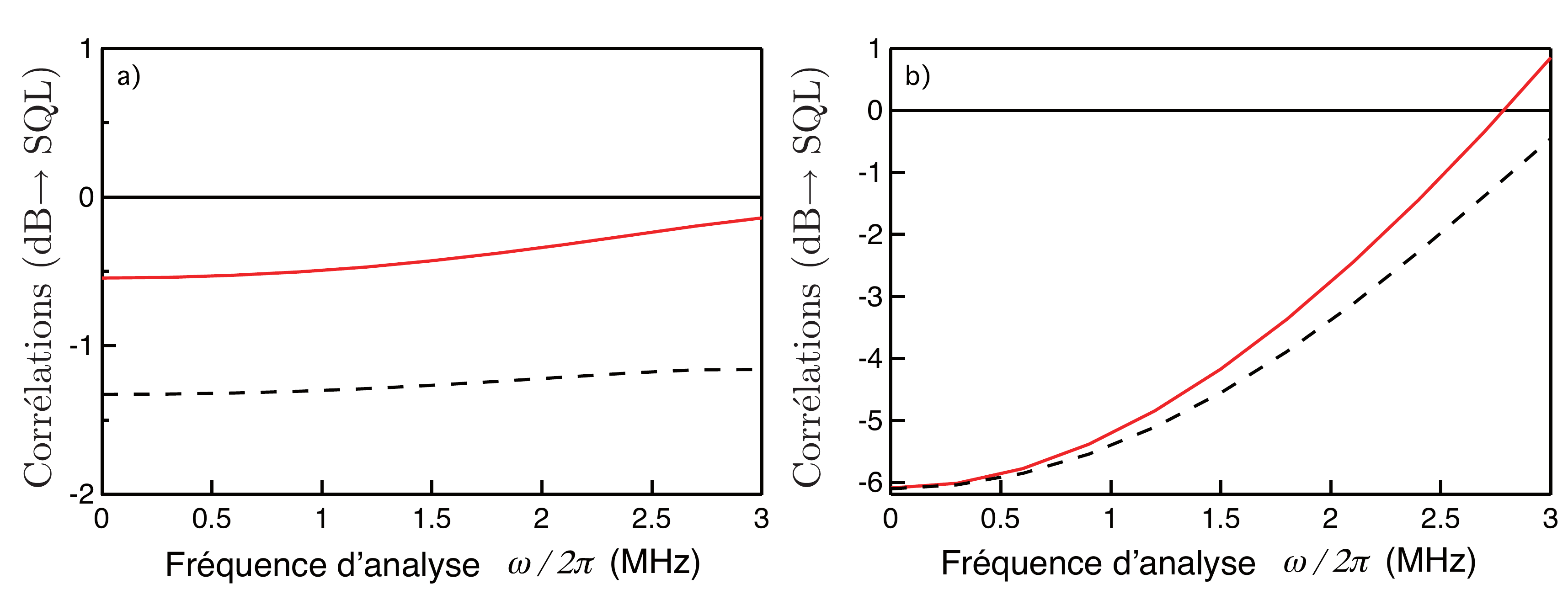} 
	\caption[Corrélations d'intensité et  anti-corrélations de phase en fonction de la fréquence d'analyse]	{ Corrélations d'intensité (noir pointillé) et anti-corrélations de phase (rouge) en dB en fonction de la fréquence d'analyse comparées à la limite quantique standard (noir).
	Paramètres : a) $\Omega/2\pi=\Delta/2\pi=0.3$~GHz  and $\delta/2\pi=$-48 MHz et b) $\Omega/2\pi=\Delta/2\pi=2$~GHz and $\delta/2\pi=$-217 MHz.}
		\label{fig3}
	\end{figure}

 
On présente maintenant les résultats du calcul concernant les fluctuations quantiques.
La figure \ref{fig2.5} présente les corrélations d'intensité à une fréquence d'analyse de 1 MHz en fonction de $\delta$.
Le profil de cette courbe montre que le processus de mélange à 4 ondes est très sensible au désaccord à deux photons $\delta$. La gamme d'accord optimale est imposée par le déplacement lumineux qui fixe la condition de résonance, elle est de l'ordre de la dizaine de MHz.
On se place dans la zone décrite précédemment comme la ``zone de mélange à 4 ondes''.
La valeur de $\delta$ correspondante est fixée par le déplacement lumineux.
La figure \ref{fig3} présente les corrélations d'intensité et les anti--corrélations de phase normalisées en fonction de la fréquence d'analyse pour deux valeurs de $\Omega$ et $\Delta$.
Pour tracer les courbes de cette figure, on utilise les valeurs de $\delta$ qui maximisent, dans chaque cas, les corrélations en compensant le déplacement lumineux.
Les spectres de corrélations d'intensité présentés sont les données typiquement accessibles expérimentalement grâce à l'analyse spectral du bruit du photocourant.
Nous avons vérifié numériquement que, dans la condition\footnote{De manière plus générale, une étude numérique en deux dimensions démontre que l'on pourra atteindre des niveaux de corrélations légèrement plus élevés en augmentant $\Omega$ à $\Delta$ fixé.}  $\Delta=\Omega$, les plus hauts taux de corrélations sont atteints pour les plus grandes valeurs de $\Delta$ et $\Omega$
Les figures \ref{fig3} a) et b) sont en accord avec cet effet.
La figure \ref{fig3} b) montre que, dans ce régime, il est possible d'atteindre 6dB de réduction du bruit sous la limite quantique standard pour la différence d'intensité et la somme des phases.
La figure \ref{fig4} présente, pour les mêmes paramètres que la figure \ref{fig3} b), l'inséparabilité en fonction de la fréquence d'analyse.
On peut voir que l'inséparabilité est inférieure à 1 jusqu'à 3 MHz environ.
Cela démontre la présence d'intrication entre les champs sonde et conjugué.
Ce résultat ouvre la perspective de la génération d'états intriqués en variables continues à l'aide du mélange à 4 ondes dans un nuage d'atomes froids.
Ce résultat est donc complémentaire des travaux du groupe de Harris dans le régime des variables discrètes, sur la génération de paires de photons dans un milieu constitué d'atomes  de $^{87}$Rb froids \cite{Kolchin:2006p9848}.
\subsection{Contribution des forces de Langevin}\label{Lange}
\begin{figure}
	\centering
	\includegraphics[width=9.5cm]{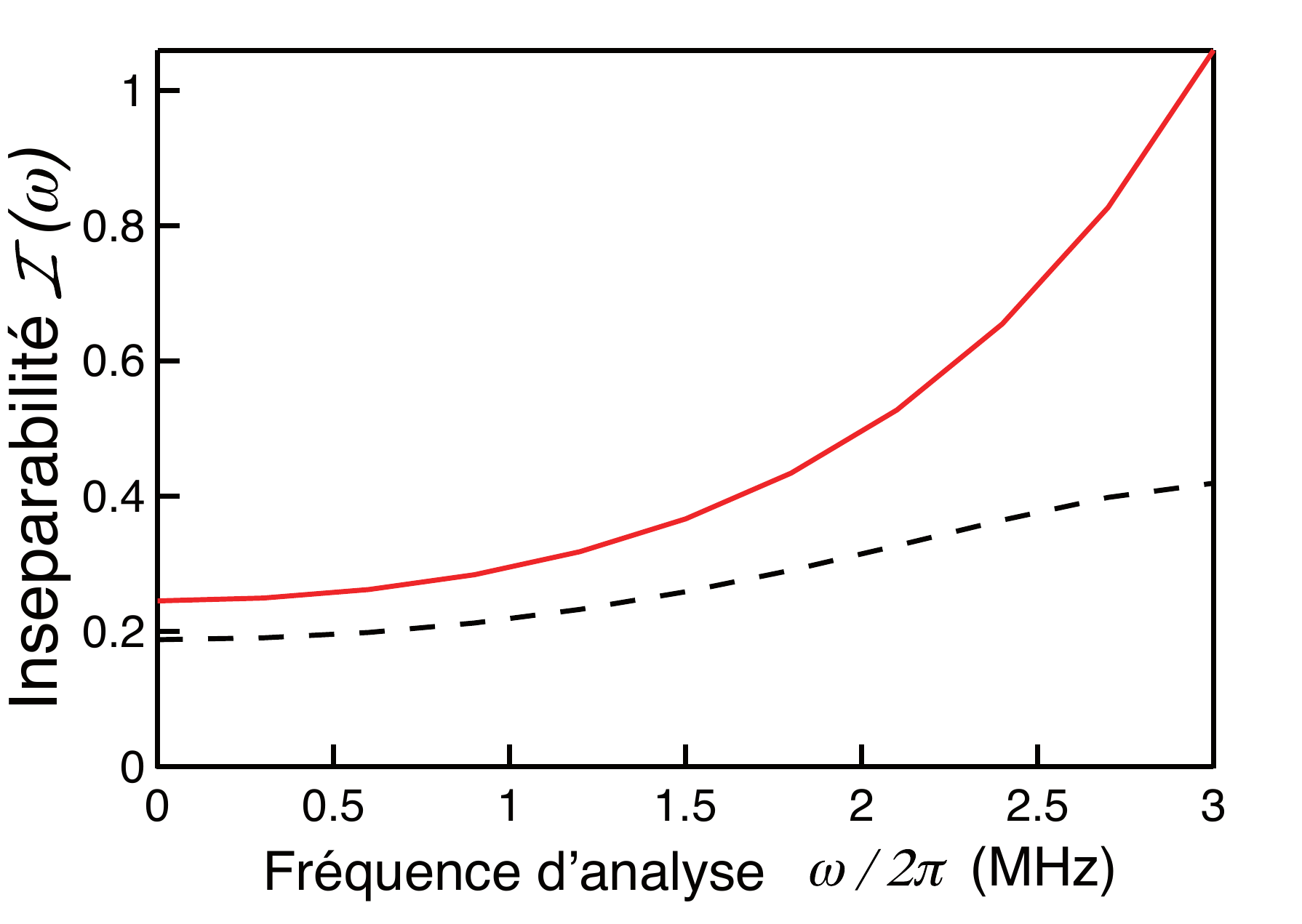} 
	\caption[Inséparabilité en fonction de la fréquence d'analyse]	{ Inséparabilité en fonction de la fréquence d'analyse. La courbe rouge correspond aux simulations numériques incluant les forces de Langevin, la courbe noire pointillés a été tracée en les négligeant. Une valeur de l'inséparabilité inférieure à 1 démontre l'intrication des champs $\hata$ et $\hat b$.
	Paramètres :  $\Omega/2\pi=\Delta/2\pi=2$~GHz et $\delta/2\pi=$-217 MHz.}
		\label{fig4}
	\end{figure} 

Nous avons vu dans les équations \eqref{SNa3} et \eqref{Sphia4} que les termes provenant des forces de Langevin introduisent un bruit supplémentaire et ne peuvent pas améliorer les corrélations.
Sur la figure \ref{fig4}, il est possible d'évaluer la contribution de ces termes en comparant les spectres d'inséparabilité obtenus en leur présence ou en leur absence.
Comme leur effet est relativement faible à basse fréquence ($\simeq 20\%$), on pourrait donc être tenté de les négliger.
Or, en les négligeant, le spectre de bruit sur la sonde seule fait apparaître des solutions non physiques.
En effet, le bruit sur l'intensité et sur la phase du mode $\hata$ peuvent passer sous la limite quantique standard simultanément, ce qui viole l'inégalité d'Heisenberg.
Au contraire, lorsque les forces de Langevin sont prises en compte, le bruit individuel en intensité sur chacun des faisceau est au dessus de la limite quantique standard comme on l'attend pour une configuration d'amplification paramétrique insensible à la phase.\\
En conclusion, même si la contribution des forces de Langevin sur les spectres de bruit de la différence d'intensité est faible, on calculera tout de même toujours ces termes pour vérifier que les solutions pour le bruit individuel sur chacun des deux modes sont physiques.
\begin{figure}[]
\centering
\includegraphics[width=10cm]{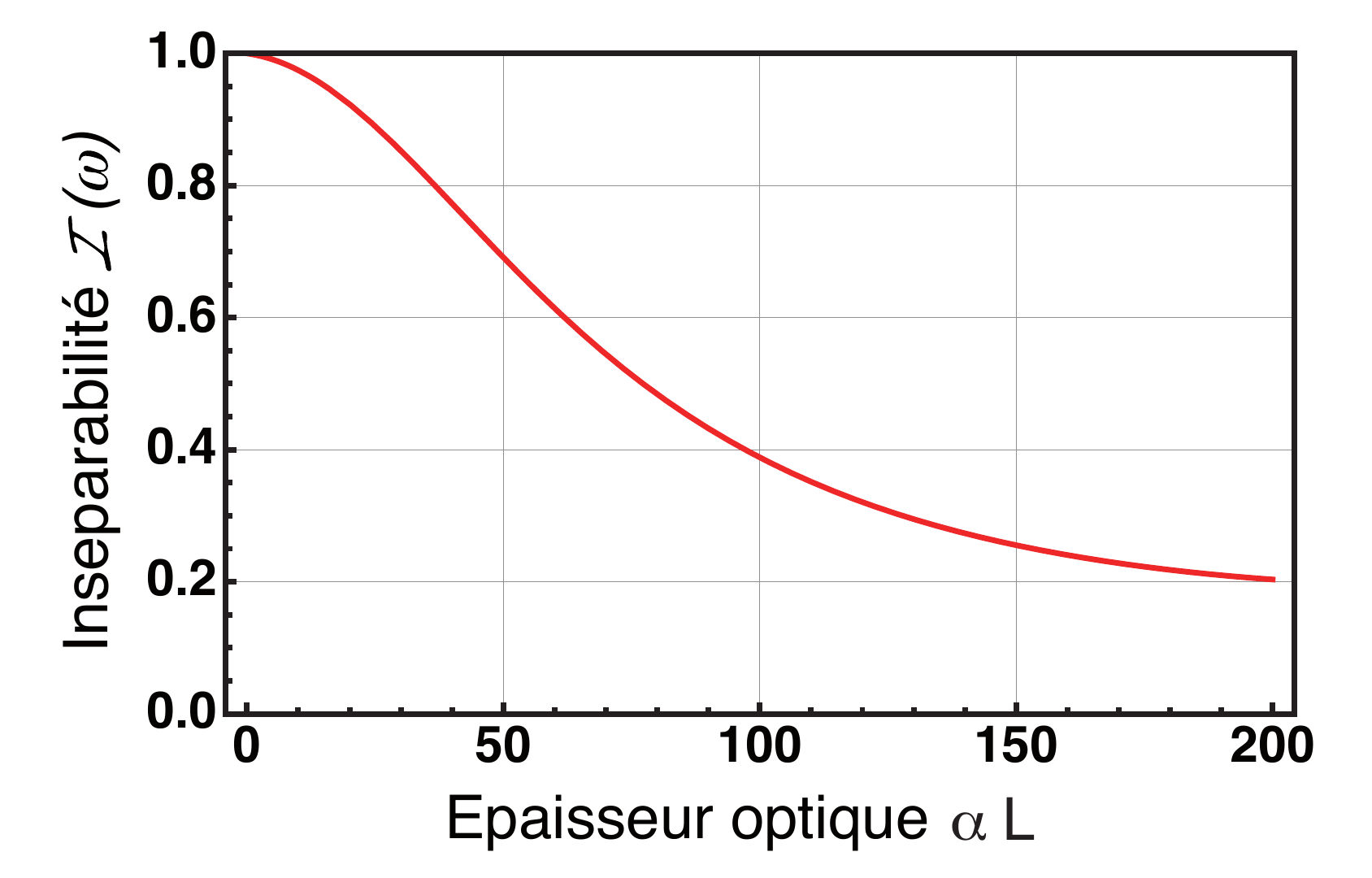} 
\caption[Inséparabilité en fonction de l'épaisseur optique]	{ Inséparabilité en fonction de l'épaisseur optique.
		Paramètres :  $\Omega/2\pi=\Delta/2\pi=2$~GHz et $\delta/2\pi=$-217 MHz.}
	\label{fig412}
\end{figure}

\subsection{Epaisseur optique}

Dans les simulations numériques que nous avons présentées jusqu'à maintenant, la valeur de l'épaisseur optique était fixée à 150.
Même s'il est envisageable d'atteindre prochainement de telles valeurs dans un nuage d'atomes froids, il peut être plus simple expérimentalement de travailler avec une épaisseur optique plus faible et il est intéressant d'évaluer la dépendance du niveau des corrélations en fonction de ce paramètre. 
Ainsi, la figure \ref{fig412} présente l'effet de l'épaisseur optique sur l'inséparabilité et démontre que l'intrication peut tout de même  être observée, bien qu'à des niveaux moins importants, pour des épaisseurs optiques plus faibles.
On peut voir que, même si elle tend vers 1, l'inséparabilité reste inférieure à cette borne pour l'épaisseur optique tendant vers zéro.

\subsection{Effet de la décohérence}	
\begin{figure}
					\centering
					\includegraphics[width=9.3cm]{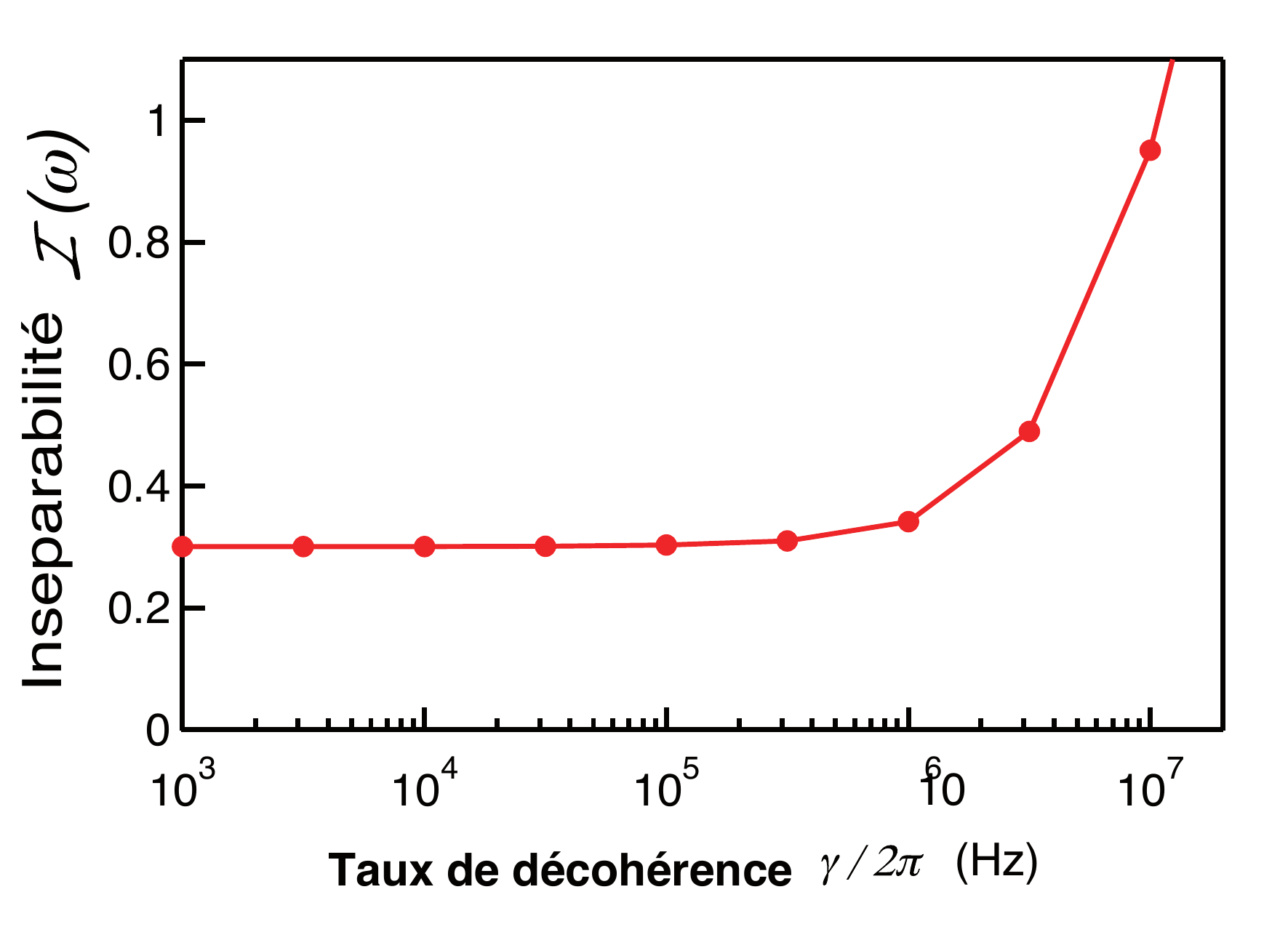} 
					\caption[Inséparabilité en fonction du taux de décohérence $\gamma$]	{ Inséparabilité à 1MHz en fonction du taux de décohérence $\gamma$.
							Paramètres :  $\Omega/2\pi=\Delta/2\pi=2$~GHz et $\delta/2\pi=$-217 MHz.}
						\label{fig5}
					\end{figure}
L'intrication est sensible aux processus de décohérence \cite{Haroche:1998p16411,Andre:2002p9491,Yu:2002p16430,Carvalho:2004p16423,Laurat:2007p11567}.
Sur la figure \ref{fig5}, nous présentons l'effet du taux de relaxation $\gamma$ (décohérence entre les niveaux $|1\ket$et $|2\ket$).
On peut constater que le système que nous avons présenté est robuste vis à vis de la décohérence jusqu'à des valeurs de $\gamma\simeq 10$ MHz.
Pour des valeurs plus élevées, l'inséparabilité devient supérieure à 1 et l'on ne peut plus garantir l'intrication.
Ces simulations mettent donc en évidence le rôle crucial que joue le taux de relaxation $\gamma$ sur la production d'états intriqués.
\subsection{Conclusion sur l'intrication en variables continues dans un milieu d'atomes froids}
Nous avons étudié dans cette section les corrélations quantiques produites par mélange à 4 ondes dans milieu constitué d'atomes froids décrit par un modèle microscopique en double-$\Lambda$.
Le principal résultat que nous avons obtenus est la possibilité de  générer des faisceaux intriqués en variables continues dans un tel système.\\
Nous avons ainsi présenté une valeur d'inséparabilité maximale de -6 dB sous la limite quantique standard pour des paramètres expérimentaux réalistes.
Dans un second temps, nous avons mis en avant le rôle joué par les termes de forces de Langevin.
Ainsi, même si la contribution de ces termes est relativement faible sur l'inséparabilité, nous avons vu que négliger ces termes pouvaient conduire à des situations non physiques.
Enfin, nous avons étudié le rôle de deux paramètres expérimentaux pour démontrer la robustesse du processus : l'effet de la décohérence d'une part et de l'épaisseur optique d'autre part.
Pour ces deux paramètres, nous avons montré que dans des situations expérimentales réalistes, il était possible de générer de l'intrication.
Ces résultats ouvrent la perspective de la réalisation d'une expérience de mélange à 4 ondes dans un milieu constitué d'atomes froids afin de générer des états non-classiques du champ.

\newpage
	\section{Extension du modèle microscopique dans le cas d'une vapeur atomique }\label{extension}
Jusqu'ici nous avons présenté le processus de mélange à 4 ondes dans un milieu décrit par un modèle d'atomes en double-$\Lambda$, tous considérés comme immobiles.
Ce cas correspond à un milieu constitué d'atomes froids.
Les expériences récentes \cite{Boyer:2007p1404,McCormick:2007p652,GlorieuxSPIE2010} montrent qu'il est possible d'observer ce processus dans une vapeur atomique ``chaude''.
Certaines hypothèses que nous avons faites dans le cas d'un milieu constitué d'atomes froids ne sont pas valables dans le cas d'une vapeur atomique.
Cette partie du manuscrit vise donc à étendre le modèle que nous venons de présenter à ce type d'expériences.
Nous allons étudier d'une part le rôle de la distribution de vitesse et par conséquent du décalage de fréquence Doppler vu par les atomes lors d'un processus de mélange à 4 ondes et d'autre part l'origine de l'absorption linéaire subie par la sonde proche de résonance observé expérimentalement.

\subsection{Distribution de vitesse}
Dans une vapeur atomique, les différents atomes ne sont pas immobiles 
La probabilité pour un atome d'avoir la norme de son vecteur vitesse égale à $\rm v$ est donnée par une loi de distribution de Maxwell-Boltzmann $\mathcal{P}(\rm v)$ :
\begin{equation}\label{maxboltz}
\mathcal{P}({\rm v})=\sqrt{\frac{m}{2 \pi k_B T}}\ \exp\left[\frac{-m \text{v}^2}{2  k_B T}\right],
\end{equation}
où $m$ est la masse d'un atome et T la température en kelvin.\\
Il s'agit d'une distribution gaussienne de valeur moyenne nulle et d'écart type $\sigma = \sqrt{\frac{k_B T}{m}}$.
On peut en déduire la valeur moyenne de la norme du vecteur vitesse $\rm v_m$  (relation \eqref{rouge})  :
\begin{equation}
\rm v_m=\sqrt{\frac{2k_B T}{m}}.
\end{equation}

\subsection{Etat stationnaire et absorption résiduelle}

\subsubsection{Préparation des atomes}
\begin{figure}
\centering
\includegraphics[width=8.5cm]{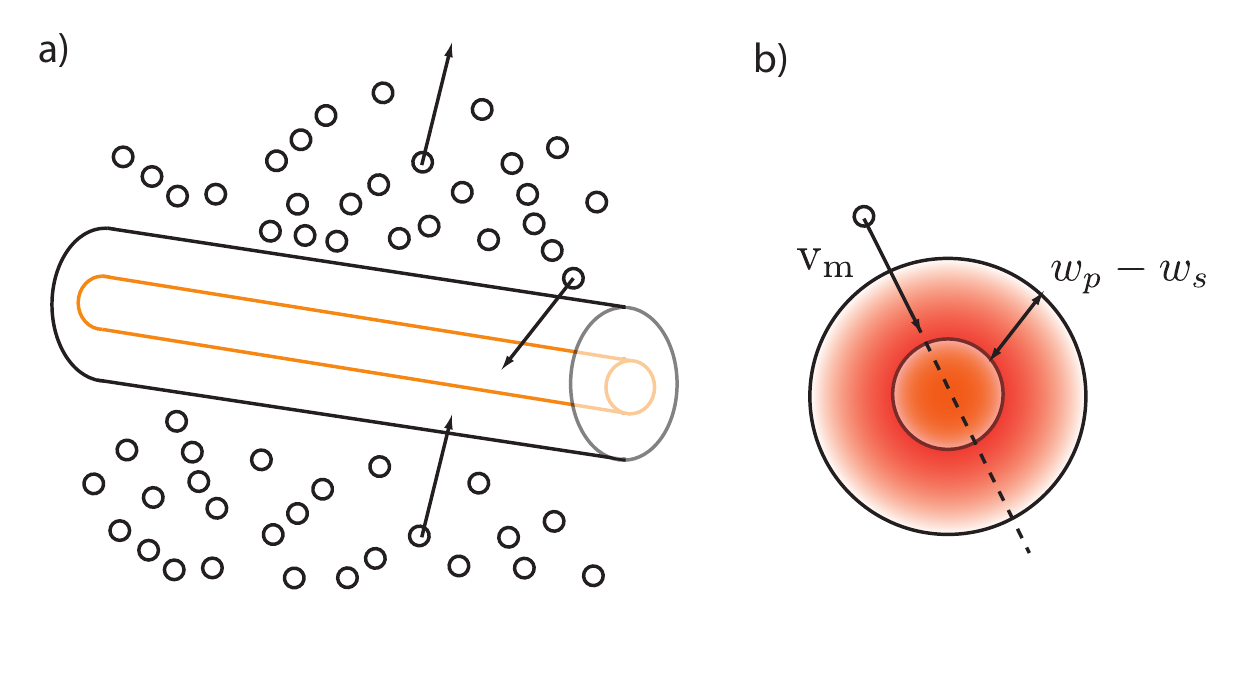} 
\caption[Géométrie des faisceaux et temps de passage des atomes]	{ Géométrie des faisceaux et temps de passage des atomes. La figure b) est une coupe transverse de la figure a). Elle correspond au modèle que nous avons choisi pour définir le temps de passage. }
	\label{tranche}
\end{figure}
\begin{figure}
\centering
\includegraphics[width=15cm]{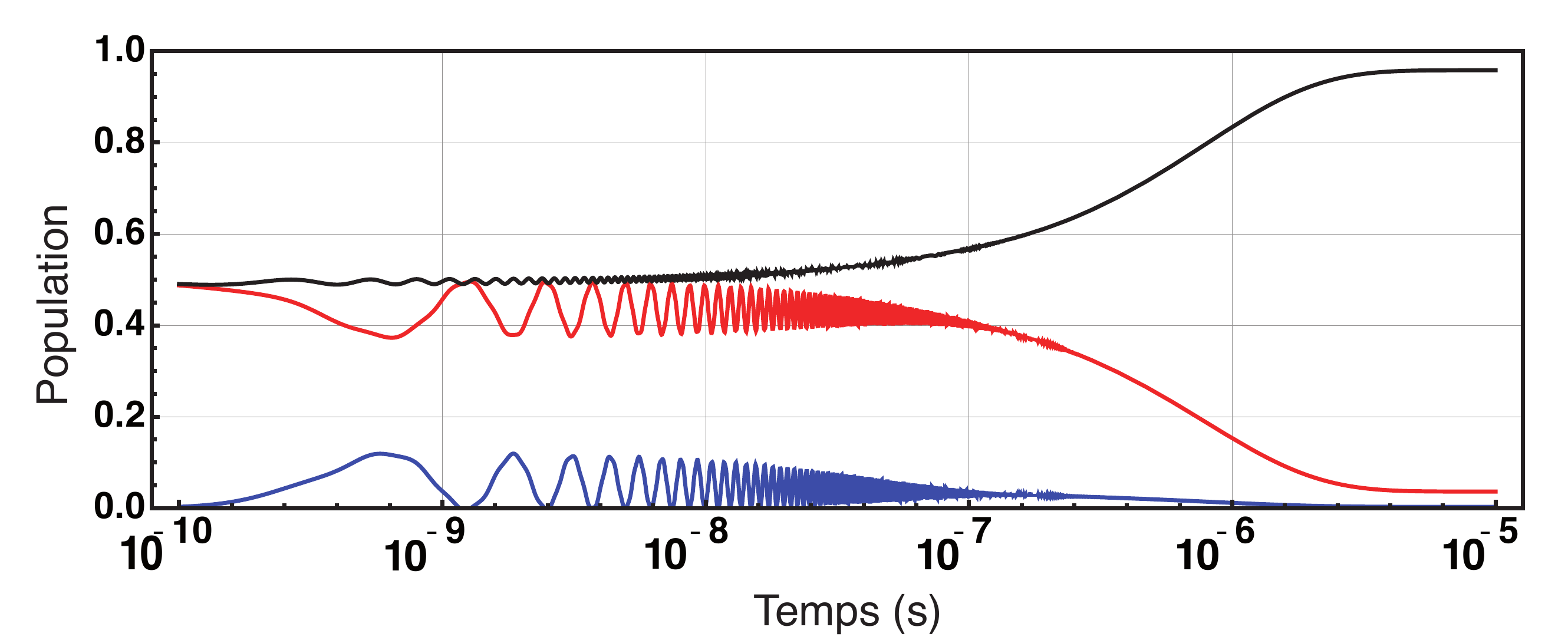} 
\caption[Evolution des populations en fonction du temps.]	{ Evolution des populations en fonction du temps. En rouge la population dans $|1\ket$, en noir la population dans $|2\ket$ et en bleu la population dans $|3\ket$. Paramètres : $\Delta /2\pi = 700$ MHz,  $\Omega /2\pi = 300$ MHz.}
	\label{evolcohe}
\end{figure}

Lors de la résolution du système d'équation de Heisenberg Langevin, nous avons fait l'hypothèse que les populations et les cohérences étaient imposées par la pompe\footnote{Il s'agit des solutions des équations \eqref{eq_evolution_pop_4WM}.}.
Cela revient à dire que $\tilde\sigma_{11}$, $\tilde \sigma_{22}$, $\tilde \sigma_{33}$, $\tilde \sigma_{44}$, $\tilde \sigma_{41}$ ,$\tilde \sigma_{32}$  ont atteint l'état stationnaire.
Pour des atomes considérés comme immobiles, le temps d'interaction entre les atomes et le faisceau pompe est infini.
Les atomes sont donc effectivement préparés  par la pompe dans l'état stationnaire décrit en \eqref{solS0}.\\
A l'inverse, si nous prenons en compte la distribution de vitesse, il n'est pas évident, a priori, de supposer  l'état stationnaire atteint.\\
On définit le temps $\tau$  par :
\begin{equation}
\tau = \frac{w_p-w_s}{\rm v_m},
\end{equation}
où l'on a introduit le waist (rayon à $1/e^2$) de la pompe $w_p$ et de la sonde $w_s$.
Pour des valeurs typique de $w_p=600$ microns et $w_s=300$ microns, on a : $\tau\simeq 1\ \mu$s pour une vapeur de $^{85}$Rb à 120$^\circ$C.
Ce temps correspond au temps de passage dans la géométrie décrite sur la figure \ref{tranche}, c'est-a-dire, des atomes à la vitesse moyenne ${\rm v_m}$ qui traversent le faisceau dans le plan normal à la direction de propagation en passant par centre du faisceau. 
Pour cette classe de vitesse, $\tau$ est donc une borne inférieure du temps de passage d'un atome dans le faisceau pompe avant de pouvoir interagir avec la sonde.\\
L'équation \eqref{415} permet de déterminer le temps typique d'évolution des populations et des cohérences $\sigma_{11},\tilde \sigma_{22},\tilde \sigma_{33},\tilde \sigma_{44},\tilde \sigma_{41},\tilde \sigma_{32}$.
En effet, cette équation montre que la dynamique du vecteur $|\Sigma_0]$ est pilotée par un terme en $e^{i[M_0]t}$.
Afin de déterminer la dynamique du système, on peut utiliser l'analyse linéaire de stabilité \cite{Haken:1983p18249}. Le système converge si et seulement si la partie imaginaire des valeurs propres de $[M_0]$ est négative.
Nous l'avons vérifié numériquement pour la gamme de paramètres correspondant aux expériences dans une vapeur atomique.
On peut voir sur la figure \ref{evolcohe}, l'évolution des populations en fonction du temps.
Dans l'état initial, la population est équi--répartie entre les états $|1\ket$ et $|2\ket$. 
Aux temps courts, on observe des oscillations puis les populations tendent vers l'état stationnaire.
\\
De plus, le temps caractéristique d'évolution d'un vecteur propre de  $M_0$ est donné par l'inverse de la partie réelle de la valeur propre associée.
Ainsi on déterminera pour un atome, la probabilité  $P(t)$ d'avoir atteint l'état stationnaire à un temps t  à partir de la relation suivante :
\begin{equation}\label{prepara}
P(t)=1-e^{-t/T_0},
\end{equation}		
où l'on a introduit $T_0$  qui est l'inverse de la plus petite (en valeur absolue) des parties réelles des valeurs propres de $M_0$, c'est-à-dire le temps caractéristique d'évolution le plus long.
\begin{figure}
\centering
\includegraphics[width=15cm]{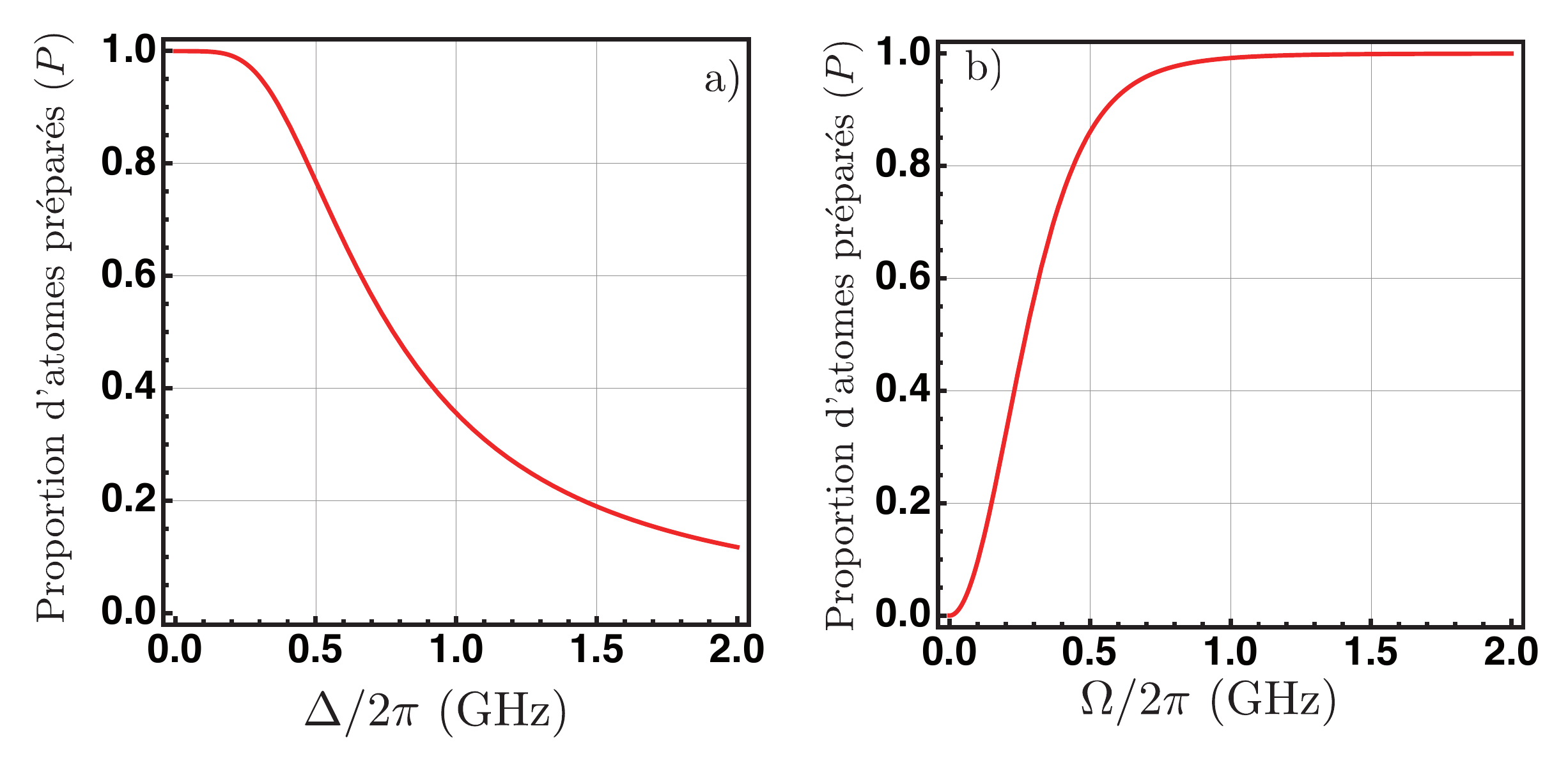} 
\caption[Proportion d'atomes préparés par la pompe.]	{Proportion $P$ d'atomes préparés par la pompe en présence d'élargissement Doppler. La figure a) présente l'effet de désaccord $\Delta$ pour $\Omega /2\pi = 300$ MHz. La figure b) présente l'effet de $\Omega$ pour $\Delta /2\pi = 700$ MHz. Paramètres :$w_p=600$ microns, $w_s=300$ microns, $\Gamma=36$ MHz, $\tau=1\mu$s .}
\label{prepa}
\end{figure} 
Ainsi, en appliquant la relation précédente au temps $t=\tau$, on obtient la proportion ``moyenne'' $P$ d'atomes préparés dans l'état stationnaire.
La figure \ref{prepa} présente des simulations numériques de $P$ pour différents paramètres expérimentaux sur la raie D1 du rubidium 85.
Sur la figure a) on peut voir qu'à pulsation de Rabi fixée, le taux de préparation augmente lorsque l'on se rapproche de résonance.
Au delà de quelques centaines de MHz de désaccord, le taux de préparation chute brutalement pour tendre vers 0 à désaccord très important.
De façon symétrique pour un désaccord de 700~MHz, on peut voir que la pulsation de Rabi doit être supérieure à 1~GHz pour préparer 95$\%$ des atomes dans l'état stationnaire que nous avons décrit précédemment.
Ceci se comprend par le fait que pour compenser le désaccord du laser de pompe par rapport à la transition atomique, il est nécessaire d'utiliser des pulsations de Rabi de l'ordre ou plus grande que ce désaccord.

\subsubsection{Pertes par absorption linéaire}
La conséquence de cette analyse est que des atomes (dans le cas d'une vapeur atomique) peuvent interagir avec la sonde sans avoir atteint l'état stationnaire, c'est-à-dire avant d'avoir été ``préparés'' dans cet état par la pompe.
Ces atomes ne vont donc pas suivre l'évolution décrite par les relations \eqref{evol_coherence}.
En première approximation, ces atomes vont être supposés dans l'état initial en l'absence de pompe.
En particulier, ils vont donc se comporter comme un milieu présentant une absorption linéaire pour les faisceaux sonde et conjugué.\\
Nous faisons ici une hypothèse forte, qui consiste à séparer les atomes en deux catégories, d'un coté ceux qui ont atteint l'état stationnaire et de l'autre ceux qui sont dans l'état initial.
Une partie des atomes (ceux qui sont préparés dans l'état stationnaire) va ainsi contribuer au processus de mélange à 4 ondes, tandis que l'autre partie sera modélisée par un terme d'absorption proportionnel à l'exponentielle de la proportion d'atomes non préparés.
Pour simplifier les calculs nous n'utiliserons pas le modèle des pertes réparties au cours de la propagation, comme nous l'avons fait au chapitre 3, mais nous ferons l'hypothèse que l'ensemble des pertes a lieu à l'entrée du milieu.
Cette simplification, ne se justifie qu'\textit{a posteriori} par le bon accord que l'on obtient entre notre modèle et les données expérimentales.
%
%
%

\subsection{Effet de l'élargissement inhomogène}
Le second effet que nous allons étudier et celui de l'élargissement inhomogène de la transition.
La dispersion de vitesse introduit un décalage en fréquence par effet Doppler pour les atomes à vitesse non nulle sur le désaccord à un photon $\Delta$.
On observe ainsi une dispersion sur la valeur de $\Delta$ et il est nécessaire d'adapter notre modèle à cet effet.

\subsubsection{Position du problème}
A l'aide de (\ref{def_alphaL}) et en négligeant la contribution des forces de Langevin pour simplifier les notations, on peut écrire l'équation (\ref{eq_evol_2}) sous la forme :
\begin{deqn}\label{eq_evol_chaud}
\frac{\partial}{\partial z}|\hat{A}(z,\omega)]= \mathcal{N} \sigma  [N(\omega)]|\hat{A}(z,\omega)],
\end{deqn}
où 
\begin{ddeqn}
[N(\omega)]=i\ \frac{\Gamma}{4} [T][M_1'(\omega)]^{-1}[S_1],
\end{ddeqn} et  $\mathcal{N}$ est la densité d'atomes et $\sigma $ la section efficace de la transition.\\
Lorsque l'on prend la solution de cette équation différentielle de propagation (entre 0 et $L$), il apparait un préfacteur de l'exposant de l'exponentielle en  $\alpha L=\mathcal{N}\sigma L$.\\
Dans le cas d'un milieu constitué d'atomes froids, $\mathcal{N}$ correspond exactement à la densité d'atomes dans le milieu.
Pour une vapeur chaude, la dispersion en vitesse va induire une dispersion en désaccord à un photon.
Or la matrice $[M_1'(\omega)]^{-1}$ dépend de $\Delta$.
Ainsi tous les atomes ne contribuent pas de la même façon au processus.
Le problème que nous allons résoudre est de déterminer comment les différentes classes de vitesses contribuent à la valeur du gain et à la réduction du bruit sur la différence des intensité des deux champs.
\subsubsection{Modèle}
Pour prendre en compte la contribution des différentes classes de vitesse lors de la propagation nous allons adopter le modèle suivant.
Le milieu $0\rightarrow L$ est décomposé en tranches d'épaisseur $dz$.
Dans chaque tranche on suppose qu'il n'y a des atomes que d'une seule classe de vitesse $\rm v(z)$, telle que le désaccord à un photon vu par les atomes dans cette tranche soit : $\Delta_z$.
La probabilité pour une tranche donnée d'être composée d'atomes de vitesse $\rm v$ est donnée par la loi de distribution de vitesse dans une vapeur.\\
Pour une tranche $dz$, l'équation (\ref{eq_evol_chaud}) se résout :
\begin{deqn}
|\hat{A}(dz)]=e^{[N(\omega,\Delta_0)].\mathcal{N}\sigma dz} |\hat{A}(0)],
\end{deqn}
et donc 
\begin{ddeqn}
|\hat{A}(z+dz)]=e^{[N(\omega,\Delta_z)].\mathcal{N}\sigma dz} |\hat{A}(z)],
\end{ddeqn}
avec $[N(\omega,\Delta_z)]$ la valeur de la matrice $[N]$ pour des atomes de vitesse $\rm v(z)$, c'est-à-dire un désaccord $\Delta_z$.
\subsubsection{Somme}
Nous devons sommer les contributions des différentes tranches pour obtenir le vecteur $|\hat{A}(L)]$ en sortie du milieu.
On peut donc écrire pour $T$ tranches :
\begin{equation}\label{496a}
|\hat{A}(L)]=  |\hat{A}(0)]\prod_{j=0}^{T-1}e^{T\mathcal{N}\sigma dz[N(\omega,\Delta_{jdz})]}.
\end{equation}
De manière générale, les matrices $[N(\omega,\Delta_{jdz})]$ et $[N(\omega,\Delta_{kdz})]$ pour $j\neq k$ ne commutent pas et le produit des exponentielles de matrice ne pourra pas s'écrire comme l'exponentielle de la somme des matrices.\\
On pourrait se demander si le résultat  dépend du choix l'ordre des tranches.
Nous avons donc simulé numériquement une distribution aléatoire de vitesse suivant la loi \eqref{maxboltz} afin de calculer $|\hat{A}(L)]$ pour $10^5$ tranches ce qui est suffisant pour avoir une distribution très proche d'une loi normale.
Nous avons comparé le résultat obtenu à celui que l'on obtient en supposant que les $[N(\omega,\Delta_{jdz})]$ et $[N(\omega,\Delta_{kdz})]$ commutent, c'est-à-dire :
\begin{equation}\label{496}
|\hat{A}(L)]_2=  |\hat{A}(0)]\ \exp\sum_{j=0}^{T-1}T\mathcal{N}\sigma dz[N(\omega,\Delta_{jdz})].
\end{equation}
Dans la gamme de paramètres étudiée, la différence entre $|\hat{A}(L)]$ et $|\hat{A}(L)]_2$ est inférieur à $10^{-4}$ sur chacune des composantes.

Nous venons donc de vérifier numériquement que l'ordre des tranches n'influence pas le résultat final.
On pourra donc choisir arbitrairement la valeur du désaccord dans une tranche  (en respectant la distribution générale des vitesses), sans perdre en généralité.
Etant donné que les relations \eqref{496a} et \eqref{496} donnent des résultats très proches, nous utiliserons donc  la relation \eqref{496} plus simple à manipuler numériquement. 
En passant à la limite de $T\rightarrow\infty$ en transformant la somme de l'équation \eqref{496} en une intégrale sur les classes de vitesses on obtient alors :
\begin{equation}\label{vitesse}
|\hat{A}(L)]=  |\hat{A}(0)]\exp\int_{-\infty}^{\infty} \mathcal{N}\sigma L\ \mathcal{P}(\rm v)\left[N\left(\omega,\Delta_{\rm v}\right)\right]d \rm v ,
\end{equation} 
où l'on a remplacé le désaccord d'une tranche $\Delta_{jdz}$ par le désaccord des atomes dans une classe de vitesse $\Delta_{\rm v}$, dont la distribution normalisée est donnée par $\mathcal{P}(v)$ définie à l'équation \eqref{maxboltz}.
On peut voir l'équation \eqref{vitesse}, comme la moyenne pondérée par la distribution de vitesse de la matrice $[N]$.
Ainsi, ce résultat est consistant avec l'approche de \cite{CohenTannoudji:1996p4732} pour l'établissement de l'équation pilote d'un petit système couplé à un réservoir.\\
Pour étudier l'effet de la dispersion de vitesse, il faut désormais comparer l'équation \eqref{vitesse} à l'équation  \ref{sol_propa}  que nous avons utilisée pour décrire un milieu constitué d'atomes froids.
Dans cette étude nous avons négligé l'effet des forces de Langevin, en vérifiant numériquement que les coefficients de diffusion sont très peu modifiés par la distribution de vitesse.
Les effets étant indépendants on pourra ajouter in fine, les deux contributions des forces de Langevin et de la distribution de vitesse.

\subsubsection{Intégration numérique de la distribution de vitesses}
\begin{figure}
\centering
\includegraphics[width=7cm]{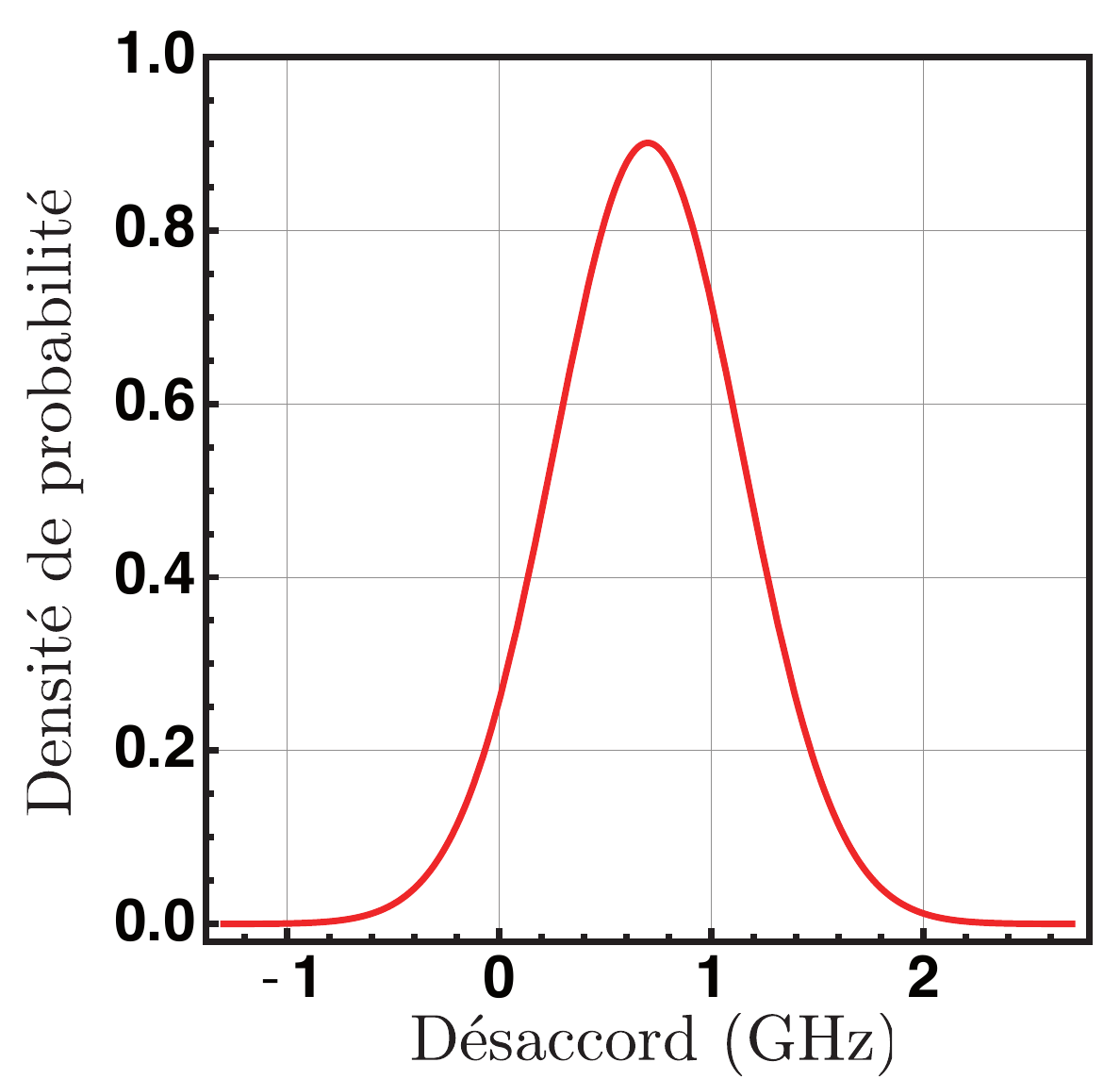} 
\caption[Distribution du désaccord réel vu par les atomes en présence d'élargissement Doppler]	{Distribution du désaccord réel vu par les atomes en présence d'élargissement Doppler pour un désaccord du laser de 700 MHz pour une vapeur atomique à 120$^\circ$C.	}
	\label{probaDelta}
\end{figure} 
On intègre numériquement la relation (\ref{vitesse}) pour obtenir les contributions de toutes les classes de vitesses au processus.
Les deux grandeurs que nous allons alors étudier sont le gain sur le mode $\hata$ et les fluctuations de la différence d'intensité.\\
Pour les températures de l'ordre de 100$^\circ$C, une valeur de $\mathcal{N}\sigma L\simeq 4500$ est typique\footnote{Il s'agit par exemple d'une cellule de 3cm chauffée à $120^\circ C$.}.
A cette température, la distribution du désaccord vu par les atomes en présence d'élargissement Doppler pour un désaccord du laser de 700 MHz est donné dans la figure \ref{probaDelta}.
Dans ces conditions, nous regardons l'effet de dispersion de ce désaccord sur le gain du faisceau sonde.

\begin{figure}
\centering
\includegraphics[width=14.5cm]{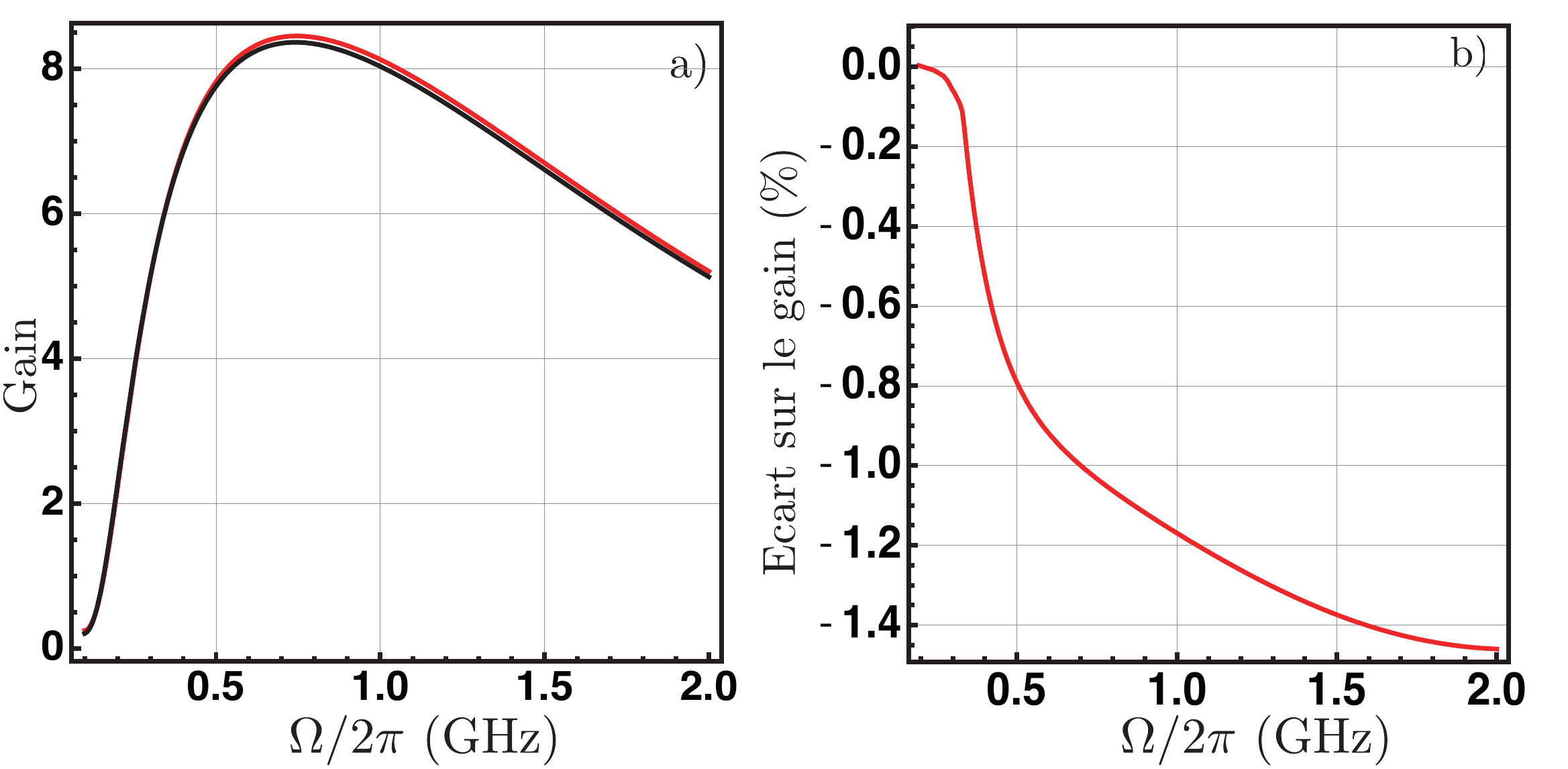} 
\caption[Effet de la distribution de vitesses sur le gain de la sonde]	{Effet de la distribution de vitesses sur le gain de la sonde pour une température de $120^\circ$C. a) la courbe rouge donne le gain pour le cas où tous les atomes sont à vitesse nulle et voient un désaccord de 700 MHz, la courbe noire donne le gain pour la situation où les atomes suivent une distribution de vitesses gaussienne.  b) Ecart en pourcentage entre les deux courbes en a).\\ Paramètres : $\gamma/2\pi =1$~MHz, $\delta/2\pi = 4$ MHz, $\alpha L=4500$	\label{ecart}}.

\includegraphics[width=14.5cm]{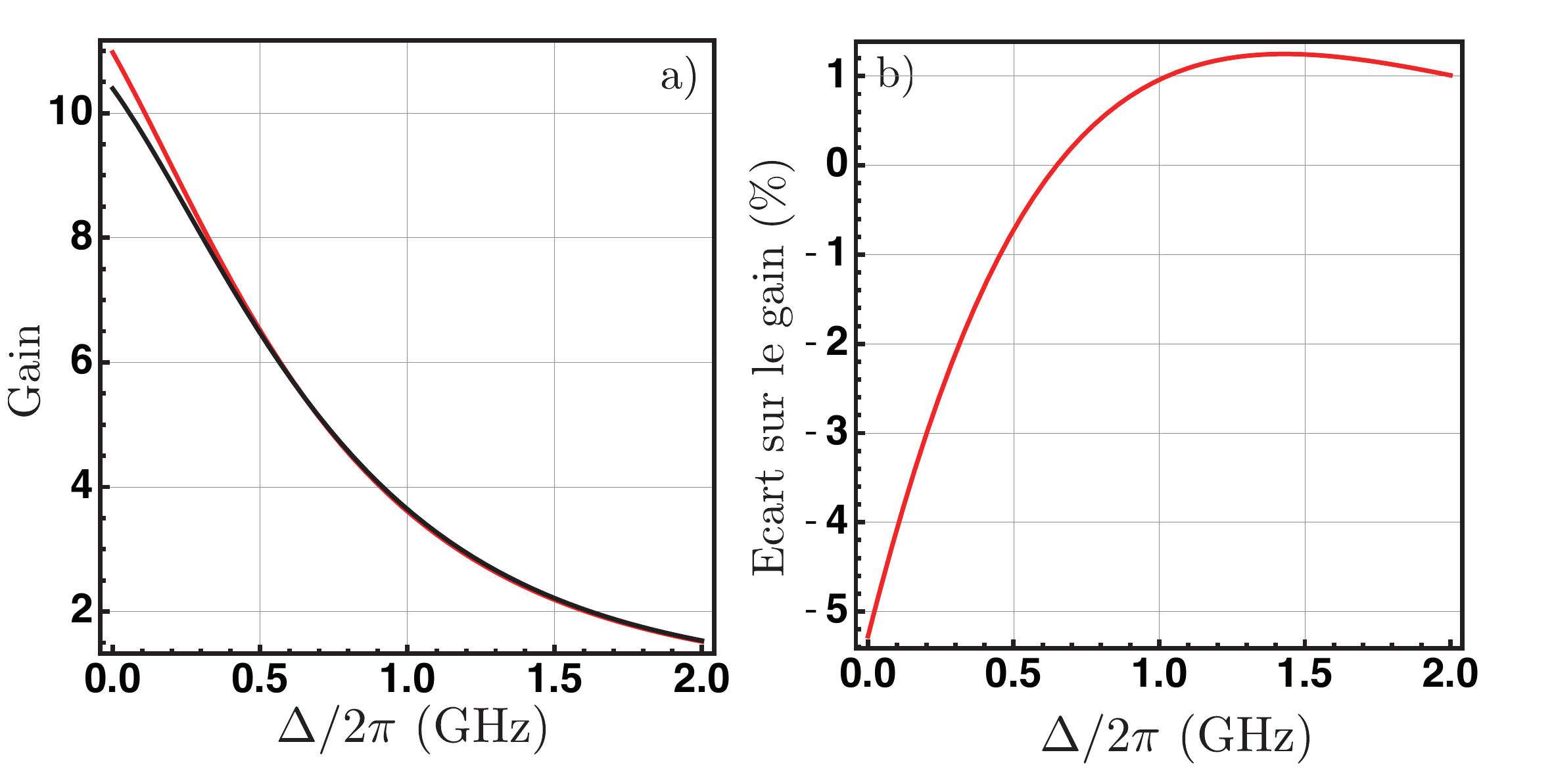} 
\caption[Effet de la distribution de vitesses sur le gain de la sonde]	{Effet de la distribution de vitesses sur le gain de la sonde pour une température de $120^\circ$C. a) la courbe rouge donne le gain pour le cas où tous les atomes sont à vitesse nulle et voient le désaccord $\Delta$ MHz, la courbe noire donne le gain pour la situation où les atomes suivent une distribution de vitesses gaussienne pour un laser fixé avec un désaccord $\Delta$. b) Ecart en pourcentage entre les deux courbes précédentes.\\ Paramètres : $\gamma/2\pi =1$~MHz, $\delta/2\pi = 4$ MHz, $\alpha L=4500$, $\Omega/2\pi=$ 0.3 GHz	\label{ecartdelta}}.
\end{figure}

On trace sur la figure \ref{ecart}, en fonction de la pulsation de Rabi du faisceau pompe, l'écart entre le gain dans le cas où tous les atomes sont à vitesse nulle et lorsque les atomes suivent une distribution de vitesse gaussienne 
De même dans la figure \ref{ecartdelta} en fonction du désaccord à un photon des atomes à vitesse nulle.
On peut noter que, cet écart est  faible (inférieur à 5 $\%$) dans la gamme de paramètres étudiée.
Sur la figure \ref{ecartdelta}, on voit que pour des désaccords faibles, l'écart augmente.
Ainsi la prise en compte de la dispersion de vitesse des atomes réduit légèrement la valeur du gain.
On peut comprendre cet effet car le gain étant plus important pour les atomes à résonance, la distribution de vitesses diminue nécessairement la valeur du gain moyenné sur la distribution de vitesse.
De manière symétrique loin de résonance, la distribution de vitesses induit une légère augmentation du gain car certains atomes vont voir des désaccord plus faibles que s'ils étaient tous froids.
De plus autour de 700 MHz, la courbe de gain de la figure \ref{ecart} peut être approximée par une droite.
Ainsi la contribution d'un atome dans la classe de vitesse $-\rm v$ sera exactement compensée par le contribution d'un atome dans la classe de vitesse $+\rm v$.
C'est la zone où l'écart entre les deux modèles s'annule.\\

\begin{figure}
\centering
\includegraphics[width=14.5cm]{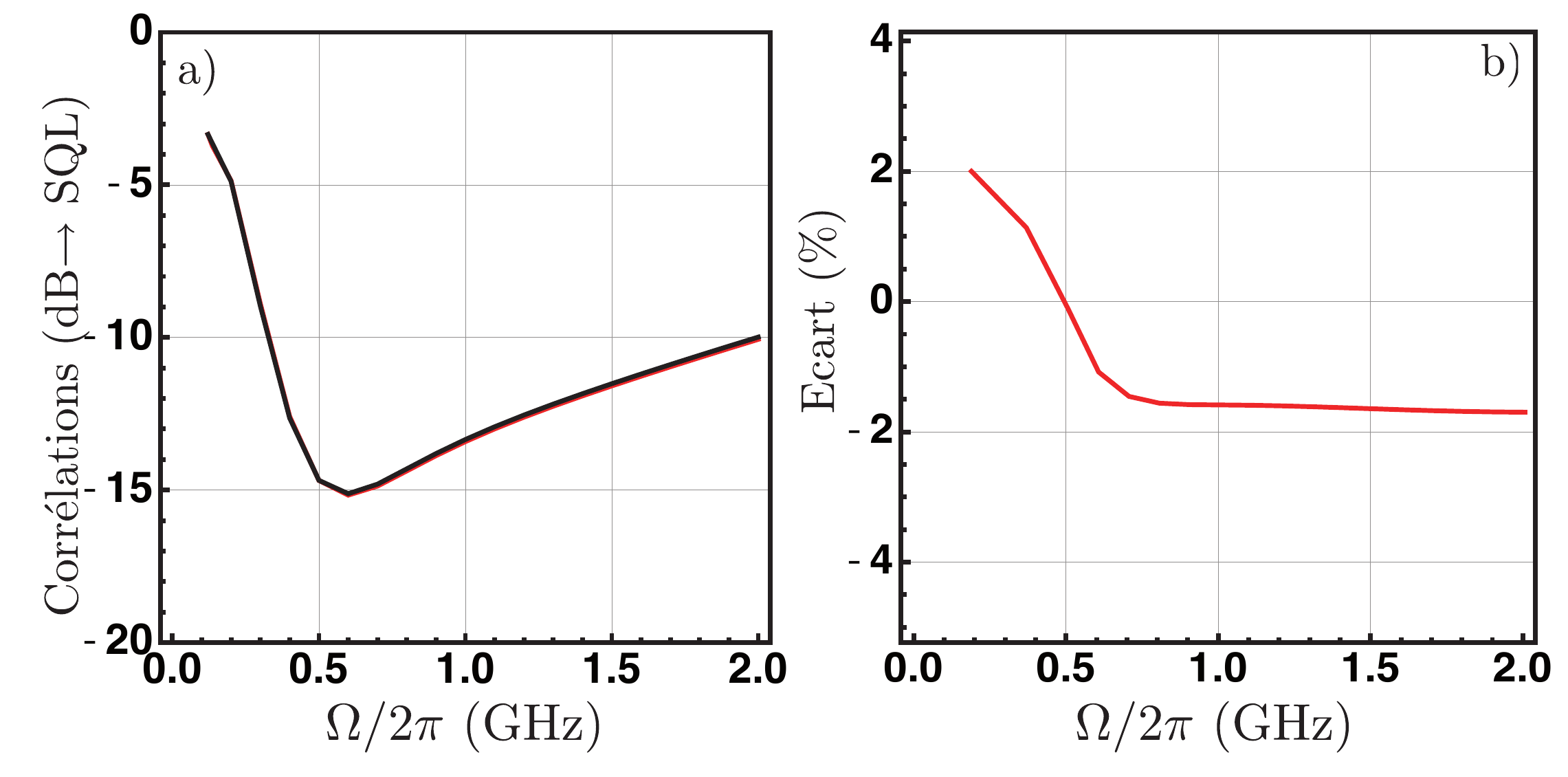} 
\caption[Effet de la distribution de vitesses sur les corrélations en intensité.]	{Effet de la distribution de vitesses sur les corrélations en intensité pour une température de $120^\circ$C. a) la courbe rouge donne les corrélations par rapport à la limite quantique standard  pour le cas où tous les atomes sont à vitesse nulle et voient un désaccord de 700 MHz, la courbe noire donne les corrélations pour la situation où les atomes suivent une distribution de vitesses gaussienne. b) Ecart en pourcentage entre les corrélations dans ces deux cas (échelle linéaire).\\ Paramètres : $\gamma/2\pi =1$~MHz, $\delta/2\pi = 4$ MHz, $\alpha L=4500$	\label{cor_Omega_effet}}.

\includegraphics[width=14.5cm]{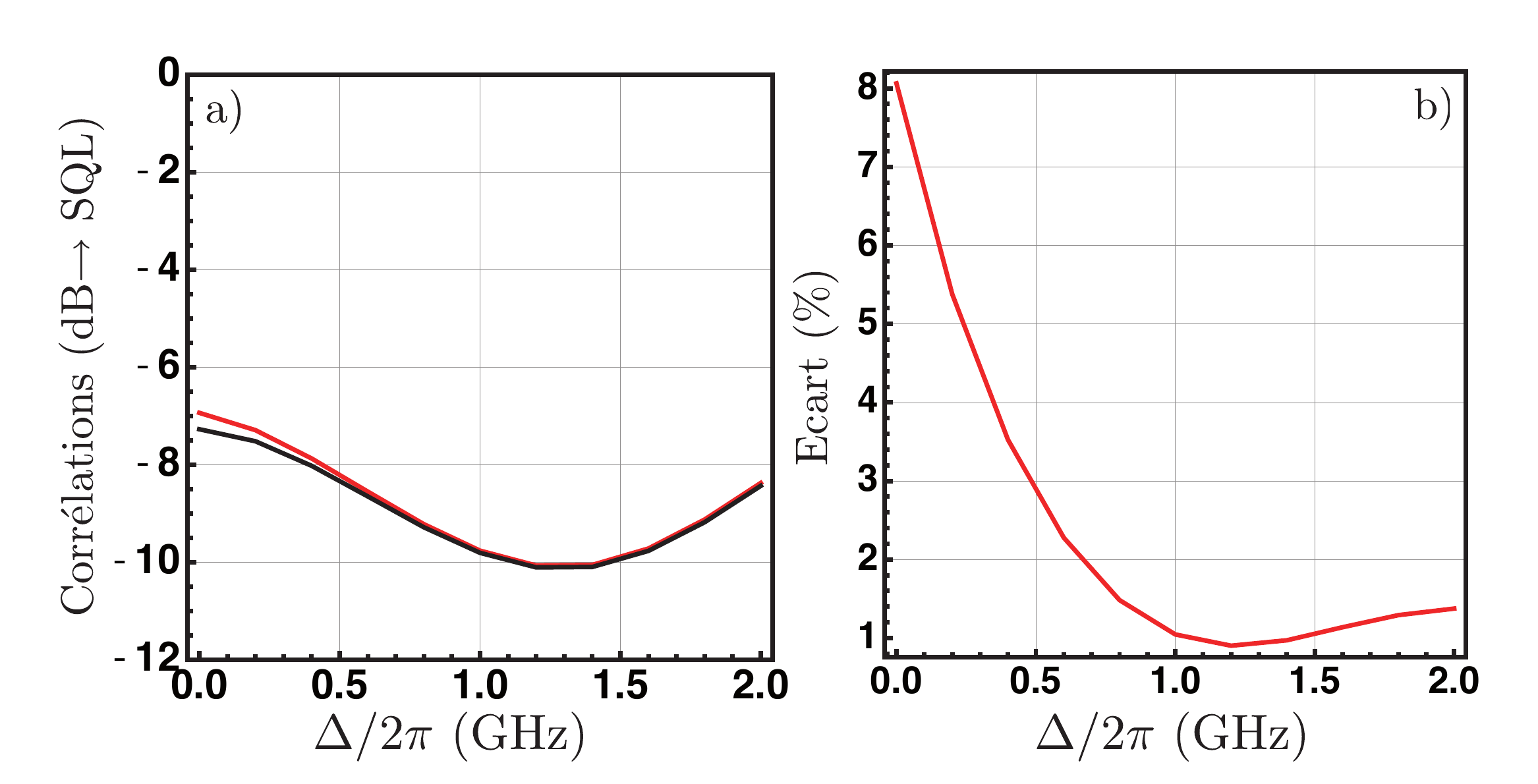} 
\caption[Effet de la distribution de vitesses sur le gain de la sonde]	{Effet de la distribution de vitesses sur  les corrélations en intensité pour une température de $120^\circ$C. a) la courbe rouge donne les corrélations par rapport à la limite quantique standard pour le cas où tous les atomes sont à vitesse nulle et voient le désaccord $\Delta$ MHz, la courbe noire donne les corrélation pour la situation où les atomes suivent une distribution de vitesses gaussienne pour un laser fixé avec un désaccord $\Delta$. b) Ecart en pourcentage entre les corrélations dans ces deux cas (échelle linéaire).\\ Paramètres : $\gamma/2\pi =1$~MHz, $\delta/2\pi = 4$ MHz, $\alpha L=4500$, $\Omega/2\pi=$ 0.3 GHz	\label{cor_Delta_effet}}.
\end{figure}

De la même manière, on étudie l'effet de la distribution de vitesses sur les corrélations d'intensité à 1MHz entre les faisceaux sonde et conjugué pour les deux paramètres de $\Omega$ et de  $\Delta$.
On peut voir sur les figures \ref{cor_Omega_effet} et \ref{cor_Delta_effet}, que cet effet est de nouveau faible (inférieur à 10 $\%$).
On peut constater que le taux de corrélations est légèrement améliorée dans le cas d'une vapeur par rapport à un modèle d'atomes immobiles pour des pulsations de Rabi faible (inférieure à 300 MHz).
L'amélioration sera d'autant plus sensible que l'on se rapprochera de résonance. En effet, comme on peut le constater sur la figure \ref{cor_Delta_effet} a) les corrélations sont plus faibles à $\Delta=0$ avec ces paramètres.
La distribution de vitesse va donc permettre à des atomes voyant un désaccord différent de 0 d'intervenir et d'augmenter les corrélations.\\

De manière générale, nous venons de voir que la distribution de vitesse modifie les équations d'évolution du système mais, qu'après intégration, l'effet sur les grandeurs mesurées expérimentalement (gain, corrélations) est faible.
\subsection{Conclusion sur l'extension du modèle à une vapeur atomique}
Le modèle simplifié que nous avons développé nous a permis d'étudier le rôle de la distribution de vitesse dans les expériences de mélange à 4 ondes.
D'une part nous avons vu, qu'il existe une fraction d'atomes non préparés dans l'état stationnaire par la pompe.
Cette fraction peut devenir significative, lorsque le laser de pompe est loin de résonance ou lorsque sa pulsation de Rabi est faible.
On appellera cet effet \textit{l'absorption résiduelle.}
Dans la suite nous traiterons cette effet par un terme correctif correspondant à des pertes sur la propagation des faisceaux.\\
D'autre part, nous avons étudié l'effet de la dispersion de désaccord vu par les atomes dans le cas d'une vapeur.
Dans cette étude, nous avons vu que cet effet était faible (typiquement inférieur à 10$\%$).
Dans la suite, nous négligerons donc cet effet.
Ainsi nous pourrons utiliser les relations du modèle dit ``atomes froids'' avec les densités (et donc les épaisseurs optiques) que l'on peut atteindre dans une vapeur atomique chaude.
Dans le cas d'une vapeur atomique, la valeur de densité $N$ est bien plus importante que celle obtenue dans un piège magnéto-optique.
On peut atteindre des valeurs de $10^{13}$ atomes par cm$^3$, soit une épaisseur optique de 10000 pour cellule de $^{85}$Rb de 1 cm de longueur.
La relation qui relie la densité d'atome en fonction de la température a été donnée au chapitre 2.
\subsection{Perspectives}
\begin{figure}[]
\centering
\includegraphics[width=10cm]{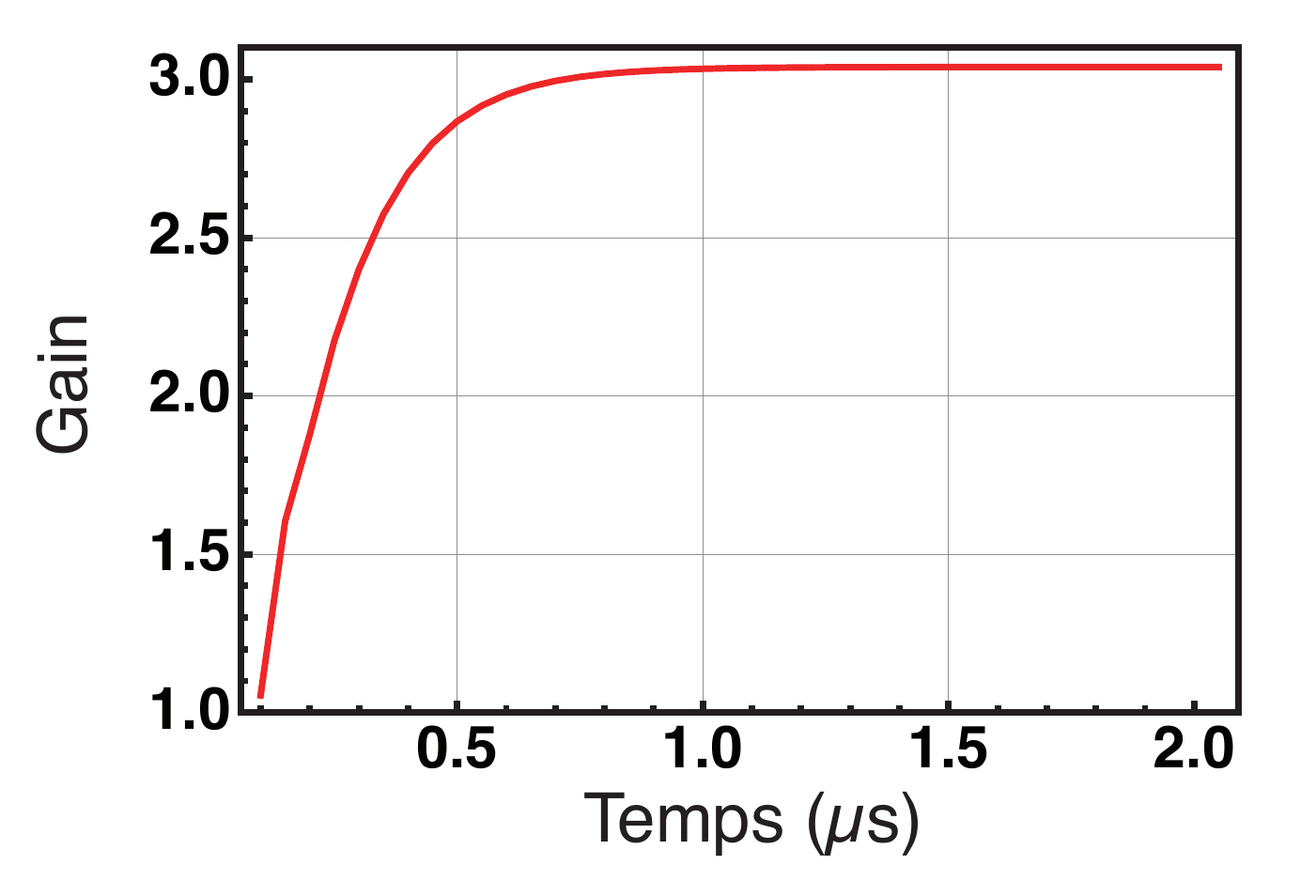} 
\caption[Etablissement du régime stationnaire.]	{Etablissement du régime stationnaire. Gain sur le faisceau sonde en fonction du temps d'interaction préalable des atomes avec le faisceau pompe. Paramètres :  $\gamma/2\pi =1$~MHz, $\delta/2\pi = 0$ MHz, $\Omega/2\pi=$ 0.9 GHz, $\Delta/2\pi=$ 1 GHz.\label{gain_f_t}}
\end{figure}

Une piste que nous n'avons pu pas aborder durant cette thèse est de résoudre le système \eqref{eq_evolution_pop_4WM} sans faire l'hypothèse de l'état stationnaire.
En connaissant la trajectoire des atomes dans les faisceaux pompe et sonde, il est possible d'injecter dans le système \eqref{evol_coherence} l'état de chaque atome lorsqu'il interagit avec le faisceau sonde 
Pour faire ces simulations, une très grande puissance de calcul est nécessaire.\\
En première approximation, il est plus simple d'utiliser, non pas l'état de chaque atome, mais un \textit{état moyen},
que l'on définit comme la solution exacte de l'équation  \eqref{eq_evolution_pop_4WM} au temps $\tau$.
Il correspond aux atomes de vitesse égale à la vitesse moyenne et passant un minimum de temps dans le faisceau pompe avant d'entrer dans le faisceau sonde.
Si l'état stationnaire n'est pas atteint, on peut alors résoudre le système \ref{evol_coherence} en injectant les solutions au temps $\tau$ du système  \eqref{eq_evolution_pop_4WM}.
Les résultats que nous avons obtenu, notamment pour le gain, sont présenté dans la figure \ref{gain_f_t}.
Pour des temps courts le gain tend vers 1 (les atomes sont encore dans l'état initial), puis il augmente vers une valeur stationnaire (ici 3) en quelques centaines de ns.

\section{Corrélations quantiques en variables continues dans une vapeur atomique chaude}
\subsection{Corrélations quantiques sur la transition $5S_{1/2}\to 5P_{1/2}$ du rubidium.}\label{sec:vapchaud}
Comme nous l'avons présenté dans la section précédente, il est possible de faire l'hypothèse que toutes les classes de vitesse d'une vapeur atomique contribuent au processus de mélange à 4 ondes à condition que l'état stationnaire soit atteint.
On va donc pouvoir étudier théoriquement le régime décrit dans \cite{Boyer:2007p1404,McCormick:2007p652,GlorieuxSPIE2010}, c'est-à-dire un régime d'épaisseur optique ($\alpha L= \mathcal{N} \sigma L$) de plusieurs milliers.
Après avoir comparé notre étude théorique aux résultats expérimentaux de \cite{Boyer:2007p1404} pour les grandeurs classiques, nous détaillerons les différents paramètres qui affectent sur la génération de corrélations quantiques en intensité.
De plus nous verrons que notre modèle prédit la génération de faisceaux intriqués dans les conditions expérimentales de \cite{GlorieuxSPIE2010}.
\subsubsection{Gain}
\begin{figure}[h]
\centering
\includegraphics[width=12.5cm]{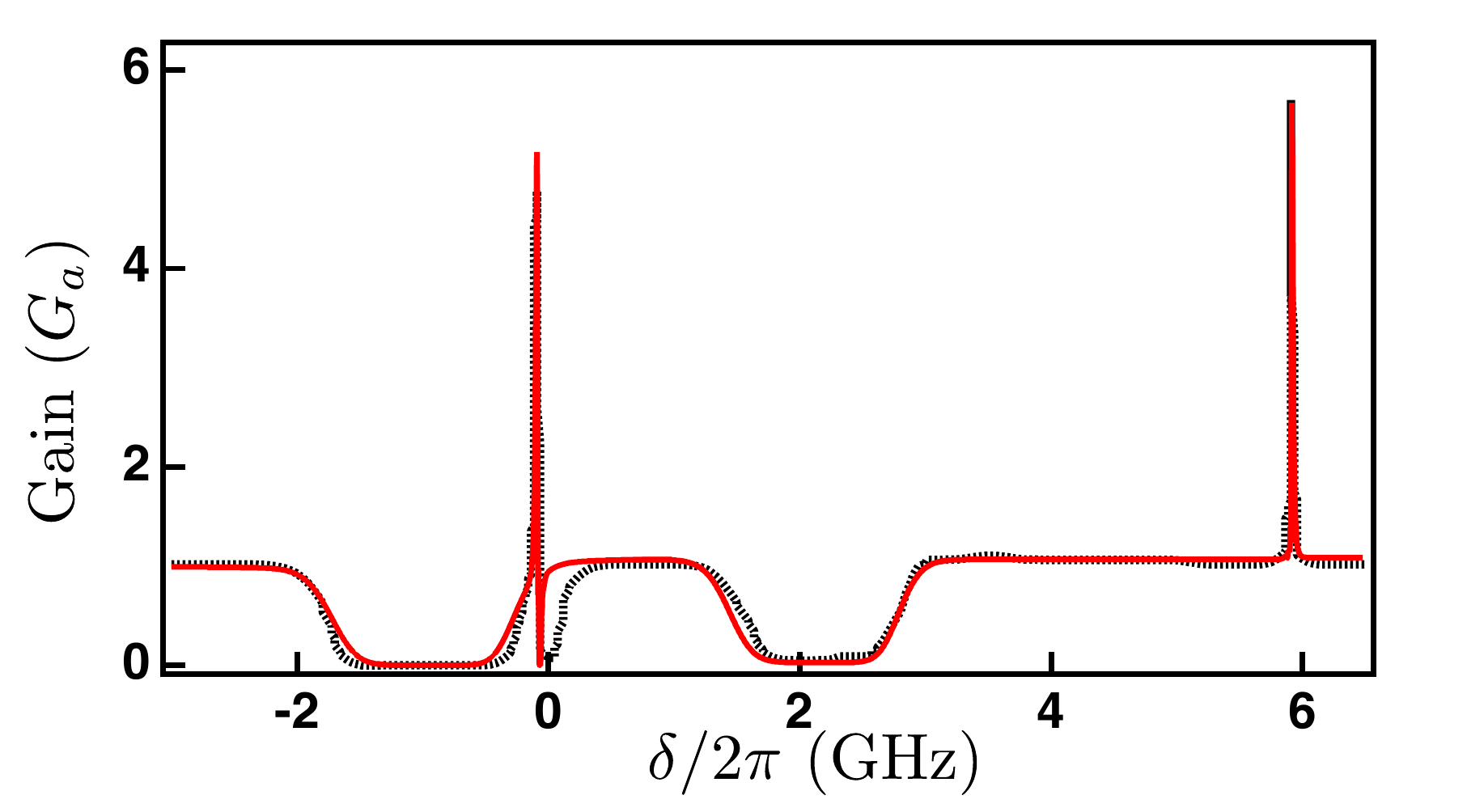} 
\caption[]	{Spectre de gain du faisceau sonde $G_a$ en fonction du désaccord à deux photons pour un désaccord à un photon de la pompe fixé. Le trait noir pointillé représente les résultats expérimentaux obtenus pour une puissance de pompe de 400 mW, focalisée sur 650 microns et une cellule de 12.5 mm de long. La courbe rouge représente la simulation numérique avec les paramètres suivants : $\gamma/2\pi =1$~MHz, $\Omega/2\pi = 0.33$ GHz, $\Delta/2\pi = 800$ MHz, $\alpha L=4500$. L'absorption résiduelle par les atomes non préparés dans l'état stationnaire et de plus pris en compte.\label{lett}}.
\end{figure}
Le gain sur le faisceau sonde en fonction du désaccord $-\delta$ est présenté sur la figure \ref{lett}.
Nous avons comparé le calcul aux résultats expérimentaux aimablement mis à disposition par P.D. Lett.
Dans ces conditions, le taux de préparation des atomes, tel que nous l'avons introduit en \eqref{prepara}, est de 97$\%$.
En première approximation on fera subir au faisceau des pertes correspondant à une absorption linéaire due à 3$\%$ de l'épaisseur optique totale.
On observe ainsi un bon accord entre les données expérimentales et les prédictions théoriques et ce sans aucun paramètre ajustable.
\subsubsection{Corrélations quantiques : effet de l'épaisseur optique}
\begin{figure}
\centering
\includegraphics[width=14.5cm]{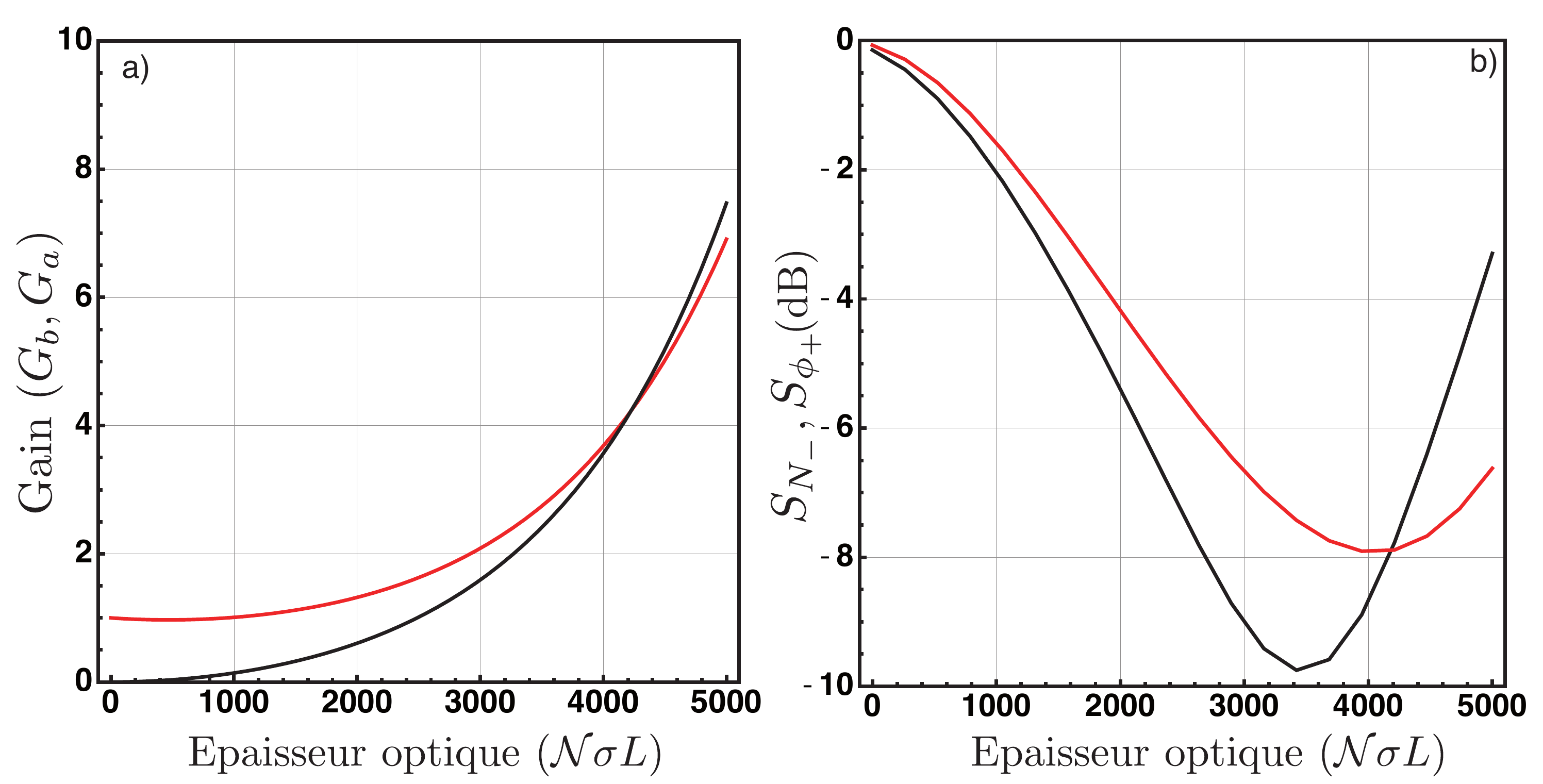} 
\caption[Effet de l'épaisseur optique ]	{Effet de l'épaisseur optique sur a) le gain sur la sonde (rouge) et la conjugué (noir) et sur b) les corrélations d'intensité (noir) et les anti-corrélations de phase (rouge).\\ Paramètres : $\gamma/2\pi =0.5$~MHz, $\delta/2\pi = 4$ MHz, $\Omega=0.3$ GHz, $\Delta=0.8$ GHz.	\label{effetalphaL}}.
\end{figure}
Nous allons nous intéresser maintenant aux corrélations quantiques.
Le premier paramètre que nous allons étudier est l'épaisseur optique.
En effet, comme nous l'avons souligné, nous pouvons atteindre dans un vapeur un tout autre régime que celui étudié pour les atomes froids.
Comme nous pouvons le voir dans la figure \ref{effetalphaL}, dans un premier temps (jusqu'à 3000) augmenter la profondeur optique permet d'amplifier les corrélations d'intensité et les anti corrélations de phase. 
Au delà de cette limite  les corrélations d'intensité diminuent puis de même les anti corrélations de phase se dégradent.\\
On peut comprendre cet effet en regardant l'effet de l'épaisseur optique sur le gain de la sonde et du conjugué.
En effet contrairement au cas d'un amplificateur insensible à la phase idéal, la différence d'intensité entre le faisceau sonde et le faisceau conjugué n'est pas indépendante du gain.
Elle diminue, puis s'annule et change de signe.
Si dans un premier temps cette diminution est un avantage pour les corrélations en intensité, en équilibrant les deux voies de la détection, dans un second temps, le taux de génération de photons conjugué devient plus grand que celui de photons sonde et les corrélations en intensité diminuent.
La figure \ref{effetalphaL} a) a été tracée en prenant en compte l'effet des forces de Langevin qui ne sont pas négligeables dans ce cas.
Pour les paramètres que nous avons utilisés (voir la légende de la figure \ref{effetalphaL}), la fraction d'atomes dans l'état stationnaire est supérieure à 85$\%$ et le désaccord à un photon est grand $\Delta=800$ MHz.
Dans ces conditions l'absorption résiduelle est négligeable.
\subsubsection{Effet de la pulsation de Rabi : $\Omega$}
\begin{figure}
\centering
\includegraphics[width=14.5cm]{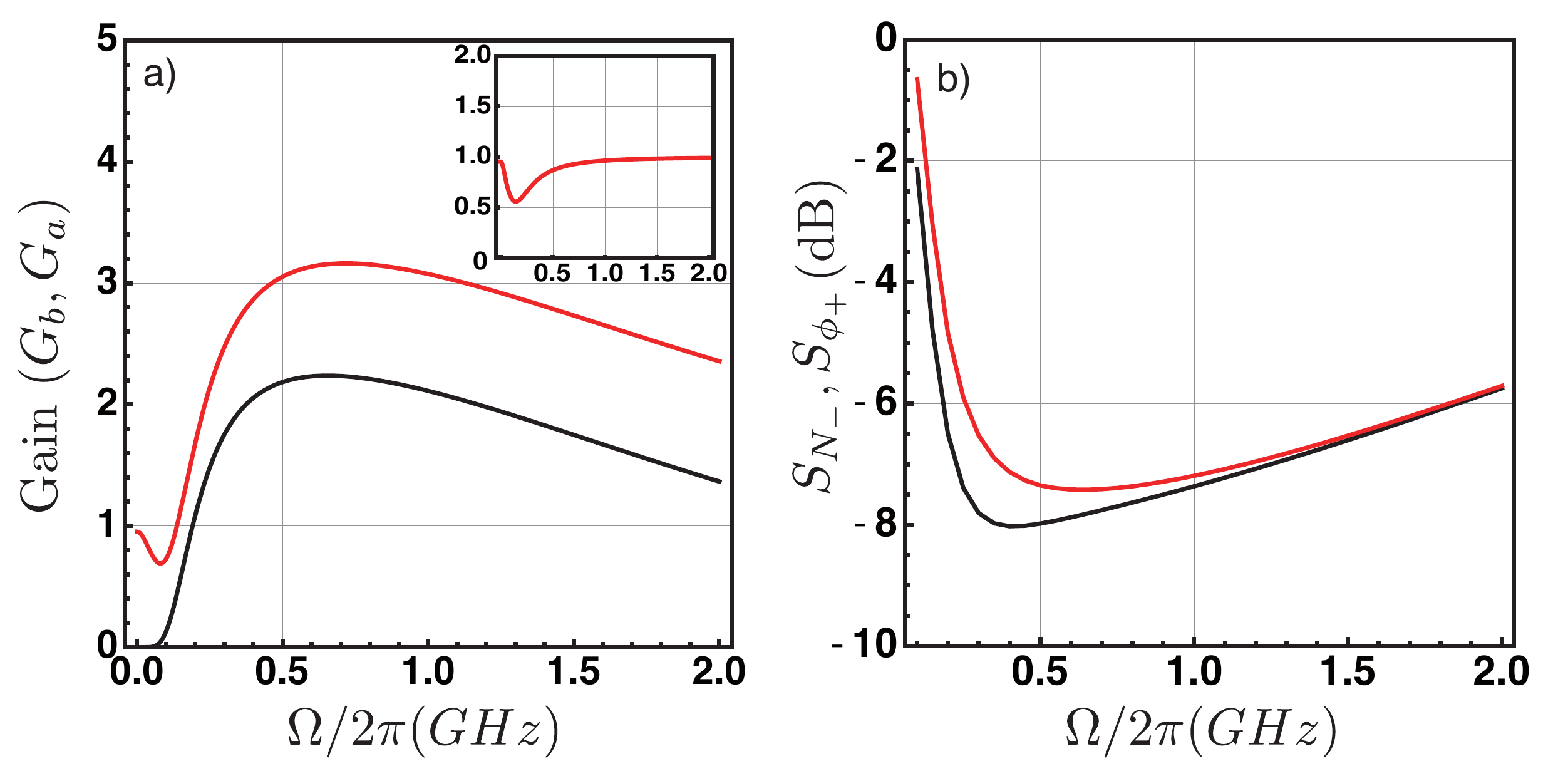} 
\caption[Effet de $\Omega$ ]	{Effet de $\Omega$ sur a) le gain sur la sonde (rouge) et la conjugué (noir) et sur b) les corrélations d'intensité (noir) et les anti-corrélations de phase (rouge). L'encart de la figure a) présente la différence  $G_a-G_b$\\ Paramètres : $\gamma/2\pi =0.5$~MHz, $\delta/2\pi = 4$ MHz, $\alpha L=3000$, $\Delta=0.7$ GHz.	\label{ch_Omega}}.
\end{figure}
La pulsation de Rabi $\Omega$ quantifie l'intensité du champ pompe.
On peut constater sur les figures \ref{ch_Omega} a) et b) que lorsque $\Omega\to 0$ le gain sur champ sonde tend vers 1, le gain sur le champ conjugué tend vers 0 et les corrélations tendent vers la limite quantique standard.
On observe ensuite un maximum dans les courbes de gain, puis l'efficacité du processus diminue.
La position de ce maximum est déterminé par les autres paramètres utilisés.
Typiquement le maximum va se déplacer vers des pulsations de Rabi plus faibles lorsque $\Delta$ diminue.\\
On peut noter aussi que le maximum des corrélations d'intensité est obtenu pour $\Omega$ plus faible que celui correspondant au maximum de gain.
On peut expliquer cela en constatant que la position du maximum des corrélations correspond au minimum de l'écart entre l'intensité de la sonde et du conjugué (encart de la figure \ref{ch_Omega} a)).
\subsubsection{Effet du désaccord à un photon : $\Delta$}
\begin{figure}
\centering
\includegraphics[width=14.5cm]{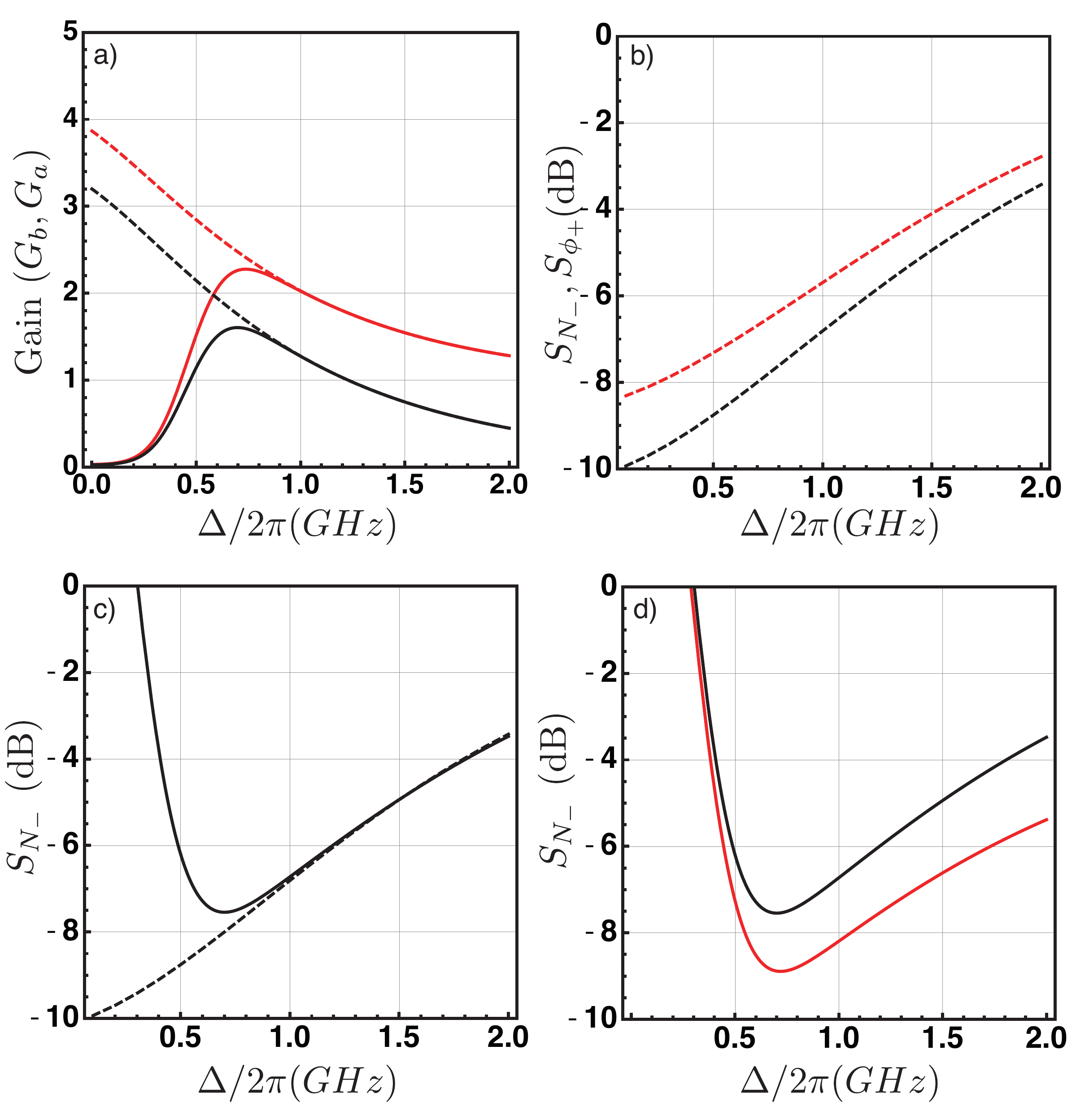} 
\caption[Effet de $\Delta$]	{a) Effet de $\Delta$ sur le gain sur la sonde (rouge) et le conjugué (noir). Les courbes en pointillés correspondent au cas idéal et les courbes en trait plein correspondent au cas prenant en compte l'absorption. b) Corrélations d'intensité (noir) et les anti-corrélations de phase (rouge) dans le cas idéal.
c) Corrélations d'intensité dans le cas idéal (pointillés) et en prenant en compte de l'absorption (plein). d) Corrélations d'intensité en prenant en compte de l'absorption en utilisant le modèle de l'amplificateur parfait (rouge) et dans notre modèle (noir).\\
Paramètres : $\gamma/2\pi =0.5$~MHz, $\delta/2\pi = 4$ MHz, $\alpha L=3000$, $\Omega=0.3$ GHz.\label{ch_Delta}}.
\end{figure}
Dans les expériences récentes de mélange à 4 ondes \cite{Boyer:2007p1404,McCormick:2007p652,GlorieuxSPIE2010}, les valeurs utilisées pour le désaccord à 1 photon $\Delta$ sont de l'ordre de plusieurs centaines de MHz, ce qui les différencient nettement du régime exploré dans \cite{Slusher:1985p3642}.
En effet, pour des lasers suffisamment intenses ($\Omega \gtrsim 0.3$ GHz), l'interaction lumière matière reste forte même à désaccord important (plusieurs centaines de MHz).
La figure \ref{ch_Delta} présente l'effet de $\Delta$ sur le gain et sur les corrélations.
On peut voir sur la  courbe rouge pointillée de figure a) que le gain de la sonde croit vers le désaccord nul.
Ce cas est le cas idéal sans prendre en compte l'absorption de la sonde par les atomes n'ayant pas atteint l'état stationnaire.
Ce cas pourrait être atteint pour des faisceaux pompes de diamètre très grand (c'est-à-dire pour lesquels la vitesse des atomes fois le temps de préparation est petit devant le rayon du faisceau pompe.)\\
Pour des faisceaux pompes plus petits, l'approximation que nous utilisons est de considérer les atomes non préparés dans l'état stationnaire comme un milieu absorbant en entrée de la cellule.
En utilisant cette approximation on obtient la courbe rouge pour le gain de la sonde.
On voit que dans ce cas le gain atteint un maximum autour de 700 MHz, qui est un compromis entre l'efficacité du processus de mélange à 4 ondes et l'absorption.
Dans ce modèle, les pertes sont appliquées sur le faisceau sonde en entrée du milieu; le gain sur le conjugué est donc affecté par le même facteur que le gain sur la sonde (courbes noires).
Cette hypothèse sera discutée dans la partie consacrée à la présentation des résultats expérimentaux (chapitre~\ref{ch5}).\\
De façon similaire, les corrélations augmentent jusqu'au désaccord nul dans le cas idéal (figure b).
En présence d'absorption, on obtient le spectre de bruit en intensité relatif présenté sur la figure c) en trait plein.
Dans le cas idéal, les spectres de bruit sont normalisés au bruit quantique standard que l'on peut déduire directement du gain.
La correction que nous avons effectuée par rapport à ce cas idéal, correspond à normaliser les spectres de bruit par le gain corrigé par l'absorption.
Dans la figure d), on compare le modèle que nous avons présenté dans ce chapitre (noir) au modèle de l'amplificateur idéal présenté au chapitre~\ref{ch3} (rouge).
On peut noter le bon accord qualitatif du modèle de l'amplificateur idéal avec notre modèle.
Le modèle de l'amplificateur idéal et des pertes réparties présentés au chapitre 3 seront donc, dans ce régime, des moyens simples de prédire une borne supérieure pour les corrélations.
En prenant en compte la structure microscopique des atomes, notre modèle permet de donner une borne supérieure plus faible pour le niveaux des corrélations attendues.

\subsubsection{Effet du désaccord à deux photons : $\delta$}
\begin{figure}
\centering
\includegraphics[width=14.5cm]{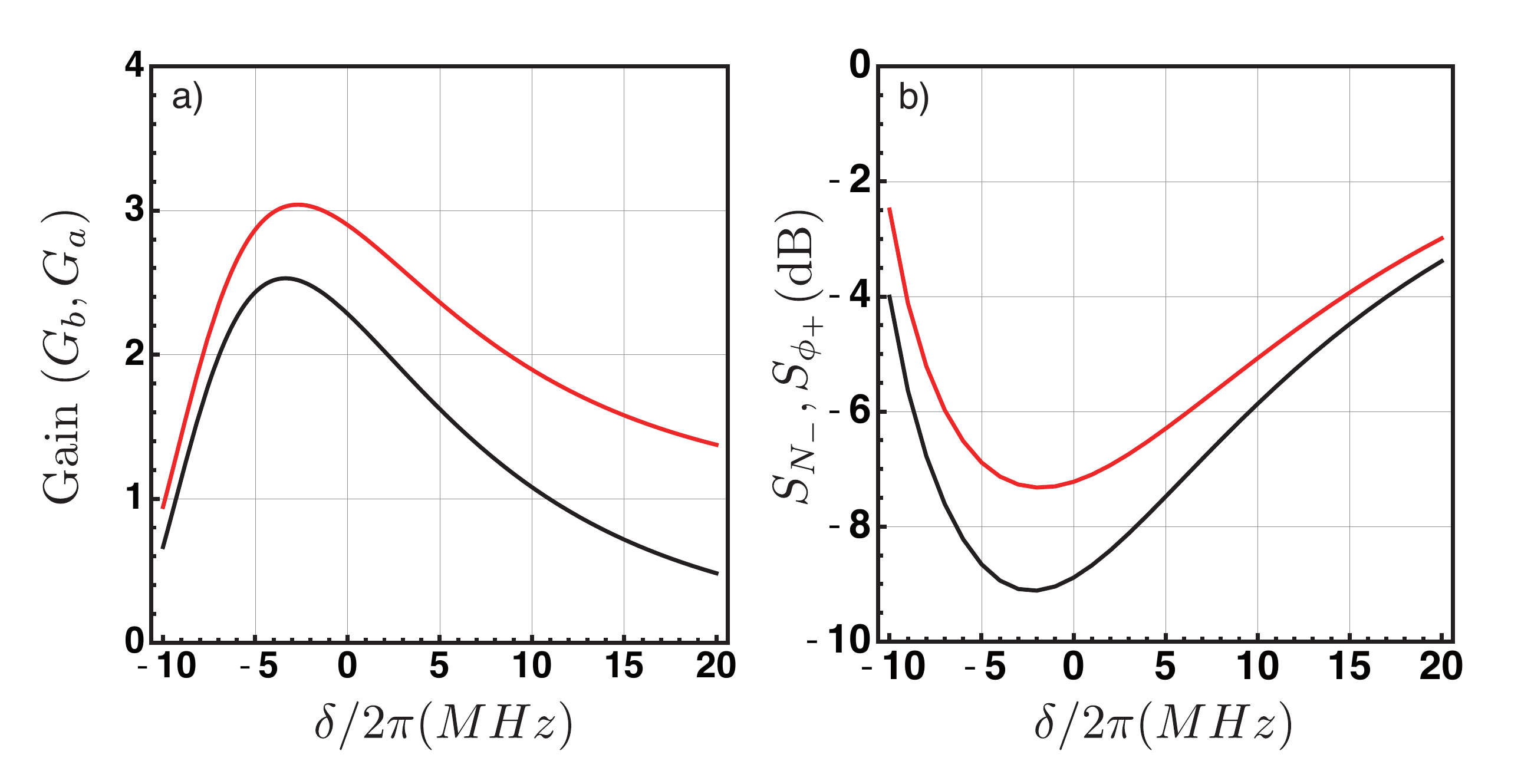} 
\caption[Effet de $\delta$]	{Effet de $\delta$ sur a) le gain de la sonde (rouge), du conjugué (noir) et sur b) les corrélations d'intensité (noir) et les anti-corrélations de phase (rouge).\\ Paramètres : $\gamma/2\pi =0.5$~MHz, $\Omega/2\pi =0.3$~GHz, $\alpha L=3000$, $\Delta=0.7$~GHz.	\label{ch_delta2}}.
\end{figure}
L'effet du désaccord à deux photons est plus difficile à analyser.
Comme nous l'avons présenté à la section \ref{parag:spec}, différents processus élémentaires contribuent au profil observé.
On peut voir sur la figure \ref{ch_delta2} a) que le gain passe par un maximum autour de $\delta=0$~MHz.
De façon analogue, on observe un maximum pour les corrélation sur la figure b) pour un désaccord similaire.
Dans les résultats présentés jusqu'ici nous avons choisi $\delta=+4$ MHz même si le maximum de la courbe de gain est plutot vers $\delta=-3$~MHz.
Ce choix a été fait pour rester proche des observation expérimentales décrites dans \cite{Boyer:2007p1404,McCormick:2007p652,GlorieuxSPIE2010} et présentées au chapitre \ref{ch5}.
\subsubsection{Spectres de bruit : effet de la fréquence d'analyse $\omega$}
\begin{figure}
\centering
\includegraphics[width=10cm]{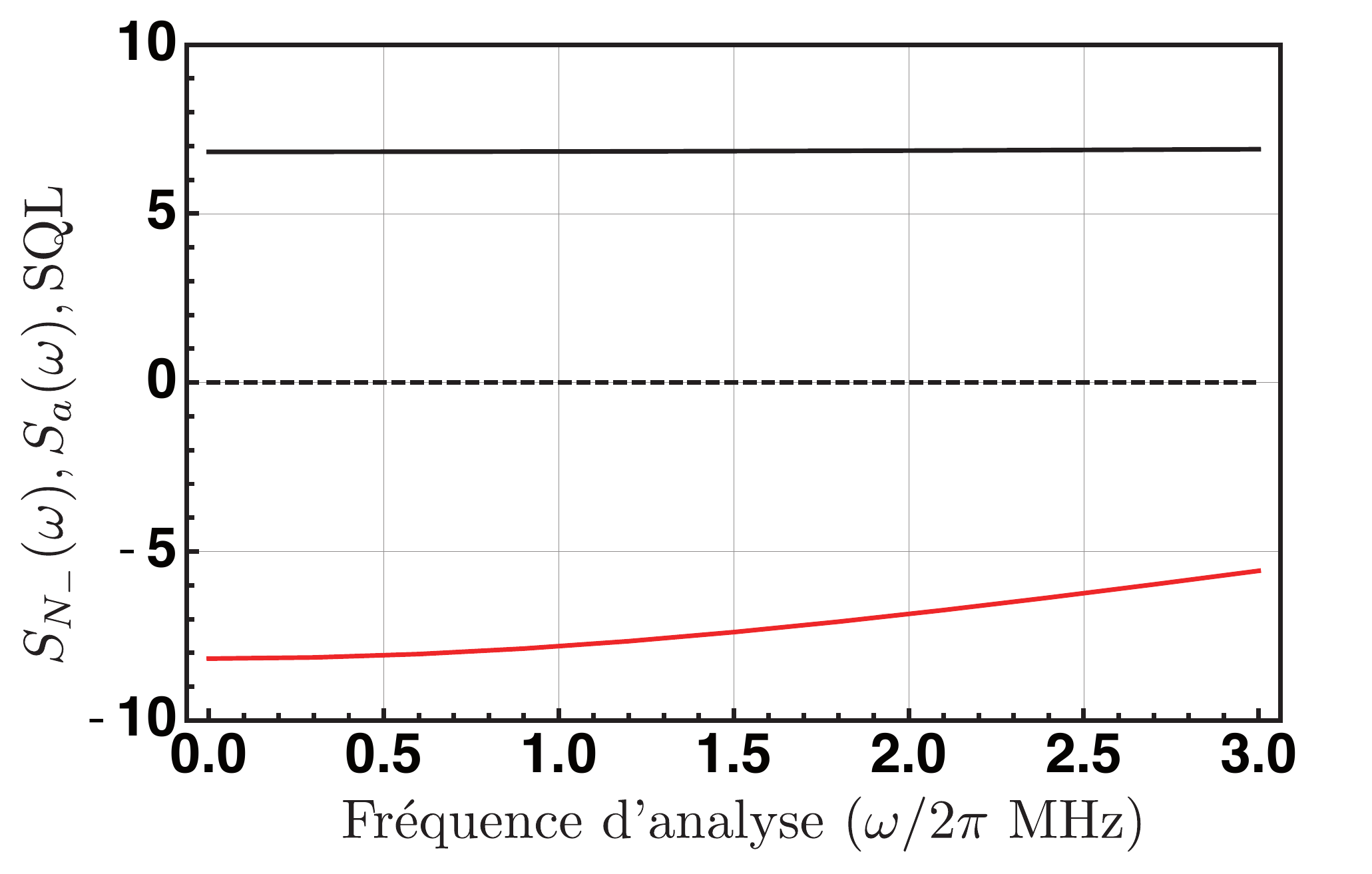} 
\caption[Spectres de bruit sur la différence d'intensité et sur le mode $\hata$.]	{Spectres de bruit en échelle logarithmique (dB par rapport à la limite quantique standard SQL) sur la différence d'intensité (rouge) et sur le mode $\hata$ (noir).\\
Paramètres : $\gamma/2\pi =0.5$~MHz, $\delta/2\pi = 4$ MHz, $\alpha L=3000$, $\Delta=0.7$~GHz	$\Omega/2\pi =0.3$~GHz	\label{ch_omega2}}.
\centering
\includegraphics[width=10cm]{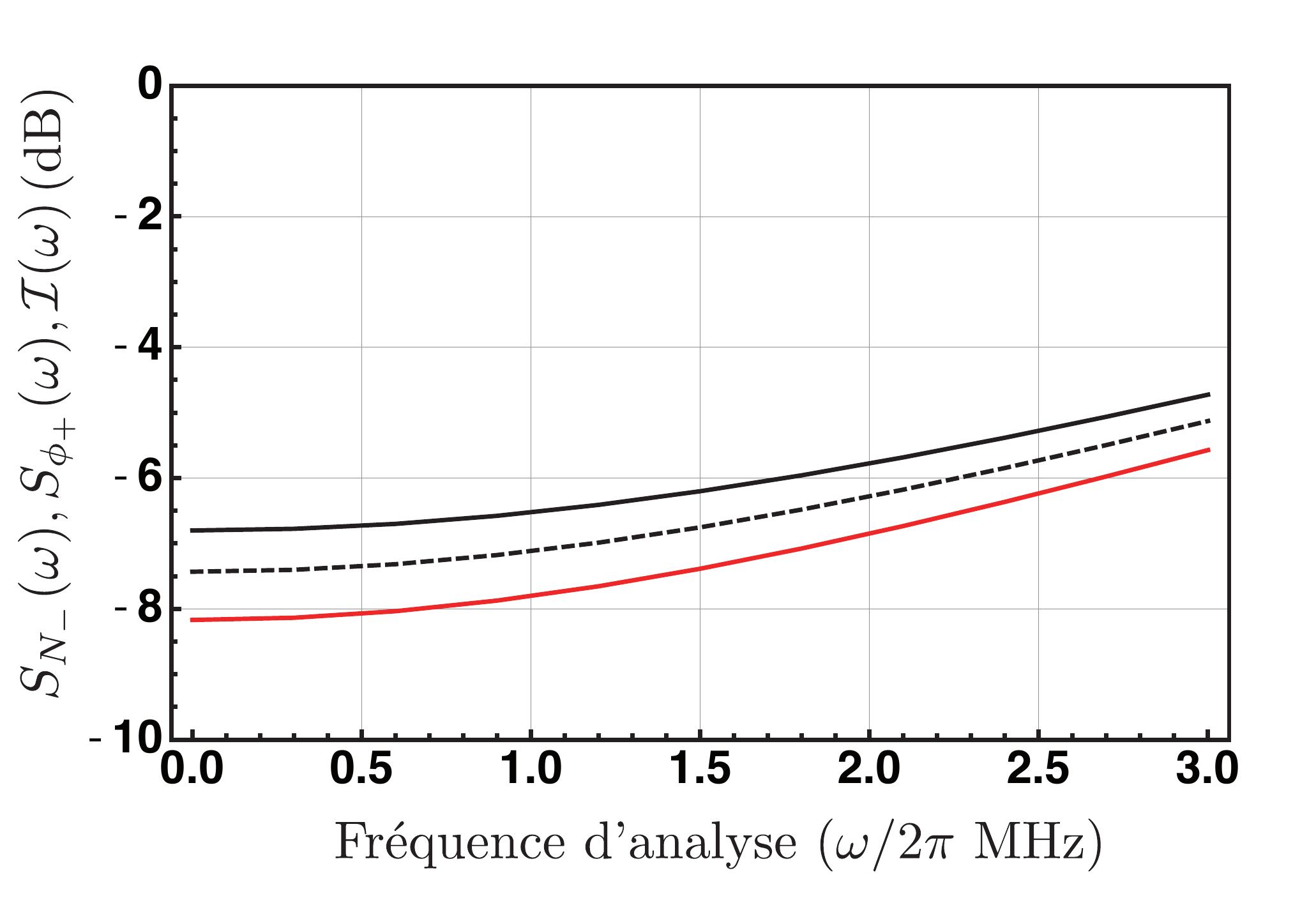} 
\caption[Spectres de bruit sur la différence d'intensité.]	{Spectres de bruit sur la différence d'intensité en rouge, des anti-corrélations de phase en noir, et de l'inséparabilité en pointillés noirs.\\ 
Paramètres : $\gamma/2\pi =0.5$~MHz, $\delta/2\pi = 4$ MHz, $\alpha L=3000$, $\Delta=0.7$~GHz	$\Omega/2\pi =0.3$~GHz. \label{ch_omega1}}.
\end{figure}
Les corrélations sont étudiées à fréquence non nulle. La fréquence d'analyse $\omega$ était jusqu'à maintenant fixée à 1~MHz.
On peut reproduire les grandeurs accessibles expérimentalement en faisant varier la fréquence d'analyse et en traçant des spectres.
Les spectres de la figure \ref{ch_omega2} présentent les corrélations quantiques sur la différence d'intensité autour de -8$dB$ sous la limite quantique standard (rouge), ainsi que l'excès de bruit de +7$dB$ dans le mode de la sonde (noir).
On peut voir que les corrélations sont maximales à basse fréquence.
Expérimentalement, ces mesures sont limitées par le bruit technique à basse fréquence.
Dans la référence \cite{McCormick:2008p6669}, une réduction du bruit sous la limite quantique standard a été observée à 4.5~kHz.

\subsubsection{Contribution des forces de Langevin}
Dans toutes les courbes présentées dans cette section, la contribution des forces de Langevin a été prise en compte.
Les termes de forces de Langevin ont une forme très compliquée et dépendent d'un grand nombre de paramètres.
Pour analyser de manière qualitative leur rôle, nous pouvons dire, de manière générale, que la contribution des forces de Langevin sera plus importante pour les pulsations de Rabi faibles.
De même, nous constatons une augmentation de leur contribution pour $\delta< 0$.
On pourra noter de plus que la contribution des forces de Langevin augmente lorsque $\gamma$ croit.

\subsection{Corrélations quantiques pour $G_a+G_b<1$}\label{sec:qbs}
\begin{figure}
\centering
\includegraphics[width=14.5cm]{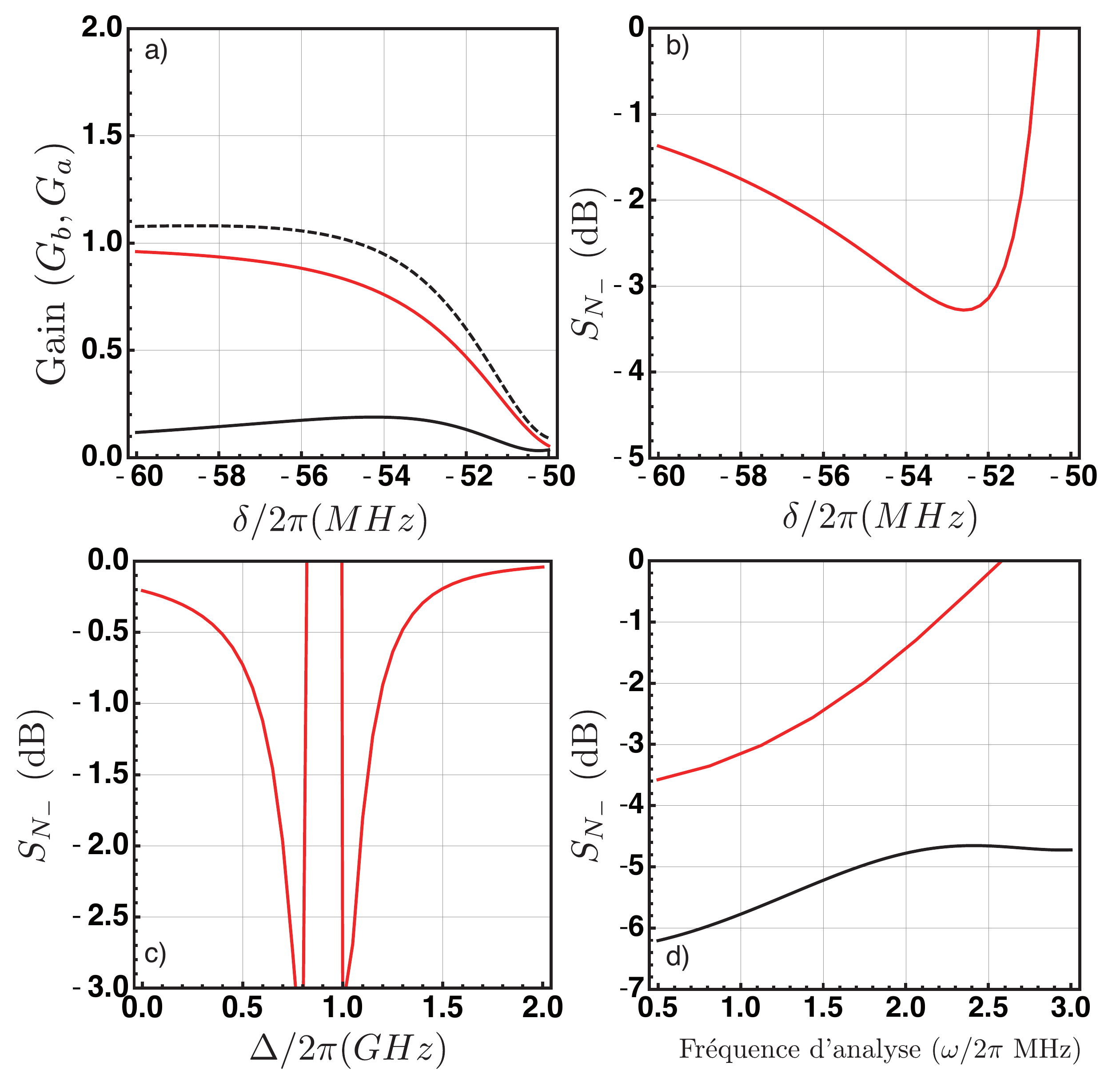} 
\caption[Corrélations quantiques pour $G_a+G_b<1$]	{ a) Effet de $\delta$ sur le gain de la sonde (rouge) et du conjugué (noir). La courbe pointillé représente $G_a+G_b$ b) Effet de $\delta$  sur les corrélations d'intensité. c) Effet de $\Delta$  sur les corrélations d'intensité. d) Spectres de bruit de la différence d'intensité en prenant en compte les forces de Langevin (rouge) et en les négligeant (noir).\\
Paramètres : $\gamma/2\pi =0.5$~MHz, $\Omega/2\pi =0.52$~GHz, $\alpha L=300$, $\Delta/2\pi=1$~GHz, $\delta/2\pi=-52$~MHz, $\omega/2\pi=1$~MHz.\label{ch_qbs}}.
\end{figure}
Nous proposons ici un nouveau régime, pour lequel on peut observer des corrélations quantiques en intensité entre les champs sonde et conjugué tout en conservant un gain total inférieur à l'unité : $G_a+G_b<1$.
On présente les résultats obtenus sans prendre en compte la dispersion de vitesses des atomes\footnote{Pour des raisons de difficulté d'intégration du profil de vitesse dans le cas où les forces de Langevin ne sont plus négligeables, nous ne sommes pas en mesure de présenter les résultats généraux en prenant en compte la distribution de vitesses. Les résultats présentés seront donc vus comme la situation d''un milieu constitué d'atomes froids ou bien comme une borne supérieure pour les corrélations dans une vapeur atomique.} sur la figure \ref{ch_qbs}.\\

En modifiant les paramètres précédents et en s'éloignant du point de fonctionnement décrit dans \cite{Boyer:2007p1404,McCormick:2007p652,GlorieuxSPIE2010}, nous avons pu mettre en évidence d'abord théoriquement (puis expérimentalement au chapitre 5), un régime très différent de l'amplificateur idéal.
En effet, comme nous pouvons le voir sur la figure \ref{ch_qbs} a), il existe des situations où le gain sur le mode $\hata$ est inférieur à 1 sur une certaine gamme.
En particulier pour $\delta=-52$ MHz qui sera la valeur que nous utiliserons par la suite.
Nous nous plaçons donc à la frontière entres les deux phénomènes décrits à la section \ref{parag:spec} (Raman et mélange à 4 ondes).\\
Sur la figure \ref{ch_qbs} b), on observe la génération de corrélations quantiques en intensité entre les champs sonde et conjugué pour un gain total ($G_a+G_b$) inférieur à 1.
Cela est en contraste avec ce que l'on avait étudié dans le cas de l'amplificateur idéal (la production de corrélations pour un gain uniquement supérieur à 1).\\
Sur la figure \ref{ch_qbs} c), on peut voir que la dépendance en $\Delta$ est plus fine dans ce régime que dans le régime présenté à la section \ref{sec:vapchaud}.
Comme nous n'avons pas pu intégrer le profil de vitesse dans ce cas, nous faisons l'hypothèse que la principale conséquence de cet effet est de réduire le nombre d'atomes qui vont contribuer au processus (et donc l'épaisseur optique effective).\\
Enfin la figure \ref{ch_qbs} d) présente un spectre de bruit de la différence d'intensité.
Il est intéressant de noter que dans ces conditions la contribution des forces de Langevin devient très grande et qu'elle ne peut pas du tout être négligée.\\

On peut donner une image physique du système que nous venons de présenter.
Pour deux faisceaux incidents (une sonde sur le mode $\hata$ et le vide sur le mode $\hat b$) l'effet de ce système sur les valeurs moyennes et de redistribuer les photon avec un gain total ($G_a+G_b$) inférieur à 1.
De plus, les deux modes qui ne sont pas corrélés en intensité en entrée, le sont en sortie.
Le système agit donc comme une lame séparatrice.
Or les corrélations que l'on observe en sortie, ne sont pas uniquement des corrélations au sens classique (comme on peut le voir avec une lame séparatrice) mais des corrélations quantiques telles que le bruit sur la différence d'intensité soit sous la limite quantique standard.
On propose donc d'appeler ce système, un \textit{''quantum beamspliter''} ou lame séparatrice quantique.

\subsection{Corrélations quantiques sur la transition $5S_{1/2}\to 6P_{1/2}$ du rubidium.}
\subsubsection{Contexte}
Le travail mené dans l'équipe IPIQ au sein du laboratoire MPQ consiste en la réalisation d'une mémoire quantique dans un nuage d'ions $^{88}$Sr$^+$.
Pour tester une telle mémoire, il peut être intéressant de disposer d'un état non-classique du champ (typiquement un état comprimé).
Sans entrer dans le détail du principe d'une mémoire quantique \cite{Julsgaard:2004p6526,Coudreau:2007p160,Choi:2008p11519} nous allons étudier la faisabilité d'états comprimés de la lumière à la longueur d'onde de la transition $5S_{1/2}\to 5P_{1/2}$ du strontium par mélange à 4 ondes dans une vapeur de rubidium 85.
En effet il existe une quasi-coïncidence (400 MHz d'écart) entre les transitions $5S_{1/2}\to 6P_{1/2}$ du rubidium 85 et $5S_{1/2}\to 5P_{1/2}$  de l'ion Sr$^+$ \cite{Madej:1998p13999}.
Nous allons donc étudier le mélange à 4 ondes sur la transition  $5S_{1/2}\to 6P_{1/2}$ du rubidium 85.
\subsubsection{Paramètres}
Nous allons lister les principales différences entre la transition $5S_{1/2}\to 5P_{1/2}$ dite ``rouge'' et la transition $5S_{1/2}\to 6P_{1/2}$ dite ``bleue'' :\\
\begin{itemize}
\item Tout d'abord la transition bleue correspond à une longueur d'onde de 421.6 nm (contre 795nm pour la transition rouge)
La largeur Doppler associée à la transition bleue sera donc beaucoup plus importante (facteur 1.9) que celle associée à la la transition rouge.
\item Pour la transition bleue $\Gamma/2\pi = 0.24$~MHz contre 5.7~MHz pour la transition rouge, soit un facteur $24$.
L'effet dominant du paramètre $\Gamma$ dans le modèle que nous avons présenté au chapitre \ref{ch4} est le préfacteur en $\Gamma \alpha L$ dans l'exposant de l'exponentiel.
On voit que pour compenser un facteur 24 sur $\Gamma$ la méthode la plus simple sera de prendre un facteur identique sur $\alpha L$.
Cette idée sera étudiée à la section \ref{bourrin}.
\item De plus, la puissance laser envisageable à cette longueur d'onde sera à prendre en compte pour déterminer la faisabilité des configurations proposées.
En effet selon la puissance disponible, on pourra faire varier la focalisation du faisceau pour conserver une pulsation de Rabi constante.
Par contre, dans ce cas les effets de décohérence et de préparation des atomes pourraient devenir critiques et seront donc détaillés.
\end{itemize}

\subsubsection{Augmenter l'épaisseur optique}\label{bourrin}
\begin{figure}
\centering
\includegraphics[width=14.5cm]{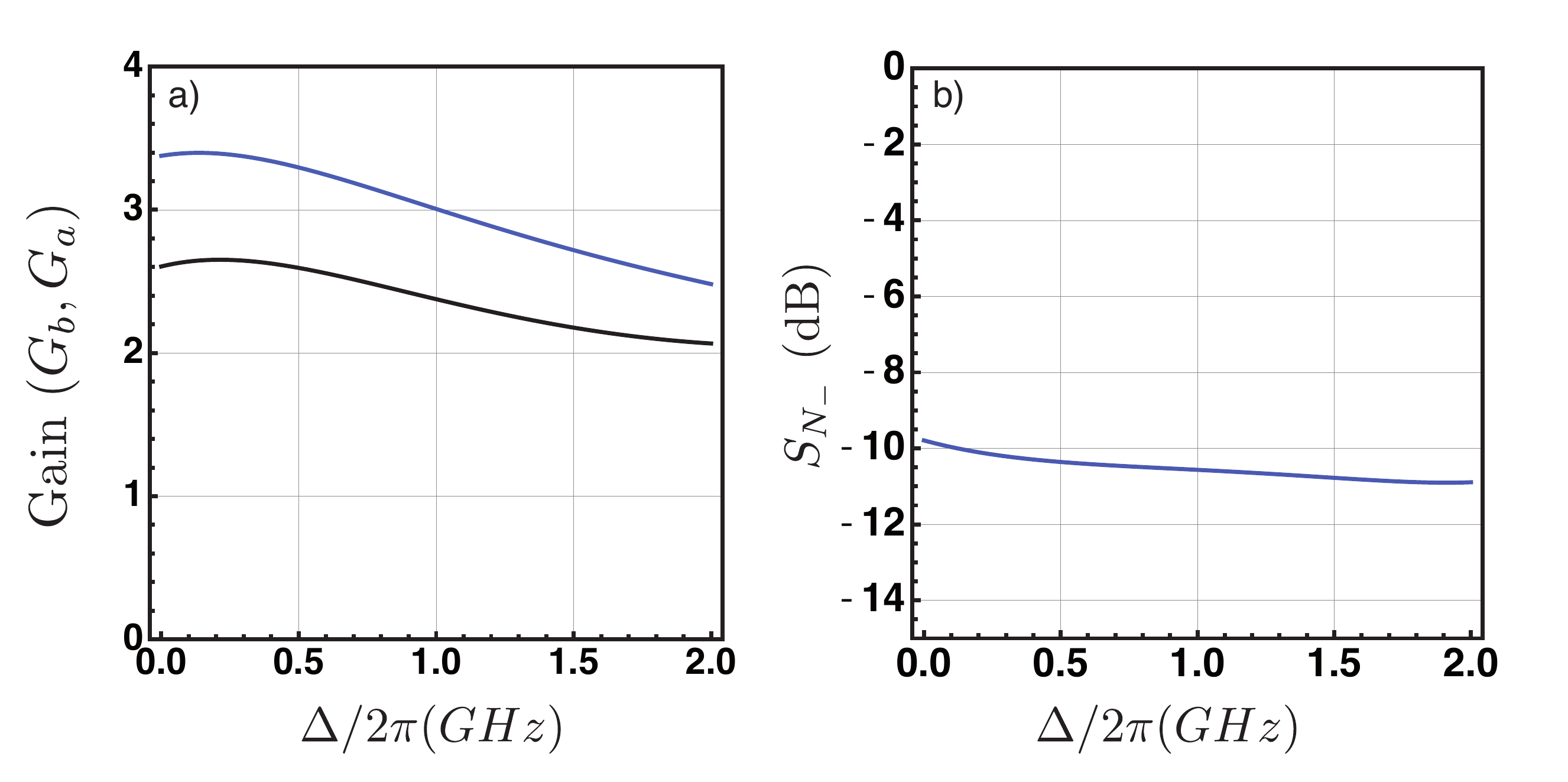} 
\caption[]	{ Effet de $\Delta$   a) sur le gain de la sonde (bleu) et du conjugué (noir) et b)  sur les corrélations d'intensité pour la transition $5S_{1/2}\to 6P_{1/2}$.\\
Paramètres : $\Gamma/2\pi =0.24$~MHz, $\gamma/2\pi =1$~MHz, $\Omega/2\pi =0.5$~GHz, $\alpha L=60000$, $\delta/2\pi=-5$~MHz, $\omega/2\pi=1$~MHz.\label{fig_bourin}}.
\end{figure}
Comme nous l'avons vu, une stratégie pour observer des corrélations quantiques consiste à essayer de compenser le facteur 24 sur $\Gamma$ sur la transition bleue en augmentant l'épaisseur optique.
La figure \ref{fig_bourin} présente les résultats dans cette situation sans prendre en compte l'absorption.
Dans ce cas, on voit que l'on retrouve sans problème les résultats de la section \ref{sec:vapchaud}, c'est-à-dire un gain supérieur à 1 sur la sonde et le conjugué et un important niveau de corrélations.\\

Or, dans cette situation, il n'est pas possible de négliger l'absorption.
En effet pour gagner un facteur 24 sur l'épaisseur optique, on doit porter le rubidium à des températures de l'ordre de 200$^\circ$C.
Nous l'avons vu, l'absorption dépend du taux de préparation des atomes dans l'état stationnaire.\\
Dans les conditions de la figure \ref{fig_bourin}, même à résonance, c'est-à-dire où le temps de préparation est le plus faible, le temps caractéristique $T_0$ est de l'ordre de 2 $\mu$s.
Ainsi pour que l'absorption à résonance soit faible (inférieure à 3$\%$) pour une épaisseur optique de $\alpha L=60000$, il faut que le temps de transit $\tau$ soit supérieur à $15\ T_0$ soit 30 $\mu$s.\\
A 200 $^\circ$C cela correspond à un parcours de l'ordre de 10 mm, des atomes dans le faisceau pompe, avant d'interagir avec le faisceau sonde.
Ainsi pour éviter l'absorption sur le faisceau sonde, il faut que le waist du faisceau pompe soit de l'ordre de 10 mm.
Pour une telle taille, des puissances de plusieurs milliers de watts sont nécessaires pour conserver la valeur de la pulsation de Rabi constante.\\

L'analyse des ordres de grandeurs nous permet donc de conclure, que la méthode naïve qui consiste à augmenter l'épaisseur optique afin de compenser la diminution de $\Gamma$, n'est pas réalisable expérimentalement.
Il faut donc essayer une autre approche.

\subsubsection{Exploration de l'espace des paramètres}
Afin de chercher une configuration envisageable expérimentalement nous avons exploré l'espace des paramètres $\Delta$, $\delta$, $\Omega$, $\alpha L$.\\
Sans donner tous les détails de ce travail afin de ne pas alourdir la lecture, nous allons présenter les principaux résultats.\\

Dans un premier temps, nous avons cherché à sortir de la zone d'absorption, c'est-à-dire en dehors du profil Doppler.
Pour ce faire, il faut aller au delà $\Delta=6$~GHz.
A ce désaccord, on n'observe du gain que pour des pulsations de Rabi très élevées $\Omega \simeq 1$ GHz et des épaisseurs optiques de $10^5$.
Pour atteindre de telles puissances optiques, il est nécessaire de focaliser le faisceau pompe sur des tailles inférieures à $50\ \mu$m.
Dans ce cas, la valeur du taux de décohérence $\gamma$ augmente inversement proportionnellement à la taille du faisceau.
Pour un waist de $50\ \mu$m, le taux de décohérence vaut $2\pi \times 6$~MHz.
Dans ces conditions, les résultats des simulations numériques ne prédisent pas de corrélations d'intensité sous la limite quantique standard.\\

D'autre part, nous avons cherché à reproduire le régime décrit dans la section \ref{sec:qbs}, c'est-à-dire un régime où $G_a+G_b<1$.
De façon similaire à ce que nous venons de voir, cette situation n'a pas pu être atteinte numériquement.
Proche de résonance, les phénomènes d'absorption dominent l'effet étudié.
Loin de résonance, l'exploration de l'espace des paramètres n'a pas permis de dégager des conditions similaires à \ref{sec:qbs}.
\subsection{Conclusion sur la production de corrélations quantiques sur la transition $5S_{1/2}\to 6P_{1/2}$}
En conclusion, nous avons étudié théoriquement la production de corrélations quantiques sur la transition $5S_{1/2}\to 6P_{1/2}$ du rubidium afin d'obtenir un source utile aux expériences d'optique quantique avec des ions Sr$^+$.
Nous avons démontré qu'il existe une situation qui prévoit des corrélations sous la limite quantique standard.
Malheureusement, cette situation ne correspond pas à des paramètres réalistes pour des expériences d'atomes froids ($\alpha L= 60000$).
De même, pour des expériences dans une vapeur atomique, l'absorption par les atomes n'ayant pas atteint l'état stationnaire rend aussi impossible l'observation de ces corrélations. 
Enfin, nous avons étudié l'espace des paramètres à l'aide de simulation numériques et nous pouvons conclure qu'il n'existe pas de solutions, réalistes expérimentalement avec un laser continu, qui permettent la génération d'états corrélés générés par mélange à 4 ondes à 422 nm.
Des résultats récents sur la réalisation d'expériences similaires en utilisant des laser pulsés semblent ouvrir la voie à des expériences dans ce régime \cite{Agha:2010p13052}.

\section{Conclusion du chapitre}
Dans ce chapitre nous avons présenté deux calculs d'optique quantique.
Dans un premier temps nous avons rappelé les résultats du phénomène de transparence électromagnétiquement induite que nous allons étudier expérimentalement dans le chapitre 6 pour un modèle basé sur des atomes à 3 niveaux en simple--$\Lambda$.
Puis dans un second temps nous avons introduit un modèle microscopique basé sur des atomes immobiles ayant une structure des niveaux d'énergie en double--$\Lambda$ pour décrire les expériences de mélange à 4 ondes.\\
Nous avons ainsi obtenu, dans le formalisme de Heisenberg-Langevin, les relations reliant les opérateurs de création et d'annihilation des champs sonde et conjugué en entrée et en sortie d'un milieu atomique constitué d'atomes immobiles.
Ce modèle a permis de prévoir la génération de faisceaux intriqués en variables continues dans un milieu constitué d'atomes froids \cite{Glorieux:2010p9415}.\\
Dans une seconde partie de ce chapitre, nous avons abordé une extension de ce modèle microscopique au cas d'une vapeur atomique ``chaude''.
Dans ce cas, nous avons démontré que les résultats obtenus pour les atomes immobiles pouvaient être appliqués en modifiant quelques paramètres :
c'est-à-dire en augmentant l'épaisseur optique afin de prendre en compte la densité que l'on observe dans des vapeurs ainsi que le taux de décohérence qui devient piloté par le temps de passage des atomes dans le faisceau.
D'autre part, pour prendre en compte le temps d'interaction non-infini des atomes avec le faisceau pompe, nous avons étudié l'établissement du régime stationnaire et nous avons démontré qu'une part significative des atomes pouvaient  ne pas avoir atteint cet état lorsqu'ils interagissaient avec le faisceau sonde.
Nous avons alors modélisé ces atomes, comme un milieu absorbant placé pour des raisons de simplicité avant le milieu de gain.
Ces hypothèses, nous ont permis d'explorer numériquement l'espace des paramètres et de prevoir un nouveau régime qui n'avait pas été étudié expérimentalement à savoir la production de corrélations quantiques pour $G_a+G_b<1$.\\
Enfin, nous avons étudié théoriquement la possibilité de générer des faisceaux corrélés à 422 nm sur la transition $5S_{1/2}\to 6P_{1/2}$ du rubidium et nous avons montré qu'il n'était pas possible de dégager des conditions réalistes expérimentalement similaires à celles de la transition 
$5S_{1/2}\to 5P_{1/2}$.

%% file: chapitre5v5.tex
\chapter{Génération d'états comprimés à 795 nm}\minitoc\label{ch5}

\vspace{1.1cm}
Dans ce chapitre, nous présentons les résultats expérimentaux que nous avons obtenus pour la génération d'états non classiques à 795 nm (corrélations en intensité sous la limite quantique standard entre les champs sonde et conjugué).
En utilisant les outils expérimentaux que nous avons introduits au chapitre \ref{ch2}, nous décrirons dans un premier temps le montage expérimental.\\
Nous présenterons ensuite les mesures que nous avons réalisées pour caractériser le milieu atomique en présence du laser de pompe.
Cette étude permet de tester le modèle théorique présenté au chapitre \ref{ch4}, en le comparant aux mesures de gain que nous avons effectué sur les faisceaux sonde et conjugué.\\
Dans une seconde partie, nous étudierons l'effet des différents paramètres expérimentaux sur la génération des faisceaux corrélés.
Cette étude, nous a permis d'optimiser notre montage expérimental et de trouver un point de fonctionnement optimal afin d'atteindre -9.2dB de corrélations sous la limite quantique standard.\\
Enfin, nous terminerons ce chapitre par la démonstration expérimentale du régime de génération de corrélations quantiques sans gain $G_a+G_b<1$ présenté à la section \ref{sec:qbs} du chapitre~4 et que l'on appelé la ``lame séparatrice quantique''.

\section{Dispositif expérimental}
	\begin{figure}
		\centering
		\includegraphics[width=14.5cm]{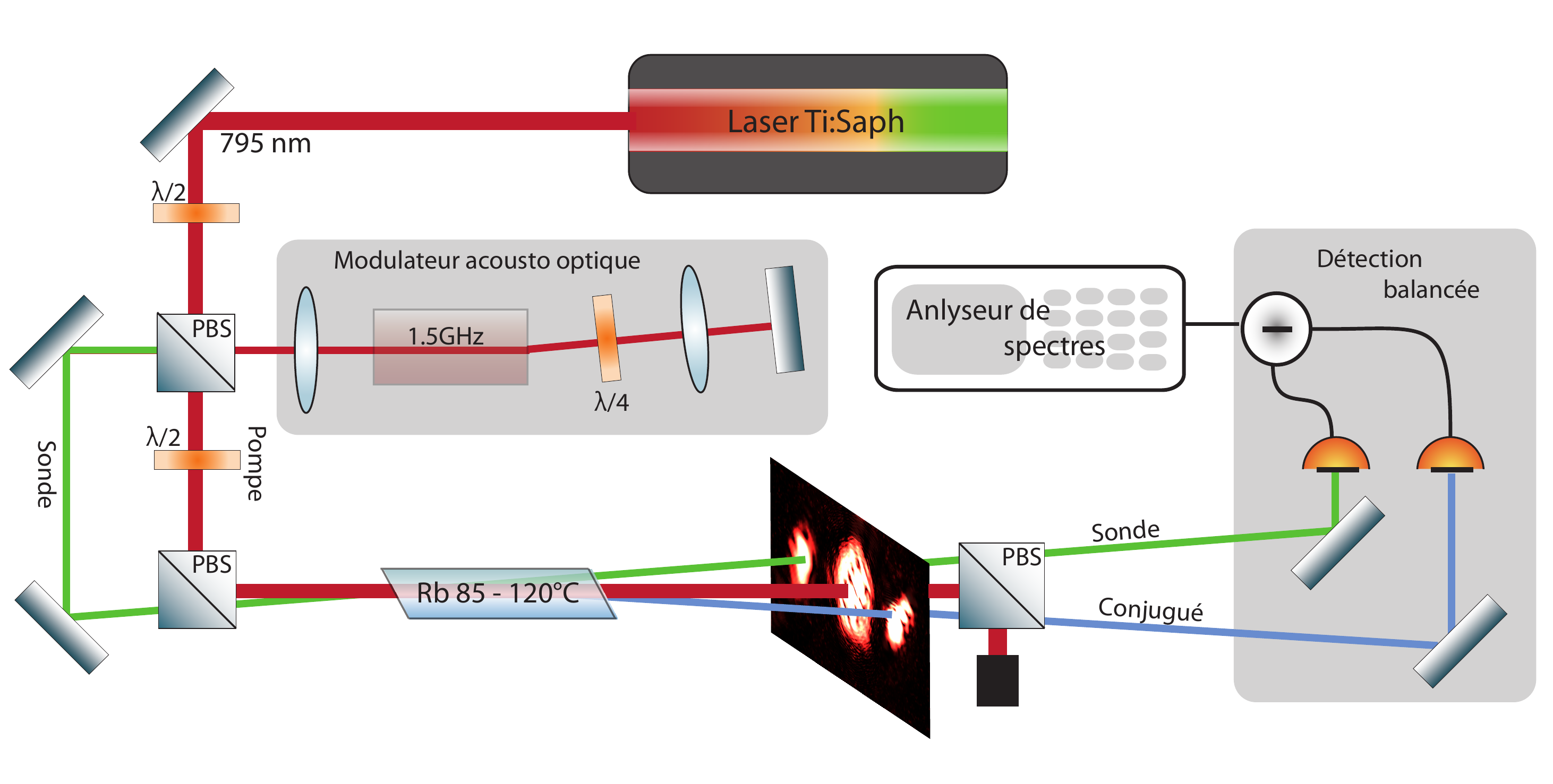} 
		\caption[Schéma du dispositif expérimental.]{Schéma du dispositif expérimental utilisé pour la génération d'états comprimés à deux modes. Un champ pompe intense (trait rouge) et un champ sonde décalé en fréquence de 3GHz (trait vert) interagissent dans une vapeur de rubidium 85. Le résultat de cette interaction est une amplification du faisceau sonde et la génération du faisceau conjugué (trait bleu).}
		\label{manip1}
		\end{figure}

Nous utilisons un montage expérimental similaire à celui utilisé par \cite{McCormick:2007p652} et détaillé dans \cite{GlorieuxSPIE2010}.
Ce montage est décrit sur le schéma de la figure \ref{manip1}.\\

\begin{figure}
\centering
\includegraphics[width=14cm]{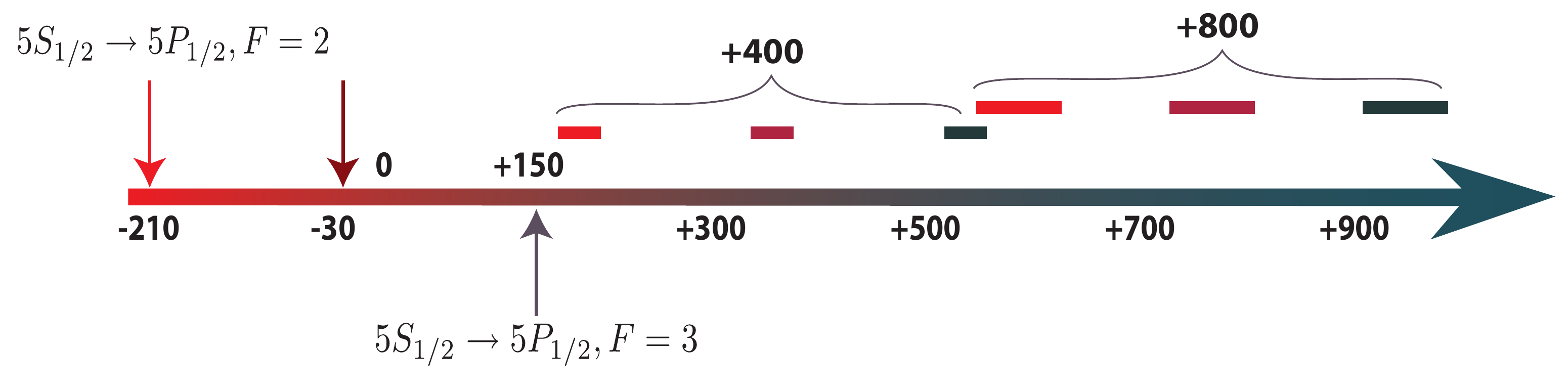} 
\caption[Échelle de fréquence pour la ligne D1 du rubidium 85.]{Échelle de fréquence pour la ligne D1 du rubidium 85. 
L'état fondamental considéré  est $5S_{1/2}, F=2$.
Les flèches indiquent la position des deux niveaux hyperfins ($F=2$, et $F=3$) de l'état excité $5P_{1/2}$.
La flèche du milieu correspond au croisement de niveaux que l'on observe dans les spectres d'absorption saturée. 
On donne ensuite les plage d'accordabilité de l'asservissement du laser à l'aide d'un acousto optique de fréquence centrale 200 MHz, utilisé en double passage dans l'ordre 1 ou dans l'ordre 2.
\label{ma0}}
\end{figure}

Un faisceau pompe intense de plusieurs centaines de mW est directement issu d'un laser titane saphir.
Ce laser est asservi en fréquence sur une référence atomique (la ligne D1 du rubidium 85) à l'aide d'un montage basée sur l'absorption saturée (non présenté sur la figure  \ref{manip1}).
Afin de pouvoir régler la fréquence de ce laser, un modulateur acousto-optique de fréquence centrale 200 MHz, permet de se décaler par rapport à la transition.
Utilisé en double passage, ce modulateur permet d'atteindre un décalage de 400 MHz $\pm$ 20 MHz en utilisant le premier ordre diffracté et 800 MHz $\pm$ 40 MHz en utilisant l'ordre 2.
Une échelle de fréquence est donnée sur la figure \ref{ma0}, afin de repérer les gammes de désaccord que nous pouvons atteindre dans ces configurations.
Une partie du faisceau pompe est prélevée à l'aide d'un cube séparateur de polarisation, et quelques dizaines de mW (typiquement 50 mW) sont focalisés sur 50 microns dans un modulateur acousto-optique de fréquence centrale 1.5 GHz.
Une configuration en oeil de chat permet d'obtenir un faisceau que l'on appelle faisceau sonde de plusieurs dizaines de $\mu$W décalé de $3$ GHz.
La plage de décalage que l'on peut atteindre dans ce montage est de 3020 à 3060 MHz.
Le modulateur est alimenté par le générateur de signaux Rhode$\&$Schwartz décrit au chapitre \ref{ch2}, lui même contrôlé par un ordinateur.\\

Les faisceaux sonde et pompe (de polarisations linéaires orthogonales) sont superposés à l'aide d'un cube séparateur de polarisation de la société Fichou.
En effet afin de filtrer le faisceau pompe en sortie du milieu, sa polarisation doit être très bien définie, les deux cubes de part et d'autre de la cellule doivent être de très bonne qualité.
Avec les cubes dont nous disposons avons mesuré un taux d'extinction (en l'absence de cellule) de $5\ 10^{-4}$.\\
Pour éviter la présence de photons de la pompe dans les faisceaux sonde et conjugué et augmenter le rapport signal à bruit, nous n'utilisons pas la géométrie colinéaire pour les faisceaux sonde et pompe.
Les faisceaux se croisent dans la cellule avec un angle de $5\ mrad$.
Selon les expériences présentées, nous utiliserons une cellule de 12.5 mm dont les faces sont traitées anti-reflet ou de 50mm dont les faces sont à l'angle de Brewster.
Une image de ce que l'on observe à l'aide d'un caméra dans le plan transverse à l'axe de propagation est présentée sur la figure \ref{manip1}.
Après le cube, un filtrage spatial (un diaphragme réglable sur le trajet de chacun des faisceaux sonde et conjugué), permet d'éliminer la présence éventuelle de photons parasites.\\
Les faisceaux sondes et conjugués sont alors focalisés sur deux photodiodes et la différence des photocourants est envoyé sur un analyseur de spectre, pour être étudié.
Une sortie basse fréquence sur chacune des photodiodes permet d'enregistrer la valeur moyenne de chacun des photocourants et ainsi connaitre le gain du processus.\\

Dans une telle expérience, il est difficile de passer d'une mesure de la différence d'intensité des faisceaux sonde et conjugué à une mesure du bruit quantique standard sur un faisceau de référence.
L'utilisation de miroir à bascule n'étant pas adaptée à une automatisation de l'expérience, nous avons rempli au préalable une base de données des spectres de bruit à la limite quantique standard, afin de calibrer le système en fonction de la puissance incidente.
Ces mesures ont été présentées au chapitre 2.
En fonctionnement normal, lors des expériences de mélange à 4 ondes, une acquisition de la valeur moyenne du photocourant est faite en temps réel sur un ordinateur.
A l'aide de cette valeur, le spectre de bruit à la limite quantique standard est généré numériquement et comparé au spectre de bruit enregistré par l'analyseur de spectre.
On obtient donc en temps réel, le niveau de corrélations par rapport à la limite quantique standard, c'est à dire exactement $S_{N_-}$\footnote{Expérimentalement, il faut corriger les valeurs mesurées, par le bruit électronique de la chaine de détection et  éventuellement par les pertes de propagation et de détection (l'efficacité quantique des photodiodes) pour obtenir $S_{N_-}$. Les résultats que nous présenterons, ne prennent uniquement en compte que la correction du bruit électronique.}.

\vspace{1cm}

\section{Caractérisation du milieu atomique en présence d'un champ pompe}
Pour caractériser le milieu atomique, nous avons fait une série d'expériences où nous avons mesurés uniquement des quantités classiques : les gains sur le champ sonde et conjugué ($G_a$ et $G_b$).
Dans un premier temps nous allons présenter un spectre de gain typique en fonction de $\Delta$ que nous comparerons aux prédictions théoriques.
Puis, dans un deuxième temps, nous étudierons comment ces spectres sont modifiés en fonction de la température de la cellule et du désaccord à 2 photons $\delta$.
\begin{figure}
\centering
\includegraphics[width=14.8cm]{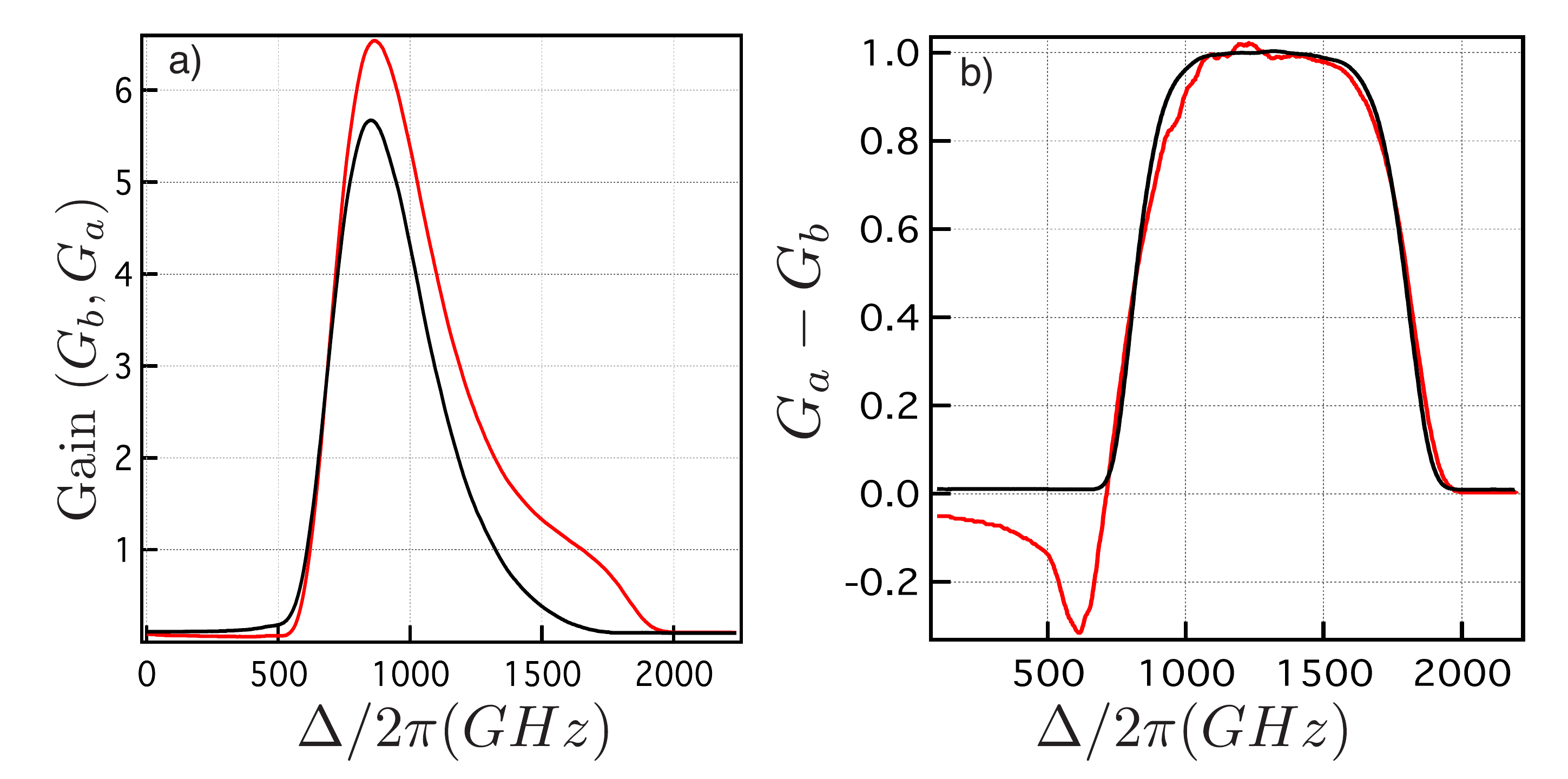} 
\caption[.]{a) Gain en fonction du désaccord à un photon, $\Delta$, pour les champs sonde (rouge) et conjugué  (noir). b) Différence entre le gain $G_a$ et $G_b$ (en rouge) comparée au spectre d'absorption du faisceau sonde en l'absence de champ pompe (noir).
Paramètres : $\delta/2\pi=2$ MHz, $T=$110$ ^\circ$C, $P_{pompe}=500$ mW, $L=$5 cm (longueur de la cellule).
\label{fig53}}
\end{figure}
\clearpage
\subsection[Gain en fonction du désaccord à un photon]{Gain en fonction du désaccord à un photon : $\Delta$}

La figure \ref{fig53} présente les courbes de gain en fonction de $\Delta$ pour des paramètres typiques (voir légende).
On peut noter sur la figure \ref{fig53} a) que le gain, aussi bien de la sonde que du conjugué, atteint un maximum pour un désaccord d'environ 900 MHz\footnote{Les désaccords $\Delta$ sont toujours donnés par rapport à la transition $5S^{1/2},F=3\to 5P^{1/2}$. La structure hyperfine du niveau excité n'est pas prise en compte.}.
L'écart entre le gain $G_a$ et $G_b$ à ce point, est proche de 1, comme dans le cas de l'amplificateur idéal décrit au chapitre 3.
Comme nous l'avons déjà expliqué, à plus grand désaccord, l'efficacité du processus de mélange à 4 ondes chute car l'interaction lumière matière s'éloigne de résonance \cite{Lukin:2000p1648}.
A l'inverse, proche de résonance, le gain devrait augmenter.
Comme on peut le constater sur la figure a), au lieu d'augmenter, le gain de la sonde chute vers zéro.
L'absorption par les atomes non préparés dans l'état stationnaire que nous avons présentée au chapitre 4 est l'effet dominant dans ce cas.\\
De plus, on peut voir que le gain du faisceau conjugué chute lui aussi vers zéro même si ce champ est loin de résonance.
Comme nous l'avons déjà proposé, le fait que la sonde soit très fortement absorbée, revient à voir le milieu en première approximation comme un milieu absorbant pour la sonde décrit par un profil Doppler d'absorption suivi, d'un milieu de gain par mélange à 4 ondes.
Dans ce cas, si le faisceau sonde est totalement absorbé, le processus de mélange à 4 ondes n'est pas injecté sur la voie sonde et le taux de génération de photons dans le mode du conjugué chute.
Il est clair que cette hypothèse simplifie grandement la complexité des processus mis en jeu, c'est à dire la compétition, au cours de la propagation, entre le gain et l'absorption.
Malgré cela, nous allons montrer qu'elle permet de décrire, de façon qualitative, les phénomènes observés.\\

Sur la figure \ref{fig53} b), on présente la différence entre le gain $G_a$ et $G_b$ (en rouge) et on la compare au spectre d'absorption du faisceau sonde en l'absence de champ pompe (noir).
Le spectre d'absorption est simplement un profil élargi par Doppler pour une épaisseur optique de plusieurs milliers (donc très plat).
Or, dans notre modèle d'un milieu absorbant suivi d'un milieu de gain, le spectre de $G_a$ en absence de champ pompe et le spectre de  $G_a-G_b$ sont égaux.
On peut constater que la différence de gain a un profil très similaire au spectre d'absorption de la sonde seule, ce qui veut dire que lorsque la sonde est absorbée d'un certain facteur, généralement le gain du conjugué chute du même facteur.\\
Autour de $500$ MHz, il y a un écart de l'ordre de 20$\%$ entre les deux courbes ; le faisceau conjugué devient même plus intense que le faisceau sonde en sortie du milieu.
Cela vient du fait que la sonde n'est pas entièrement absorbée dès l'entrée du milieu et qu'au début de sa propagation elle permet la génération d'un faisceau conjugué qui, lui, ne sera pas absorbé par la suite. \\

\begin{figure}
\centering
\includegraphics[width=14.5cm]{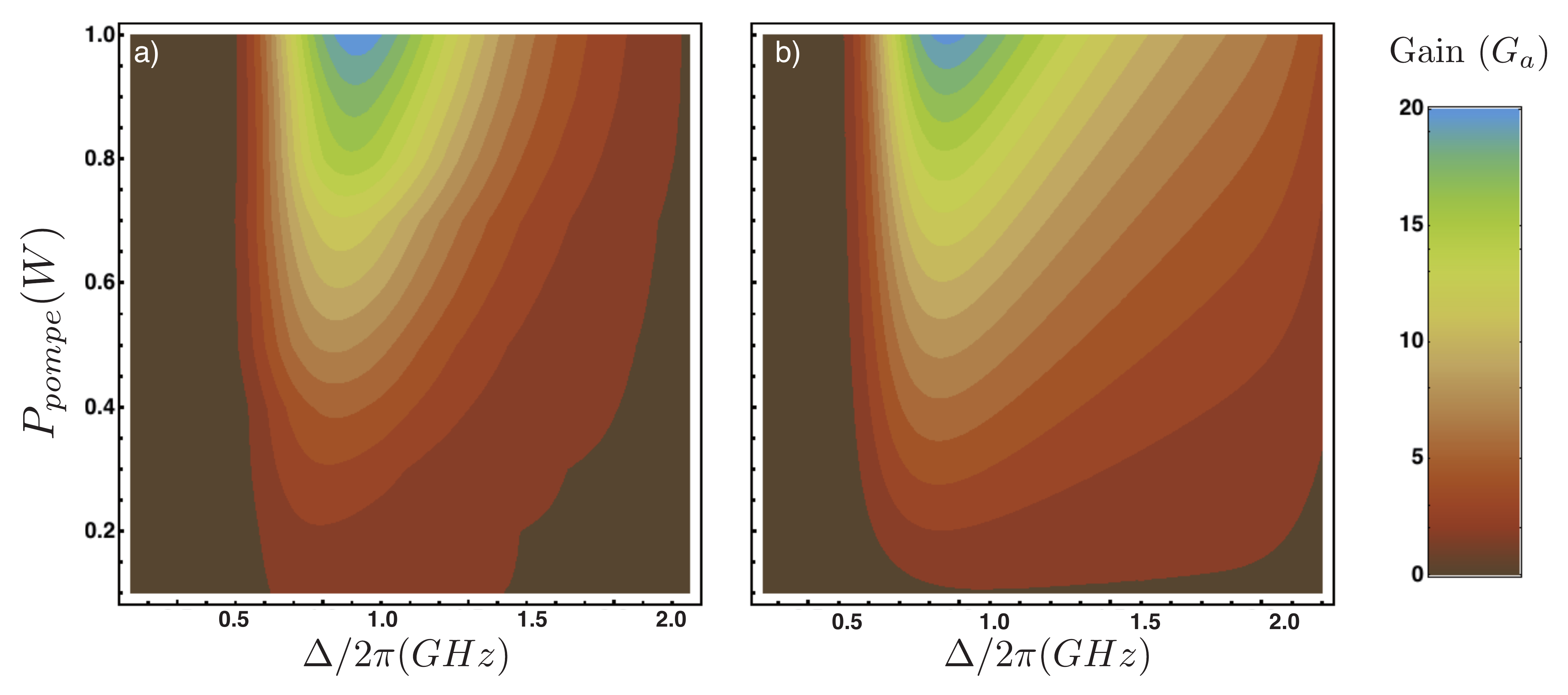} 
\caption[Cartographie du gain $G_a$ en fonction de la puissance du laser de pompe et du désaccord $\Delta$.]{\label{map1}Cartographie du gain $G_a$ en fonction de la puissance du laser de pompe et du désaccord $\Delta$. a) Résultats expérimentaux, b) simulations numériques.
Paramètres expérimentaux :  $\delta/2\pi=5$ MHz, $T$ entre 110$ ^\circ$C et 114$ ^\circ$C, $L=$5 cm (longueur de la cellule).
Données numériques : $\gamma/2\pi=500$ kHz, $\alpha L\simeq 5000$.\\
 }
\end{figure}

Afin d'étudier le rôle de l'intensité du champ pompe, on présente sur la figure \ref{map1}, une carte 2D, du gain $G_a$.
En coupe horizontale, on retrouve les spectres en fonction de $\Delta$ que nous venons d'étudier.
On peut voir sur la figure \ref{map1} a) que la position du maximum de gain se décale très légèrement vers les désaccord plus élevé à mesure que l'on augmente la pulsation de Rabi du champ pompe (le rayon du faisceau pompe dans la cellule est fixé et vaut 650$\mu$m, la puissance est donc directement proportionnelle à la pulsation de Rabi).
La figure \ref{map1} b) présente les résultats issues des simulations numériques présentées au chapitre 4.
On peut constater le bon accord avec les résultats expérimentaux. 
On notera cependant que les spectres théoriques sont plus larges que les spectres expérimentaux, notamment le gain est surestimé pour les désaccords supérieurs à 1.5 GHz.\\
Il est important de rappeler que les simulations numériques que nous présentons ici ne contiennent aucun paramètre ajustable, et que les valeurs numériques utilisées sont identiques aux paramètres expérimentaux mesurés. 

\subsection[Effet de l'épaisseur optique]{Effet de l'épaisseur optique : $\alpha$ L}
\begin{figure}
\centering
\includegraphics[width=14.5cm]{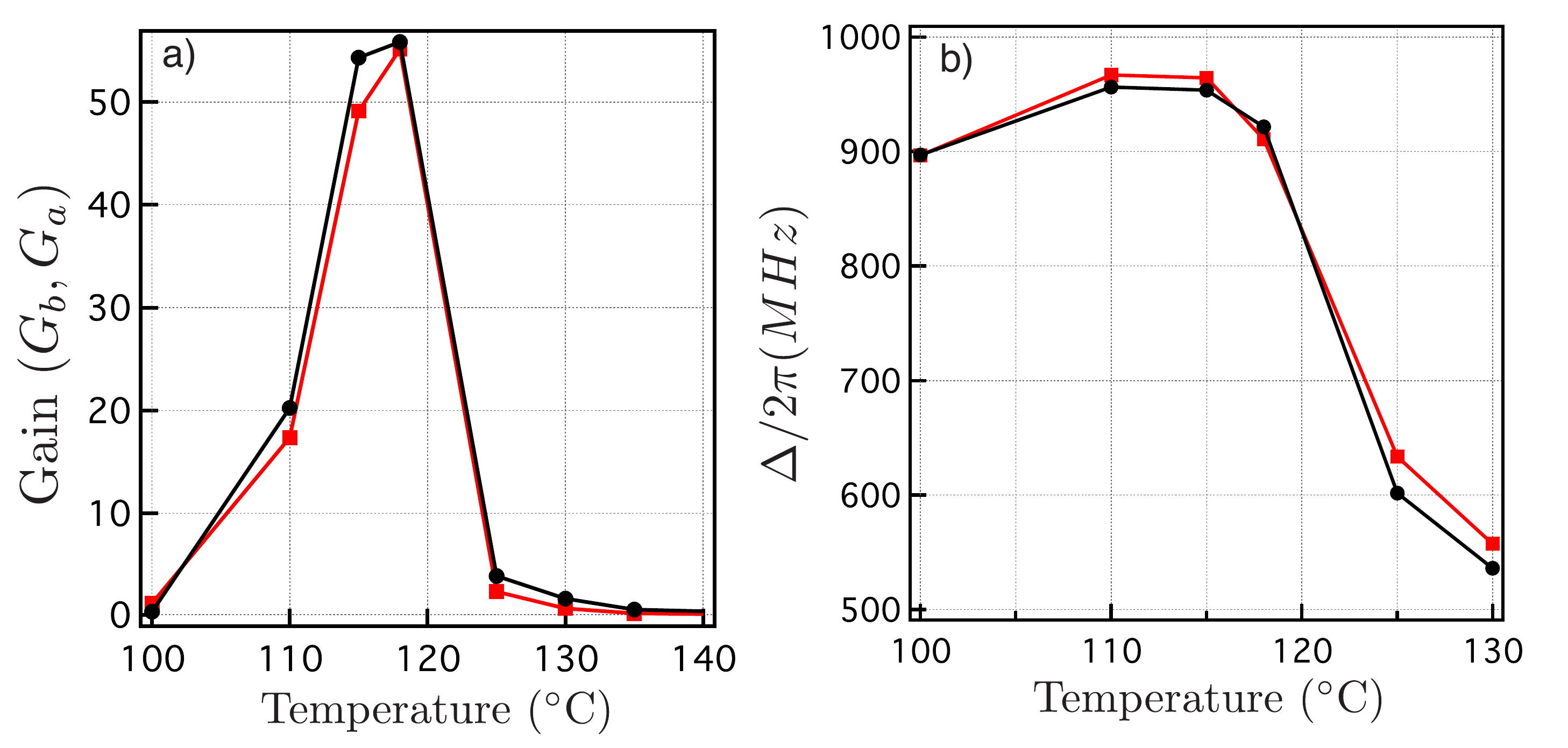} 
\caption[Effet de l'épaisseur optique sur le gain.]{a) Effet de la température sur le maximum (en fonction de $\Delta$) du gain $G_a$ (rouge) et $G_b$ (noir). b) Position (désaccord à 1 photon) du maximum de gain en fonction de la température pour la sonde (rouge) et le conjugué (noir).
Paramètres expérimentaux :  $\delta/2\pi=5$ MHz, $P=$ 400 mW.
\label{fig55}}
\end{figure}
La température est un paramètre de contrôle très important pour les expériences que nous présentons dans ce manuscrit.
Son rôle est double, d'une part modifier l'épaisseur optique en modifiant la densité d'atomes, et d'autre part augmenter la largeur du profil d'absorption Doppler en élargissant la distribution de vitesse.
Cependant, dans la gamme de température explorée (90$ ^\circ$C à 140$ ^\circ$C), le second effet est négligeable (de l'ordre de 5$\%$).
Par contre la densité atomique (et donc l'épaisseur optique) est modifiée d'un facteur 20 sur la même gamme de température.
Dans la suite, on parlera donc de l'épaisseur optique ou de le température comme du même paramètre.\\

Sur la figure \ref{fig55} a), on peut voir qu'il existe une plage de températures relativement étroite pour laquelle, le gain devient très important.
Cette plage se situe (pour une cellule de 5 cm) autour de 115$ ^\circ$C, ce qui correspond à une épaisseur optique de l'ordre de 6000.
Au dela, on peut voir que l'efficacité chute brusquement.
Cela s'explique par les phénomènes d'auto-focalisation qui apparaissent à ces températures (voir chapitre 2).
En dessous de cette plage, l'épaisseur optique est trop faible (trop peu d'atomes en interaction) pour observer l'amplification du champ sonde avec les paramètres utilisés.\\
La figure b) donne la position (désaccord à 1 photon) du maximum de gain en fonction de la température.
On voit que dans la plage où l'amplification est grande, le gain est maximum autour de $900$ MHz.
Dans la suite on se placera donc autour de cette valeur de gain entre 110$ ^\circ$C  et 120$ ^\circ$C.

\subsection[Effet du désaccord à 2 photons]{Effet du désaccord à deux photons : $\delta$}

\begin{figure}
\centering
\includegraphics[width=14.5cm]{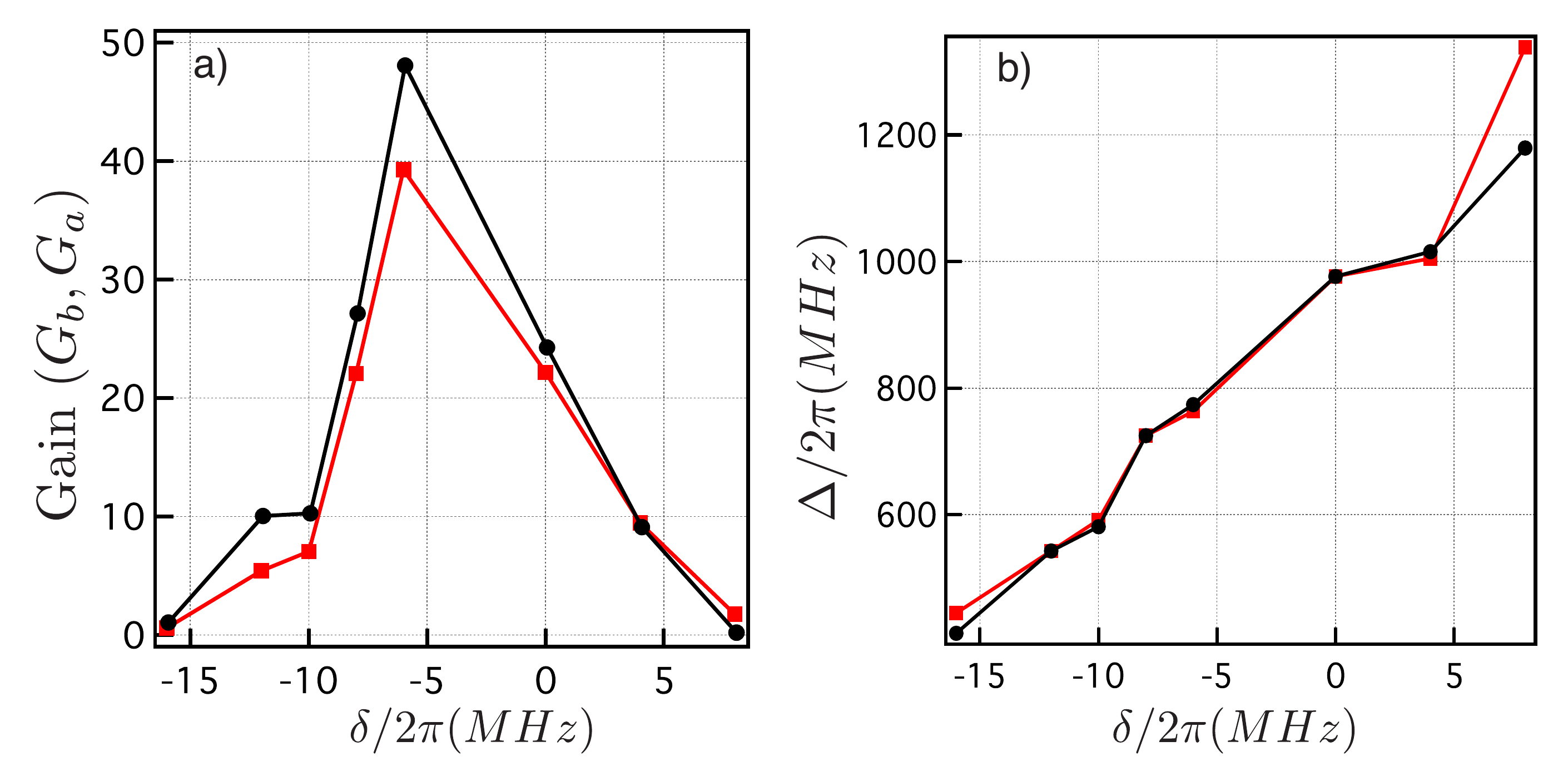} 
\caption[Effet du désaccord à 2 photons sur le gain.]{a) Effet du désaccord à deux photons, $\delta$ sur le maximum du gain $G_a$ (rouge) et $G_b$ (noir). b) Position (désaccord à 1 photon) du maximum de gain en fonction de $\delta$ pour la sonde (rouge) et le conjugué (noir).
Paramètres expérimentaux :  $T=$ 110$ ^\circ$C, $P=$ 400 mW.
\label{fig56}}
\end{figure}
A une température fixée, on fait varier le désaccord à deux photons et on cherche le maximum de gain en fonction de $\Delta$.
La figure \ref{fig56} présente les résultats obtenus.
Des gains très importants (supérieur à 50) peuvent ainsi être atteints pour des désaccords légèrement négatifs.
Dans ces régimes le gain sur le faisceau conjugué devient supérieur au gain sur le faisceau sonde.\\
On le vérifiera dans la section suivante, mais il semble clair que ce régime n'est pas le plus favorable à la génération d'un fort taux de corrélation quantiques.
En effet, il parait préférable de travailler dans un régime, où l'intensité de la sonde et du conjugué sont égales, afin de maximiser le taux de  de ``photons corrélés'' détectables.\\
Sur la figure \ref{fig56} b), nous avons tracé la position du maximum de gain en fonction de $\delta$. 
On peut noter que le désaccord à un photon qui maximise la gain augmente avec $\delta$, comme nous pouvons le vérifier numériquement à l'aide de notre modèle.

\section{Corrélations quantiques à 795 nm }
L'étude du gain dans les expériences de mélange à 4 ondes, nous a permis de caractériser le milieu atomique utilisé.
Nous présentons maintenant les mesures de corrélations quantiques en variables continues obtenues dans cette configuration.
Ce travail constitue le coeur des résultats expérimentaux présentés dans ce manuscrit.
Il s'agit d'une étude approfondie des paramètres influant sur le taux de corrélations quantiques observées.
Cette étude a été réalisée avec les photodiodes de série du montage amplificateur Thorlabs PDB-150, décrites en détail au chapitre 2.
Ces photodiodes ont une efficacité quantique de 85$\%$ à 795 nm.
Si pour une étude qualitative de l'effet des différents paramètres, l'efficacité quantique des photodiodes n'est pas crucial, elle le devient lorsque l'on souhaite mesurer des forts taux de corrélations. 
Nous présenterons donc, dans un deuxième temps, les résultats que nous avons obtenus après changement des photodiodes et optimisation de la photodétection.
Enfin le régime de génération de corrélations sans gain sera étudié expérimentalement.
\subsection{Mesures des corrélations quantiques}
\begin{figure}
\centering
\includegraphics[width=9.5cm]{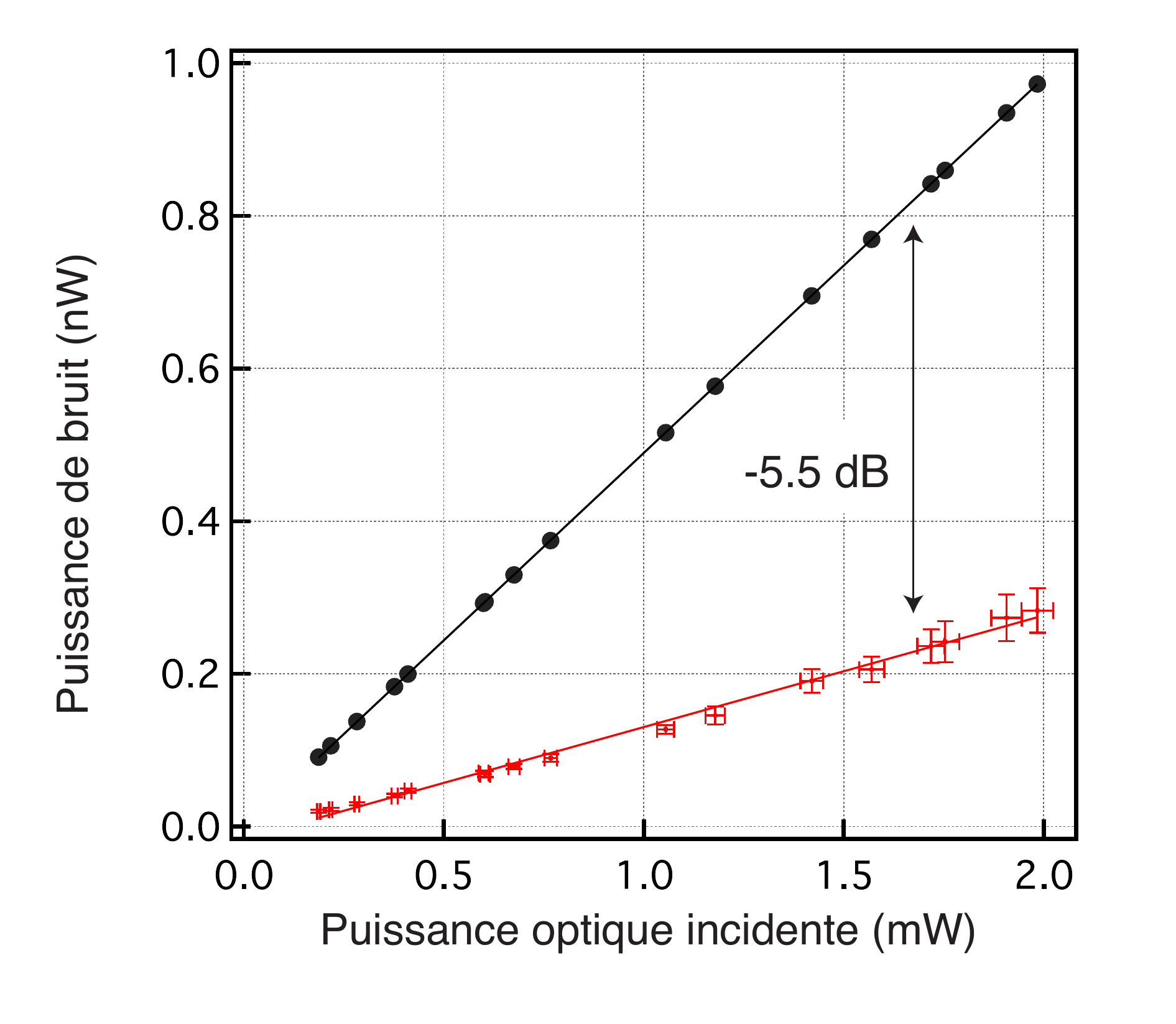} 
\caption[Puissance de bruit mesurée par détection balancée.]{Puissance de bruit à 1.5 MHz mesurée par détection balancée pour un faisceau cohérent (noir) et pour les faisceaux sonde et conjugué (rouge). La puissance incidente correspond à la somme de la puissance des deux voies. Le rapport de la pente des ajustements linéaires donne le taux de compression.\\
Paramètres expérimentaux : $\delta/2\pi=10$ MHz, $\Delta/2\pi=900$ MHz, $P=$~1,08~W, $T=$~114$ ^\circ$C, $L=$5 cm, RBW = 100 kHz, VBW = 10 Hz.
\label{fig57}}
\end{figure}
Comme nous l'avons présenté au chapitre 2, les corrélations quantiques sont mesurées à l'aide de deux photodiodes et d'une détection balancée.
Dans un premier temps, nous avons fait varier la puissance incidente du faisceau sonde afin de vérifier la linéarité de la réponse.
Les résultats obtenus sont présentés sur la figure \ref{fig57}.
Le bruit quantique standard est mesuré pour différentes puissances incidentes sur les photodiodes (courbe noire) et il est comparé au bruit de la différence d'intensité entre les faisceaux sonde et conjugué (généré par mélange à 4 ondes) pour différentes valeurs de puissance optique en sortie du milieu.\\
Ces deux séries de données sont décrites à l'aide d'un ajustement linéaire dont le rapport entre les deux pentes permet de déterminer le taux de compression sous la limite quantique standard.
Sur la figure \ref{fig57}, on observe une réduction du bruit sur la différence d'intensité de $-5.5$ dB par rapport à la limite quantique standard.
On peut bien sûr évaluer le taux de compression à l'aide d'une seule valeur de puissance incidente mais cette méthode nous permet d'augmenter la précision.
\subsection[Effet du désaccord à deux photons]{Effet du désaccord à deux photons : $\delta$}
De la même manière que nous l'avons étudié pour le gain, nous allons nous intéresser à l'effet du désaccord à deux photons sur les corrélations.
\begin{figure}
\centering
\includegraphics[width=13.5cm]{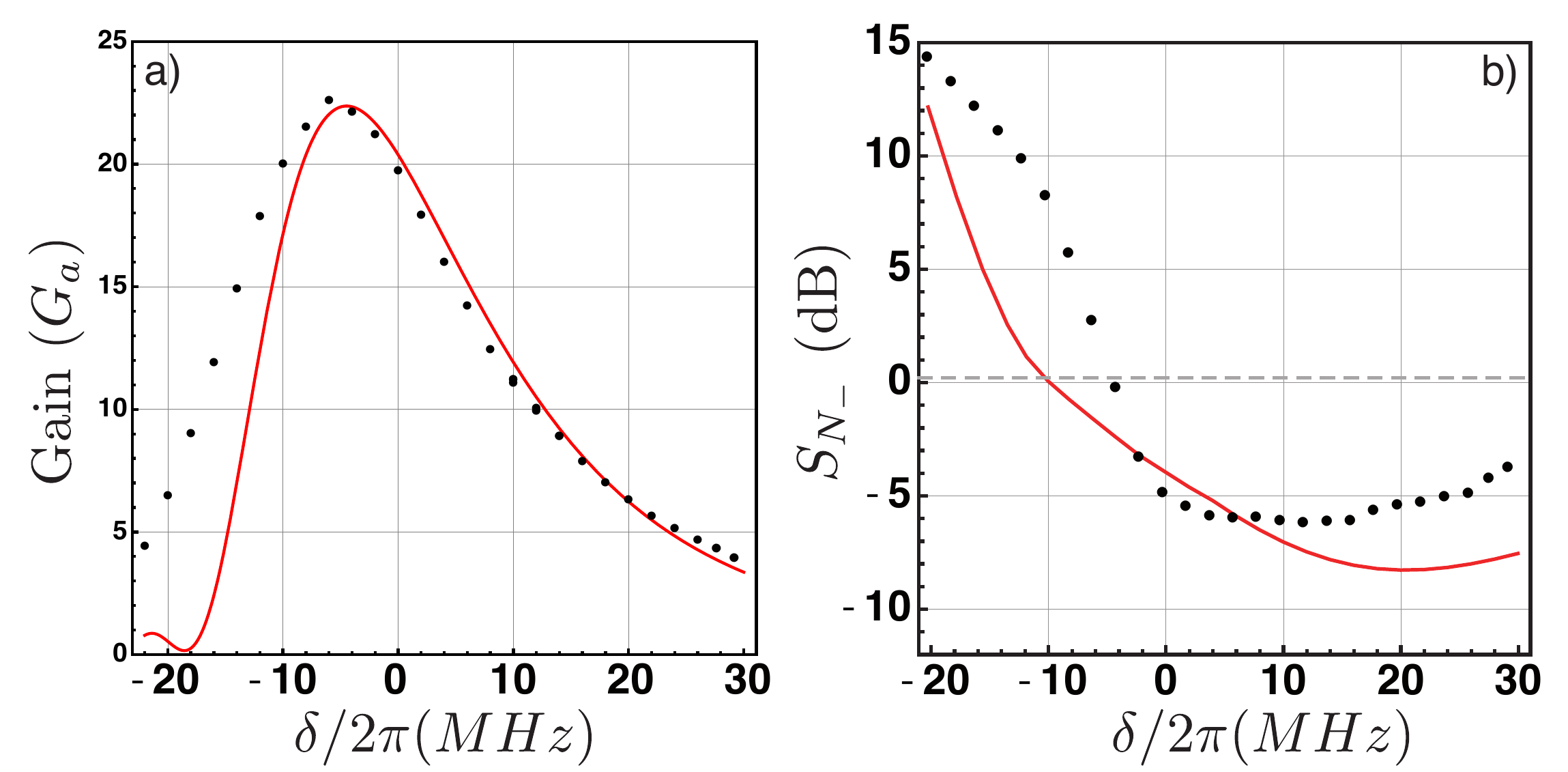} 
\caption[Effet du désaccord à 2 photons.]{Effet du désaccord à deux photons sur a) le gain de la sonde et sur b) les corrélations entre les champs sonde et conjugué. En noir, les données expérimentales et en rouge le modèle théorique.\\
 Paramètres expérimentaux : $\Delta/2\pi=1$ GHz, $P=$~1,08~W, $T=$~114$ ^\circ$C, $L=$5 cm, RBW = 100 kHz, VBW = 10 Hz.
Données numériques correspondantes : $\alpha L=5900$, $\gamma/2\pi=500$ kHz, $\Omega/2\pi$=0.47 GHz.
\label{fig58}}
\end{figure}
La figure \ref{fig58} présente les résultats expérimentaux que nous avons obtenus comparés aux simulations numériques dans le modèle qui décrit un milieu constitué ``d'atomes chauds''.
La figure \ref{fig58} a) donne le gain pour le champ sonde en fonction de $\delta$, tandis que la figure \ref{fig58} b) présente le bruit de la différence d'intensité à 2 MHz par rapport à la limite quantique standard en fonction de $\delta$.
Comme nous l'avons déjà signalé, le gain est maximal autour de $\delta= -5$ MHz, ce qui se comprend en prenant en compte le déplacement lumineux engendré par le faisceau sonde.
Par contre les corrélations entre les faisceaux sonde et conjugué sont plus importantes et passent sous la limite quantique standard pour des désaccord positif et donc des gains plus faibles.
On peut voir sur la figure \ref{fig58} b) que le maximum de corrélations est obtenu pour $\delta \simeq 5$ MHz, mais que la courbe est très plate entre 2 et 20 MHz.
Le modèle théorique prévoit aussi, un excès de bruit pour des désaccords inférieur à $\delta =-10$ MHz et une zone de $\delta$ qui permet la génération de corrélation sous la limite quantique standard.
Même si l'accord quantitatif n'est pas parfait\footnote{Il faut rappeler qu'aucun paramètre ajustable n'est introduit dans le modèle et que les paramètres sont fixés par ceux utilisés dans l'expérience.}, le modèle permet de déterminer la gamme dans laquelle, il est souhaitable de travailler et le niveau de corrélations atteignable.
Les données expérimentales présentées sont corrigées du bruit électronique, et les données numériques sont corrigées par l'efficacité quantique des photodiodes utilisées (85$\%$)
Dans ce cas les pertes sur la propagation qui sont de l'ordre de 3$\%$ sont négligées par rapport à l'efficacité quantique.

\subsection[Effet du désaccord à un photon]{Effet  du désaccord à un photon : $\Delta$}
\begin{figure}
\centering
\includegraphics[width=14cm]{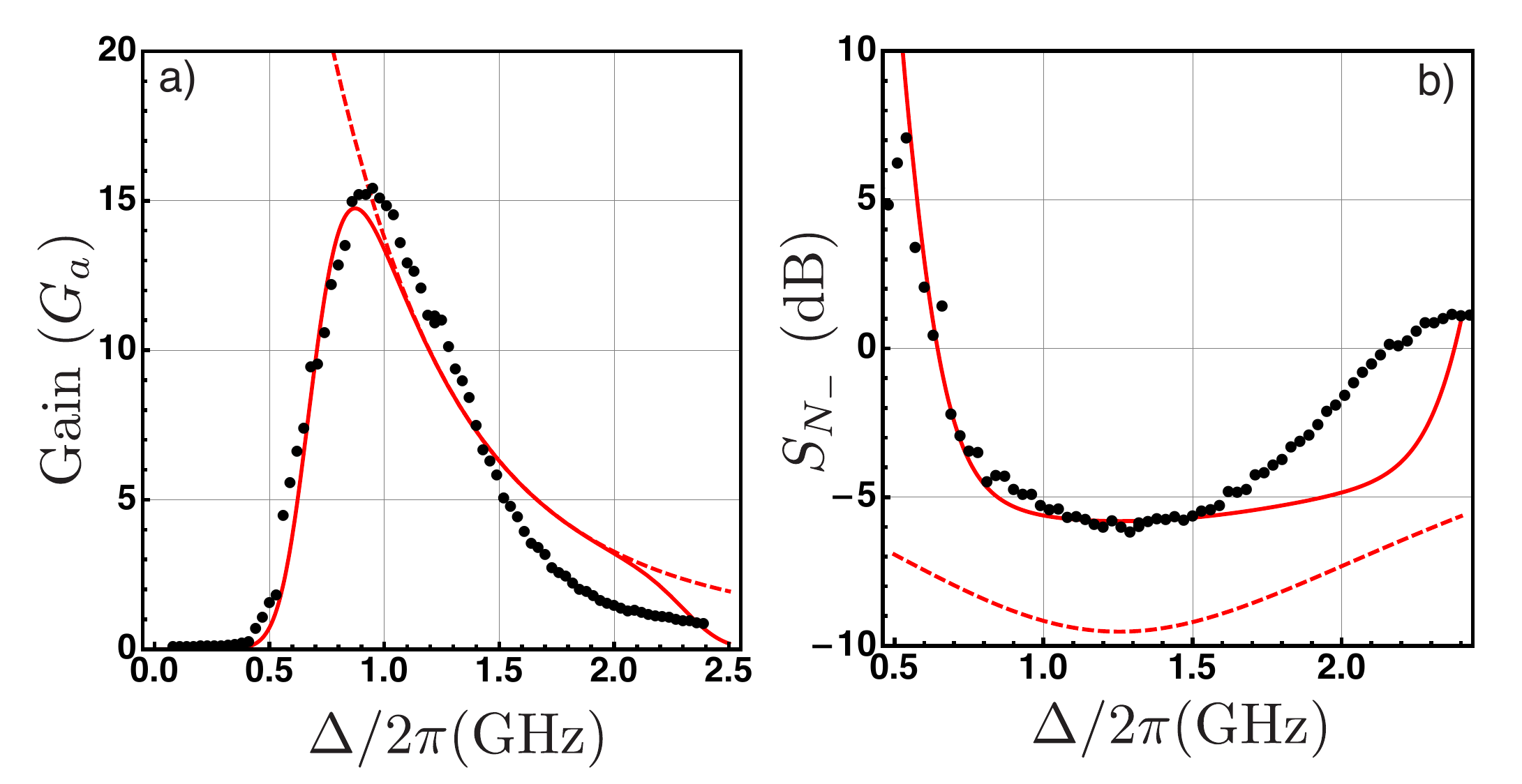} 
\caption[Effet du désaccord à 1 photon.]{Effet du désaccord à un photon sur a) le gain de la sonde et sur b) les corrélations entre les champs sonde et conjugué. En noir, les données expérimentales et en rouge le modèle théorique prenant en compte de l'absorption, en rouge pointillé le modèle théorique non corrigé par l'absorption (ni l'efficacité quantique).\\
 Paramètres expérimentaux : $\Delta/2\pi=1$ GHz, $P=$~1,08~W, $T=$~114$ ^\circ$C, $L=$5 cm, RBW = 100 kHz, VBW = 10 Hz.
Données numériques correspondantes : $\alpha L=5900$, $\gamma/2\pi=500$ kHz, $\Omega/2\pi$=0.47 GHz.
\label{fig59}}
\end{figure}
La figure \ref{fig59} présente l'effet du désaccord à un photon $\Delta$.
Sur la figure \ref{fig59} a), nous comparons les résultats expérimentaux avec deux calculs différents.
Dans un premier cas (pointillé rouge) nous ne prenons pas en compte l'absorption par les atomes non préparés dans l'état stationnaire.
Dans ce cas, on voit que le gain augmente vers la résonance, ce qui n'est pas le cas expérimentalement.
Dans une seconde approche (en rouge), nous avons pris en compte l'effet de l'absorption.
On voit dans ce cas que l'accord avec les résultats expérimentaux est bien meilleur.
Il reste cependant, un écart pour des désaccords au delà de 1.5 GHz.
Cet écart peut s'expliquer par une prise en compte incomplète du rôle de la transition $5S^{1/2},F=2\to5P^{1/2}$ pour le faisceau sonde.
En effet au delà de $\Delta=1.5$ GHz, le faisceau sonde est plus proche de la résonance pour la transition $5S^{1/2},F=2\to5P^{1/2}$ que pour la transition $5S^{1/2},F=3\to5P^{1/2}$.
Or dans le modèle que nous avons développé, seule la transition $5S^{1/2},F=3\to5P^{1/2}$ est prise en compte pour le faisceau sonde.
Ainsi un effet qui était négligeable pour $\Delta<1.5$ GHz, cesse de l'être et notre modèle microscopique ne permet plus de décrire correctement les processus mis en jeu.\\

Sur la figure \ref{fig59} b), on présente l'effet de $\Delta$ sur les corrélations générées.
On peut noter que le maximum de corrélations est atteint pour des valeurs de $\Delta$ légèrement plus élevées que le maximum de gain.
En effet, comme la valeur du gain dépend à la fois de l'amplification et de l'absorption, alors le maximum est un équilibre entre les deux et il se situe dans une zone où l'on observe de l'absorption.
Comme les pertes vont détruire les corrélations quantiques entre nos deux faisceaux, il est normal de trouver l'optimum des corrélations décalés vers une zone où l'absorption est plus faible, c'est à dire vers $\Delta$ plus élevé.\\
Par contre pour des valeurs de $\Delta$ encore plus grandes, le gain devient faible et le taux de photons corrélés générés diminuent.
L'optimum est donc à nouveau un compromis entre l'absorption et le gain.
Sa position sera en général (cela dépend des différents paramètres) autour de 1 GHz.\\
On peut constater que le modèle théorique simplifié que nous présentons, consistant à en compte l'absorption (courbe rouge), uniquement pour  la normalisation des spectres de bruit par le bruit quantique standard, permet de rendre compte de ces différents effets.
Comme pour le gain, nous pouvons noter que ce modèle surestime la valeur des corrélations pour des désaccord autour de 2 GHz.

\subsection[Effet de la pulsation de Rabi]{Effet de la pulsation de Rabi : $\Omega_R$}
\begin{figure}
\centering
\includegraphics[width=14.5cm]{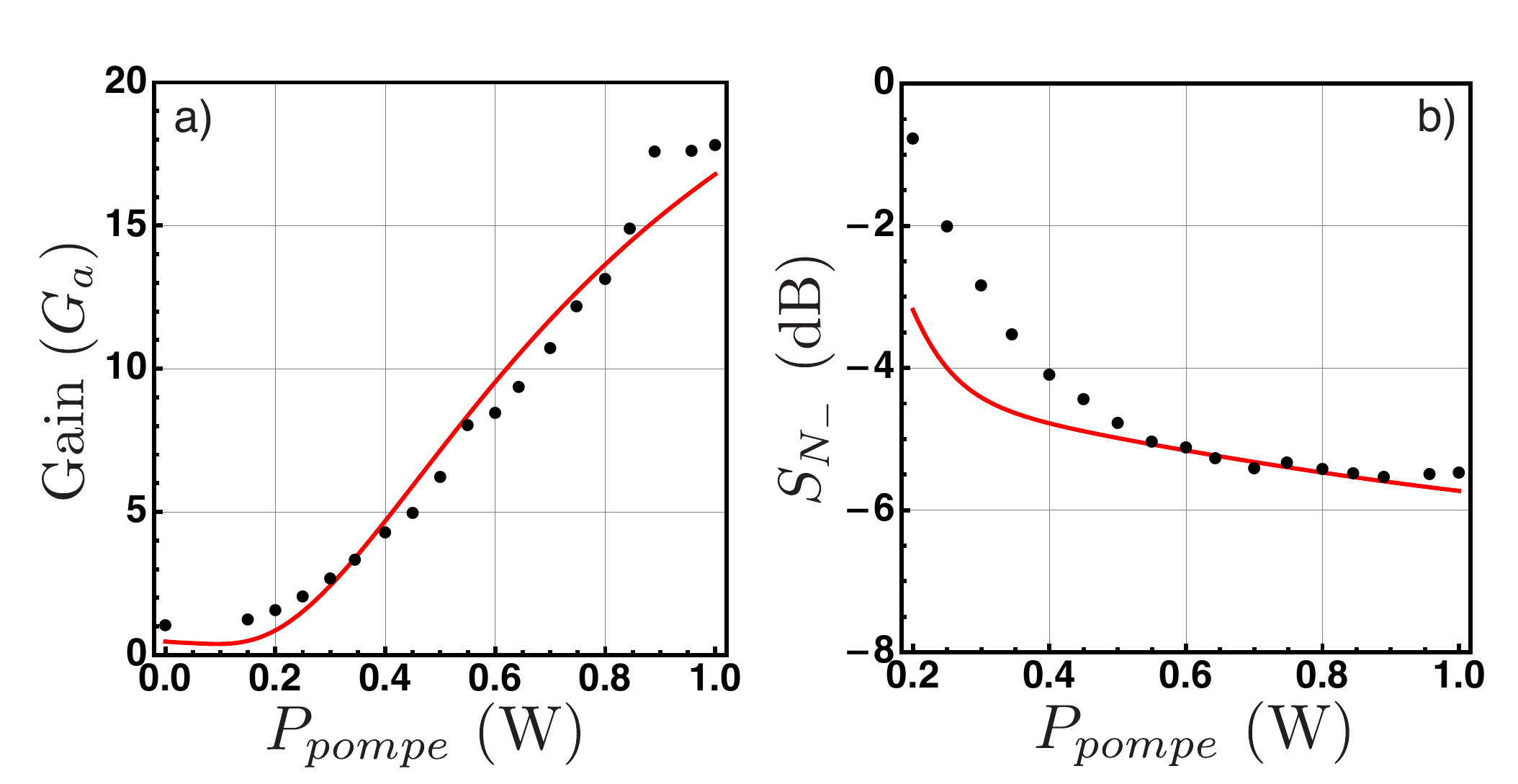} 
\caption[Effet de la pulsation de Rabi.]{Effet de la puissance de pompe sur a) le gain de la sonde et sur b) les corrélations entre les champs sonde et conjugué. En noir, les données expérimentales et en rouge le modèle théorique.\\
 Paramètres expérimentaux : $\Delta/2\pi=1$ GHz, $\delta=$~10~MHz, $T=$~116$ ^\circ$C, $L=$5 cm, RBW = 100 kHz, VBW = 10 Hz.
Données numériques correspondantes : $\alpha L=6600$, $\gamma/2\pi=500$ kHz.
\label{fig510}}
\end{figure}
La puissance du faisceau laser de pompe est limitée par la puissance délivrée par la source laser.
On pourrait penser modifier l'intensité du laser de pompe et donc la pulsation de Rabi, en le focalisant plus au sein de la cellule.
Cette voie n'a pas été explorée car une réduction de la taille du faisceau pompe réduirait aussi le taux de préparation des atomes et augmenterait le taux de décohérence $\gamma$, ce que l'on ne souhaite pas a priori.
On a donc étudié, l'effet de la puissance du laser de pompe pour une focalisation donnée (waist de 650 microns).\\
La figure \ref{fig510} présente les résultats de cette étude.
Sur la figure \ref{fig510} a), on voit que la valeur du gain augmente avec la puissance de pompe, comme le prédit le modèle théorique.
Sur la figure \ref{fig510} b), on peut voir que les corrélations augmentent aussi avec la puissance de pompe.\\
On notera qu'au delà d'un certain seuil (environ 600 mW), le taux de compression ne varie que faiblement avec la puissance de pompe.
A cette température, pour des puissances supérieures à 1.5W, d'autres effets non--linéaires indésirables (autofocalisation par exemple) viennent parasiter le processus de mélange à 4 ondes et fait diminuer drastiquement les corrélations.
Ainsi, on travaillera généralement autour d'une puissance de 1W de pompe.
\subsection[Effet de l'épaisseur optique]{Effet de l'épaisseur optique : $\alpha$ L}
\begin{figure}
\centering
\includegraphics[width=14.5cm]{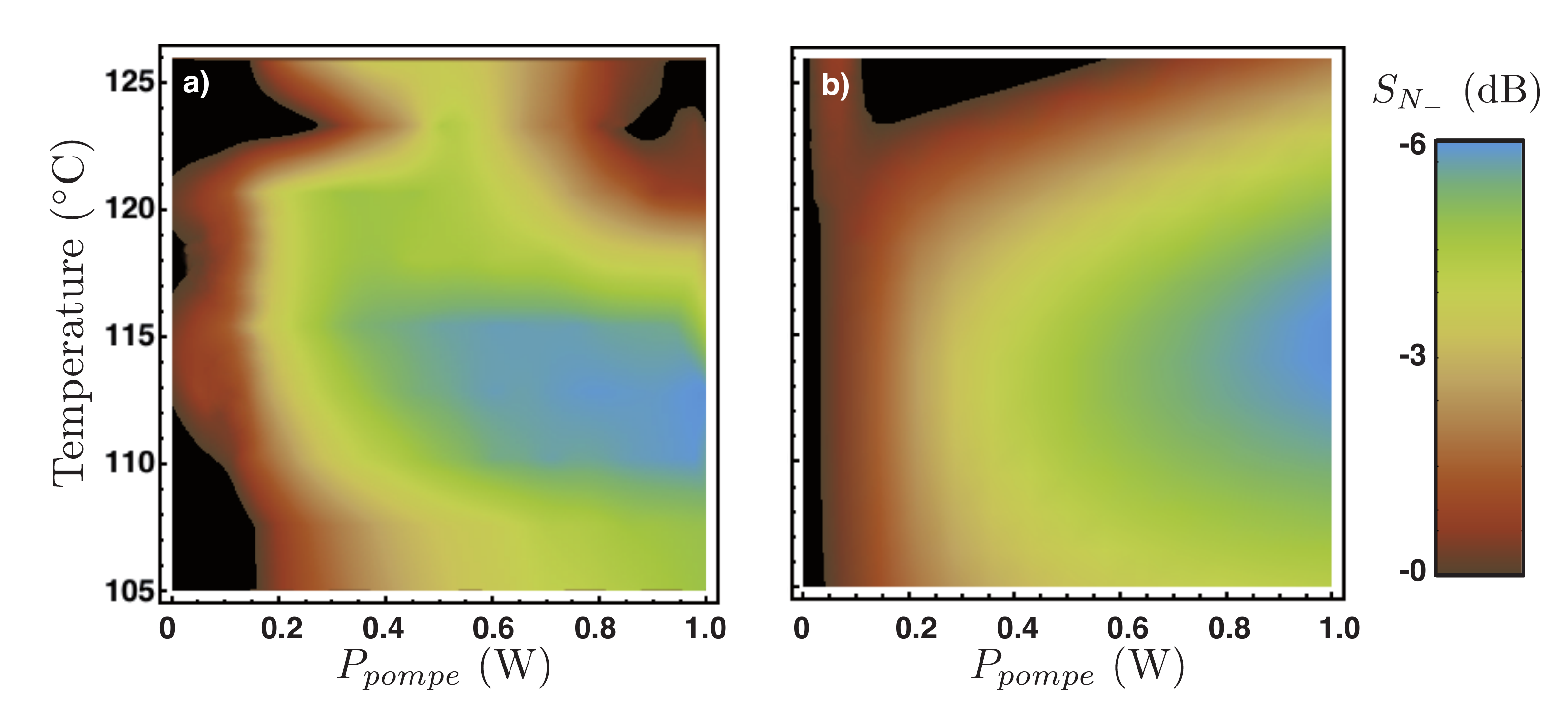} 
\caption[Effet de l'épaisseur optique.]{Effet de la température (de l'épaisseur optique) sur les corrélations entre les champs sonde et conjugué. a) les données expérimentales et b) le modèle théorique.\\
 Paramètres expérimentaux : $\Delta/2\pi=1$ GHz, $\delta=$~10~MHz, $L=$5 cm, RBW = 100 kHz, VBW = 10 Hz.
Données numériques correspondantes : $\gamma/2\pi=500$ kHz.
\label{fig511}}
\end{figure}
Nous avons étudié le rôle de l'épaisseur optique sur le niveau des corrélations générés.
Comme nous l'avons vu dans le paragraphe précédent, à 116$ ^\circ$C, augmenter la puissance de pompe permet d'accroitre les corrélations.
Les résultats présentés sur la figure \ref{fig511}, montrent que ce n'est pas le cas à toutes les températures.
On peut voir sur la figure \ref{fig511} a), que pour des températures plus élevées (125$ ^\circ$C), il existe un maximum dans le niveau des corrélations pour une puissance de pompe autour de 500 mW.
Bien que les simulations numériques présentées sur la figure \ref{fig511} b) ne prennent pas en compte les effets parasites (tel que l'auto-focalisation par exemple), on observe aussi sur cette figure, pour des densité élevées, un maximum vers 100 mW puis une diminution des corrélations.
Ainsi, on peut faire l'hypothèse que l'excès de bruit a deux origines physiques distinctes : d'une part, un effet qui n'est pas pris en compte par notre modèle et qui correspond aux effets d'auto-focalisation que l'on a pu observer expérimentalement.
Et d'autre part, un effet que l'on observe dans notre modèle et qui est donc liée à la structure microscopique du système en double lambda d'autre part, mais nous ne connaissons pas l'origine physique de ce maximum.\\
A l'aide de cette étude, on peut conclure qu'il existe une température optimale (autour de 115$ ^\circ$C  pour notre cellule de 1.25 cm de long) pour la génération de faisceaux corrélés par mélange à 4 ondes dans une vapeur atomique.

\subsection{Effet d'un champ magnétique transverse}
Dans la configuration expérimentale que nous avons décrite, le champ magnétique n'est pas contrôlé : le champ magnétique terrestre, non écranté, affecte les niveaux atomiques.
Afin d'étudier son rôle, nous avons réalisé des expériences en ajoutant un champ magnétique au niveau de la cellule.
Notre modèle théorique, décrit au chapitre 4, ne prend pas en compte les sous niveaux Zeeman, mais simplement un modèle simplifié à quatre niveaux.
Nous allons donc présenter les résultats expérimentaux obtenus, sans rentrer dans leur analyse théorique qui dépasse le cadre de ce travail de thèse. 
Le travail que nous avons réalisé ici est une étude préliminaire de l'effet d'un champ magnétique externe.
Nous rendons compte des résultats uniquement dans le cas d'un champ appliqué de manière transverse par rapport à l'axe propagation des lasers.
En effet, il serait naturel d'appliquer un champ longitudinal pour voir l'effet de la levée de dégénérescence des sous niveaux Zeeman, mais cette situation s'avère difficile expérimentalement.
En appliquant un champ longitudinal, on observe une fuite importante du faisceau pompe à travers le cube de sortie (qui traduit une rotation de la polarisation de la pompe), de telle sorte que le rapport signal à bruit sur les intensités détectées des faisceaux sonde et conjugué se détériore très fortement.\\
Ainsi, nous avons utilisé des bobines dans la configuration de Helmoltz, afin de générer un champ magnétique au centre de la cellule approximativement uniforme.
Les bobines sont constituées de 50 tours de fils de cuivre sur des anneaux en plastique de diamètre central 5 cm.
Avec le générateur de courant dont nous disposons nous pouvons atteindre un champ de 15 Gauss au centre de la cellule.\\
\begin{figure}
\centering
\includegraphics[width=14cm]{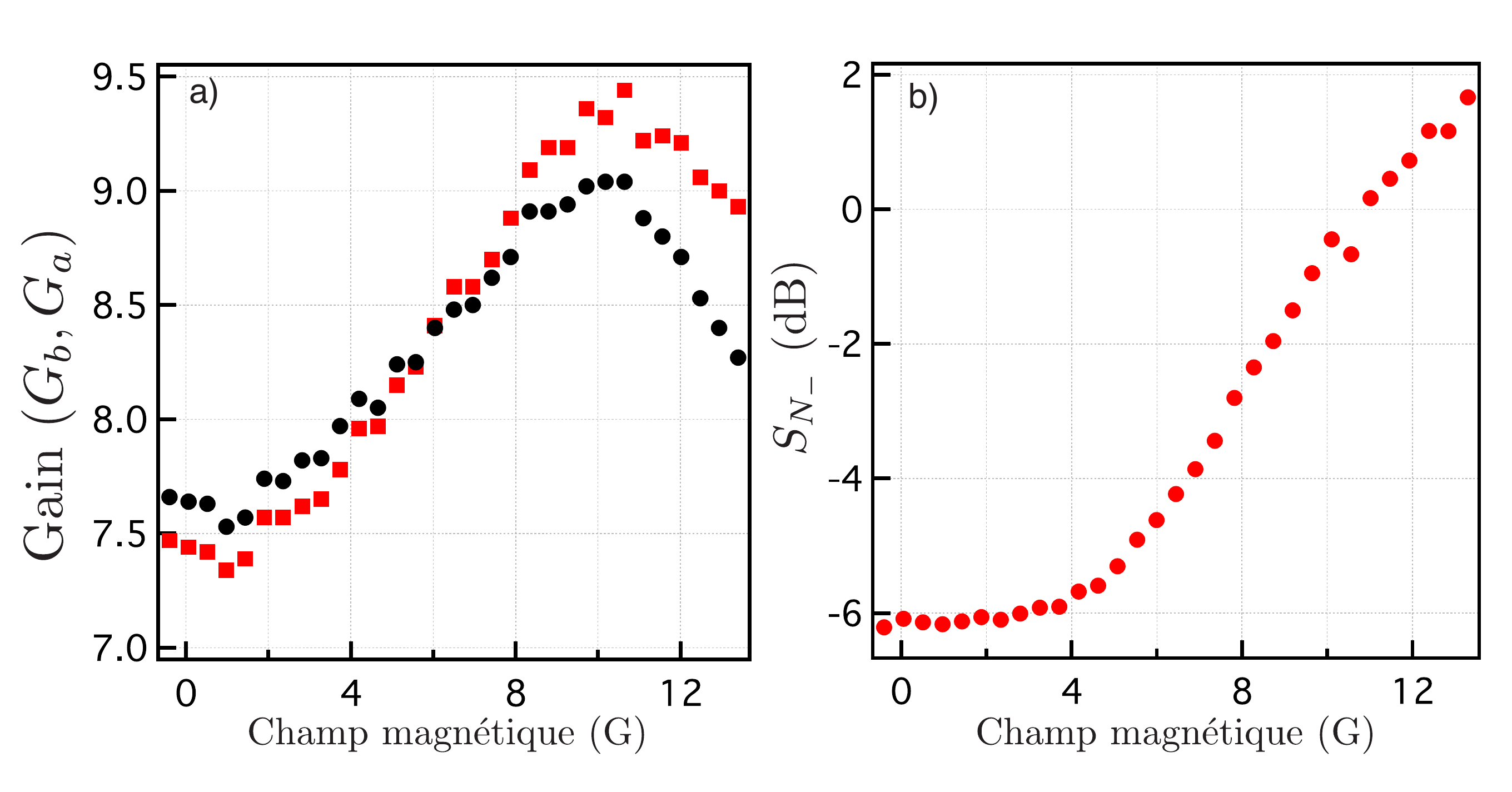} 
\caption[Effet d'un champ magnétique transverse.]{Effet d'un champ magnétique transverse sur a) le gain de la sonde (noire) et du conjugué (rouge) et sur b) les corrélations entre les champs sonde et conjugué. \\
 Paramètres expérimentaux : $\Delta/2\pi=1$ GHz, $\delta=$~4~MHz, $P=$~1,1~W, $T=$~114$ ^\circ$C, $L=$5 cm, RBW = 100 kHz, VBW = 10 Hz.
\label{champB}}
\end{figure}

Nous présentons sur la figure \ref{champB}, l'effet du champ en fixant tous les autres paramètres (voir légende).
Sur la figure \ref{champB} a), on peut constater que l'augmentation du champ magnétique fait initialement augmenter le gain sur les champs sonde et conjugué.
Il est important de noter le croisement entre les deux séries de points de la figure \ref{champB} a).
En effet, sans champ magnétique le gain sur la faisceau sonde est légèrement supérieur au gain sur le faisceau conjugué ($G_a> G_b$) pour les paramètres choisis.
Alors que, autour de 6 Gauss, les gains s'égalisent.
Le gain sur le conjugué augmentant plus vite que celui sur la sonde pour des champs entre 6 Gauss et 10 Gauss, on a alors $G_a< G_b$.\\
Bien que le régime ou $G_a\simeq G_b$ semblerait être le plus favorable pour observer des corrélations, on constate sur la figure \ref{champB} b), qu'au delà de 4 Gauss, et sans modifier le désaccord à 2 photons $\delta$, les corrélations se détériorent fortement et qu'au delà de 10 Gauss, nous n'observons plus de corrélations sous la limite quantique standard.
Les autres paramètres étant fixés, l'ajout d'un champ magnétique supérieur à 10 Gauss, va détruire les corrélations.
La plage sur laquelle l'effet du champ magnétique transverse est faible est relativement large (4 Gauss) par rapport au champ magnétique terrestre.
Ainsi, dans la plupart des cas, il semble qu'il n'est pas nécessaire de faire une protection autour de la cellule pour l'isoler des champs magnétiques parasites.\\

Nous avons réalisé de plus, une série de spectres en faisant varier $\delta$ pour différents champs magnétiques.
\begin{figure}
\centering
\includegraphics[width=14.5cm]{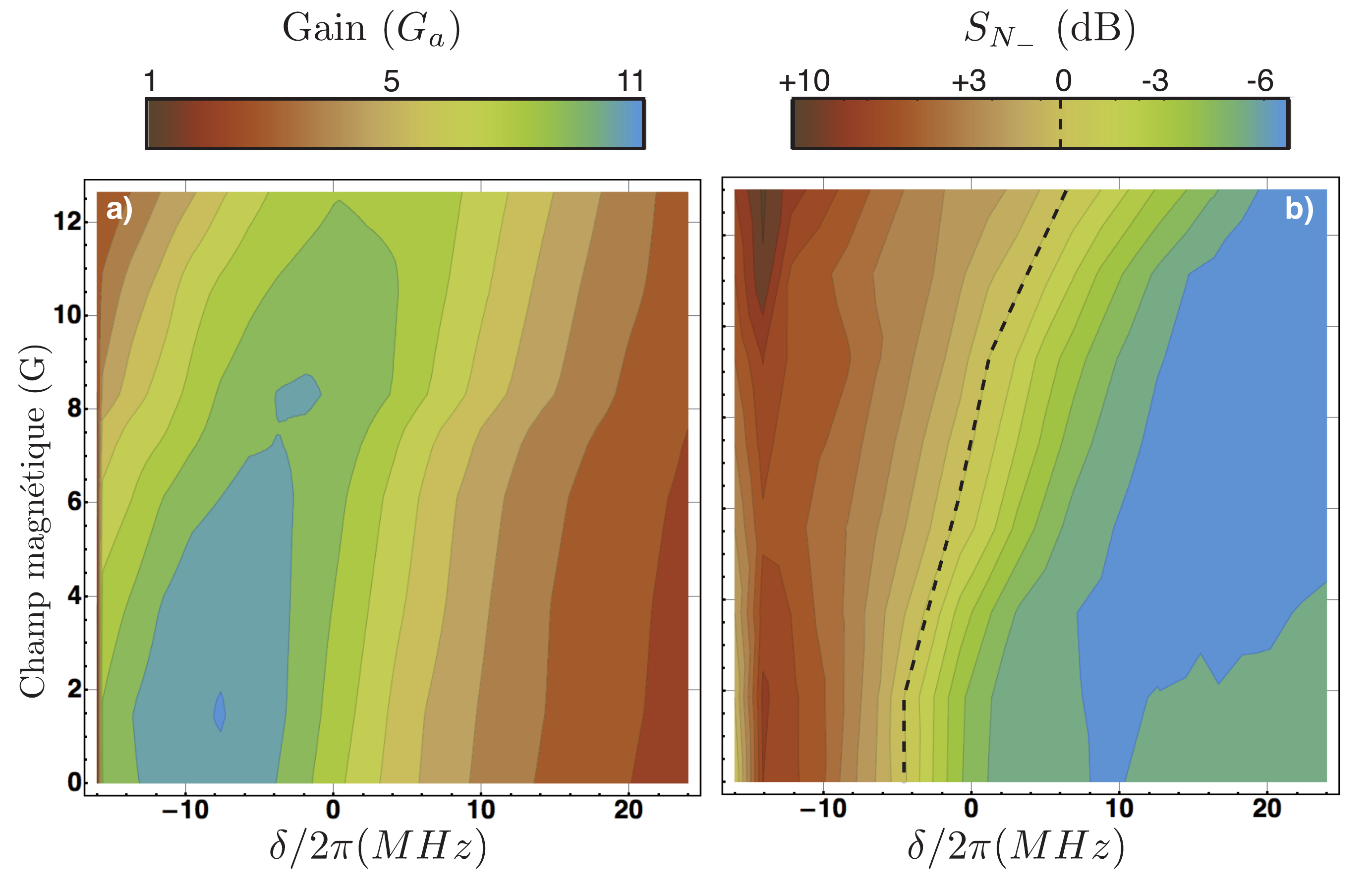} 
\caption[Effet d'un champ magnétique transverse.]{Effet d'un champ magnétique transverse sur a) le gain de la sonde et sur b) les corrélations entre les champs sonde et conjugué. \\
 Paramètres expérimentaux : $\Delta/2\pi=1$ GHz, $P=$~1,1~W, $T=$~114$ ^\circ$C, $L=$5 cm, RBW = 100 kHz, VBW = 10 Hz.
\label{B2D}}
\end{figure}
La figure \ref{B2D} présente les résultats obtenus.
Sur la figure \ref{B2D} a), on peut voir qu'appliquer un champ magnétique faible, décale le maximum de gain vers des valeurs plus grandes de $\delta$.
En effet le maximum de gain est autour de $\delta=-9$~MHz pour un champ nul et atteint $\delta=1$ MHz pour un champ de 12 Gauss.
On observe donc un décalage de l'ordre de 0.8 MHz par Gauss  .
On peut noter de plus, que le maximum diminue légèrement (10$\%$) avec le champ sur la gamme étudié.\\

Sur la figure \ref{B2D} b), on a représenté $S_{N_-}$ à 1 MHz en fonction de $\delta$ pour différentes valeurs de B.
On peut voir, de façon similaire aux courbes de gain, que les spectres sont décalés vers les valeurs plus élevées de $\delta$ à mesure que l'on augmente le champ magnétique.
Le désaccord pour lequel le bruit passe sous la limite quantique standard (tracée en pointillé), varie d'une valeur de $\delta=-5$ MHz pour un champ nul, à $\delta=+5$ MHz pour un champ de 12 Gauss.
Le décalage est le même que précédemment, c'est à dire de l'ordre de 0.8 MHz par Gauss.\\
On peut noter enfin, que l'effet d'un champ magnétique faible (entre 2 et 10 Gauss) élargi légèrement la gamme de désaccord $\delta$ permettant d'observer un haut niveau de corrélation (zone bleu sur la figure  \ref{B2D} b)).
De plus, et c'est le résultat principal de cette étude, il est possible de compenser  l'effet d'un champ magnétique transverse uniforme en décalant de 0.8 MHz par Gauss le désaccord à 2 photons appliqué.
Ce résultat suggère donc qu'il n'y a pas de processus du type ``piégeage cohérent de population'' qui joue un rôle sur la production de corrélations dans ce type d'expérience.
En effet, dans un tel cas, un champ magnétique transverse déstabiliserait fortement les états noirs du système en mélangeant les sous-niveaux Zeeman et par conséquent on observerait une diminution importante des corrélations \cite{Berkeland:2002p18576}.

\subsection{Effet du bruit en entrée sur le champ sonde}
\begin{figure}[h]
\centering
\includegraphics[height=6.6cm]{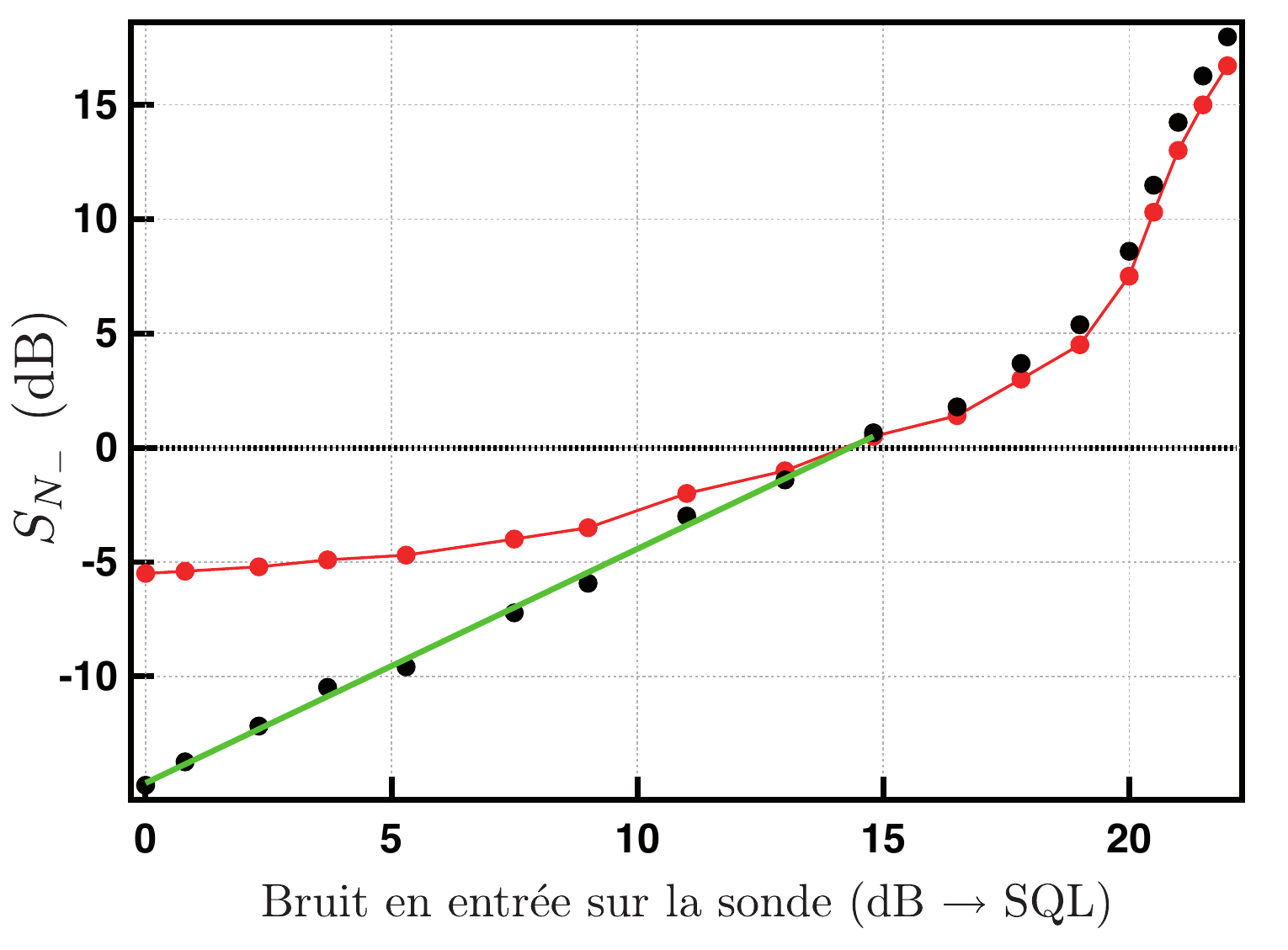} 
\caption[Effet du bruit en entrée sur le champ sonde.]{Effet du bruit en entrée sur le champ sonde sur les corrélations entre les champs sonde et conjugué. En rouge les données brutes. En noir les données corrigées des pertes de façon a minimiser les résidus lors d'un ajustement linéraire (en vert). On obtient un coefficient de pertes de 0,257. \\
 Paramètres expérimentaux : $\Delta/2\pi=1$ GHz, $P=$~1,1~W, $\delta=+5$ MHz, $T=$~114$ ^\circ$C, $L=$5 cm, RBW = 100 kHz, VBW = 10 Hz.
\label{bruitin}}
\end{figure}

Comme nous l'avons présenté au chapitre 2, il est très important de contrôler le bruit sur le faisceau sonde \cite{McKenzie:2004p18635}.
Pour démontrer son rôle crucial dans la génération de corrélations quantiques par mélange à 4 ondes, nous avons réalisé une série de mesures en ajoutant volontairement du bruit sur le faisceau sonde entrée du milieu en utilisant une source RF bruitée (figure \ref{bruitin}).\\
On peut voir qu'il devient impossible de mesurer des corrélations sous la limite quantique standard pour un excès de bruit en entrée de 15dB.
Par contre tant que l'excès de bruit reste faible (inférieur à 3dB), le niveaux de corrélations mesurées reste inchangé. 
On peut avancer l'hypothèse suivante pour comprendre cet effet.
La valeur de compression maximale mesurée dans les conditions de cette expérience sont sous-estimées par rapport à la valeur réelle, du fait des pertes (importantes dans ce montage préliminaire) dans la chaine de détection.
Ainsi la valeur mesurée est ``saturée'' par ces pertes et la dégradation de la valeur réelle (due à l'augmentation du bruit en entrée) n'affecte que très peu la valeur mesurée.
Pour rendre compte de cet effet nous avons corrigé les données mesurées à l'aide d'un coefficient de pertes et ajusté ces données corrigées par une droite.
Le coefficient de pertes qui permet le meilleur ajustemen linéaire est 25.7$\%$, ce qui correspond à une compression maximale de 14dB sous la limite quantique standard (ce qui est probablement sur-évaluée).\\
Par ailleurs, lorsque le bruit dépasse la limite quantique standard, on observe une forte dégradation des corrélations avec l'augmentation du bruit en entrée.
L'augmentation n'étant alors plus linéaire, il pourrait être intéressant dans une ouverture de ce travail de thèse d'étudier théoriquement plus en détail ce phénomène.

\subsection{Optimisation}
\begin{figure}[]
\centering
\includegraphics[width=12cm]{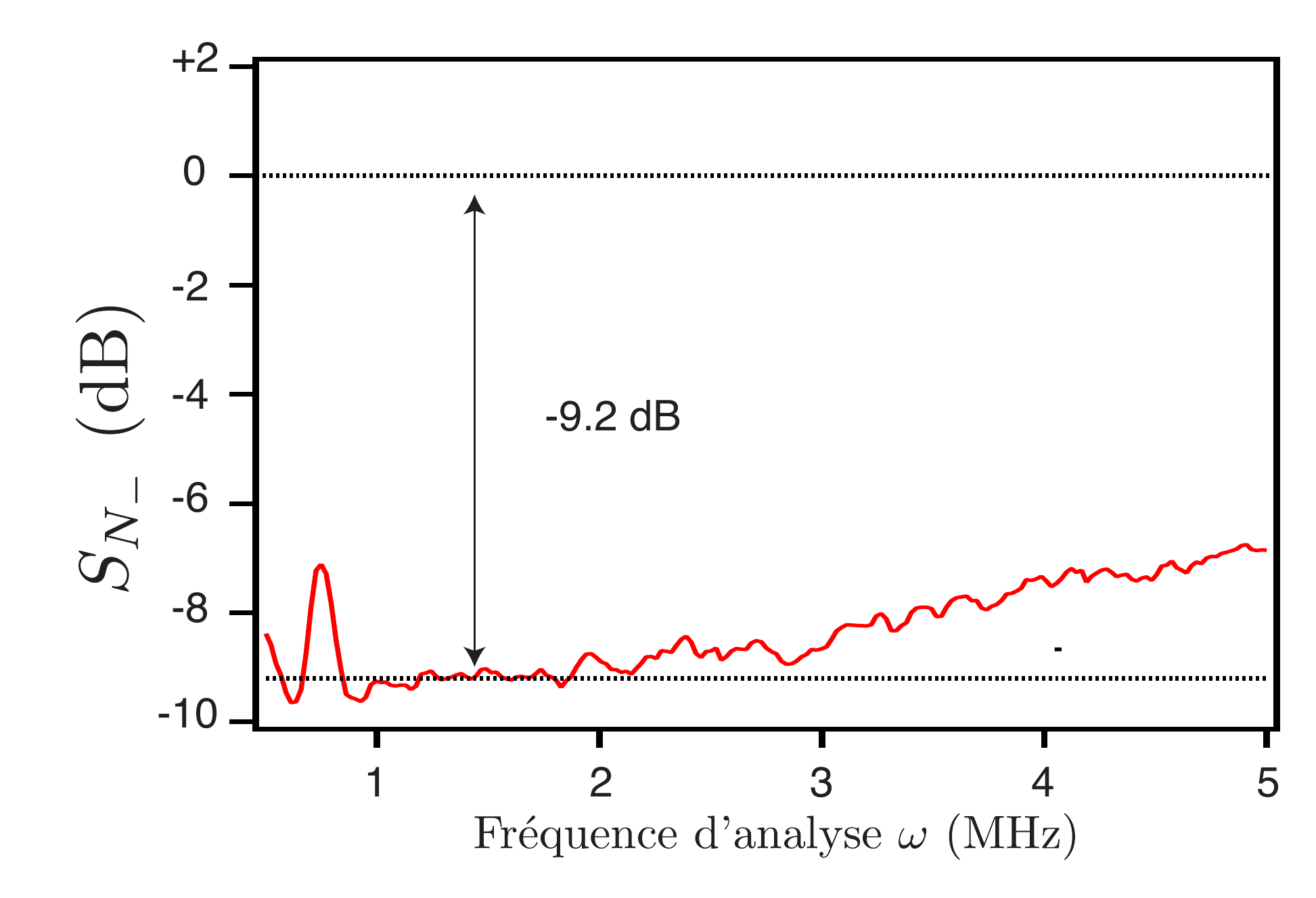} 
\caption[Spectre de bruit de la différence d'intensité.]{Spectre de bruit de la différence d'intensité entre les champs sonde et conjugué comparé au bruit quantique standard. \\
 Paramètres expérimentaux : $\Delta/2\pi=0.8$ GHz, $P=$~1,2~W, $\delta=+6$ MHz, $T=$~118$ ^\circ$C, $L=$1.25 cm, RBW = 100 kHz, VBW = 10 Hz.
\label{record}}
\end{figure}

\begin{figure}[]
\centering
\includegraphics[width=7cm]{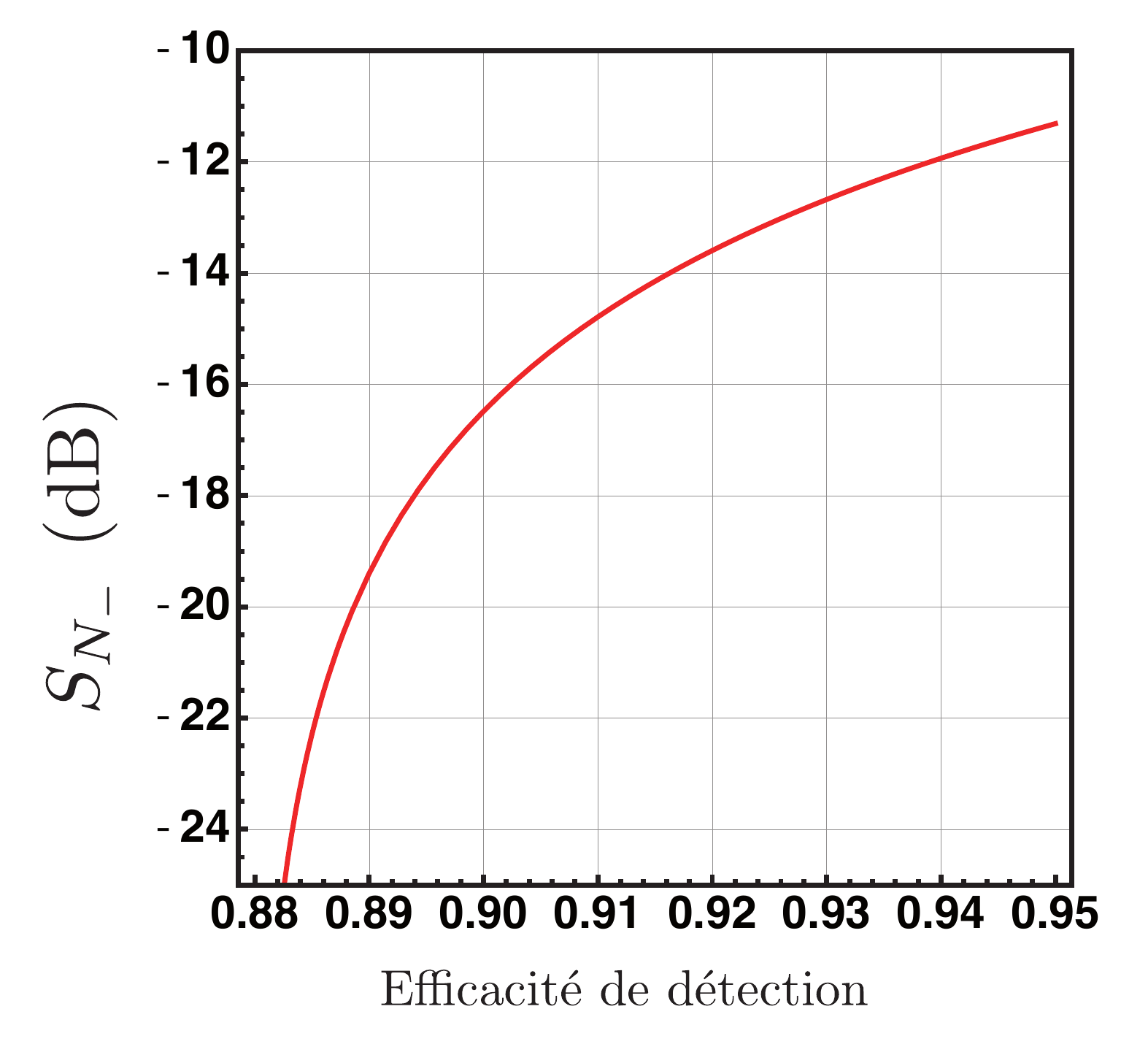} 
\caption[Valeur de $S_{N_-}$ vraie pour une valeur mesurée de -9.2dB.]{Correction de la valeur mesurée de $S_{N_-}$. Valeur de $S_{N_-}$ vraie pour une valeur mesurée de -9.2dB en fonction de l'efficacité de détection.
\label{correc}}
\end{figure}

Nous avons maintenant présenté l'effet de l'ensemble des paramètres que nous pouvons contrôler lors des expériences de mélange à 4 ondes sur la générations de corrélations quantiques.
En se plaçant dans des conditions optimales et en choisissant des photodiodes d'efficacité quantique optimale, nous avons pu mesurer jusqu'à $-9.2$ dB de corrélations sous la limite quantique standard.
Cette valeur est une valeur mesurée, uniquement corrigée du bruit électronique (qui ne tient pas compte des pertes liées à l'efficacité de détection).
Le spectre en fonction de la fréquence d'analyse, c'est-à-dire la grandeur mesurée à l'analyseur de spectre, est présenté sur la figure \ref{record}.
Comme nous l'avons dit, la réduction du bruit sous la limite quantique standard qui est détectée vaut -9.2dB.
Pour des taux de compression aussi important, même de faibles pertes lors de la détection vont réduire sensiblement la valeur détectée par rapport à la compression réelle.
La figure \ref{correc} présente le taux de compression réel en prenant en compte les pertes de la chaine de détection pour le spectre présenté sur la figure \ref{record}.
Dans la gamme d'incertitude de l'efficacité de détection (90$\% \pm 3\%$), on voit que les corrélations générées en sortie de cellule sont à des niveaux très importants.
Même dans l'hypothèse où les pertes sont les plus faibles ($\eta=93 \%$), la réduction du bruit est supérieur à -10 dB sous la limite quantique standard.
Un des pistes qui restent à explorer sur ce système est donc l'amélioration du système de détection.
Ainsi des corrélations supérieures à 90 $\%$ sont envisageables expérimentalement par mélange à 4 ondes dans une vapeur atomique.\\

En conclusion cette étude détaillée de l'ensemble des paramètres nous a permis de trouver le point de fonctionnement optimal et d'atteindre des niveaux de corrélations les plus haut de la littérature, dans ce type de système.
Cette expérience démontre clairement, une nouvelle fois \cite{Boyer:2007p1404}, l'intérêt du mélange à 4 ondes dans une vapeur atomique pour la génération d'état quantiques du champ à 2 modes.
On obtient ainsi une source de lumière non-classique avec une efficacité comparables aux meilleures sources basées sur des milieux $\chi^{(2)}$ \cite{Eberle:2010p18613}.
On peut noter de plus, qu'une telle source est naturellement bien adaptée à une interaction ultérieure avec un milieu atomique car elle est automatiquement proche de résonance sur la transition utilisée.
\section{Mise en évidence du régime $G_a+G_b<1$}
Le régime que nous avons présenté théoriquement à la section \ref{sec:qbs} du chapitre 4, qui consiste à observer des corrélations sans amplification ($G_a+G_b<1$) a pu être mis en évidence expérimentalement pour la première fois au cours de cette thèse.
Nous présentons ici les résultats obtenus lors d'expériences préliminaires visant a démontrer l'existence de ce nouveau régime.\\

\subsection{Etude du gain}
\begin{figure}[]
\centering
\includegraphics[width=14cm]{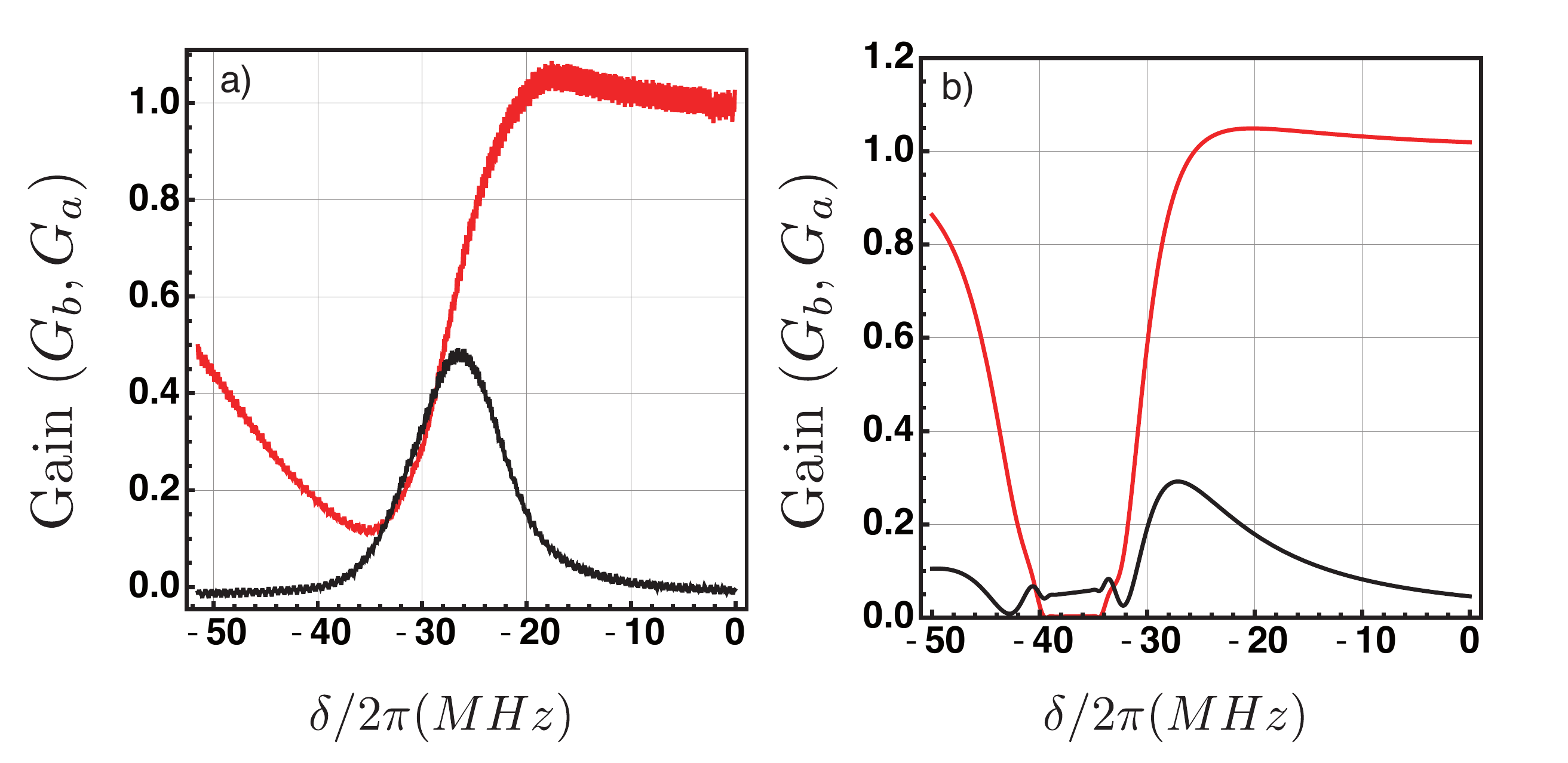} 
\caption[Spectres de $G_a$ (rouge) et $G_b$ (noir) en fonction du désaccord à 2 photons.]{Spectres de $G_a$ (rouge) et $G_b$ (noir) en fonction du désaccord à 2 photons. a) Données expérimentales, b) simulation numérique. \\
 Paramètres expérimentaux : $\Delta/2\pi=0.8$ GHz, $P=$~0.4~W, $T=$~90$ ^\circ$C, $L=$1.25 cm.
 Données numériques : $\gamma/2\pi=500$ kHz, $\alpha L$= 500, $\Omega=0.4$ GHz.
\label{qbsgain}}
\end{figure}

Dans un premier temps, nous allons étudier le gain sur le faisceau sonde et conjugué à l'aide des paramètres que nous avons trouvés dans les simulations numériques.
Après avoir baissé la température à 90$ ^\circ$C, nous avons étudié le gain en fonction du désaccord à 2 photons.
La figure \ref{qbsgain} présente les résultats que nous avons obtenus dans cette situation.
Sur la figure \ref{qbsgain} a), on peut voir qu'autour de $\delta=-30$ MHz, le régime $G_a+G_b<1$ est atteint.
On peut remarquer de plus que expérimentalement, on peut aussi obtenir dans ce régime $G_a\simeq G_b$.
Sur la figure \ref{qbsgain} b), nous présentons les simulations numériques dans les mêmes conditions.
Même si l'accord entre les deux figures n'est pas parfait, on peut noter l'intérêt de ce modèle, qui nous a permis de mettre en évidence ce nouveau régime en utilisant directement les valeurs prédites théoriquement.
\subsection{''Lame séparatrice quantique''}
Nous allons maintenant démontrer que nous avons pu réaliser expérimentalement un outil qui serait l'équivalent d'un séparateur de faisceaux ($G_a+G_b<1$) produisant des corrélations quantiques entre ses deux sorties.
C'est ce que nous avons appelé au chapitre 4 la ''lame séparatrice quantique''.\\
A l'entrée de notre dispositif, nous avons donc deux voies : le champ sonde d'une part et le vide d'autre part (dans le mode du conjugué).
Dans un premier temps, nous devons vérifier qu'en sortie l'énergie totale dans les deux voies n'est pas supérieure à l'entrée. 
Cela pourrait être le cas, car de l'énergie est apporté dans le système par la voie de la pompe.
\begin{figure}[]
\centering
\includegraphics[width=14cm]{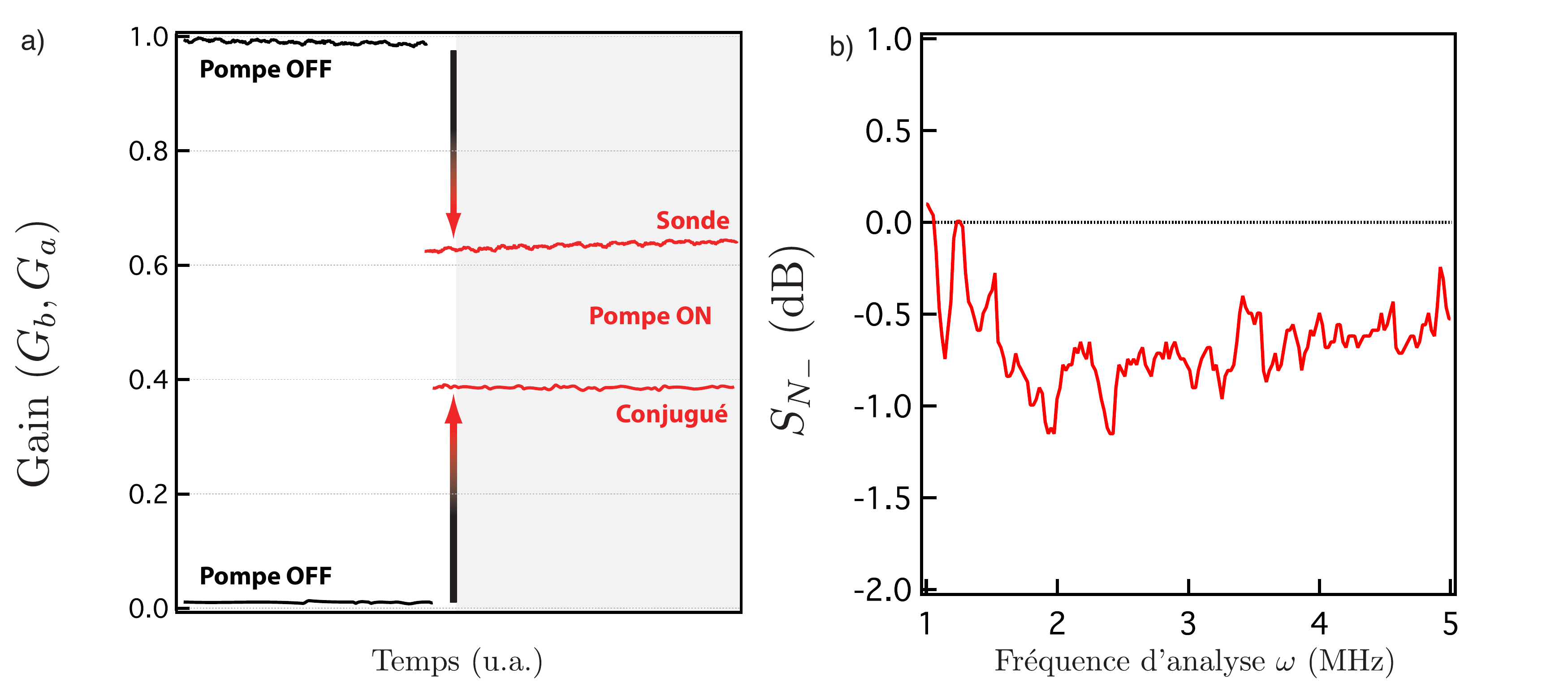} 
\caption[Démonstration expérimentale d'une ''lame séparatrice quantique''.]{Démonstration expérimentale d'une ''lame séparatrice quantique''. a) Gains sur les faisceaux sonde et conjugué en l'absence (noir) et en présence (rouge) du champ pompe. b) Spectres de bruit de la différence d'intensité comparé à la limite quantique standard.\\
 Paramètres expérimentaux : $\Delta/2\pi=1$ GHz, $\delta=-52$ MHz, $P=$~0.4~W, $T=$~126$ ^\circ$C, $L=$1.25 cm.
\label{sq_qbs}}
\end{figure}
La figure \ref{sq_qbs} a) donne les amplitudes normalisées (gains) $G_a$ et $G_b$ en entrée et en sortie du milieu.
En entrée, comme nous l'avons dit le vide est injecté sur le mode du conjugué, ainsi $G_a+G_b=1$ (on peut le voir sur les courbes noires qui correspondent à l'absence de champ pompe et donc d'interaction non-linéaire).
En sortie du milieu, le gain $G_a$ vaut 0.62 et le gain $G_b$ 0.39.
On a donc $G_a+G_b$ très proche de 1 ($1\%$ d'écart).
Aux incertitudes de mesures près, nous avons donc démontré qu'il n'y avait pas de gain d'énergie pour l'ensemble des faisceaux sonde et conjugué dans cette configuration.\\

Sur la figure \ref{sq_qbs} b), nous présentons le spectre de bruit de la différence d'intensité, pour les deux champs sonde et conjugué en sortie du milieu, comparé au bruit quantique standard.
Ainsi, nous pouvons voir que les deux modes en sortie du milieu possèdent des corrélations au delà de la limite quantique standard (environ 0.8$dB$ sous cette limite).
Le dispositif que nous présentons ici, se comporte donc bien comme un séparateur de faisceau pour les grandeurs classiques (gain) mais il va au delà car il génère des corrélations entre les deux voies de sortie.
\subsection{Perspectives}
Si nous avons démontré le fonctionnement en tant que lame séparatrice quantique du mélange à 4 ondes dans ce régime, nous n'avons pas fait une étude approfondie du phénomène.
Par exemple, il pourrait être intéressant d'étudier ce qui se passe en injectant non pas le vide sur le mode du conjugué mais un état cohérent.
Si dans cette configuration un tel outil permet de produire des corrélations quantiques entre les deux modes de sortie, il serait alors possible d'utiliser ce dispositif comme l'élément de base d'un interféromètre de Mach--Zehnder.

\section{Conclusion du chapitre}

Dans ce chapitre nous avons présenté les résultats expérimentaux que nous avons obtenu sur une expérience de mélange à 4 ondes dans une vapeur atomique.
Nous avons décrit le dispositif utilisé puis caractérisé le milieu atomique utilisé.
Nous avons ensuite étudié en détail l'espace des paramètres qui influent sur ce processus.
Au cours de cette étude, nous avons pu valider le modèle théorique que nous avons présenté au chapitre 4 en le comparant aux résultats expérimentaux.
A l'aide de ce travail, nous avons pu trouver un point de fonctionnement optimal et démontrer la génération de -9.2dB de corrélations sous la limite quantique standard.\\
Enfin, à l'aide des prévisions théoriques, nous avons mis en évidence un nouveau régime permettant la générations de faisceaux corrélés dans une vapeur atomique.
Ce régime se caractérisant par $G_a+G_b<1$, nous avons appelé le dispositif ainsi créé une ``lame séparatrice quantique''.

%% file: chapitre6v5.tex
\chapter{Transparence électromagnétiquement induite à 422 nm}\label{ch6}
\setcounter{minitocdepth}{2}
\minitoc
\vspace{1 cm}
\normalsize \textit{Comme nous l'avons vu au chapitre 4, les expériences de mélange à 4 ondes à 795 nm sur la raie D1 du rubidium 85, dans les conditions que nous avons présentées dans ce manuscrit, ne sont pas transposables sur la transition $5S^{1/2}\to 6 P^{1/2}$ (à 422 nm).
D'autres régimes pourraient être envisagés, notamment en s'éloignant encore de la résonance pour éviter l'absorption du faisceau sonde ($\Delta> 4.5$GHz).
Dans une telle configuration, les puissances optiques nécessaires pour mesurer un gain supérieur à 1 sont très élevées (plus de 5 Watts).
Nous ne disposons malheureusement pas d'une source laser à 422 nm produisant une telle puissance.
Par contre, comme nous l'avons présenté au chapitre 2, nous pouvons réaliser le doublage de fréquence du laser titane saphir afin de générer jusqu'à 200 mW à 422 nm.
De plus, l'équipe possède une diode laser en cavité étendue de la société Toptica à cette même longueur d'onde.
Grâce à ces deux sources, nous avons pu réaliser les premières expériences de transparence électromagnétiquement induite sur la transition  $5S^{1/2}\to 6 P^{1/2}$ du rubidium 85.
Ce chapitre est dédié à la présentation de ces résultats.}
\\

\large L'intérêt de la transparence électromagnétiquement induite (Electromagnetically Induced Transparency ou EIT) afin de réaliser des mémoires quantiques a été démontré dans les expériences de lumière lente \cite{Hau:1999p19999} puis de lumière arrêtée \cite{Liu:2001p20008}.
Plus récemment, deux équipes ont montré que le phénomène d'EIT permettait de stocker et de restituer des états vides comprimés \cite{Appel:2008p6196} et \cite{Arikawa:2010p19756,Honda:2008p19767}
Enfin les travaux de thèse Jean Cviklinski \cite{Cviklinski:2008p1266} et Jérémie Ortalo \cite{Ortalo09} sont consacrés à la réalisation et à l'étude d'une mémoire quantique via le phénomène d'EIT.
De plus, il a été souligné par de nombreux auteurs qu'à l'aide des effets d'EIT \cite{Harris:1997p9203}, il était possible d'amplifier fortement les interactions non-linéaires par rapport à un milieu modélisé par un système résonant à deux niveaux \cite{Li:1996p10474,Lu:1998p10458,Li:2007p10766}.

\begin{sidewaysfigure}
\centering
\includegraphics[width=20cm]{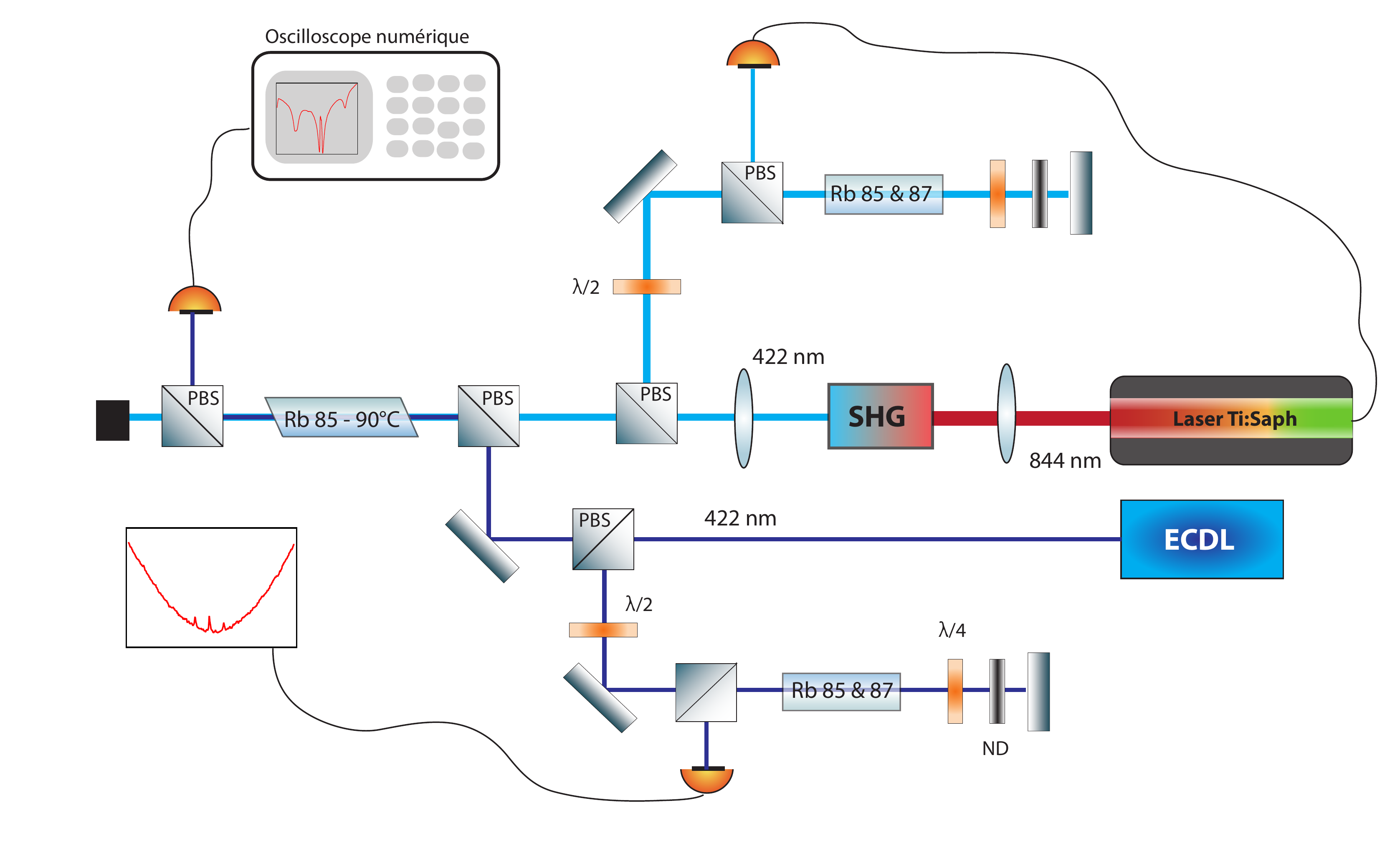} 
\caption[Schéma du dispositif expérimental.]{Schéma du dispositif expérimental utilisé dans les expériences de transparence électromagnétiquement induite sur la transition  $5S^{1/2}\to 6 P^{1/2}$ du $^{85}$Rb . Un champ de contrôle, à 422 nm, généré par doublage de fréquence (SHG) d'un laser titane saphir continu interagit dans une vapeur de $^{85}$Rb avec un champ sonde à la même longueur d'onde, produit par une diode laser en cavité étendue (ECDL). Un montage d'absorption saturé permet de disposer d'une référence de fréquence pour le faisceau sonde.  }
\label{manipeit}
\end{sidewaysfigure}

Pour réaliser une expérience d'EIT, il est nécessaire de disposer de deux sources laser accordables à résonance avec les transitions atomiques considérées (voir figure \ref{niv}).
Dans le cas que nous présentons ici, il s'agit de la transition $5S^{1/2}\to 6 P^{1/2}$ du $^{85}$Rb, dont les détails sont donnés à l'annexe \ref{Annexe_Rb}.
Les deux niveaux hyperfins (F=2 et F=3) de l'état $5S^{1/2}$, ainsi qu'un des états excités $6 P^{1/2}$ sont utilisé pour former une configuration en $\Lambda$ sur cette transition à 422nm.\\
Nous  utilisons deux sources différentes qui sont représentées sur la figure \ref{manipeit}.
Pour générer le champ de contrôle, il s'agit du laser titane saphir à 844 nm doublé (voir chapitre 2).
Ce laser est asservi sur le pic de croisement des transitions atomiques du rubidium $5S^{1/2},F=2 \to 6 P^{1/2}, F=2$ et $6 P^{1/2}, F=3$ à l'aide d'un montage d'absorption saturée.
Pour le champ sonde, nous avons utilisé une diode laser en cavité étendue (ECDL) DL-100 de la société Toptica\footnote{www.toptica.com} décrite en détail dans la thèse de Sébastien Rémoville \cite{Removille:2009p19653}.
Un second montage d'absorption saturée utilisant une partie du faisceau sonde permet de disposer d'une référence en fréquence lorsque l'on balaye la fréquence de la diode laser.\\
Ces deux faisceaux sont superposés dans une cellule de rubidium 85 isotopique à une température comprise entre 90 et 135 $ ^\circ$C.
Le waist du faisceau de contrôle mesure 250 $\mu$m et celui de faisceau sonde 125 $\mu$m.
Les positions du waist des deux faisceaux se situent au milieu de la cellule de longueur $L=$5 cm.\\
Le faisceau de contrôle est filtré en sortie de la cellule et le faisceau sonde est collecté par une photodiode dont le photocourant est enregistré par un oscilloscope numérique.

\clearpage

\section[Transparence électromagnétiquement induite à 422 nm]{Transparence électromagnétiquement induite sur la transition  $5S^{1/2}\to 6 P^{1/2}$ du $^{85}$Rb}
La transparence électromagnétiquement induite est un phénomène largement étudié dans la littérature \cite{Harris:1990p9025,Imamoglu:1991p9024,Arimondo:1996p9015} qui a été observé sur de nombreuses transitions atomiques en particulier les raies D1 et D2 du rubidium 85.
Une étude extensive du phénomène d'EIT dans une vapeur de rubidium peut être trouvée dans \cite{Fulton:1995p9269}.
Aucune des références que nous avons pu trouver dans la littérature ne reporte l'observation du phénomène d'EIT sur la transition $5S^{1/2}\to 6 P^{1/2}$ du rubidium 85.
C'est donc la première mise en évidence expérimentale de cet effet sur cette transition.\\

\subsection{Spectre de transmission}

\begin{figure}
\centering
\includegraphics[width=9.5cm]{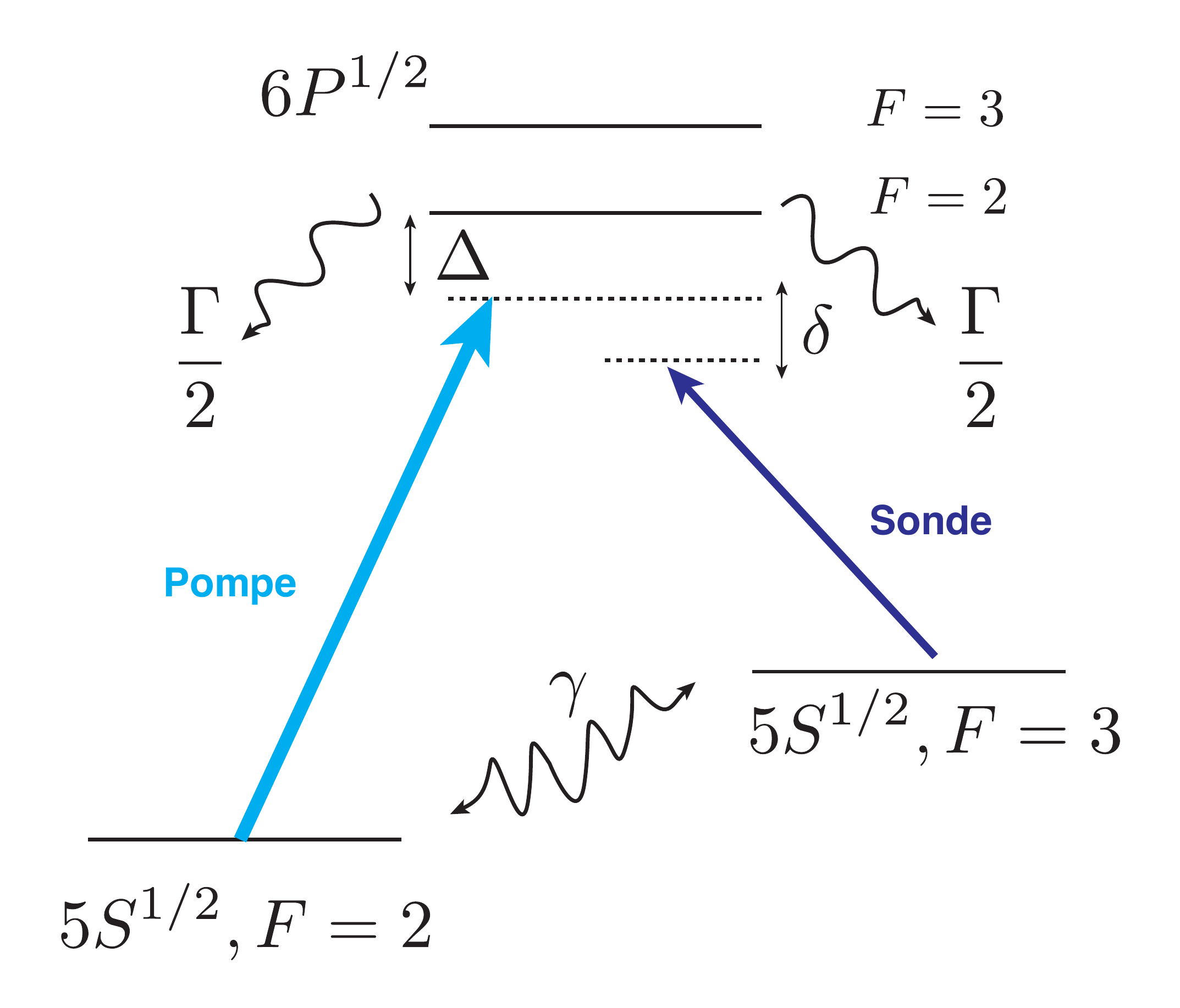} 
\caption[Schéma de niveaux de la transition $5S^{1/2}\to 6 P^{1/2}$.]{Schéma de niveaux de la transition $5S^{1/2}\to 6 P^{1/2}$. Les deux niveaux $5S^{1/2}$ $F=2$ et $F=3$ sont les niveaux fondamentaux séparés de 3.036 GHz. Les deux niveaux $6P^{1/2}$ $F=2$ et $F=3$ sont les niveaux excité séparés de 117 MHz.
On appelle $\Delta$ le désaccord du champ de contrôle par rapport à la transition $5S^{1/2},F=2\to 6 P^{12},F=2$. Le laser de contrôle est stabilisé sur le pic de croisement de niveaux de l'absorption saturée soit : $\Delta=117/2$ MHz. $\delta$ est le désaccord à 2 photons entre le champ de contrôle et le champ sonde.
Le taux de desexcitation du niveaux excité vers chacun des niveaux fondamentaux vaut $\Gamma/2$. Le taux de décohérence des deux niveaux hyperfins du fondamental vaut $\gamma$.
\label{niv}}
\centering
\includegraphics[width=13cm]{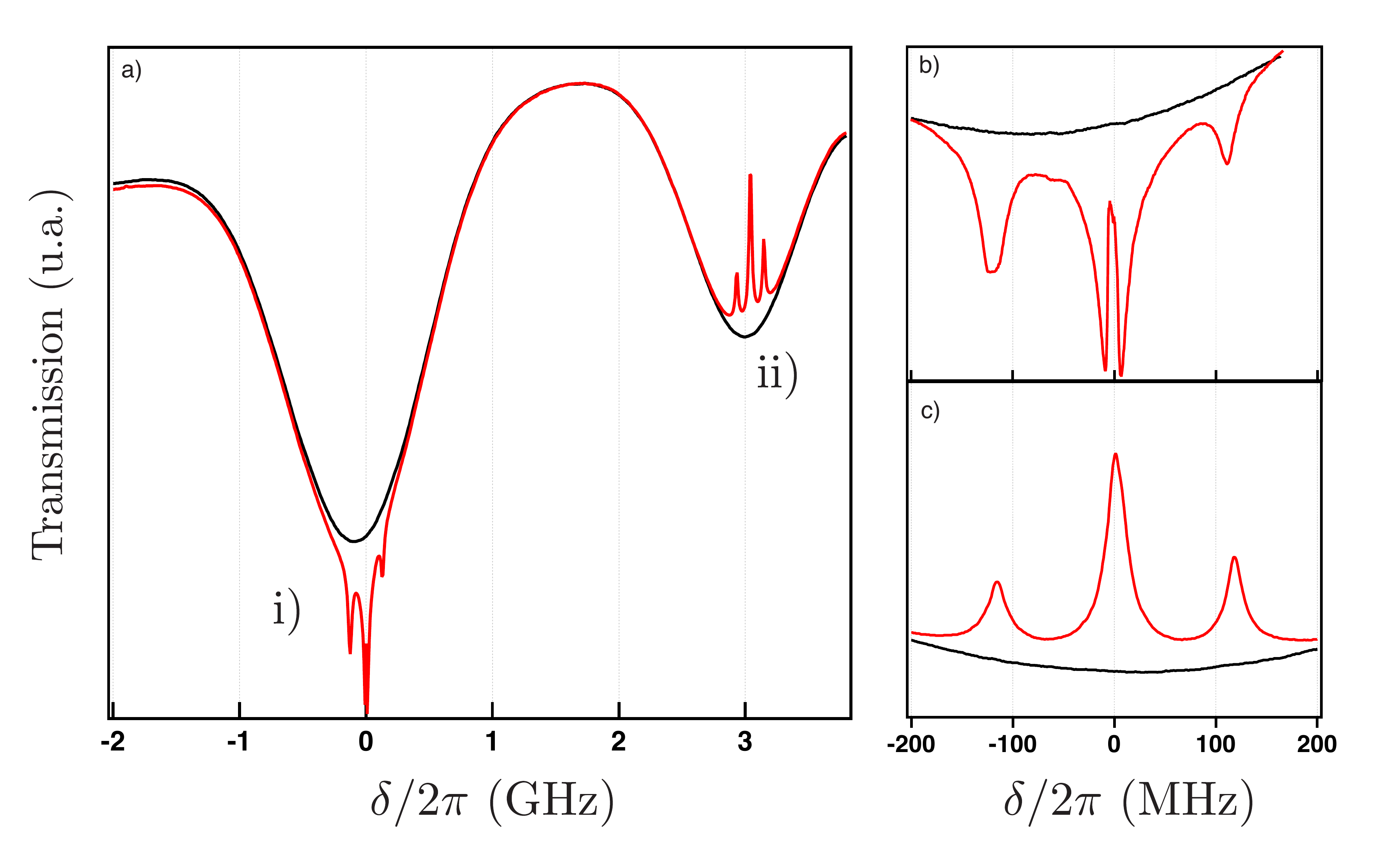} 
\caption[Spectres d'absorption de la sonde en présence d'un champ de contrôle.]{Spectres d'absorption de la sonde. La figure a) représente l'ensemble du spectre mesuré en présence (rouge) et en l'absence (noir) du champ de contrôle. La figure b) correspond à la zone notée i) dans laquelle, on peut observer le phénomène d'EIT, la figure c) correspond à la zone notée ii) (sur cette figure l'échelle horizontale est décalée de 3036 MHz).\\
Paramètres expérimentaux : $T=130^\circ$C, $L=$5cm, $P_{controle}$=120 mW.}
\label{eit1}
\end{figure}

Les 4 niveaux atomiques que nous allons prendre en compte pour expliquer les observations sont représentés sur la figure \ref{niv}.
A l'aide du laser sonde nous pouvons balayer les transitions $5S^{1/2},F=2\to 6 P^{1/2}$ et $5S^{1/2},F=3\to 6 P^{1/2}$ qui sont séparées de 3.036 GHz.
Nous avons choisi d'asservir la fréquence du champ de contrôle à $\Delta=+63.5$ MHz de la transition $5S^{1/2},F=2\to 6 P^{1/2},F=2$, à l'aide du pic de croisement de niveaux du montage d'absorption saturée.
Dans ces conditions, nous avons obtenus les spectres présentés sur la figure \ref{eit1}\footnote{Notons que le choix de la valeur exacte du désaccord $\Delta$ ne va que très faiblement influencer la forme des spectres observés dans une large gamme autour de $\Delta=0$ (plusieurs centaines de MHz).  En effet la largeur de la distribution Doppler est de l'ordre de 1 GHz à cette température, ce qui est largement supérieur à l'écart entre les deux niveaux excités (117 MHz).   }.
On peut distinguer sur la figure a), deux zones qui ont été notées i) et ii).
Chacune de ces zones a pour forme générale, un profil d'absorption élargie par effet Doppler.
Sur ces profils relativement large, on peut observer autour de $\delta=0$ MHz et $\delta=3036$ MHz, une variation rapide de la transmission du faisceau sonde.
Le détail de ces deux zones est présenté sur les figures b) et c).\\
Commençons par commenter la zone ii) décrite sur la figure \ref{eit1} c).
L'échelle horizontale de cette figure est décalée de 3036 MHz, ce qui signifie que $\delta=0$ correspond à la situation où le champ de contrôle et le champ sonde sont exactement à la même fréquence.
L'élargissement inhomogène (effet Doppler) est à l'origine des deux séries de trois pics observées. 
En effet, il existe une classe de vitesse que l'on notera $v$ qui va entrainer un désaccord $kv=+63.5$ MHz sur les faisceaux sonde et contrôle et une classe de vitesse que l'on notera $-v$ qui va entrainer un désaccord $-kv=-63.5$ MHz.
Ces désaccords sont utilisés dans les tableaux qui suivent pour décrire les différentes configurations.
Nous donnons ainsi dans chaque cas, le schéma de niveaux  équivalent et une description du processus.
Notons que la configuration symétrique des deux désaccords n'est pas une condition nécessaire pour expliquer les phénomènes que nous allons détailler, mais qu'elle permet de simplifier la compréhension.
\begin{changemargin}{-0.3cm}{-1cm}

~\\
\begin{tabular}{|c|c|p{5.7cm}|}
\hline
Schéma de niveaux & Schéma de niveaux équivalent & \hspace{2cm} Explication\\ & prenant en compte l'effet Doppler  &   \\ \hline
\begin{minipage}{3.5cm}\includegraphics[width=3.5cm]{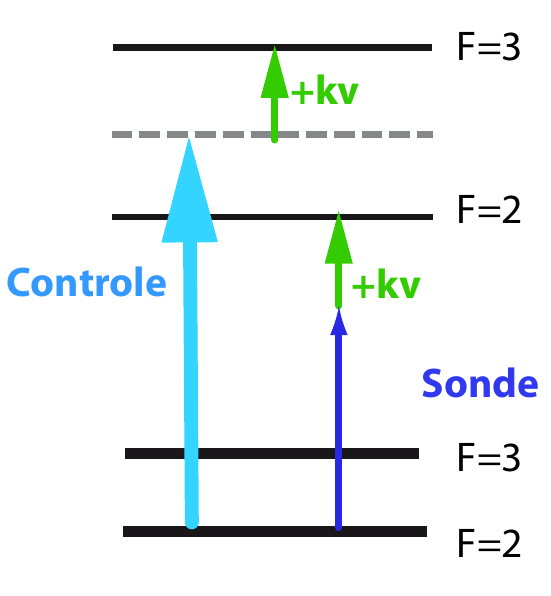}  \end{minipage}
& \begin{minipage}{3.5cm}\includegraphics[width=3.5cm]{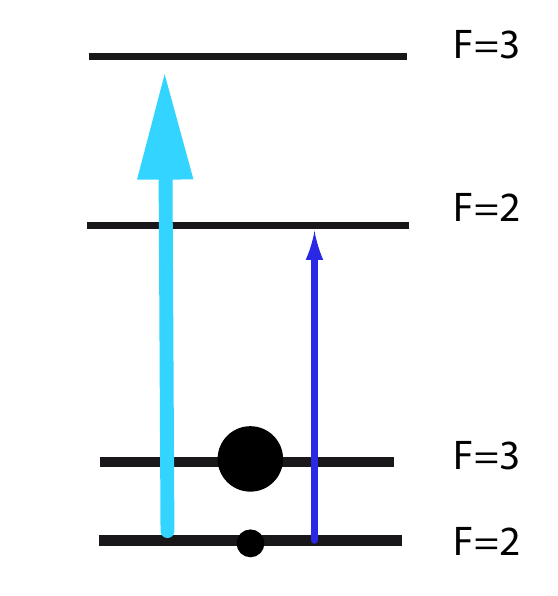}  \end{minipage} 
&  \begin{minipage}{5.7cm} \vspace{0.3cm} Le faisceau de contrôle réalise le pompage optique de la majorité des atomes dans l'état $5S^{1/2},F=3$. Le niveau $5S^{1/2},F=2$ étant par conséquent moins peuplé, le faisceau sonde à résonance sur la transition $F=2\to F=2$ est moins fortement absorbé. On observe le pic de transmission noté (a) sur la figure \ref{eit2}.\vspace{0.3cm}\end{minipage}\\ \hline
\begin{minipage}{3.5cm}\includegraphics[width=3.5cm]{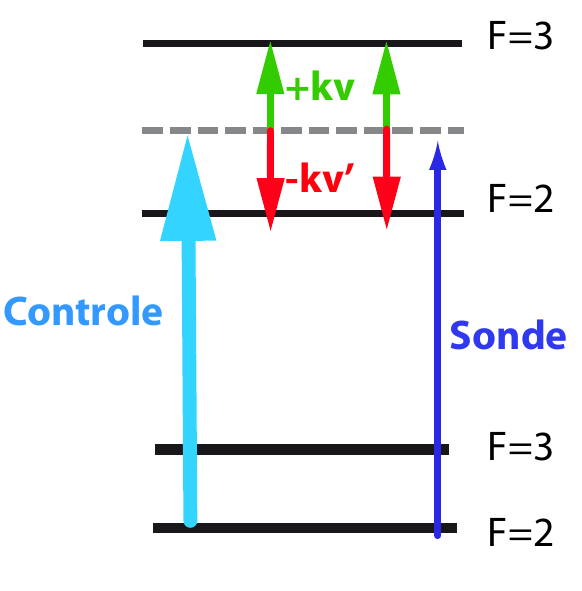}  \end{minipage}
&\begin{minipage}{4.7cm}\includegraphics[width=4.7cm]{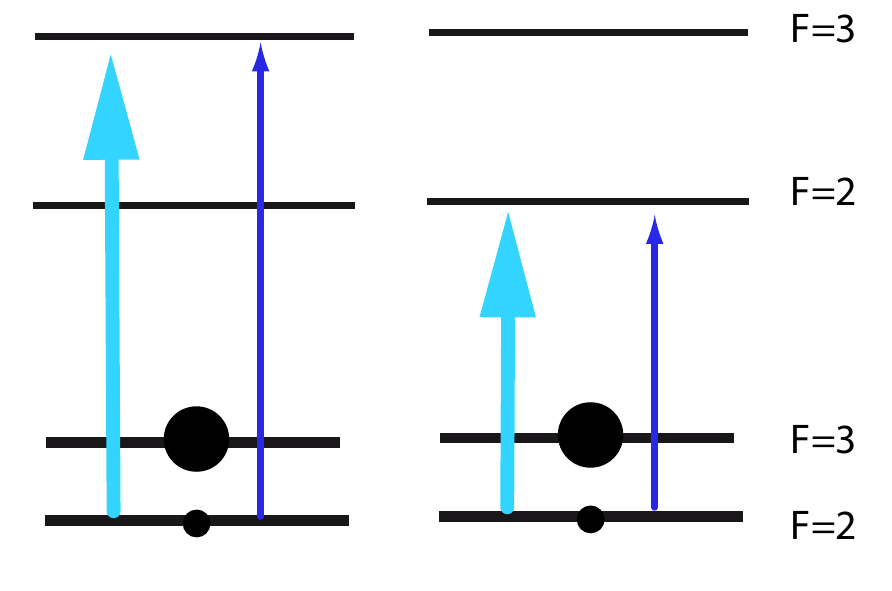}  \end{minipage}
&  \begin{minipage}{5.7cm} \vspace{0.3cm} Le faisceau de contrôle réalise le pompage optique de façon similaire à ce qui est décrit dans la situation précédente. On observe de même un pic de transmission noté (b) sur la figure \ref{eit2}. Bien que le niveau excité avec lequel interagit les deux laser est le même, il n'y a pas de phénomène d'EIT dans ce cas car les niveaux fondamentaux considérés sont aussi les mêmes. Il ne s'agit donc pas d'un système en $\Lambda$. \vspace{0.3cm}\end{minipage} \\ \hline
\begin{minipage}{3.5cm}\includegraphics[width=3.5cm]{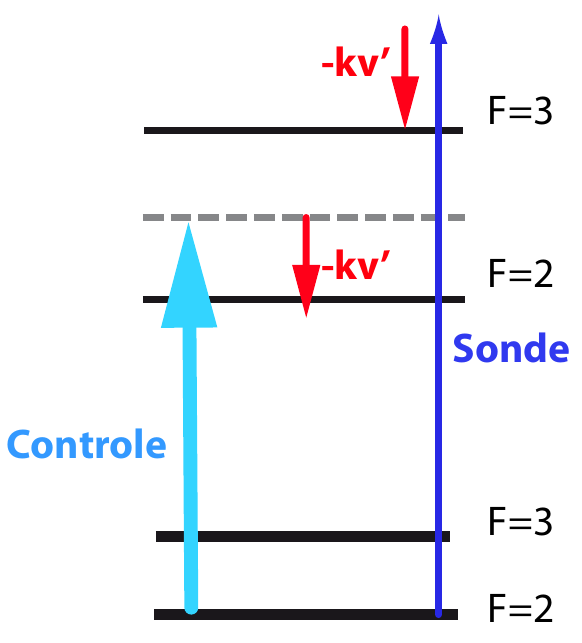}  \end{minipage}
&\begin{minipage}{3.5cm}\includegraphics[width=3.5cm]{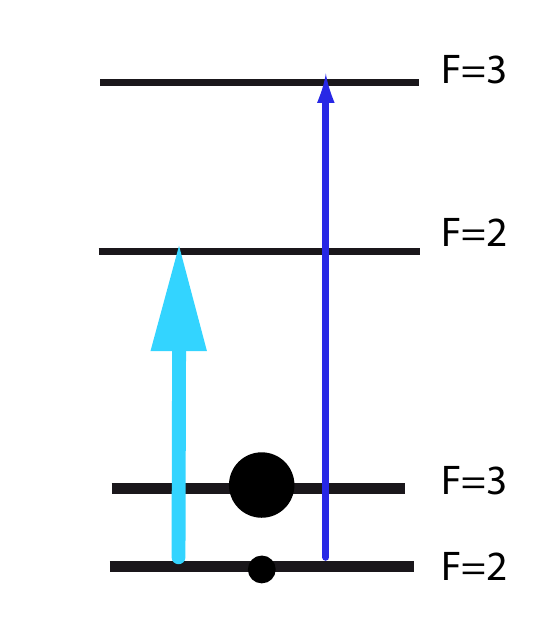}  \end{minipage}
&  \begin{minipage}{5.7cm} \vspace{0.3cm} Le faisceau de contrôle réalise le pompage optique de la majorité des atomes dans l'état $5S^{1/2},F=3$. Le niveau $5S^{1/2},F=2$ étant par conséquent moins peuplé, le faisceau sonde à résonance sur la transition $F=2\to F=3$ est moins fortement absorbé. On observe un pic de transmission noté (c) sur la figure \ref{eit2}. \vspace{0.3cm}\end{minipage} \\ \hline
\end{tabular}\vspace{0.3cm}

Sur ces figures le faisceau pompe ou faisceau de contrôle est représenté en bleu ciel et le faisceau sonde en bleu foncé.
Les différents désaccords introduits par l'effet Doppler sont représenté en vert pour les atomes de vitesse $v$ et en rouge pour les atomes de vitesse $-v'$.
Sur les schémas de niveaux équivalents, le déséquilibre de population entre les niveaux fondamentaux dans l'état stationnaire est représenté par la taille des disques noirs.\\
\end{changemargin}

\begin{changemargin}{-1.5cm}{0cm}

\begin{tabular}{|c|c|p{6.5cm}|}
\hline
Schéma de niveaux & Schéma de niveaux équivalent &\hspace{2cm} Explication\\ & prenant en compte l'effet Doppler  &   \\ \hline
\begin{minipage}{3.5cm}\includegraphics[width=3.5cm]{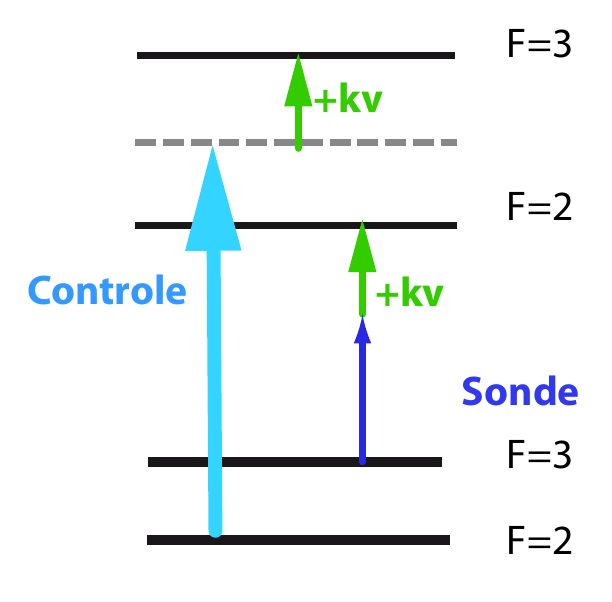}  \end{minipage}
& \begin{minipage}{3.5cm}\includegraphics[width=3.5cm]{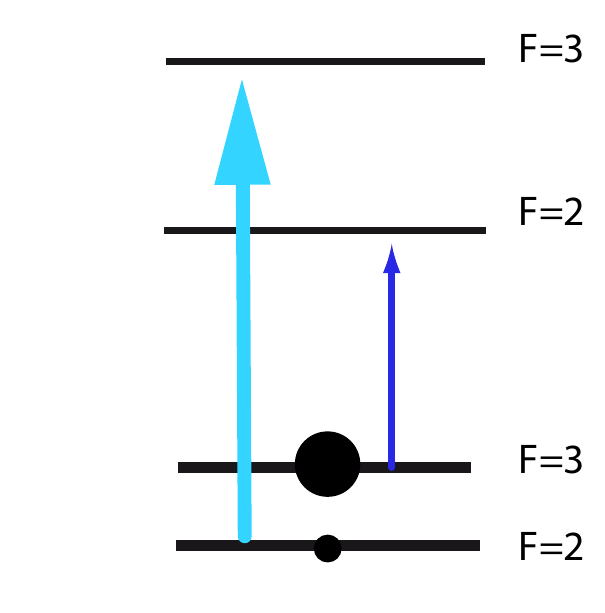}  \end{minipage} 
&  \begin{minipage}{6.5cm} \vspace{0.3cm} Le faisceau de contrôle réalise le pompage optique de la majorité des atomes dans l'état $5S^{1/2},F=3$. Le niveau $5S^{1/2},F=3$ étant par conséquent plus peuplé, le faisceau sonde à résonance sur la transition $F=3\to F=2$ est plus fortement absorbé. On observe le pic d'absorption noté (d).\vspace{0.3cm}\end{minipage}\\ \hline
\begin{minipage}{3.5cm}\includegraphics[width=3.5cm]{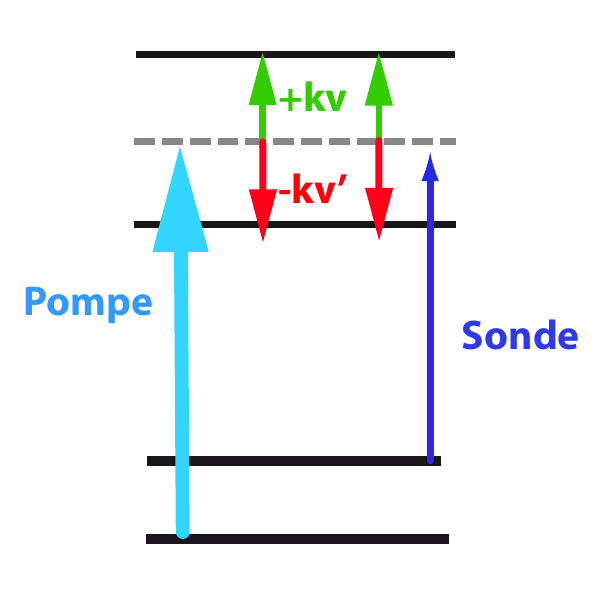}  \end{minipage}
&\begin{minipage}{4.7cm}\includegraphics[width=4.7cm]{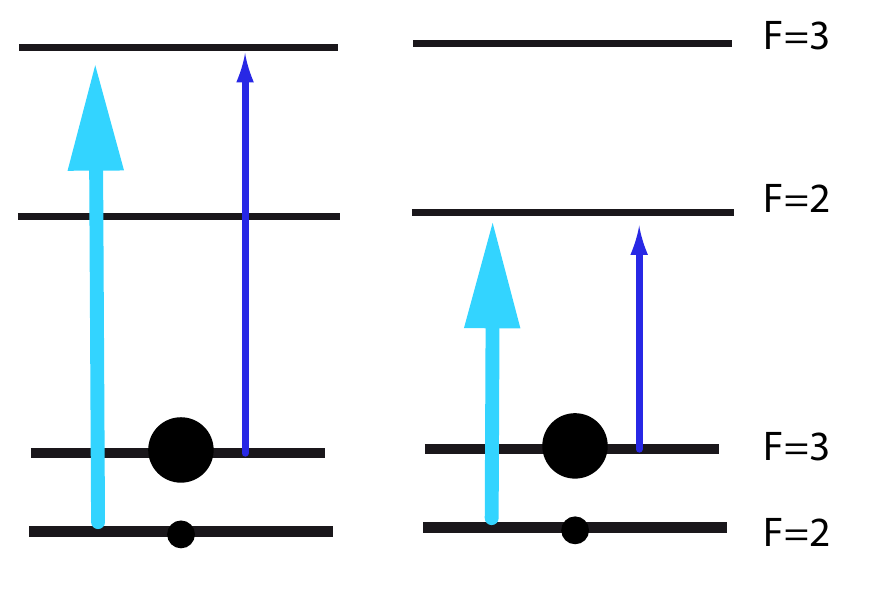}  \end{minipage}
&  \begin{minipage}{6.5cm} \vspace{0.3cm} Le faisceau de contrôle réalise le pompage optique de façon similaire à ce qui est décrit dans la situation précédente. On observe de même un pic d'absorption. Au sein de ce pic d'absorption, on observe un pic de transmission qui correspond au phénomène d'EIT dans une configuration en $\Lambda$. Ce profil est noté (e) sur la figure \ref{eit2}. \vspace{0.3cm}\end{minipage} \\ \hline
\begin{minipage}{3.5cm}\includegraphics[width=3.5cm]{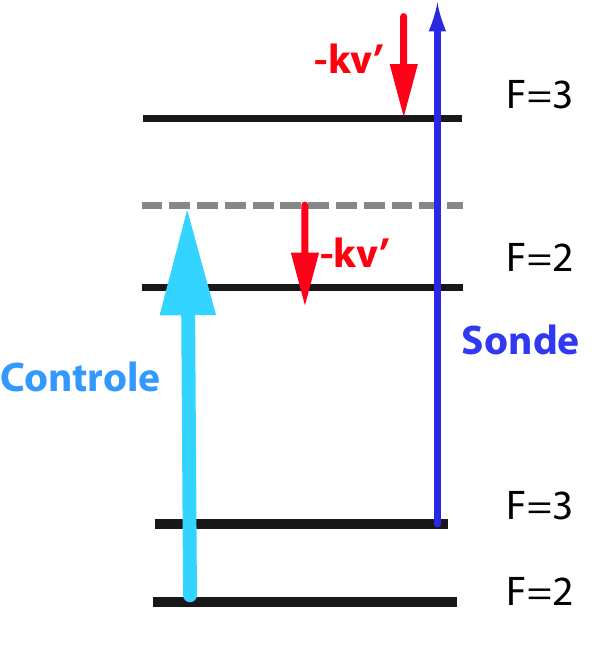}  \end{minipage}
&\begin{minipage}{3.5cm}\includegraphics[width=3.5cm]{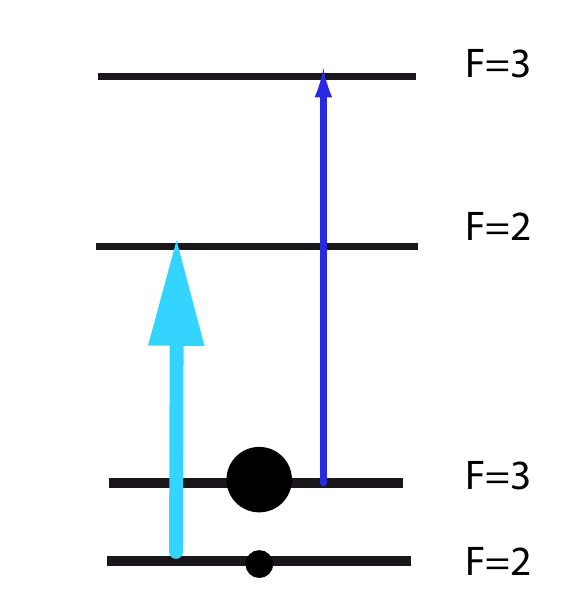}  \end{minipage}
&  \begin{minipage}{6.5cm} \vspace{0.3cm} Le faisceau de contrôle pompe optiquement le niveau $5S^{1/2},F=3$. De manière similaire à ce qui est décrit dans la situation d). On observe un pic d'absorption noté (f). \vspace{0.3cm}\end{minipage} \\ \hline
\end{tabular}
\begin{figure}[h!]\centering
\includegraphics[width=11cm]{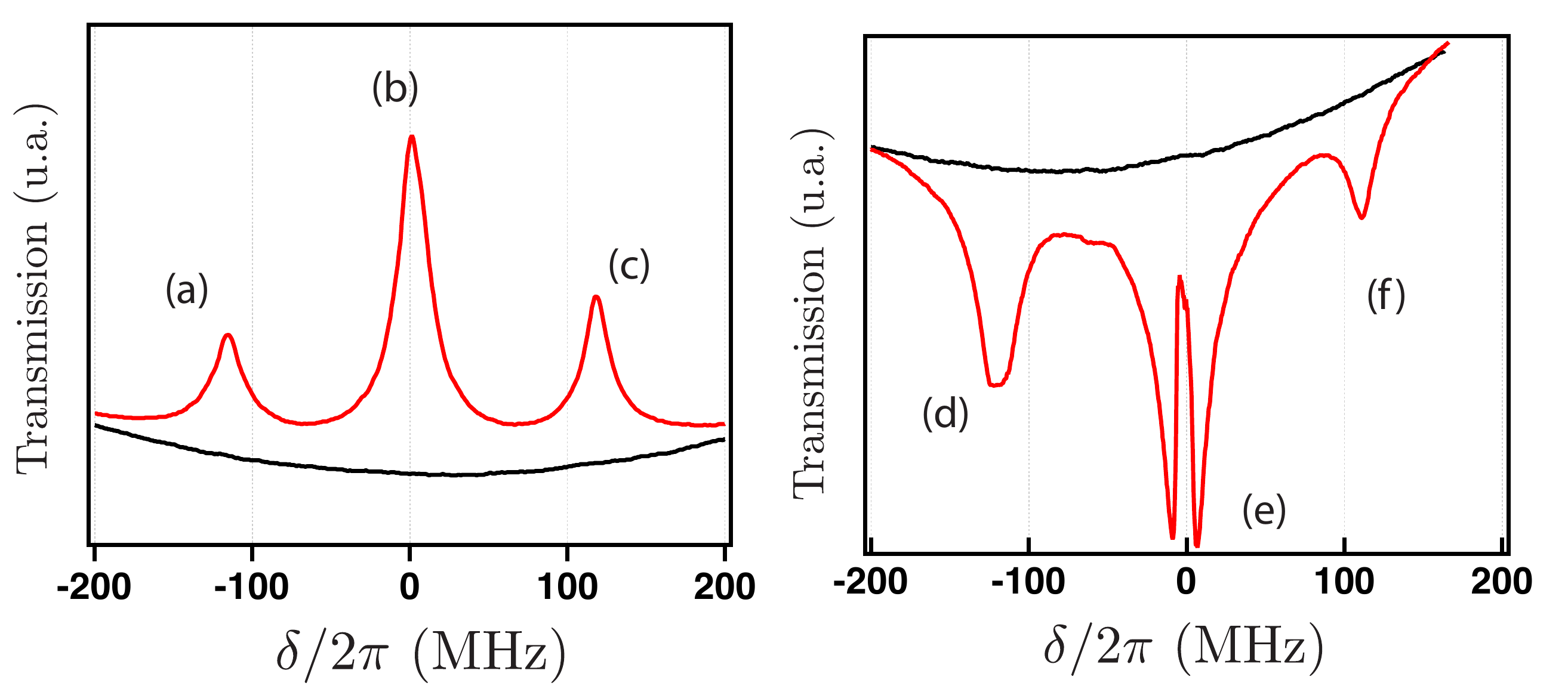} 
\caption[Schéma du dispositif expérimental.]{Spectres d'absorption de la sonde. Détails de la figure \ref{eit1}\label{eit2}.\\}
\end{figure}
\end{changemargin}
\clearpage
Pour résumer, le profil complet de transmission est donc donné par trois phénomènes physiques distincts :
\begin{itemize} 
\item l'absorption linéaire élargie par effet Doppler par une vapeur atomique\\
\item la modification des populations par pompage optique qui va imposer des pics d'absorption ou de transmission selon la transition adressée par le champ sonde.\\
\item le phénomène d'EIT qui ajoute un pic de transmission dans une zone d'absorption lorsque les champs sonde et contrôle adressent le même niveau excité pour deux niveaux fondamentaux distincts.
\end{itemize}
\subsection{Etude de la fenêtre de transparence}
\begin{figure}[]
\centering
\includegraphics[width=10cm]{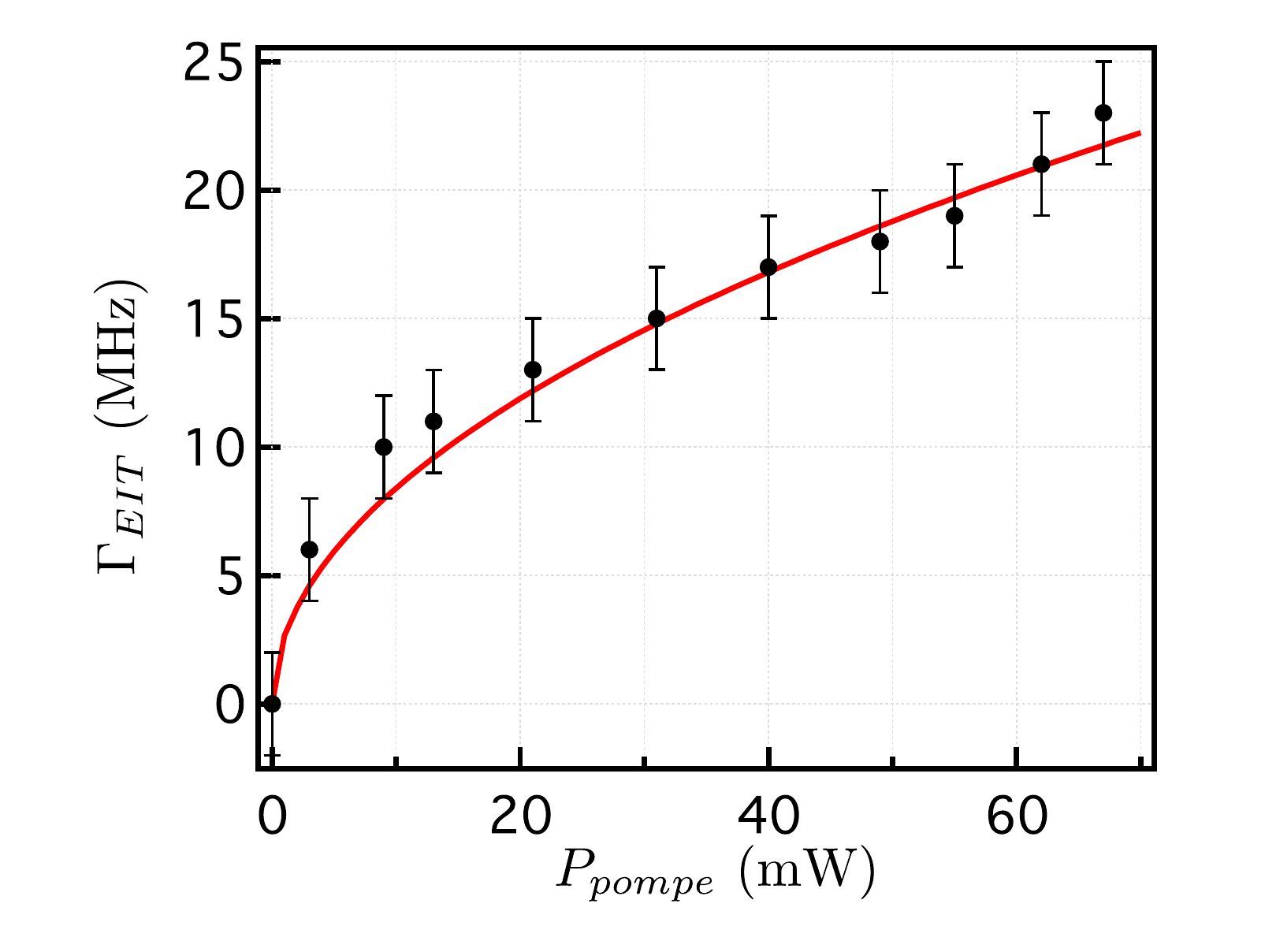} 
\caption[Schéma de niveaux de la transition $5S^{1/2}\to 6 P^{1/2}$.]{Largeur de la fenêtre d'EIT en fonction de la puissance du faisceau de contrôle. L'ajustement est donné  par le modèle décrit dans \cite{Javan2002}.
\label{EITfit}}
-\end{figure}
Depuis les travaux pionniers de \cite{Feld:1969p9016} sur la réduction de la largeur de raie à l'aide d'un champ de contrôle, le rôle de l'effet Doppler dans les expériences d'EIT réalisées à l'aide d'une vapeur atomique, a été largement étudié, notamment par \cite{Li:1995p9295,Vemuri:1996p9371,Vemuri:1996p9330,Vemuri:1996p9336}.
Plus récemment les travaux de \cite{Li:2004p8962} ont porté sur la largeur de la fenêtre d'EIT en fonction du temps de passage des atomes dans le faisceaux sonde.
Dans les conditions expérimentales que nous avons présentées, la largeur de la fenêtre d'EIT, $\Gamma_{EIT}$, est donnée par la relation suivante \cite{Javan2002} :
\begin{equation}
\Gamma_{EIT}=\sqrt{\frac{2\gamma}{\Gamma}}\Omega
\end{equation}
Nous avons réalisé une série de mesures pour des puissances du laser de contrôle variant entre 0 et 70 mW.
Le faisceau de contrôle est focalisé sur 450 microns et le faisceau sonde sur 400 microns (rayon à $1/e^2.$).
Ainsi, à 130$^\circ$C, le taux de décohérence des niveaux hyperfins vaut $\gamma\simeq0.6$ MHz.\\
Les résultats expérimentaux sont présentés sur la figure \ref{EITfit} et nous les comparons au modèle décrit dans \cite{Javan2002}.
Il est intéressant de noter que le seul paramètre ajustable dans ce modèle est le taux de décohérence $\gamma$.
Notons de plus que l'ajustement présenté sur la figure \ref{EITfit} utilise une valeur de $\gamma=0.64$ MHz, ce qui est consistant avec la valeur estimée (par le temps de passage des atomes dans le faisceau) de $\gamma\simeq0.6$ MHz.
L'écart entre les deux valeurs est inférieur à 10$\%$, ce qui doit être comparé avec l'incertitude sur la mesure de la taille du faisceau laser de contrôle dont dépend la valeur estimée de $\gamma$ et qui est aussi de l'ordre de 10$\%$.

\section{Conclusion du chapitre}
Dans ce chapitre, nous avons étudié expérimentalement le phénomène de transparence électromagnétiquement induite sur la transition $5S^{1/2}\to 6 P^{1/2}$ du rubidium 85.
Bien que le processus d'EIT dans une vapeur atomique de rubidium soit largement traité dans la littérature, nous n'avons trouvé aucune mention d'expériences entièrement réalisées à 422 nm, c'est à dire pour le champ de contrôle et le champ sonde.
Nous présentons des spectres de transmission du faisceau sonde en présence et en absence de champ de contrôle.
Les différents pics observés ont été interprétés, puis nous avons étudié spécifiquement le pic d'EIT et sa largeur en fonction de la puissance du champ de contrôle.
On trouve un bon accord entre les valeurs mesurées expérimentalement et la littérature \cite{Javan2002}.\\